\pdfoutput=1

\newif\ifprintfig
\printfigtrue

\documentclass{emulateapj}
\usepackage{graphicx}

\newcommand\degree{{^\circ}}
\newcommand\pc{{\rm\,pc}}

\newcommand\Mpc{{\rm\,Mpc}}

\newcommand\Myr{{\rm\,Myr}}
\newcommand\Myrs{{\rm\,Myrs}}
\newcommand\Gyr{{\rm\,Gyr}}

\newcommand\kmsec{{\rm\,km\,s^{-1}}}
\newcommand\kms{\kmsec}

\newcommand\Msun{{\rm\,M_\odot}}

\newcommand\clock{\count0=\time \divide\count0 by 60
     \count1=\count0 \multiply\count1 by -60 \advance\count1 by \time
     \number\count0:\ifnum\count1<10{0\number\count1}\else\number\count1\fi}

\shortauthors{Dalcanton et al.}
\shorttitle{NIR Stellar Populations}

\begin{document}

\title{Resolved Near-Infrared Stellar Populations in Nearby Galaxies}

\author{Julianne J.\ Dalcanton\altaffilmark{1}, 
  Benjamin F.\ Williams\altaffilmark{1}, 
  Jason L.\ Melbourne\altaffilmark{2},
  L\'eo Girardi\altaffilmark{3},
  Andy Dolphin\altaffilmark{4},
  Philip A.\ Rosenfield\altaffilmark{1},
  Martha L.\ Boyer\altaffilmark{5},
  Roelof S.\ de Jong\altaffilmark{6},
  Karoline Gilbert\altaffilmark{1},
  Paola Marigo\altaffilmark{7},
  Knut Olsen\altaffilmark{8},
  Anil C.\ Seth\altaffilmark{9},
  Evan Skillman\altaffilmark{10}
}

\altaffiltext{1}{Department of Astronomy, Box 351580, University of Washington, Seattle, WA 98195; jd@astro.washington.edu, ben@astro.washington.edu, philrose@astro.washington.edu, kgilbert@astro.washington.edu}
\altaffiltext{2}{Caltech Optical Observatories, Division of Physics, Mathematics and Astronomy, Mail Stop 301-17, California Institute of Technology, Pasadena, CA 91125, USA; jmel@caltech.edu}
\altaffiltext{3}{Obsservatorio Astronomico di Padova - INAF, Vicolo
dellœôòùOsservatorio 3, I-35122 Padova, Italy; leo.girardi@oapd.inaf.it}
\altaffiltext{4}{Raytheon, 1151 E. Hermans Road, Tucson, AZ 85706, USA; adolphin@raytheon.com}
\altaffiltext{5}{Space Telescope Science Institute, Baltimore, MD 21218, USA; mboyer@stsci.edu}
\altaffiltext{6}{Astrophysikalisches Institut Potsdam (AIP), An der Sternwarte 16, 14482 Potsdam, Germany; rdejong@aip.de}
\altaffiltext{7}{Dipartimento di Astronomia, Universit\'a di Padova, Vicolo
dellœôòùOsservatorio 2, I-35122 Padova, Italy; paola.marigo@unipd.it}
\altaffiltext{8}{National Optical Astronomy Observatory, 950 North Cherry Avenue, Tucson, AZ 85719, USA; kolsen@noao.edu}
\altaffiltext{9}{Center for Astrophysics Fellow, Harvard-Smithsonian Center for Astrophysics, 60 Garden Street, Cambridge, MA 02138, USA; aseth@cfa.harvard.edu}
\altaffiltext{10}{Department of Astronomy, University of Minnesota, 116 Church St. SE, Minneapolis, MN 55455, USA; skillman@astro.umn.edu}
  
\begin{abstract}
  We present near-infrared (NIR) color-magnitude diagrams (CMDs) for
  the resolved stellar populations within 26 fields of 23 nearby
  galaxies ($\lesssim4\Mpc$), based on images in the $F110W$ and
  $F160W$ filters taken with Wide Field Camera 3 (WFC3) on the Hubble
  Space Telescope (HST).  The CMDs are measured in regions spanning a
  wide range of star formation histories, including both old dormant
  and young star-forming populations.  We match key NIR CMD features
  with their counterparts in more familiar optical CMDs, and identify
  the red core Helium burning (RHeB) sequence as a significant
  contributor to the NIR flux in stellar populations younger than a
  few $100\Myrs$ old.  The strength of this feature suggests that the
  NIR mass-to-light ratio can vary significantly on short timescales
  in star forming systems.  The NIR luminosity of star forming
  galaxies is therefore not necessarily proportional to the stellar
  mass.  We note that these individual RHeB stars may also be
  misidentified as old stellar clusters in images of nearby galaxies.
  For older stellar populations, we discuss the CMD location of
  asymptotic giant branch (AGB) stars in the HST filter set, and
  explore the separation of AGB subpopulations using a combination of
  optical and NIR colors.  We empirically calibrate the magnitude of
  the NIR tip of the red giant branch (TRGB) in $F160W$ as a function of
  color, allowing future observations in this widely adopted filter
  set to be used for distance measurements. We also analyze the
  properties of the NIR RGB as a function of metallicity, showing a
  clear trend between NIR RGB color and metallicity.  However, based
  on the current study, it appears unlikely that the slope of the NIR
  RGB can be used as an effective metallicity indicator in
  extragalactic systems with comparable data.  Finally, we highlight
  issues with scattered light in the WFC3, which becomes significant
  for exposures taken close to a bright earth limb.
\end{abstract}
\keywords{galaxies: stellar content --- galaxies: abundances --- galaxies:
  distances and redshifts --- galaxies: dwarf --- galaxies: irregular ---
  stars: AGB and post-AGB --- stars: carbon --- infrared: stars}

\vfill
\clearpage

\section{Introduction}  \label{introsec}

Near-infrared (NIR) observations have become increasingly important
for studies of galaxies and their evolution over cosmic time
\citep[e.g.,][]{conselice05,dahlen05,saracco06,cirasuolo10}.  This
trend will no doubt continue during the coming decade, thanks to
continued improvements in NIR detectors, the growing maturity of
adaptive optics, the installation of WFC3 on HST, and the upcoming
launch of the James Webb Space Telescope (JWST).

The importance of NIR observations for galaxy evolution studies rests
on the reduced sensitivity of the NIR mass-to-light ratio to dust and to the
age or metallicity of the underlying stellar population, particularly
compared to the optical or ultraviolet.  The NIR luminosity of a
galaxy is therefore thought to be a robust indicator of stellar mass 
\citep[e.g.][]{thronson88,brinchmann00,bundy05}.
However, the ability to correctly interpret NIR observations relies on
accurate stellar population modeling, which in turn requires accurate
isochrones, stellar lifetimes, and spectra for the evolving stars
which dominate the flux at NIR wavelengths (see \cite{conroy2010}, for
an estimate of current uncertainties).  

Recently, there has been some concern that the mass-to-light ratio in
the NIR is not nearly as stable as has been assumed.  On the
theoretical side, \citet{maraston2006} have suggested that the NIR flux
from asymptotic giant branch (AGB) stars may lead to drastically lower
mass-to-light ratios when intermediate-age populations are present, as
must be the case for {\emph{in situ}} observations of young galaxies
at high redshift (although see \citet{kriek2010}).  Lower NIR
mass-to-light ratios could also help to explain why some elliptical
galaxy progenitors at high redshift appear to have stellar densities
higher than seen in the local universe \citep{vandokkum2008}.

The most accurate constraints on evolving stars' contribution to the
NIR come from resolving the stellar populations directly \citep[e.g.][
most recently]{rejkuba2006,gullieuszik07,gullieuszik08,melbourne10}.
With WFC3's IR channel on HST, we are now able to resolve large
numbers of individual stars in nearby galaxies.  There are well over a
hundred galaxies within the Local Volume ($D\lesssim4\Mpc$) that are
close enough to be targeted for resolved stellar population studies.
These galaxies have already been observed extensively with HST in the
optical \citep{dalcanton2009,gilbert2011}, and their color-magnitude
diagrams have revealed the galaxies to have a wide range of
metallicities and star formation histories.  These systems therefore
form an ideal set for probing the properties of the same stellar
populations in the NIR.

To this end, we undertook an HST ``snapshot'' survey of nearby
galaxies with existing high-quality optical data.  We took advantage
of the high throughput and resolving power of the WFC3 IR channel to
produce CMDs of individual fields in 23 galaxies, spanning a variety
of stellar populations.  In this paper, we present our sample, and
characterize the resulting color-magnitude diagrams.  These data will
be used in subsequent papers to place quantitative constraints on the
contribution of AGB and RHeB stars to the total luminosity
\citep{melbourne2011}, on the lifetimes of AGB stars, and on the
population of carbon stars.

The outline of the paper is as follows.  In
Section~\ref{observationsec} we present our sample selection,
observations, reductions, and matched optical photometry.  We also
discuss a previously uncharacterized transient scattered light feature
in the Wide Field Camera 3 infrared channel (hereafter WFC3/IR). In
Section~\ref{cmdsec} we discuss the origin of different features in
the NIR CMDs and optical-NIR color-color diagrams, and highlight the
importance of core red Helium-burning (RHeB) stars to the overall
luminosity, for populations younger than $\sim0.5\Gyr$.  We discuss
the properties of the tip of the red giant branch (TRGB) in the NIR
(Section~\ref{trgbsec}), and the AGB and red giant branch (RGB)
luminosity function in Section~\ref{LFsec}.  Finally, we discuss the
metallicity dependence of the NIR RGB morphology and color on
metallicity in Section~\ref{rgbstructsec}.

\section{Data}  \label{observationsec}

\subsection{Sample Selection}  \label{samplesec}

To characterize NIR CMDs as a function of stellar age and metallicity,
we require targets whose stellar populations have been constrained
using optical observations in tandem with well-calibrated stellar
isochrones.  We therefore chose SNAP targets from among those nearby
galaxies with archival high-quality multi-color imaging (typically
from ACS, but in some cases from WFPC2).  Of these, we selected a
subset of 61 fields in 51 galaxies from the ACS Nearby Galaxy Survey
Treasury (ANGST), a volume-limited sample of non-Local Group galaxies
out to $\sim\!4\Mpc$ \citep{dalcanton2009}, and from the archival
legacy program ANGRRR: Archival Nearby Galaxies: Reduce, Reuse,
Recycle \citep{gilbert2011}.  Based on their HST stellar photometry,
we have verified that these target fields have (1) sufficient numbers
of stars to contain hundreds of candidate AGB stars within a single
WFC3/IR FOV; (2) uncrowded stellar photometry of sufficient quality to
provide useful constraints on the star formation history and
metallicity distribution\footnote{Galaxies beyond $4\Mpc$ are
  sufficiently crowded that the deep optical photometry needed for
  deriving SFHs is not possible.  Galaxies closer than $1\Mpc$ are
  sufficiently extended that few AGB stars actually fall in a typical
  FOV and contamination from Galactic foreground sources is severe,
  making this study more effective in more distant galaxies.}; (3)
indications from the range of red giant branch colors that the stellar
populations host a non-negligible fraction of stars whose
metallicities fall outside of the range of well-studied LMC/SMC
metallicities; and (4) a sufficiently broad range of star formation
histories to allow sampling of evolving stars of different masses.
Our target list included multiple pointings in the more massive
systems, whenever the optical data suggested that there are
significant variations in stellar age and metallicity at different
locations within the galaxy.

During Cycle 17 we obtained observations for 26 of the 61 possible
SNAP targets.  Two galaxies (Holmberg~II and NGC~2403) had multiple
pointings (2 and 3 pointings, respectively), isolating regions with
different star formation histories and metallicities.  The properties
of the 23 observed galaxies are listed in Table~\ref{sampletable}.
Galaxy names, positions, apparent blue magnitudes ($B_{\rm T}$), diameters,
morphological $T$-types, and H{\sc i} line widths ($W_{50}$) have been
adopted from the primary name in the \citet{karachentsev04b} Catalog
of Neighboring Galaxies.  Distance moduli are based on the $F814W$ TRGB
from \citet{dalcanton2009} for most galaxies,
and from \citet{karachentsev03} for NGC~7793.  Group memberships are
from \citet{karachentsev05} or \citet{tully06}.  Foreground
extinctions ($A_{\rm V}$) are from \citet{schlegel98}, as reported by the
on-line Galactic Dust Extinction Service at the Infrared Science
Archive\footnote{\tt{http://irsa.ipac.caltech.edu/applications/DUST/}}.

\subsection{Observations}  \label{obssec}

Observations for this program were carried out during Cycle 17 as
SNAP-11719. Some fraction of these were taken during the period of
instrument commissioning after the Hubble repair mission.  All targets
were observed in both $F110W$ and $F160W$ with HST's WFC3 IR channel.
These filters offer the greatest depth in a given exposure time, and
thus are likely to become the workhorse filters for the WFC3/IR
camera.  Observations were carried out in a 3-point
``WFC3-IR-DITHER-LINE'' pattern, with exposures in both $F110W$ and
$F160W$ taken at each pointing.  We adopted the {\tt STEP50} exposure
sequence, which accommodates data with a large dynamic range by using
both short and long non-destructive reads.  We used ${\tt NSAMP}=9$ in
$F110W$ and ${\tt NSAMP}=11$ in $F160W$, giving total exposure times
of 597.7\,s and 897.7\,s respectively.  The exposure sequence was
interleaved to allow buffers to dump during the exposures, with no
latency.  To maximize the schedulability of our observations, we did
not specify a roll-angle orientation constraint.

\begin{figure}
\centerline{
\includegraphics[width=2.25in]{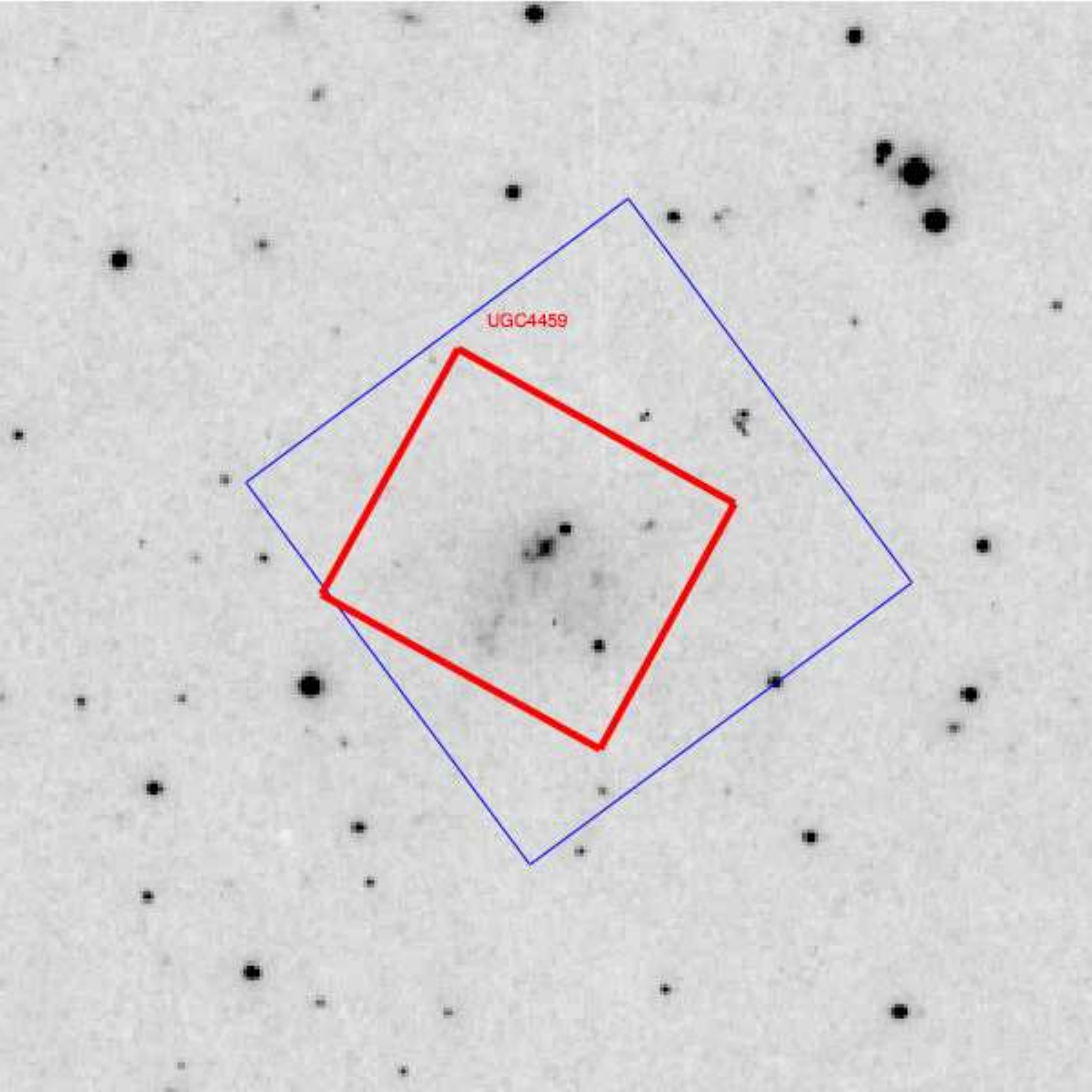}  
\includegraphics[width=2.25in]{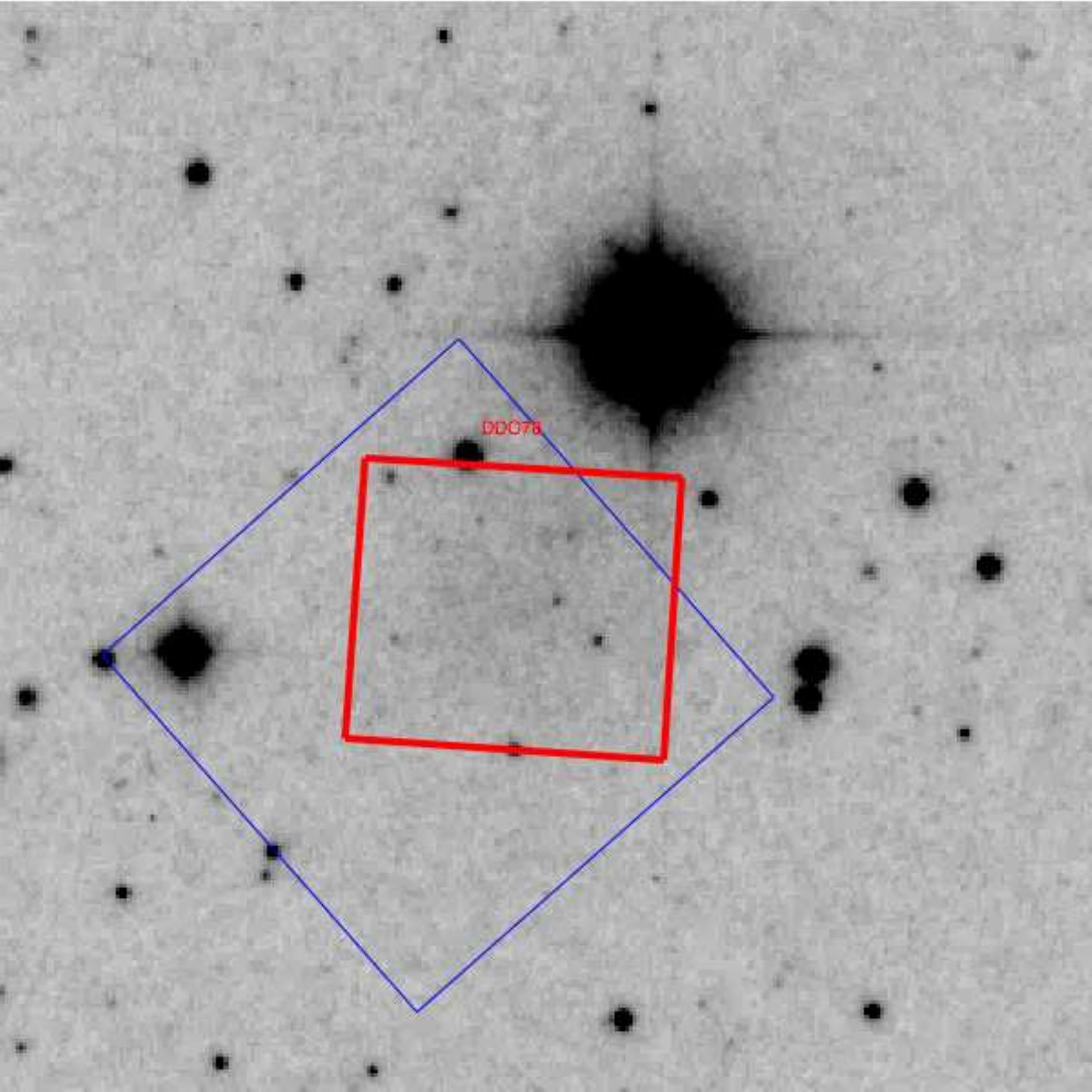}  
\includegraphics[width=2.25in]{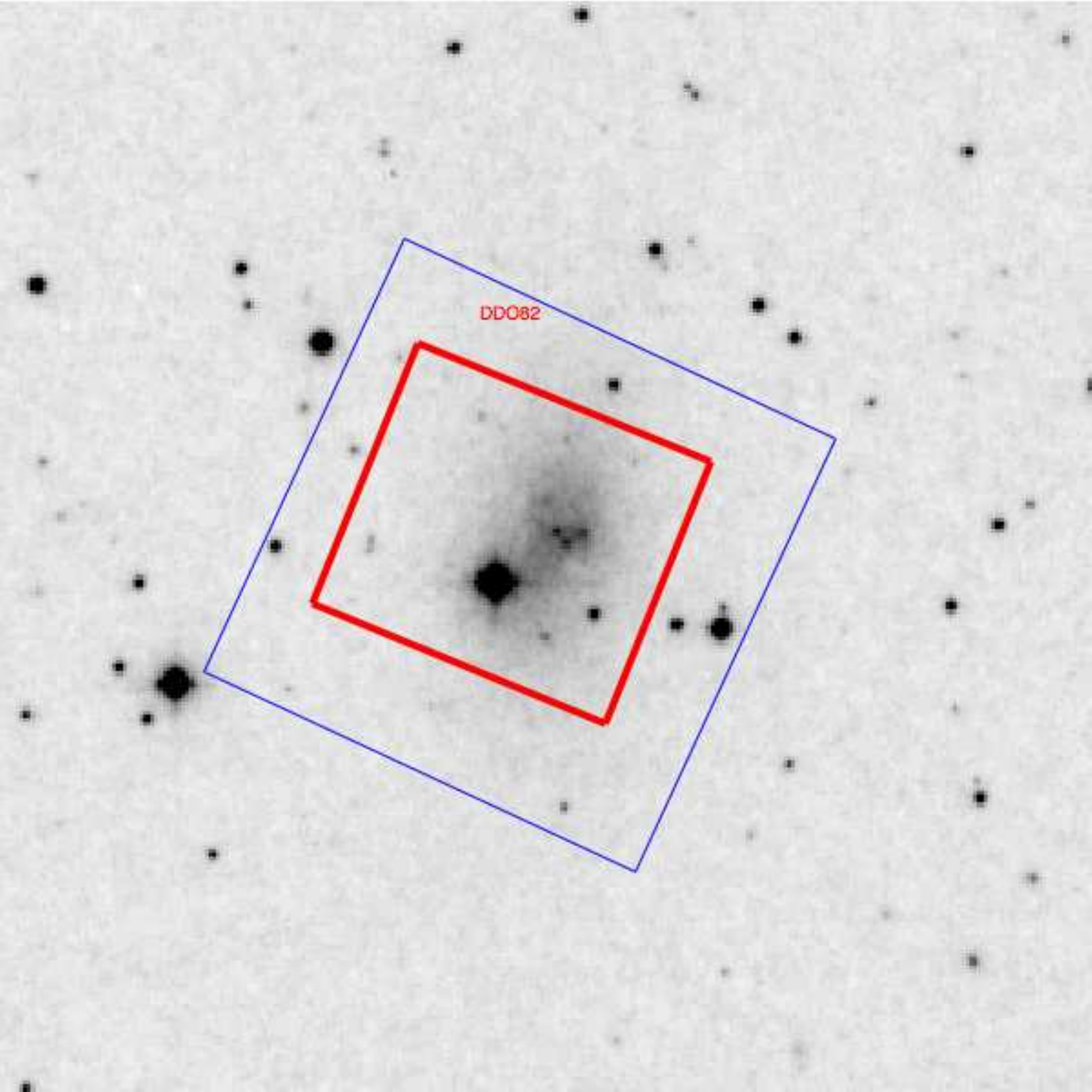}  
}
\centerline{
\includegraphics[width=2.25in]{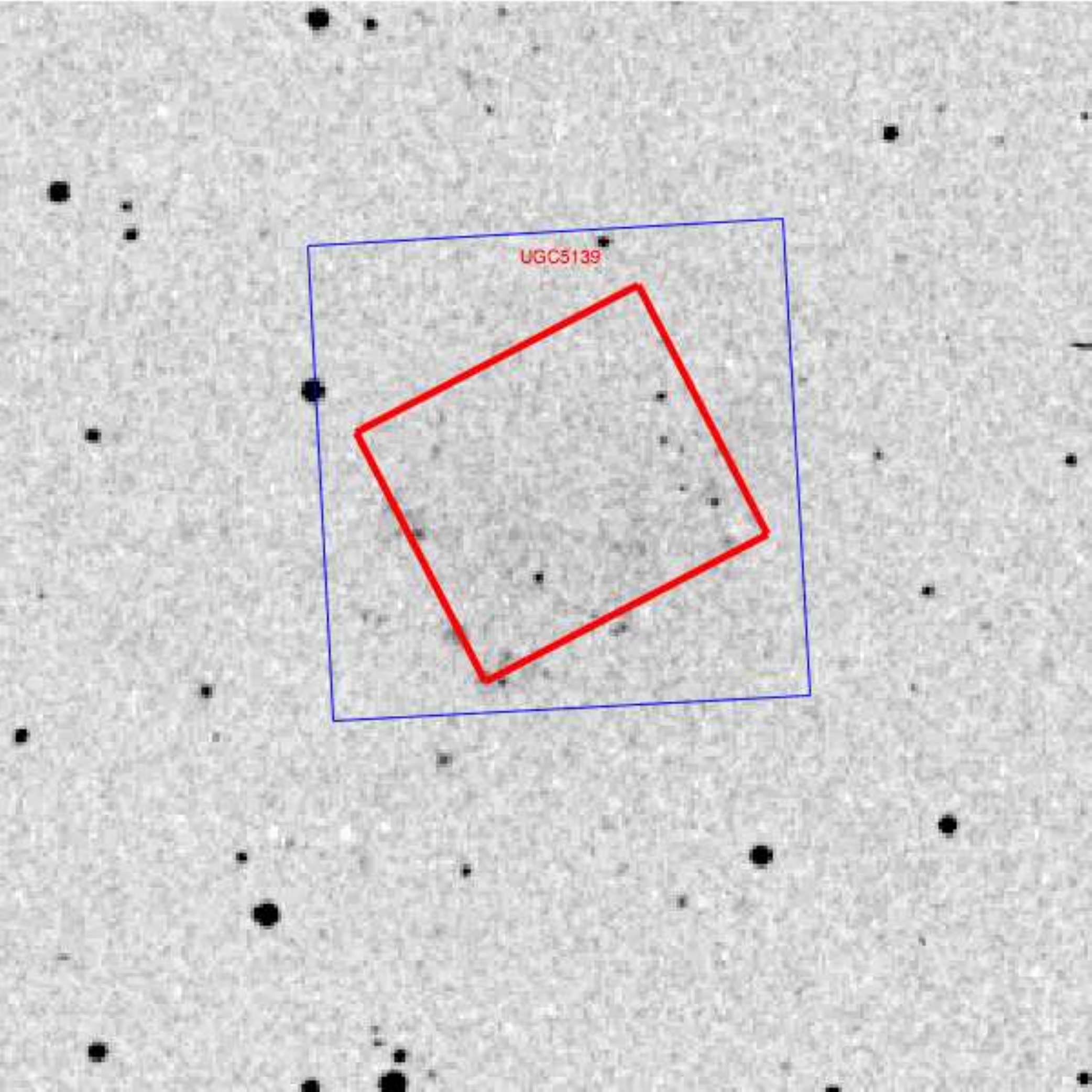}  
\includegraphics[width=2.25in]{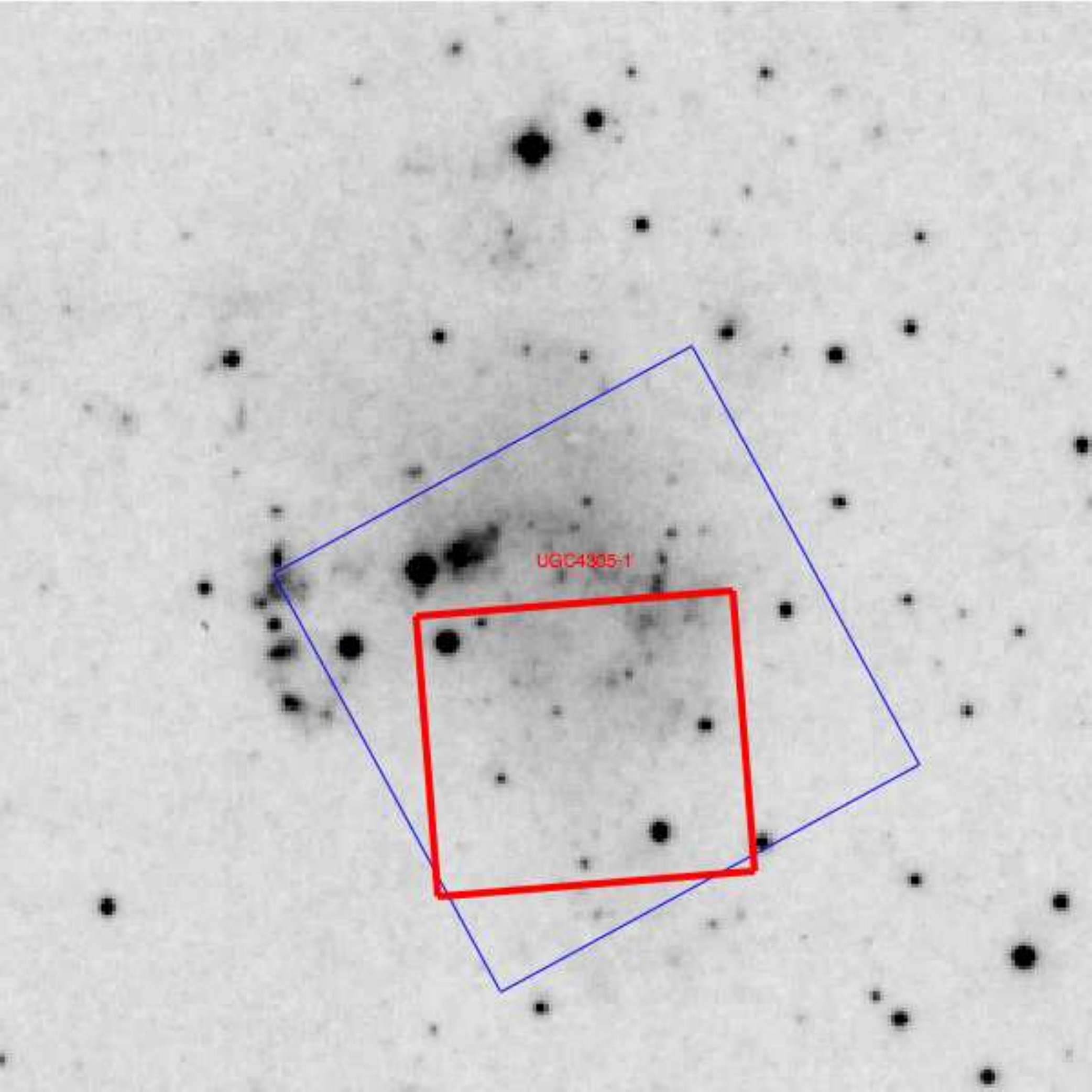}  
\includegraphics[width=2.25in]{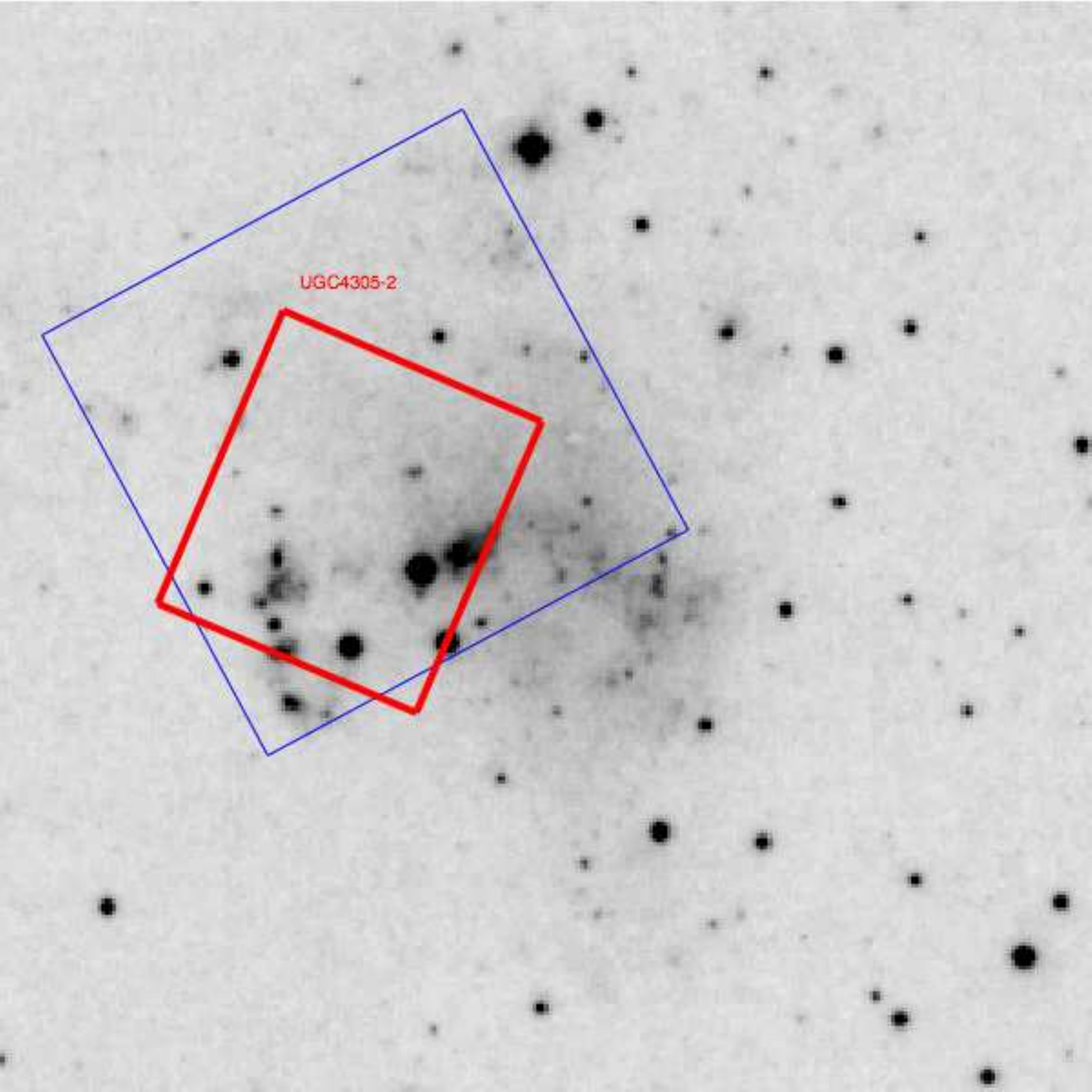}  
}
\centerline{
\includegraphics[width=2.25in]{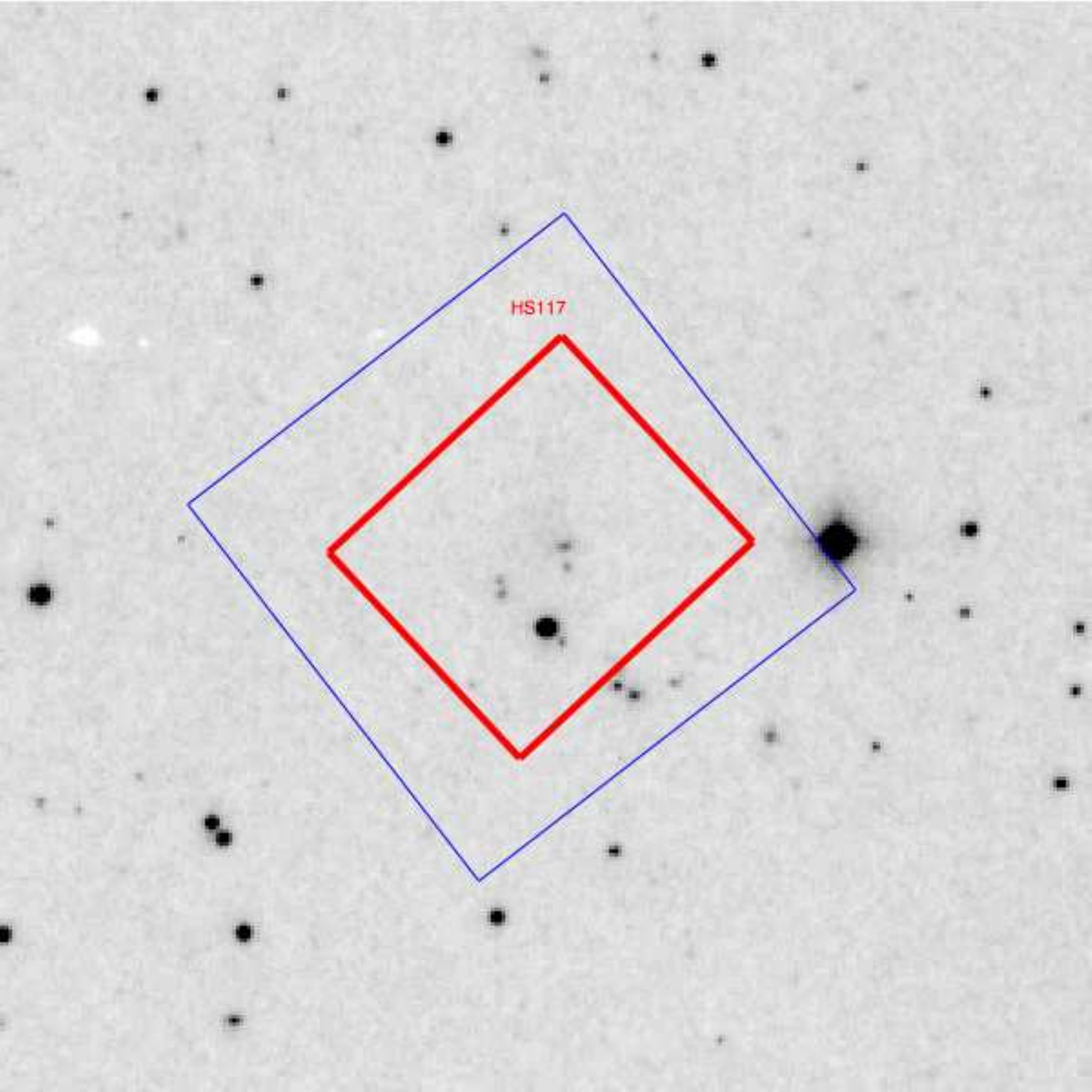}  
\includegraphics[width=2.25in]{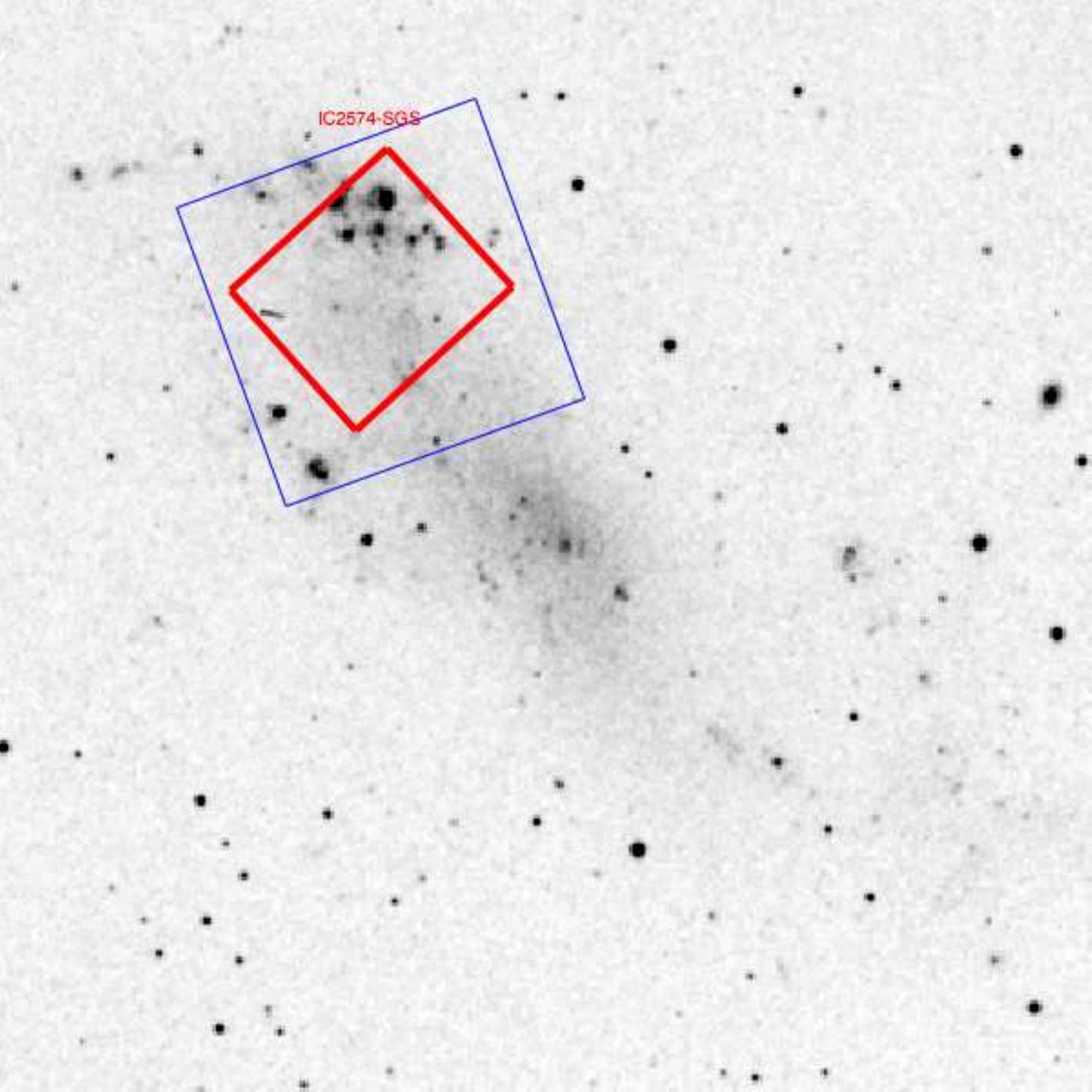}  
\includegraphics[width=2.25in]{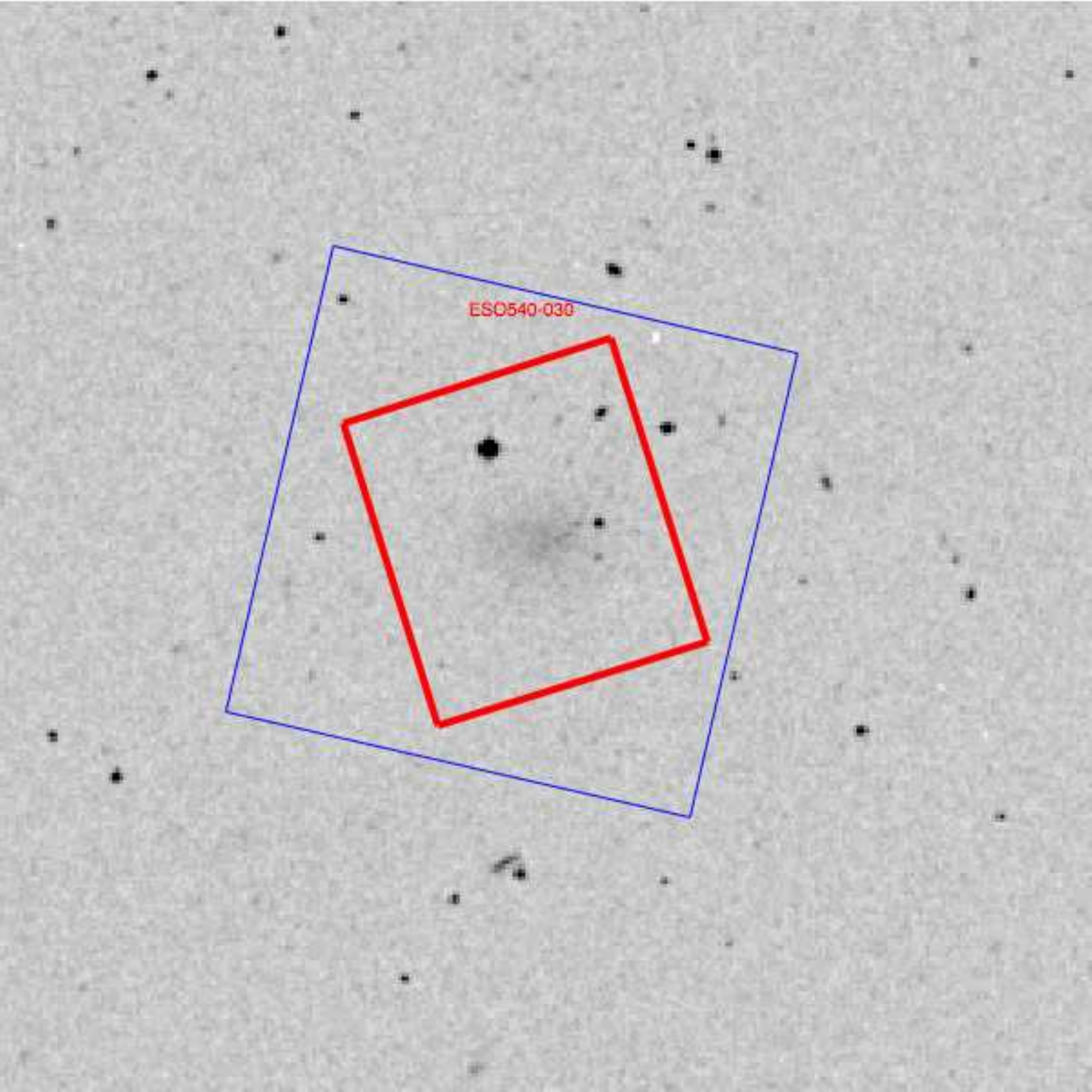}  
}
\caption{ Position of the WFC3/IR field of view (red), overlayed on an
  optical image from the Digitized Sky Survey.  Blue region shows the
  area covered by optical HST data.  (Target names from upper left to
  lower right: [a] UGC4459; [b] DDO78; [c] DDO82; [d] UGC5139; [e]
  UGC4305-1; [f] UGC4305-2; [g] HS117; [h] IC2574-SGS; [i]
  ESO540-030;).
\label{overlayfig}}
\end{figure}

Properties of the actual observations can be found in
Table~\ref{obstable}, where we include both the name of the galaxy
from Table~\ref{sampletable} and the name of the specific target
within the galaxy.  Target names were chosen to match the names of
existing optical data sets at the same position; these associated data
sets are also listed in Table~\ref{obstable}.  The centers of the WFC3
field-of-view (FOVs) were chosen to maximize the overlap with the
optical data.  The locations of the WFC3/IR footprints are shown in
red in Figure~\ref{overlayfig}, superimposed on Digitized Sky Survey
images of each galaxy.  Blue regions show the locations of the
associated optical imaging from ACS or WFPC2.  Because we did
not specify an orientation, occasionally a small corner of the WFC3/IR
FOV fell off the area covered by the optical data.  This mismatch
occurred in cases where the optical FOV was not centered on the most
desirable part of the galaxy, and the scheduled roll angle happened to
be unfavorable; however, the fraction of the WFC3/IR FOV lacking
optical coverage is less than 10\% in almost all cases.

In Figure~\ref{gridfig} we show false color images of the 
observations.  There are a number of features to note.  First, in many
cases it is clear that our short exposures have already reached the
``crowding limit'', where stars are sufficiently close on the sky that
fainter stars could not be detected, even with longer exposures; this effect 
can also be seen as spatially-varying depth in the CMDs.
Second, background galaxies are far more prevalent than in optical HST
images, suggesting that unresolved background galaxies are likely to
be a more significant contaminant of the NIR CMDs.  Third, the images
are marked with a number of circles, located at positions of ``IR
blobs'', that are flagged as bad data by the WFC3/IR pipeline (see the
WFC3-2010-06 ISR by N. Pirzkal).  Fourth, extended line emission is
visible in $F110W$ images of young star forming regions, as can be
seen in the close-up of a star forming region in IC2574-SGS, shown in
Figure~\ref{iroptrgbfig}. The line emission in this filter is
dominated by the [S{\sc iii}]9069\AA,9532\AA\ doublet, with
additional contributions from HeI at 10830\AA\ and 10833\AA\  (from 2
photon emission) and Paschen $\beta$ at 12818\AA.

Finally, and perhaps most concerning, a number of the images show
significant scattered light, particularly in the lower left region of
the image (see UGC4305-1 or NGC2403-HALO-6, for example).  We now
discuss the origin of this scattered light component.

\subsubsection{Scattered Light} \label{scatteredlightsec}

Figure~\ref{scatteredlightfig} shows the series of 6 combined {\tt
  *FLT} images taken during our 1 orbit visit, after all detected
sources have been fit and subtracted from the images.  The series
shows images in both $F110W$ (exposures 1, 3, \& 5 in the sequence),
and $F160W$ (exposures 2, 4, \& 6). The images show two types of
structures in the ``sky'' background.  The first is persistent
structure due to unresolved stars in the image; this component does
not change throughout the visit, but varies from galaxy-to-galaxy.
The second structure (falling typically on the left-hand side of the
image, and peaking in the bottom left corner), varies rapidly in
amplitude during a single orbit but has a consistent shape when
present.  The pattern of the time variation is not consistent,
however, such that the brightness of the scattered light feature peaks
at different times during an orbit.  This variability leads to an
apparent variation of the color of the scattered light; if the
scattered light peaks during a $F110W$ exposure, the scattered light
can appear blue, but if it peaks during an $F160W$ exposure, the light
can appear red.

The rapid variation in the amplitude of the scattered light places
strong constraints on its origin, as there are few telescope pointing
attributes that vary significantly during an orbit.  The angles
between the principal {\tt V1} axis and the Sun or the moon (FITS
header keywords {\tt SUNANGLE} and {\tt MOONANGL}) vary little during
an orbit, nor does the amount of zodiacal light.  The only quantities
which do vary on that timescale are the angle between the Sun and the
Earth's limb (keyword {\tt SUN\_ALT}) and, more directly, the angle
between the spacecraft pointing and the Earth's limb (keyword {\tt
  LOS\_LIMB} in the {\tt *.JIT} or {\tt *.JIF} files).  We thank Ben
Weiner (private communication) for pointing out this possibility,
based on his analysis of similar effects seen in WFC3/IR grism
data\footnote{See also ISR-ACS 2003-05 by Biretta et al for an
  analysis of ACS background light as a function of limb angle.}.

To examine the variation of the scattered light as a function of the
angle to the Earth's limb, we analyzed the sky levels in each
star-subtracted {\tt *.FLT} image.  We assume that there are three
contributors to the star-subtracted ``sky'' image: (1) a mean uniform
background; (2) a spatially-structured, time-invariant background due
to unresolved stars; and (3) a spatially-structured, time-variable
background due to scattered light, which peaks in the lower left
quadrant.  To assess these different components, we recorded the mean
level in the upper right quadrant of each image, which is assumed to
be free of the scattered light feature, and in the lower left region
dominated by the feature ([0:308,0:507], in image coordinates).  Note
that we do not expect these two sky levels to be identical, even in
the absence of the scattered light feature, because of the
contribution of light from unresolved stars.  However, they both
should share a common ``floor'', representative of the mean sky
background during the observation.

In the left hand panel of Figure~\ref{limbangfig}, we show the mean
sky level in the upper right quadrant of each exposure, plotted as a
function of the limb angle between the spacecraft and the earth, for
the $F110W$ and $F160W$ filters, respectively.  If the earth's limb
was dark at the time of observation, the sky level is plotted with an
open circle.  Observations taken within a single orbit are connected
with a line.  

From the left hand side of Figure~\ref{limbangfig}, it is clear that
the sky brightness depends strongly on the limb angle for the $F110W$
filter, at least when the earth's limb is bright.  Below angles of
$40\degree$, the mean sky level increases by roughly a factor of two
or more.  We expect that this correlation would be significantly
tighter if this test were repeated for empty fields, where there would
be little contribution from unresolved sources to the measurement of
the sky.

In the right hand panel of Figure~\ref{limbangfig}, we attempt to
capture the amplitude of the scattered light feature visible in the
first two panels of Figure~\ref{scatteredlightfig}.  Unfortunately, we
cannot simply track the variation of the sky level in that region,
since we expect some limb angle dependent variation in the overall sky
level, as seen in the left hand plot of Figure~\ref{limbangfig}.
Instead, we look at the limb angle dependence of the excess light in
the lower left, compared to the upper right (${\rm Gradient} \equiv
{\rm Sky}_{lower\,left} - {\rm Sky}_{upper\,right}$).  We assume that
when the scattered light feature is absent, the difference in the sky
level between the upper right and lower left regions reflects the
structure in the sky due to unresolved sources.  This difference
should be constant with time, no matter what the mean sky background
level is.  For each set of observations (3 per filter in a single
orbit), we then look at the ``excess'' in the difference between the
lower left and the upper right, compared to the minimum during the
orbit.  We then scale the resulting excess by the minimum sky level in
the four quadrants of the image, to give an indication of the
fractional strength of the feature, compared to the typical sky level
when unresolved sources are not present.

The right hand side of Figure~\ref{limbangfig} shows that for most of
the exposures, the relative sky brightness of the lower left and upper
right varies by less than 1\% between exposures, as long at the angle
to the bright earth limb is greater than $30\degree$.  If the angle is
less than $30\degree$, however, a strong scattered light feature is
produced in the majority of cases.  The onset is quite sharp, and is
present in both filters, unlike the increase in the uniform sky
brightness with limb angle, which is gradual and much stronger in
$F110W$.  Based on this analysis, the galaxies with the most prominent
scattered light features ($>$4\% in either filter) are UGC4305-1,
IGC2574-SGS, DDO71, SN-NGC2403-PR, NGC2403-DEEP, NGC2403-HALO-6, and
KDG73, in order of strongest to weakest.  This ranking agrees with the
visual impression seen in the color images in Figure~\ref{gridfig}.

For our subsequent analysis, we make no additional attempt to
compensate for the structure of the scattered light feature in these
images.  Since our photometry uses a very localized estimate of the
sky, our results should be insensitive to this feature, except through
having to accept more noise in the affected region.  Given the
brightness of the stellar sources, including observations with low
limb angle observations produces more benefits through increased
integration time than are lost through increased noise levels.  For
many other possible WFC3/IR programs, however, this may not be the
case, and observers should consider specifying a minimum earth limb
angle for their observations.

There are two additional complications that may result from the
presence of rapidly varying scattered light, however.  The first
complication is that some reads affected by scattered light may have
pixels that are erroneously flagged as cosmic rays, if flux in the
scattered light feature appears as an outlier compared to the large
number of unaffected reads; however, we do not see in any obvious
increase in the number of masked cosmic rays in the limb brightened
regions in Figure~\ref{scatteredlightfig}.  The second complication
results from the fact that the final IR image is produced by fitting a
linear function to a series of non-destructive reads taken during the
course of the exposure.  However, while the flux in a pixel from a
star will rise linearly during the exposure, the flux from the
background will {\emph{not}} rise linearly when the angle to the limb
or the limb's brightness changes during an observation.  The
assumption of a linear fit may therefore be invalid when the flux in
the background is comparable to the flux of the source in a given
pixel.  One should further note that this effect could be present even
when the strong scattered light feature is not obvious, since, as one
sees in Figure~\ref{limbangfig}, the uniform sky background can vary
strongly during the observation, particularly in $F110W$.
Unfortunately, it is not straightforward for us to evaluate this
effect in our data.  First, it is difficult to separate the true
time-variable sky background, from the spatially-variable
time-constant background from unresolved stars.  Second, we cannot use
spatial variations in the color magnitude diagram to identify
systematic errors in the photometry associated with the scattered
light feature, since galaxies typically exhibit significant gradients
in the their stellar populations.


\subsection{Data Reduction \& Photometry}  \label{photometrysec}

Photometry was performed using the DOLPHOT package\footnote{{\tt
    http://purcell.as.arizona.edu/dolphot/}} \citep{dolphin2000},
including a new WFC3-specific module generated in support of this HST
program.  The core functionality of DOLPHOT remains as in previous
versions (i.e., simultaneous PSF fitting across a stack of images
aligned to sub-pixel accuracy), but the WFC3 update includes several
new capabilities: a preprocessing routine that uses the data quality
extensions to mask the image and then applies the appropriate WFC3
pixel area maps, a WFC3 point spread function (PSF) library, WFC3
distortion corrections, and new zero points for WFC3.  Pixel area
maps, distortion terms, zero points, and encircled energy corrections
were obtained from the WFC3 documentation\footnote{WFC3 PSFs from
  release of 15-Nov-2009; WFC3 zero points and corrections to infinite
  aperture from J. Kalirai 2009 (ISR 2009-30); WFC3 IDC tables use for
  distortion correction from t20100519\_ir\_idc.fits}.  The PSF
library was computed using the preliminary version of Tiny Tim 7.0
with WFC3 support\footnote{http://www.stecf.org/instruments/TinyTim/};
note that the current version of the WFC3/IR Tiny Tim PSF library is
based on pre-flight data.  For WFC3/IR, PSFs are computed with
10$\times$10 subsampling and a 3$\arcsec$ radius, for 64 regions on
the chip.  DOLPHOT also adopts small aperture corrections calculated
from uncrowded stars in the image; these corrections were always
$\le$0.01 mag.

All PSF photometry was performed by minimizing residuals on all
individual {\tt{flt}} exposures, and then combined into final
magnitudes for each star in each filter.  Non-detections are
considered to be stars with signal-to-noise ratios of less than 4, for
a given filter.  Quality parameters are computed and included in the
combined photometry as well.  Two catalogs were then produced for each
field using the DOLPHOT signal-to-noise, sharpness, and crowding
parameters.  The first ``{\tt{*.st}}'' catalog contains all sources
detected in at least one band with S/N$>$4, where the sharpness of the
source does not exclude the possibility that it is stellar
({\tt{sharpness}}$^2\,<$0.1).  The second ``{\tt{*.gst}}'' catalog
only contains the highest quality photometry (including detections
with S/N$>$4 in both filters, ({\tt{sharpness}}$_{F110W}$ +
{\tt{sharpness}}$_{F160W}$)$^2\,<$0.12, and {\tt{crowding}}$_{F110W}$
+ {\tt{crowding}}$_{F160W}\!<$0.48).  The {\tt{*.gst}} catalog
typically contains $\sim$50$\pm$10\% of the sources from the
{\tt{*.st}} catalog. These final parameter cuts were chosen to produce
the cleanest CMD features (few stars outside of the known features)
without culling large numbers of sources from the features themselves;
the culled sources are almost entirely valid stellar detections, but
have more uncertain photometry.  The number of stars detected varies
between 6,000 and 40,000 per pointing, with a median of $\sim$10,000
(Table~\ref{obstable}).  We also include the maximum and minimum
stellar surface density of the {\tt{*.st}} detections, calculated
within 100 subregions defined by a 10$\times$10 grid in the
image\footnote{Empirically, the data in Table~\ref{obstable} indicates
  that one cannot expect to detect more than $\sim$5.5 stars per
  square arcsecond in an WFC3/IR frame.}; observations with a higher
stellar surface density will be more affected by crowding, and
observations with larger differences between the maximum and minimum
surface density will have stronger spatial variations in the depth of
the CMD and the accuracy of the photometry.

Finally, we assessed the biases in our photometry using artificial
star tests.  False stars were placed into the image stack 100,000
times, such that the artificial stars sampled the full range of
colors, magnitudes, and locations of our real photometry.  We
recovered the artificial stars using identical photometry routines and
post-processing cuts as those applied to the data.

Table~\ref{obstable} includes the resulting ``50\% completeness''
magnitude for each filter, as determined by the limit where 50\% of
artificial stars inserted into the image are successfully recovered by
our photometry (described below) in both filters (as in the
{\tt{*.gst}} catalog).  The 50\% completeness limits of our least
crowded fields are $\sim$26.2 mag in $F110W$ and $\sim$25.1 mag in
$F160W$; these limits were significantly fainter than originally
expected for photon-limited sources, based on the pre-flight WFC3/IR
Exposure Time Calculator.  Note, however, that this depth is
frequently position dependent, as can be seen from the
position-dependent CMDs in the right hand panels of
Figure~\ref{gridfig}. In crowded regions, the proximity of adjacent
stars compromises the recovery of faint stars, producing brighter
limiting magnitudes than one would expect based on photon counting
statistics alone.  The field for SN-NGC2403-PR has been particularly
affected by crowding, and has a 50\% completeness limit that is more
than 2 magnitudes brighter than the median of the sample.

Figure~\ref{errfig} shows the magnitude uncertainties reported by
DOLPHOT for the $F110W$ and $F160W$ filters, for the full stellar
``{\tt *.st}'' catalog.  The small number of points that scatter to
larger uncertainties typically come either from regions that were not
covered in all three dithered exposures, or that have been flagged as
being poorly photometered.  The number of stars with higher than
average uncertainties are reduced in the culled ``{\tt *.gst}''
catalog, which produces somewhat sharper features in the CMD, at the
expense of a modest drop in completeness.  The effect of this culling
can be seen in Figure~\ref{st_vs_gst_fig}, where we plot the CMDs for
the {\tt *.st} and the {\tt *.gst} photometry catalogs for one of our
fields.  We will describe the many features in these CMDs in
Section~\ref{cmdsec} below.

In Figure~\ref{fakestarfig} we show the distribution of $F160W$
magnitude errors for artificial stars inserted into our images, in
bins of 1 magnitude, for a crowded and an uncrowded field (left and
right plots, respectively).  At the extremes, the distributions show a
tail towards brighter recovered output magnitudes, due to the blending
of faint undetected stars with brighter stars.  In the median,
however, there is very little bias in the recovered magnitude
($<\!0.02\,{\rm mag}$ in the median for the faintest bin, which is
nearly an order of magnitude smaller than the photometric uncertainty
at the same magnitude).  Recovered colors are even less biased, as
crowding typically offsets the magnitudes of both filters in similar
directions.  The amplitudes of these effects are of comparable
amplitude in $F110W$, but are not quite identical due to the different
level of crowding and sky brightness in the two filters.

\subsection{Matching NIR and Optical Catalogs}  \label{matchedsec}

We match stellar catalogs from the distortion corrected optical and
WFC3 images as follows.  We calculate the astrometric transformation
by first requiring $\sim$150 stars that are bright in both the optical
and NIR data sets, and that spatially span the entire overlap region
between the WFC3 and ACS (or WFPC2) images.  We produce a list of
candidate alignment stars by first culling the optical and NIR star
catalogues from DOLPHOT to only stars that are in the spatial overlap
region.  We then select all reasonably bright, red stars in each
dataset (optical color $>\!0.7$ mags and F814W $<\!26$ mags; IR color
$F110W - F160W\!>\!  0.5$ mags and F160W $<\!24$ mags), producing
lists of $\sim$1000 stars in the optical and in the NIR.  We sort the
resulting two lists by luminosity to preferentially select luminous
stars.  Rather than selecting the 150 brightest stars, which are
frequently spatially clustered and do not span the full chip, we
insist that as long as we have more than one star to choose from, we
will select stars with an average separation of $\sim5\arcsec$,
ensuring broader areal coverage of the alignment stars.

We derive the transformation between the optical and IR star lists by
first visually identifying a roughly linear shift between the two
coordinate systems.  We then iteratively calculate a final
transformation using the ``method of triangles'' described in
\citet{valdes1995}, as implemented in the routine
\emph{MATCH}\footnote{\tt{http://spiff.rit.edu/match/match-0.8/match.html}}
by Michael Richmond.  First we run a linear fit between the two
coordinate systems and then we use that solution as a starting point
for a quadratic solution.  We find that a cubic solution is generally
unnecessary for the transformation between the distortion corrected
WFC3 and ACS images, but we do use it for the transformation of the
two data sets of WFPC2 observations.  We apply the transformation to
the entire NIR dataset bringing it into the optical coordinate system
used for the ANGST and ANGRRR data releases.  Note that we make no
attempt to tie the images to the global astrometric frame, due to the
lack of appropriate astrometric standards in the small WFC3 fields of
view.

Once the WFC3/IR catalog has been astrometrically aligned to the
optical catalog, we match individual stars in the two catalogs.  We
consider stars to be a match if the angular separation between the
stars is less than $0.07\arcsec$ (i.e., $\sim$0.5 WFC3/IR pixel).
Typically 90\% of the stars in the NIR catalogue are well matched to a
star in the optical catalogue. Unmatched stars typically fall in the
ACS chip gap or in the diffraction spikes of optically saturated
bright stars.

\subsection{Characterizing the Age of the Stellar Population}  \label{SFHsec}

The morphology of the CMDs that result from our NIR and optical
photometry reflect the age and metallicity of the underlying stellar
population.  To provide an intial constraint of these parameters, we
have analyzed the star formation history (SFH) of these galaxies using
optical CMD fitting, similar to the procedures described in
\citet{williams2011} and \citet{weisz2011}, but restricted to the
optical data that overlap the WFC3/IR FOV.

Specifically, we measured the star formation rate and metallicity as a
function of stellar age, by fitting the optical CMDs using the
software package MATCH \citep{dolphin2002}.  We adopted magnitude cuts
set to the 50\% completeness limits, and then fit the CMDs using
linear combinations of the stellar evolution models of \citet[][with
updates in \citealp{marigo2008} and \citealp{girardi2008}]{girardi02},
populated with stars following an IMF with a slope of -2.3 and a
binary fraction of 0.35 (with random mass sampling), convolved with
the photometric error and completeness statistics derived from
artificial star tests.  We first fit the data assuming a single
foreground reddening and distance, adopting values used in the ANGST
survey \citep{dalcanton2009} based on the \citet{schlegel98} Galactic
dust maps and the magnitude of the TRGB, respectively.  The best fit
provides the relative contribution of stars of each age and
metallicity in each field, which is then converted into cumulative
stellar mass produced as a function of time.  For consistency with
\citet{weisz2011}, we restricted the metallicity evolution to be constant or
increasing with time, for galaxies in common with the \citet{weisz2011}
dwarf sample.  This choice provides more accurate recent star
formation histories, at the expense of introducing occasional biases
against SF in the oldest bin for the few galaxies with very deep data

To assess the uncertainty of the best fit, we ran extensive Monte
Carlo tests. These tests assess two types of uncertainties: (1) random
errors due to Poisson sampling of the CMD and errors in photometry;
and (2) systematic errors due to deficiencies in the stellar evolution
models, or due to offsets in distance, reddening,
and/or magnitude zero-points.  The Poisson errors are accounted for by
generating artificial CMDs from the best-fit convolved model 100
times, and refitting the resulting CMD.  While fitting each of these
realizations, the systematic errors are assessed by introducing small
random shifts in log($T_{\rm{eff}}$) and $M_{\rm{bol}}$.
The random values were drawn from a Gaussian distribution
with a width of 0.03 in log($T_{\rm{eff}}$) and 0.41 in $M_{\rm{bol}}$.
The size of these shifts are set by differences between models in the
literature, and therefore serve as a proxy of the effects of our
particular choice of stellar evolution models.  Our final
uncertainties are the 68\% confidence intervals of the results of all
of our Monte Carlo test fits.  These total uncertainties are used as
the error bars in all subsequent plots and analysis.  Note however
that adjacent time bins in the SFH are covariant to some degree, such
that when one time bin fluctuates high, the adjacent bins fluctuate
low.  As a result, the plots of the cumulative SFH offer a truer
representation of the uncertainties.

The resulting SFH and cumulative age distributions are shown in the
bottom panels of Figure~\ref{cmdfig}.  The lower left panel shows the
SFR as a function of time, and the lower right hand panel shows the
cumulative star formation history, measured from the present back to
$14\Gyr$ ago.  A detailed discussion of the SFH analysis of the NIR
CMDs will be discussed in an upcoming paper.

\section{Color Magnitude Diagrams}  \label{cmdsec}

In this section we describe the properties of stellar populations in
the $F110W\!+\!F160W$ filter set using both models
(Section~\ref{modelcmdsec}) and observations
(Section~\ref{cmdoverviewsec}).  We discuss the qualitative dependence
of these properties on the age of the stellar population, as
derived from optical CMDs in Section~\ref{SFHsec}.

\subsection{Overview of Key Features in Model CMDs} \label{modelcmdsec}

To elucidate interpretation of the CMDs, in Figure~\ref{fakecmdfig} we
show simulated NIR color magnitude diagrams, color-coded either by age
or stellar mass (for a constant star formation rate at a fixed
metallicity of [Fe/H]=-1.45), or by metallicity (for a burst of star
formation between $8.9$ and $11.2\Gyr$). These simulations include the
updated models for AGB mass-loss at low metallicity described in
\citet{girardi2010}, and use the artifical stars from the M81-DEEP
field to calculate typical photometric uncertainties and biases.
Comparable figures for optical CMDs can be found in
\citet{dalcanton2009}.

In both the optical and the NIR, the most prominent feature is the red
giant branch (RGB), found at colors of $F110W\!-\!F160W\!\sim\!0.8$ in
the NIR and $F606W\!-\!F814W\!\sim\!1$ or $F475W\!-\!F814W\!\sim\!2$
in the optical.  This color depends on metallicity
(Section~\ref{rgbstructsec}), and becomes bluer when the metallicity
is low \citep{aaronson1978}.  At low metallicities, the optical RGB
also becomes more vertical and exhibits less curvature. In the NIR,
however, the slope of the RGB is nearly vertical, with only modest
variations with metallicity \citep{ferraro2000}.  The ages of stars in
the RGB can span a wide range ($\gtrsim\!1\Gyr$).  For the galaxies in
our sample, early star formation is particularly vigorous
\citep{weisz2011,williams2011}, which will weight the population of
RGB stars towards older ages, compared to the constant SFR shown in
Figure~\ref{fakecmdfig}.

One of the most prominent features along the RGB is the ``red clump'',
typically found at 3-4 magnitudes fainter than the TRGB in the
optical. The stars in the red clump are burning Helium in their cores
in the same way as horizontal branch stars, but appear red due to
either their young ages (large envelope mass) or their high
metallicities (see \citet[][]{girardi2001,salaris2002,castellani2000}
for theoretical models and
\citet[][]{ivanov2002,grocholsky2002,valenti2004} for observational
constraints from globular clusters).  Unfortunately, our NIR data does
not reach this information-rich feature, due to the shortness of our
NIR exposures and the bright crowding limit resulting from the
larger WFC3/IR pixel scale.  The absence of this feature reduces the
utility of using our NIR CMDs to constrain the relative amounts of
ancient and intermediate age SF.

Ancient star formation also produces a population of low mass
thermally pulsing AGB stars (TP-AGB; $<1.5\Msun$), found just above
the TRGB\footnote{AGB stars are also present in the same region
  occupied by the RGB, but are greatly outnumbered.  They cannot be
  distinguished cleanly as a separate population fainter than the tip
  of the red giant branch at $M_{\rm{F814W}}\!\sim\!-4$.} at
$M_{\rm{F814W}}\!\sim\!-4$.  At younger ages, TP-AGB stars are more
massive and have much brighter magnitudes \citep{marigo2008}.  As we
discuss below in Section~ \ref{rgbagbsec}, the colors of AGB stars
vary little in the NIR, and thus produce nearly vertical sequences for
all but the smaller population of red extreme AGB stars
\citep[e.g.][]{nikolaev2000,gullieuszik08,melbourne10,davidge10}.

Figure~\ref{fakecmdfig} shows that younger stars are expected to
dominate the stellar populations blueward and brightward of the RGB.
The youngest stars are main sequence stars, found at the bluest edge
of the CMD.  While the main sequence is quite well defined in the
optical, it is not nearly as distinct in the NIR.  This difference
must result in part from the fact that only the most luminous high
mass main sequence stars are detectable in these NIR observations.  In
the optical CMDs, we typically detect main sequence stars with masses
of $4.5\Msun$ and above, whereas in the NIR, we can typically only
detect main sequence stars with $\gtrsim20\Msun$ (assuming a magnitude
limit of 25 mag in $F160W$ versus 28 mag in $F814W$, for a median
distance of $3.5\Mpc$ and metallicity of $0.1Z_\odot$).  Such stars
are typically rare, and thus the detectable main sequence will not be
sufficiently well-populated to appear as a distinct sequence unless there
has been ample very recent star formation.

Complicating the clear detection of the main sequence in some cases is
the presence of slightly older blue helium core burning (BHeB) stars.
These are evolving post-main sequence $5\!-\!20\Msun$ stars, with ages
of typically $10-500\Myr$.  These stars, along with their red
counterparts (RHeB stars), are formed following the exhaustion of core
hydrogen burning, when the core rapidly collapses and the stars'
effective temperature decreases\footnote{There is some overlap in the
  literature between the brightest stars in the HeB sequences and
  classical red and blue supergiants.  However, the former extends to
  much lower masses and luminosities.}.  The core soon reaches an
equilibrium, when core helium burning and inner shell hydrogen burning
commence, expanding the redder outer layers. A star at this phase is
known as a red helium burning star. As helium in the core is converted
into carbon, the RHeBs become visibly hotter and enter the blue helium
burning (BHeB) phase of evolution
\citep[e.g.,][]{langer1995}. Although the BHeB phase occupies a large
fraction of the total core helium burning lifetime
\citep[e.g.,][]{bertelli1994}, stars are capable of alternating
between BHeB and RHeB stages, with the precise lifetimes determined by
the intricate relationships between underlying physical parameters
\citep[e.g.,][]{chiosi1992}.

The red and blue core helium burning sequences are most obvious at
optical wavelengths.  They appear as two sequences emerging brightward
of the red clump, where low mass $\lesssim2\Msun$ core helium
burning stars are found.  The BHeB sequence emerges diagonally, heads
to bluer colors at higher luminosities, then becomes vertical and
parallel to the main sequence at high luminosities and high stellar
masses The RHeB sequence emerges nearly vertically from the red clump,
and extends steadily brightwards, at somewhat bluer optical colors
than the RGB.  Along these sequences, stars of a given mass appear at
a single luminosity, so that the number of stars along the sequences
indicate the numbers of evolving stars of different masses.  This
connection allows the star formation rate between $\sim$5 and
$\sim$400 Myr to be essentially read off the sequence, as was
pioneered by \citet{dohm2002b}, and more recently discussed in
\citet{mcquinn2011}.  More massive, and thus younger, stars appear at
very bright magnitudes along both the BHeB and RHeB sequences.
However, we note that the models for these stars are currently
uncertain, and are known not to produce the correct colors or relative
number of stars on the red and blue sequences at low metallicities
\citep{dohm2002a,mcquinn2011}.

\subsection{Observed Properties of NIR CMDs}  \label{cmdoverviewsec}

We now turn to the observed NIR and optical CMDs for the fields
covered by our WFC3/IR observations, shown in the top panels of
Figure~\ref{cmdfig}.  In the upper left panel of Figure~\ref{cmdfig},
we show the NIR CMD of the cleaned {\tt *.gst} photometry catalogs,
uncorrected for foreground extinction or reddening.  In the adjacent
panel, we show the highest quality optical CMD that is available from
overlapping archival imaging.  The stars in the optical CMD have been
restricted to those that overlap the area covered by the WFC3/IR FOV.
These observed CMDs can be compared with the models in
Figure~\ref{fakecmdfig} to understand which phases of stellar
evolution are populating various features in the CMD.  We now discuss
the principal features of these CMDs.

\subsubsection{Old and Intermediate Age Populations: The AGB \& RGB} \label{rgbagbsec}

In a broad qualitative sense, the NIR CMDs in Figure~\ref{cmdfig} show
the features expected from the stellar models, particularly at old
ages.  In fields dominated by old stellar populations (as judged from
the optical CMD; see KDG~63, or NGC~404 for examples), the NIR CMDs
are quite simple.  They show roughly 3 magnitudes of a well-defined
upper red giant branch terminating at $M_{\rm{F160W}}\!\sim\!-5.7$.
The red giant branch is more vertical and exhibits less curvature than
in the optical (see M81-DEEP for an example), as expected from
Figure~\ref{fakecmdfig}.  The color at the tip of the RGB varies from
between 0.7 and 1.1, as we discuss in more detail in
Section~\ref{trgbsec} below.  This range of colors agrees well with
the range expected from the models in Figure~\ref{fakecmdfig}.
Unfortunately, the data are not deep enough to reveal the red clump in
the NIR; this feature is expected to appear $\sim\!4$ mag below the
TRGB.

Galaxies dominated by old stars also host a sparse population of
TP-AGB stars located above the tip of the RGB
($M_{\rm{F160W}}\!\lesssim\!-5.7$).  The NIR population of TP-AGB stars
falls in a vertical sequence spanning a narrow range of color, in
contrast to the broad ``fan'' of TP-AGB stars seen in the optical.  At
optical wavelengths, the TP-AGB stars span a wide range of colors, due
to a combination of dust in their circumstellar envelopes, large
variations in molecular line spectra with photospheric temperature,
the distance between the optical and the NIR peak of an AGB's star
bolometric flux \citep{Frogel90}, and long-period variability (which
affects all passbands, but only produces significant color spreads in
the optical). 

The NIR TP-AGB sequence typically extends $\lesssim$1 magnitude in
$F160W$ above the TRGB.  However, the TP-AGB reaches even brighter
magnitudes in populations with more recent intermediate age SF (see
NGC~3741, for an example).  The existence of bright, massive TP-AGB
stars at younger stellar ages is predicted by the models as well (see
Figure~\ref{fakecmdfig}, and luminosity functions in
\citet{olsen2006}).  

At lower luminosities in the TP-AGB sequence, CMDs sometimes show a
noticeable ``gap'' between the brightest RGB stars and the faintest
TP-AGB stars above the TRGB (see NGC2403-HALO-6).  This localized drop
in the luminosity function seems to have a counterpart in evolutionary
tracks of TP-AGB stars \citep{marigo2007}.  The minimum likely
reflects the steeper rate of brightening ($-dM_{\rm{bol}}/dt$) that
characterises the initial stages of the TP-AGB phase, when the first
thermal pulses have still not reached the full-amplitude regime and
TP-AGB stars are expected to be fainter than predicted by the core
mass-luminosity relation.  
%
%
As a result, the brighter late stages of TP-AGB evolution are slower,
such that luminosity bins appear relatively more populated than the
fainter early stages.  Simulated CMDs based on the \citet{marigo2007}
TP-AGB models do show a local minimum in the luminosity function right
above the TRGB, as expected.  However, the gap is not apparent in the
more recent models described in \citet{girardi2010} (lower right panel
of Figure~\ref{fakecmdfig}).  The observation of this features opens
the possibility that it could help to calibrate the models at the
early stages of the TP-AGB.

The mean color at the base of the TP-AGB sequence is comparable to the
mean color at the tip of the RGB, such that the TP-AGB sequence appears
to emerge vertically from the RGB.  However, the cumulative distributions
of colors immediately above and below the TRGB\footnote{We define the
  RGB stars ``below'' the TRGB as being the 0.25 magnitude interval
  starting 0.05 mag fainter than the TRGB, and the AGB stars ``above''
  the TRGB as being the 0.5 magnitude interval starting 0.05 mag
  brighter than the TRGB.  This selection produces roughly equal
  numbers of stars in each subsample.} suggests that the AGB is
slightly redder by 0.02-0.04 magnitudes for galaxies where there is
minimal contamination from RHeB stars (e.g., KKH37, HS117, DDO82,
DDO78, N2976).  Visually, the TP-AGB sequence appears to span a smaller
range of color than the RGB in some cases. However, there is no
statistical evidence that this is the case, as both the RGB and TP-AGB
have statistically indistiguishible widths, when the fewer number of
stars in the AGB is taken into account.  

We note that our sample does not immediately show a large obvious
population of carbon-rich AGB stars.  In the corresponding $J+H$
ground-based NIR filter set, such stars dominate a roughly diagonal
sequence that starts near $M_{\rm{H}}\!\sim\!-7$ and extends to redder colors
and fainter magnitudes. To illustrate this sequence, we show empirical
data in Figure~\ref{lmcfig}, using $J-H$ vs $H$ CMD for 2MASS
observations of the the LMC.  This filter combination is the closest
match to the $F110W\!+\!F160W$ filter set used in our
observations\footnote{For reference, at $J-H=1$, magnitudes in $F160W$
  are 0.28 magnitudes fainter than in $H$, and colors in $F110W-F160W$
  are 0.08 magnitudes redder than in $J-H$, as shown below in
  equations~\ref{magtranseqn}~\&~\ref{colortranseqn}.}.  In the LMC
data the sequence of carbon-rich and/or extreme AGB stars is roughly horizontal at
$M_{\rm{H}}\sim-7.3$, and extends redward from the sequence of more numerous
oxygen-rich AGB stars.  There are a few cases where a comparable
red sequence of extreme AGB stars is hinted at in our data (e.g., UGC4305,
NGC4163, NGC2403-HALO-6, NGC0300-WIDE1, DDO82).  However, in none of
these is the sequence as noticeable as in the LMC.  

The absence of a strong carbon star tail is initially suprising given
that our sample of galaxies is dominated by lower metallicity galaxies than
the LMC.  The dredge-up of carbon is believed to be more efficient at
lower metallicities, both because of more favourable conditions for
the occurence of strong helium shell flashes
\citep{karakas2002,stancliffe2006}, and because of the lesser carbon
dredge-up needed to reach a carbon-rich condition (with C/O$>$1 at the
surface).  Thus, one expects carbon-rich stars to be more numerous in
low metallicity galaxies.  Empirically, this trend has been seen in
several studies
\citep[e.g.,][]{battinelli2005,groenewegen1999,groenewegen2007,boyer2011}.

On the other hand, carbon-rich stars of low metallicity are
significantly hotter than at solar metallicities, which inhibits the
formation of the molecules that cause the spectral features typical of
carbon stars \citep{marigo2007}.  The resulting reduced opacity in low
metallicity carbon-rich stars \citep[e.g.,][]{marigo2009} may
move the carbon stars back to bluer NIR colors than their high
metallicity counterparts \citep[see, for instance, Figure~7
of][]{marigo2007}.  Thus, while carbon stars may be more numerous in
low metallicity systems, they may be less obvious outliers in NIR
CMDs.  The difficulty of identifying carbon stars in the NIR has been
demonstrated empirically by \citet{battinelli2009} for galaxies within
the Local Group.

We believe that the lack of obvious carbon stars is further exacerbted
by use of the $F110W$+$F160W$ filter set.  Simulations with
TRILEGAL \citep{Girardi_etal05, GirardiMarigo07} shown in
Figure~\ref{carbonfig} indicate that the majority of carbon-rich AGB stars
occupy similar locations as oxygen-rich AGB stars in the WFC3/IR
CMDs. These simulations were performed using both the
\citet{loidl2001} (top row) and \citet{aringer2009} (bottom row)
synthetic spectra for C stars.  Whereas carbon stars are distinctly
redder in $J-H$ than the oxygen-rich AGB (left column), in
$F110W-F160W$ they are (if anything) slightly {\emph{bluer}} (middle
column), particularly for the \citet{aringer2009} atmosphere models.
The properties of the carbon-rich AGB population in our sample will be
explored in more detail in an upcoming paper.

\subsubsection{Young populations: The Importance of Red Core Helium
  Burning Stars} \label{RHeBsec}

In contrast to older stellar populations, young stellar populations
have surprisingly complex NIR CMDs.  Fields with the most active star
formation show far more features than just the RGB and AGB (see
targets IC2574-SGS (in I2574), NGC0300-WIDE1 (in N300), UGC4305-1 (in
HoII), and UGC5139 (in HoI)).  We now identify and briefly discuss
these various features.

Star forming NIR CMDs host a vertical sequence at
$F110W-F160W\!\sim\!0$ on the blue side of the CMD, made up of both
main sequence and BHeB stars.  They also show a broad cloud of stars
with colors intermediate between the main sequence and the RGB.  These
stars are most likely to be BHeB stars and other evolving massive
stars. Note that only the most massive blue stars are detectable in
the NIR.  Such stars are rare, making it less likely that a
``sequence'' will be sufficiently well-populated to appear as a
distinct feature (see Figure~\ref{fakecmdfig}), which reduces the
clarity of the main sequence.  In fields with on-going but low
intensity star formation (e.g., targets NGC3741, NGC7793-HALO-6,
NGC3077-PHOENIX), the ``main sequence'' manifests more as a blue edge
to the cloud of points, rather than the narrow sequence seen in the
optical.

To elucidate the correspondence between features in the optical and
NIR CMDs, in Figure~\ref{matchedfig} we show CMDs for stars that have
been matched between the NIR and optical photometry catalogs
(\S\ref{matchedsec}) for IC2574-SGS, after correcting for foreground
extinction and distance.  The stars have been color-coded by their
likely evolutionary phase, as deduced from the optical.  For this
exercise, we use optical catalogs involving the $F555$ filter, which
provides good separation of CMD features, particularly between main
sequence and BHeB stars, and between RHeB and RGB stars.  Stars that
did not have high quality positional matches (coincident within
0.07$\arcsec$), or that were faint in $F814W$, are plotted in black.
We note that the matching is unlikely to be perfect, and that some
modest fraction of stars in the NIR catalog may have been incorrectly
matched with stars in the optical catalog.  This mismatching would be
most likely in the crowded star-forming regions in this particular
field, where slight errors in the alignment could still leave stars
within the error radius for matching.

Figure~\ref{matchedfig} reveals a number of features.  First, as
expected, the main sequence stars (in blue) crowd along the blue edge
of the NIR CMD.  However, the blue core helium burning stars (in
green) also contribute the blue vertical sequence, as well as filling
the region between the NIR vertical blue sequence and the RGB.  Thus,
the lack of a clear BHeB sequence in the NIR is due primarily to its
merging with the MS.  On the red side of the NIR CMD, we see the RGB
(red) and AGB (cyan) features discussed in \S\ref{rgbagbsec}, along
with a modest horizontal tail of likely carbon-rich AGB stars at
$M_{\rm{F160W}}\!\sim\!-7$.  In addition to the AGB stars seen in the
CMDs of older stellar populations there is a brighter plume of
AGB stars extending brightward of $M_{\rm{F160W}}\!\sim\!-7.5$ and
$M_{\rm{F814W}}\!\sim\!-5$.  This plume is likely due to the presence
of more massive stars on the AGB, as a result of the strong SF at
$t<2\Gyr$ in the IC2574-SGS field.  Note also that the AGB stars fall
on a well-defined sequence in the NIR CMD, but spread over a wide
swath of color in the optical, as discussed above in
Section~\ref{rgbagbsec}.

The most remarkable feature in Figure~\ref{matchedfig} is the strong
sequence of luminous RHeB stars.  The color of the sequence is only
slightly bluer than the AGB and RGB (by $\sim\!0.2$ mag), and extends
to far brighter magnitudes.  Indeed, the red core Helium burning
evolutionary phase is responsible for {\emph{all}} of the most NIR
luminous stars in IC2574.  As can be seen in Figure~\ref{iroptrgbfig},
these luminous RHeB stars are tightly clustered within an active star
forming region that features several young clusters spanning a
$\sim\!100\Myr$ range of ages.  Looking at the optical image (left),
the youngest clusters in the region ($\lesssim10\Myr$) are still
embedded in luminous HII regions, while the slightly older clusters
are somewhat fainter, more diffuse, and lack extended line emission,
although they are still quite blue, due to large concentrations of
luminous BHeB stars.  However, in the NIR (right), the relative
luminosities of these clusters are reversed, such that the youngest
clusters have negligible NIR emission, but the slightly older clusters
have large concentrations of extremely luminous RHeB stars.

Thus, although much recent work has focused on the importance of the
AGB phase to setting the NIR colors and luminosities of young stellar
populations \citep{maraston2006,Henriques10}, it appears that RHeB
stars can potentially be equally important.  We demonstrate this in
Figure~\ref{ic2574LFfig}, where we show the integrated luminosity of
stars in different evolutionary phases.  At the limit where our data
becomes incomplete, RHeB stars are the dominant contribution, followed
by AGB stars.  The RGB, which is canonically assumed to dominate the
NIR light, is only the third most important contributor to the
luminosity at the completeness limit of our data. Note, however, that
the fractional contribution of RGB stars will increase when the full
range of stellar luminosities is considered; corrections for the
missing stars are included in a companion paper by
\citet{melbourne2011}, where the fraction of integrated light
contributed by RHeB and AGB stars is quantified, showing that the RHeB
can contribute up to 25\% of the total light in $F160W$, and that
current models underpredict the luminosity contribution of RHeB stars
by up to a factor of 4.

The contribution of RHeB stars to the total luminosity will be
strongest when the SFR has been elevated between 25 and
$\sim\!100\Myrs$ ago, which produces RHeB stars brighter than
$M_{\rm{F814W}}\!\sim\!-5$.  This timescale is the same as the one over
which UV emission is expected to be significant, and thus surveys that
select for galaxies with high UV flux may also be selecting for
galaxies with a significant RHeB population.  Failure to account for
the luminous, somewhat blue RHeB population would lead one to
overestimate the mass-to-light ratio in the NIR, and to infer lower
metallicities from the broad-band colors.  Unfortunately, the models
of RHeB stars are even more uncertain than for AGB stars, and are
known to produce erroneous colors and relative numbers of BHeB and
RHeB at some metallicities \citep{langer1995, dohm2002a, gallart05,
  mcquinn2011}

Our conclusions about the importance of RHeB stars are not
significantly affected by uncertainties from foreground contamination.
Unlike studies in the Magellanic Clouds, our fields cover small areas
on the sky, which minimizes contributions from bright Milky Way stars.
The lack of luminous contaminants can be seen empirically in
Figure~\ref{gridfig} (presented below), where we show the CMD of
different subregions in the chip.  We see essentially no luminous
stars in the outer regions of small galaxies, where we are presumedly
dominated by foreground stars and background galaxies (see HS117, for
example).  Predictions from the default TRILEGAL Milky Way model
\citep{Girardi_etal05} also suggest that we expect fewer than one
contaminating star in each of the NIR CMDs plotted in
Figure~\ref{cmdfig}.

We also note that individual RHeB stars could easily be confused as
individual stellar clusters, in images of galaxies that are nearby,
but that are not sufficiently close to resolve individual stars.  Due
to their high luminosity and red colors, it may be difficult to
distinguish single IR-luminous RHeB stars from older stellar clusters
with larger numbers of fainter RHeB stars.

\subsection{Color-Color Diagrams}  \label{colorcolorsec}

We can further explore the separation of different stellar evolutionary
phases using the optical-NIR color-color diagram.  We adopt the target
IC1574-SGS as a test case, and use the matched catalogs from
Figure~\ref{matchedfig} to generate an optical-NIR color-color
diagram, color-coded as in Figure~\ref{matchedfig}.  The resulting
color-color diagram is shown in Figure~\ref{colorcolorfig}.

Figure~\ref{colorcolorfig} shows that the optical and NIR colors are
highly correlated for colors bluer than $F555W-F814W\!\lesssim1.7$.
In this regime, the optical and NIR colors track each other
extremely well, and are thus of little utility for separating phases of
stellar evolution in data of this quality.
At redder colors, however, there is noticably decreased correlation
between the optical and NIR colors, driven almost entirely by the
behavior of the AGB population (cyan points in
Figure~\ref{colorcolorfig}).  Although the bluest AGB stars follow a
narrow sequence of optical-NIR colors, the optically redder AGB stars
are far less well-behaved.  To first order, the NIR color saturates at
$F110W-F160W\!\sim\!1$ for a wide range of optical colors, with the
exception of a likely population of extreme AGB stars \citep{blum2006}.
Qualitatively, it appears that the narrower blue AGB sequence bends
over to fixed NIR color with increasing optical color.  Beyond
$F555W-F814W\!\gtrsim\!1.7$, the narrow sequence appears to become
embedded in a larger swath of red AGB stars, for which there is little
correlation between optical and NIR color, and wide dispersion.  There
is no qualitative evidence for the narrower sequence continuing
redwards of $F555W-F814W\!\gtrsim\!2.2$.

We can further explore these different regimes of AGB behavior by
qualitatively separating the AGB sequence into two regions and
examining the behavior of the stars in the NIR CMD.
Figure~\ref{colorcoloragbfig} shows the adopted division of the AGB
population in color-color space (upper left), and the resulting NIR
CMD (upper right).  Although the division into two classes is somewhat
arbitrary, and not motivated by a particular choice of stellar model,
the adopted separation does appear to break the AGB population into
two luminosity classes.  As shown in the histograms in the lower left
panels, the redder, high-dispersion sub-population is a constant
fraction of the AGB stars brighter than $M_{\rm{F160W}}\!<\!-6.8$, it
also lacks lower luminosity AGB stars, and makes up fewer than 10\% of
the stars fainter than $M_{\rm{F160W}}\!>\!-6.2$.  In light of our
selection criteria, this luminosity difference suggests that the
higher luminosity, more massive AGB stars have systematically redder
optical colors (lower right panel) and/or bluer NIR colors (with the
exception of the extreme AGB stars redward of
$F110W-F160W\!\gtrsim\!1$).  Brighter than $M_{\rm{F160W}}\!<\!-6.8$,
however, we see no noticible difference in the NIR luminosities or
median NIR colors of the two sub-populations (beyond the larger
dispersion in the NIR color expected by the adopted division of the
two populations in color-color space).

Although Figure~\ref{colorcoloragbfig} suggests that there are
systematic luminosity-dependent differences in the optical-NIR colors
of AGB stars, we currently lack a physical model that would allow us
to make better motivated choices for dividing the AGB population on
the basis of optical-NIR colors.  It is tempting to think that the
division we have made is helping to isolate carbon stars, based on the
behavior seen in Figure~\ref{carbonfig}.  On the other hand, that same
figure shows the large current uncertainties in the modeling at the
relevant wavelengths.  A more definitive approach to isolating and
modeling carbon stars will likely require observations at somewhat
longer wavelengths than are currently possible with WFC3's IR channel,
or the addition of narrow-band imaging.

\subsection{Spatial variations}

In addition to the field-to-field variations of the NIR CMDs discussed
above, we also find strong spatial variations within individual
fields.  Such variation is naturally expected if young and
intermediate age populations contribute significantly to the NIR CMD.
In the right hand panels of Figure~\ref{gridfig} we show the NIR CMDs
in 16 subfields within each WFC3/IR frame, next to a color image of
the same frame.

These grids show a number of features.  First, one can see the effects
of crowding.  In the most well-populated fields, the depth of the CMD
is strongly dependent on position, such that the densest subregions
have the shallowest depth.  Second, one can see significant spatial
variations in the underlying stellar populations, particularly in
fields that have significant amounts of recent star formation.
Features due to the main sequence, RHeB, and AGB vary in strength and
luminosity within individual fields.  As an example, in NGC0300-WIDE1
the stellar populations from the left side of the frame host a
prominent main sequence and luminous RHeB stars, whereas these
populations are nearly absent from the lower right of the frame.
Finally, CMD features due to young stellar populations typically
appear with more clarity in the subregions than in the field as a
whole (for example, note the sharp localization of the RHeB and BHeB
in the upper left quadrants of IC2754-SGS, or in the relative
population of the AGB and RGB across the field).

\section{Tip of the Red Giant Branch in the NIR}  \label{trgbsec}

In the optical, the magnitude of the tip of the red giant branch has
become a widely used distance indicator for galaxies with resolved
stellar populations \citep{Lee93,Sakai96,Mendez02, Karachentsev06,
  Makarov06}.  Its utility as a distance indicator rests on the
relative insensitivity of the TRGB magnitude to age (for
age$\gtrsim\!3\Gyr$) or metallicity (for [Fe/H]$\lesssim\!-0.5$) in
the $I$ band, where the bolometric magnitude of RGB stars typically
peak \citep{Lee93}.

In the NIR, however, the behavior of the TRGB magnitude is quite
different.  Theoretical models indicate that the magnitude of the TRGB
should vary significantly with the properties of the underlying
stellar population \citep[Figure 2 of][]{salaris2005}, complicating
the use of the NIR TRGB as a distance indicator.
The variation in the NIR TRGB magnitude is known to be
metallicity-dependent for sub-solar metallicities
($\lesssim0.5Z_\odot$), as has been demonstrated conclusively in data
for globular clusters in the 2MASS filter set \citep{valenti2004}.
One can therefore potentially correct for the $\lesssim1$ magnitude
variation in the NIR TRGB absolute magnitude if the metallicity is
known, allowing the TRGB to be used as a distance indicator in the
NIR.

Unfortunately, this basic approach may not be effective in
extragalactic systems, where the mean metallicity is uncertain, and
the underlying stellar population is more complex.  As shown in
Figure~10 of \citet{melbourne2010}, the presence of intermediate age
stars ($1-2.5\Gyr$) shifts the NIR TRGB to significantly
fainter magnitudes.  However, one could potentially diagnose the
presence of this population through other means (for example, through
a larger population of bright AGB stars), and correct for it.  The
metallicity dependence could likewise be corrected for as well; since
metallicity directly affects the color of the RGB, it should be
possible to empirically calibrate a relationship between the magnitude
and color of the TRGB.

In this section, we derive the magnitude and color of the TRGB for our
sample.  We assume that the distances in Table~\ref{sampletable}
(based on $F814W$ TRGB measurements) are correct, and then derive the
absolute magnitude of the NIR TRGB.  In what follows, all magnitudes
and colors have been corrected for the foreground extinction given in
Table~\ref{sampletable}, assuming that
$A_{\rm{F110W}}/A_{\rm{V}}=0.33669$ and
$A_{\rm{F160W}}/A_{\rm{V}}=0.20443$ \citep[][, updated to included the
latest in-flight calibrations for the WFC3 filter sets]{girardi2008}.

As a first step, in Figure~\ref{LFfig} we plot the observed luminosity
functions of the red extinction-corrected stars (dark black
histogram), and their median color in magnitude bins (red line,
plotted only for bins with more than 12 stars).  Red stars are
selected using a magnitude-dependent color cut to suppress the
contribution of bluer young BHeB and MS stars; brighter than the TRGB,
we keep all stars redward of $F110W-F160W=0.5$, and fainter than the
TRGB, we keep all stars redward of the diagonal line connecting
$m=m_{\rm{TRGB}}$, $F110W-F160W=0.5$ with $m=m_{\rm{TRGB}}+4$,
$F110W-F160W=-0.2$.  This cut has no appreciable effect on any feature
visible in the luminosity function.  The luminosity functions have not
been corrected for incompleteness, and we therefore expect them to
roll over at faint magnitudes due solely to observational effects; we
carry out a full correction for these effects in
\citet{melbourne2011}, where we analyze the NIR luminosity contributed
by different evolutionary phases.

Figure~\ref{LFfig} shows that there is a clear variation in the
absolute magnitude of the TRGB, relative to a fiducial TRGB absolute
magnitude of $M_{\rm{F160W}}=-5.7$, plotted as a vertical dotted line.
Visual inspection shows that redder RGBs typically terminate at
brighter absolute magnitudes; for these presumably metal rich RGB
stars, the bolometric flux peaks at redder wavelengths, increasing the
flux in the NIR.  In some cases one can also see the slight dip in the
number of stars just brightward of the TRGB that was discussed in
Section~\ref{rgbagbsec} (for example, see UGC4459, NGC4163, NGC3741,
NGC2976-DEEP, and IC2574-SGS for particularly obvious cases).

To quantify the variation in the TRGB, we measure the TRGB magnitude
in $F160W$ using the edge-detection filter described in
\citet{Mendez02} applied to a Gaussian-smoothed luminosity function as
in \citet{Sakai96} and \citet{seth05}.  Although more sophisticated
techniques exist \citep[e.g.,][]{Makarov06,frayn03}, the TRGB of our
sample is typically well-populated and falls well above the
photometric limit of the data, making our use of the widely used and
calibrated edge-detection technique adequate for an initial distance
measurement.  We use the identical procedure as was used to measure
the TRGB for the optical data \citep{dalcanton2009}, making the
$F814W$ and NIR TRGB directly comparable.  Specifically, after
extinction correcting all magnitudes and colors, we restrict our
analysis to stars on the RGB by iteratively fitting a line to stars
that are less than 1 magnitude fainter than the estimated TRGB and
that have colors consistent with potential RGB stars
($0.6<F110W-F160W<1.1$). We retain all stars that are within $2\sigma$
of the fit to the RGB sequence towards the red, and within $1.5\sigma$
to the blue; we use the more restrictive cut towards the blue to
suppress the contribution from RHeB stars.  The candidate RGB stars
were then used to construct a Gaussian-smoothed luminosity function,
which was then passed through an edge-detection filter.

The final TRGB magnitude and uncertainty was measured by executing 750
Monte Carlo bootstrap resampling trials.  In each trial, additional
Gaussian random errors were added to the stars' photometry, based on
the magnitude of each star's photometric error.  The TRGB for each
trial was taken to be the magnitude corresponding to the peak of the
edge-detection response filter within a 1 magnitude interval around
the likely TRGB.  We then fit the histogram of the returned TRGB
magnitudes with a Gaussian at $m_{\rm{TRGB}}$, taking the mean and
width of the Gaussian to be the magnitude of the TRGB and its
uncertainty.  These quantities are converted to absolute magnitudes
using the distance moduli in Table~\ref{sampletable}, adopted from the
TRGB analysis in $F814W$.

Once the magnitude of the TRGB has been measured, we characterize the
color of the TRGB using stars within 0.05 magnitudes fainter than the
TRGB; in the few cases where there are fewer than 15 stars within the
adopted magnitude range, we increase the range to 0.1 magnitudes.  We
calculate the mean and standard deviation of these stars' colors using
a biweight, iteratively clipping stars within 2$\sigma$ of the
biweight mean, which visual inspection suggests captures all the width
of the TRGB with acceptable rejection of RHeB stars.

Figure~\ref{trgbfig} shows the resulting absolute TRGB magnitude in
$F160W$, as a function of the color of the TRGB.  The data show the
expected trend of brighter TRGB magnitudes for redder colors.  This
correlation can be well approximated as

\begin{equation}      \label{Mtrgbeqn}
M_{\rm{TRGB}}(F160W) = -2.576 (F110W-F160W) - 3.496
\end{equation}

\noindent with the absolute magnitudes showing a scatter of $0.050$ magnitudes
around the mean.  

Also shown on Figure~\ref{trgbfig} are expectations for two sets of
theoretical models.  The red line shows the magnitude and color of the
TRGB for $10\Gyr$ old Padova isochrones from the WFC3 extension of
\citet{girardi2008}, spanning a range of metallicities.  The blue line
shows a fit to the magnitude and color of the TRGB for a series of
isochrones supplied by Aaron Dotter, spanning a range of ages
(6-$14\Gyr$, shown as asteri with increasing point sizes for older
TRGBs) and metallicities (from [Fe/H]$\!=\!-2.49$ (magenta) to
[Fe/H]$\!=\!-0.99$ (green)).  In general, the models show the same
trend between TRGB color and magnitude seen in our data, suggesting
that the bulk of the TRGB luminosity variation is driven by variations
in metallicity.  Similar behavior has been hinted at longer
wavelengths for Spitzer IRAC observations of galaxies within the Local
Group \citep{boyer2009}.

Although the slope of the fit to the Dotter isochrones is
nearly identical to a weighted (in $X$ and $Y$) least squares fit to
the data (Eqn.~\ref{Mtrgbeqn}), the measured TRGBs are shifted to
brighter magnitudes by a median of $0.099\pm0.027$ magnitudes (or
equivalently, to bluer colors by $0.043\pm0.012$), compared to
the models.  These offsets between the data and models are
statistically significant, as judged by whether or not the mean
residual in magnitude or color is significantly different from zero.

We have considered a number of possible origins for the offset between
data and models in the left hand plot of Figure~\ref{trgbfig}.  First,
as a broad check on the consistency of our measurements, on the right
side of Figure~\ref{trgbfig} we plot the correlation between our
measured NIR TRGB color ($F110W-F160W$) and the optical-NIR color of
the TRGB inferred from the difference between our measurements of
$m_{\rm{TRGB}}$ in $F814W$ and $F160W$.  These two colors appear to be highly correlated, 
as expected (Figure~\ref{colorcolorfig}).  However, they also show a small offset
from the Dotter isochrones, such that the median observed 
optical-NIR color of the TRGB is 0.04 magnitudes redder than the Dotter
isochrones, with a semi-interquartile range of $\pm0.037$.  If the
source of the offset in the $M_{\rm{TRGB}}(F160W)$ vs $F110W-F160W$
relation were primarily a bias towards bluer measurements of the mean
color of the TRGB, then the data should be offset to bluer colors in
this diagram as well, rather than to redder $F110W-F160W$ colors as is
observed.  Thus, the offset between data and models does not appear to
be due a mismeasurement of the TRGB color.

Alternatively, if the source of the offset
were primarily a bias towards brighter measurements of the $F160W$
TRGB magnitude, then the data would be shifted to redder values of
$m(F814W)-m(F160W)$ in this diagram.  We do indeed see a redward shift
between the data and the models for $m(F814W)-m(F160W)$. However, the
amplitude of this shift is a factor of 2.5 times smaller than needed
to explain the offset observed in $M_{\rm{TRGB}}(F160W)$ vs $F110W-F160W$.

The final explanation we consider is if the offset results from a
mismatch between the filter calibrations and the filter throughputs
adopted by the models (such that measured F160W magnitudes are
brighter than would be inferred for the models).  If so, then both the
$m(F814W)-m(F160W)$ and $F110W-F160W$ values would be shifted to
redder colors compared to the models.  However, this shift would move
points along the mean relation, rather than perpendicular to it, and
thus is unlikely to produce significant offsets.  This possibility is
consistent with the direction and the magnitudes of the offsets in
both diagrams shown in Figure~\ref{trgbfig}.

As a second approach to assessing the origin of the offset, we have
included measurements of the TRGB from globular clusters in
Figure~\ref{trgbfig}.  These measurements are from \citet{ivanov2002}
(solid green diamonds) and the \citet{valenti2004} and
\citet{valenti2007} samples (open yellow diamonds\footnote{Including
  updates from {\tt{http://www.bo.astro.it/$\sim$GC/ir\_archive/}}},
restricting the sample to those with $A_H < 0.5$).  Both data sets
were originally taken in $J$ and $H$. We transformed these data to
$F160W$ and $F110W$ using the following relations

\begin{equation}   \label{magtranseqn}
F160W-H = 0.2031 + 0.401\,(J-H - 0.9) + 0.3498\,(J-H - 0.9)^2
\end{equation}

\noindent and

{\scriptsize{
\begin{equation}   \label{colortranseqn}
J-H = 0.9418 + 0.841\,(F110W-F160W - 1.0) - 0.9053\,(F110W-F160W - 1.0)^2,
\end{equation}
}}

\noindent which were derived by fitting the magnitudes and colors of the TRGB for
colors redder than $F110W-F160W<1.15$ in the Padova models; these transformations are good
to $<\!0.001$ magnitudes across this range.

The globular cluster data tend to be shifted to brighter magnitudes
and/or bluer colors than our measurements, and are even more offset
from the models (in both the WFC3/IR and native $J+H$ filter sets).
The offsets are more pronounced for the Valenti sample,
which appears to be biased somewhat brightwards compared to other
data, including two clusters in common with the \citet{ivanov2002}
sample\footnote{The two clusters are NGC~6441, which is 0.07 mag
  brighter in $F160W$ and 0.08 mag bluer in the \citet{valenti2004}
  sample, and NGC~6624, which is 0.12 magnitudes brighter and has an
  identical color in the \citet{valenti2004} sample.  Both of these
  clusters have relatively high degrees of foreground extinction,
  which increases the likelihood of differences between independent
  analyses.}, and three bulge clusters in common with the sample of
\citet{chun2010}.  We note, however, that measurements of the TRGB can
be difficult in globular clusters, which frequently have uncertain
distances, sparsely populated RGBs at bright
magnitudes, and ambiguous distinctions between bright
RGB and faint AGB stars.  Furthermore, the extinction corrections are
typically quite large for the comparison globular cluster sample; even with our
restriction on the foreground extinction, the median extinction is
larger than 0.3 magnitudes in $H$ for the Valenti sample.

We have also considered that photometric biases may be responsible for
the small offsets in our TRGB magnitudes and/or colors.  For example,
the Monte Carlo process used to evaluate our uncertainties
artificially increases the photometric error (during randomization of
magnitudes) and potentially biases $m_{\rm{TRGB}}$ by scattering stars
preferentially above the tip.  In practice, however, the effect of the
added noise is negligible, since the photometric uncertainties are
extremely small at $m_{\rm{TRGB}}$.  The effect of these biases have
been explored extensively in \citet{madore1995}, and at our typical
signal-to-noise level of $SNR\sim50$, we expect negligible magnitude
biases due to photometric errors.  We likewise have considered the
effect of crowding on our measurements, and find that it too is
unlikely to explain the offset.  As shown in Figure~\ref{fakestarfig},
our photometric measurements are essentially unbiased in the median,
with less than an 0.005 magnitude shift at faint magntidues.  Crowding
does produce a slight tail to brighter magnitudes, due to undetected
stars contaminating the flux of the resolved stars.  However, at the
typical magnitude of the TRGB, fewer than 3\% of fake stars have
magnitudes that would be brighter than expected from photometric
uncertainty alone.  In addition, we see no correlation between the
degree of offset from the models and the number of stars in the field.

We also have explored population differences as being a source of the
offset between the data and the models.  Variations in the mean
stellar age can affect the magnitude of the TRGB, while contamination
from RHeB and AGB stars can change both the color and magnitude of the
TRGB.  However, we found no significant impact from these effects,
based on the lack of correlation between the the amplitude of the
offset and the fraction of stars formed in different age ranges.  

We are left without any satisfying explanation for this small offset
between the data and the models.  At this point, our best guess is
that it results from some combination of small uncertainties in the
current WFC3/IR calibrations (although these improved significantly
over the course of this program), in the filter throughputs adopted by
the models, or the models themselves.  Indeed, \citet{cassisi2010}
Figure 2 shows significant systematic variations among models for the
predicted bolometric magnitudes of the TRGB.  For now, any attempt to
use $F160W$ measurements for the TRGB should acknowledge that there
may be an underlying 10\% systematic uncertainty in the calibration.

\section{Luminosity Functions}  \label{trgbstructsec}\label{LFsec}

In the above analysis, we identify the TRGB as the point where there
is a sharp peak in the edge-detection algorithm that also corresponds
with the largest drop in the number of stars.  However, the
edge-detection algorithm frequently identifies other sharp
transitions in the luminosity function that correspond to smaller changes
in the absolute number of stars with magnitude.

To explore these effects, in Figure~\ref{allLFfig} we plot all galaxy
luminosity functions, relative to the observed magnitude of the TRGB.
The light lines are color coded according to each galaxy's rank when
sorted by the fraction of star formation the galaxy has experienced in
the most recent Gyr.  The heavy lines show the average luminosity
function when the galaxies are sorted by rank into 4 bins of recent
star formation, with equal numbers of galaxies per bin; red, magenta,
blue, and black lines go from lowest to highest fraction of recent
star formation ($\langle f_{0-1\Gyr} \rangle = 0.001, 0.04, 0.07,
0.09$, respectively).  All luminosity functions have been normalized
to have the same total number of stars in the bins between 0.25 and
0.75 magnitudes fainter than the TRGB.  We do not
analyze the luminosity function at fainter magnitudes, where the
varying depths, photometric uncertainties, and distances among the
galaxy population make the averaging procedure invalid.  These
luminosity functions include only stars that lie within the
extrapolated slope and width of the RGB, and may miss the reddest, most
luminous RHeB stars.

The average luminosity functions in Figure~\ref{allLFfig} show clear
systematic deviations that correlate with the fraction of star
formation in the most recent gigayear.  These deviations are most
obvious brighter than the TRGB, where elevated recent star formation
produces a larger population of RHeB and luminous AGB stars, extending
to brighter magnitudes.  As noted previously in Section~\ref{RHeBsec}
and as quantified in \citet{melbourne2011}, this population of luminous
stars can lead to significant reductions in the NIR mass-to-light
ratio, on much shorter timescales than discussed for AGB stars alone.

There are also subtle features close to the TRGB itself, some of which
correlate with the recent SFR.  In particular, the amplitude of the
discontinuity at the TRGB is less pronounced when recent star
formation is more prominent, most likely due to contamination from
younger stellar populations all along the RGB sequence and its
extension to brighter magnitudes.  This contamination will tend to
reduce the reliability of the NIR TRGB distance measurements for
actively star-forming galaxies.  


\section{Metallicity and the Structure of the RGB}  \label{rgbstructsec}

For two decades it has been known that the slope of the NIR RGB
correlates strongly with metallicity for globular clusters
\citep[e.g.,][]{davidge1992,cohen1995,kuchinski1995,ferraro2000}.
While this correlation allows the slope of the NIR RGB to be used as a
reddening-free diagnostic of metallicity for globular clusters
\citep[for an example, see][]{ferraro2006}, it is not clear if the
relationships used for uni-aged globular clusters can be routinely
applied to the complex stellar populations found in galaxies, where
age variations can also affect the structure of the RGB, and where
observations typically produce lower quality photometry due to
unavoidable crowding errors.

We now evaluate whether the NIR RGB slope can be used effectively as a
metallicity indicator for complex stellar populations.  We first
characterize the RGB using the measured median color as a function of
magnitude (red lines from Figure~\ref{LFfig}; see
Section~\ref{trgbsec}).  We presume that the observed variation in
color is dominated by variations in metallicity, based on the greater
sensitivity of RGB color to metallicity compared to age (upper right
of Figure~\ref{fakecmdfig}).  Thus, while we cannot directly compare
the RGB slope to metallicity, we can use the color of the TRGB as a
proxy.  

\subsection{TRGB Color and the Luminosity-Metallicity Relationship}

To demonstrate the connection between metallicity and NIR color of the
TRGB, in Figure~\ref{magcolorfig}, we plot the color of the TRGB as a
function of each galaxy's extinction-corrected $B_T$ magnitude (upper
left) and NIR luminosity (upper right; 3.6$\mu$ luminosity taken from
\citep{dale2009}, assuming that the Sun has a flux of 14.71\,Jy at a
distance of $10\pc$).  The data show a clear relationship between
galaxy luminosity and TRGB color,such that $F110W-F160W = 0.605 -
0.0181\,M_{\rm{B_T}}$ and $F110W-F160W = 0.431 +
0.0518\,\log_{10}{L_{3.6\mu}}$, with an rms of 0.046 and 0.039,
respectively.  Because of the well-known mass-metallicity
relationship \citep[e.g.,][and references therein]{lee2006}, one would
expect higher metallicities in more massive, luminous galaxies, which
should then manifest itself as more luminous galaxies having redder
TRGBs.  Figure~\ref{magcolorfig} shows this expected behavior,
suggesting that the TRGB color does indeed correlate with metallicity.

Further support for the connection between TRGB color and metallicity
comes from the lower left panel, where we plot the NIR TRGB color as a
function of the metallicity inferred from the oldest age bin of the
star formation history \citep[derived from the optical photometry
using MATCH;][]{melbourne2011}.  We see a tight correlation between
NIR TRGB color and metallicity, following the relationship
$F110W-F160W = 1.093 + 0.192\,{\rm{[Fe/H]}}$ with an rms of 0.050.
Note, however, that one expects this correlation to be tight, given
that the metallicity sensitivity of the CMD fits comes largely from
the color of the TRGB.

\subsection{RGB Slope as a Metallicity Indicator}

Having verified the relationship between TRGB color and metallicity,
we now investigate the correlation between RGB slope and the TRGB
color.  On the left side of Figure~\ref{rgbfig} we plot linear fits to
the RGB, restricted to two magnitudes fainter than the
TRGB\footnote{The RGB slope in globular clusters is typically measured
  down to near the horizontal branch \citep[e.g.,][]{kuchinski1995}.
  However, in extragalactic systems one typically cannot resolve more
  than 2-3 magnitudes of the RGB, due to crowding.}.  The plot shows
the expected variation of the magnitude of the TRGB with color, such
that redder RGBs extend to brighter magnitudes (Figure~\ref{trgbfig}).

On the right side of Figure~\ref{rgbfig} we plot the slope of the
linear fit compared to the color of the fit at the magnitude of the
TRGB.  In general, there are no systematic statistically-significant
trends across the whole sample.  There is a $\sim$25\% dispersion in
the mean slope for colors between $0.8\!\lesssim\!
F110W-F160W\!\lesssim\!1$, with no apparent correlation.  This
suggests that the observed RGB slope is unlikely to be a useful
metallicity indicator in the presence of the dispersion in age,
metallicity, and photometric errors within individual extragalactic
systems.

The only possible manifestations of the known correlation between NIR
RGB slope and metallicity can be seen at the extremes.  Of the four
galaxies with the most vertical RGBs, three also have the bluest
colors.  Two of these three (UA292 and Sc22) do indeed to have very
low metallicity populations, based upon the lack of curvature in the
optical RGB.  In contrast, the remaining galaxy (KDG73) does not have
a particularly old globular-like stellar population, nor does the
optical CMD suggest a particularly low metallicity (based upon the
observed curvature in the optical RGB).  Instead, this galaxy appears
to have a substantial younger AGB population which may be pulling the RGB to
bluer colors and altering its morphology (see Figure~\ref{fakecmdfig}).

The other extreme outliers are HoII and IC2574, both of which have
particularly shallow RGB slopes.  However, both of these galaxies also have
dramatic RHeB populations, which are likely shifting the base of the
RGB fit to bluer colors, artificially flattening the RGB slope.

In summary, there appears to be little chance that the slope of the
NIR RGB can be used to assess the metallicity of extragalactic systems
with current data quality; the existing trends appear to be weak at
best, and outliers due to contributions from AGB and RHeB stars are
not uncommon.  One may have better success by combining the NIR with
the optical to produce an optical-NIR CMD with a wide color baseline
(i.e., similar to the widely used $V-K$ color), or with using a more
sophisticated method to characterize the RGB slope.

\section{Conclusions}   \label{conclusionsec}

This work represents the first step in characterizing the NIR
properties of complex stellar populations for galaxies spanning a wide
range of metallicity and star formation history.  We have presented
CMDs in one of the most efficient and commonly used WFC3/IR filter
pairs, allowing one to assess the contribution of individual stars
to the integrated light in these bandpasses.

In all cases, the CMDs show a dominant population of RGB stars, as
expected.  They also show clear sequences of AGB stars, with only a
modest contribution of extremely red AGB stars.  These features are
characteristic of old and intermediate age star formation.  In
galaxies with more recent star formation, we also find a dramatic
sequence of red core Helium burning stars, extending to much brighter
magnitudes than the AGB sequence.  These stars are associated with
recent star formation (20-500$\Myr$), and in some cases are a major
contributor to the NIR flux.  They also have colors that are similar
to the underlying AGB and RGB population, making their presence
difficult to diagnose from broad-band colors alone.  These RHeB stars
therefore present a significant uncertainty in adopting a NIR
mass-to-light ratio when analyzing unresolved galaxies or interpreting
their NIR colors.  We quantify the contribution of both RHeB and AGB
stars in \citet{melbourne2011}.

We have used the observed CMDs to empirically calibrate the NIR TRGB
magnitude as a function of the metallicity-sensitive $F110W-F160W$ color, allowing these
standard HST filters to be used as distance indicators in the WFC3/IR
and JWST era.  We find that there is a strong correlation between
color and the absolute magnitude of the TRGB, as expected from models
and previous observations of globular clusters.  However, we find that
the relationship is offset by $\sim$0.05-0.1 magnitudes from the
prediction of current isochrone models.  Analysis of the origin of
this offset points to a residual uncertainty in either the WFC3/IR
zero points or the theoretical models.

We also explore the structure of the RGB and AGB sequences,
highlighting possible variations with the age and metallicity of the
underlying stellar population.  We see a clear age-dependent variation
in the luminosity function of red stars, due to increasing numbers of
luminous AGB stars with increasing intermediate age star formation.
We are unable to detect any metallicity dependence in the slope of the
RGB, however, which prevents the slope of the NIR RGB from being used
as a reddening-free metallicity indicator.

We also present an analysis of scattered light in the WFC3/IR
detector.  We have found a rapidly-varying scattered light component
that affects images taken at low angles to a bright Earth limb.  We
discuss how this scattered light may affect photometry, particularly
when fitting a series of non-destructive reads with a linear function.

In a series of subsequent papers, we will be using the data presented
here to refine theoretical models of RHeB and AGB stars, to assess
contributions of carbon stars, to empirically calibrate the fractional
luminosity due to AGB and RHeB stars, and to improve star formation histories
at intermediate ages.





\acknowledgements 

The authors are very happy to acknowledge helpful discussions with Ben
Weiner about scattered light in WFC3/IR.  Aaron Dotter is warmly
thanked for supplying theoretical RGB sequences, as is Jason Kalirai
for providing information helpful for WFC3/IR calibration and
processing.  As always, we thank Alison Vick for her support in
scheduling and executing these observations.  The anonymous referee is
thanked for providing useful, detailed comments on this very long
paper.  JJD thanks the Max Planck Institute f\"ur Astronomie for their
hospitality while writing part of this paper.  LG and PM acknowledge
support from contract ASI-INAF I/009/10/0.  This work was supported by
the Space Telescope Science Institute through SNAP-11719, and used
additional data products produced through GO-10915 and AR-10945.


\bibliographystyle{apj}  

\begin{thebibliography}{87}
\expandafter\ifx\csname natexlab\endcsname\relax\def\natexlab#1{#1}\fi

\bibitem[{{Aaronson} {et~al.}(1978){Aaronson}, {Cohen}, {Mould}, \&
  {Malkan}}]{aaronson1978}
{Aaronson}, M., {Cohen}, J.~G., {Mould}, J., \& {Malkan}, M. 1978, \apj, 223,
  824

\bibitem[{{Aringer} {et~al.}(2009){Aringer}, {Girardi}, {Nowotny}, {Marigo}, \&
  {Lederer}}]{aringer2009}
{Aringer}, B., {Girardi}, L., {Nowotny}, W., {Marigo}, P., \& {Lederer}, M.~T.
  2009, \aap, 503, 913, 0905.4415

\bibitem[{{Battinelli} \& {Demers}(2005)}]{battinelli2005}
{Battinelli}, P., \& {Demers}, S. 2005, \aap, 434, 657

\bibitem[{{Battinelli} \& {Demers}(2009)}]{battinelli2009}
------. 2009, \aap, 493, 1075

\bibitem[{{Bertelli} {et~al.}(1994){Bertelli}, {Bressan}, {Chiosi}, {Fagotto},
  \& {Nasi}}]{bertelli1994}
{Bertelli}, G., {Bressan}, A., {Chiosi}, C., {Fagotto}, F., \& {Nasi}, E. 1994,
  \aaps, 106, 275

\bibitem[{{Blum} {et~al.}(2006){Blum}, {Mould}, {Olsen}, {Frogel}, {Werner},
  {Meixner}, {Markwick-Kemper}, {Indebetouw}, {Whitney}, {Meade}, {Babler},
  {Churchwell}, {Gordon}, {Engelbracht}, {For}, {Misselt}, {Vijh}, {Leitherer},
  {Volk}, {Points}, {Reach}, {Hora}, {Bernard}, {Boulanger}, {Bracker},
  {Cohen}, {Fukui}, {Gallagher}, {Gorjian}, {Harris}, {Kelly}, {Kawamura},
  {Latter}, {Madden}, {Mizuno}, {Mizuno}, {Nota}, {Oey}, {Onishi}, {Paladini},
  {Panagia}, {Perez-Gonzalez}, {Shibai}, {Sato}, {Smith}, {Staveley-Smith},
  {Tielens}, {Ueta}, {Van Dyk}, \& {Zaritsky}}]{blum2006}
{Blum}, R.~D. {et~al.} 2006, \aj, 132, 2034

\bibitem[{{Boyer} {et~al.}(2009){Boyer}, {Skillman}, {van Loon}, {Gehrz}, \&
  {Woodward}}]{boyer2009}
{Boyer}, M.~L., {Skillman}, E.~D., {van Loon}, J.~T., {Gehrz}, R.~D., \&
  {Woodward}, C.~E. 2009, \apj, 697, 1993

\bibitem[{{Boyer} {et~al.}(2011){Boyer}, {Srinivasan}, {van Loon}, {McDonald},
  {Meixner}, {Zaritsky}, {Gordon}, {Kemper}, {Babler}, {Block}, {Engelbracht},
  {Hora}, {Indebetouw}, {Meade}, {Misselt}, {Robitaille}, {Sewilo}, {Shiao}, \&
  {Whitney}}]{boyer2011}
{Boyer}, M.~L. {et~al.} 2011, \aj, in preparation

\bibitem[{{Brinchmann} \& {Ellis}(2000)}]{brinchmann00}
{Brinchmann}, J., \& {Ellis}, R.~S. 2000, \apjl, 536, L77

\bibitem[{{Bundy} {et~al.}(2005){Bundy}, {Ellis}, \& {Conselice}}]{bundy05}
{Bundy}, K., {Ellis}, R.~S., \& {Conselice}, C.~J. 2005, \apj, 625, 621

\bibitem[{{Cassisi}(2010)}]{cassisi2010}
{Cassisi}, S. 2010, in IAU Symposium, Vol. 262, IAU Symposium, ed. {G.~Bruzual
  \& S.~Charlot}, 13--22

\bibitem[{{Castellani} {et~al.}(2000){Castellani}, {Degl'Innocenti}, {Girardi},
  {Marconi}, {Prada Moroni}, \& {Weiss}}]{castellani2000}
{Castellani}, V., {Degl'Innocenti}, S., {Girardi}, L., {Marconi}, M., {Prada
  Moroni}, P.~G., \& {Weiss}, A. 2000, \aap, 354, 150

\bibitem[{{Chiosi} {et~al.}(1992){Chiosi}, {Bertelli}, \&
  {Bressan}}]{chiosi1992}
{Chiosi}, C., {Bertelli}, G., \& {Bressan}, A. 1992, \araa, 30, 235

\bibitem[{{Chun} {et~al.}(2010){Chun}, {Kim}, {Shin}, {Chung}, {Lim}, {Park},
  {Kim}, {Han}, \& {Sohn}}]{chun2010}
{Chun}, S. {et~al.} 2010, \aap, 518, A15+

\bibitem[{{Cirasuolo} {et~al.}(2010){Cirasuolo}, {McLure}, {Dunlop}, {Almaini},
  {Foucaud}, \& {Simpson}}]{cirasuolo10}
{Cirasuolo}, M., {McLure}, R.~J., {Dunlop}, J.~S., {Almaini}, O., {Foucaud},
  S., \& {Simpson}, C. 2010, \mnras, 401, 1166

\bibitem[{{Cohen} \& {Sleeper}(1995)}]{cohen1995}
{Cohen}, J.~G., \& {Sleeper}, C. 1995, \aj, 109, 242

\bibitem[{{Conroy} \& {Gunn}(2010)}]{conroy2010}
{Conroy}, C., \& {Gunn}, J.~E. 2010, \apj, 712, 833

\bibitem[{{Conselice} {et~al.}(2005){Conselice}, {Bundy}, {Ellis}, {Brichmann},
  {Vogt}, \& {Phillips}}]{conselice05}
{Conselice}, C.~J., {Bundy}, K., {Ellis}, R.~S., {Brichmann}, J., {Vogt},
  N.~P., \& {Phillips}, A.~C. 2005, \apj, 628, 160

\bibitem[{{Dahlen} {et~al.}(2005){Dahlen}, {Mobasher}, {Somerville},
  {Moustakas}, {Dickinson}, {Ferguson}, \& {Giavalisco}}]{dahlen05}
{Dahlen}, T., {Mobasher}, B., {Somerville}, R.~S., {Moustakas}, L.~A.,
  {Dickinson}, M., {Ferguson}, H.~C., \& {Giavalisco}, M. 2005, \apj, 631, 126

\bibitem[{{Dalcanton} {et~al.}(2009){Dalcanton}, {Williams}, {Seth}, {Dolphin},
  {Holtzman}, {Rosema}, {Skillman}, {Cole}, {Girardi}, {Gogarten},
  {Karachentsev}, {Olsen}, {Weisz}, {Christensen}, {Freeman}, {Gilbert},
  {Gallart}, {Harris}, {Hodge}, {de Jong}, {Karachentseva}, {Mateo}, {Stetson},
  {Tavarez}, {Zaritsky}, {Governato}, \& {Quinn}}]{dalcanton2009}
{Dalcanton}, J.~J. {et~al.} 2009, \apjs, 183, 67

\bibitem[{{Dale} {et~al.}(2009){Dale}, {Cohen}, {Johnson}, {Schuster},
  {Calzetti}, {Engelbracht}, {Gil de Paz}, {Kennicutt}, {Lee}, {Begum},
  {Block}, {Dalcanton}, {Funes}, {Gordon}, {Johnson}, {Marble}, {Sakai},
  {Skillman}, {van Zee}, {Walter}, {Weisz}, {Williams}, {Wu}, \&
  {Wu}}]{dale2009}
{Dale}, D.~A. {et~al.} 2009, \apj, 703, 517

\bibitem[{{Davidge}(2010)}]{davidge10}
{Davidge}, T.~J. 2010, \apj, 718, 1428

\bibitem[{{Davidge} {et~al.}(1992){Davidge}, {Harris}, {Bridges}, \&
  {Hanes}}]{davidge1992}
{Davidge}, T.~J., {Harris}, W.~E., {Bridges}, T.~J., \& {Hanes}, D.~A. 1992,
  \apjs, 81, 251

\bibitem[{{Dohm-Palmer} \& {Skillman}(2002)}]{dohm2002a}
{Dohm-Palmer}, R.~C., \& {Skillman}, E.~D. 2002, \aj, 123, 1433

\bibitem[{{Dohm-Palmer} {et~al.}(2002){Dohm-Palmer}, {Skillman}, {Mateo},
  {Saha}, {Dolphin}, {Tolstoy}, {Gallagher}, \& {Cole}}]{dohm2002b}
{Dohm-Palmer}, R.~C., {Skillman}, E.~D., {Mateo}, M., {Saha}, A., {Dolphin},
  A., {Tolstoy}, E., {Gallagher}, J.~S., \& {Cole}, A.~A. 2002, \aj, 123, 813

\bibitem[{{Dolphin}(2000)}]{dolphin2000}
{Dolphin}, A.~E. 2000, \pasp, 112, 1383

\bibitem[{{Dolphin}(2002)}]{dolphin2002}
------. 2002, \mnras, 332, 91

\bibitem[{{Ferraro} {et~al.}(2000){Ferraro}, {Montegriffo}, {Origlia}, \& {Fusi
  Pecci}}]{ferraro2000}
{Ferraro}, F.~R., {Montegriffo}, P., {Origlia}, L., \& {Fusi Pecci}, F. 2000,
  \aj, 119, 1282

\bibitem[{{Ferraro} {et~al.}(2006){Ferraro}, {Valenti}, \&
  {Origlia}}]{ferraro2006}
{Ferraro}, F.~R., {Valenti}, E., \& {Origlia}, L. 2006, \apj, 649, 243

\bibitem[{{Fluks} {et~al.}(1994){Fluks}, {Plez}, {The}, {de Winter},
  {Westerlund}, \& {Steenman}}]{fluks}
{Fluks}, M.~A., {Plez}, B., {The}, P.~S., {de Winter}, D., {Westerlund}, B.~E.,
  \& {Steenman}, H.~C. 1994, \aaps, 105, 311

\bibitem[{{Frayn} \& {Gilmore}(2003)}]{frayn03}
{Frayn}, C.~M., \& {Gilmore}, G.~F. 2003, \mnras, 339, 887

\bibitem[{{Frogel} {et~al.}(1990){Frogel}, {Mould}, \& {Blanco}}]{Frogel90}
{Frogel}, J.~A., {Mould}, J., \& {Blanco}, V.~M. 1990, \apj, 352, 96

\bibitem[{{Gallart} {et~al.}(2005){Gallart}, {Zoccali}, \&
  {Aparicio}}]{gallart05}
{Gallart}, C., {Zoccali}, M., \& {Aparicio}, A. 2005, \araa, 43, 387

\bibitem[{{Gilbert} {et~al.}(2011){Gilbert}, {Williams}, {Dalcanton}, \& {et
  al}}]{gilbert2011}
{Gilbert}, K.~A., {Williams}, B.~F., {Dalcanton}, J.~J., \& {et al}. 2011, \aj,
  in prep.

\bibitem[{{Girardi} {et~al.}(2002){Girardi}, {Bertelli}, {Bressan}, {Chiosi},
  {Groenewegen}, {Marigo}, {Salasnich}, \& {Weiss}}]{girardi02}
{Girardi}, L., {Bertelli}, G., {Bressan}, A., {Chiosi}, C., {Groenewegen},
  M.~A.~T., {Marigo}, P., {Salasnich}, B., \& {Weiss}, A. 2002, \aap, 391, 195

\bibitem[{{Girardi} {et~al.}(2008){Girardi}, {Dalcanton}, {Williams}, {de
  Jong}, {Gallart}, {Monelli}, {Groenewegen}, {Holtzman}, {Olsen}, {Seth},
  {Weisz}, \& {the ANGST/ANGRRR Collaboration}}]{girardi2008}
{Girardi}, L. {et~al.} 2008, \pasp, 120, 583

\bibitem[{{Girardi} {et~al.}(2005){Girardi}, {Groenewegen}, {Hatziminaoglou},
  \& {da Costa}}]{Girardi_etal05}
{Girardi}, L., {Groenewegen}, M.~A.~T., {Hatziminaoglou}, E., \& {da Costa}, L.
  2005, \aap, 436, 895

\bibitem[{{Girardi} \& {Marigo}(2007)}]{GirardiMarigo07}
{Girardi}, L., \& {Marigo}, P. 2007, in Astronomical Society of the Pacific
  Conference Series, Vol. 378, Why Galaxies Care About AGB Stars: Their
  Importance as Actors and Probes, ed. F.~{Kerschbaum}, C.~{Charbonnel}, \&
  R.~F. {Wing}, 20--+

\bibitem[{{Girardi} \& {Salaris}(2001)}]{girardi2001}
{Girardi}, L., \& {Salaris}, M. 2001, \mnras, 323, 109

\bibitem[{{Girardi} {et~al.}(2010){Girardi}, {Williams}, {Gilbert},
  {Rosenfield}, {Dalcanton}, {Marigo}, {Boyer}, {Dolphin}, {Weisz},
  {Melbourne}, {Olsen}, {Seth}, \& {Skillman}}]{girardi2010}
{Girardi}, L. {et~al.} 2010, \apj, 724, 1030

\bibitem[{{Grocholski} \& {Sarajedini}(2002)}]{grocholsky2002}
{Grocholski}, A.~J., \& {Sarajedini}, A. 2002, \aj, 123, 1603

\bibitem[{{Groenewegen}(1999)}]{groenewegen1999}
{Groenewegen}, M.~A.~T. 1999, in IAU Symposium, Vol. 191, Asymptotic Giant
  Branch Stars, ed. {T.~Le Bertre, A.~Lebre, \& C.~Waelkens}, 535--+

\bibitem[{{Groenewegen}(2007)}]{groenewegen2007}
{Groenewegen}, M.~A.~T. 2007, in Astronomical Society of the Pacific Conference
  Series, Vol. 378, Why Galaxies Care About AGB Stars: Their Importance as
  Actors and Probes, ed. {F.~Kerschbaum, C.~Charbonnel, \& R.~F.~Wing}, 433--+

\bibitem[{{Gullieuszik} {et~al.}(2008){Gullieuszik}, {Held}, {Rizzi},
  {Girardi}, {Marigo}, \& {Momany}}]{gullieuszik08}
{Gullieuszik}, M., {Held}, E.~V., {Rizzi}, L., {Girardi}, L., {Marigo}, P., \&
  {Momany}, Y. 2008, \mnras, 388, 1185

\bibitem[{{Gullieuszik} {et~al.}(2007){Gullieuszik}, {Held}, {Rizzi},
  {Saviane}, {Momany}, \& {Ortolani}}]{gullieuszik07}
{Gullieuszik}, M., {Held}, E.~V., {Rizzi}, L., {Saviane}, I., {Momany}, Y., \&
  {Ortolani}, S. 2007, \aap, 467, 1025

\bibitem[{{Henriques} {et~al.}(2010){Henriques}, {Maraston}, {Monaco},
  {Fontanot}, {Menci}, {De Lucia}, \& {Tonini}}]{Henriques10}
{Henriques}, B., {Maraston}, C., {Monaco}, P., {Fontanot}, F., {Menci}, N., {De
  Lucia}, G., \& {Tonini}, C. 2010, ArXiv e-prints, 1009.1392

\bibitem[{{Ivanov} \& {Borissova}(2002)}]{ivanov2002}
{Ivanov}, V.~D., \& {Borissova}, J. 2002, \aap, 390, 937

\bibitem[{{Karachentsev}(2005)}]{karachentsev05}
{Karachentsev}, I.~D. 2005, \aj, 129, 178

\bibitem[{{Karachentsev} {et~al.}(2006){Karachentsev}, {Dolphin}, {Tully},
  {Sharina}, {Makarova}, {Makarov}, {Karachentseva}, {Sakai}, \&
  {Shaya}}]{Karachentsev06}
{Karachentsev}, I.~D. {et~al.} 2006, \aj, 131, 1361

\bibitem[{{Karachentsev} {et~al.}(2003){Karachentsev}, {Grebel}, {Sharina},
  {Dolphin}, {Geisler}, {Guhathakurta}, {Hodge}, {Karachentseva}, {Sarajedini},
  \& {Seitzer}}]{karachentsev03}
------. 2003, \aap, 404, 93

\bibitem[{{Karachentsev} {et~al.}(2004){Karachentsev}, {Karachentseva},
  {Huchtmeier}, \& {Makarov}}]{karachentsev04b}
{Karachentsev}, I.~D., {Karachentseva}, V.~E., {Huchtmeier}, W.~K., \&
  {Makarov}, D.~I. 2004, \aj, 127, 2031

\bibitem[{{Karakas} {et~al.}(2002){Karakas}, {Lattanzio}, \&
  {Pols}}]{karakas2002}
{Karakas}, A.~I., {Lattanzio}, J.~C., \& {Pols}, O.~R. 2002, PASA, 19, 515

\bibitem[{{Kriek} {et~al.}(2010){Kriek}, {Labbe}, {Conroy}, {Whitaker}, {van
  Dokkum}, {Brammer}, {Franx}, {Illingworth}, {Marchesini}, {Muzzin}, {Quadri},
  \& {Rudnick}}]{kriek2010}
{Kriek}, M. {et~al.} 2010, ArXiv e-prints, 1008.4357

\bibitem[{{Kuchinski} {et~al.}(1995){Kuchinski}, {Frogel}, {Terndrup}, \&
  {Persson}}]{kuchinski1995}
{Kuchinski}, L.~E., {Frogel}, J.~A., {Terndrup}, D.~M., \& {Persson}, S.~E.
  1995, \aj, 109, 1131

\bibitem[{{Langer} \& {Maeder}(1995)}]{langer1995}
{Langer}, N., \& {Maeder}, A. 1995, \aap, 295, 685

\bibitem[{{Lee} {et~al.}(2006){Lee}, {Skillman}, {Cannon}, {Jackson}, {Gehrz},
  {Polomski}, \& {Woodward}}]{lee2006}
{Lee}, H., {Skillman}, E.~D., {Cannon}, J.~M., {Jackson}, D.~C., {Gehrz},
  R.~D., {Polomski}, E.~F., \& {Woodward}, C.~E. 2006, \apj, 647, 970

\bibitem[{{Lee} {et~al.}(1993){Lee}, {Freedman}, \& {Madore}}]{Lee93}
{Lee}, M.~G., {Freedman}, W.~L., \& {Madore}, B.~F. 1993, \apj, 417, 553

\bibitem[{{Loidl} {et~al.}(2001){Loidl}, {Lan{\c c}on}, \&
  {J{\o}rgensen}}]{loidl2001}
{Loidl}, R., {Lan{\c c}on}, A., \& {J{\o}rgensen}, U.~G. 2001, \aap, 371, 1065

\bibitem[{{Madore} \& {Freedman}(1995)}]{madore1995}
{Madore}, B.~F., \& {Freedman}, W.~L. 1995, \aj, 109, 1645

\bibitem[{{Makarov} {et~al.}(2006){Makarov}, {Makarova}, {Rizzi}, {Tully},
  {Dolphin}, {Sakai}, \& {Shaya}}]{Makarov06}
{Makarov}, D., {Makarova}, L., {Rizzi}, L., {Tully}, R.~B., {Dolphin}, A.~E.,
  {Sakai}, S., \& {Shaya}, E.~J. 2006, \aj, 132, 2729

\bibitem[{{Maraston} {et~al.}(2006){Maraston}, {Daddi}, {Renzini}, {Cimatti},
  {Dickinson}, {Papovich}, {Pasquali}, \& {Pirzkal}}]{maraston2006}
{Maraston}, C., {Daddi}, E., {Renzini}, A., {Cimatti}, A., {Dickinson}, M.,
  {Papovich}, C., {Pasquali}, A., \& {Pirzkal}, N. 2006, \apj, 652, 85

\bibitem[{{Marigo} \& {Aringer}(2009)}]{marigo2009}
{Marigo}, P., \& {Aringer}, B. 2009, \aap, 508, 1539

\bibitem[{{Marigo} \& {Girardi}(2007)}]{marigo2007}
{Marigo}, P., \& {Girardi}, L. 2007, \aap, 469, 239

\bibitem[{{Marigo} {et~al.}(2008){Marigo}, {Girardi}, {Bressan}, {Groenewegen},
  {Silva}, \& {Granato}}]{marigo2008}
{Marigo}, P., {Girardi}, L., {Bressan}, A., {Groenewegen}, M.~A.~T., {Silva},
  L., \& {Granato}, G.~L. 2008, \aap, 482, 883

\bibitem[{{McQuinn} {et~al.}(2011){McQuinn}, {Skillman}, {Dalcanton},
  {Dolphin}, {Holtzman}, {Weisz}, \& {Williams}}]{mcquinn2011}
{McQuinn}, K.~B.~W., {Skillman}, E.~D., {Dalcanton}, J., {Dolphin}, A.,
  {Holtzman}, J., {Weisz}, D., \& {Williams}, B.~F. 2011, \aj, submitted

\bibitem[{{Melbourne} {et~al.}(2010{\natexlab{a}}){Melbourne}, {Williams},
  {Dalcanton}, {Ammons}, {Max}, {Koo}, {Girardi}, \& {Dolphin}}]{melbourne10}
{Melbourne}, J., {Williams}, B., {Dalcanton}, J., {Ammons}, S.~M., {Max}, C.,
  {Koo}, D.~C., {Girardi}, L., \& {Dolphin}, A. 2010{\natexlab{a}}, \apj, 712,
  469

\bibitem[{{Melbourne} {et~al.}(2010{\natexlab{b}}){Melbourne}, {Williams},
  {Dalcanton}, {Ammons}, {Max}, {Koo}, {Girardi}, \& {Dolphin}}]{melbourne2010}
------. 2010{\natexlab{b}}, \apj, 712, 469

\bibitem[{{Melbourne} {et~al.}(2011){Melbourne}, {Williams}, {Dalcanton},
  {Rosenfield}, {Girardi}, {Marigo}, {Skillman}, {Olsen}, {Weisz}, \&
  {Dolphin}}]{melbourne2011}
{Melbourne}, J. {et~al.} 2011, \apj, submitted

\bibitem[{{M{\'e}ndez} {et~al.}(2002){M{\'e}ndez}, {Davis}, {Moustakas},
  {Newman}, {Madore}, \& {Freedman}}]{Mendez02}
{M{\'e}ndez}, B., {Davis}, M., {Moustakas}, J., {Newman}, J., {Madore}, B.~F.,
  \& {Freedman}, W.~L. 2002, \aj, 124, 213

\bibitem[{{Nikolaev} \& {Weinberg}(2000)}]{nikolaev2000}
{Nikolaev}, S., \& {Weinberg}, M.~D. 2000, \apj, 542, 804

\bibitem[{{Olsen} {et~al.}(2006){Olsen}, {Blum}, {Stephens}, {Davidge},
  {Massey}, {Strom}, \& {Rigaut}}]{olsen2006}
{Olsen}, K.~A.~G., {Blum}, R.~D., {Stephens}, A.~W., {Davidge}, T.~J.,
  {Massey}, P., {Strom}, S.~E., \& {Rigaut}, F. 2006, \aj, 132, 271

\bibitem[{{Rejkuba} {et~al.}(2006){Rejkuba}, {da Costa}, {Jerjen}, {Zoccali},
  \& {Binggeli}}]{rejkuba2006}
{Rejkuba}, M., {da Costa}, G.~S., {Jerjen}, H., {Zoccali}, M., \& {Binggeli},
  B. 2006, \aap, 448, 983

\bibitem[{{Sakai} {et~al.}(1996){Sakai}, {Madore}, \& {Freedman}}]{Sakai96}
{Sakai}, S., {Madore}, B.~F., \& {Freedman}, W.~L. 1996, \apj, 461, 713

\bibitem[{{Salaris}(2002)}]{salaris2002}
{Salaris}, M. 2002, in Astronomical Society of the Pacific Conference Series,
  Vol. 274, Observed HR Diagrams and Stellar Evolution, ed. {T.~Lejeune \&
  J.~Fernandes}, 50--+

\bibitem[{{Salaris} \& {Girardi}(2005)}]{salaris2005}
{Salaris}, M., \& {Girardi}, L. 2005, \mnras, 357, 669

\bibitem[{{Saracco} {et~al.}(2006){Saracco}, {Fiano}, {Chincarini}, {Vanzella},
  {Longhetti}, {Cristiani}, {Fontana}, {Giallongo}, \& {Nonino}}]{saracco06}
{Saracco}, P. {et~al.} 2006, \mnras, 367, 349

\bibitem[{{Schlegel} {et~al.}(1998){Schlegel}, {Finkbeiner}, \&
  {Davis}}]{schlegel98}
{Schlegel}, D.~J., {Finkbeiner}, D.~P., \& {Davis}, M. 1998, \apj, 500, 525

\bibitem[{{Seth} {et~al.}(2005){Seth}, {Dalcanton}, \& {de Jong}}]{seth05}
{Seth}, A.~C., {Dalcanton}, J.~J., \& {de Jong}, R.~S. 2005, \aj, 130, 1574

\bibitem[{{Stancliffe}(2006)}]{stancliffe2006}
{Stancliffe}, R.~J. 2006, \mnras, 370, 1817

\bibitem[{{Thronson} \& {Greenhouse}(1988)}]{thronson88}
{Thronson}, Jr., H.~A., \& {Greenhouse}, M.~A. 1988, \apj, 327, 671

\bibitem[{{Tully} {et~al.}(2006){Tully}, {Rizzi}, {Dolphin}, {Karachentsev},
  {Karachentseva}, {Makarov}, {Makarova}, {Sakai}, \& {Shaya}}]{tully06}
{Tully}, R.~B. {et~al.} 2006, \aj, 132, 729, astro-ph/0603380

\bibitem[{{Valdes} {et~al.}(1995){Valdes}, {Campusano}, {Velasquez}, \&
  {Stetson}}]{valdes1995}
{Valdes}, F.~G., {Campusano}, L.~E., {Velasquez}, J.~D., \& {Stetson}, P.~B.
  1995, \pasp, 107, 1119

\bibitem[{{Valenti} {et~al.}(2004){Valenti}, {Ferraro}, \&
  {Origlia}}]{valenti2004}
{Valenti}, E., {Ferraro}, F.~R., \& {Origlia}, L. 2004, \mnras, 354, 815

\bibitem[{{Valenti} {et~al.}(2007){Valenti}, {Ferraro}, \&
  {Origlia}}]{valenti2007}
------. 2007, \aj, 133, 1287

\bibitem[{{van Dokkum} {et~al.}(2008){van Dokkum}, {Franx}, {Kriek}, {Holden},
  {Illingworth}, {Magee}, {Bouwens}, {Marchesini}, {Quadri}, {Rudnick},
  {Taylor}, \& {Toft}}]{vandokkum2008}
{van Dokkum}, P.~G. {et~al.} 2008, \apjl, 677, L5

\bibitem[{{Weisz} {et~al.}(2011){Weisz}, {Dalcanton}, {Williams}, {Gilbert},
  {Skillman}, {Seth}, {Dolphin}, {McQuinn}, {Gogarten}, {Holtzman}, {Rosema},
  {Cole}, {Karachentsev}, \& {Zaritsky}}]{weisz2011}
{Weisz}, D.~R. {et~al.} 2011, ArXiv e-prints, 1101.1093

\bibitem[{{Williams} {et~al.}(2011){Williams}, {Dalcanton}, {Johnson}, {Weisz},
  {Seth}, {Dolphin}, {Gilbert}, {Skillman}, {Rosema}, {Gogarten}, {Holtzman},
  \& {de Jong}}]{williams2011}
{Williams}, B.~F. {et~al.} 2011, \apj , submitted

\end{thebibliography}


\begin{deluxetable}{llccccccccccl}
\tabletypesize{\scriptsize}
\tablecaption{Sample Galaxies}
\tablehead{
    \colhead{Galaxy} &
    \colhead{Alt.} &
    \colhead{RA} &
    \colhead{Dec} &
    \colhead{Diam.} &
    \colhead{$B_{\rm{T}}$} &
    \colhead{$A_{\rm{V}}$} &
    \colhead{$m-M$} &
    \colhead{T} &
    \colhead{$W_{50}$} &
    \colhead{Group} &
    \colhead{} &
    \colhead{} \\
    \colhead{}&
    \colhead{Names} &
    \colhead{(J2000)} &
    \colhead{(J2000)} &
    \colhead{($\prime$)} &
    \colhead{} &
    \colhead{} &
    \colhead{} &
    \colhead{} &
    \colhead{$\kms$} &
    \colhead{} &
    \colhead{} &
    \colhead{} 
}
\startdata                                                              
DDO53	    & U4459        &  08:34:06.5 &  66:10:45   &  1.6  & 14.55  & 0.118  & 27.79 &   10 &  25  &    M81 &   &   \\
DDO78	    &              &  10:26:27.9 &  67:39:24   &  2.0  & 15.8   & 0.066  & 28.18 &   -3 &      &    M81 &   &   \\
DDO82	    & U5692        &  10:30:35.0 &  70:37:10   &  3.4  & 13.57  & 0.133  & 27.90 &   9  &      &    M81 &   &   \\
HoI	    & U5139,DDO63  &  09:40:28.2 &  71:11:11   &  3.6  & 13.64  & 0.153  & 27.95 &   10 &  29  &    M81 &   &   \\
HoII	    & U4305        &  08:19:05.9 &  70:42:51   &  7.9  & 11.09  & 0.098  & 27.65 &   10 &  66  &    M81 &   &   \\
HS117	    &              &  10:21:25.2 &  71:06:58   &  1.5  & 16.5   & 0.359  & 27.91 &   10 &  13  &    M81 &   &   \\
I2574	    & U5666,DDO81  &  10:28:22.4 &  68:24:58   &  13.2 & 10.84  & 0.112  & 27.90 &   9  &  115 &    M81 &   &   \\
KDG2	    & E540-030,KK9 &  00:49:21.1 &  -18:04:28  &  1.2  & 16.37  & 0.072  & 27.61 &   -1 &      &    Scl &   &   \\
KDG63	    & U5428,DDO71  &  10:05:07.3 &  66:33:18   &  1.7  & 16.01  & 0.303  & 27.74 &   -3 &  19  &    M81 &   &   \\
KDG73	    &              &  10:52:55.3 &  69:32:45   &  0.6  & 17.09  & 0.056  & 28.03 &   10 &  18  &    M81 &   &   \\
KKH37	    &              &  06:47:45.8 &  80:07:26   &  1.2  & 16.4   & 0.231  & 27.56 &   10 &  20  &        &   &   \\
M81	    & N3031,U5318  &  09:55:33.5 &  69:04:00   &  26.9 &  7.69  & 0.249  & 27.77 &   3  &  422 &    M81 &   &   \\
N300	    &              &  00:54:53.5 &  -37:40:57  &  21.9 &  8.95  & 0.039  & 26.5  &   7  &  149 &  14+13 &   &   \\
N404	    & U718         &  01:09:26.9 &  35:43:03   &  2.5  & 11.21  & 0.181  & 27.42 &   -1 &  78  &        &   &   \\
N2403	    & U3918        &  07:36:54.4 &  65:35:58   &  21.9 &  8.82  & 0.124  & 27.5  &   6  &  231 &    M81 &   &   \\
N2976	    & U5221        &  09:47:15.6 &  67:54:49   &  5.9  & 11.01  & 0.224  & 27.76 &   5  &  97  &    M81 &   &   \\
N3077	    & U5398        &  10:03:21.0 &  68:44:02   &  5.4  & 10.46  & 0.208  & 27.92 &   10 &  65  &    M81 &   &   \\
N3741	    & U6572        &  11:36:06.4 &  45:17:07   &  2.0  & 14.38  & 0.077  & 27.55 &   10 &  81  &  14+07 &   &   \\
N4163       & U7199        &  12:12:08.9 &  36:10:10   &  1.9  & 13.63  & 0.062  & 27.29 &   10 &  18  &  14+07 &   &   \\
N7793       &              &  23:57:49.4 &  -32:35:24  &  9.3  &  9.70  & 0.060  & 27.96 &   7  &  174 &    Scl &   &   \\
Sc22	    & Sc-dE1       &  00:23:51.7 &  -24:42:18  &  0.9  & 17.73  & 0.046  & 28.11 &   10 &      &    Scl &   &   \\
U8508	    & IZw60        &  13:30:44.4 &  54:54:36   &  1.7  & 14.12  & 0.047  & 27.06 &   10 &  49  &  14+07 &   &   \\
UA292	    & CVnI-dwA     &  12:38:40.0 &  32:46:00   &  1.0  & 16.10  & 0.048  & 27.79 &   10 &  27  &        &   &   \\
\enddata
\tablecomments{Distances, $B_{\rm{T}}$, $W_{50}$, and T-type taken from CNG,
  with updates from \citep{dalcanton2009} and \citep{karachentsev03} for NGC~7793; Group membership from
  \citet{karachentsev05} or \citet{tully06}; $A_V$ from IRSA}
\label{sampletable}
\end{deluxetable}

%
\begin{deluxetable}{lllccccccc}
\tabletypesize{\scriptsize}
\tablecaption{Observations \& Photometry}
\tablehead{
   \colhead{Catalog} &
    \colhead{Target} &
   \colhead{Obs.Date} &
    \colhead{$N_{stars}$} &
    \colhead{$\Sigma_{max}$} &
    \colhead{$\Sigma_{min}$} &
   \multicolumn{2}{c}{50\% Completeness} &
   \colhead{Optical} &
    \colhead{Optical} \\
    \colhead{Name} &
    \colhead{Name} &
    \colhead{} &
    \colhead{} &
   \multicolumn{2}{c}{(\#/sqr-arcsec)} &
   \colhead{($F110W$)} &
   \colhead{($F160W$)} &
   \colhead{PropID} &
    \colhead{Filters} 
}
\startdata
DDO53  & UGC4459	  & 2010-04-23 11:15:02	&  9366  &   2.24 &   0.07 &  26.02 &  25.02& GO-10605  & F555W,F814W            \\
DDO78  & DDO78	          & 2010-04-20 14:41:53	& 11536  &   2.07 &   0.28 &  25.97 &  25.04& GO-10915  & F475W,F814W            \\
DDO82  & DDO82	          & 2010-05-07 06:56:06	& 25590  &   5.51 &   0.35 &  25.71 &  24.69& GO-10915  & F475W,F606W,F814W      \\
HoI    & UGC5139	  & 2009-08-21 22:55:14	& 13425  &   2.13 &   0.60 &  25.98 &  25.00& GO-10605  & F555W,F814W            \\
HoII   & UGC4305-1        & 2010-02-26 09:38:51	& 19328  &   3.53 &   0.39 &  25.72 &  24.76& GO-10605  & F555W,F814W            \\
HoII   & UGC4305-2        & 2010-01-04 07:01:20	& 21869  &   4.85 &   0.16 &  25.79 &  24.79& GO-10605  & F555W,F814W            \\
HS117  & HS117	          & 2010-02-24 02:05:52	&  3467  &   3.24 &   0.01 &  26.00 &  25.12& GO-9771   & F606W,F814W            \\
I2574  & IC2574-SGS       & 2010-02-25 03:03:05	& 27521  &   3.81 &   0.86 &  25.62 &  24.68& GO-9755   & F435W,F555W,F814W      \\
KDG2   & ESO540-030       & 2009-12-17 12:00:32	&  3890  &   2.43 &   0.02 &  26.12 &  25.01& GO-10503  & F606W,F814W            \\
KDG63  & DDO71	          & 2010-04-21 16:01:31	&  7316  &   2.78 &   0.02 &  25.98 &  25.03& GO-9884   & F606W,F814W            \\
KDG73  & KDG73	          & 2010-06-09 17:46:07	&  2140  &   0.98 &   0.01 &  26.13 &  25.16& GO-10915  & F475W,F814W            \\
KKH37  & KKH37	          & 2009-09-29 10:41:04	&  5097  &   3.77 &   0.01 &  26.05 &  25.02& GO-10915  & F475W,F814W            \\
       &                  &                     &   $-$  &   $-$  &   $-$  &  $-$   &  $-$  & GO-9771   & F606W,F814W            \\
M81    & M81-DEEP         & 2010-06-13 00:21:05	&  6159  &   1.43 &   0.13 &  26.10 &  25.12& GO-10915  & F475W,F606W,F814W      \\
N300   & NGC0300-WIDE1    & 2010-04-19 17:45:59	& 27898  &   3.66 &   1.29 &  25.36 &  24.47& GO-10915  & F475W,F606W,F814W      \\
N404   & NGC404	          & 2009-12-31 00:24:37	& 23159  &   4.00 &   0.34 &  25.71 &  24.73& GO-10915  & F606W,F814W (WFPC2)    \\
N2403  & NGC2403-DEEP     & 2010-02-28 19:16:27	&  9492  &   2.77 &   0.27 &  25.88 &  25.01& GO-10915  & F606W,F814W (WFPC2)    \\
N2403  & NGC2403-HALO-6   & 2010-04-25 04:26:22	&  5691  &   2.03 &   0.15 &  26.07 &  25.11& GO-10523  & F606W,F814W            \\
N2403  & SN-NGC2403-PR    & 2010-04-22 07:56:14	& 38792  &   5.77 &   1.73 &  23.40 &  22.35& GO-10182  & F475W,F606W,F814W      \\
N2976  & NGC2976-DEEP     & 2010-02-25 02:03:27	& 15392  &   4.05 &   0.25 &  25.84 &  24.92& GO-10915  & F475W,F606W,F814W      \\
N3077  & NGC3077-PHOENIX  & 2010-02-21 22:49:04	&  8813  &   3.14 &   0.20 &  26.00 &  25.14& GO-9381   & F435W,F555W,F814W      \\
N3741  & NGC3741	  & 2009-11-07 01:33:13	&  6819  &   4.44 &   0.00 &  25.98 &  24.86& GO-10915  & F475W,F814W            \\
N4163  & NGC4163	  & 2010-03-23 17:39:56	& 19105  &   4.77 &   0.07 &  25.90 &  24.71& GO-10915  & F475W,F606W,F814W      \\
N7793  & NGC7793-HALO-6   & 2010-06-14 19:11:39	&  6578  &   2.14 &   0.02 &  26.11 &  25.07& GO-10523  & F606W,F814W            \\
Sc22   & SCL-DE1	  & 2009-09-08 00:45:13	&  2440  &   3.17 &   0.02 &  26.19 &  25.09& GO-10503  & F606W,F814W            \\
U8508  & UGC8508	  & 2009-10-14 19:41:41	& 12664  &   4.38 &   0.03 &  25.97 &  24.83& GO-10915  & F475W,F814W            \\
UA292  & UGCA292	  & 2010-05-18 12:38:27	&  2373  &   2.53 &   0.01 &  26.19 &  25.14& GO-10915  & F475W,F814W            \\
\enddata
\tablecomments{All star counts and surface densities are based on the numbers of stars in the {\tt{*.gst}} catalogs; the {\tt{*.st}} catalogs
typically contain twice as many stars, though with larger photometric uncertainties.  The maximum and minimum stellar surface densities were calculated
by calculating the density of sources in a 10$\times$10 grid on the image.}
\label{obstable}
\end{deluxetable}

%

%
\begin{deluxetable}{llccccccc}
\tabletypesize{\scriptsize}
\tablecaption{TRGB Measurements}
\tablehead{
    \colhead{Catalog} &
    \colhead{Target} &
    \colhead{$m-M$} &
    \colhead{$A_V$} &
    \colhead{$N_{stars}$} &
    \colhead{Mean} &
    \colhead{$m_{TRGB}$} &
    \colhead{$m_{TRGB}$} &
    \colhead{$M_{TRGB}$} \\
    \colhead{Name} &
    \colhead{Name} &
    \colhead{($F814W$)} &
    \colhead{} &
    \colhead{(TRGB fit)} &
    \colhead{Color} &
    \colhead{(raw)} &
    \colhead{($F160W$)} &
    \colhead{($F160W$)} 
}
\startdata
    KDG63 &                 DDO71 & 27.74 & 0.303 &   1024 & 0.853 & 22.142 & $22.080\pm0.064$  & $-5.660\pm0.064$ \\
    DDO78 &                 DDO78 & 27.82 & 0.066 &   1438 & 0.901 & 22.045 & $22.031\pm0.027$  & $-5.789\pm0.027$ \\
    DDO82 &                 DDO82 & 27.90 & 0.133 &   5268 & 0.909 & 22.089 & $22.062\pm0.028$  & $-5.838\pm0.028$ \\
     KDG2 &            ESO540-030 & 27.61 & 0.072 &    481 & 0.829 & 22.105 & $22.090\pm0.040$  & $-5.520\pm0.040$ \\
    HS117 &                 HS117 & 27.91 & 0.359 &    480 & 0.836 & 22.340 & $22.266\pm0.025$  & $-5.644\pm0.025$ \\
    I2574 &            IC2574-SGS & 27.90 & 0.112 &   3967 & 0.859 & 22.232 & $22.209\pm0.019$  & $-5.691\pm0.019$ \\
    KDG73 &                 KDG73 & 28.03 & 0.056 &    273 & 0.771 & 22.464 & $22.452\pm0.045$  & $-5.578\pm0.045$ \\
    KKH37 &                 KKH37 & 27.56 & 0.231 &    659 & 0.848 & 21.957 & $21.910\pm0.026$  & $-5.650\pm0.026$ \\
      M81 &              M81-DEEP & 27.77 & 0.249 &    594 & 0.986 & 21.744 & $21.693\pm0.043$  & $-6.077\pm0.043$ \\
     N300 &         NGC0300-WIDE1 & 26.50 & 0.039 &   1410 & 0.972 & 20.561 & $20.553\pm0.026$  & $-5.947\pm0.026$ \\
    N2403 &          NGC2403-DEEP & 27.50 & 0.124 &    751 & 0.927 & 21.701 & $21.675\pm0.036$  & $-5.825\pm0.036$ \\
    N2403 &        NGC2403-HALO-6 & 27.50 & 0.124 &    356 & 0.901 & 21.635 & $21.609\pm0.016$  & $-5.891\pm0.016$ \\
    N2976 &          NGC2976-DEEP & 27.76 & 0.224 &   2495 & 0.952 & 21.851 & $21.805\pm0.034$  & $-5.955\pm0.034$ \\
    N3077 &       NGC3077-PHOENIX & 27.92 & 0.208 &    945 & 1.000 & 21.931 & $21.888\pm0.025$  & $-6.032\pm0.025$ \\
    N3741 &               NGC3741 & 27.55 & 0.077 &    801 & 0.814 & 21.974 & $21.958\pm0.023$  & $-5.592\pm0.023$ \\
     N404 &                NGC404 & 27.42 & 0.181 &   2691 & 0.967 & 21.498 & $21.461\pm0.021$  & $-5.959\pm0.021$ \\
    N4163 &               NGC4163 & 27.29 & 0.062 &   2594 & 0.875 & 21.571 & $21.558\pm0.014$  & $-5.732\pm0.014$ \\
    N7793 &        NGC7793-HALO-6 & 27.96 & 0.060 &    648 & 0.949 & 21.930 & $21.917\pm0.020$  & $-6.043\pm0.020$ \\
     Sc22 &               SCL-DE1 & 28.11 & 0.046 &    377 & 0.793 & 22.570 & $22.560\pm0.037$  & $-5.550\pm0.037$ \\
     HoII &             UGC4305-1 & 27.65 & 0.098 &   1845 & 0.850 & 21.937 & $21.917\pm0.071$  & $-5.733\pm0.071$ \\
     HoII &             UGC4305-2 & 27.65 & 0.098 &   2258 & 0.835 & 21.947 & $21.927\pm0.025$  & $-5.723\pm0.025$ \\
    DDO53 &               UGC4459 & 27.79 & 0.118 &    932 & 0.805 & 22.142 & $22.118\pm0.031$  & $-5.672\pm0.031$ \\
      HoI &               UGC5139 & 27.95 & 0.153 &   1742 & 0.826 & 22.351 & $22.320\pm0.029$  & $-5.630\pm0.029$ \\
    U8508 &               UGC8508 & 27.06 & 0.047 &   1110 & 0.847 & 21.412 & $21.402\pm0.024$  & $-5.658\pm0.024$ \\
    UA292 &               UGCA292 & 27.79 & 0.048 &    184 & 0.722 & 22.459 & $22.450\pm0.022$  & $-5.340\pm0.022$ \\
\enddata
\tablecomments{Absolute magnitudes are determined using the distance
  modulus from Table~\ref{sampletable}, which were originally
  determined from the $F814W$ TRGB.  Mean colors are for the stars
  used to measure the TRGB, which are not necessarily all RGB stars,
  and include only stars within 0.05$^m$ of the TRGB.  $A_{\rm{V}}$
  values are as reported by IRSA for coordinates in Table 1.  
  Extinction corrections from $A_{\rm{V}}$ to the observed filters are
  adopted from models of \citet{girardi2008}, as described in the
  text. Listed uncertainties are dominated by photometric
  uncertainties and by stochasticity in the number of stars near the
  tip; systematic uncertainties (due to uncertainties in the assumed
  TRGB absolute magnitudes and extinction) are likely to be much
  larger, but are not included in the listed uncertainties.
  SN-NGC2403-PR is not included due to large crowding errors.
  }
\label{trgbtable}
\end{deluxetable}

\clearpage

\vfill
\clearpage
 
\begin{figure}
\figurenum{\ref{overlayfig} continued}
\centerline{
\includegraphics[width=2.25in]{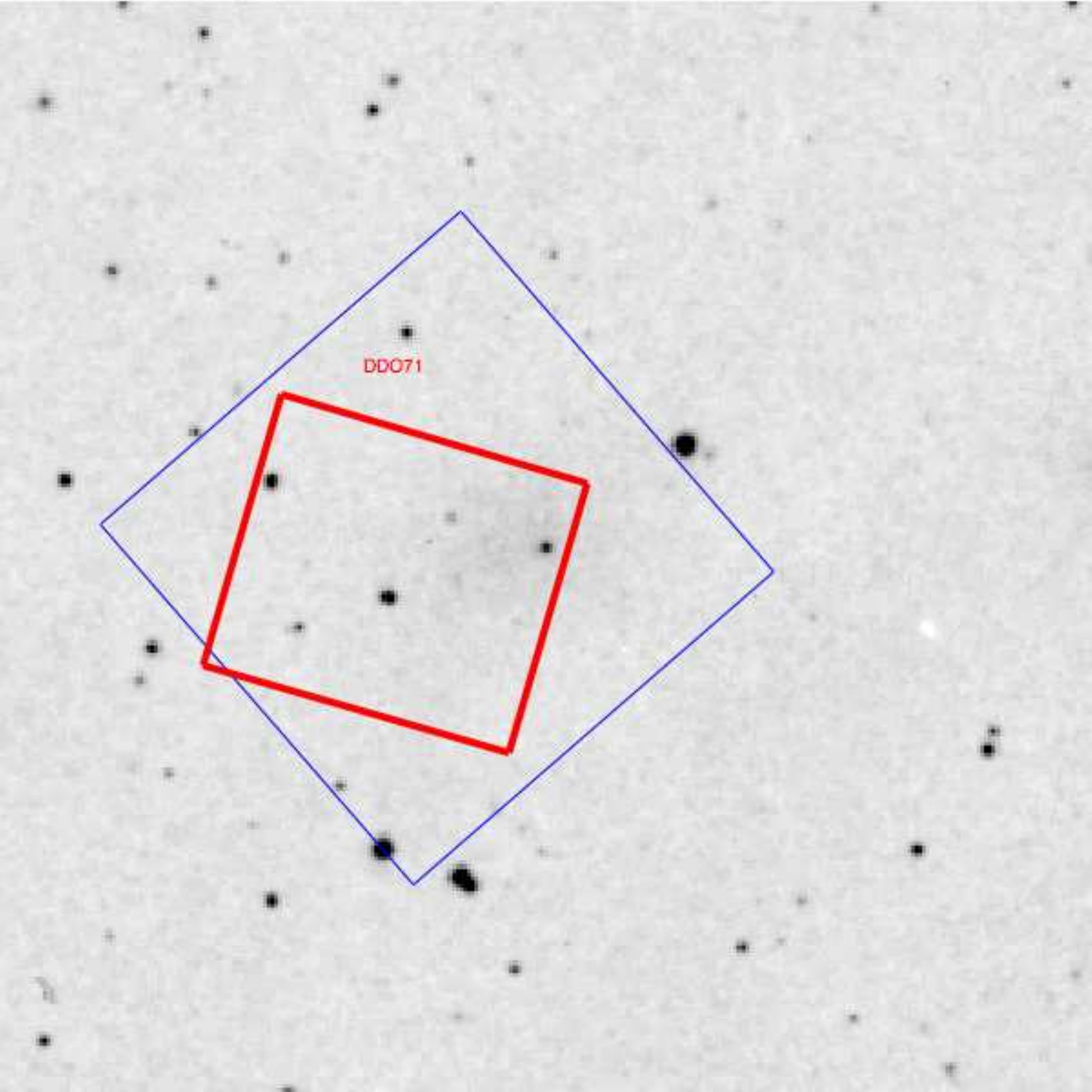}  
\includegraphics[width=2.25in]{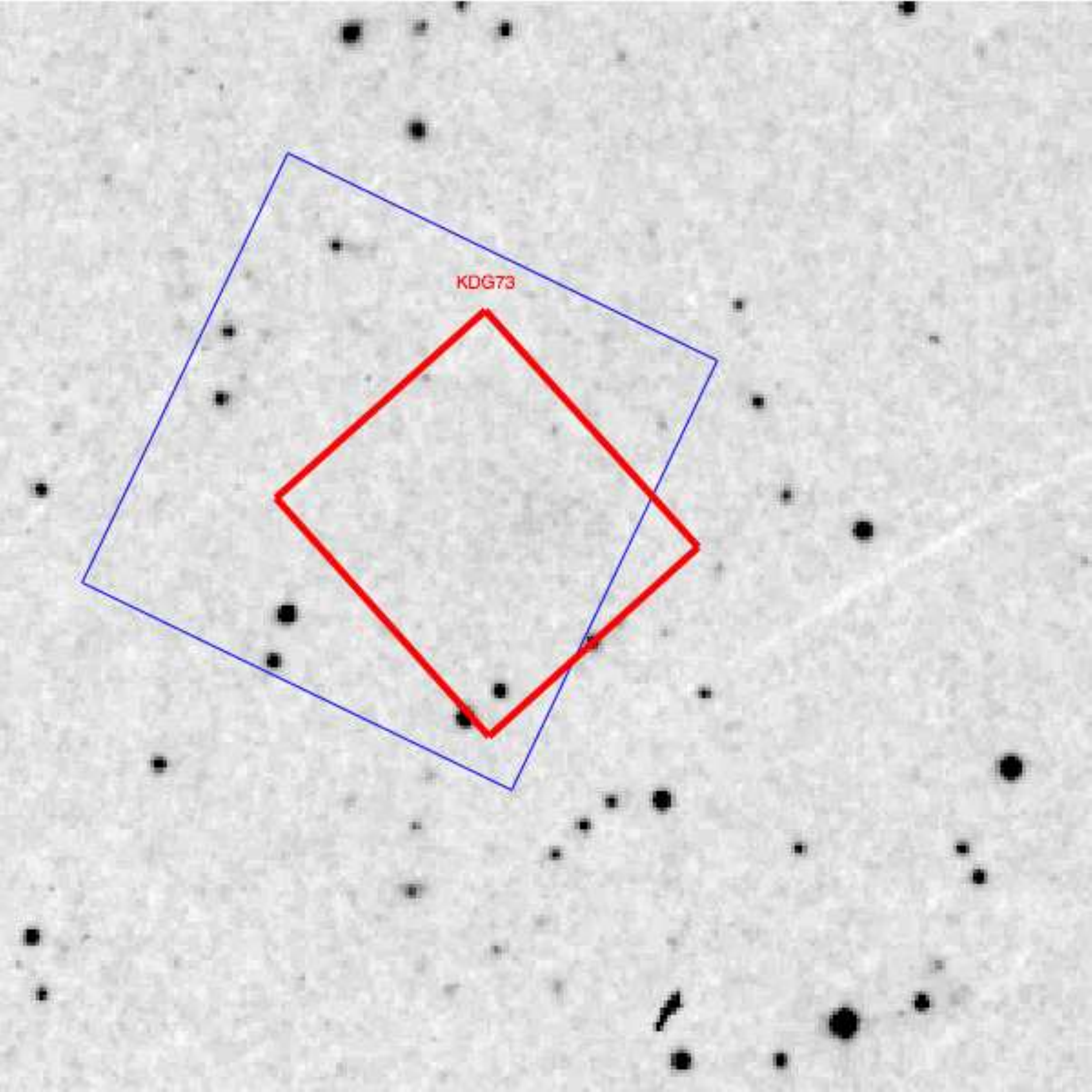}  
\includegraphics[width=2.25in]{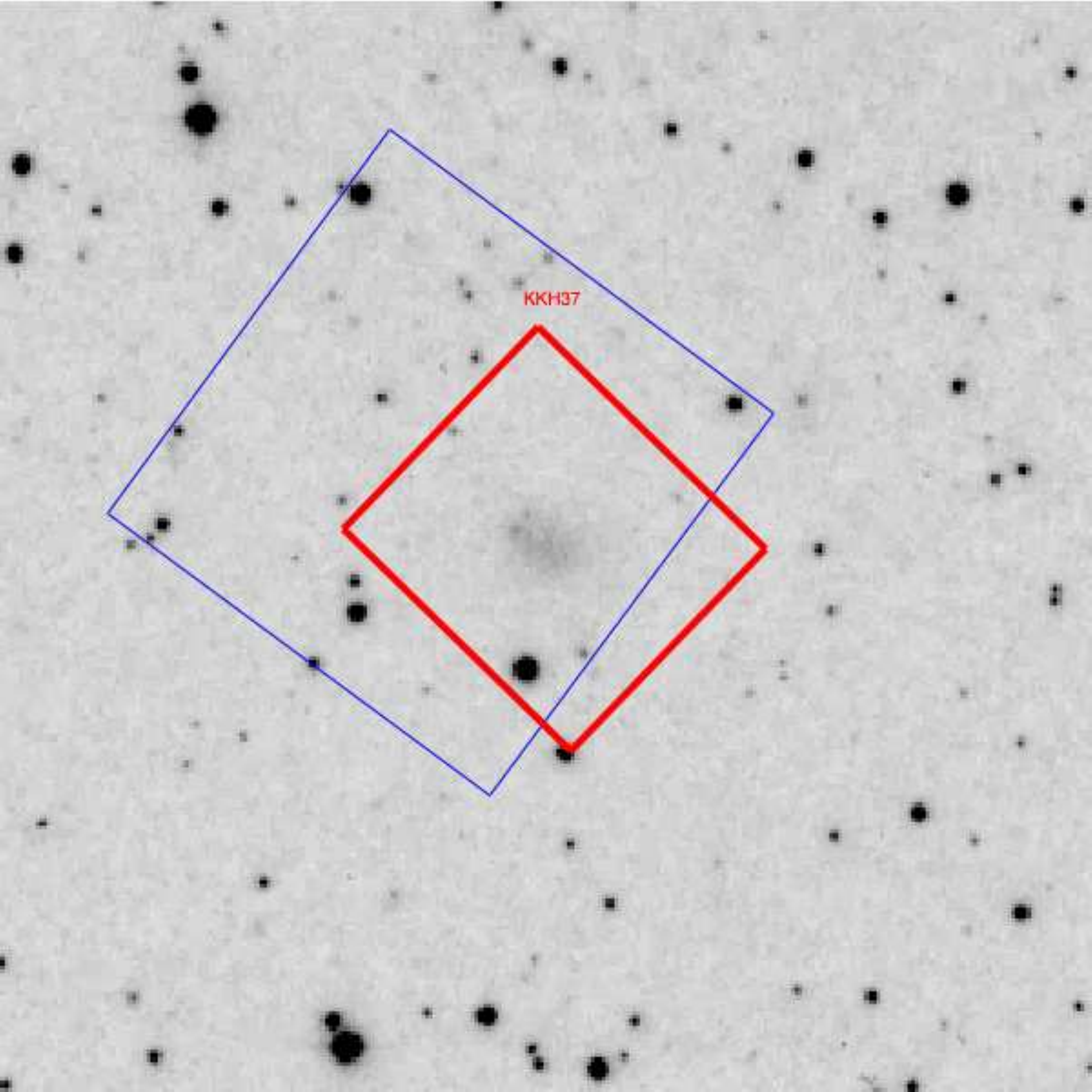}  
}
\centerline{
\includegraphics[width=2.25in]{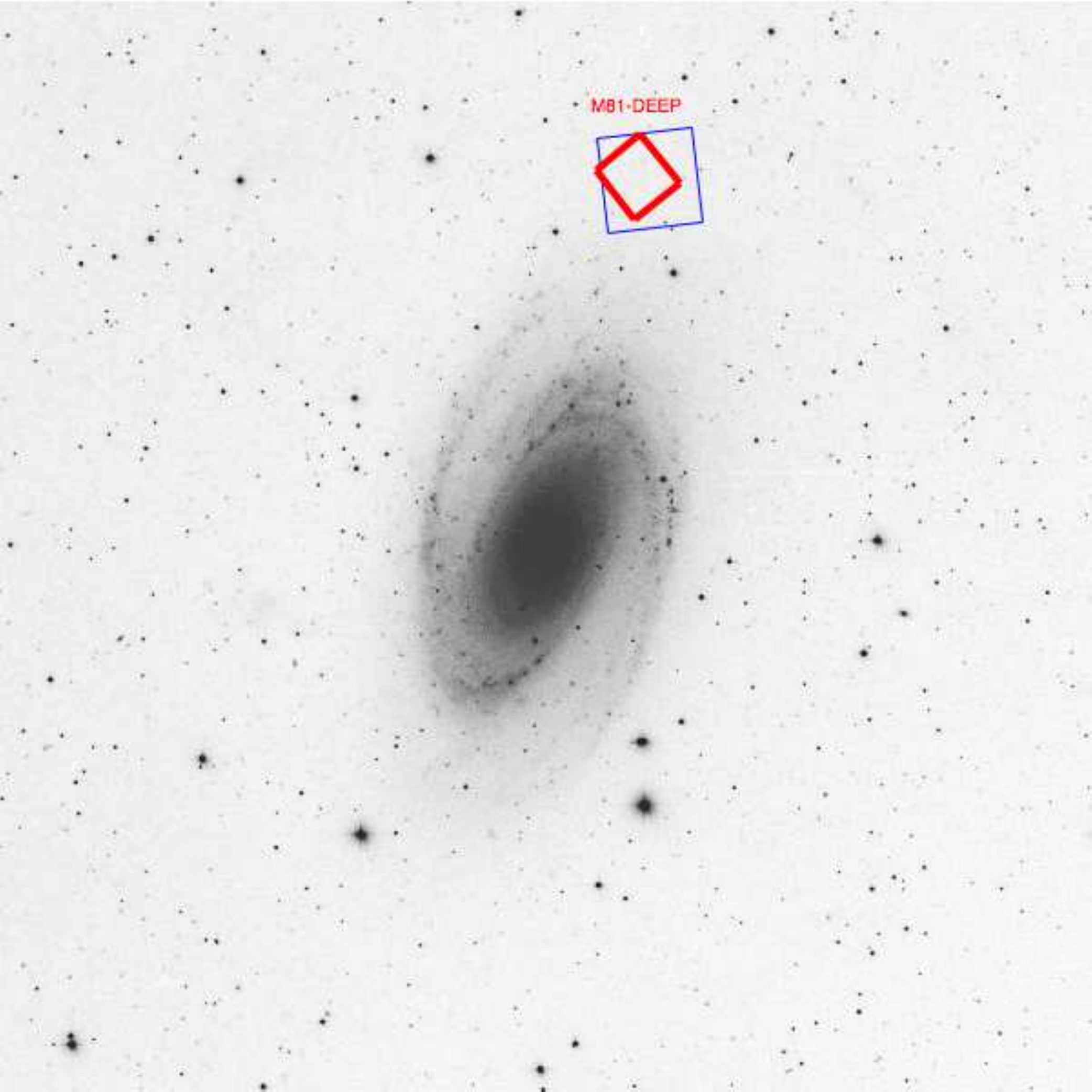}  
\includegraphics[width=2.25in]{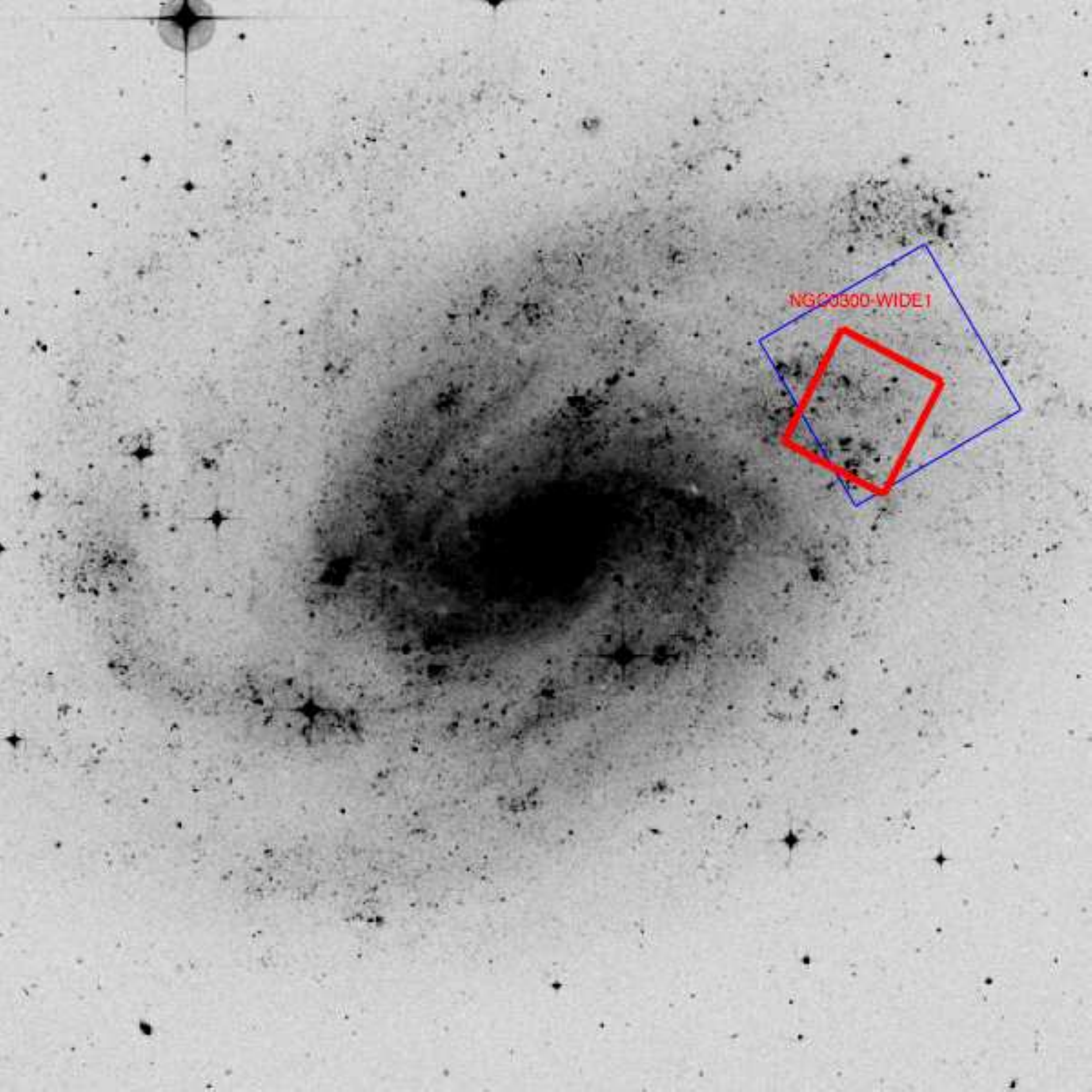}  
\includegraphics[width=2.25in]{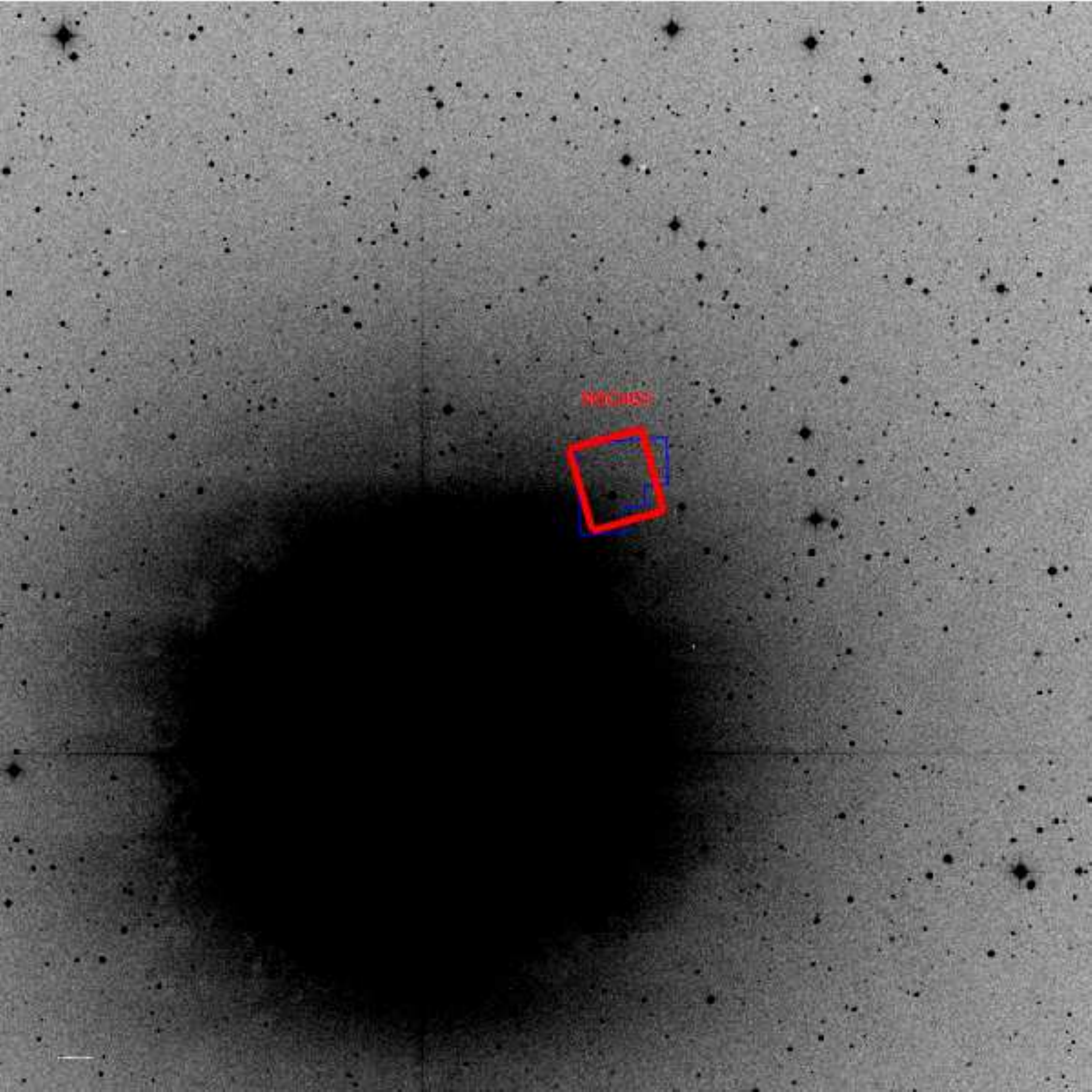}  
}
\centerline{
\includegraphics[width=2.25in]{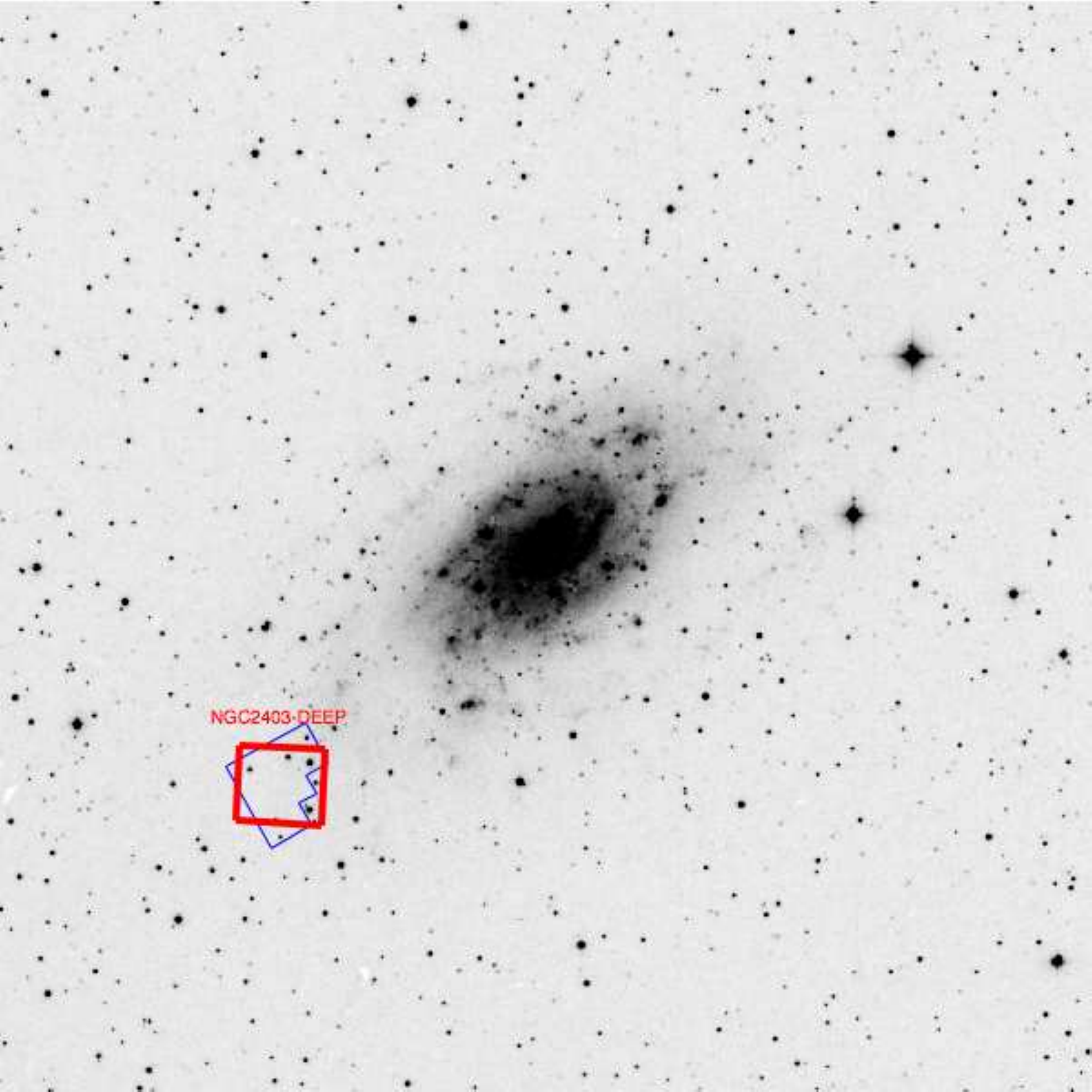}  
\includegraphics[width=2.25in]{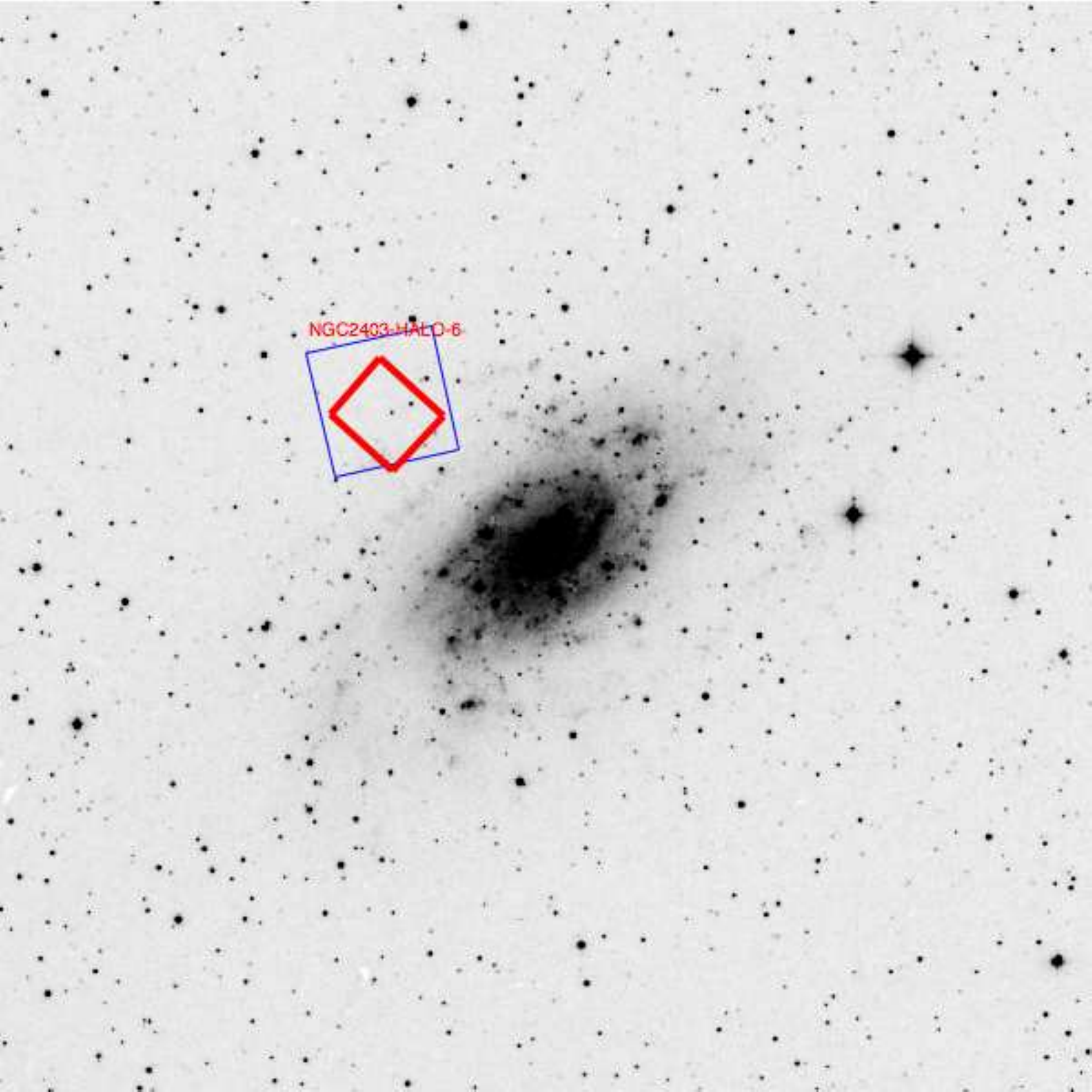}  
\includegraphics[width=2.25in]{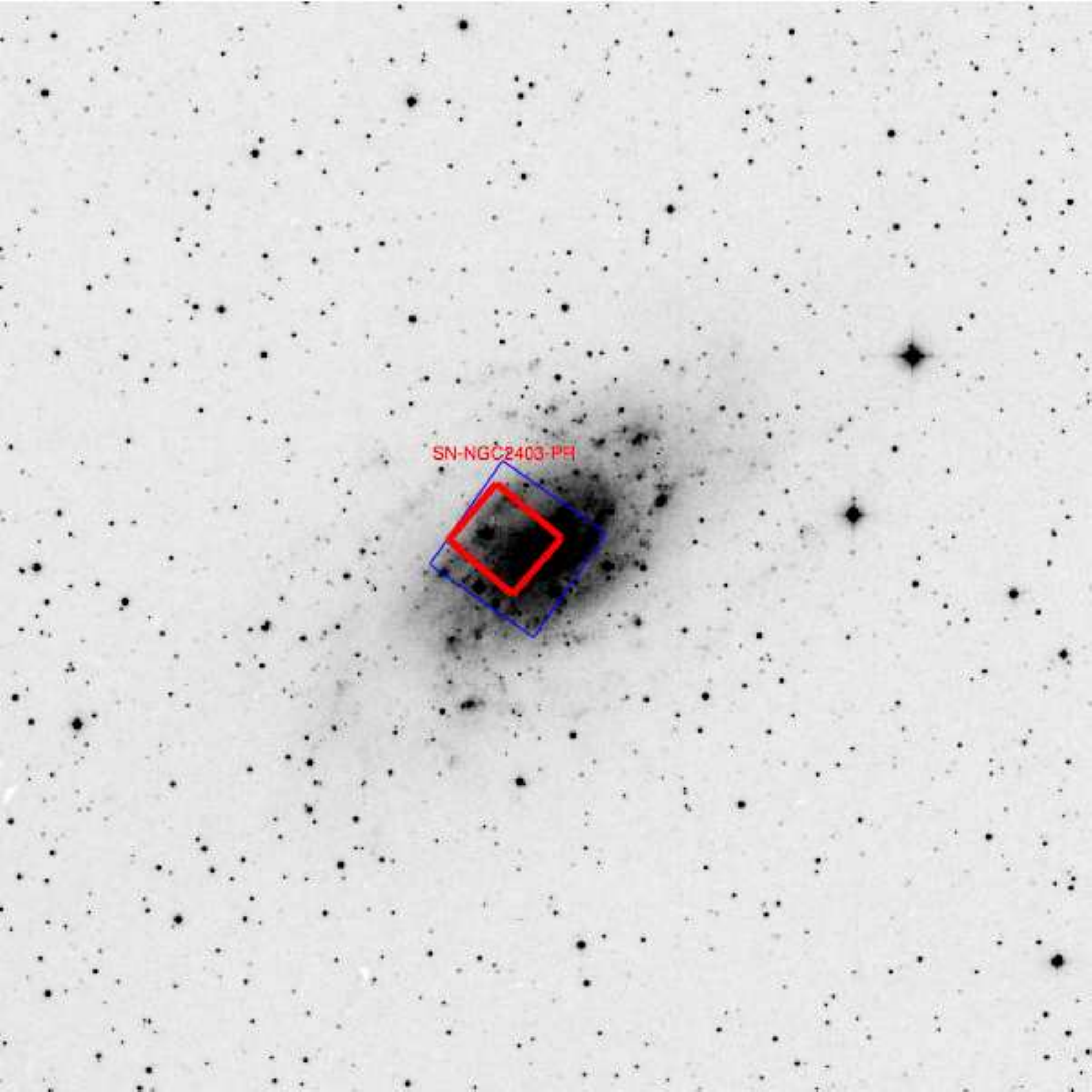}  
}
\caption{ Position of the WFC3/IR field of view (red), overlayed on an
  optical image from the Digitized Sky Survey.  Blue region shows the
  area covered by optical HST data.  (Target names from upper left to
  lower right: [j] DDO71; [k] KDG73; [l] KKH37; [m] M81-DEEP; [n]
  NGC0300-WIDE1; [o] NGC404; [p] NGC2403-DEEP; [q] NGC2403-HALO-6; [r]
  SN-NGC2403-PR;).  }
\end{figure}
\vfill
\clearpage
 
\begin{figure}
\figurenum{\ref{overlayfig} continued}
\centerline{
\includegraphics[width=2.25in]{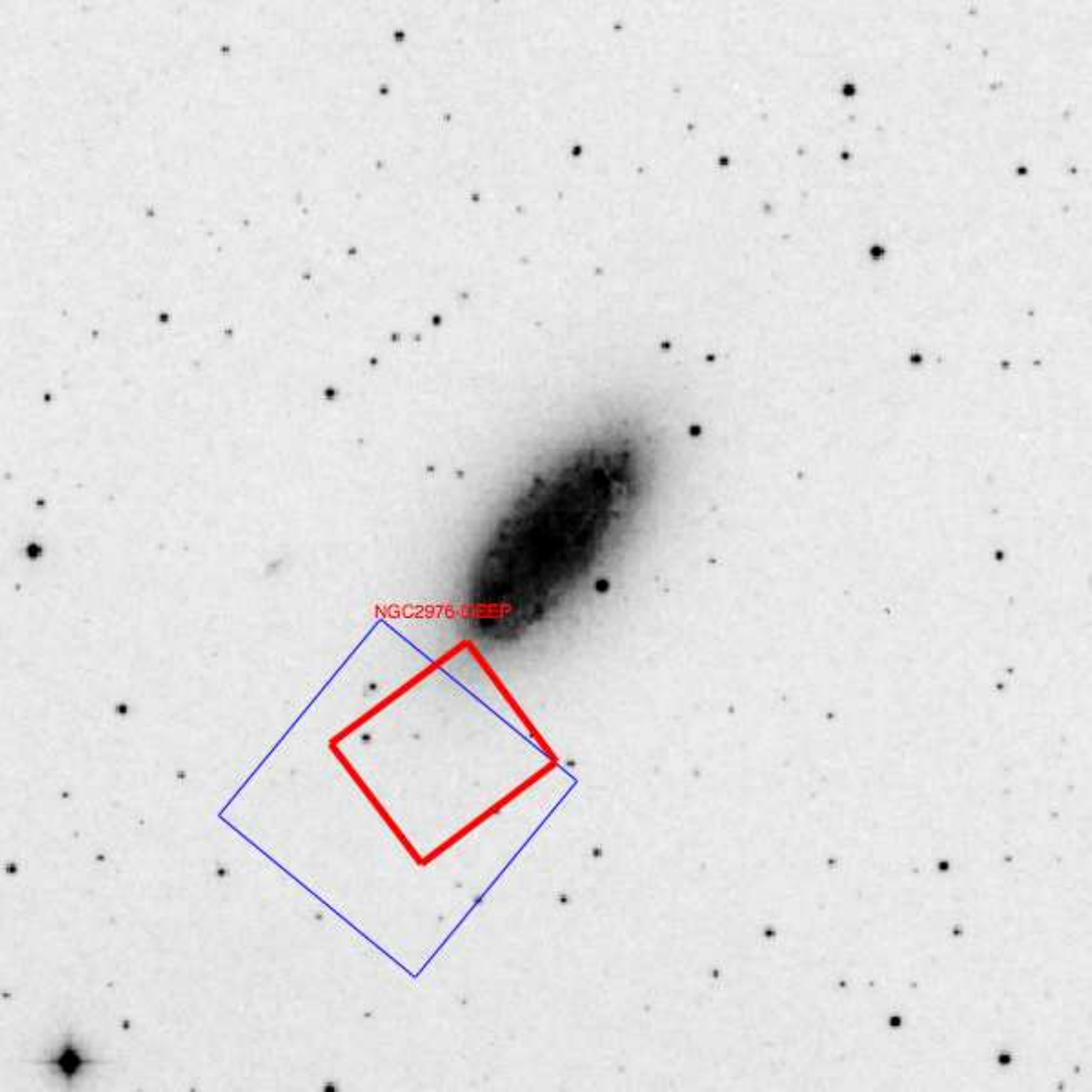}  
\includegraphics[width=2.25in]{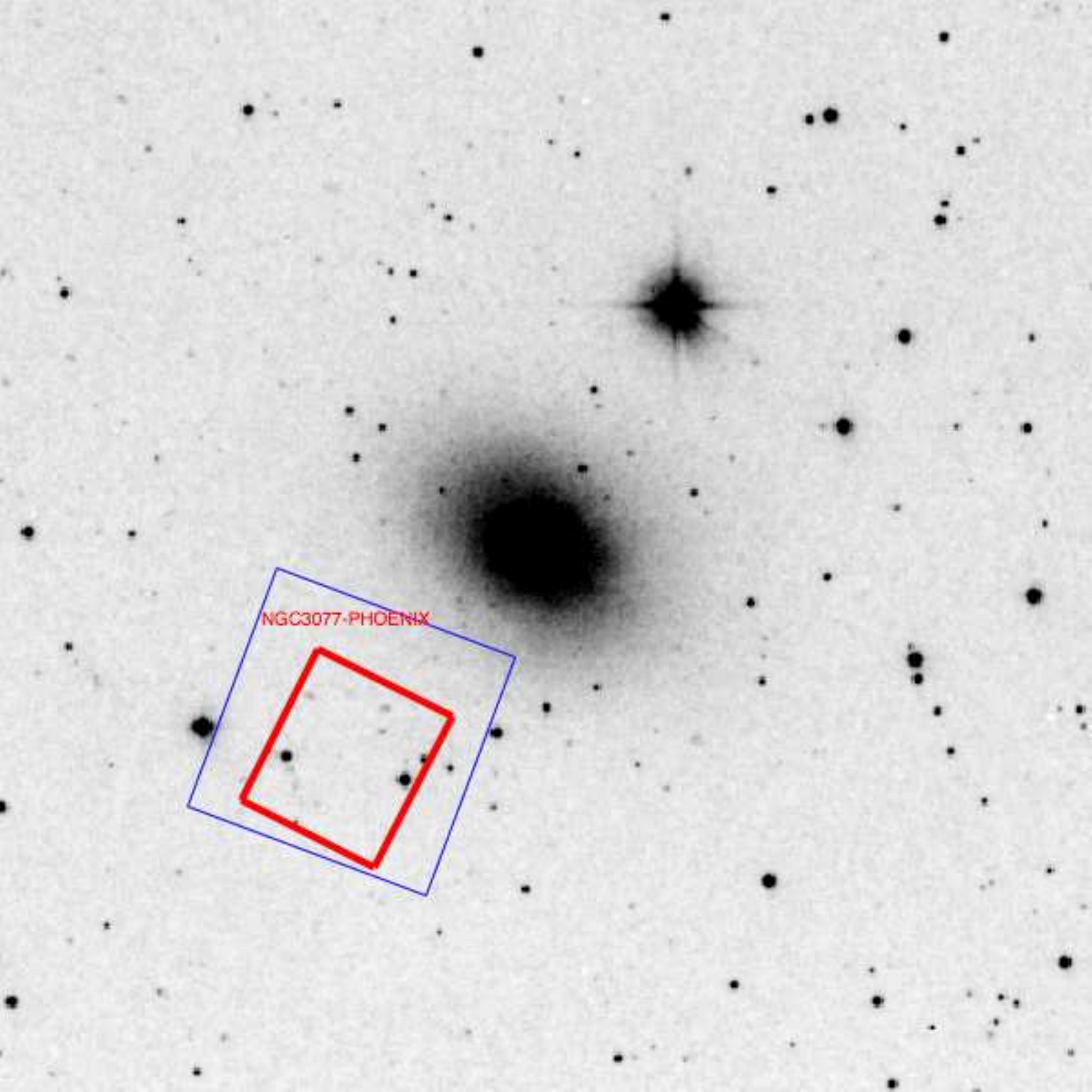}  
\includegraphics[width=2.25in]{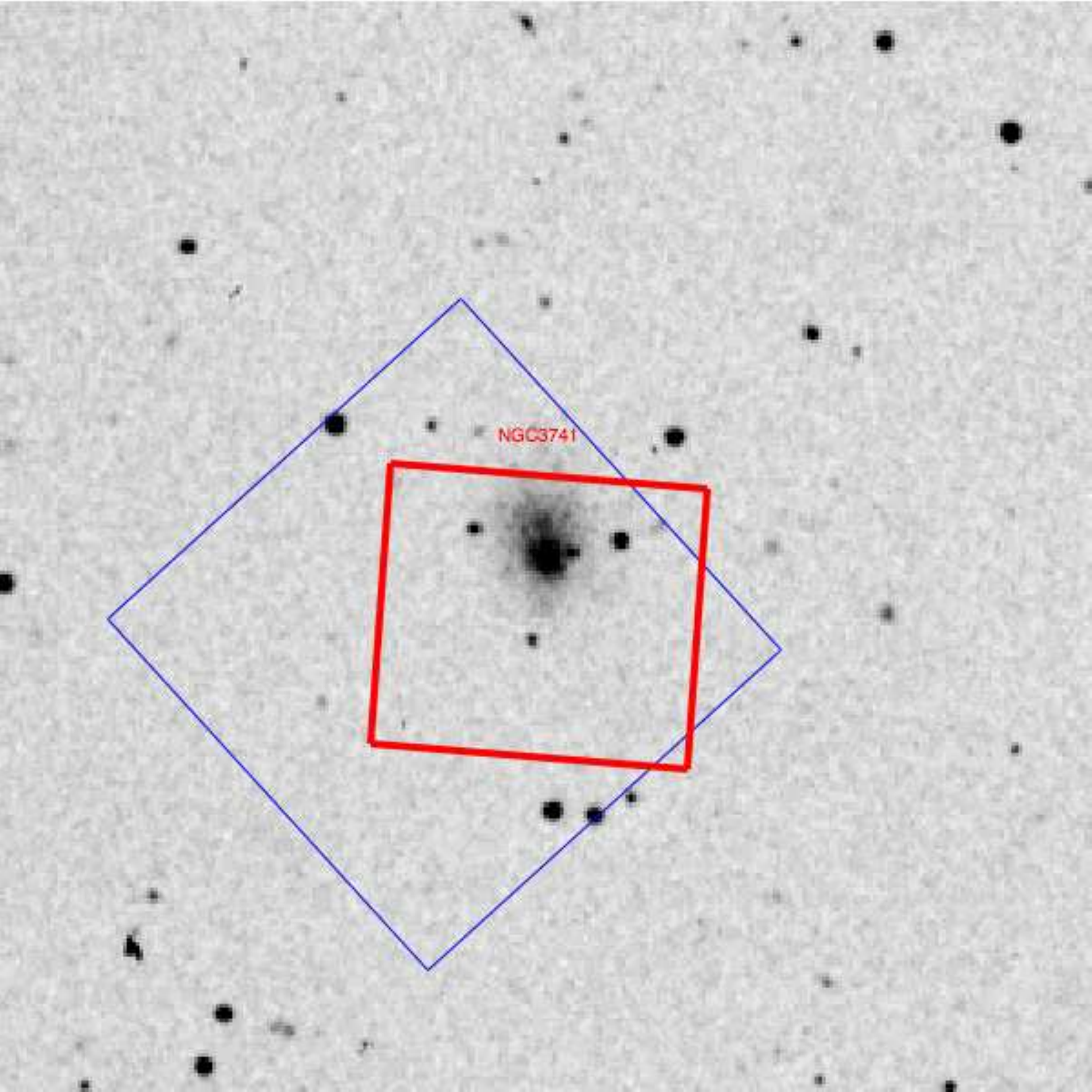}  
}
\centerline{
\includegraphics[width=2.25in]{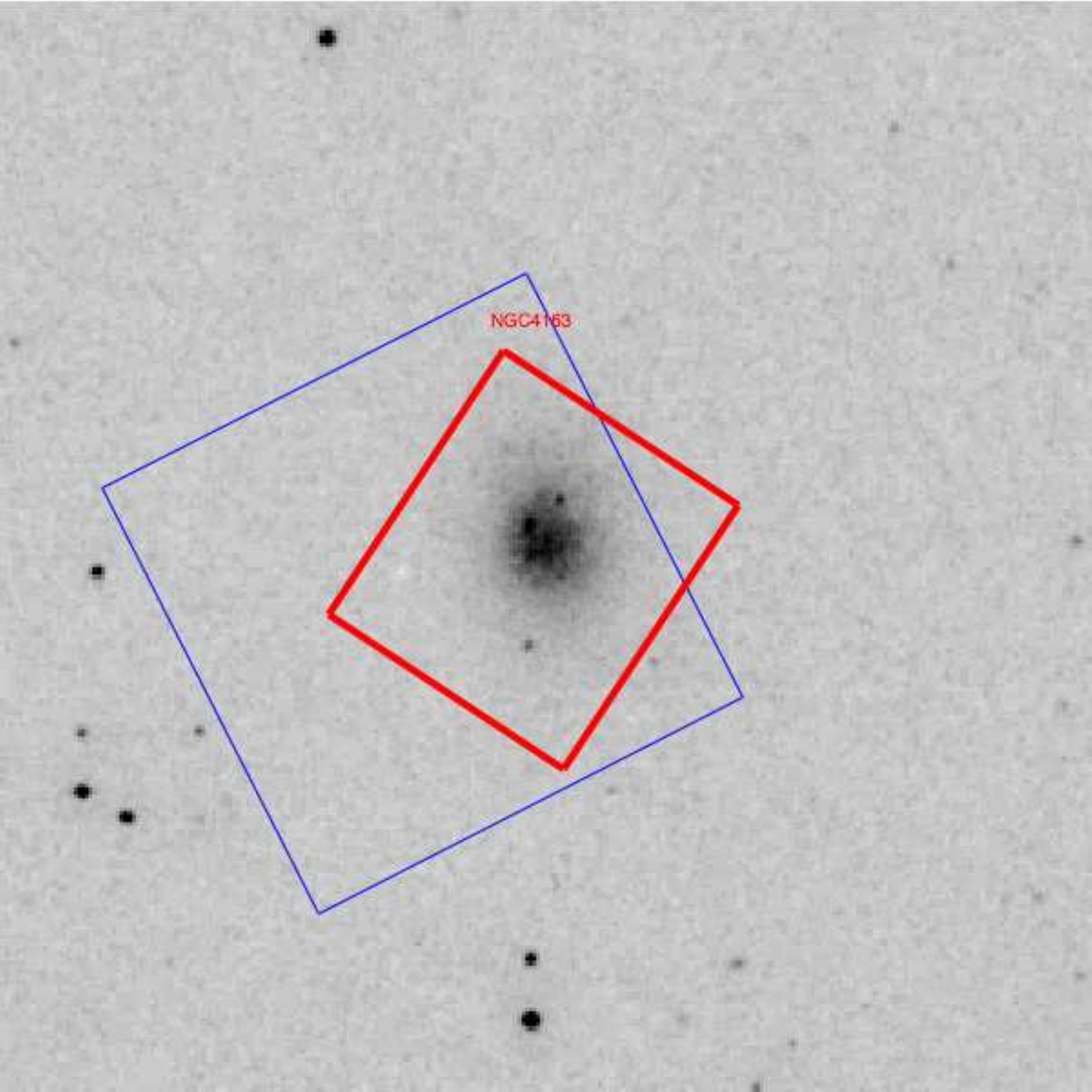}  
\includegraphics[width=2.25in]{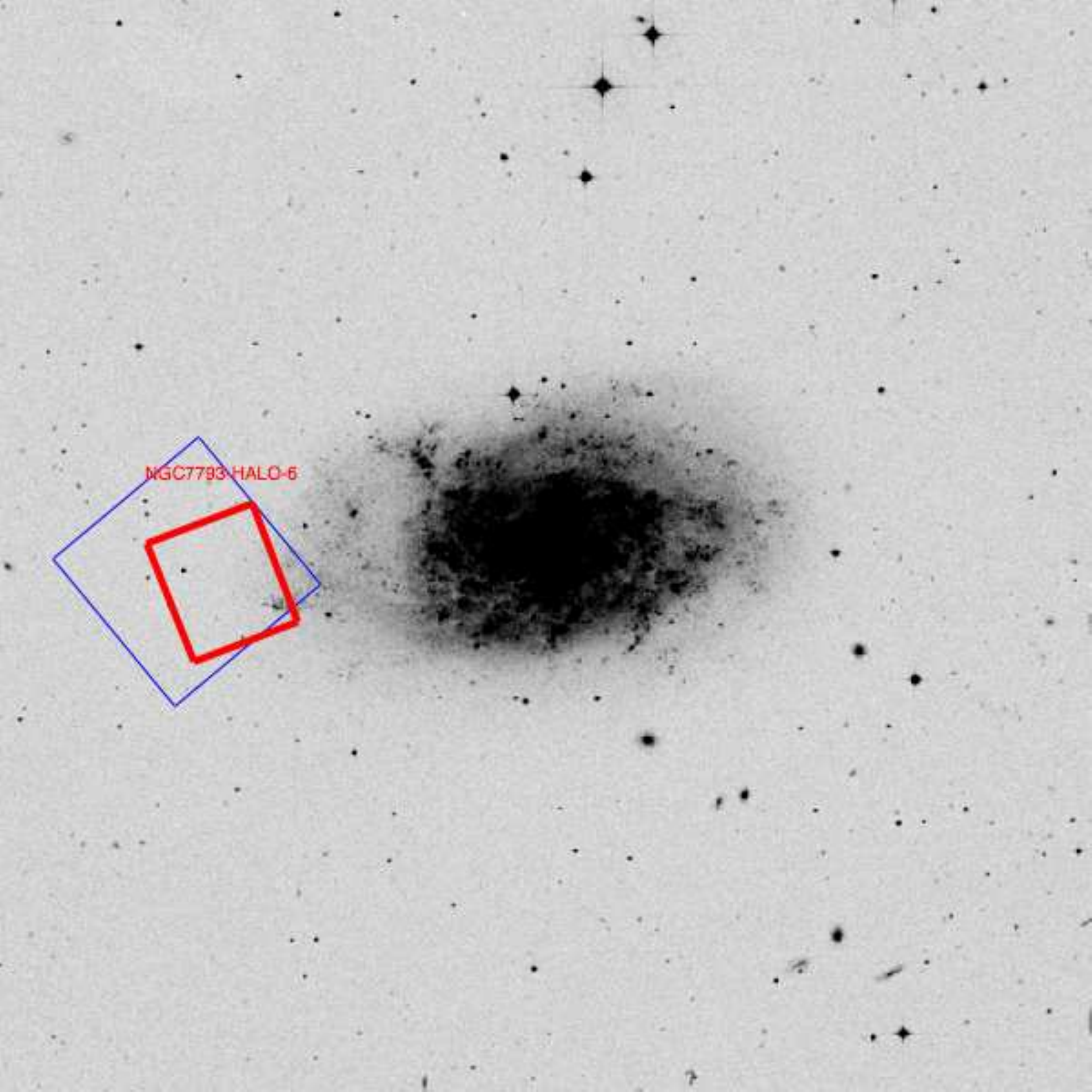}  
\includegraphics[width=2.25in]{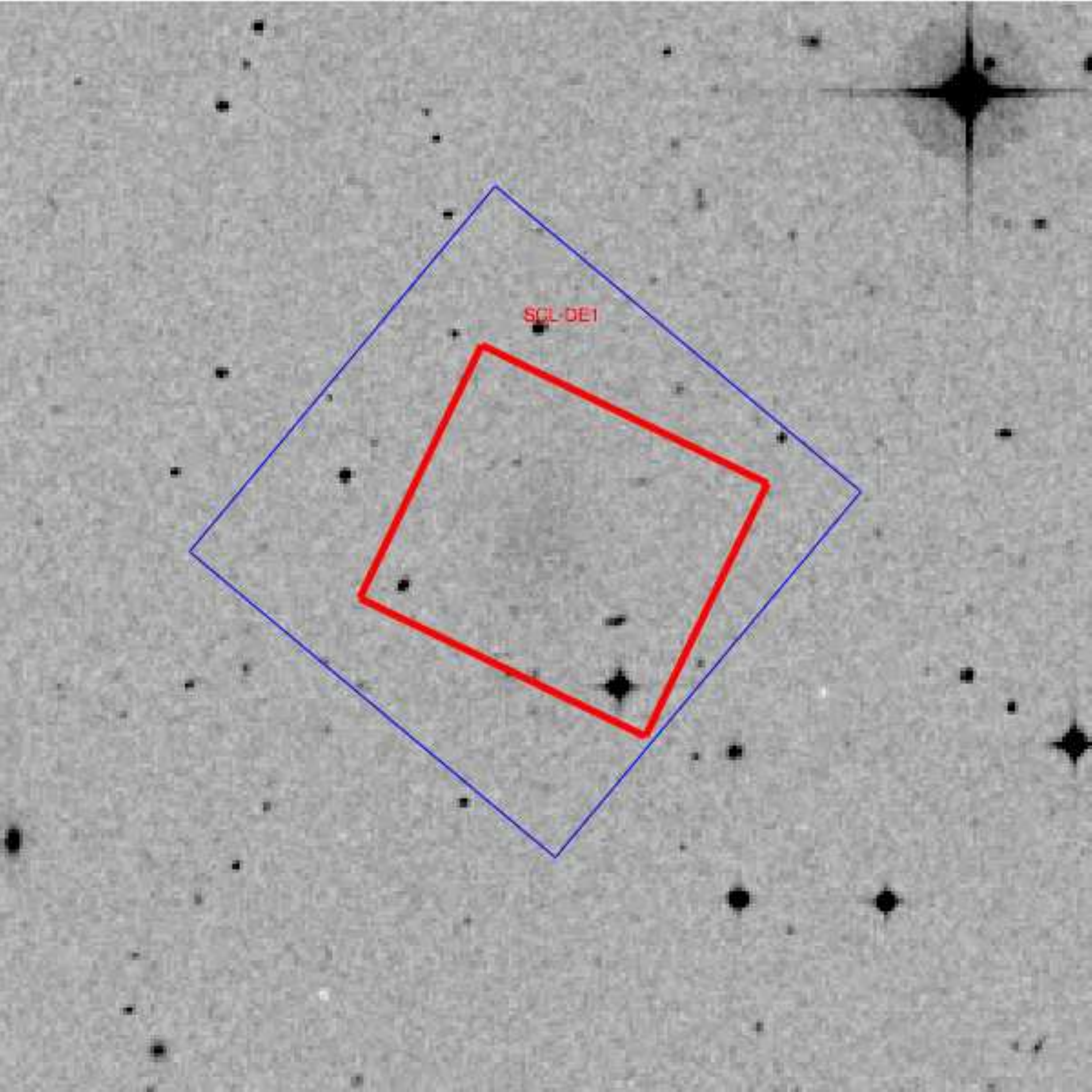}  
}
\centerline{
\includegraphics[width=2.25in]{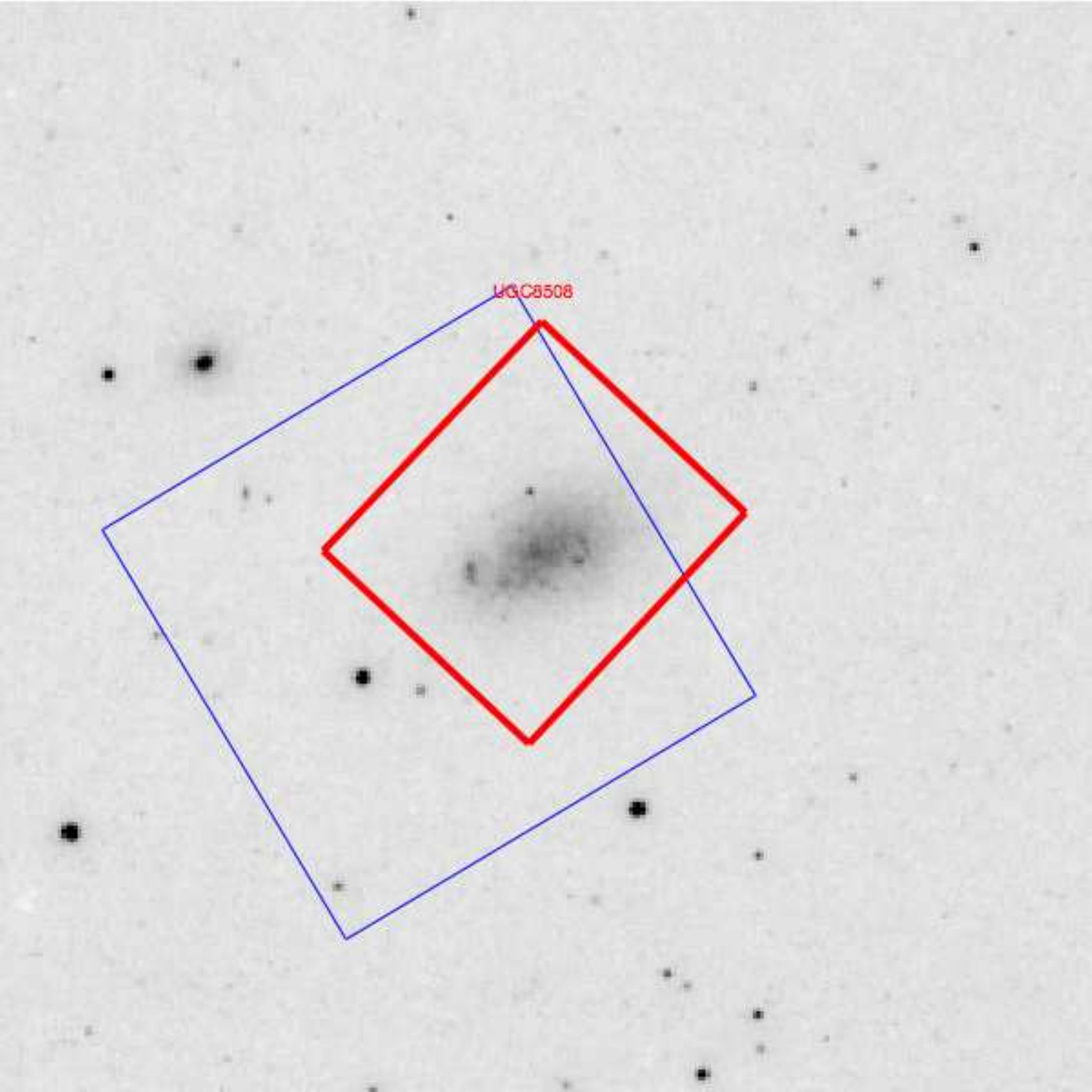}  
\includegraphics[width=2.25in]{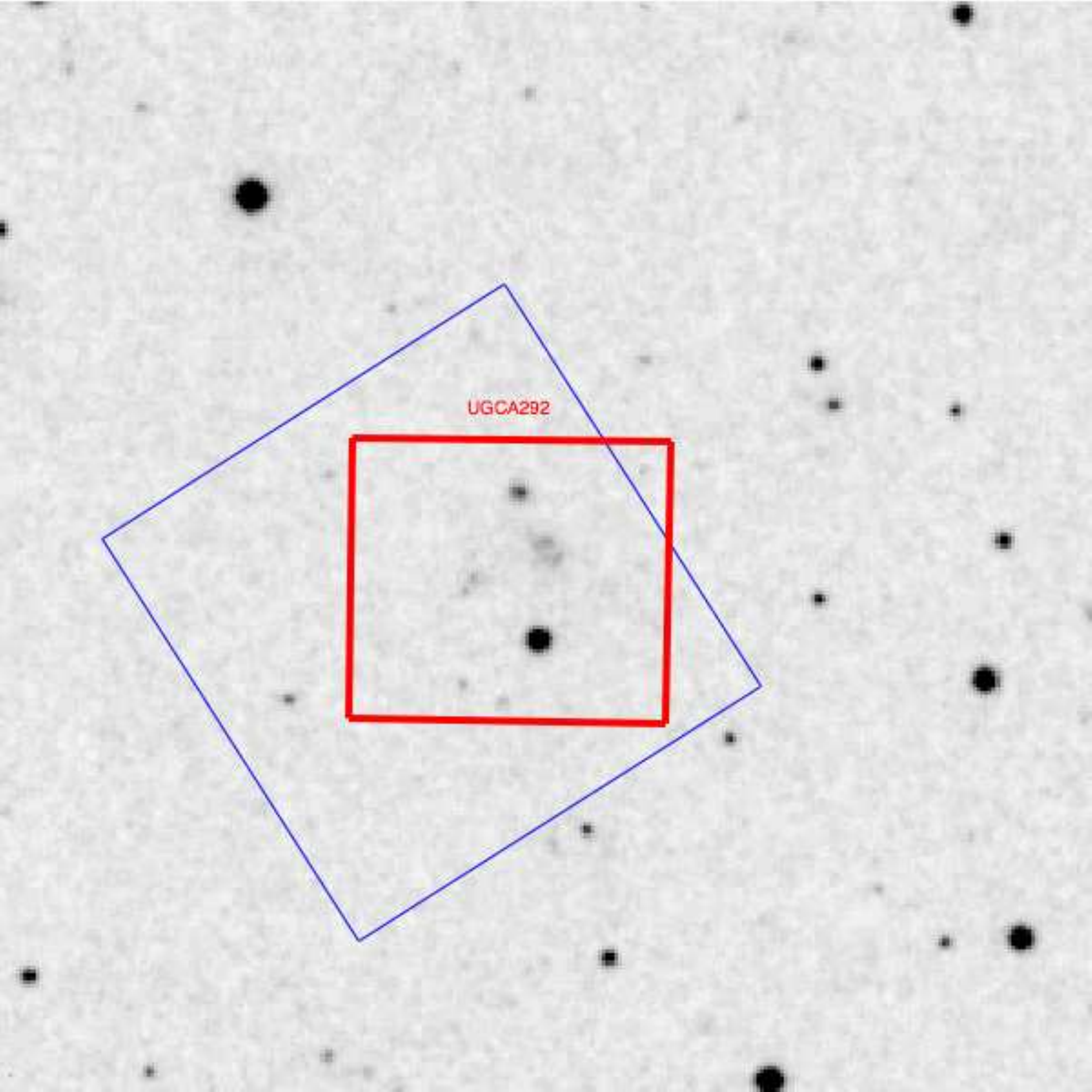}  
}
\caption{ Position of the WFC3/IR field of view (red), overlayed on an
  optical image from the Digitized Sky Survey.  Blue region shows the
  area covered by optical HST data.  (Target names from upper left to
  lower right: [s] NGC2976-DEEP; [t] NGC3077-PHOENIX; [u] NGC3741; [v]
  NGC4163; [w] NGC7793-HALO-6; [x] SCL-DE1; [y] UGC8508; [z]
  UGCA292;).  }
\end{figure}
\vfill
\clearpage

\begin{figure}
\centerline{
\includegraphics[width=3.25in]{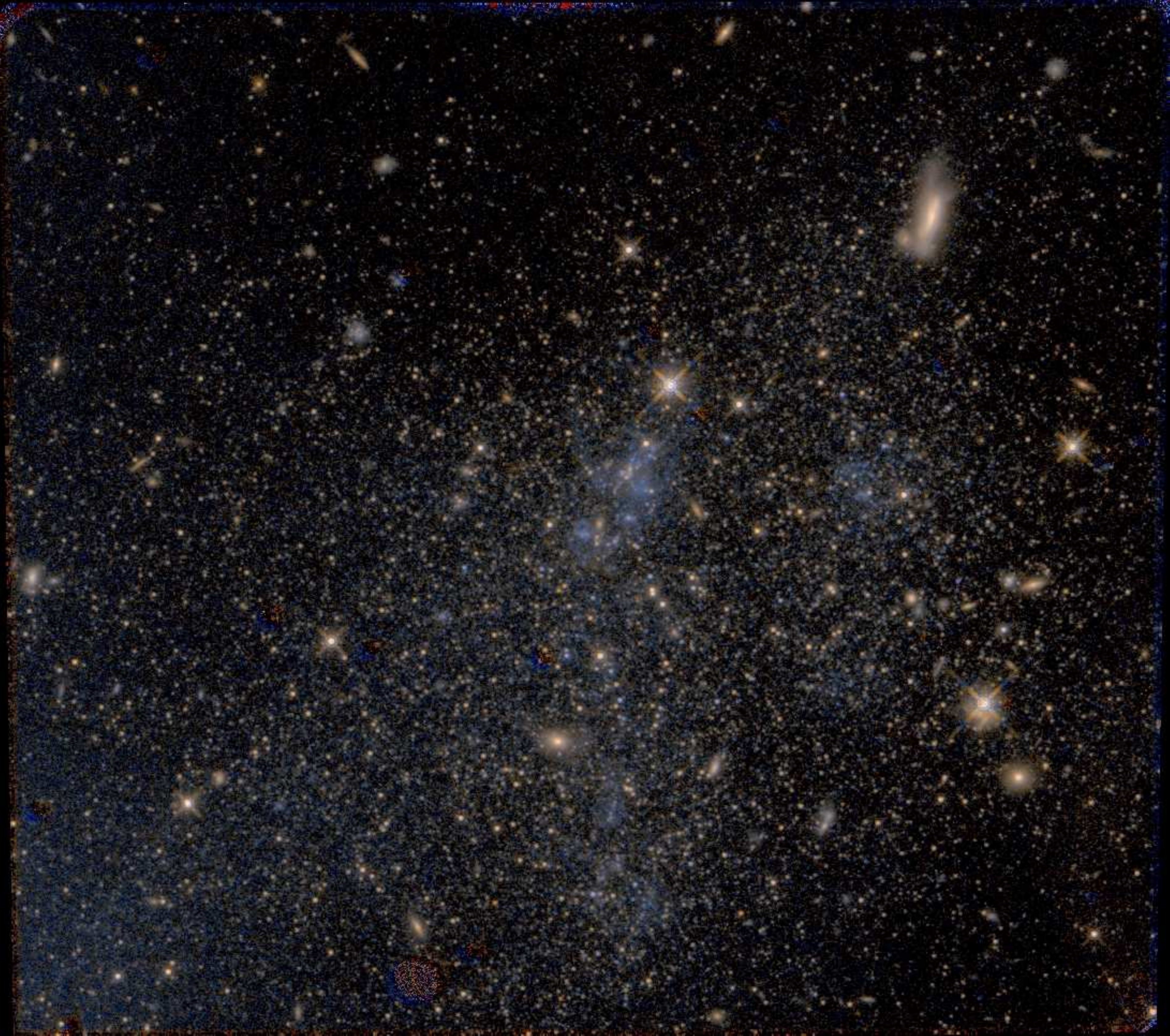}  
\includegraphics[width=3.25in]{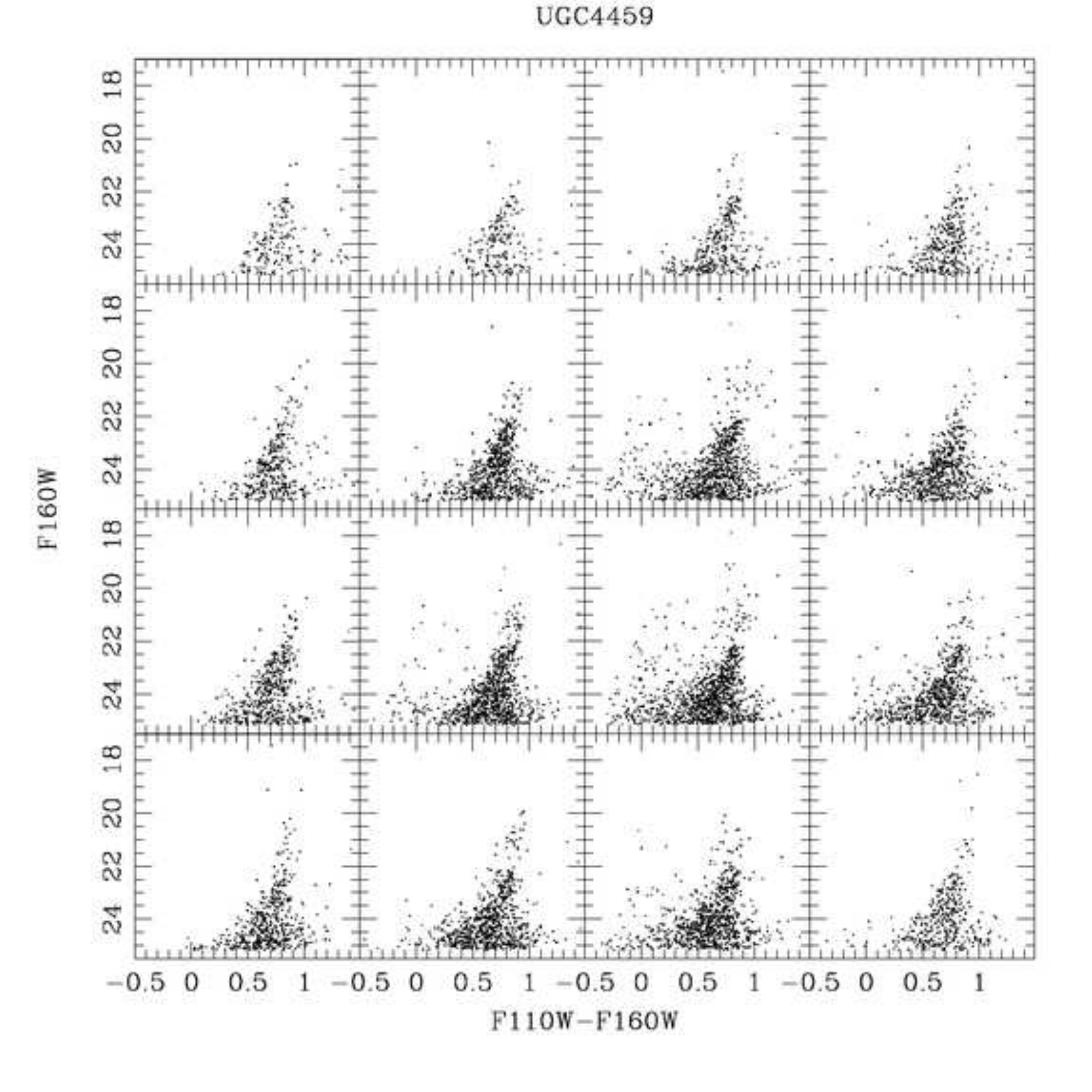}  
}
\centerline{
\includegraphics[width=3.25in]{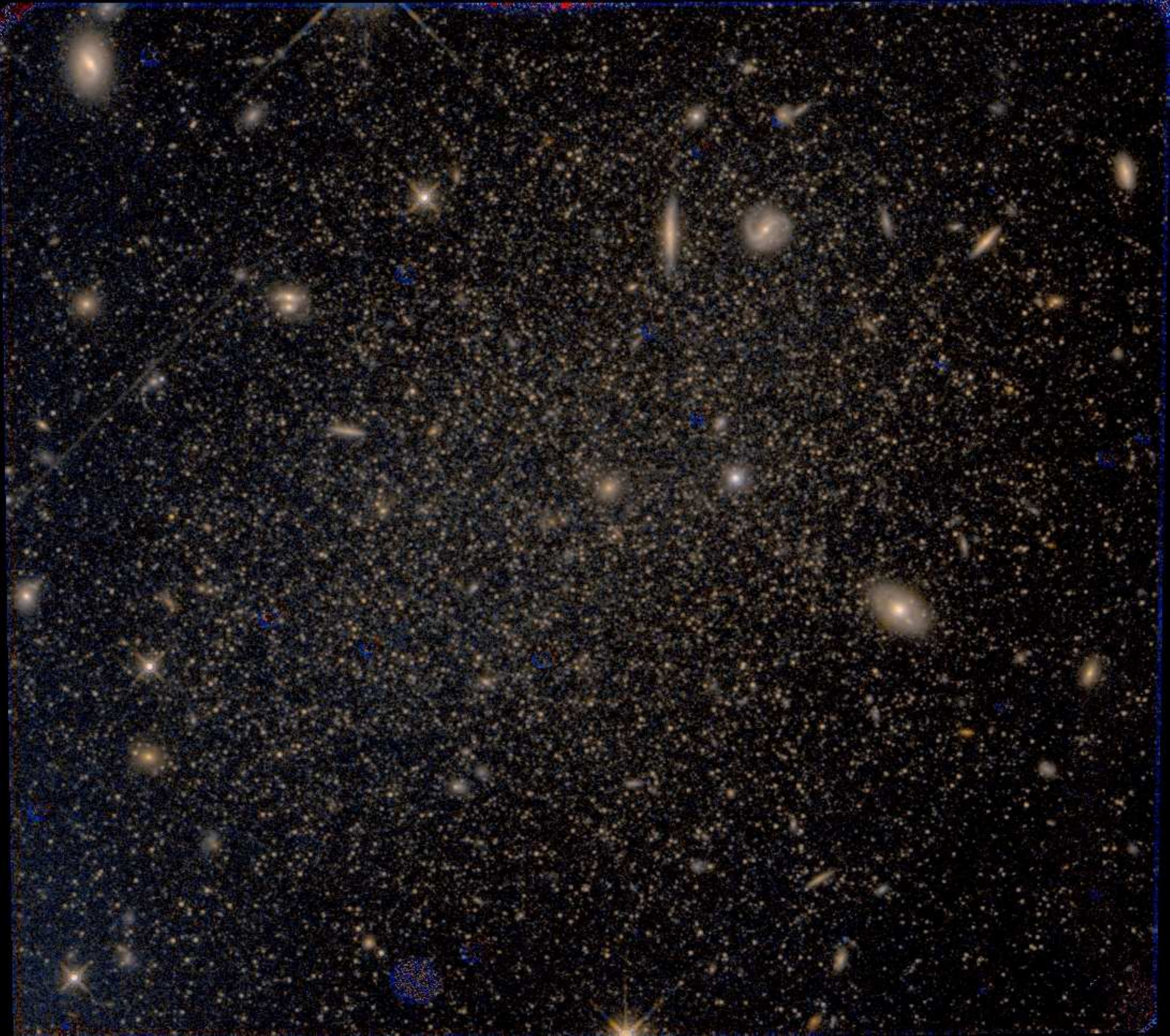}  
\includegraphics[width=3.25in]{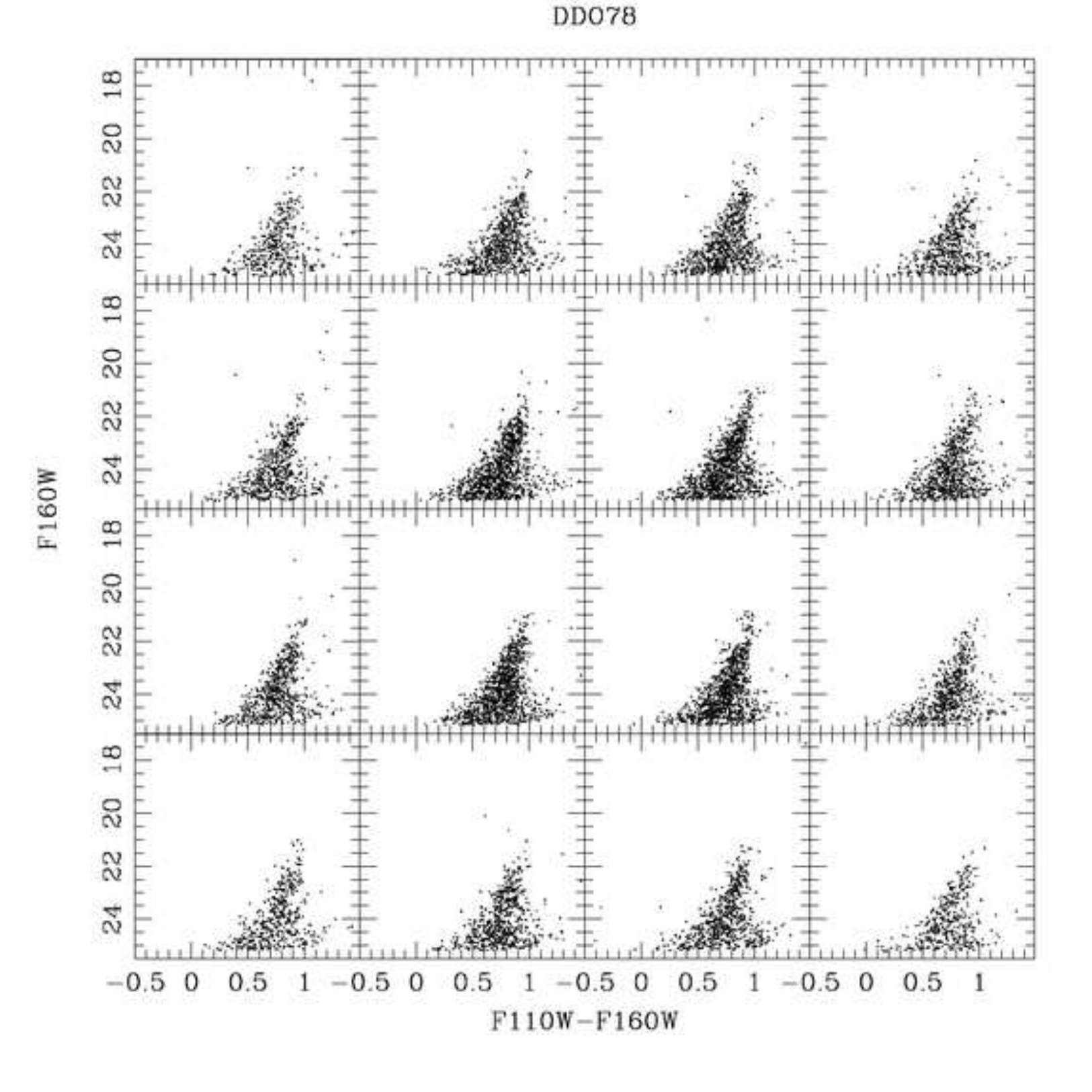}  
}
\caption{
(Left) False color $F110W+F160W$ image of the WFC3/IR field for the
target UGC4459 within the galaxy DDO53 [Top] and within the galaxy
DDO78 [Bottom].  (Right) Color-magnitude diagrams generated for a grid
of subregions, so that the upper left CMD corresponds to the upper
left of the adjacent image.
\label{gridfig}}
\end{figure}
\vfill
\clearpage
 
\begin{figure}
\figurenum{\ref{gridfig} continued}
\centerline{
\includegraphics[width=3.25in]{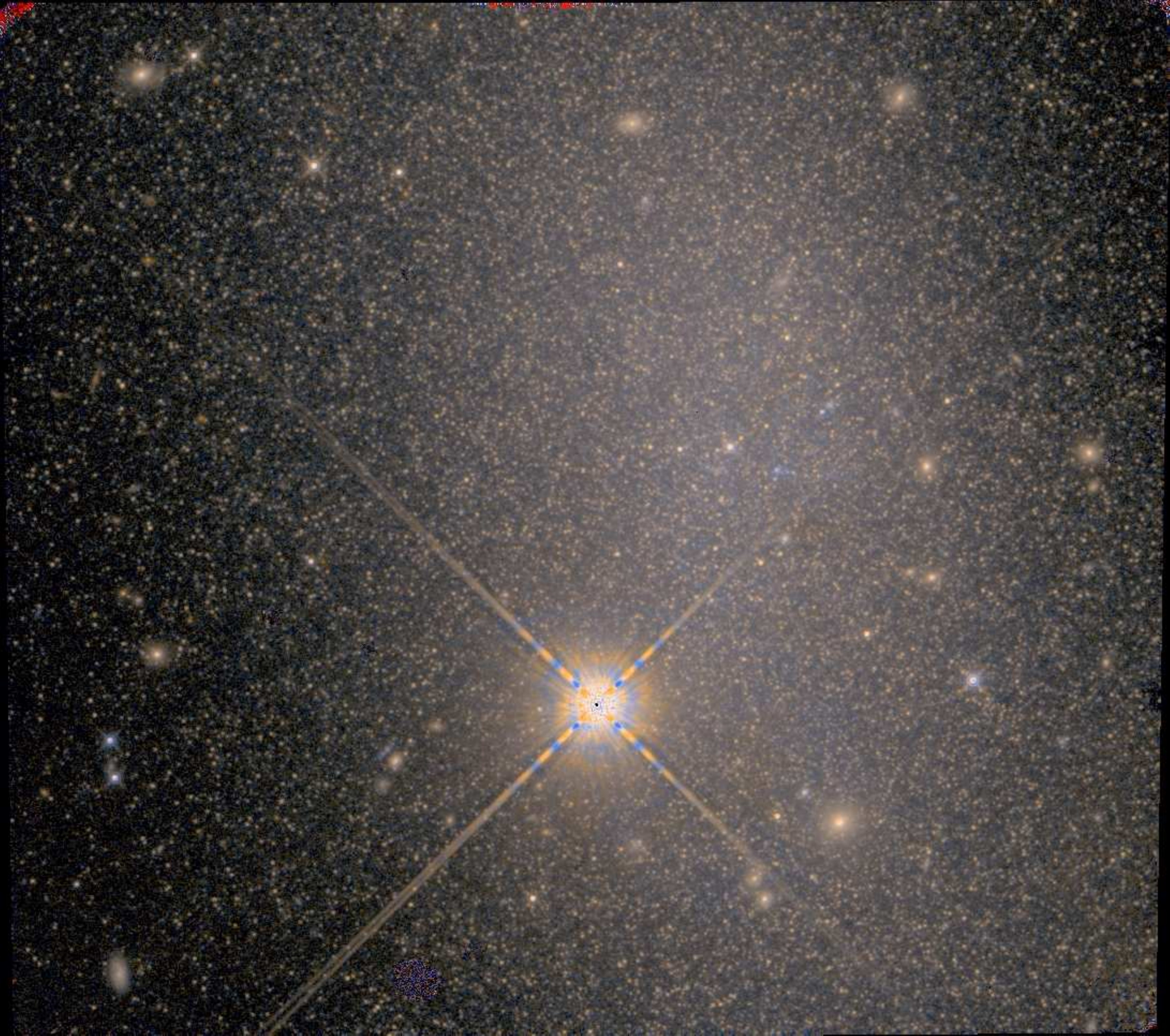}  
\includegraphics[width=3.25in]{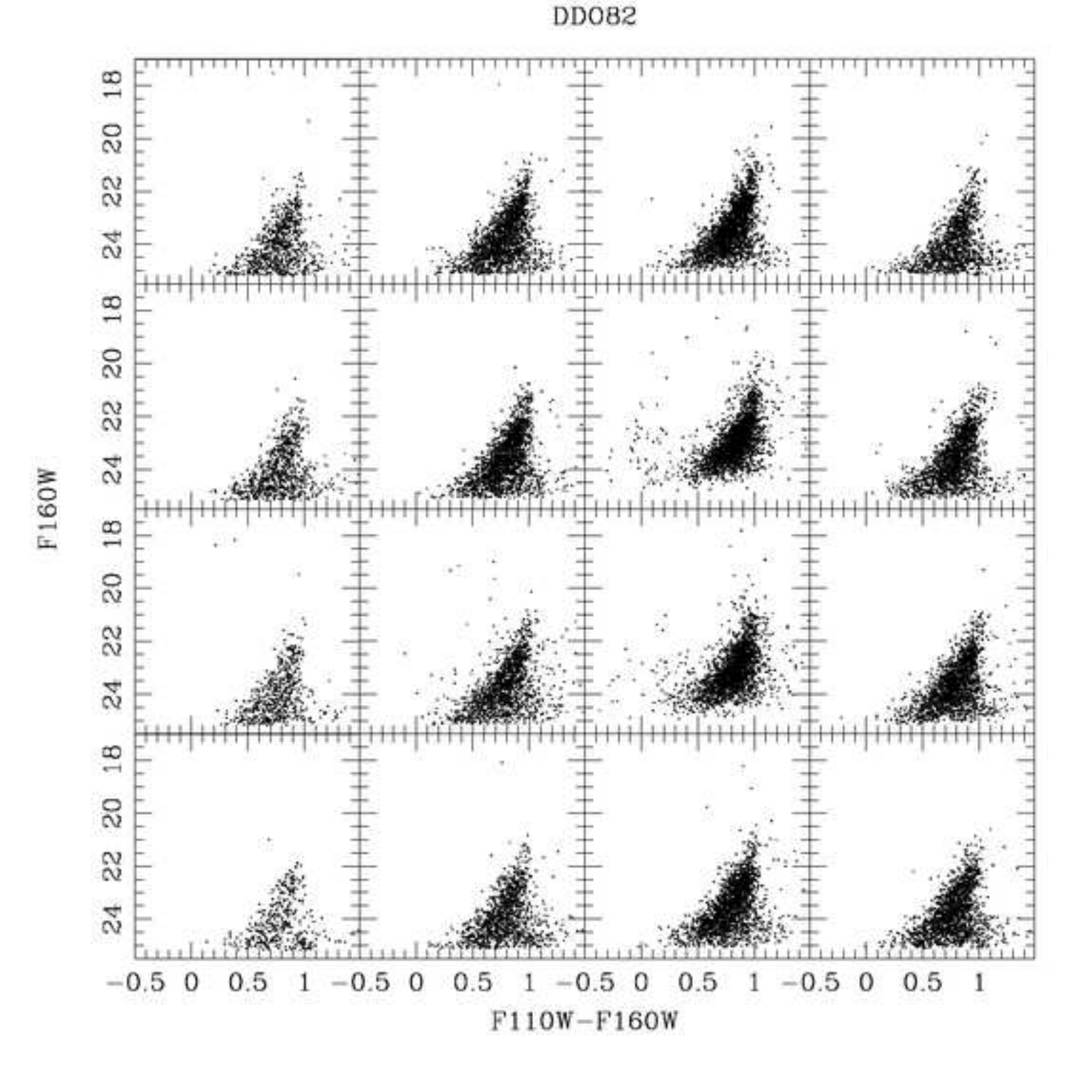}  
}
\centerline{
\includegraphics[width=3.25in]{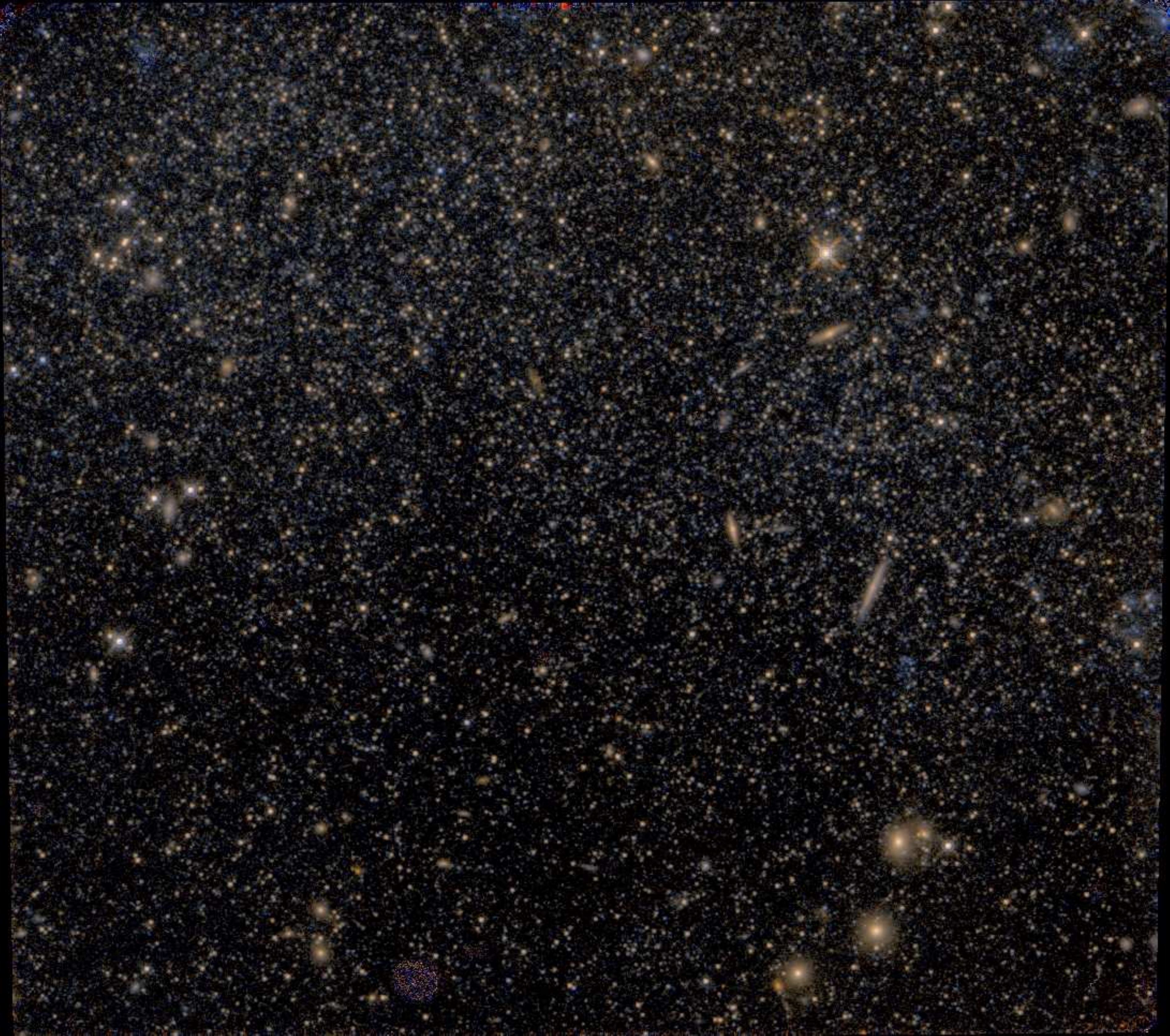}  
\includegraphics[width=3.25in]{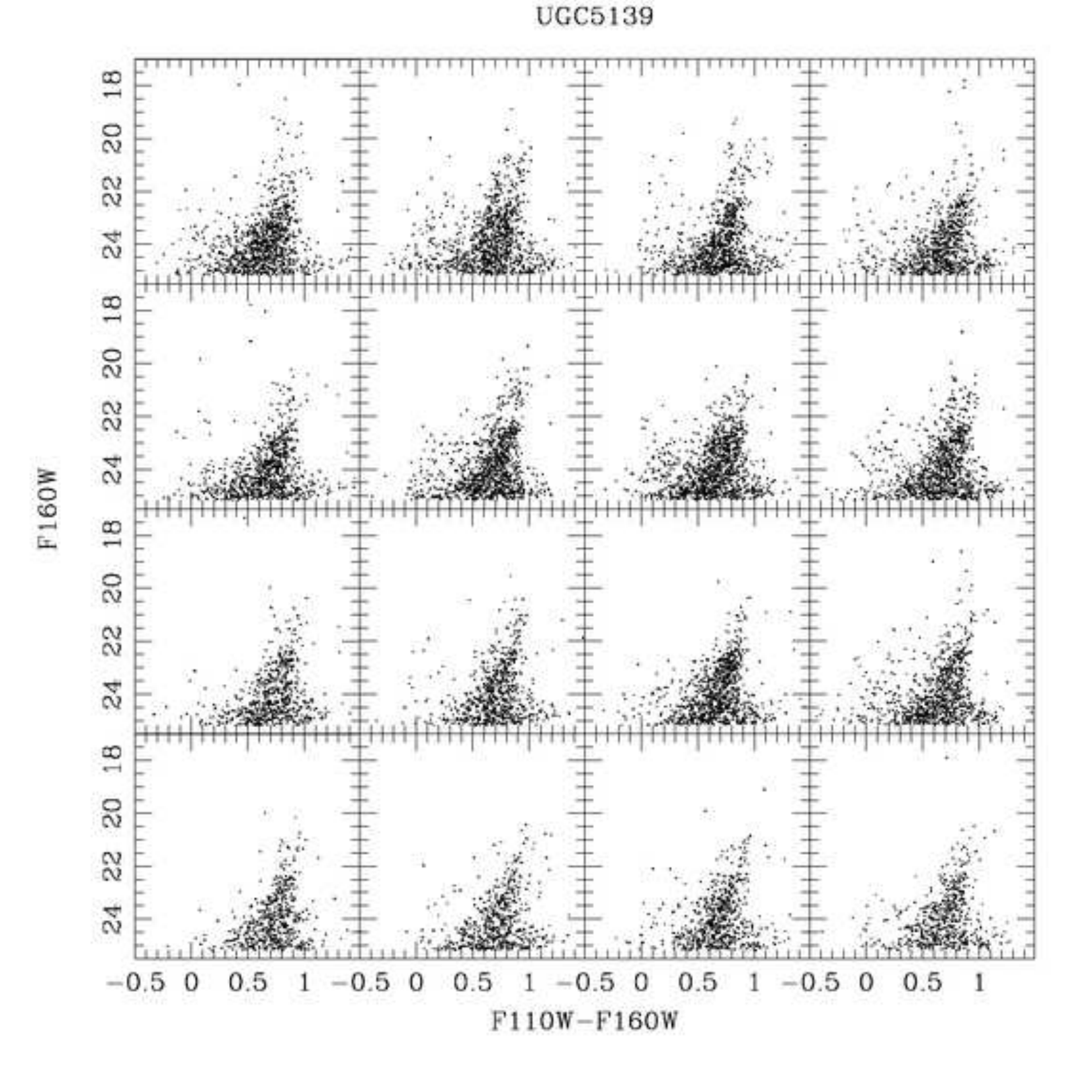}  
}
\caption{ (Left) False color $F110W+F160W$ image of the WFC3/IR field
  for within the galaxy DDO82 [Top] and the target UGC5139 within the
  galaxy HoI [Bottom].  (Right) Color-magnitude diagrams generated for
  a grid of subregions, so that the upper left CMD corresponds to the
  upper left of the adjacent image.  }
\end{figure}
\vfill
\clearpage
 
\begin{figure}
\figurenum{\ref{gridfig} continued}
\centerline{
\includegraphics[width=3.25in]{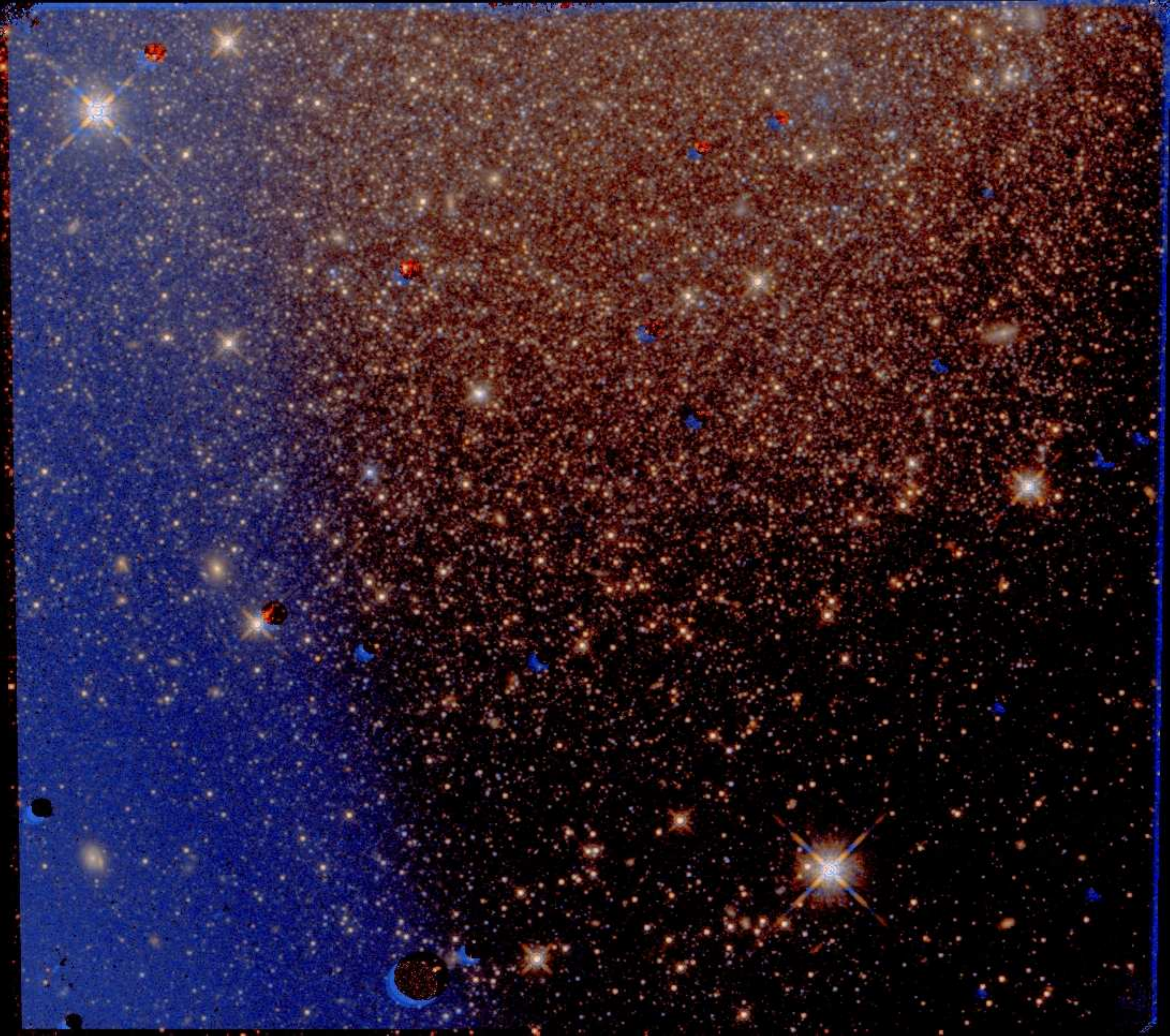}  
\includegraphics[width=3.25in]{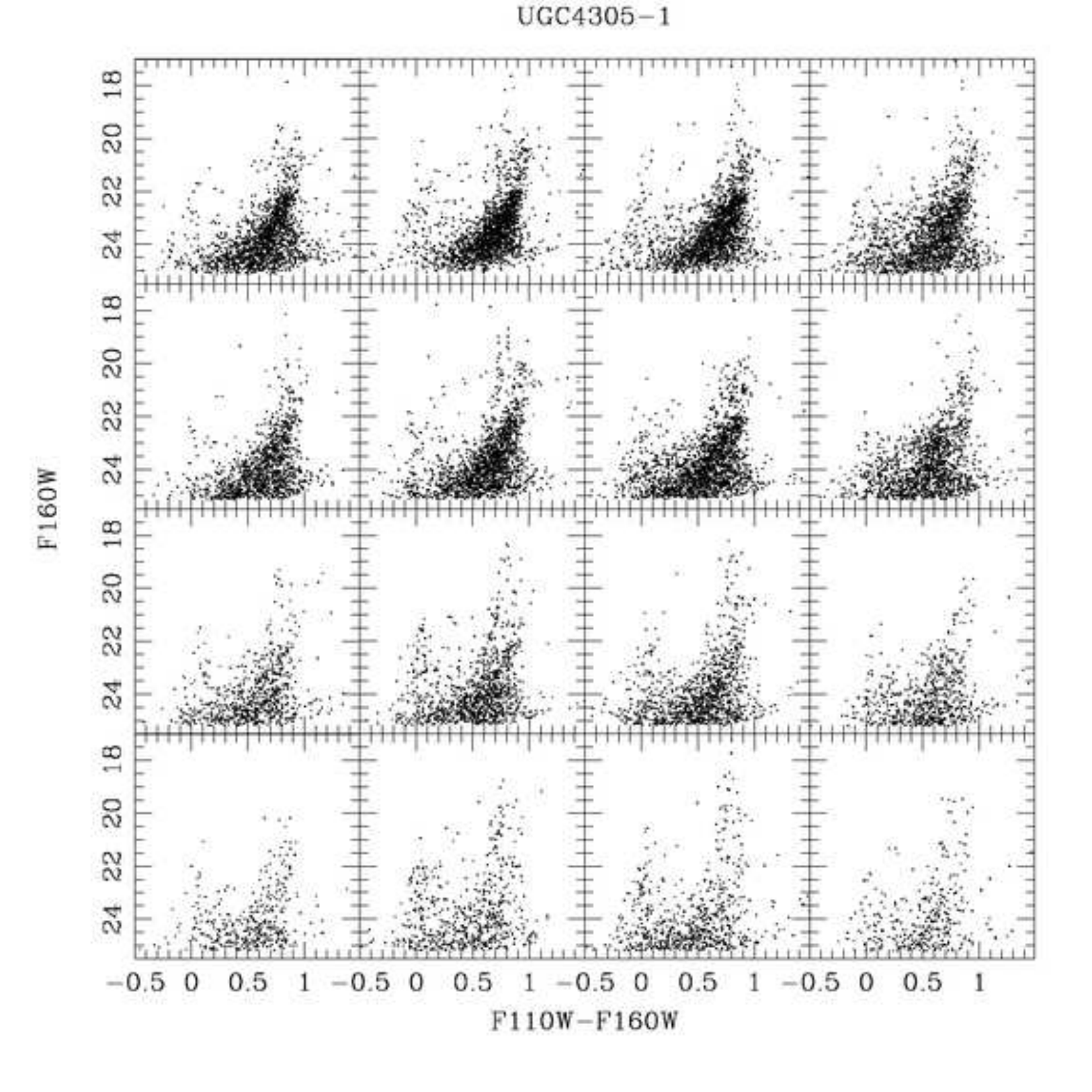}  
}
\centerline{
\includegraphics[width=3.25in]{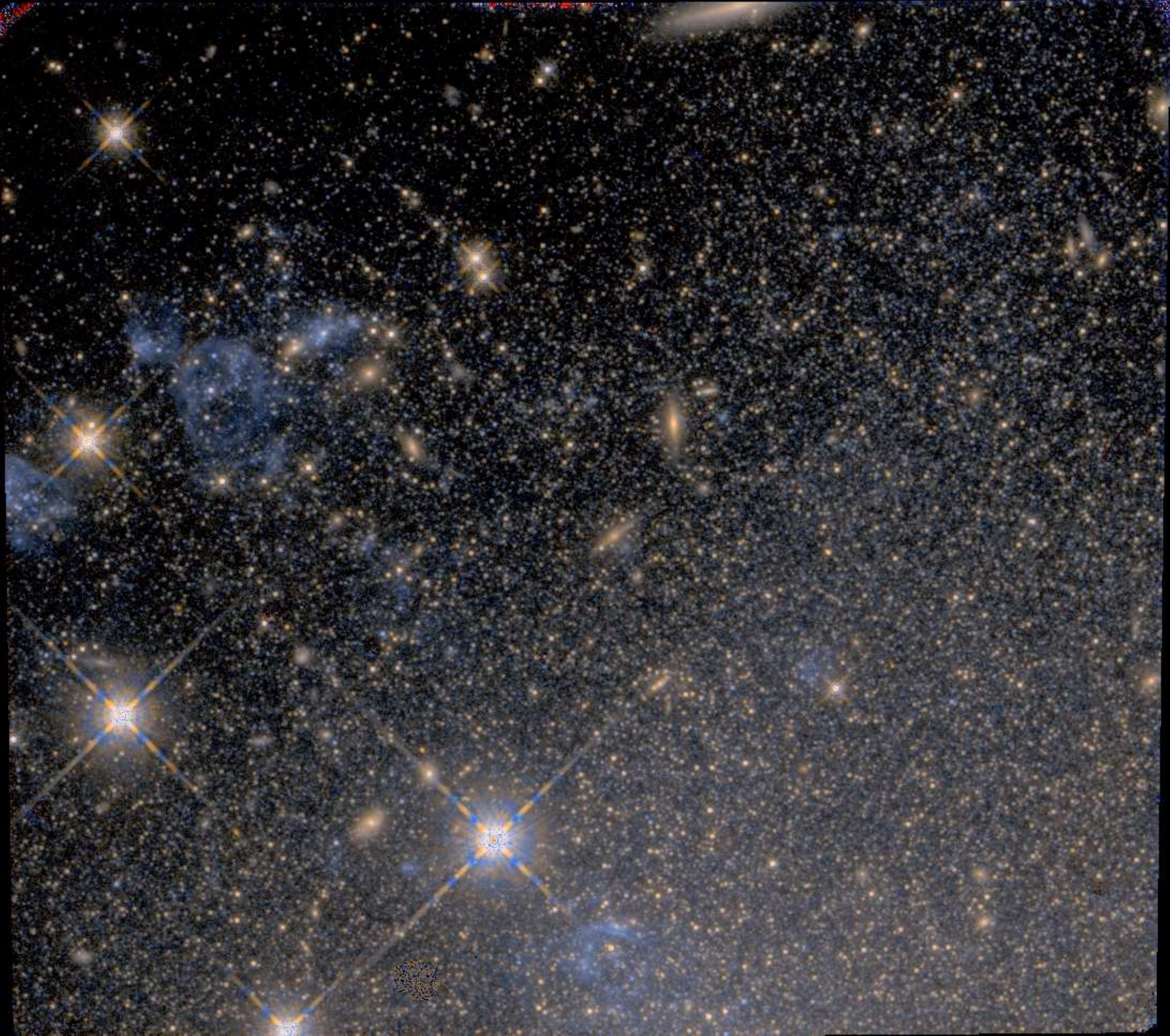}  
\includegraphics[width=3.25in]{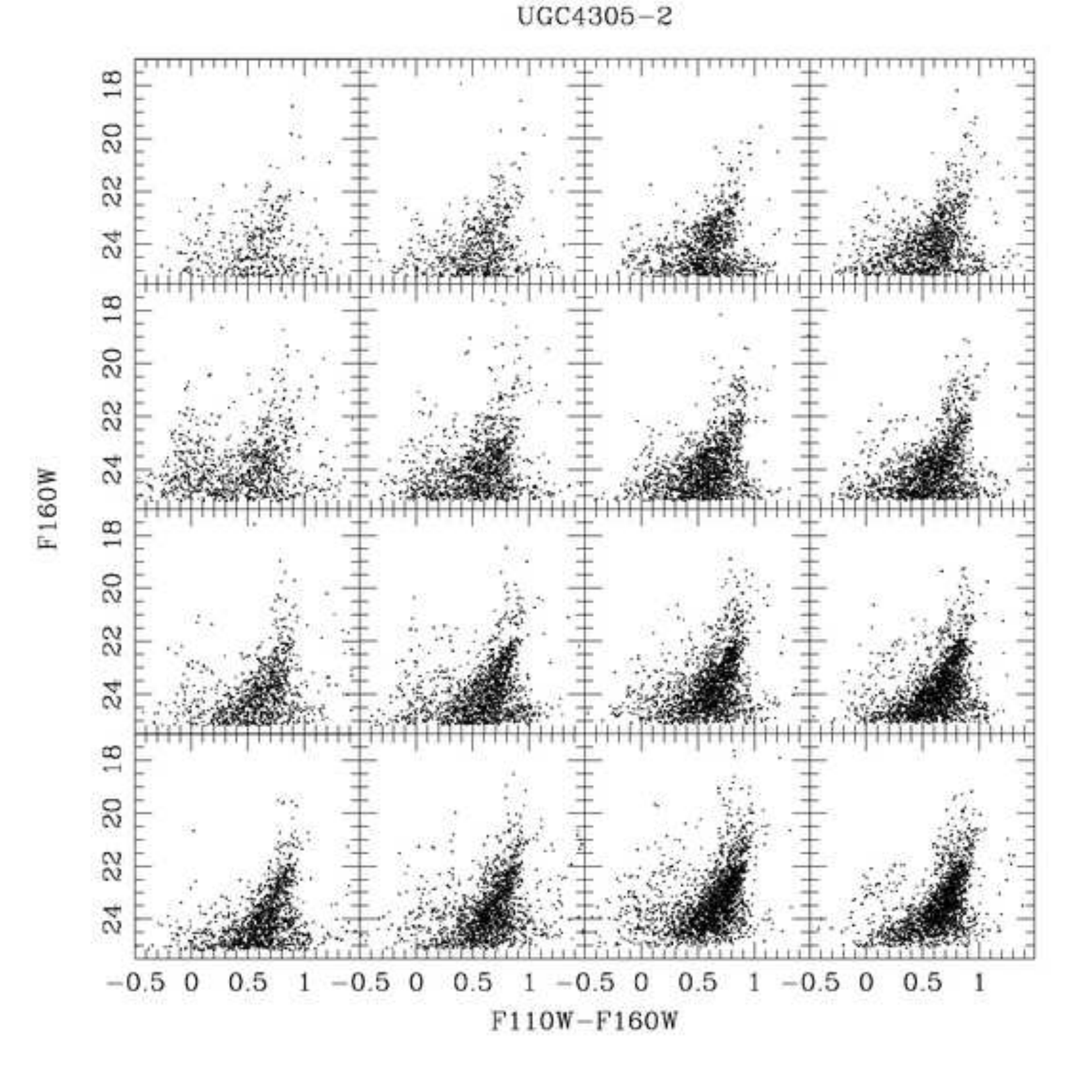}  
}
\caption{ (Left) False color $F110W+F160W$ image of the WFC3/IR field
  for the target UGC4305-1 within the galaxy HoII [Top] and the target
  UGC4305-2 within the galaxy HoII [Bottom].  (Right) Color-magnitude
  diagrams generated for a grid of subregions, so that the upper left
  CMD corresponds to the upper left of the adjacent image.  }
\end{figure}
\vfill
\clearpage
 
\begin{figure}
\figurenum{\ref{gridfig} continued}
\centerline{
\includegraphics[width=3.25in]{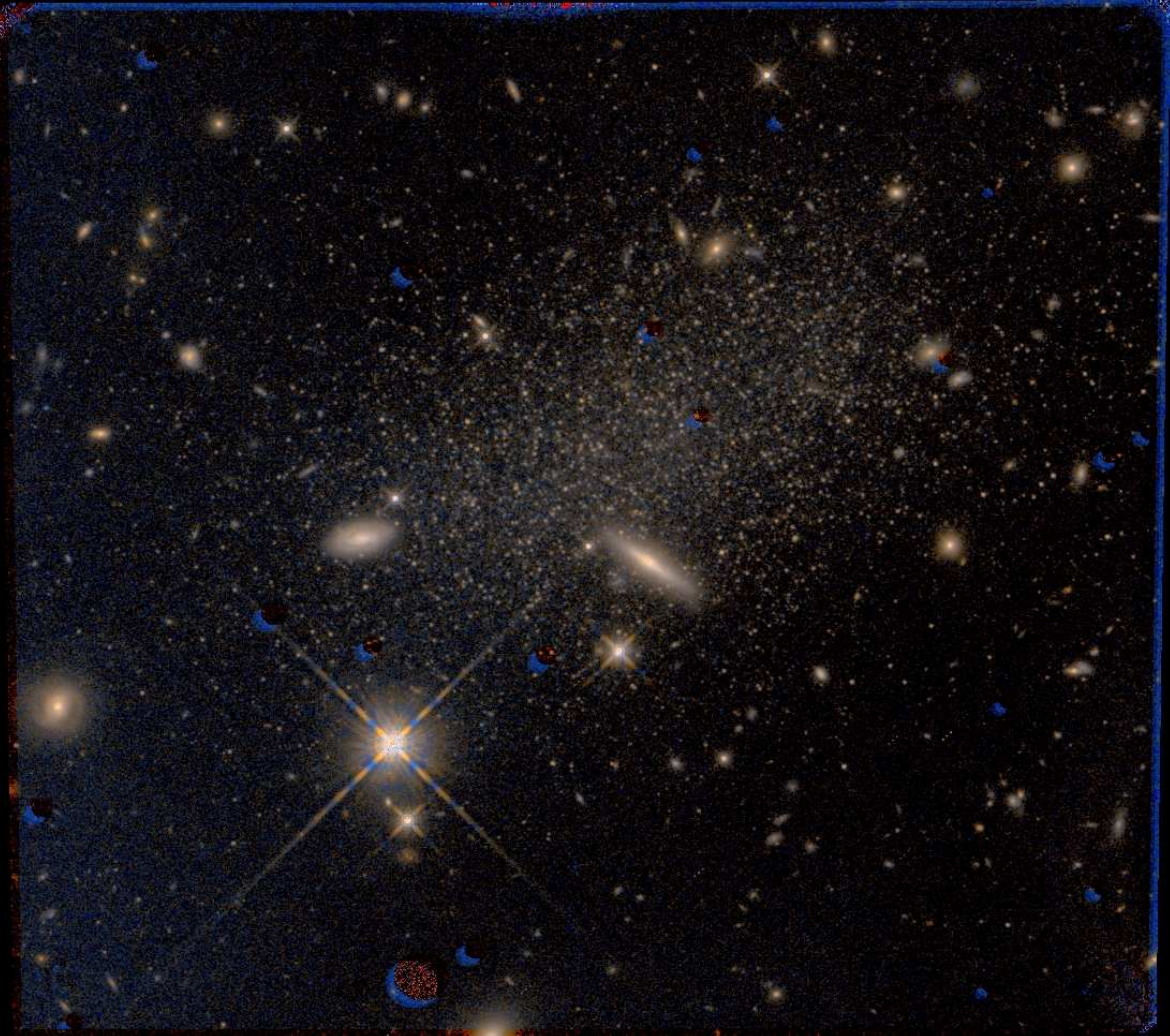}  
\includegraphics[width=3.25in]{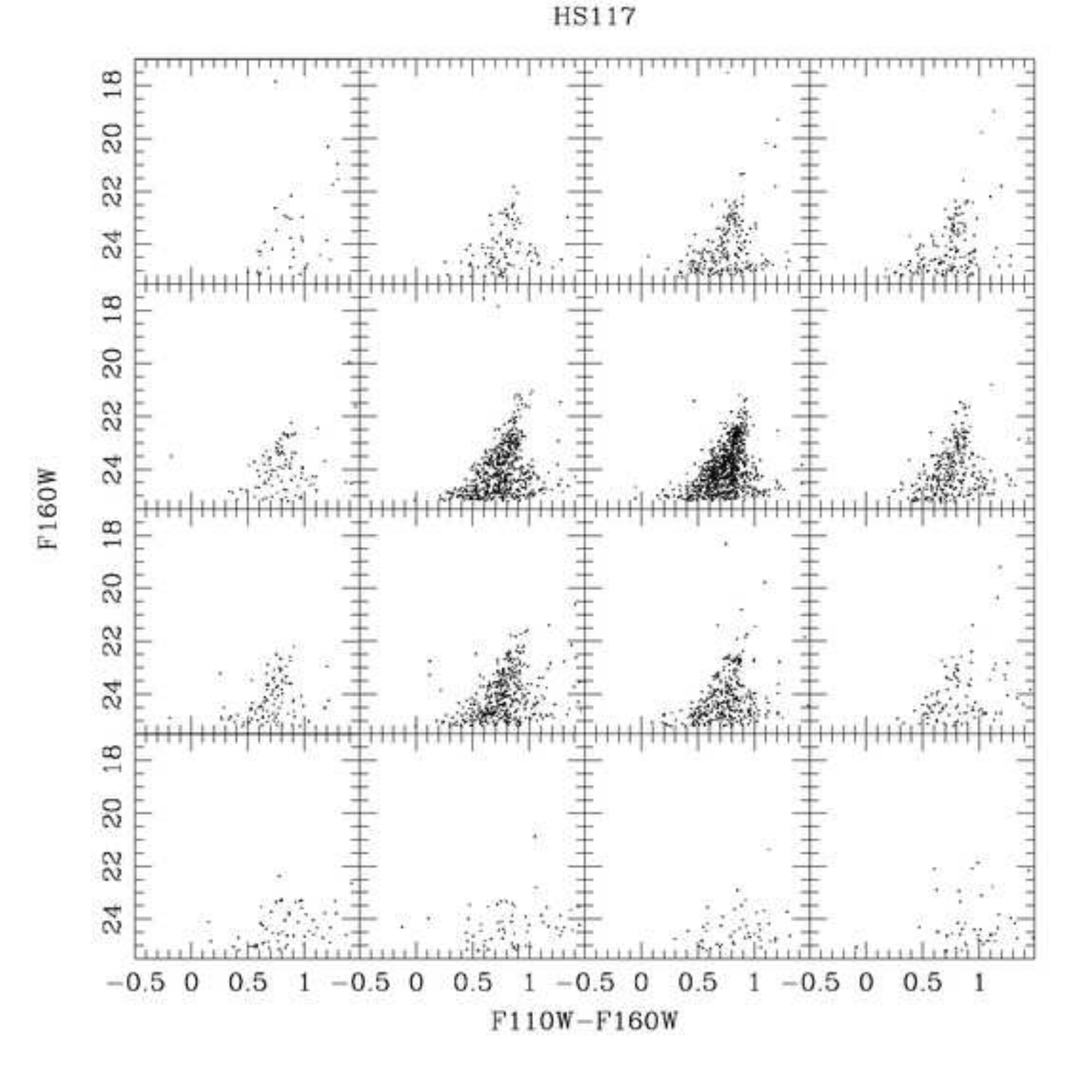}  
}
\centerline{
\includegraphics[width=3.25in]{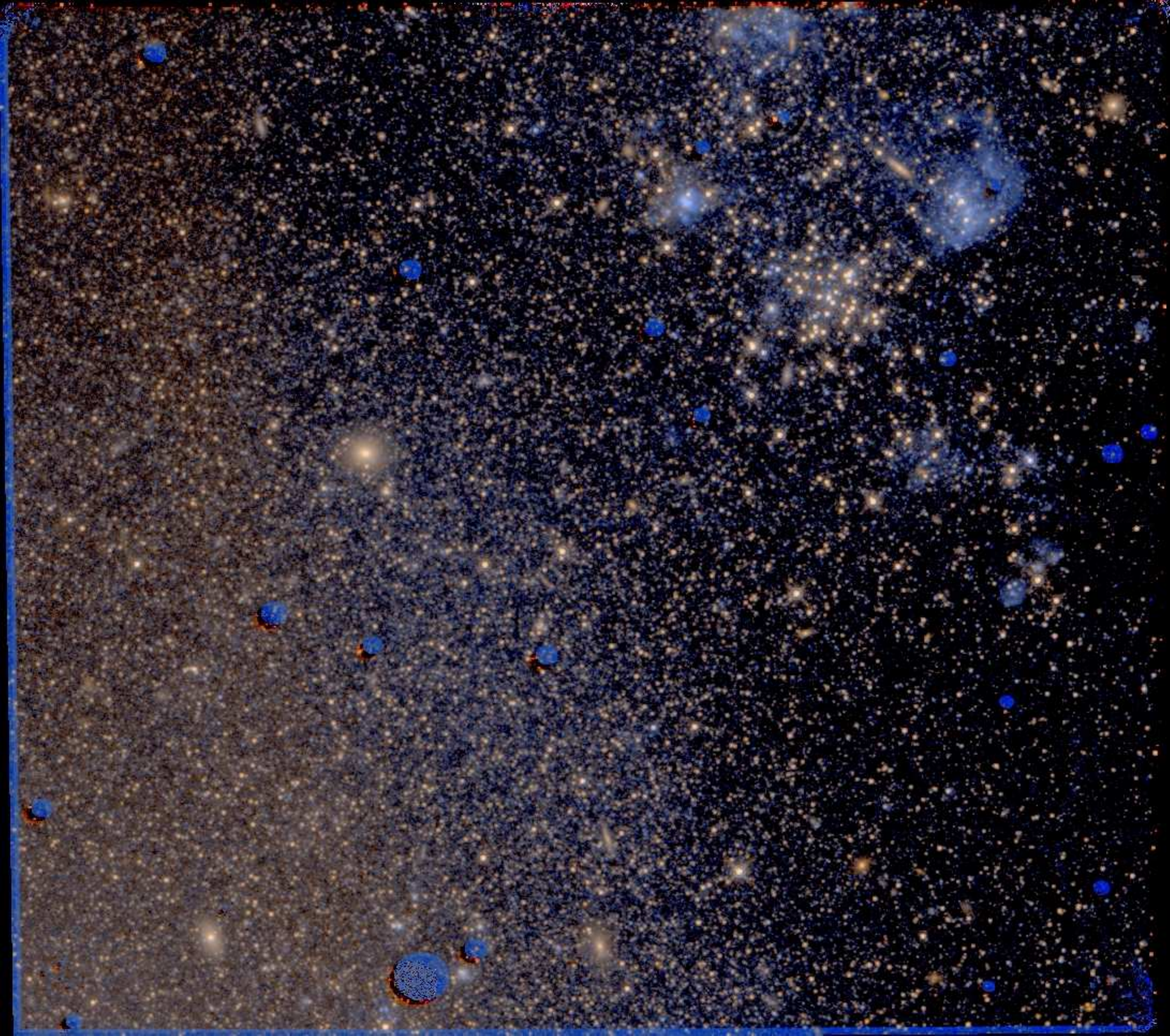}  
\includegraphics[width=3.25in]{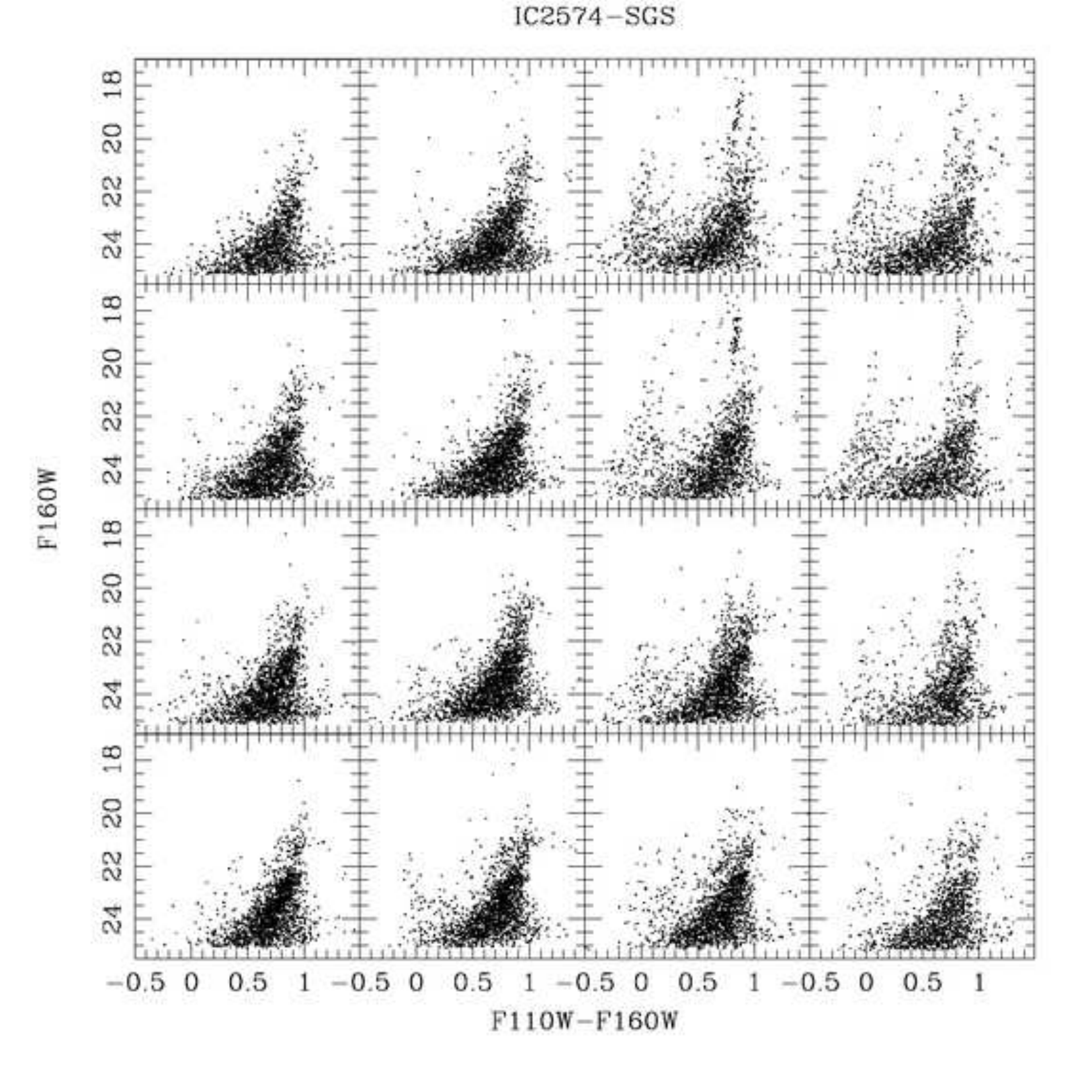}  
}
\caption{ (Left) False color $F110W+F160W$ image of the WFC3/IR field
  for within the galaxy HS117 [Top] and the target IC2574-SGS within
  the galaxy I2574 [Bottom].  (Right) Color-magnitude diagrams
  generated for a grid of subregions, so that the upper left CMD
  corresponds to the upper left of the adjacent image.  }
\end{figure}
\vfill
\clearpage
 
\begin{figure}
\figurenum{\ref{gridfig} continued}
\centerline{
\includegraphics[width=3.25in]{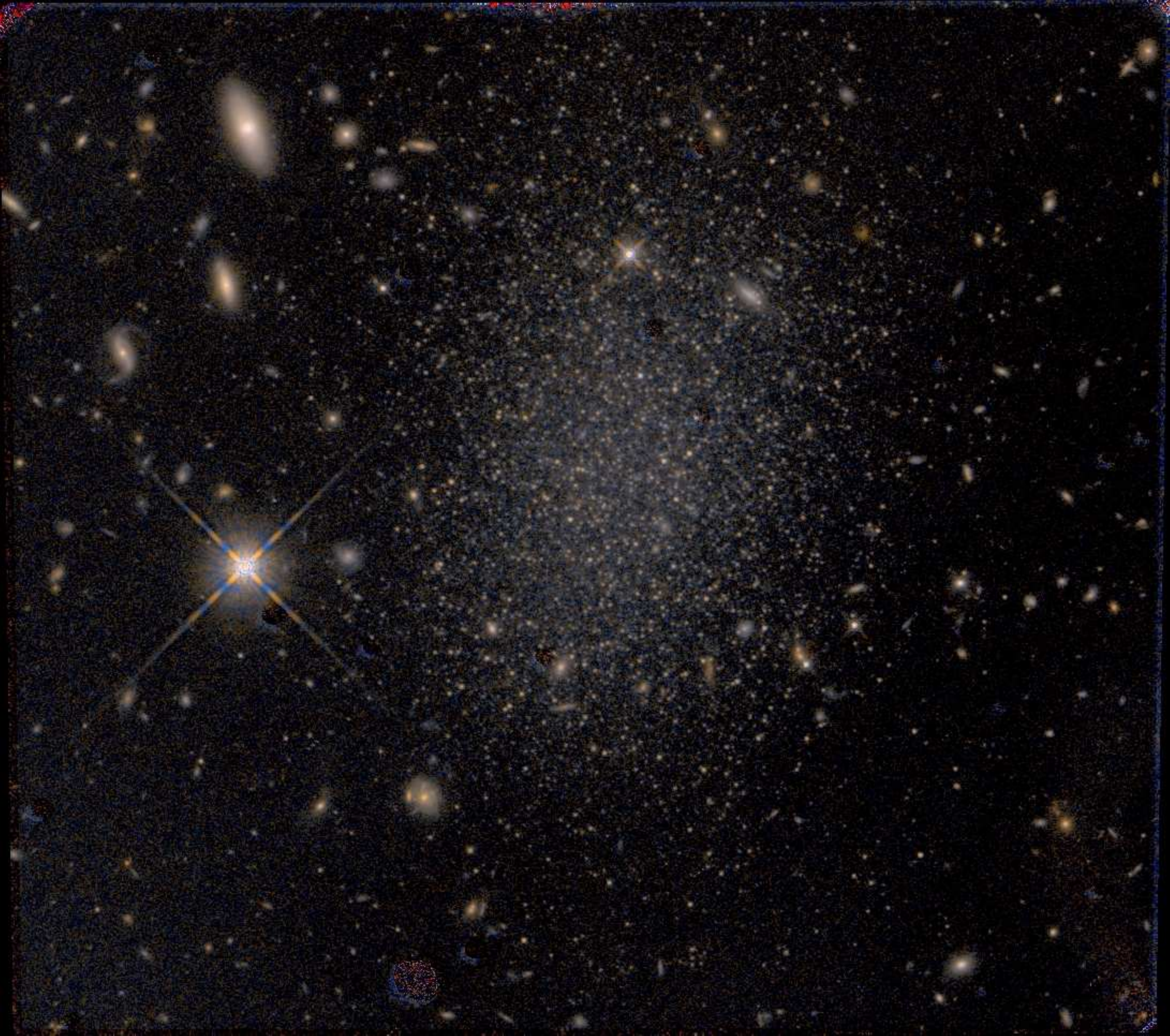}  
\includegraphics[width=3.25in]{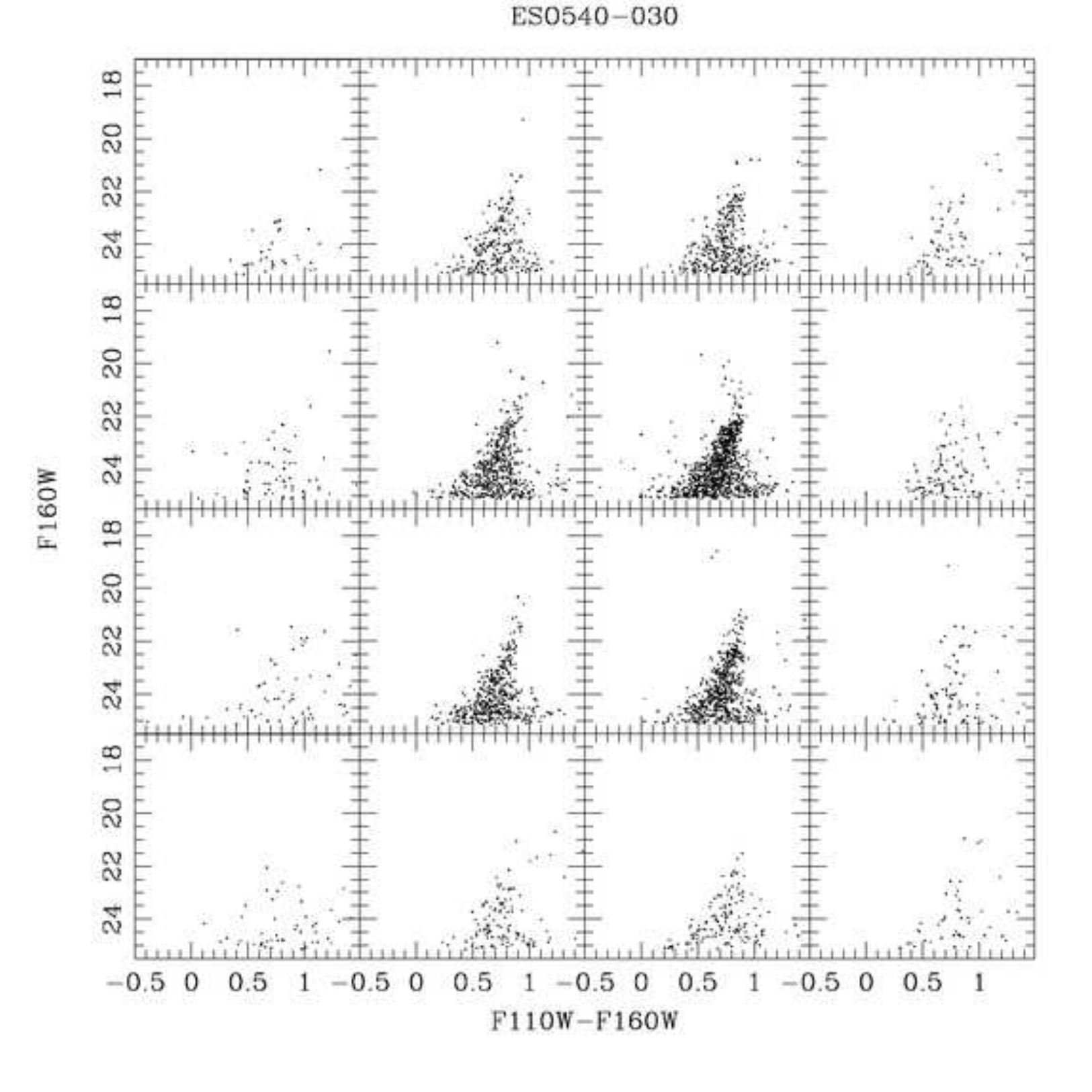}  
}
\centerline{
\includegraphics[width=3.25in]{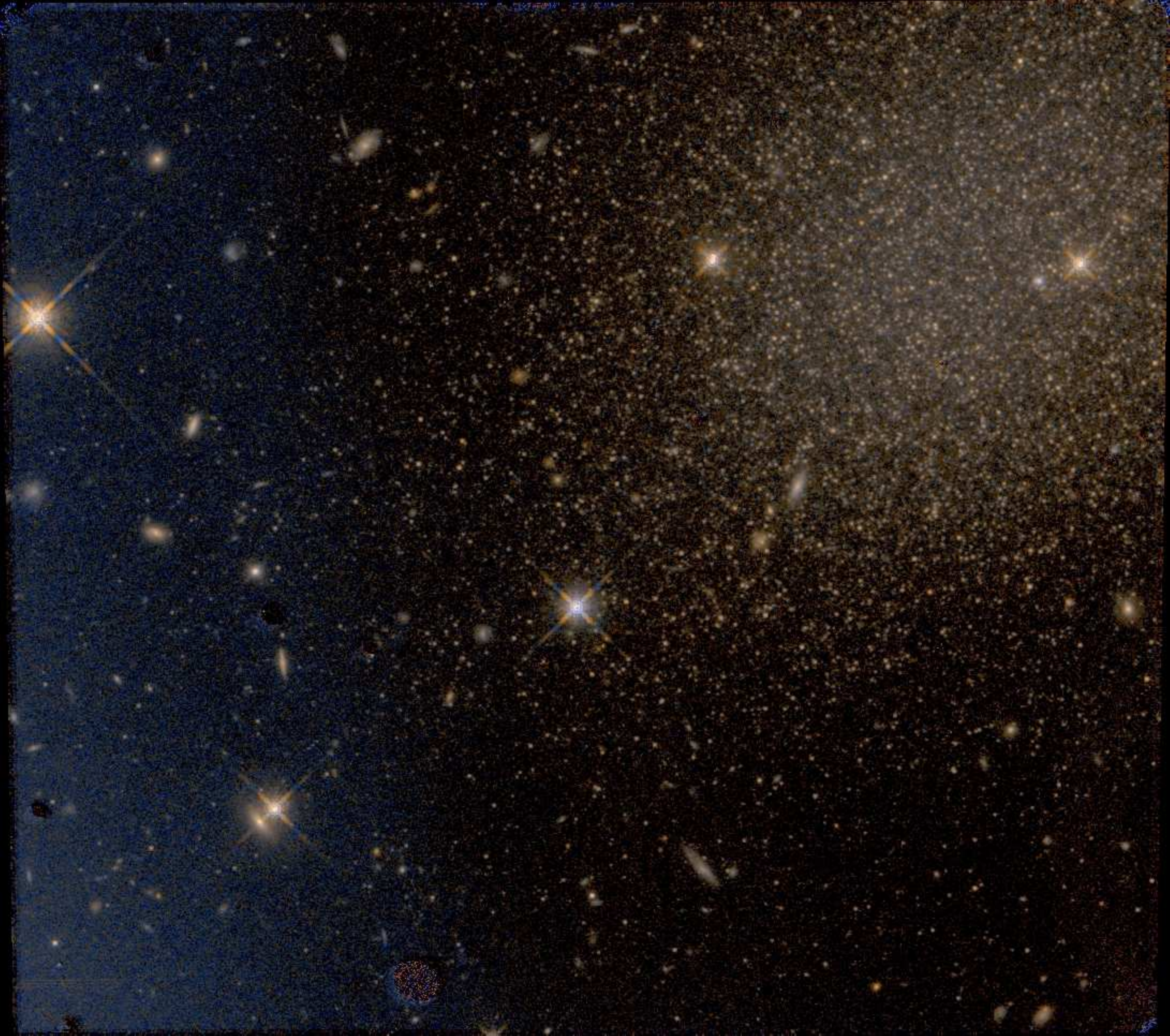}  
\includegraphics[width=3.25in]{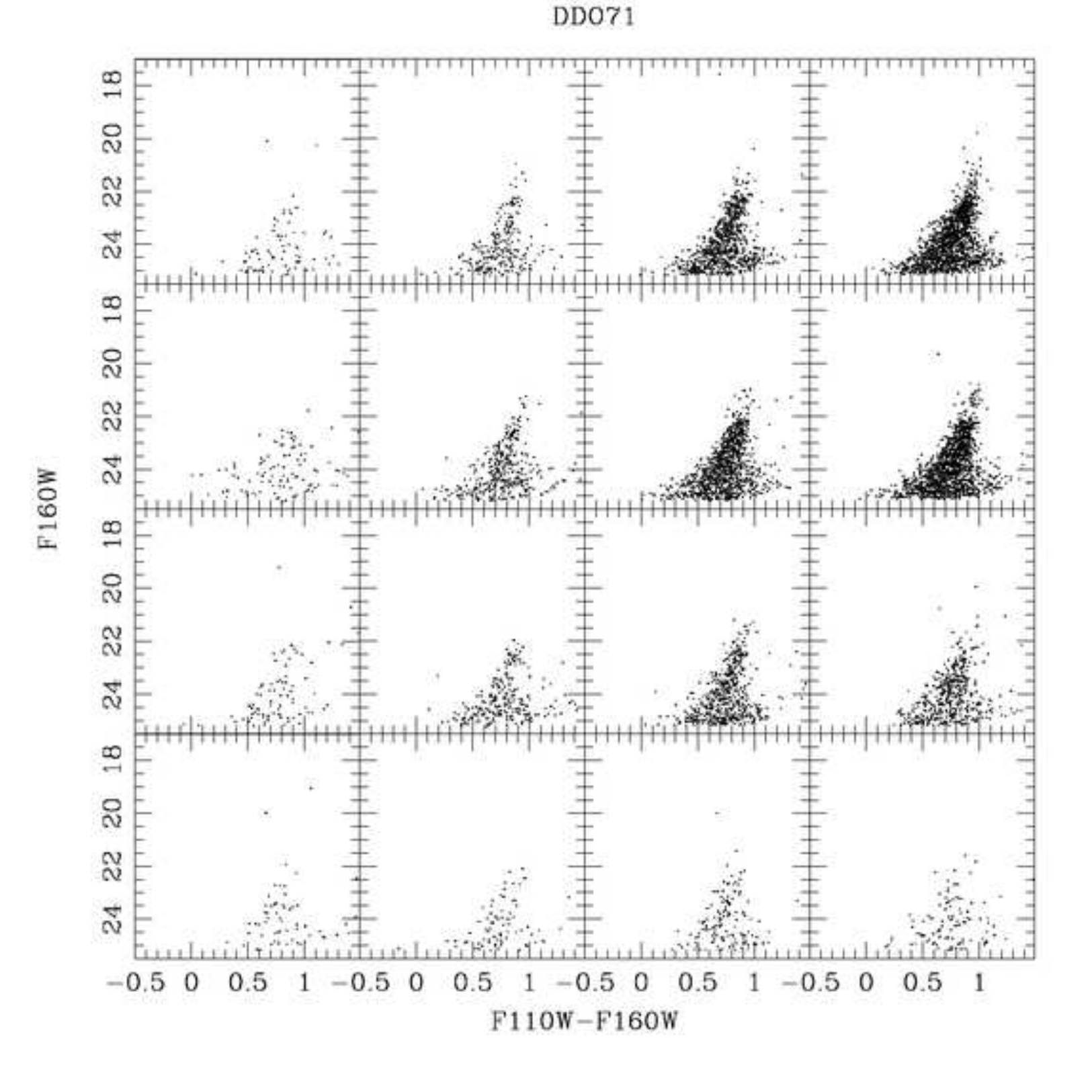}  
}
\caption{ (Left) False color $F110W+F160W$ image of the WFC3/IR field
  for the target ESO540-030 within the galaxy KDG2 [Top] and the
  target DDO71 within the galaxy KDG63 [Bottom].  (Right)
  Color-magnitude diagrams generated for a grid of subregions, so that
  the upper left CMD corresponds to the upper left of the adjacent
  image.  }
\end{figure}
\vfill
\clearpage
 
\begin{figure}
\figurenum{\ref{gridfig} continued}
\centerline{
\includegraphics[width=3.25in]{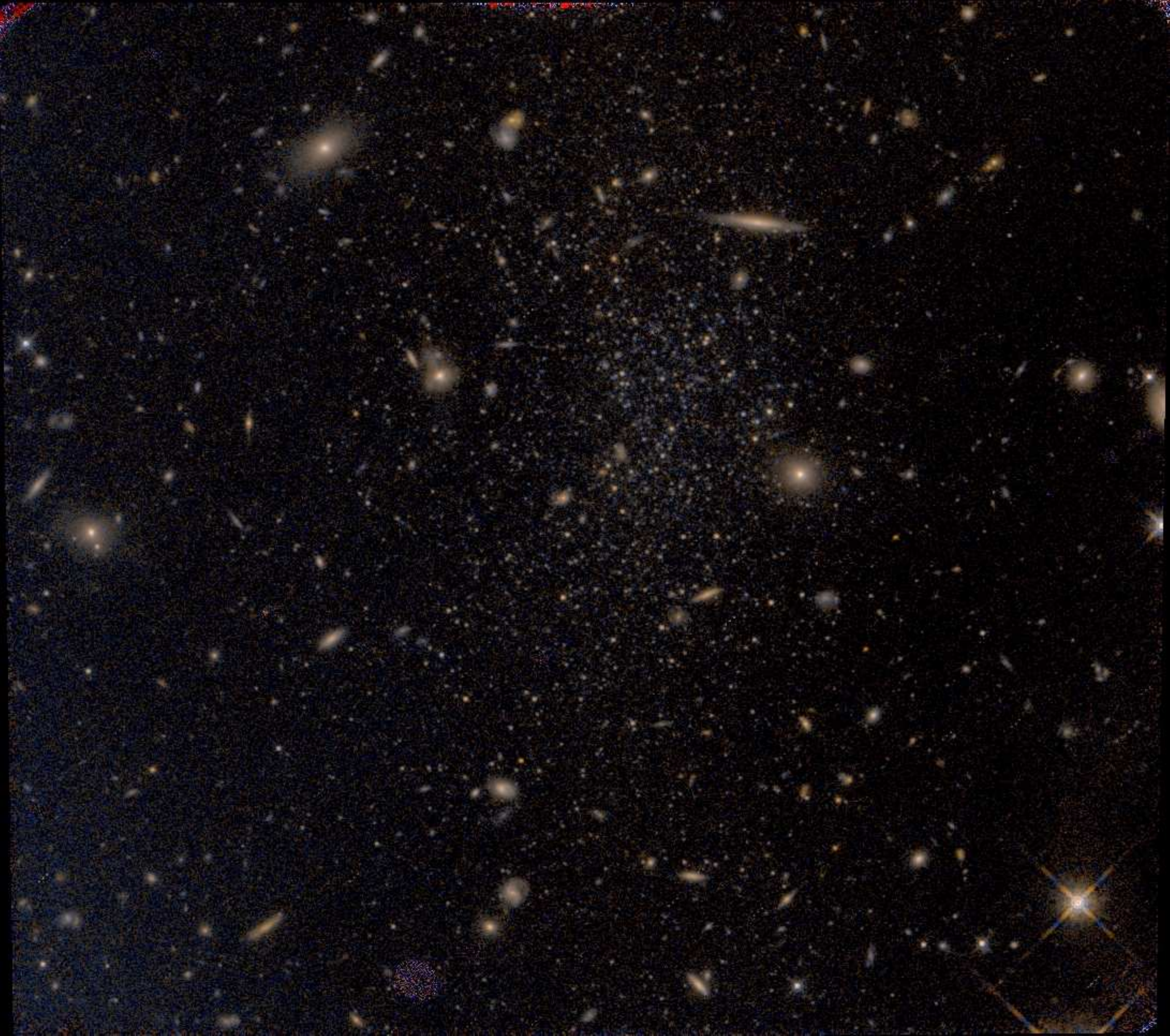}  
\includegraphics[width=3.25in]{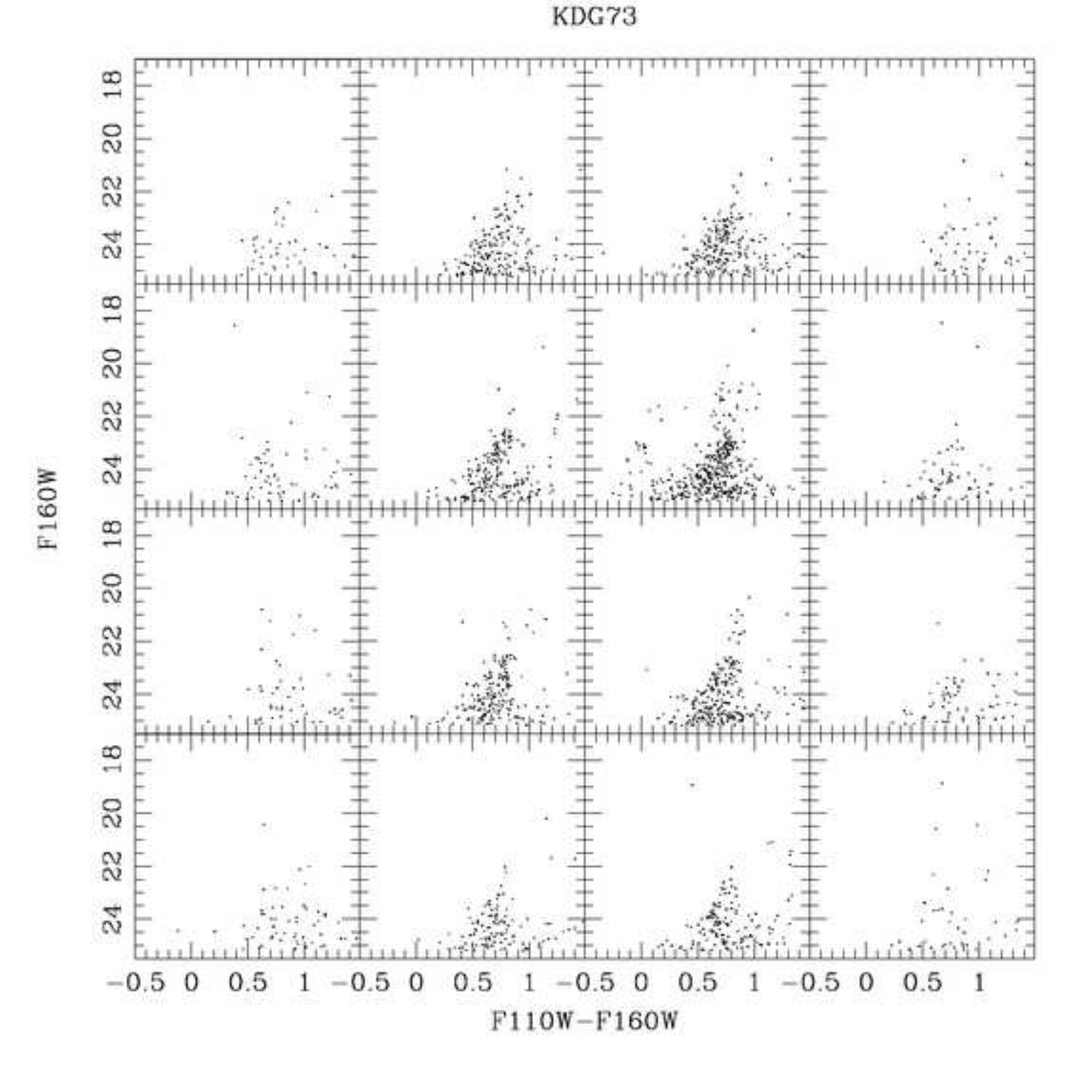}  
}
\centerline{
\includegraphics[width=3.25in]{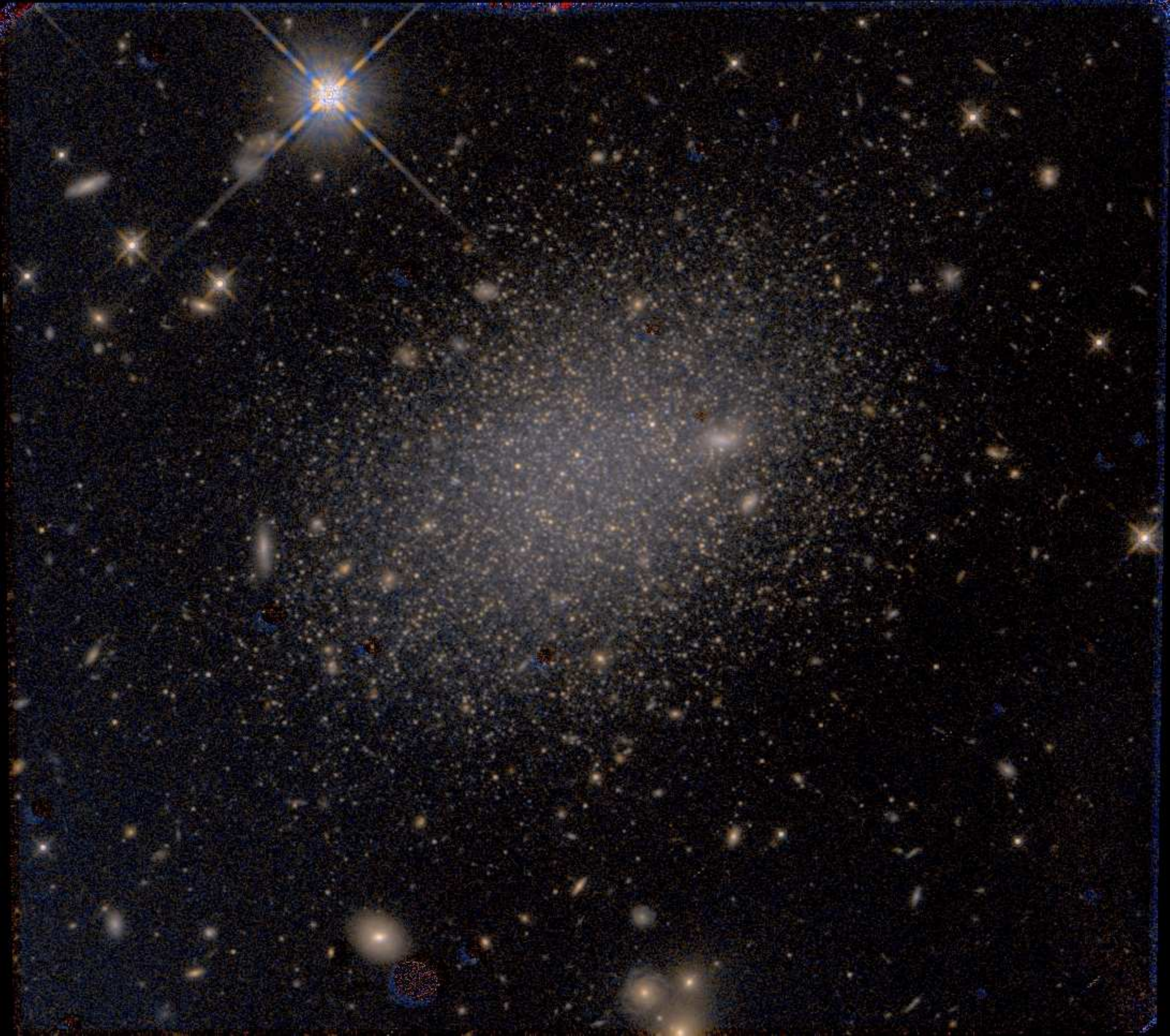}  
\includegraphics[width=3.25in]{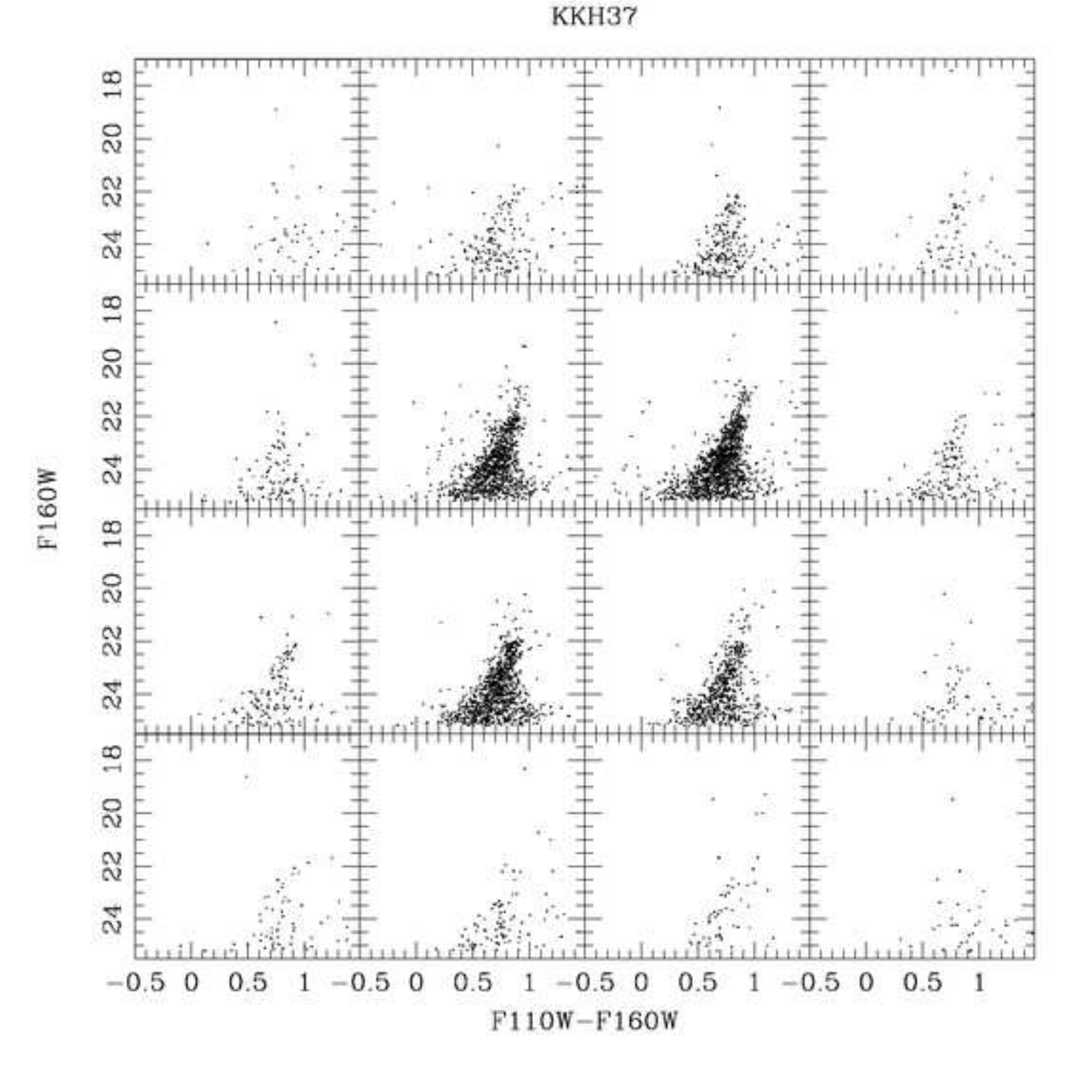}  
}
\caption{ (Left) False color $F110W+F160W$ image of the WFC3/IR field
  for within the galaxy KDG73 [Top] and within the galaxy KKH37
  [Bottom].  (Right) Color-magnitude diagrams generated for a grid of
  subregions, so that the upper left CMD corresponds to the upper left
  of the adjacent image.  }
\end{figure}
\vfill
\clearpage
 
\begin{figure}
\figurenum{\ref{gridfig} continued}
\centerline{
\includegraphics[width=3.25in]{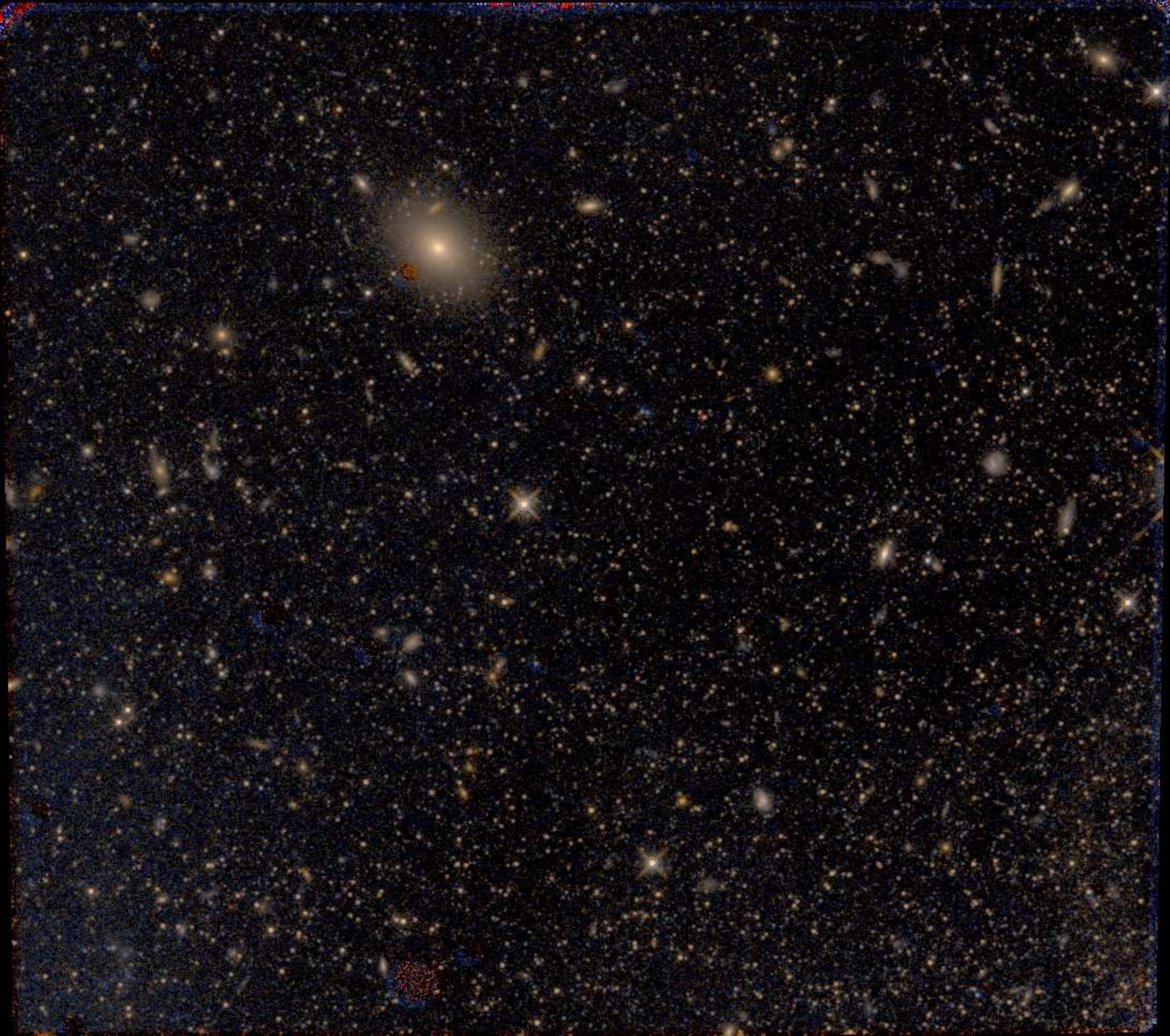}  
\includegraphics[width=3.25in]{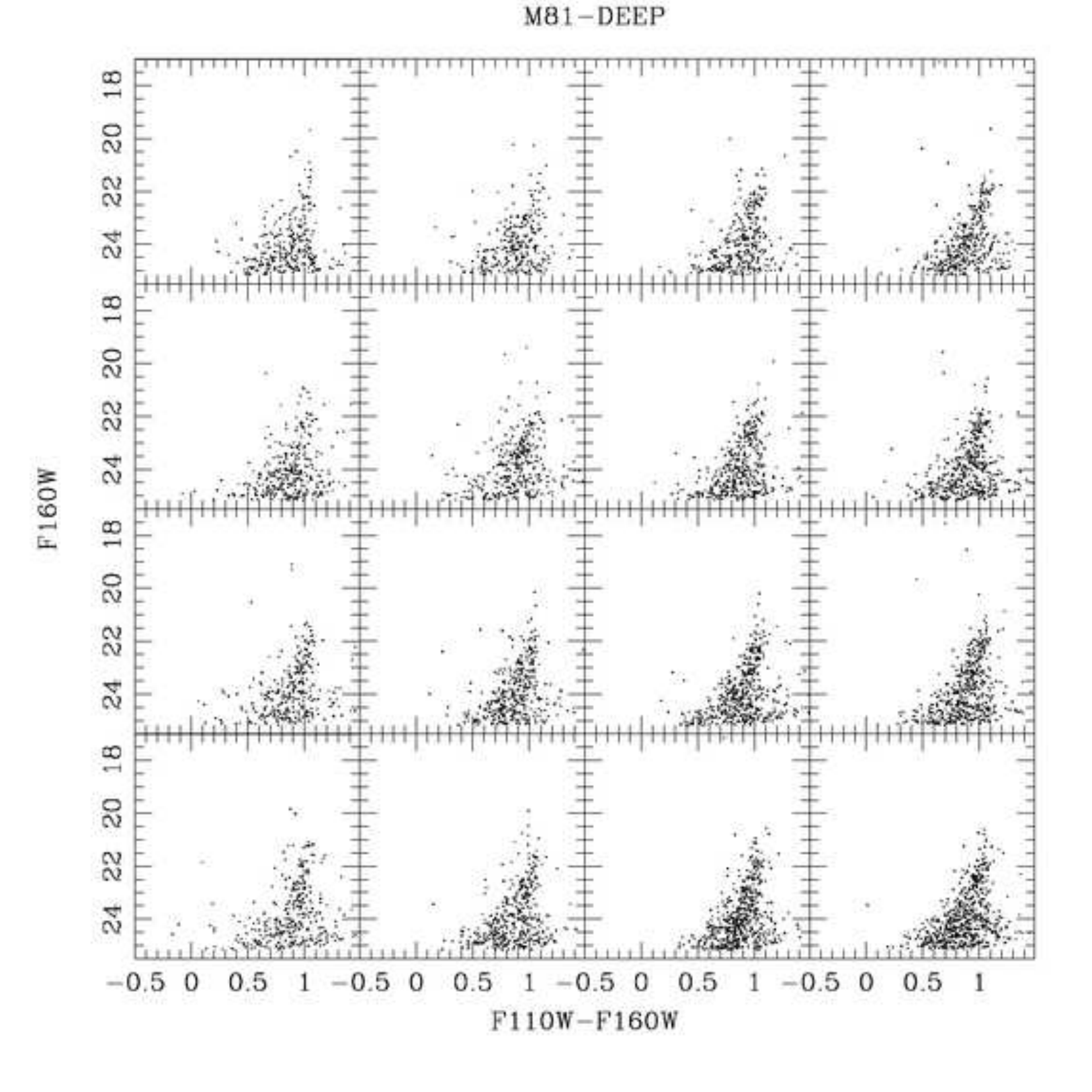}  
}
\centerline{
\includegraphics[width=3.25in]{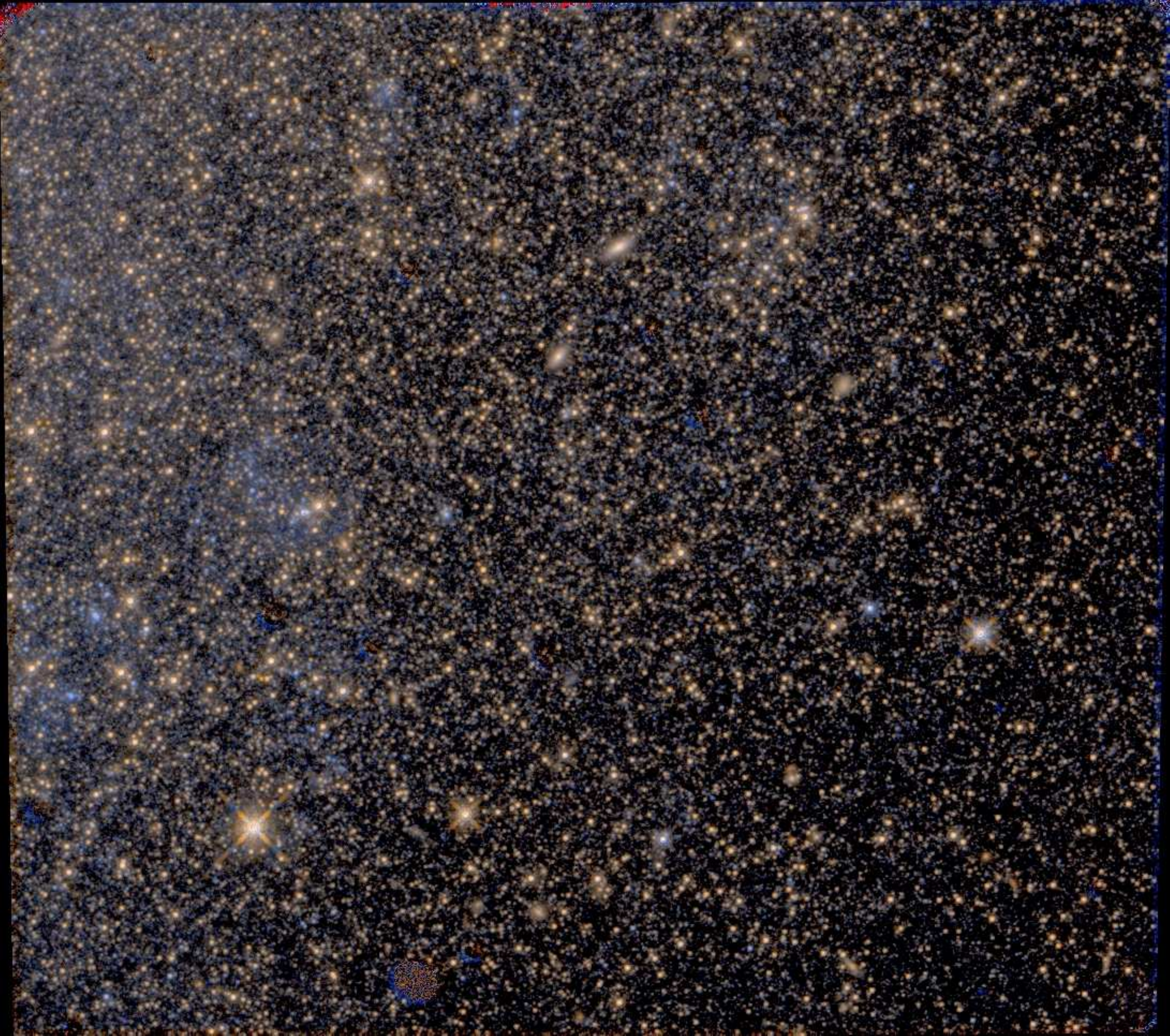}  
\includegraphics[width=3.25in]{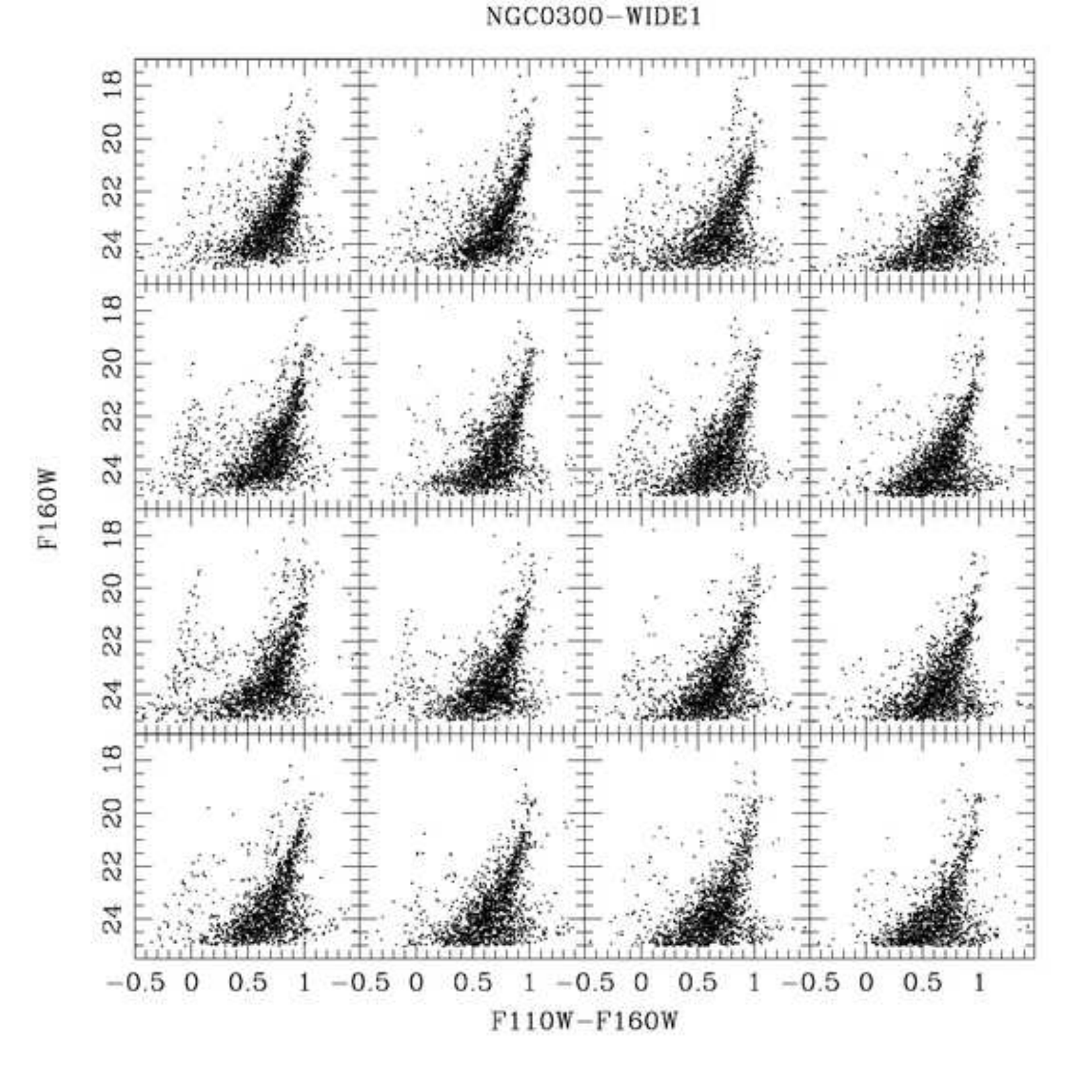}  
}
\caption{ (Left) False color $F110W+F160W$ image of the WFC3/IR field
  for the target M81-DEEP within the galaxy M81 [Top] and the target
  NGC0300-WIDE1 within the galaxy N300 [Bottom].  (Right)
  Color-magnitude diagrams generated for a grid of subregions, so that
  the upper left CMD corresponds to the upper left of the adjacent
  image.  }
\end{figure}
\vfill
\clearpage
 
\begin{figure}
\figurenum{\ref{gridfig} continued}
\centerline{
\includegraphics[width=3.25in]{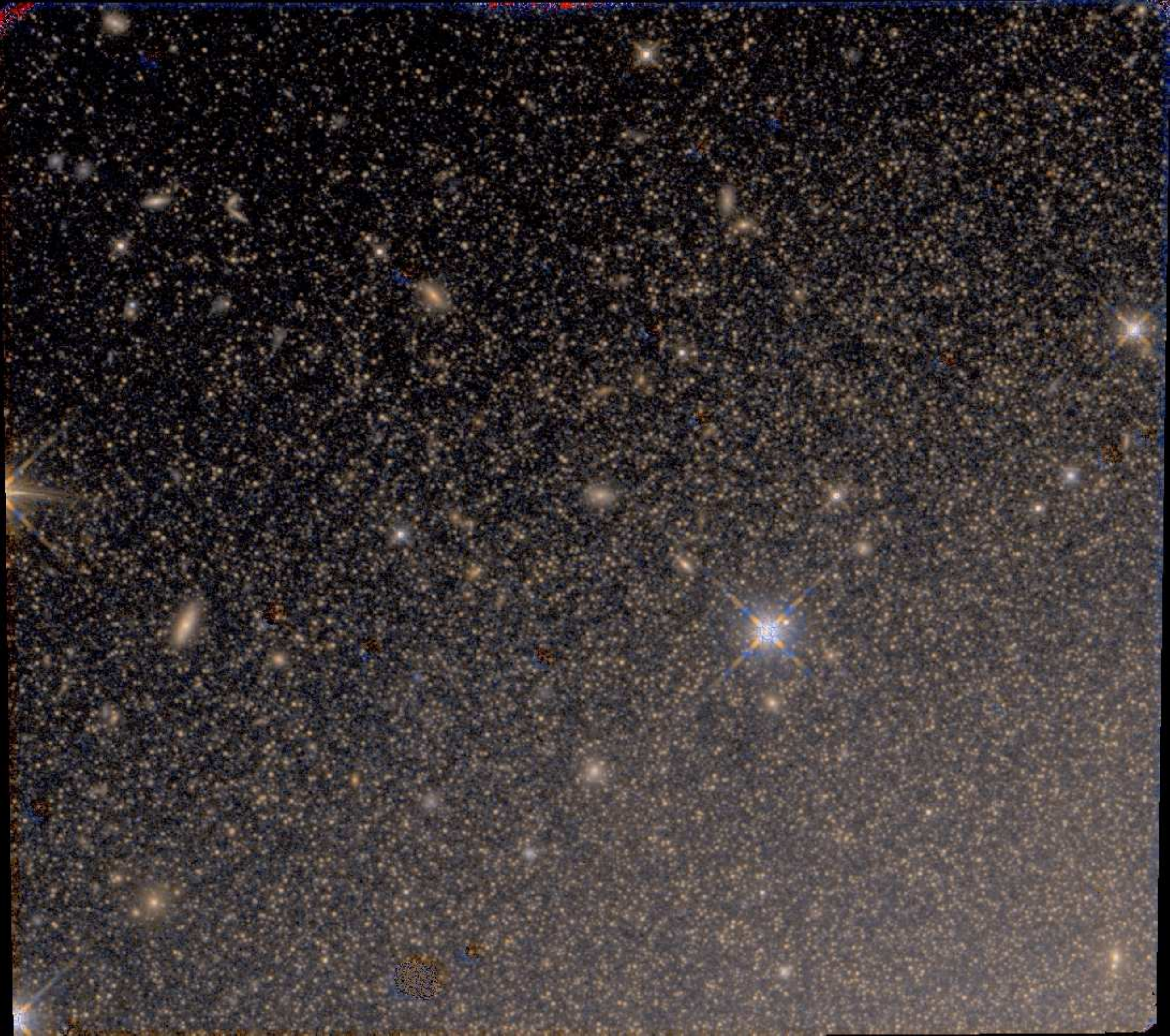}  
\includegraphics[width=3.25in]{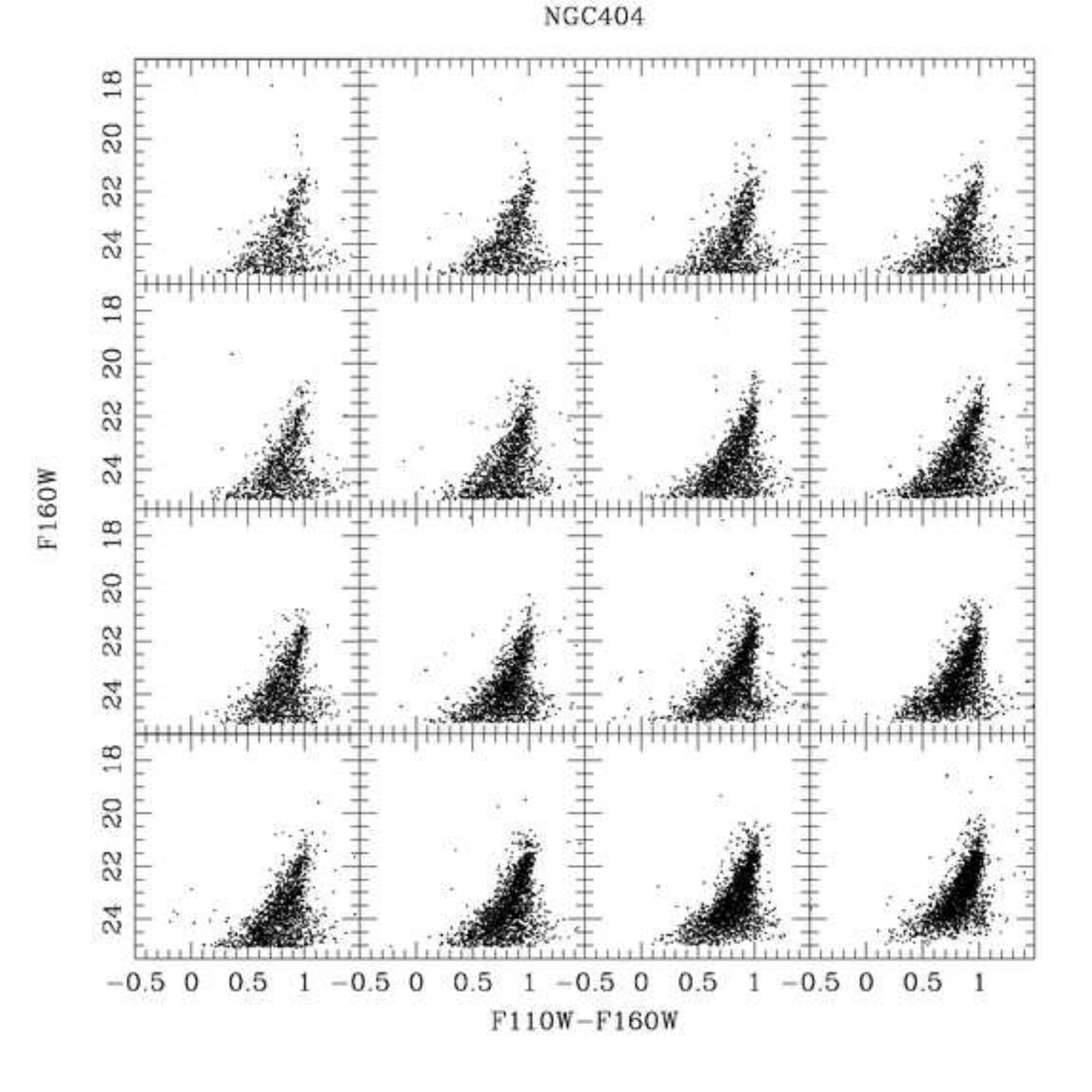}  
}
\centerline{
\includegraphics[width=3.25in]{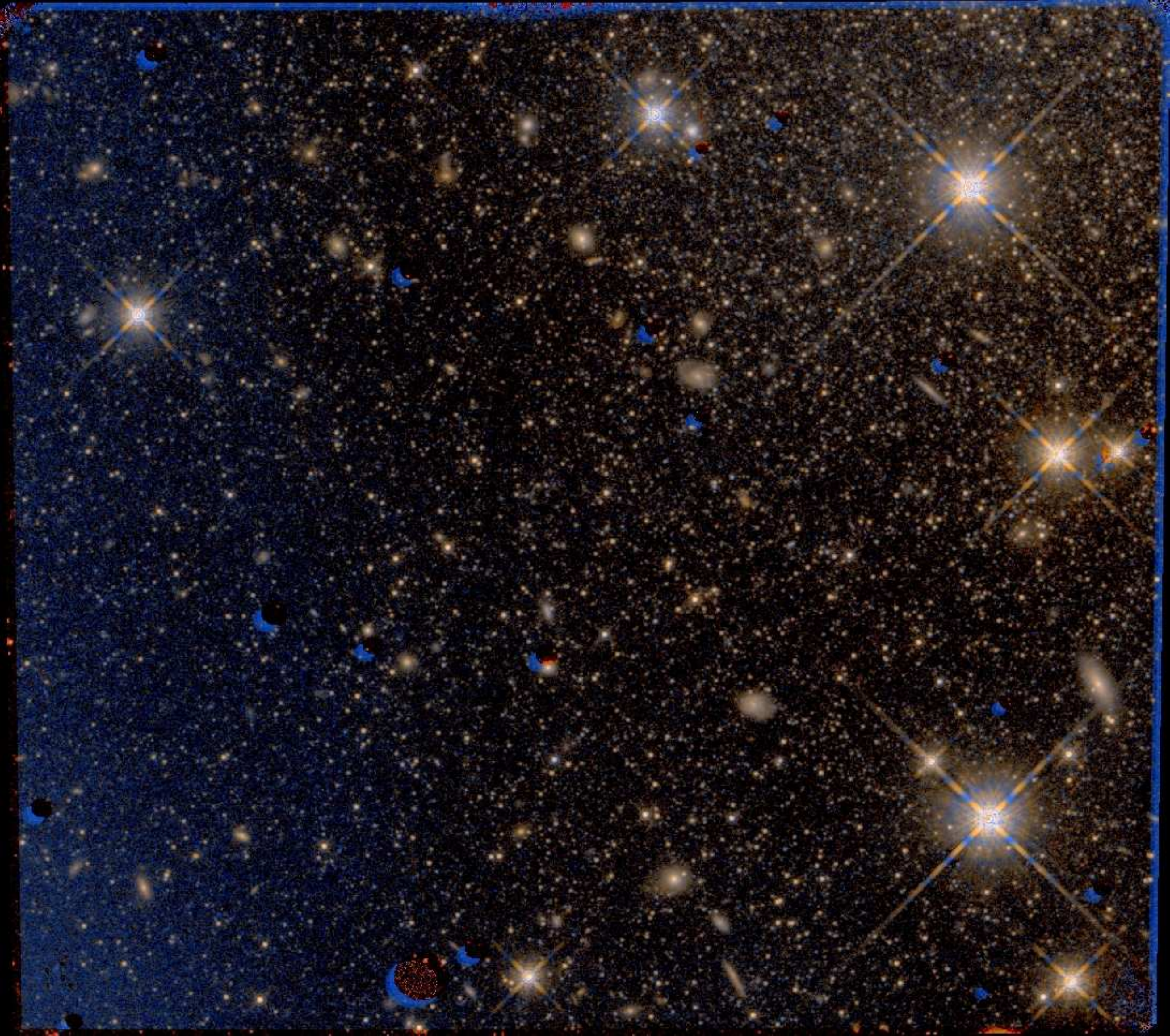}  
\includegraphics[width=3.25in]{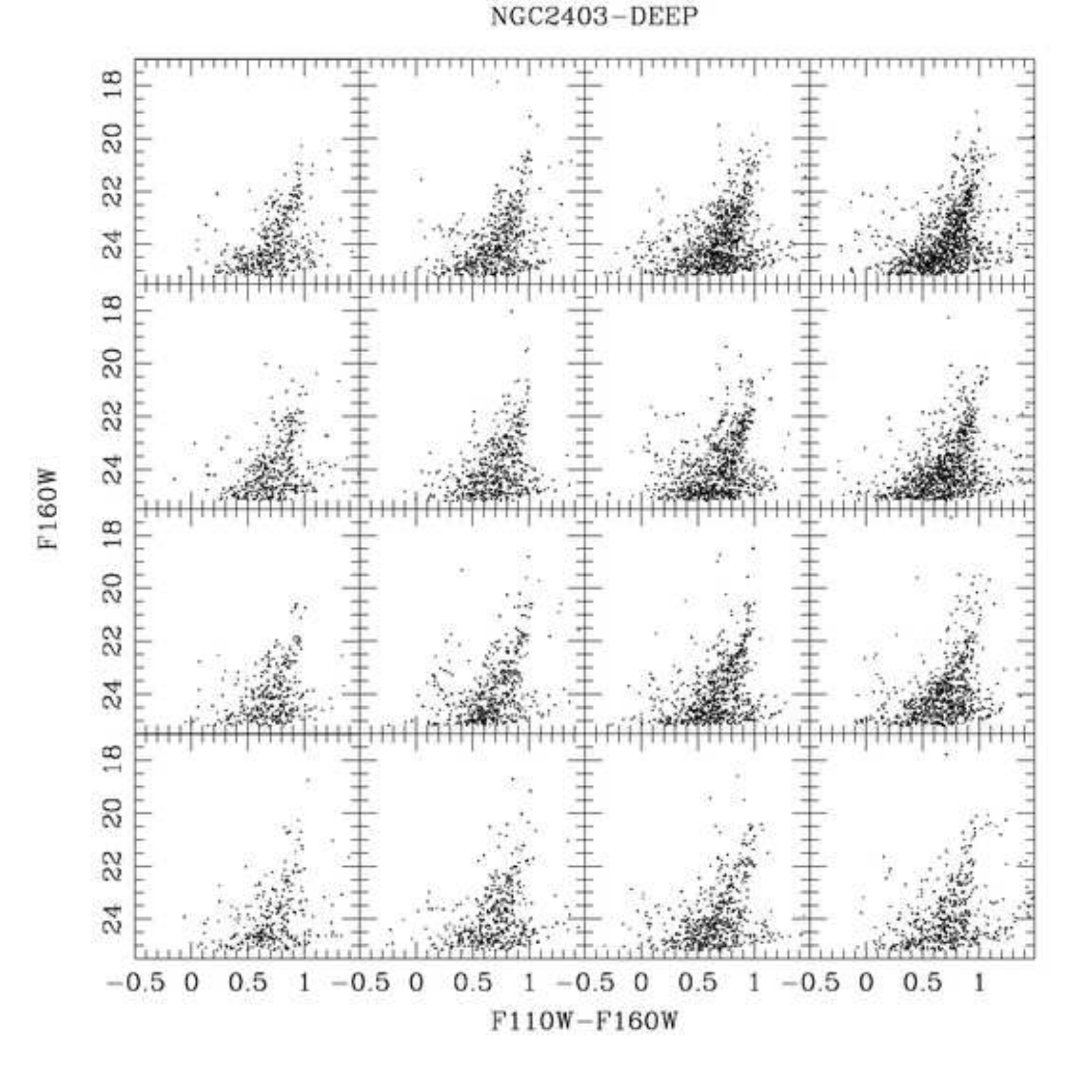}  
}
\caption{ (Left) False color $F110W+F160W$ image of the WFC3/IR field
  for the target NGC404 within the galaxy N404 [Top] and the target
  NGC2403-DEEP within the galaxy N2403 [Bottom].  (Right)
  Color-magnitude diagrams generated for a grid of subregions, so that
  the upper left CMD corresponds to the upper left of the adjacent
  image.  }
\end{figure}
\vfill
\clearpage
 
\begin{figure}
\figurenum{\ref{gridfig} continued}
\centerline{
\includegraphics[width=3.25in]{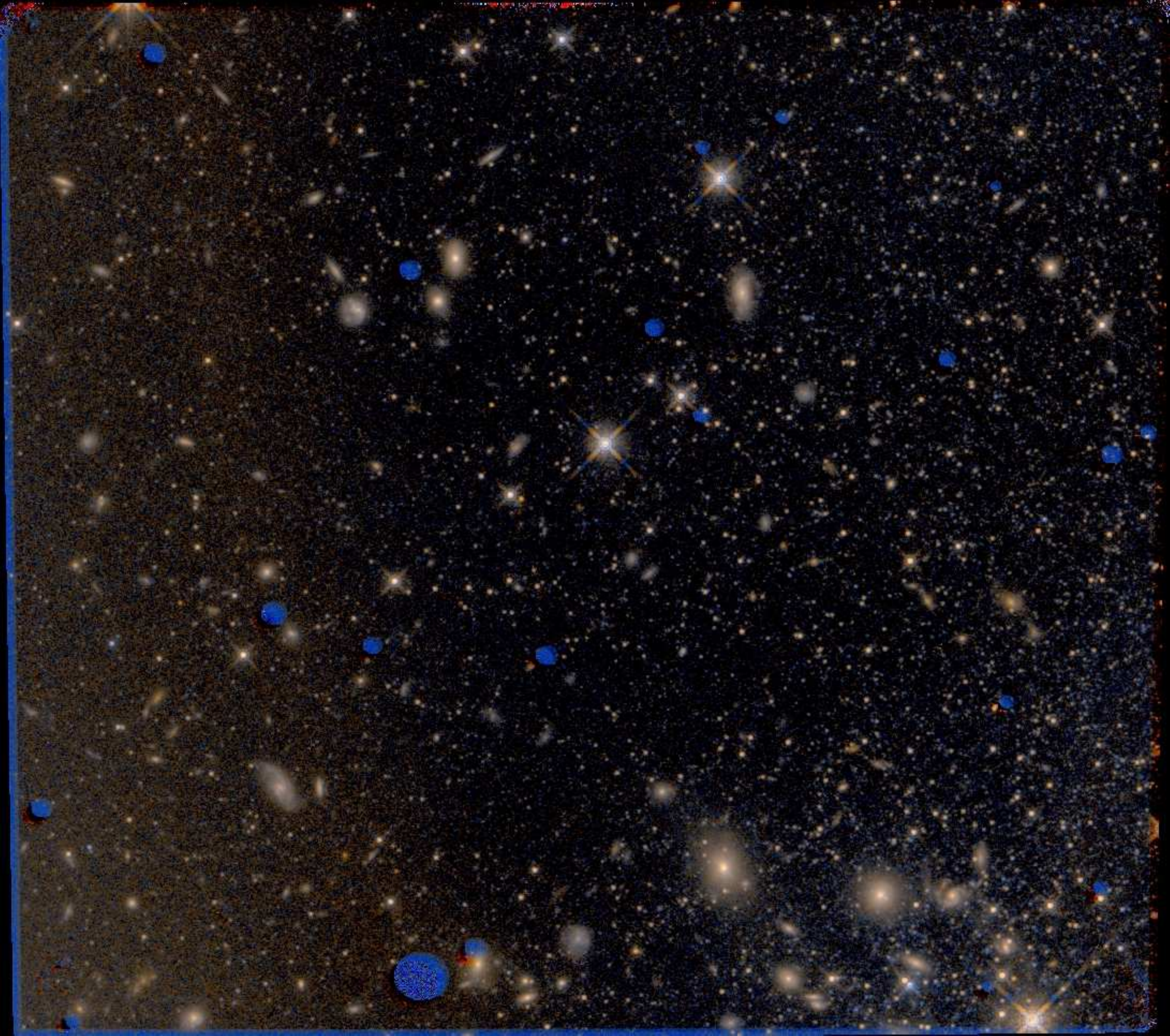}  
\includegraphics[width=3.25in]{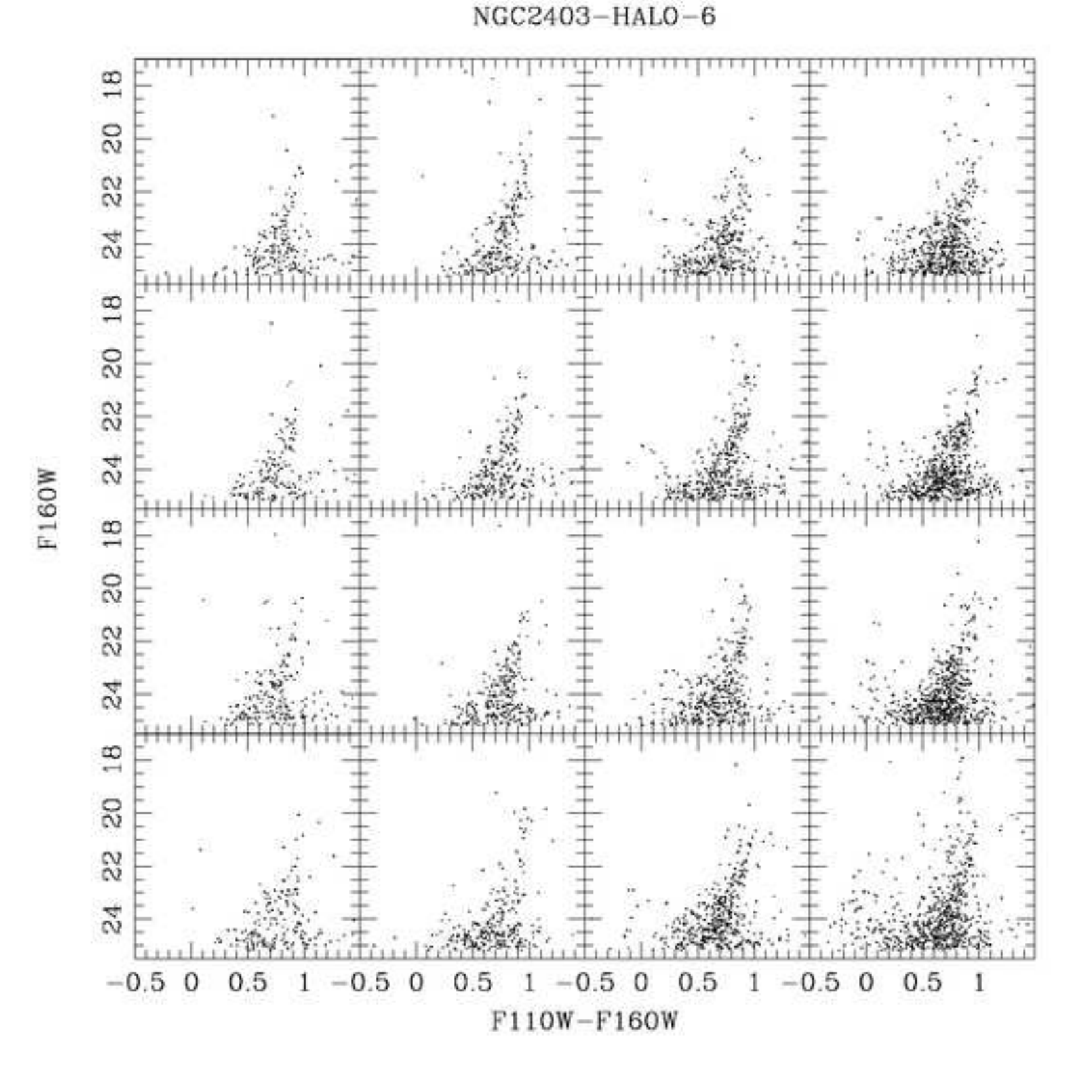}  
}
\centerline{
\includegraphics[width=3.25in]{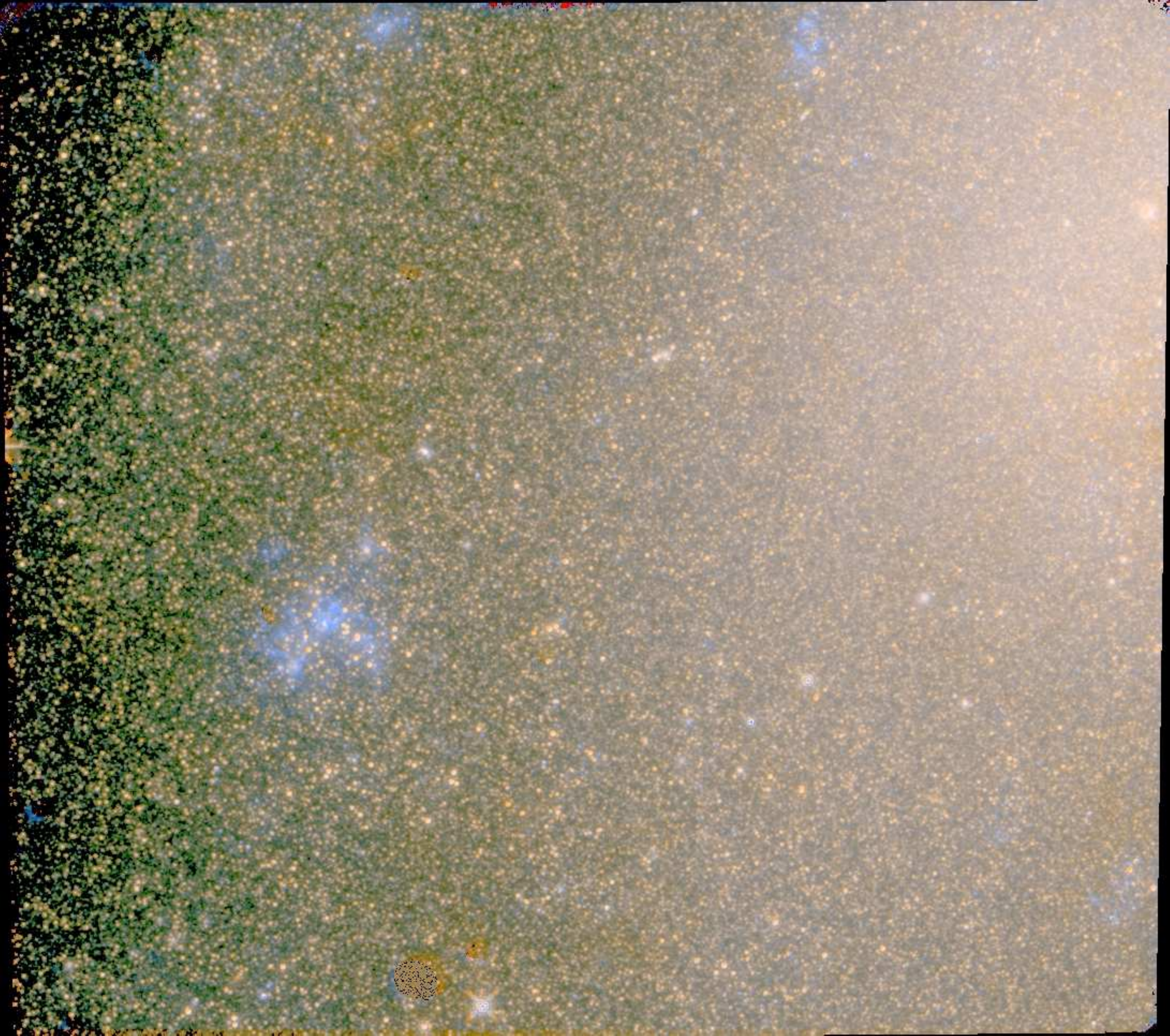}  
\includegraphics[width=3.25in]{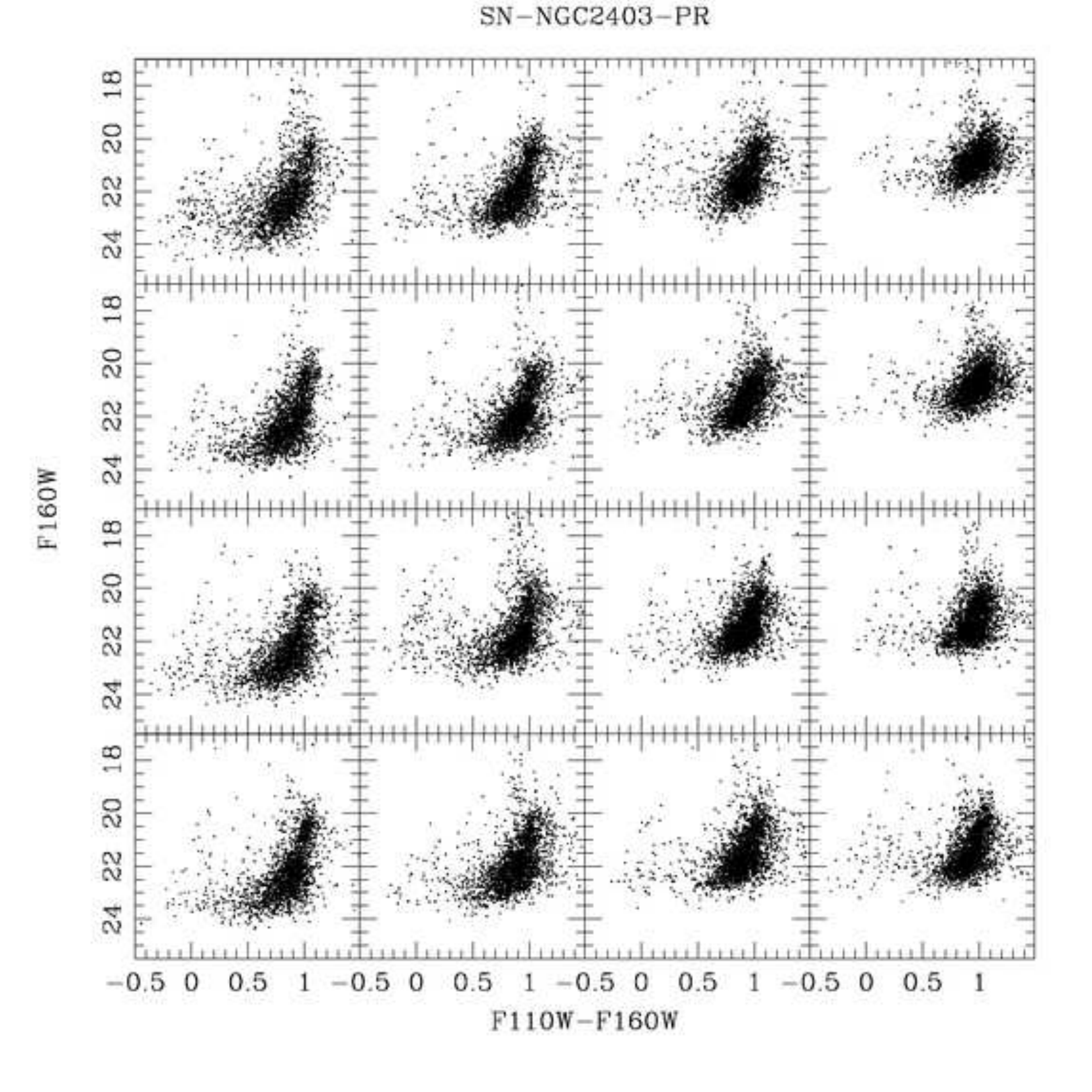}  
}
\caption{ (Left) False color $F110W+F160W$ image of the WFC3/IR field
  for the target NGC2403-HALO-6 within the galaxy N2403 [Top] and the
  target SN-NGC2403-PR within the galaxy N2403 [Bottom].  (Right)
  Color-magnitude diagrams generated for a grid of subregions, so that
  the upper left CMD corresponds to the upper left of the adjacent
  image.  }
\end{figure}
\vfill
\clearpage
 
\begin{figure}
\figurenum{\ref{gridfig} continued}
\centerline{
\includegraphics[width=3.25in]{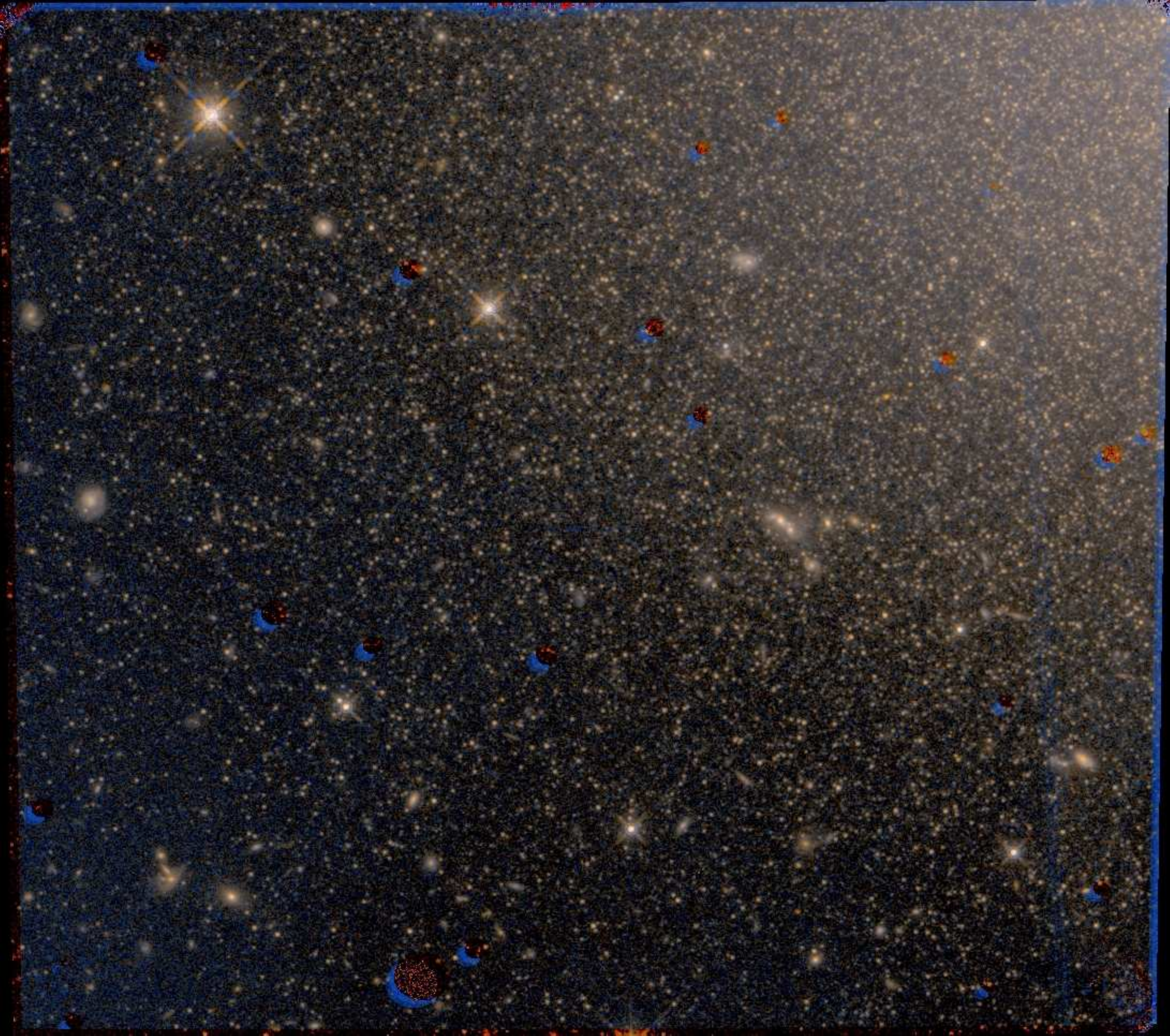}  
\includegraphics[width=3.25in]{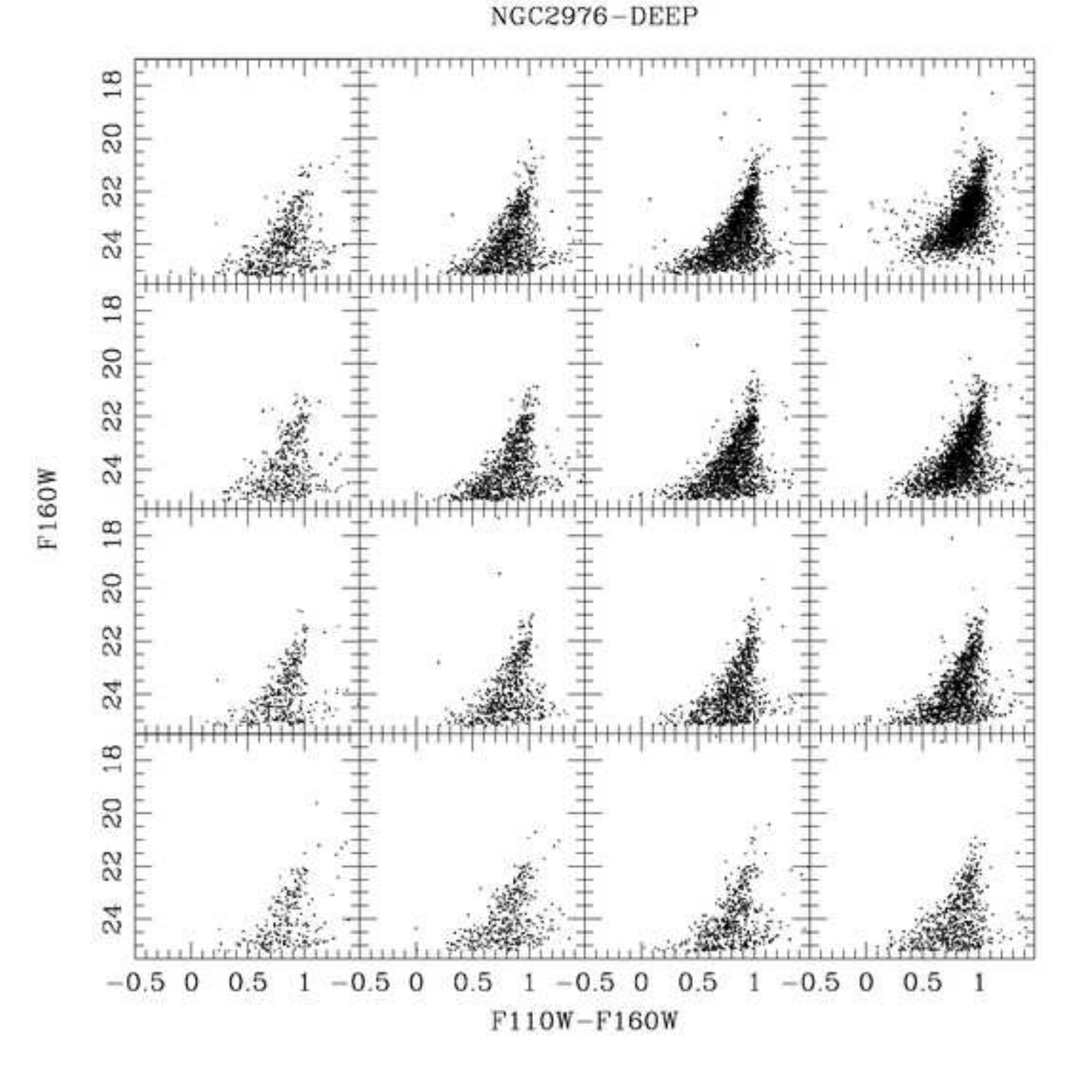}  
}
\centerline{
\includegraphics[width=3.25in]{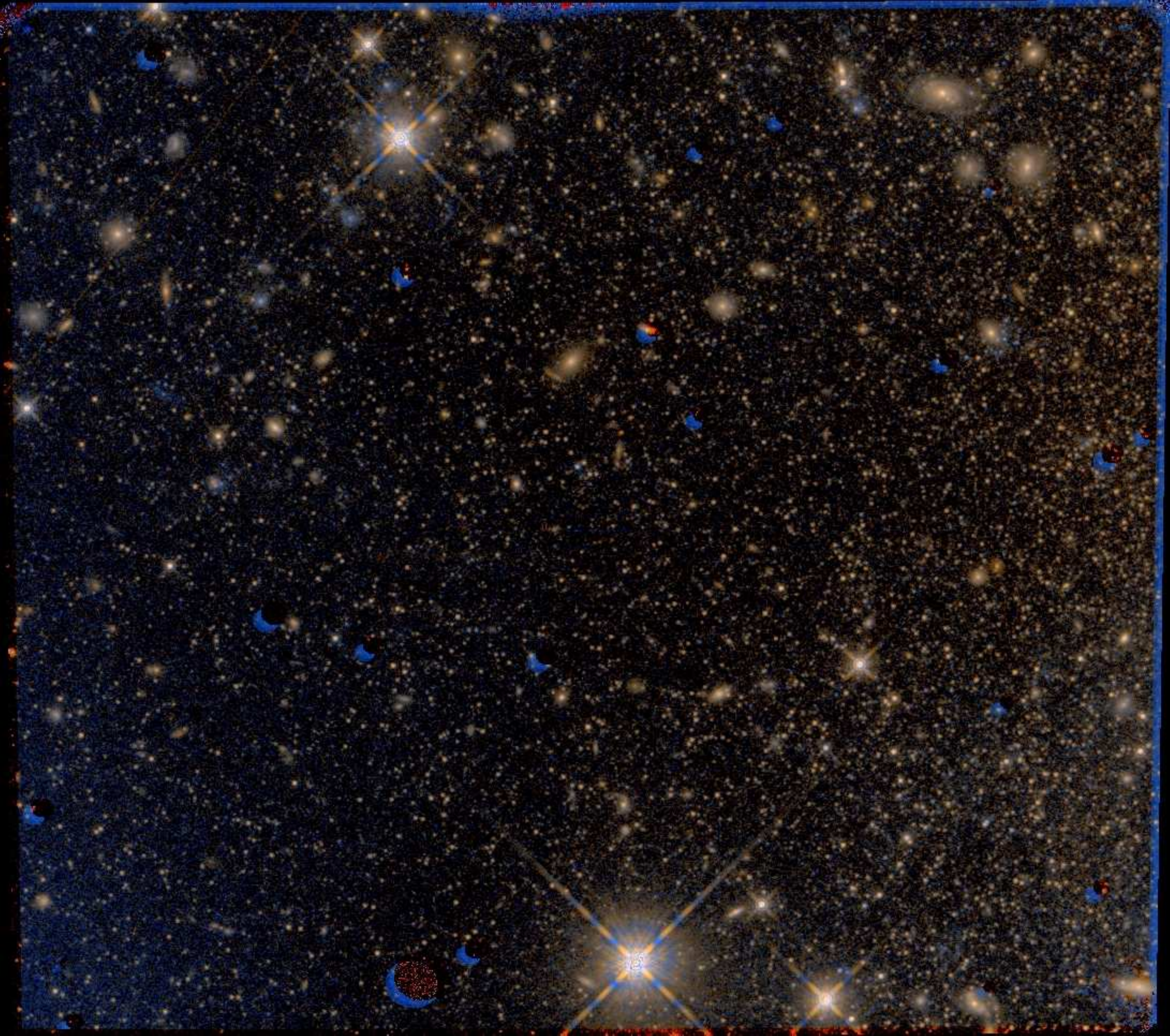}  
\includegraphics[width=3.25in]{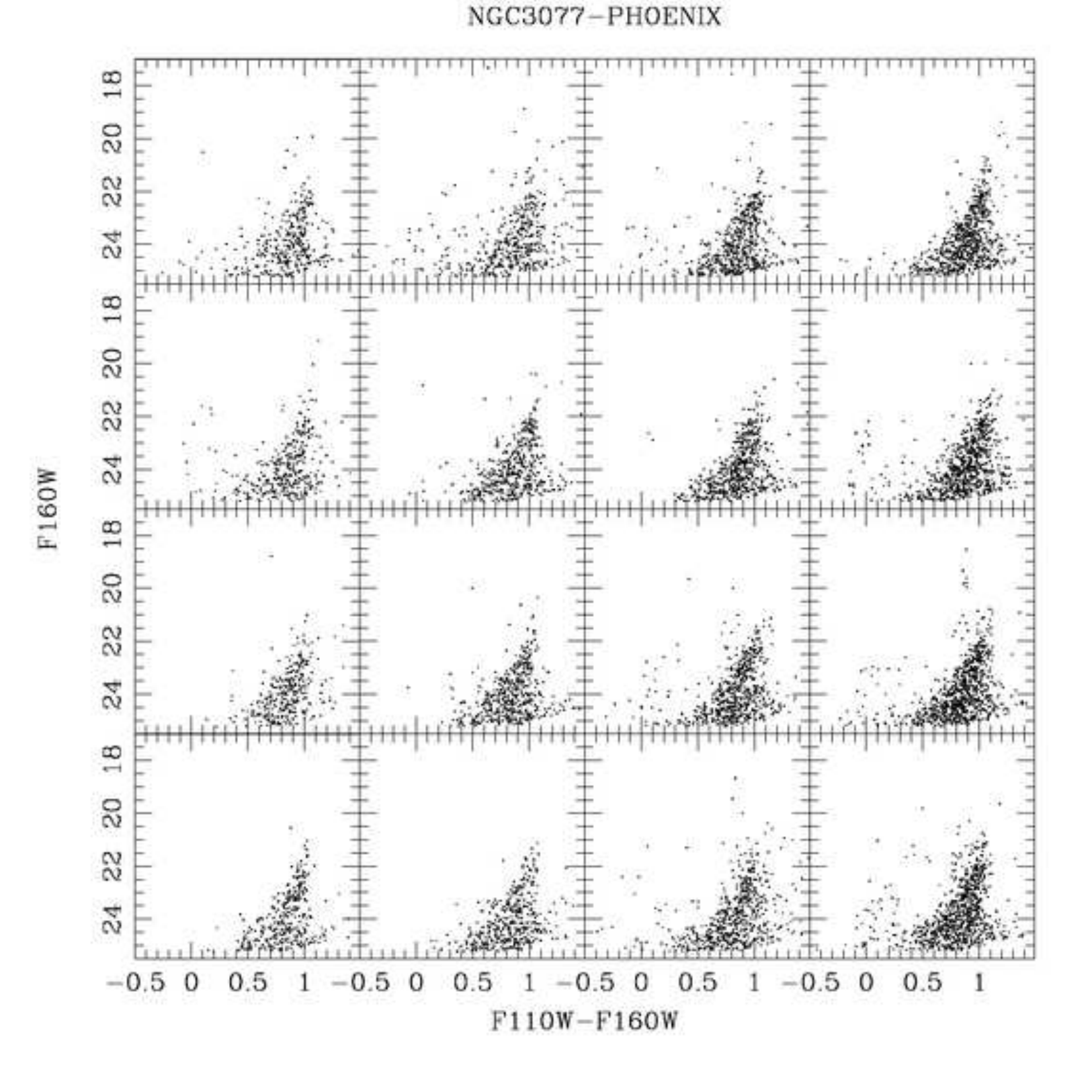}  
}
\caption{ (Left) False color $F110W+F160W$ image of the WFC3/IR field
  for the target NGC2976-DEEP within the galaxy N2976 [Top] and the
  target NGC3077-PHOENIX within the galaxy N3077 [Bottom].  (Right)
  Color-magnitude diagrams generated for a grid of subregions, so that
  the upper left CMD corresponds to the upper left of the adjacent
  image.  }
\end{figure}
\vfill
\clearpage
 
\begin{figure}
\figurenum{\ref{gridfig} continued}
\centerline{
\includegraphics[width=3.25in]{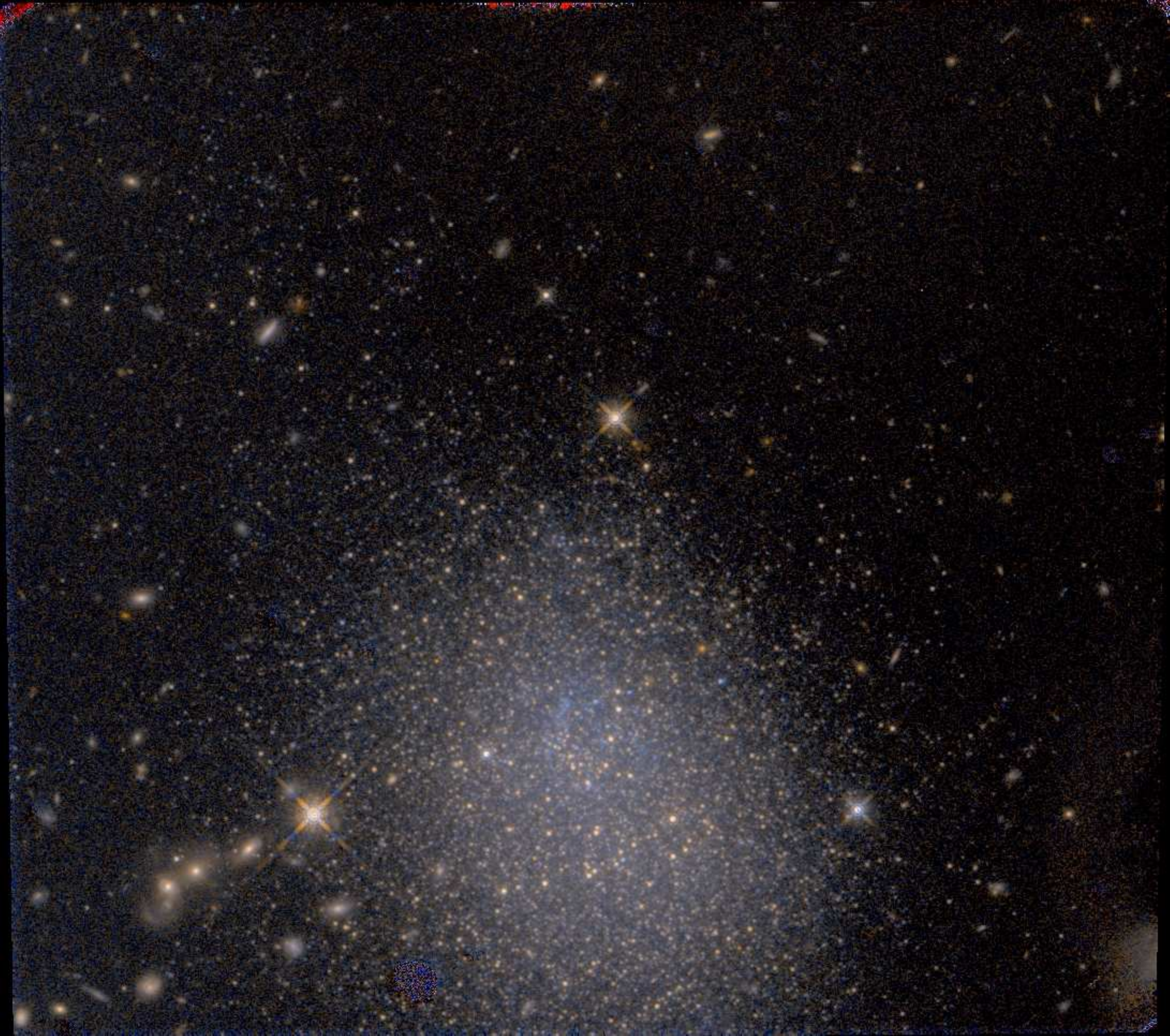}  
\includegraphics[width=3.25in]{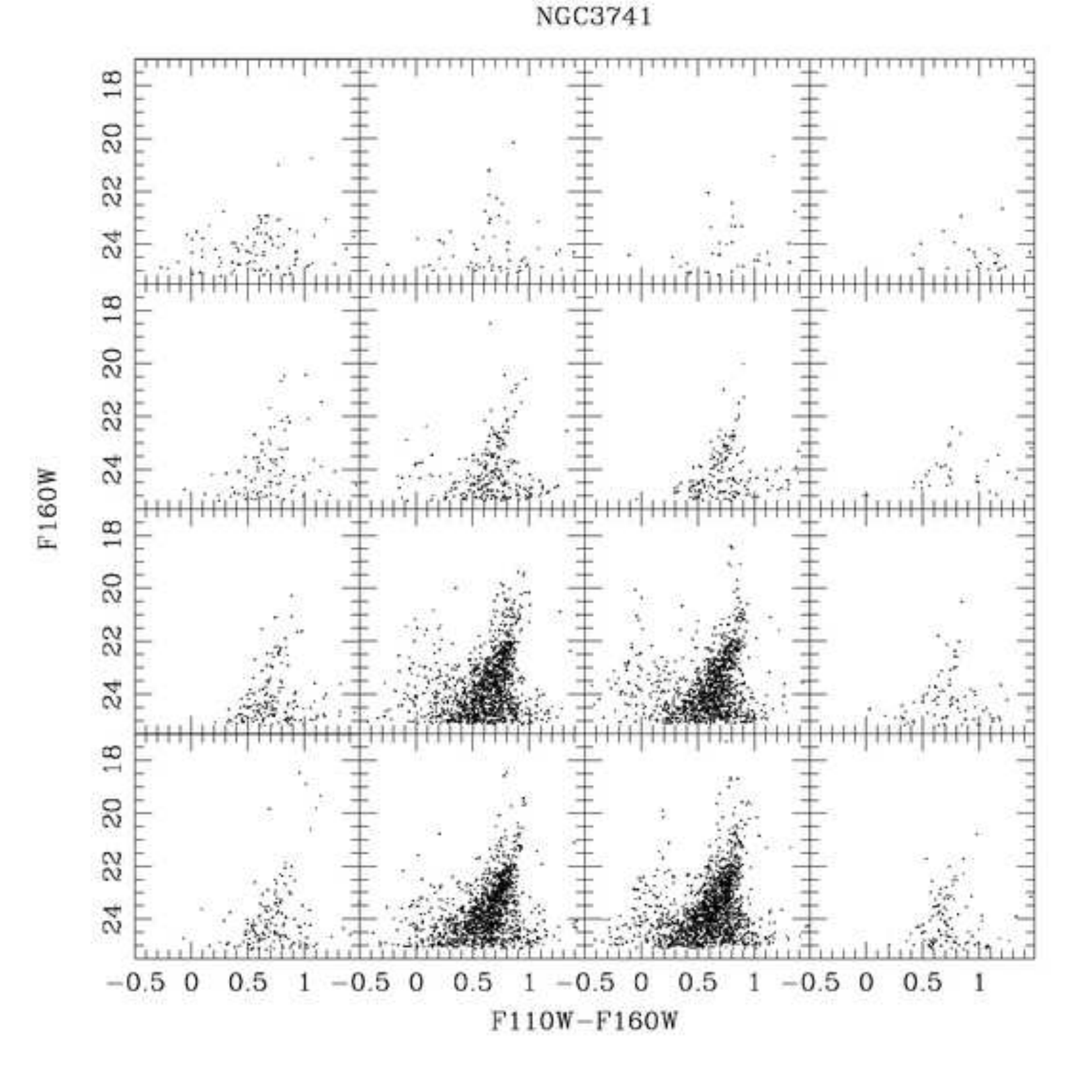}  
}
\centerline{
\includegraphics[width=3.25in]{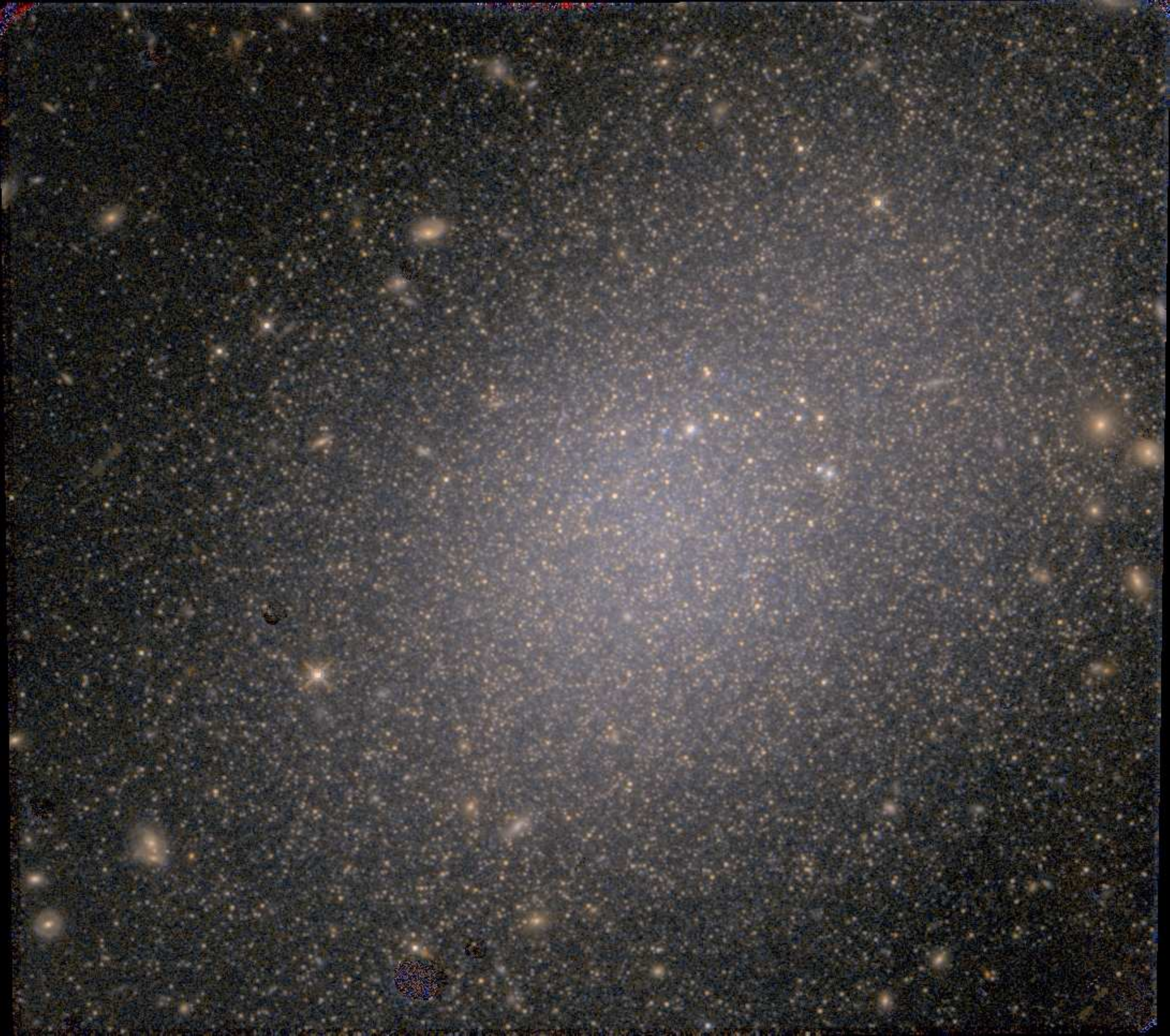}  
\includegraphics[width=3.25in]{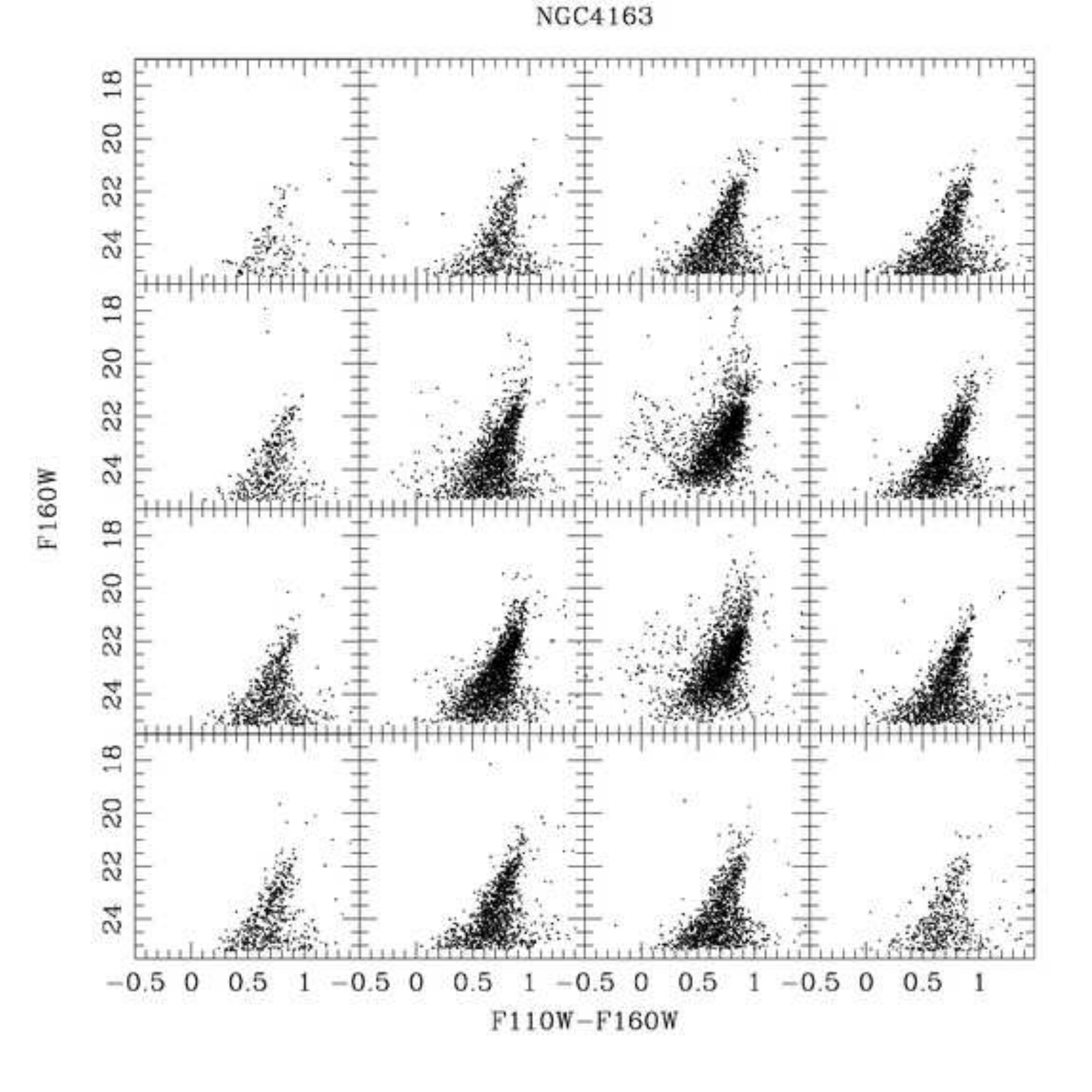}  
}
\caption{ (Left) False color $F110W+F160W$ image of the WFC3/IR field
  for the target NGC3741 within the galaxy N3741 [Top] and the target
  NGC4163 within the galaxy N4163 [Bottom].  (Right) Color-magnitude
  diagrams generated for a grid of subregions, so that the upper left
  CMD corresponds to the upper left of the adjacent image.  }
\end{figure}
\vfill
\clearpage
 
\begin{figure}
\figurenum{\ref{gridfig} continued}
\centerline{
\includegraphics[width=3.25in]{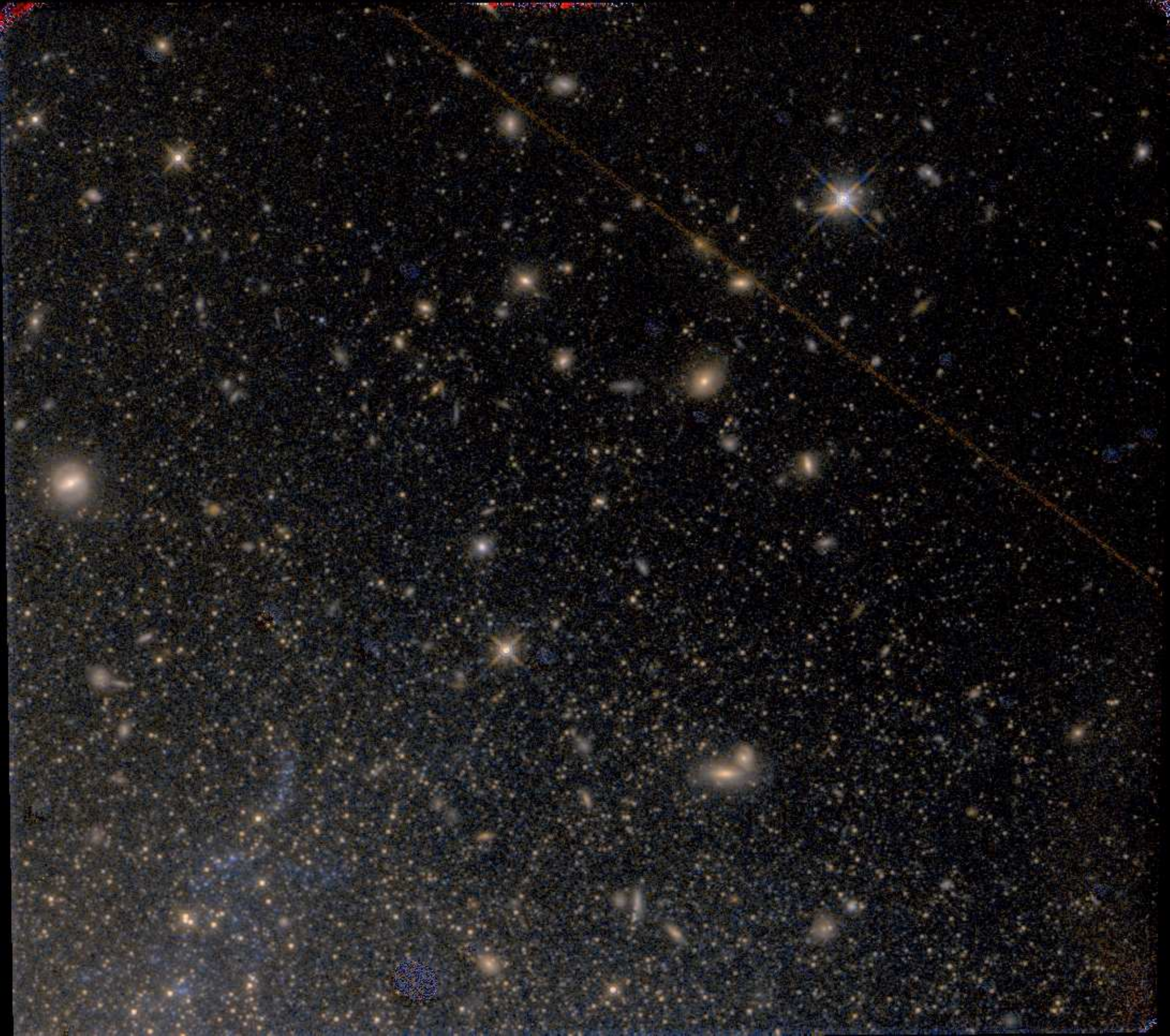}  
\includegraphics[width=3.25in]{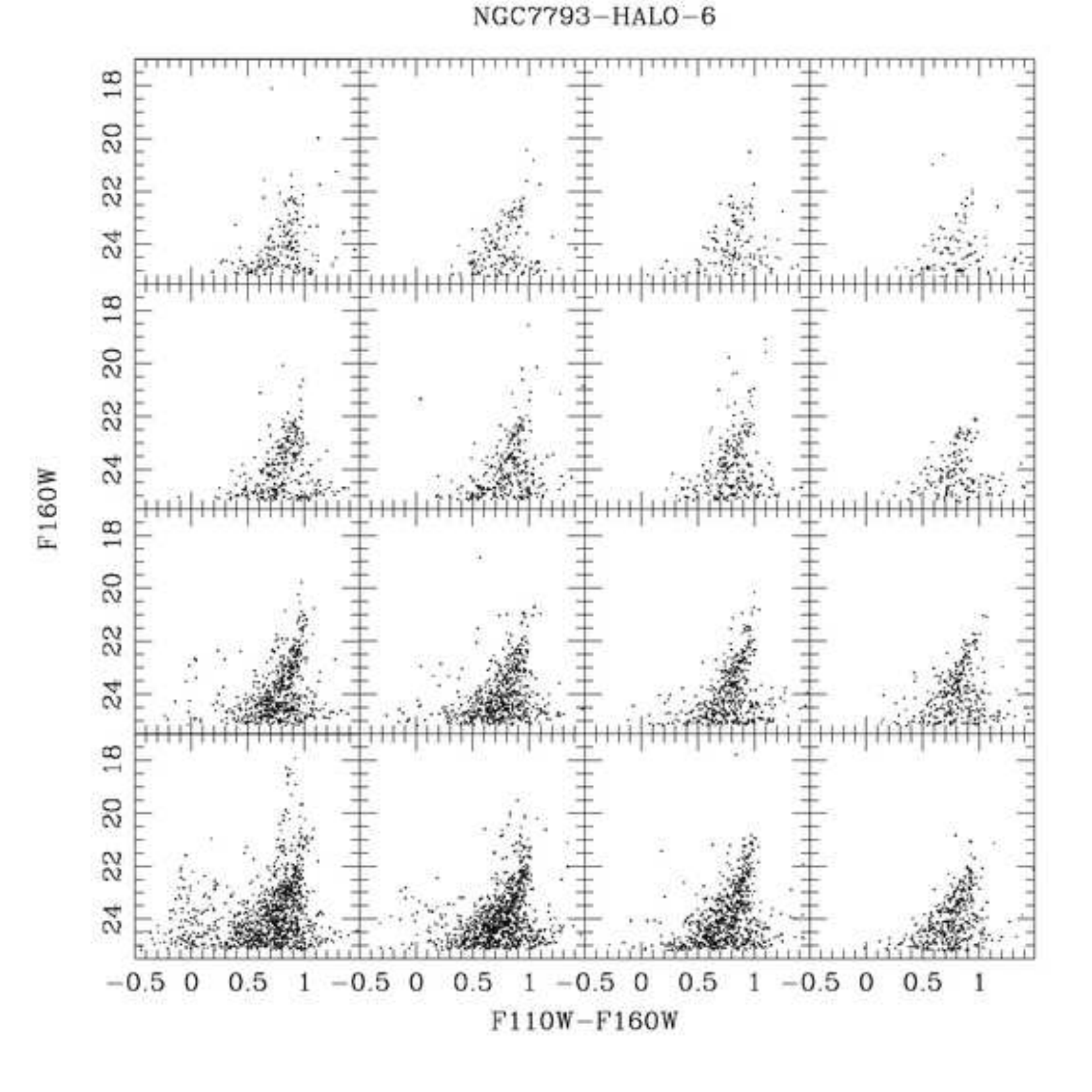}  
}
\centerline{
\includegraphics[width=3.25in]{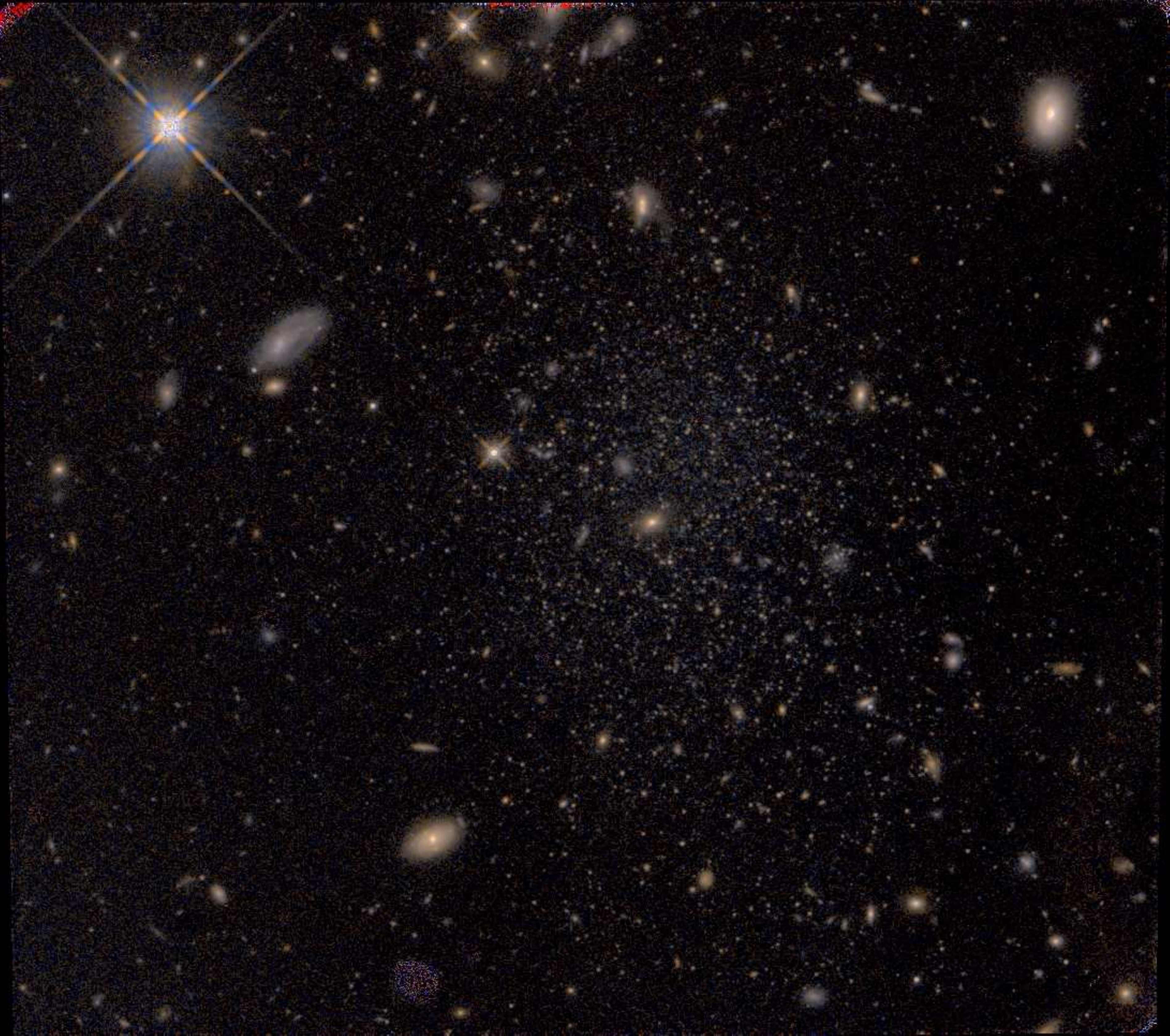}  
\includegraphics[width=3.25in]{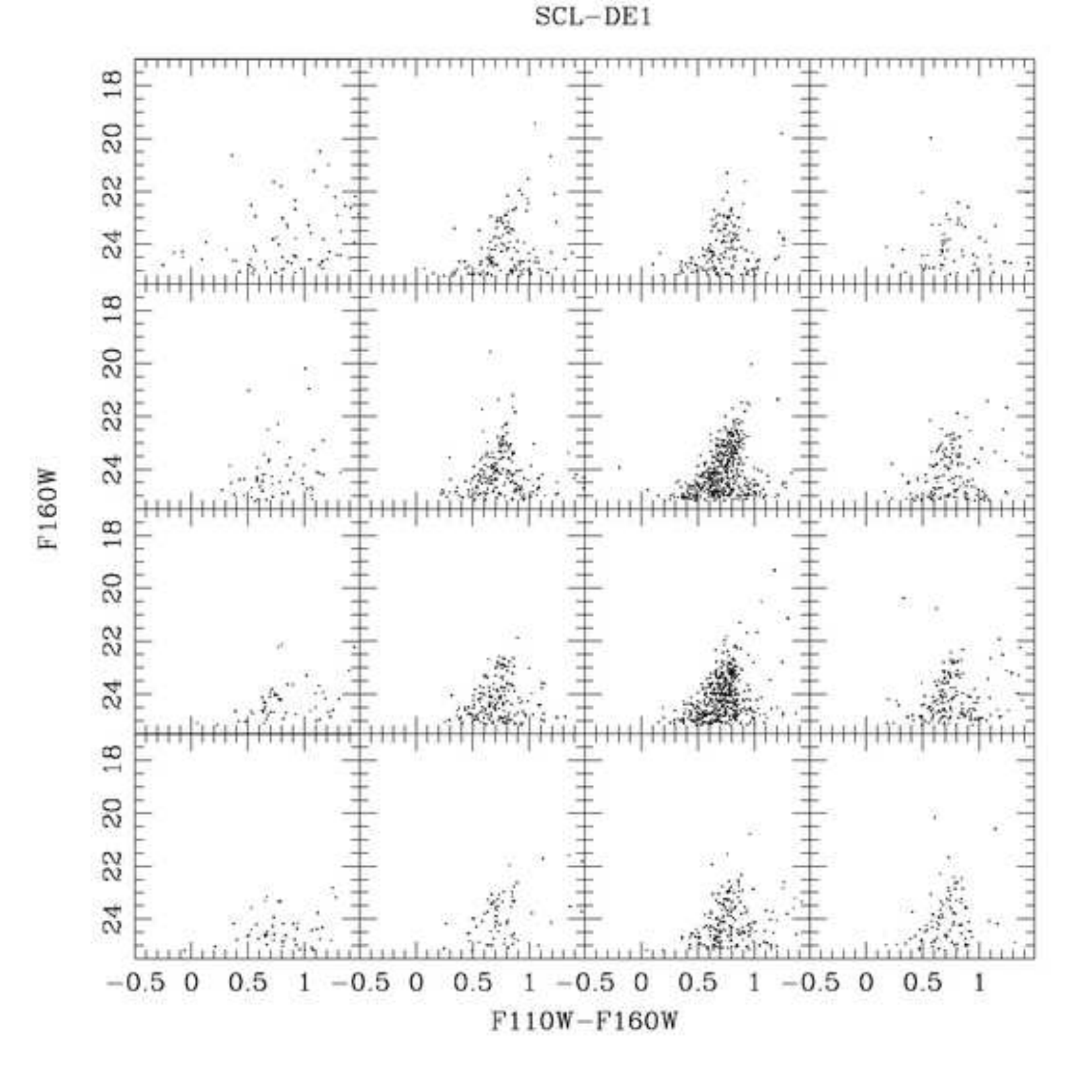}  
}
\caption{ (Left) False color $F110W+F160W$ image of the WFC3/IR field
  for the target NGC7793-HALO-6 within the galaxy N7793 [Top] and the
  target SCL-DE1 within the galaxy Sc22 [Bottom].  (Right)
  Color-magnitude diagrams generated for a grid of subregions, so that
  the upper left CMD corresponds to the upper left of the adjacent
  image.  }
\end{figure}
\vfill
\clearpage
 
\begin{figure}
\figurenum{\ref{gridfig} continued}
\centerline{
\includegraphics[width=3.25in]{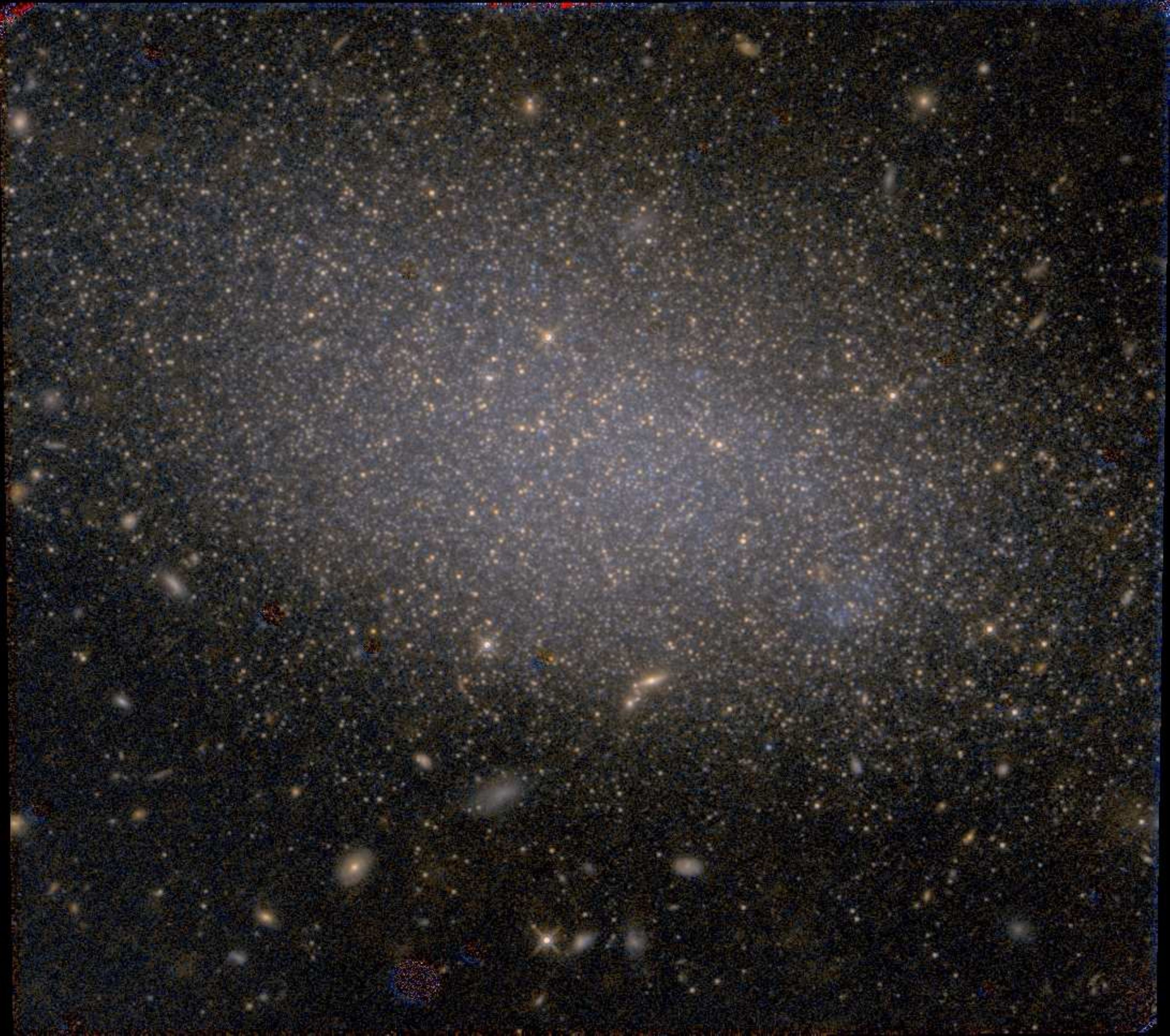}  
\includegraphics[width=3.25in]{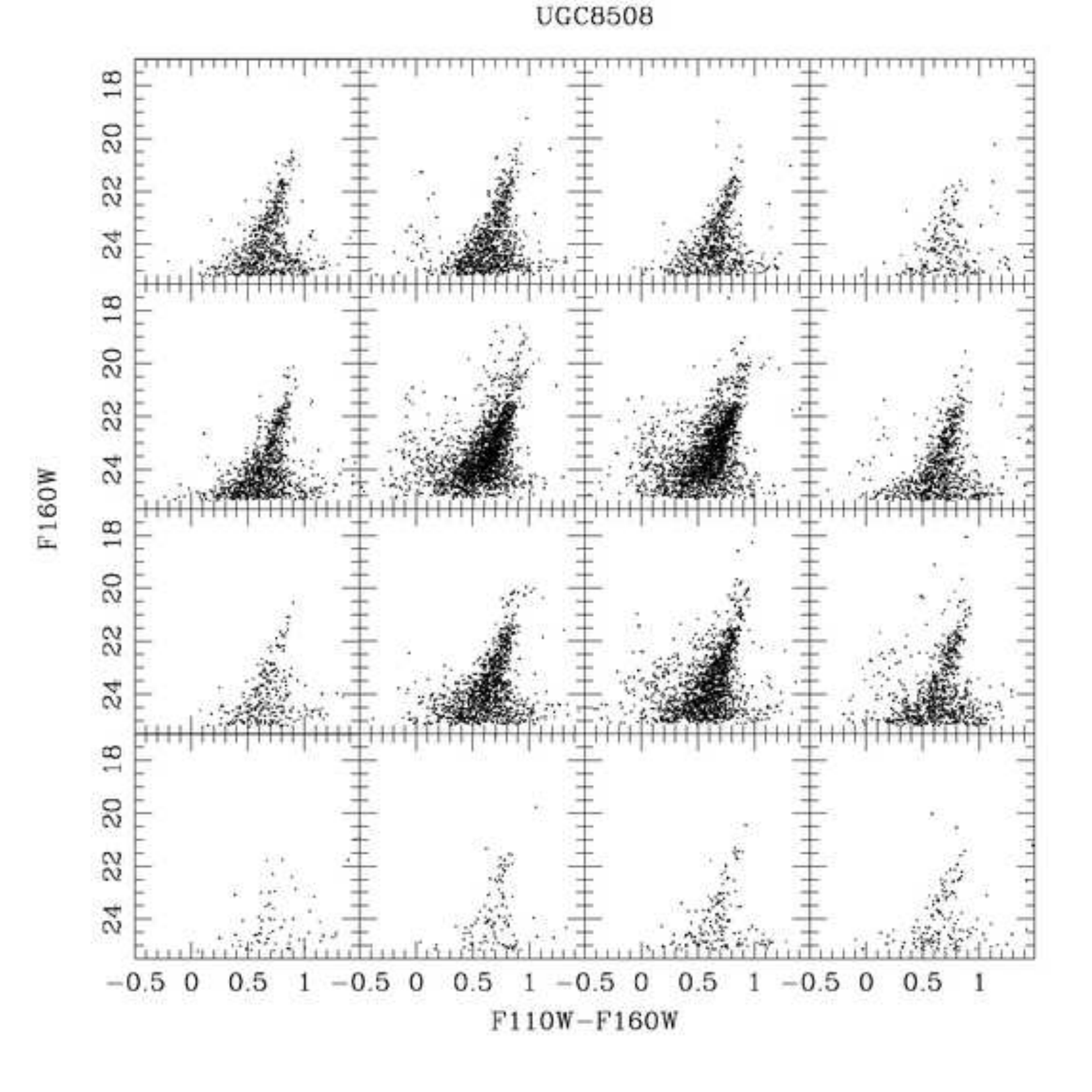}  
}
\centerline{
\includegraphics[width=3.25in]{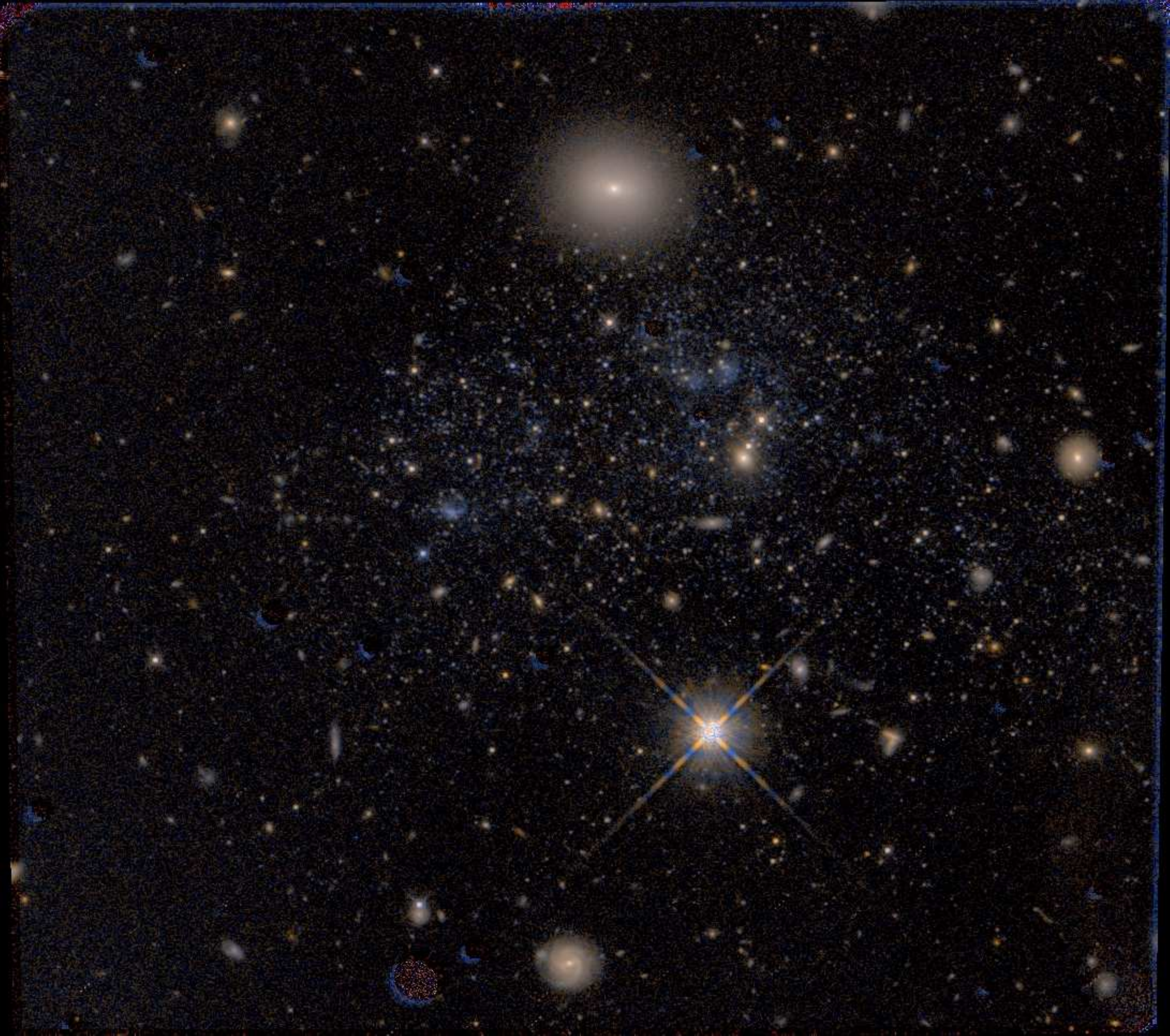}  
\includegraphics[width=3.25in]{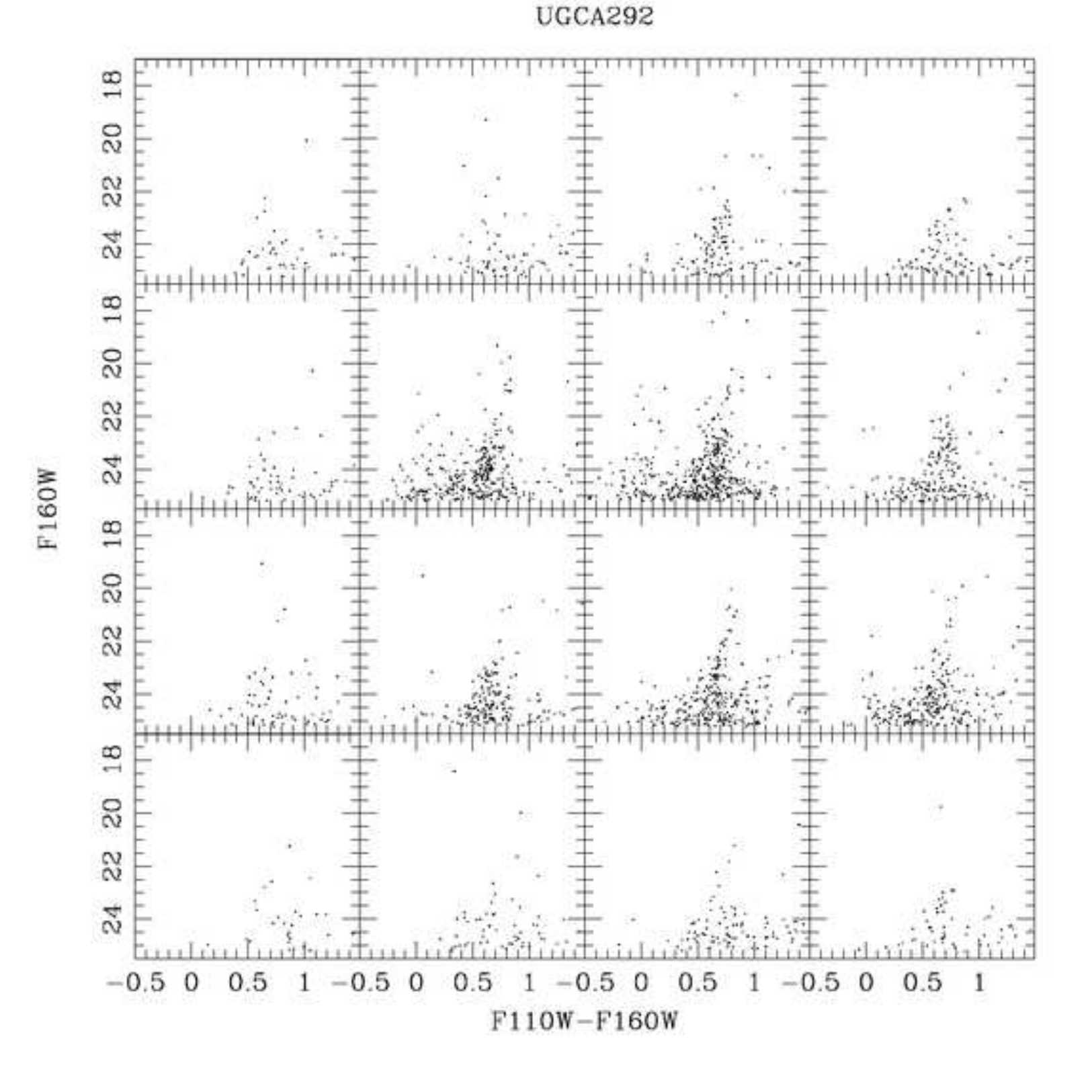}  
}
\caption{ (Left) False color $F110W+F160W$ image of the WFC3/IR field
  for the target UGC8508 within the galaxy U8508 [Top] and the target
  UGCA292 within the galaxy UA292 [Bottom].  (Right) Color-magnitude
  diagrams generated for a grid of subregions, so that the upper left
  CMD corresponds to the upper left of the adjacent image.  }
\end{figure}
\vfill
\clearpage

\begin{figure}
\centerline{
\includegraphics[width=6.25in]{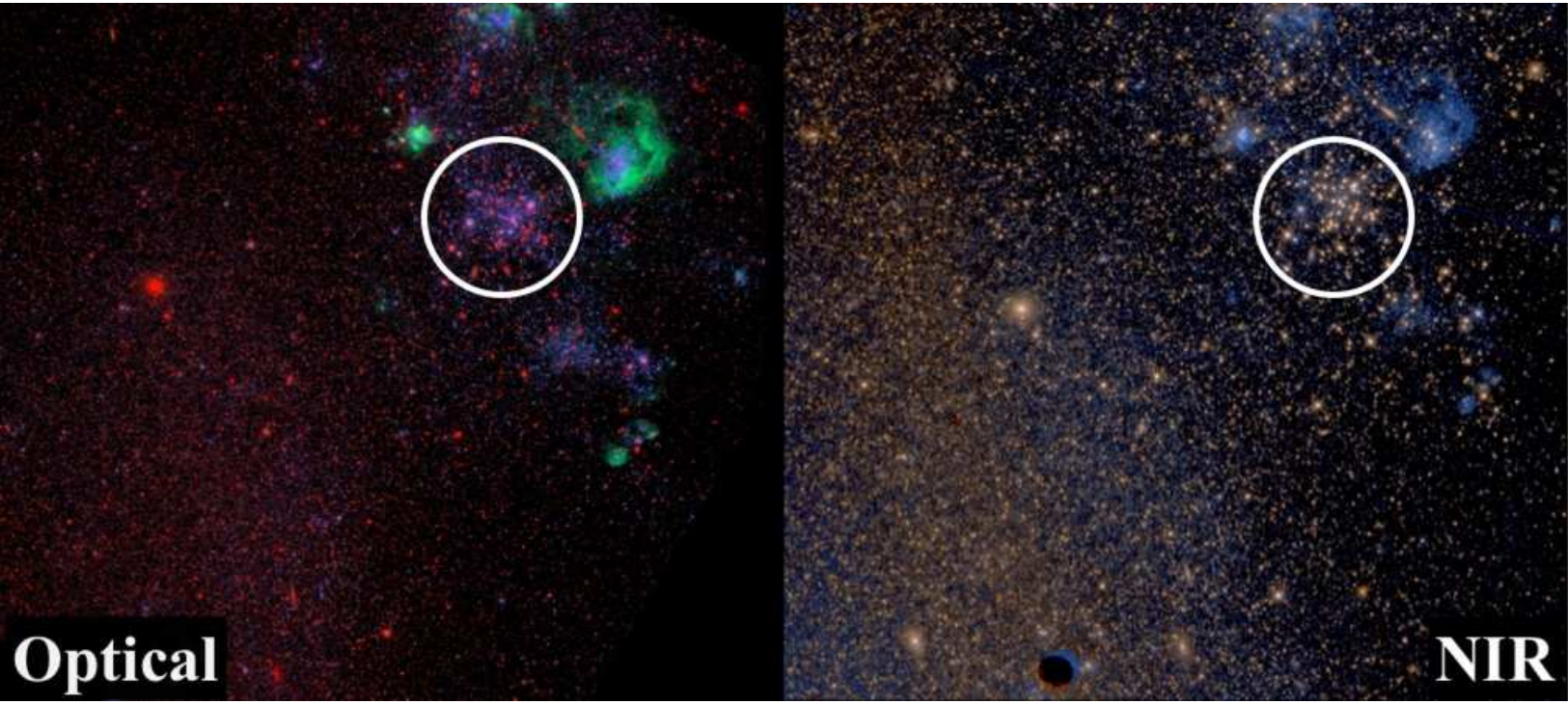}
}
\caption{
Comparison between a false-color image in the optical (left:
$F435W+F555W+F814W$) and the NIR (right: $F110W+F160W$) in a star
forming region of I2574.  Nebulosity associated with line emission
from ionized [S{\sc{iii}}] and netural He is visible near H{\sc{ii}}
regions in the NIR image.  The circle indicates a compact star
formation region dominated by luminous red core-Helium burning (RHeB)
stars.  This SF region is old enough that the photoionized H{\sc ii}
has recombined ($\gtrsim10\Myrs$), and therefore lacks the diffuse
emission seen in the younger neighboring HII regions.  However, it
also hosts all of the most NIR luminous sources (see discussion in
Section~\ref{RHeBsec}).
\label{iroptrgbfig}}
\end{figure}
\vfill

\begin{figure}
\centerline{
\includegraphics[width=6.25in]{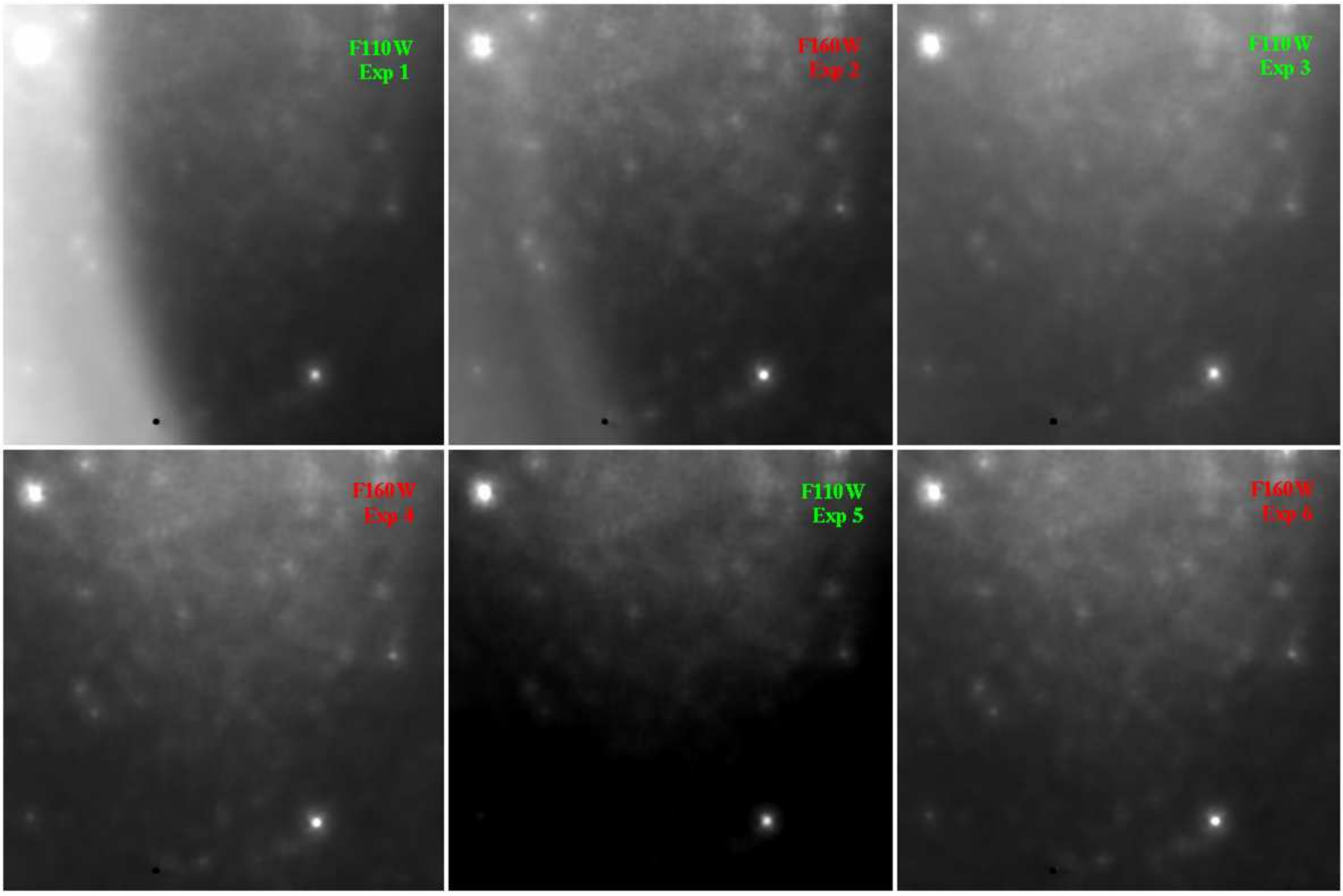}
}
\caption{
DOLPHOT's estimate of the local sky for a sequence of exposures taken during
a single orbit visit for UGC4305-1, based on DOLPHOT's PSF fitting.
The first exposure in the orbit is in the upper left, and the final
exposure is in the lower right.  Exposures alternate between $F110W$
and $F160W$, and all images for a single filter have been displayed
with a common scale and stretch.  There is an overall gradient from
top to bottom in all images, due to a larger fraction of unresolved
sources in the upper half of the image.  There is also a
time-dependent feature on the left hand side of the image, due to
scattered light. We show in Figure~\ref{limbangfig} that this feature
appears when the telescope is pointed to within $30\degree$ of a
bright earth limb; the first two exposures in the orbit had average
limb angles of $22\degree$ and $26\degree$, respectively, while the
third through sixth exposures have limb angles of $31\degree$,
$34\degree$, $36\degree$, and $37\degree$, respectively.  Even after the bright
limb feature disappears, there are still temporal variations in the overall 
sky brightness level.
\label{scatteredlightfig}}
\end{figure}
\clearpage

\begin{figure}
\centerline{
\includegraphics[width=3.25in]{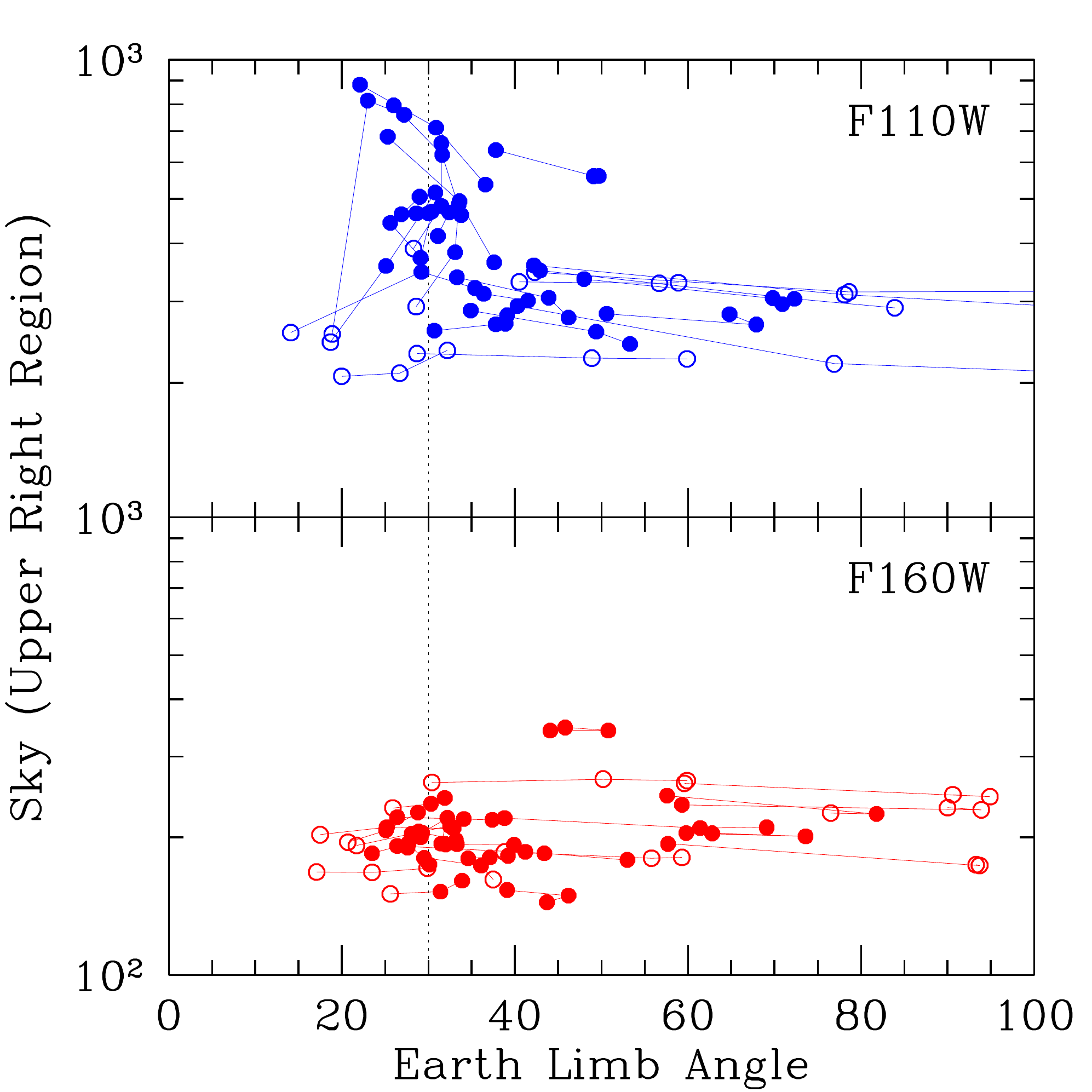}
\includegraphics[width=3.25in]{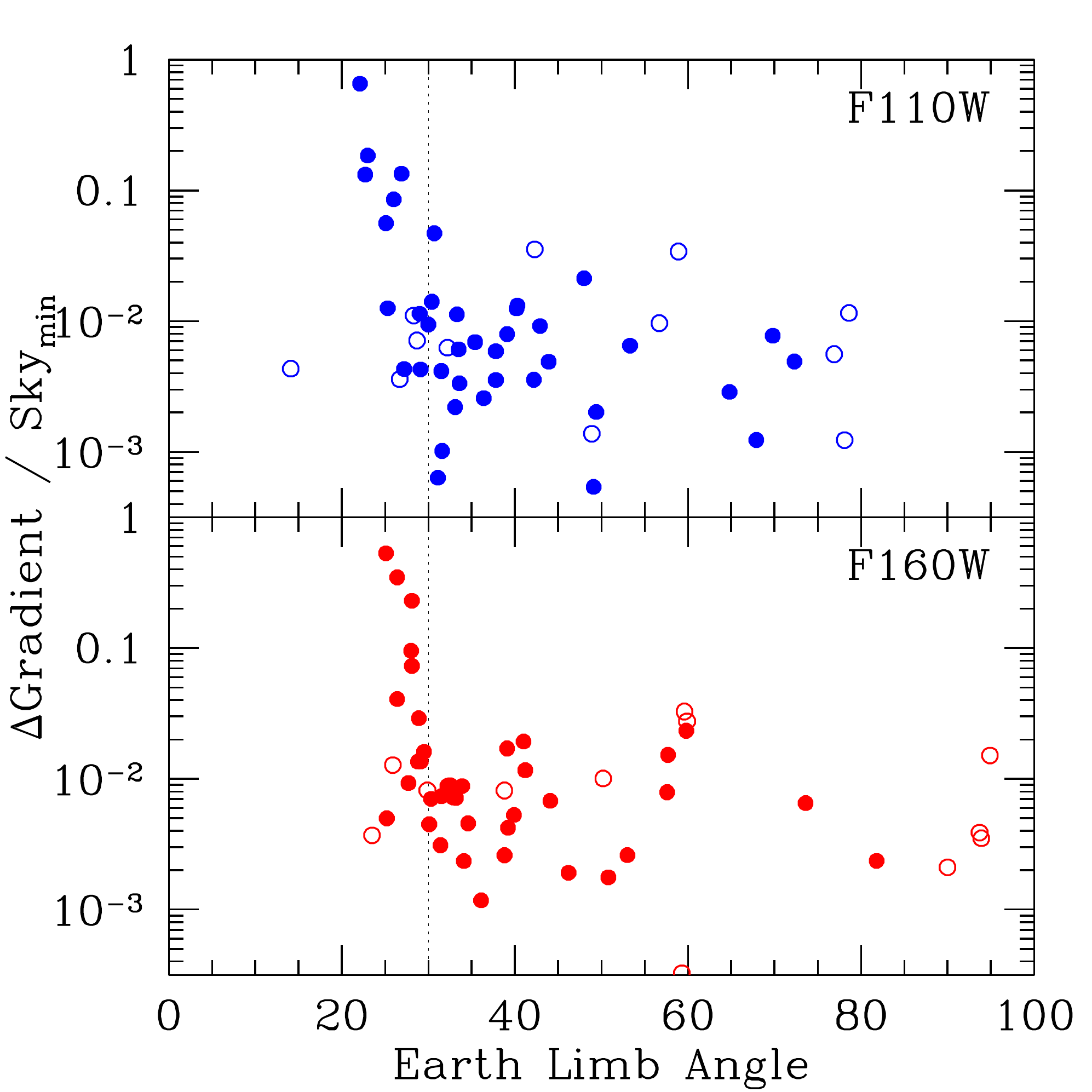}
}
\caption{ 
(Left) Sky level (counts per pixel) in the upper right quadrant of all $F110W$ and
$F160W$ exposures (top and bottom, respectively), as a function of the
average angle between the spacecraft's V1 axis and the earth's limb
during the exposure; open circles indicate that the limb was dark at
the time of observation. Successive exposures in a single orbit are
connected by lines, but intermediate non-destructive reads are
not shown.  In $F110W$, the mean sky level is clearly
correlated with proximity to the earth's limb, when the limb is
bright.  (Right) The amplitude of the scattered light feature seen in
Figure~\ref{scatteredlightfig} (as a fraction of the sky brightness),
as a function of the average angle to the earth's limb during each
observation.  The amplitude of the scattered light feature is taken to
be the excess counts in the affected region, compared to the variation
in the mean sky level seen in the unaffected upper right quadrant.  It
is calculated by first calculating a gradient across the image using
the difference between the amplitude of the sky in the affected region
and the unaffected region in the upper right of the image.  The
smallest gradient for the exposures in a given orbit is then taken to
be the expected true spatial variation in the sky background, in the
absence of scattered light.  The amplitude of the scattered light
feature is then scaled to the level of the sky in the upper right
quadrant of each image, giving the approximate amplitude of the
feature as a percent of sky. The scattered light feature turns on
abruptly at earth limb angles less than 30$\degree$.
\label{limbangfig}}
\end{figure}
\vfill
\clearpage

\begin{figure}
\centerline{
\includegraphics[width=6.25in]{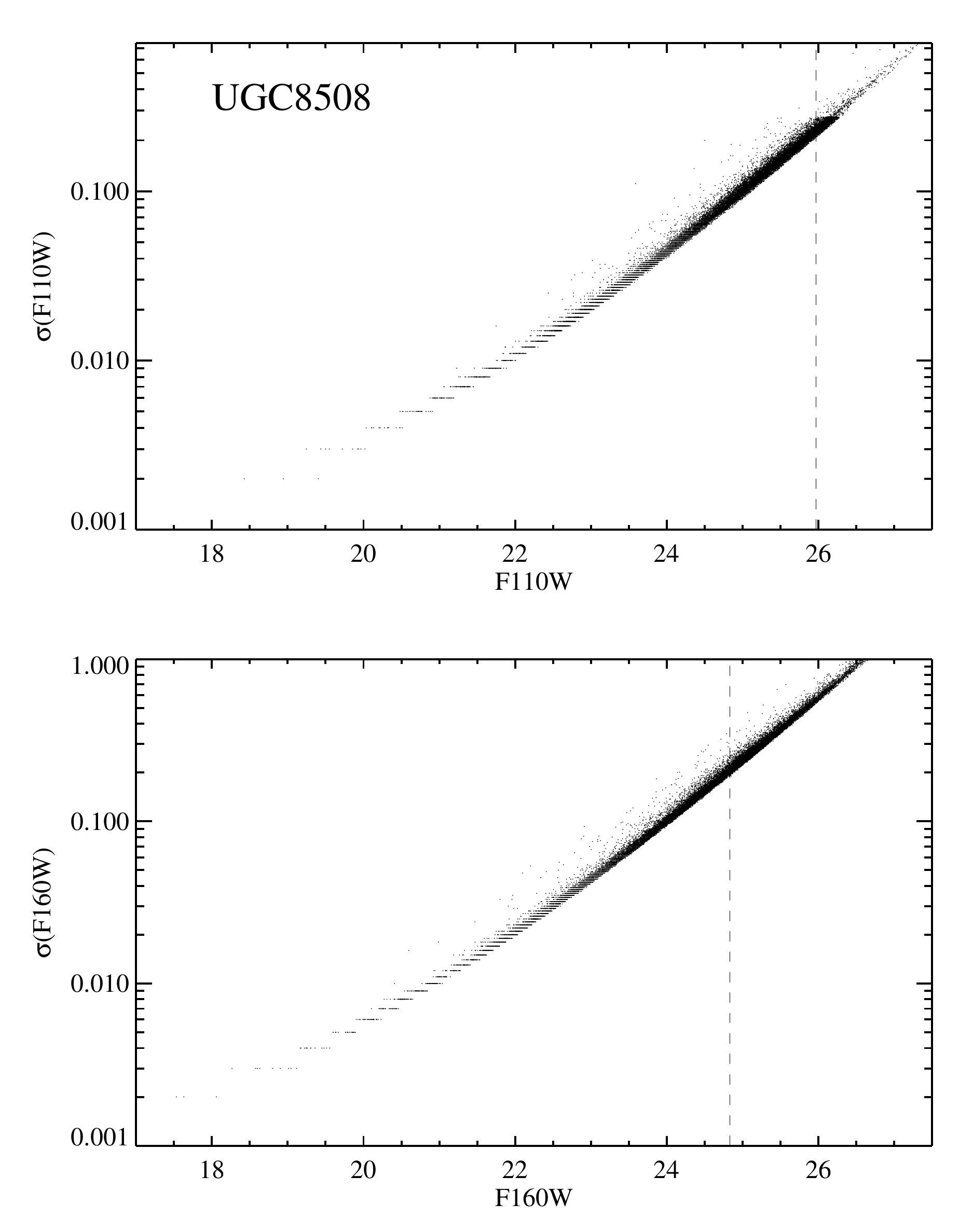}
}
\caption{
Photometric uncertainties as a function of magnitude in the $F110W$
(top) and $F160W$ filters, for a representative field with moderate
crowding (UGC8508), using the more complete {\tt *.st} catalog.  The
vertical lines indicate the average 50\% completeness level determined
from artificial star tests (Table~\ref{obstable}).  Photometric uncertainties
are those reported by DOLPHOT, and do not fully capture the uncertainties
due to crowding (see Figure~\ref{fakestarfig}).
\label{errfig}}
\end{figure}
\vfill
\clearpage

\begin{figure}
\centerline{
\includegraphics[width=3.25in]{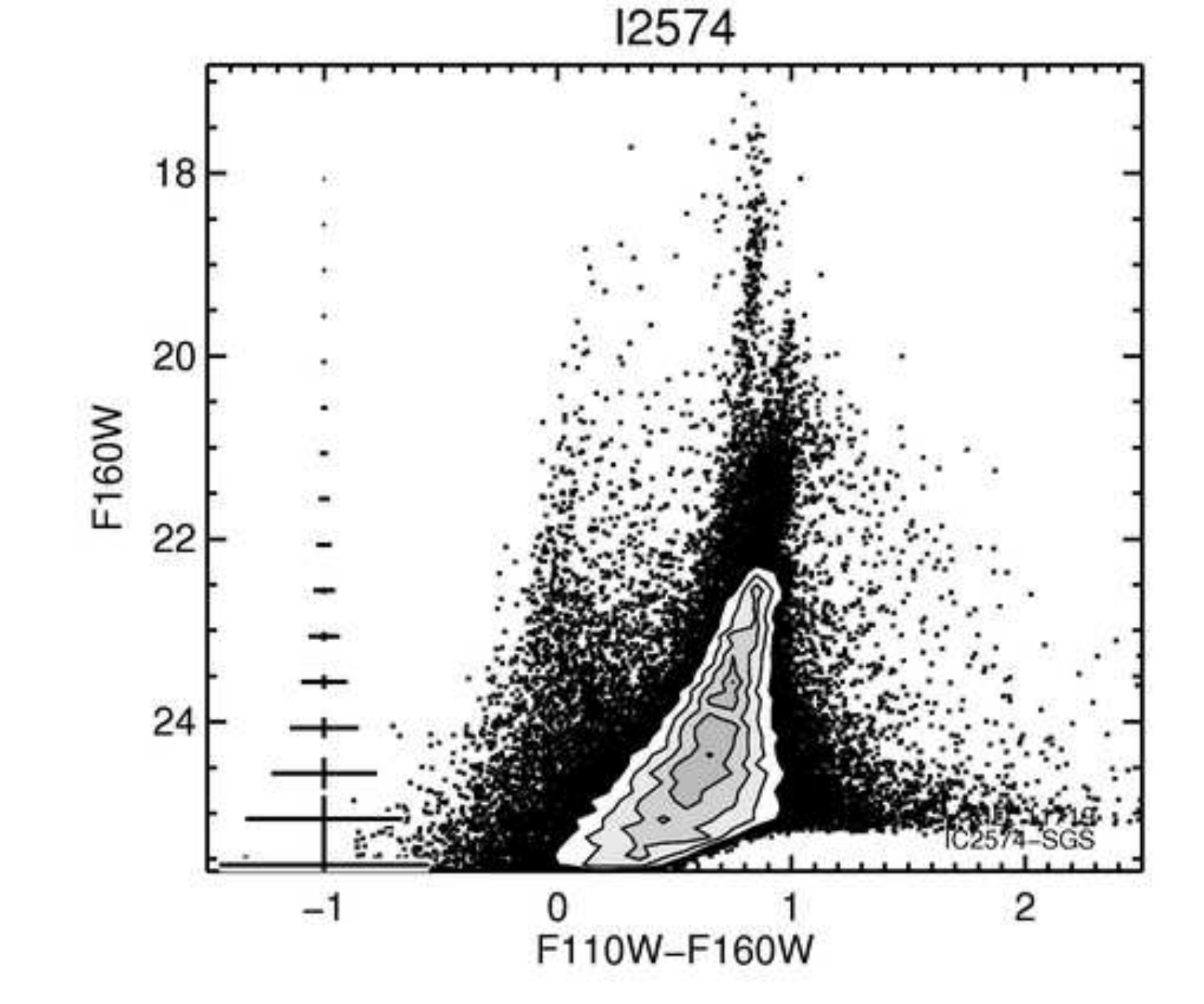}
\includegraphics[width=3.25in]{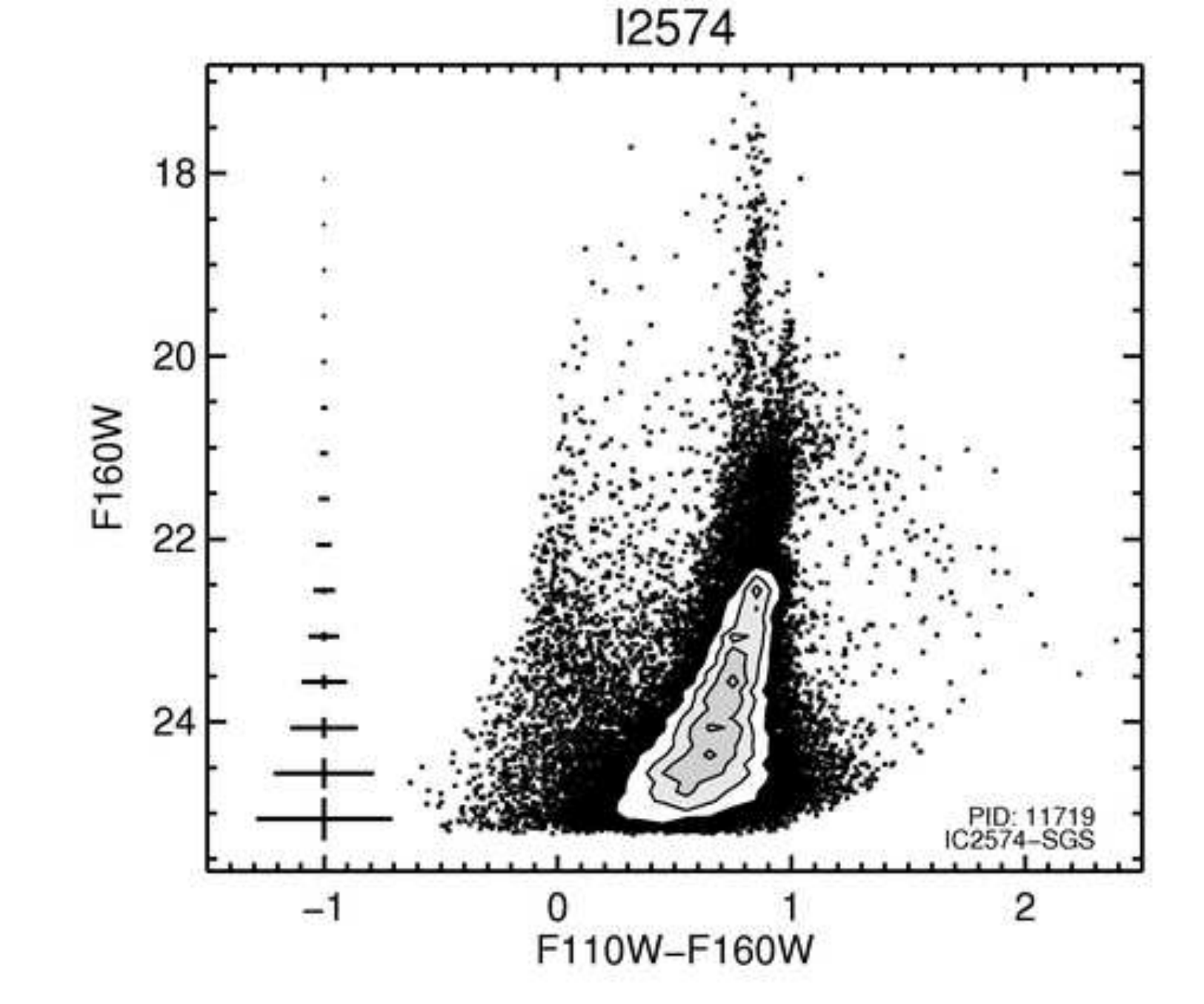}
}
\caption{
Comparison of the color magnitude diagrams of the original stellar
{\tt *.st} photometry (left) and the high quality ``cleaned'' {\tt
  *.gst} photometry (right) for target IC2574-SGS.  The original photometry has higher
completeness, but at the expense of increased errors due to crowding,
producing less sharp features in the CMD.  The {\tt *.st} photometry
includes stars that have a signal-to-noise of greater than 4 in only
one filter, making the colors unreliable at faint
magnitudes; the {\tt *.gst} photometry requires high signal-to-noise in
both filters.\label{st_vs_gst_fig}}
\end{figure}
\vfill

\begin{figure}
\centerline{
\includegraphics[width=3.25in]{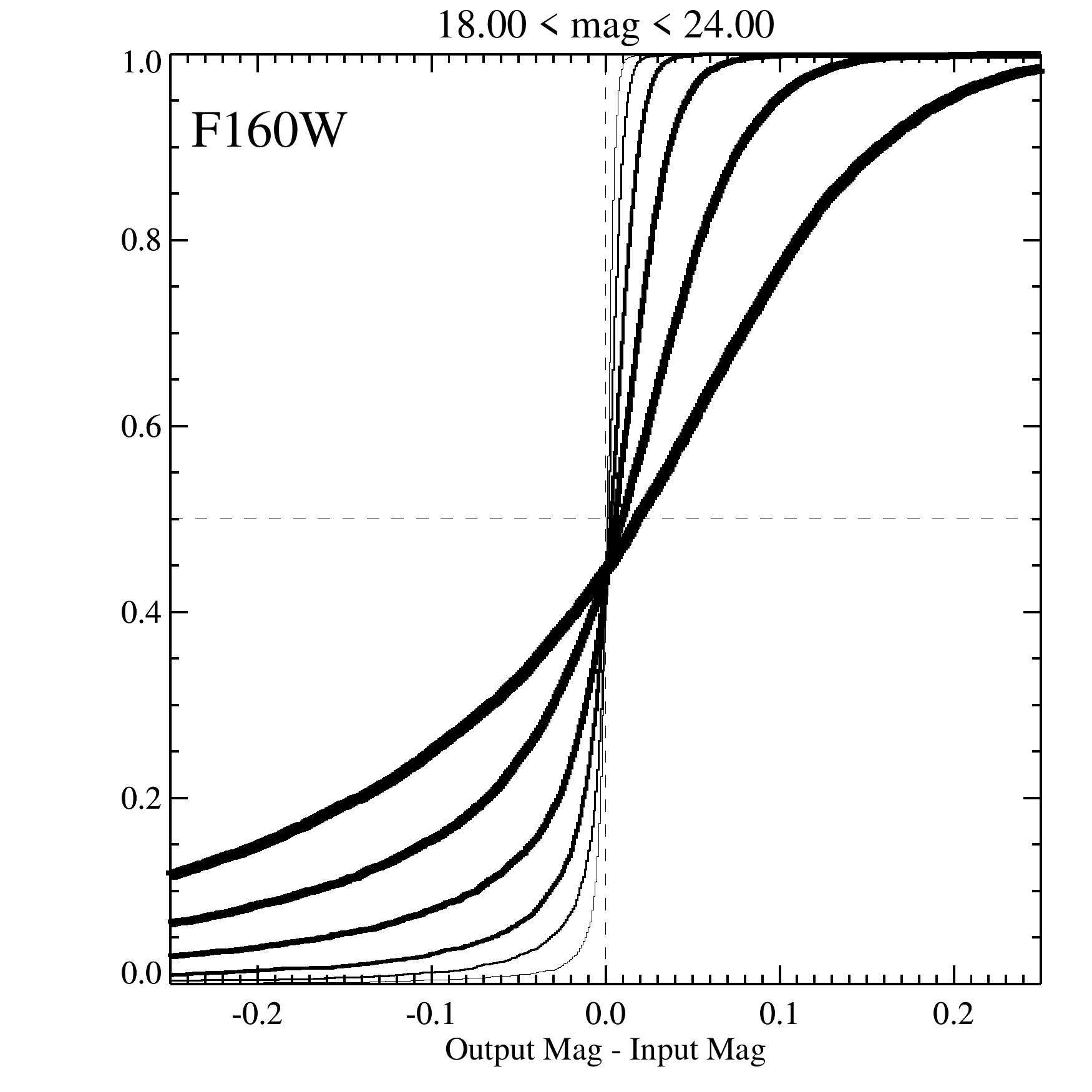}
\includegraphics[width=3.25in]{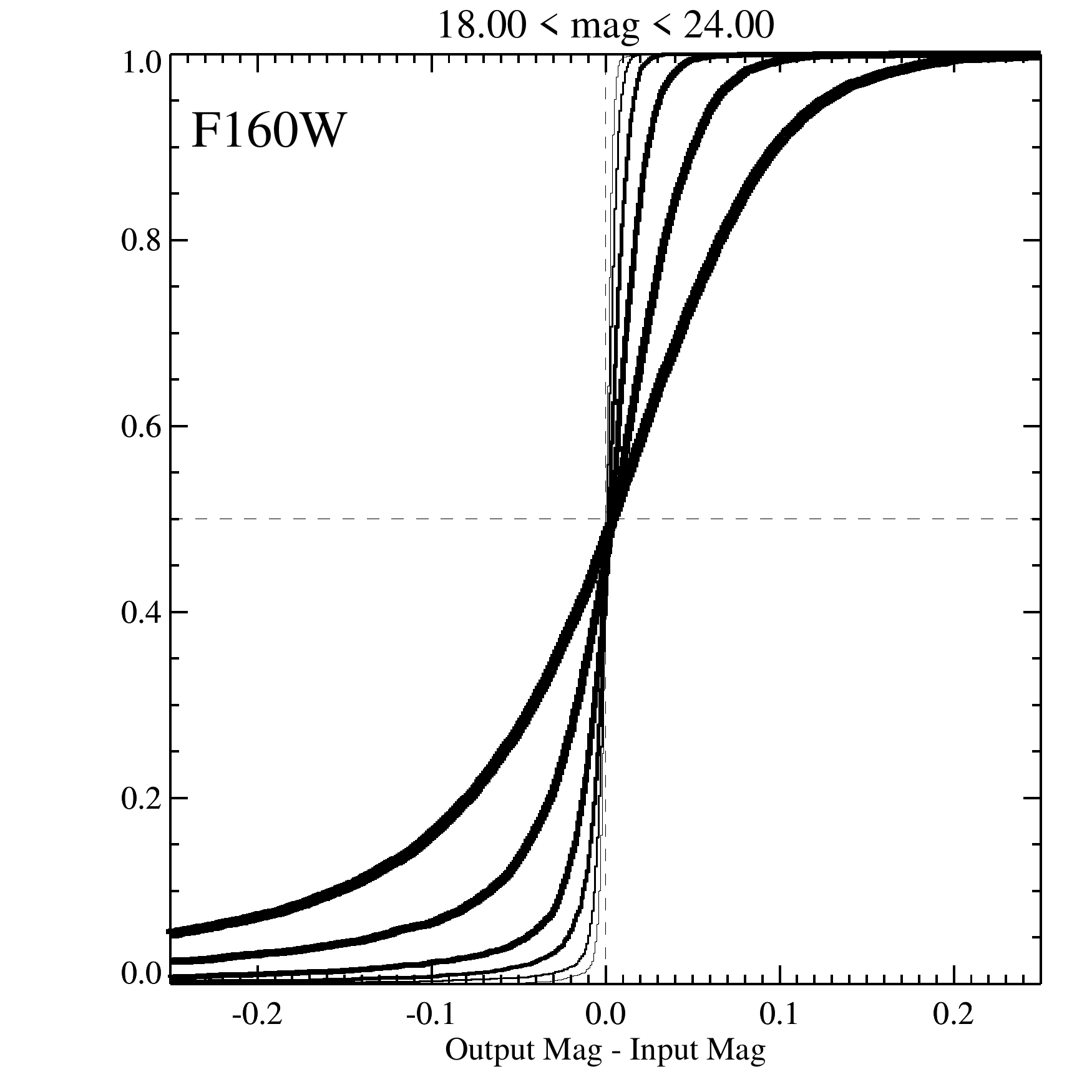}
}
\caption{Distribution of differences between the true and the
  recovered $F160W$ magnitudes for artificial stars inserted into
  images of IC2574-SGS (left; highly crowded) and UGC~4459 (right;
  uncrowded).  Distributions are calculated in bins 1 magnitude wide,
  with increasing line thicknesses indicating fainter magnitudes; the
  heaviest line includes stars with $23<m_{\rm{F160W}}\le 24$.
  Crowding biases the photometry very slightly ($<0.004\,$mag)
  towards fainter observed magnitudes at the median, with larger biases
  at fainter magnitudes.  However, these biases are much smaller than
  the photometric errors ($<\!0.05\sigma_m$ for IC2574-SGS) in all
  magnitude ranges.  The uncertainties in the magnitude of
  individual stars is frequently dominated by crowding, however (see
  Figure~\ref{errfig} for typical photometric errors as a function of
  magnitude.)
  \label{fakestarfig}}
\end{figure}
\clearpage

\begin{figure}
\centerline{
\includegraphics[width=3.25in]{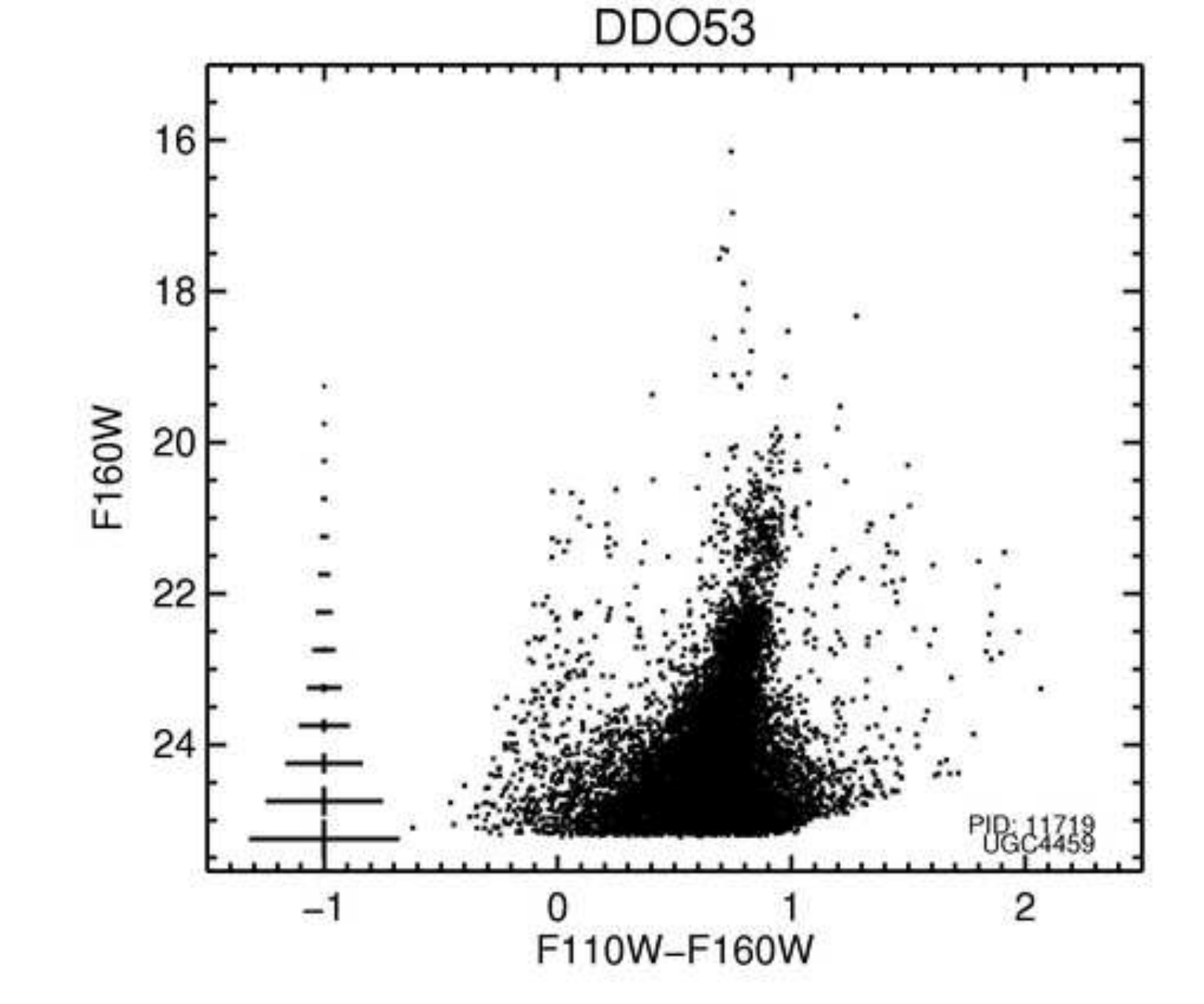}
\includegraphics[width=3.25in]{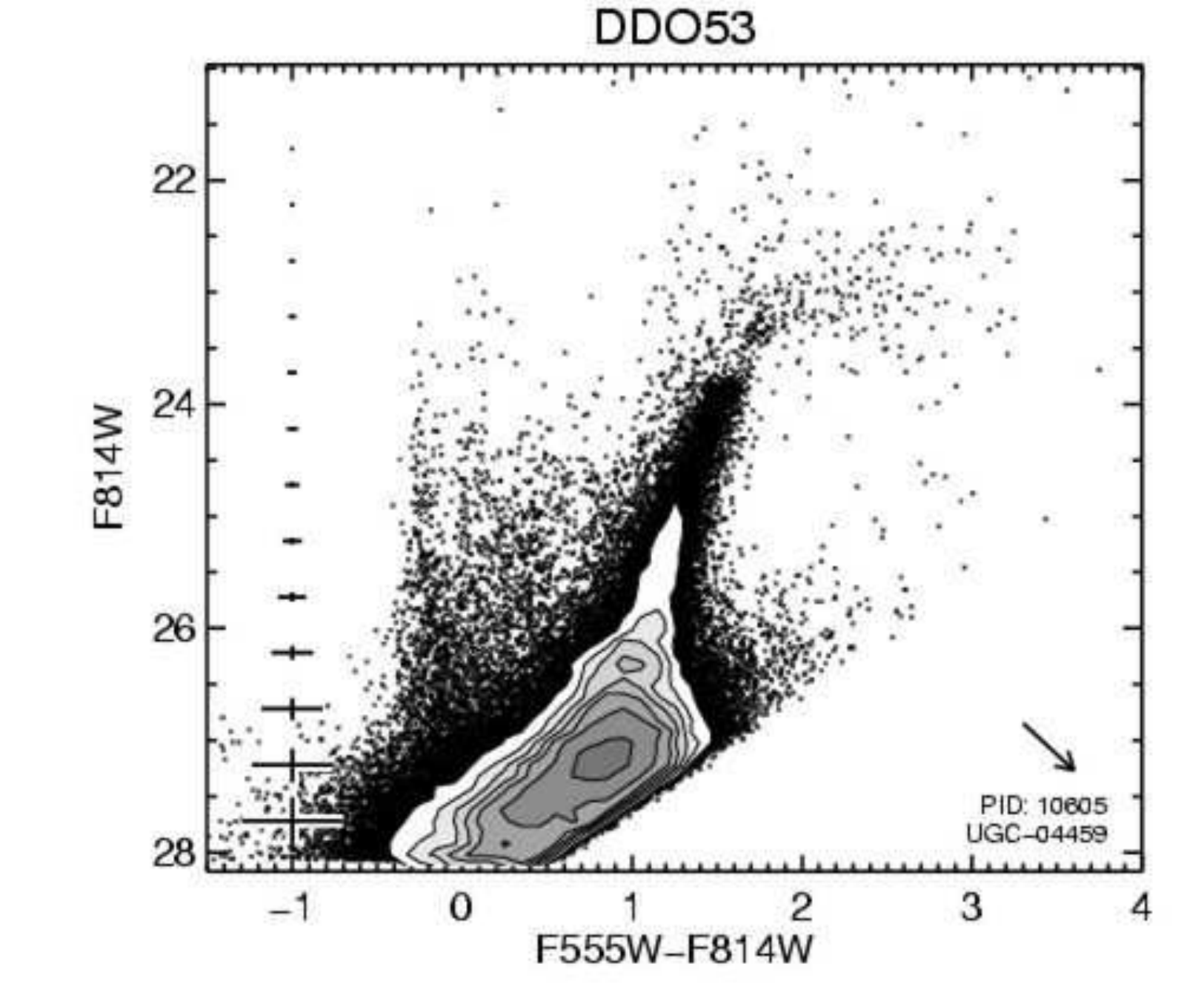}
}
\centerline{
\includegraphics[width=3.25in]{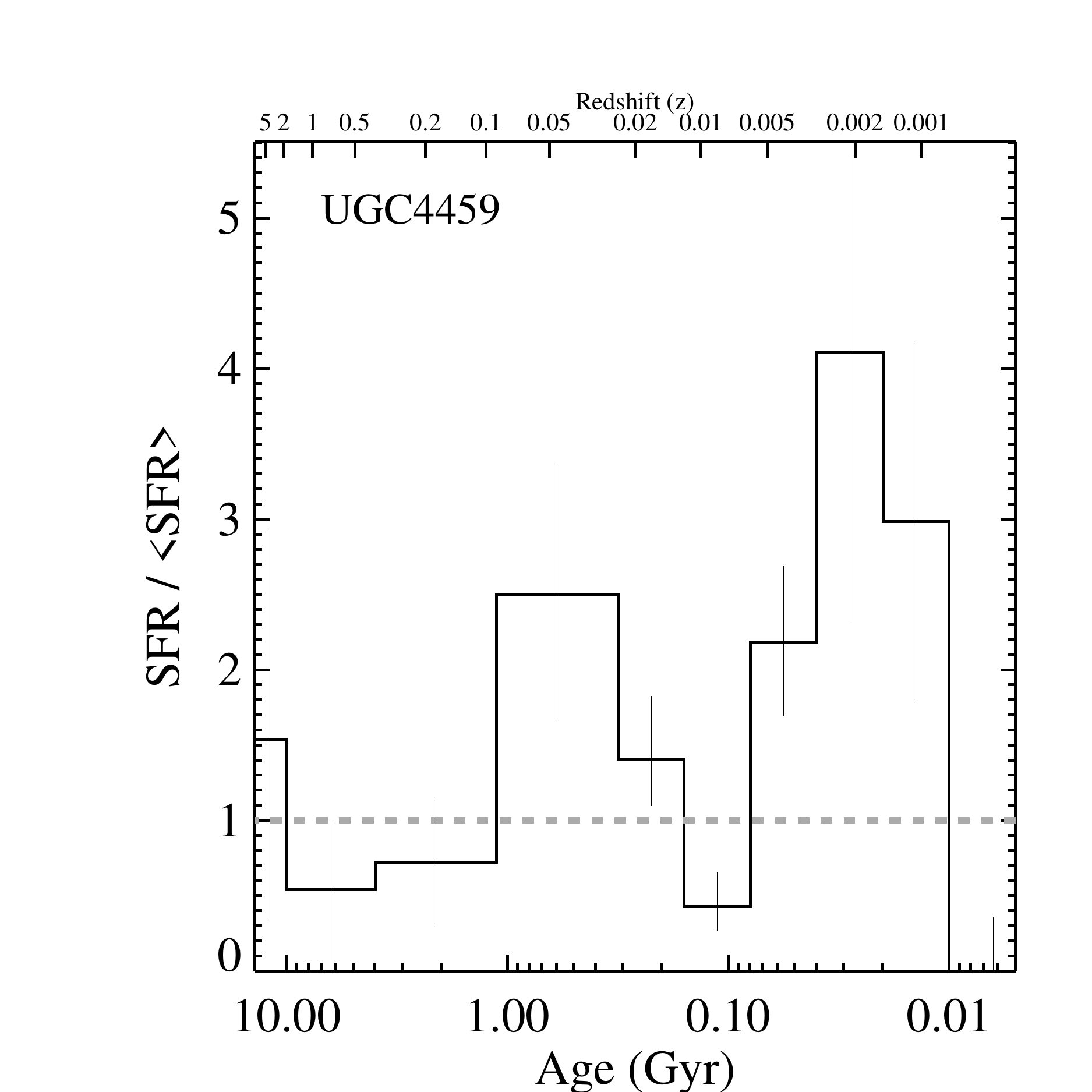}
\includegraphics[width=3.25in]{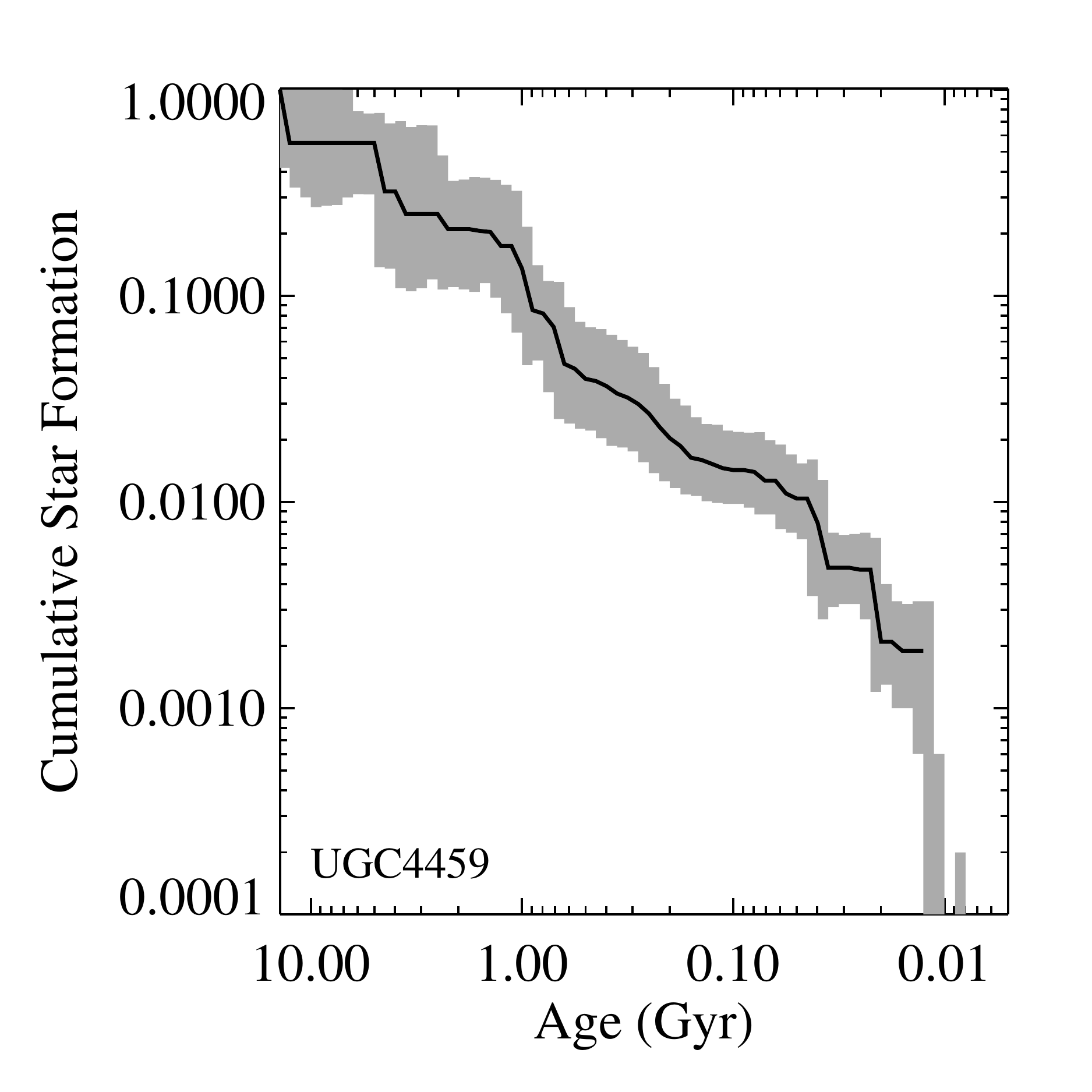}
}
\caption{ Color magnitude diagrams of the WFC3/IR (upper left) and
  optical (upper right) for the target UGC4459 within galaxy DDO53.
  Lower panels show the star formation history derived from the
  optical data, for both the differential (left, with horizontal
  dotted line indicating the past average SFR) and cumulative (right)
  star formation histories.  The cumulative star formation history is
  calculated from the present back to 14\,Gyrs.  Uncertainties in the
  lower two panels are the 68\% confidence intervals, calculated from
  Monte Carlo tests including random and systematic uncertainties.
  Optical CMDs are restricted to the area covered by the WFC3
  FOV.\label{cmdfig}}
\end{figure}
\vfill
\clearpage
 
\begin{figure}
\figurenum{\ref{cmdfig} continued}
\centerline{
\includegraphics[width=3.25in]{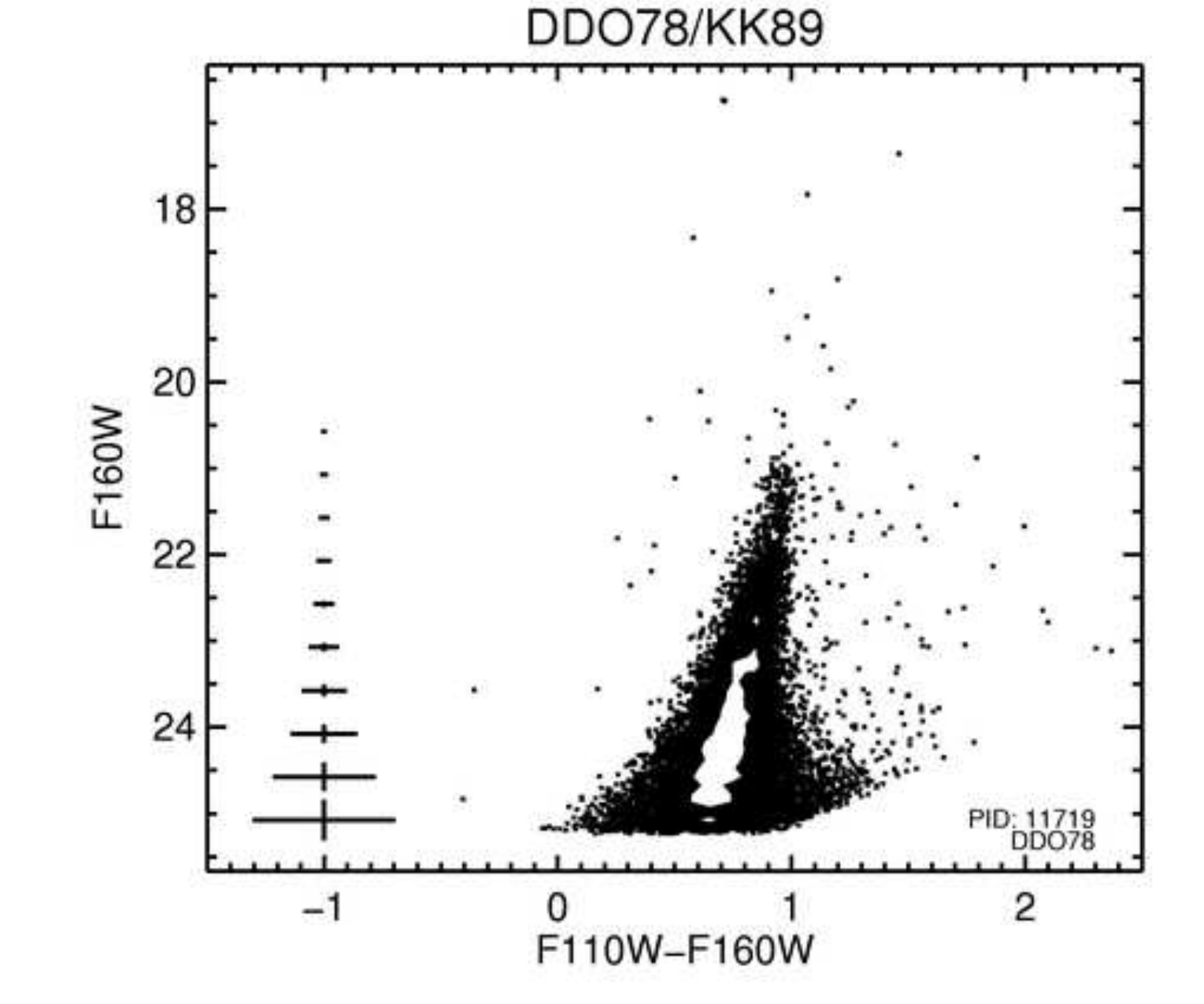}
\includegraphics[width=3.25in]{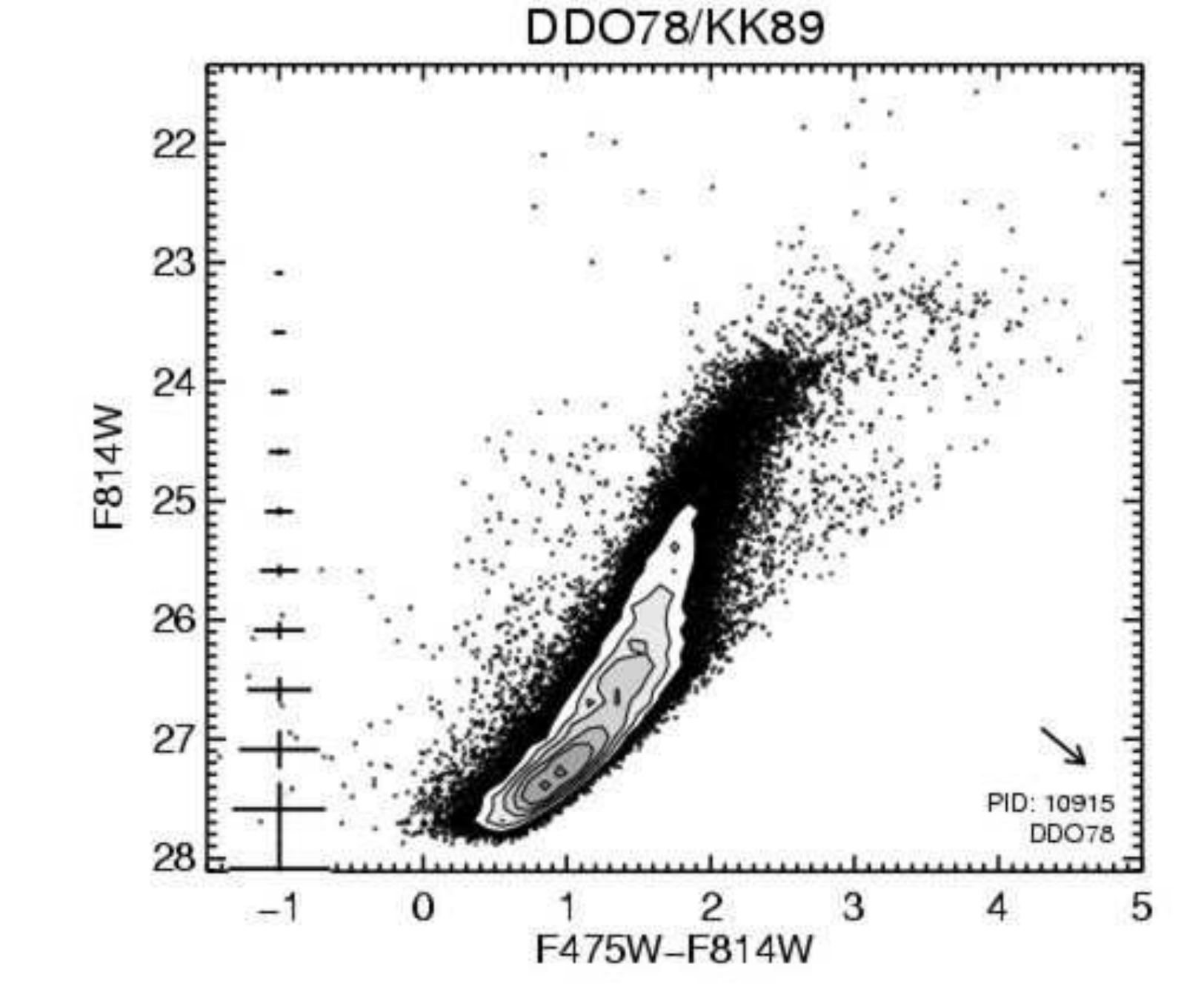}
}
\centerline{
\includegraphics[width=3.25in]{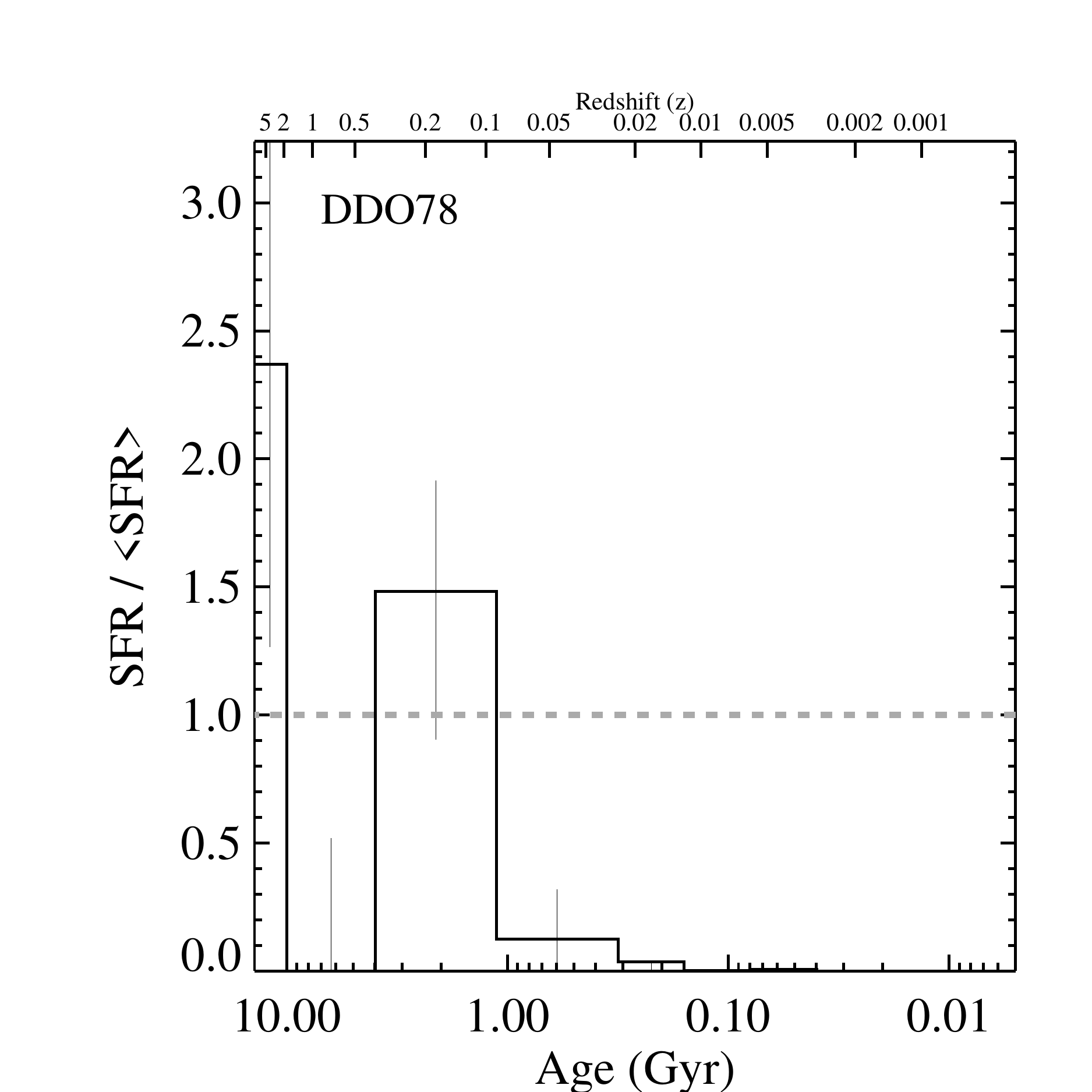}
\includegraphics[width=3.25in]{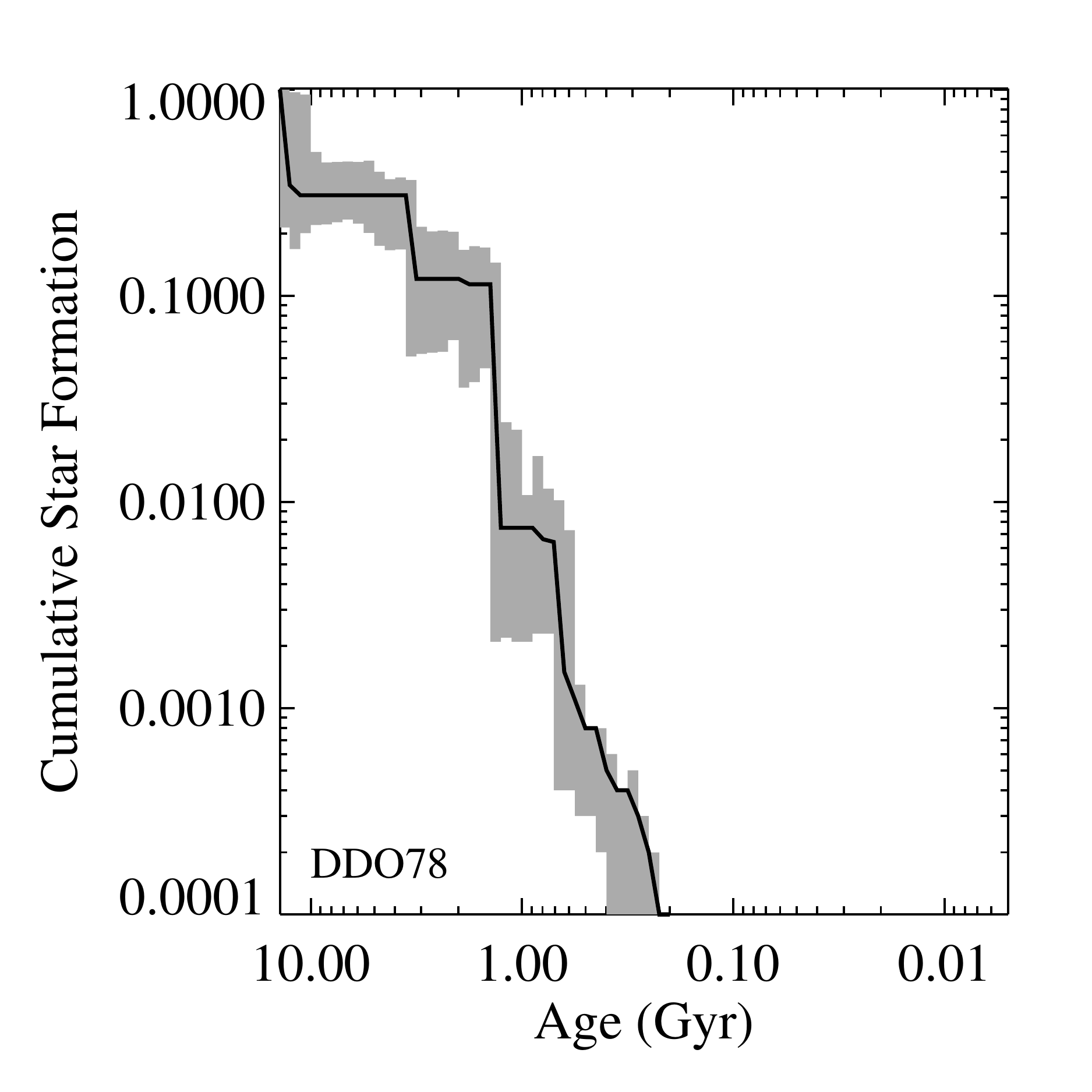}
}
\caption{ Color magnitude diagrams of the WFC3/IR (upper left) and
  optical (upper right) for DDO78.  Lower panels show the star
  formation history derived from the optical data, for both the
  differential (left, with horizontal dotted line indicating the past
  average SFR) and cumulative (right) star formation histories.  The
  cumulative star formation history is calculated from the present
  back to 14\,Gyrs.  Uncertainties in the lower two panels are the
  68\% confidence intervals, calculated from Monte Carlo tests
  including random and systematic uncertainties.  Optical CMDs are
  restricted to the area covered by the WFC3 FOV. }
\end{figure}
\vfill
\clearpage
 
\begin{figure}
\figurenum{\ref{cmdfig} continued}
\centerline{
\includegraphics[width=3.25in]{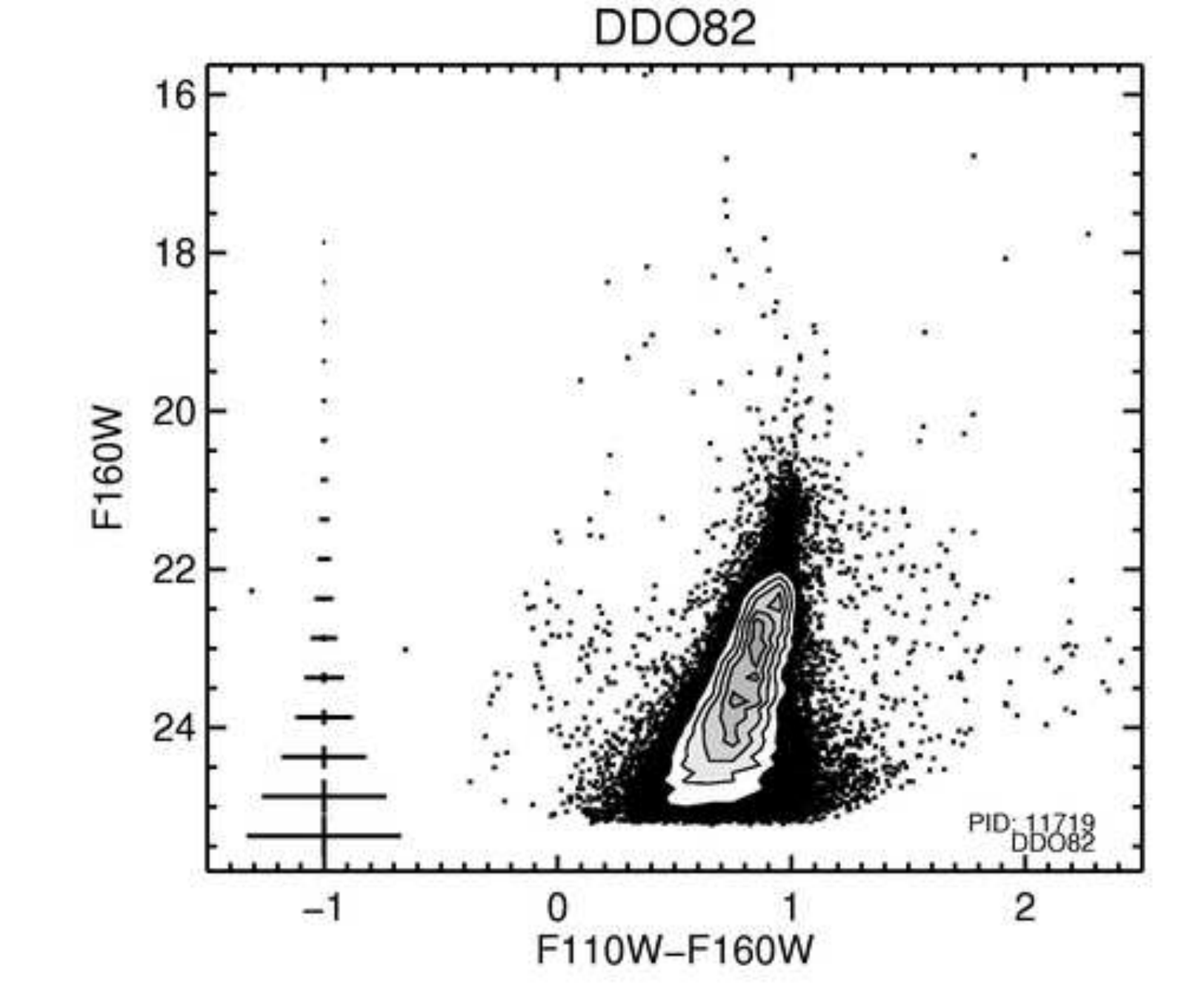}
\includegraphics[width=3.25in]{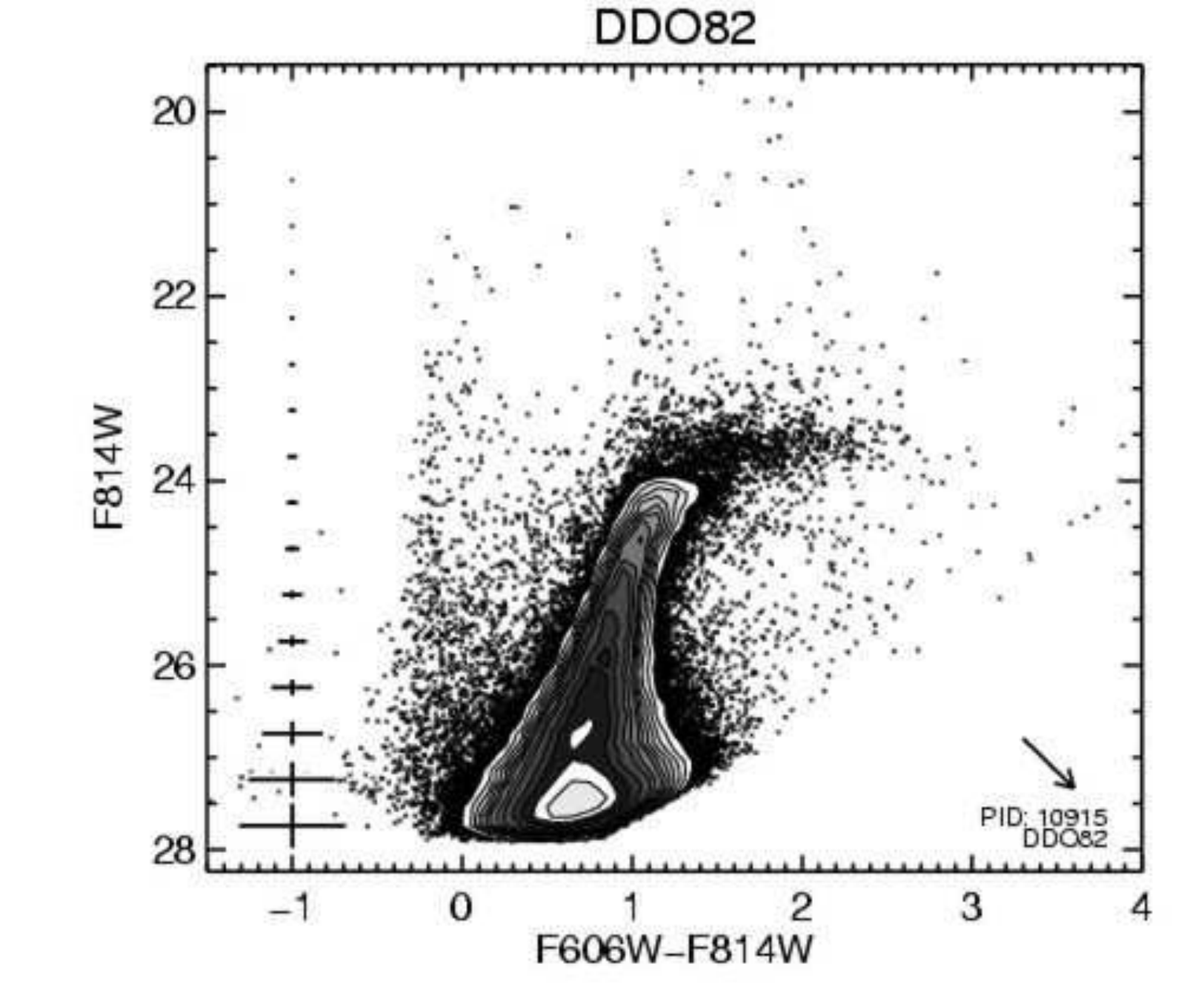}
}
\centerline{
\includegraphics[width=3.25in]{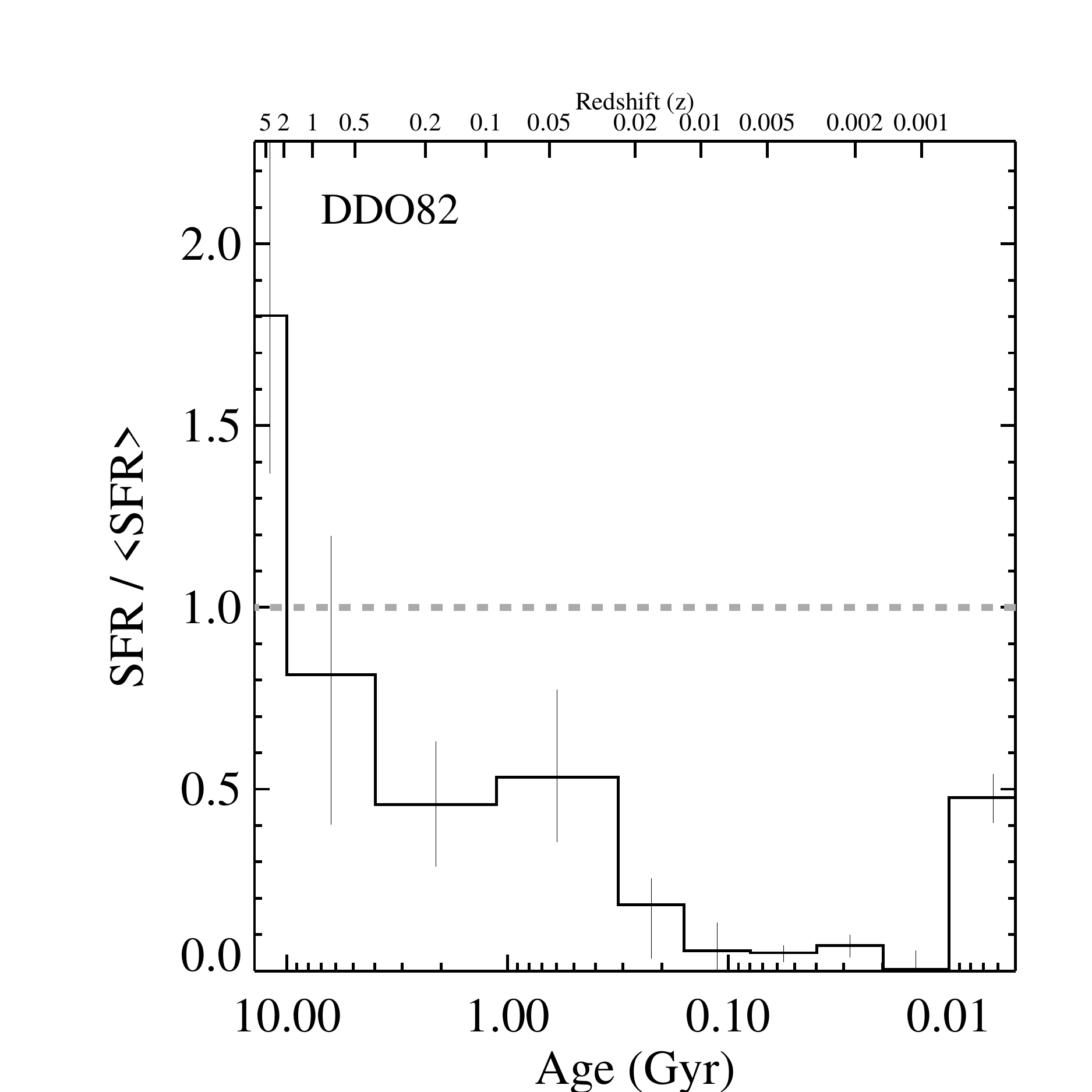}
\includegraphics[width=3.25in]{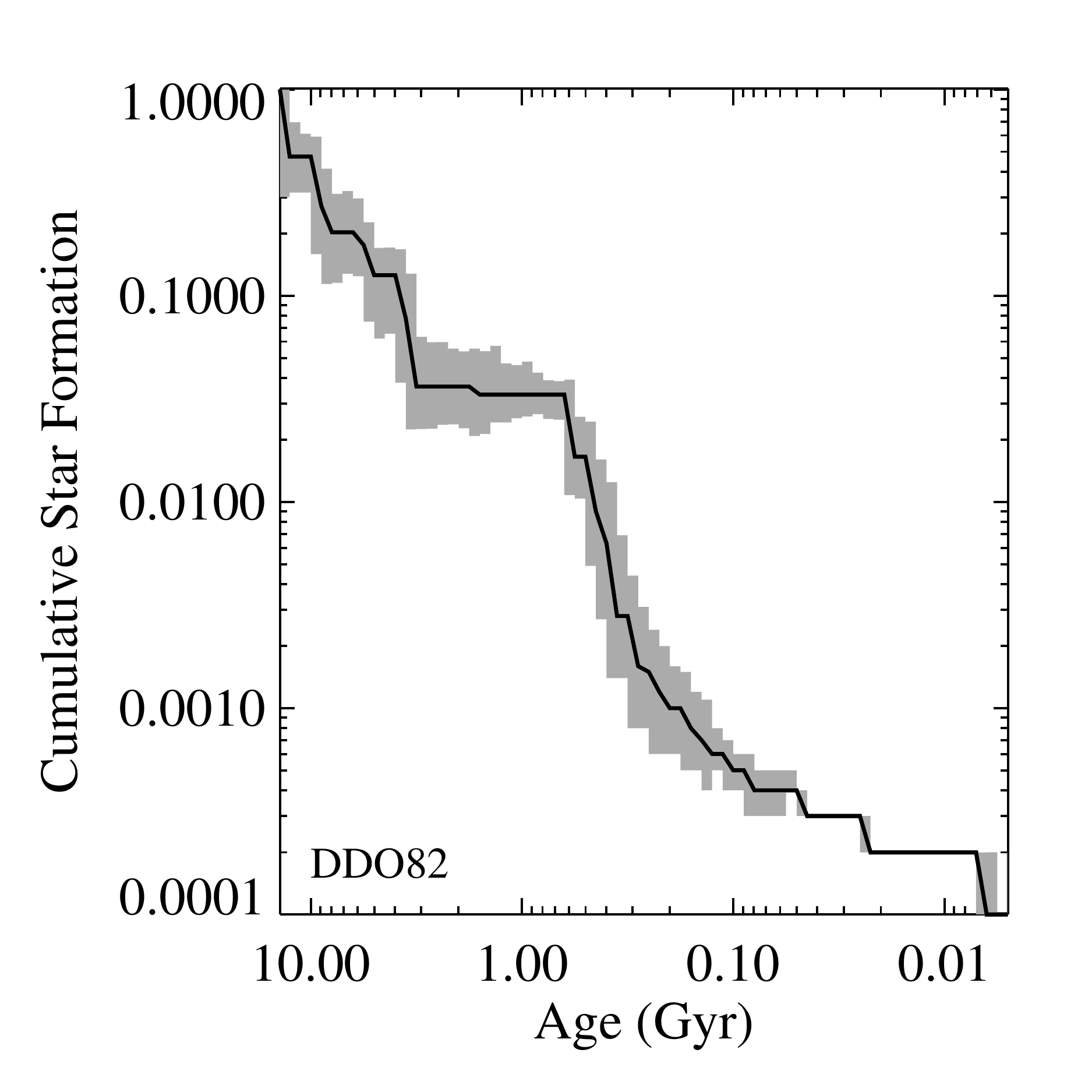}
}
\caption{ Color magnitude diagrams of the WFC3/IR (upper left) and
  optical (upper right) for DDO82.  Lower panels show the star
  formation history derived from the optical data, for both the
  differential (left, with horizontal dotted line indicating the past
  average SFR) and cumulative (right) star formation histories.  The
  cumulative star formation history is calculated from the present
  back to 14\,Gyrs.  Uncertainties in the lower two panels are the
  68\% confidence intervals, calculated from Monte Carlo tests
  including random and systematic uncertainties.  Optical CMDs are
  restricted to the area covered by the WFC3 FOV. }
\end{figure}
\vfill
\clearpage
 
\begin{figure}
\figurenum{\ref{cmdfig} continued}
\centerline{
\includegraphics[width=3.25in]{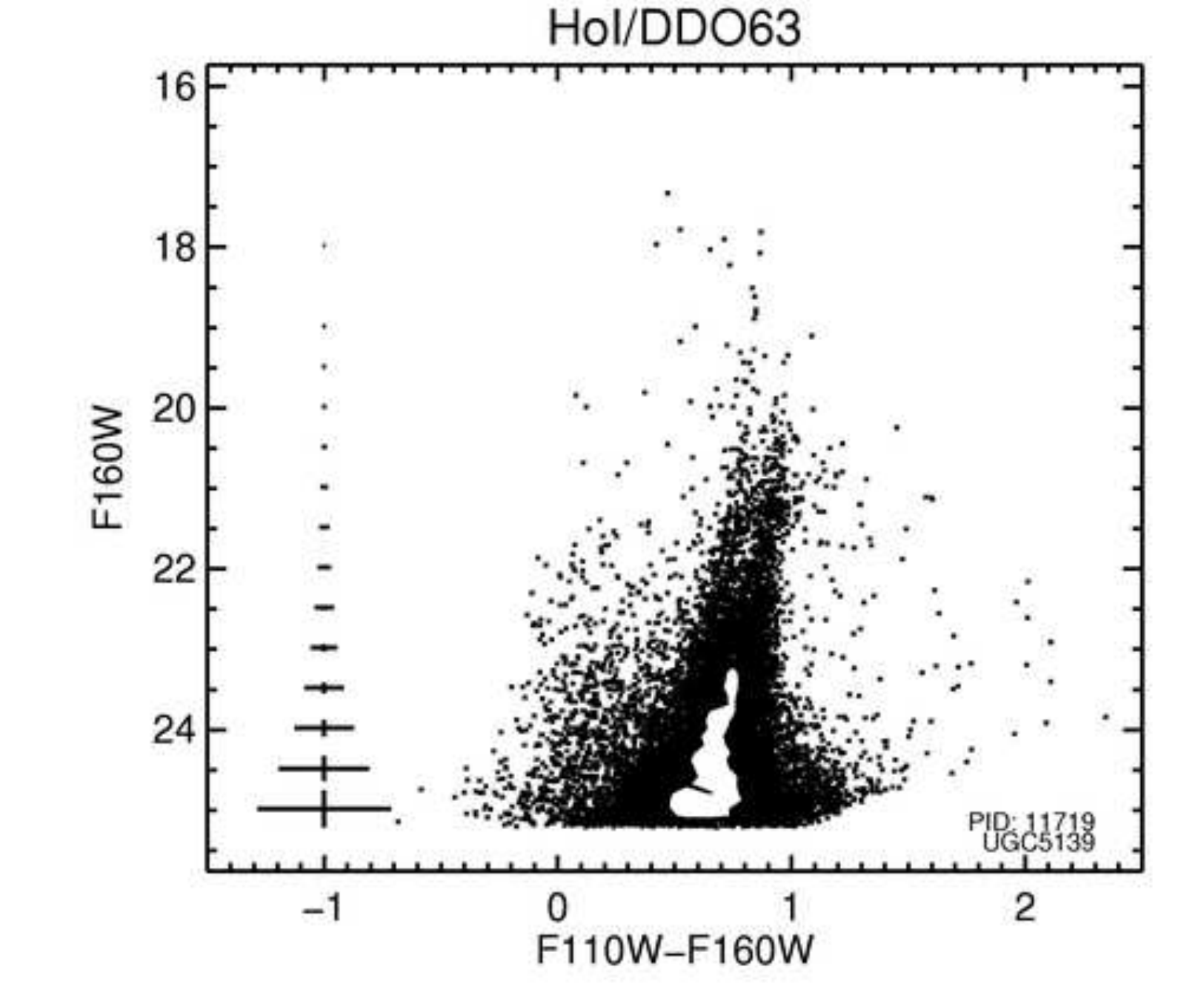}
\includegraphics[width=3.25in]{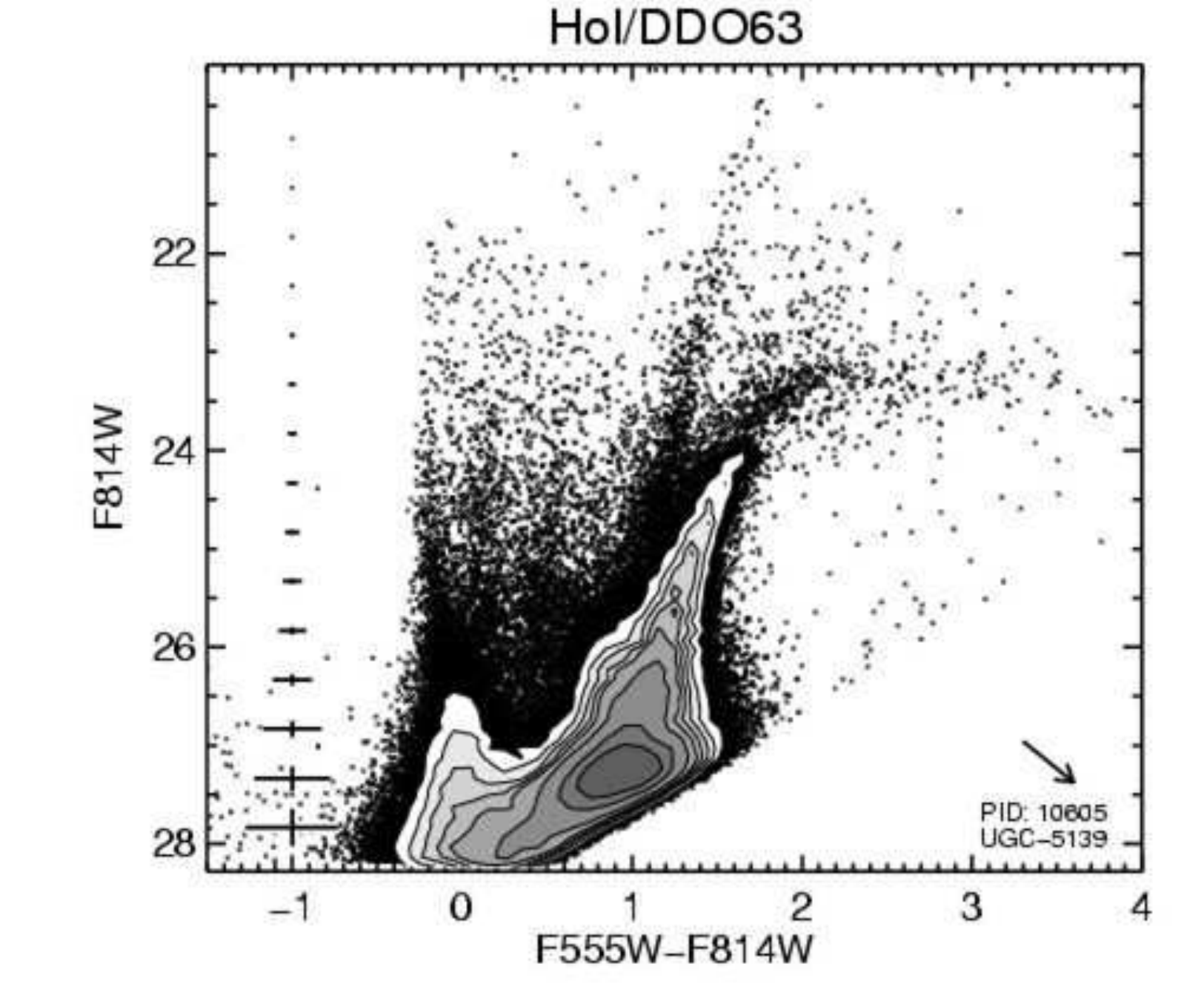}
}
\centerline{
\includegraphics[width=3.25in]{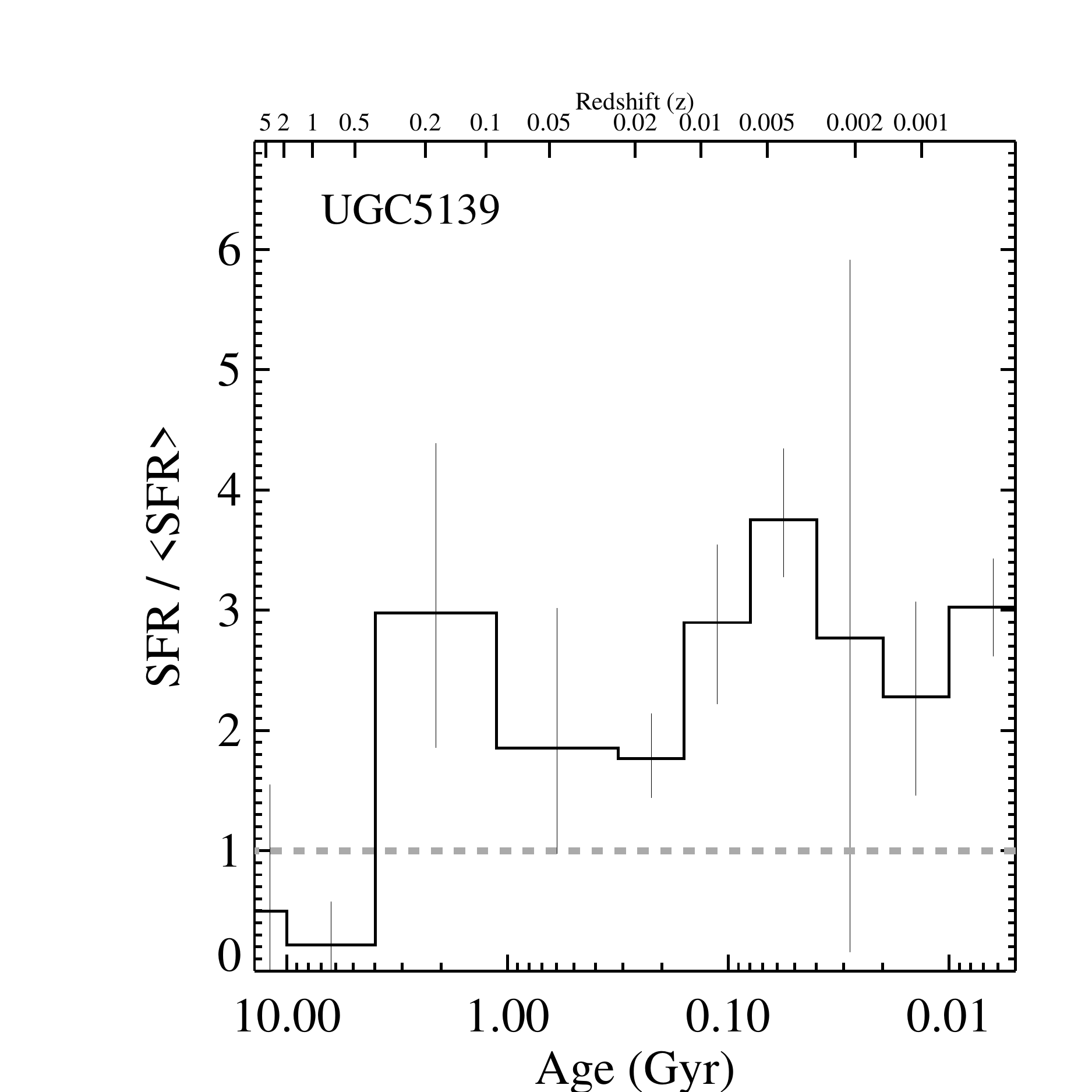}
\includegraphics[width=3.25in]{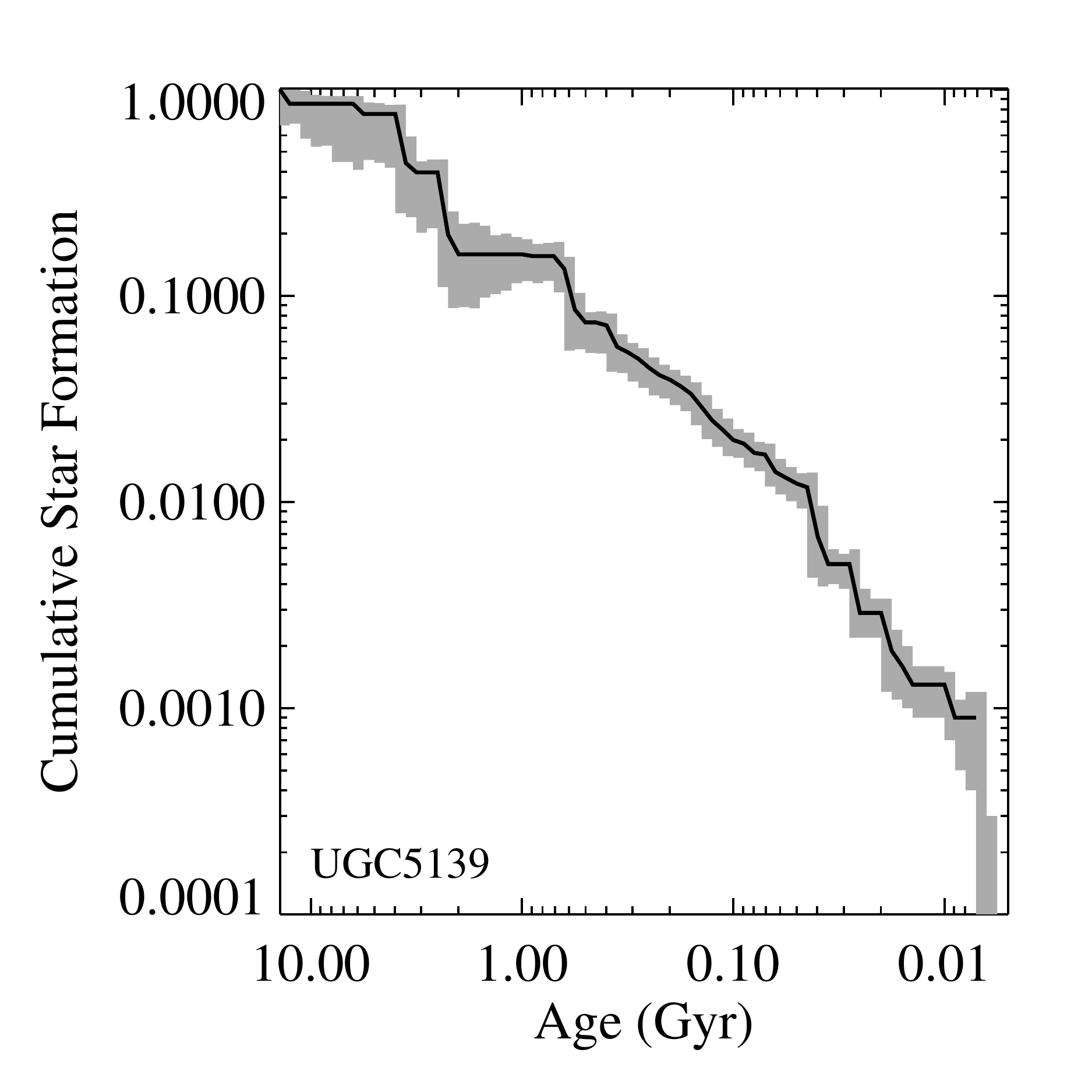}
}
\caption{ Color magnitude diagrams of the WFC3/IR (upper left) and
  optical (upper right) for the target UGC5139 within galaxy HoI.
  Lower panels show the star formation history derived from the
  optical data, for both the differential (left, with horizontal
  dotted line indicating the past average SFR) and cumulative (right)
  star formation histories.  The cumulative star formation history is
  calculated from the present back to 14\,Gyrs.  Uncertainties in the
  lower two panels are the 68\% confidence intervals, calculated from
  Monte Carlo tests including random and systematic uncertainties.
  Optical CMDs are restricted to the area covered by the WFC3 FOV. }
\end{figure}
\vfill
\clearpage
 
\begin{figure}
\figurenum{\ref{cmdfig} continued}
\centerline{
\includegraphics[width=3.25in]{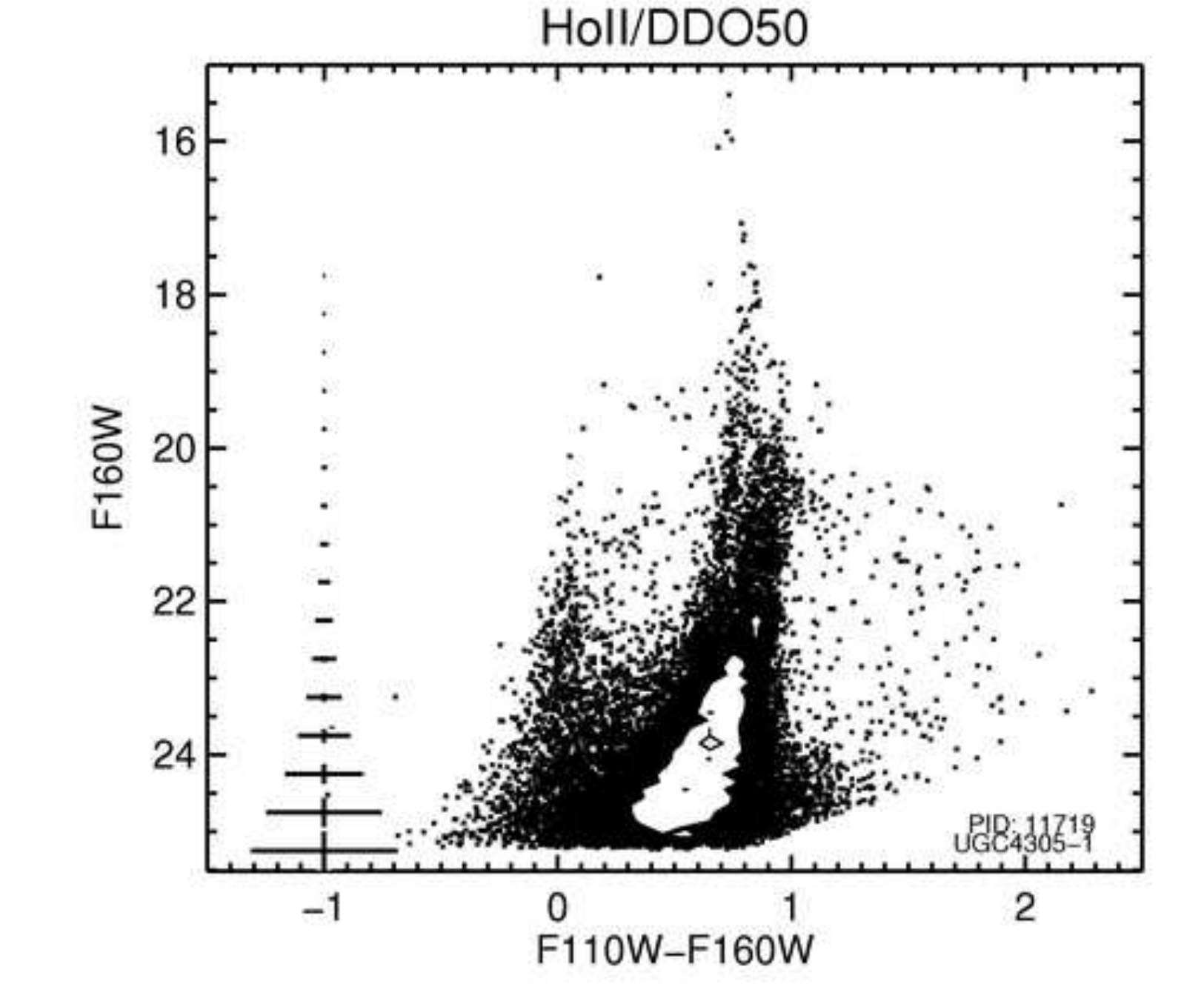}
\includegraphics[width=3.25in]{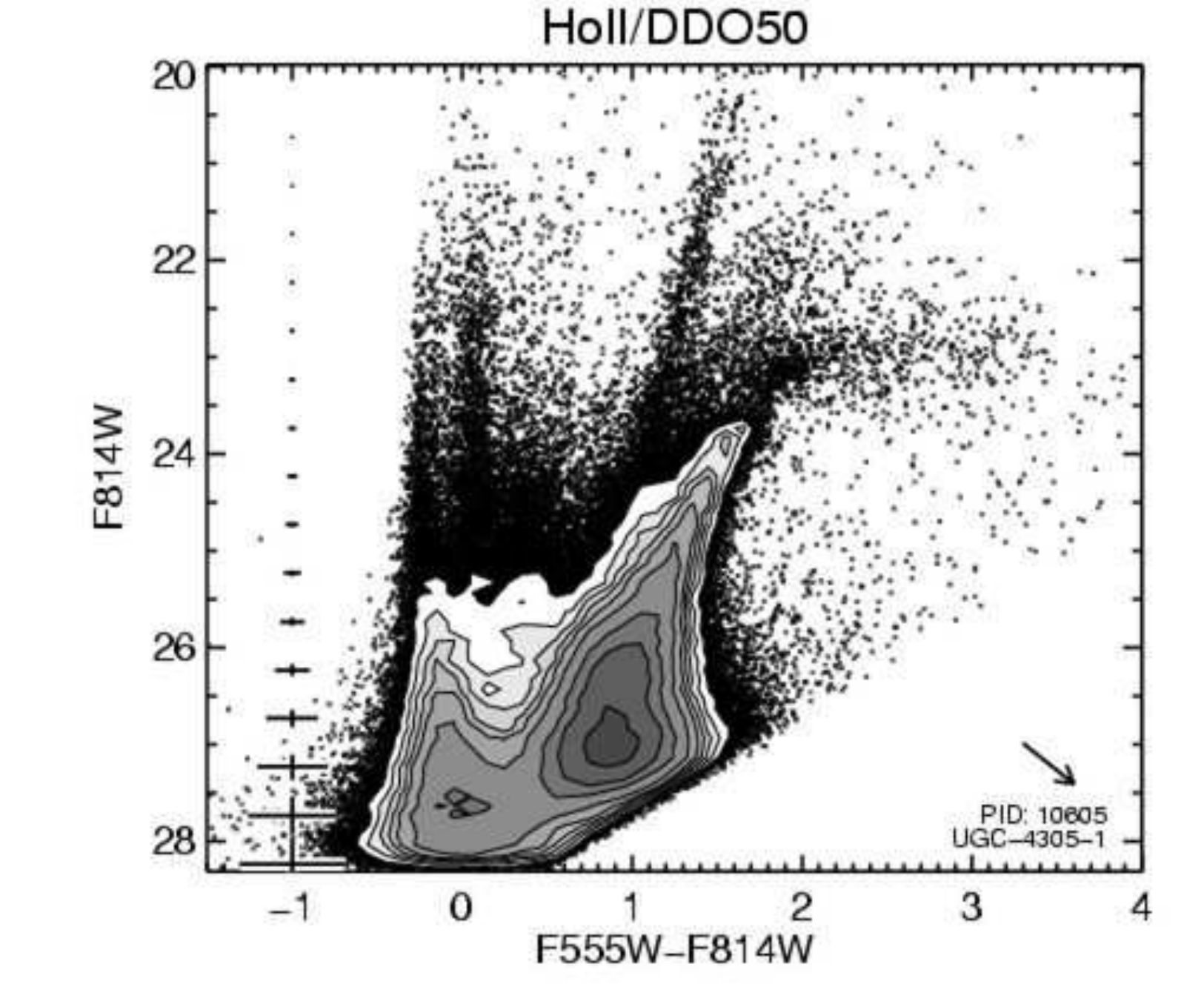}
}
\centerline{
\includegraphics[width=3.25in]{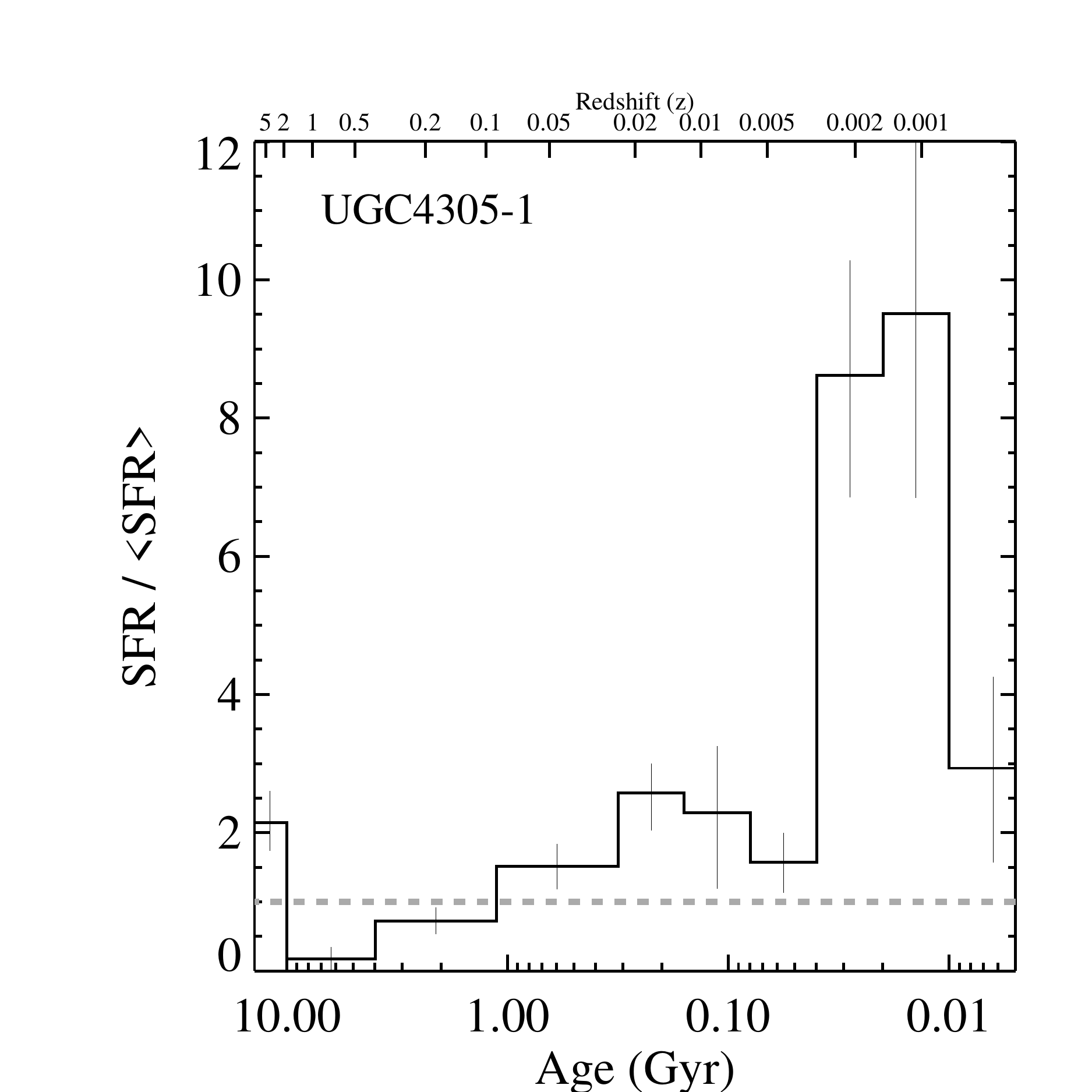}
\includegraphics[width=3.25in]{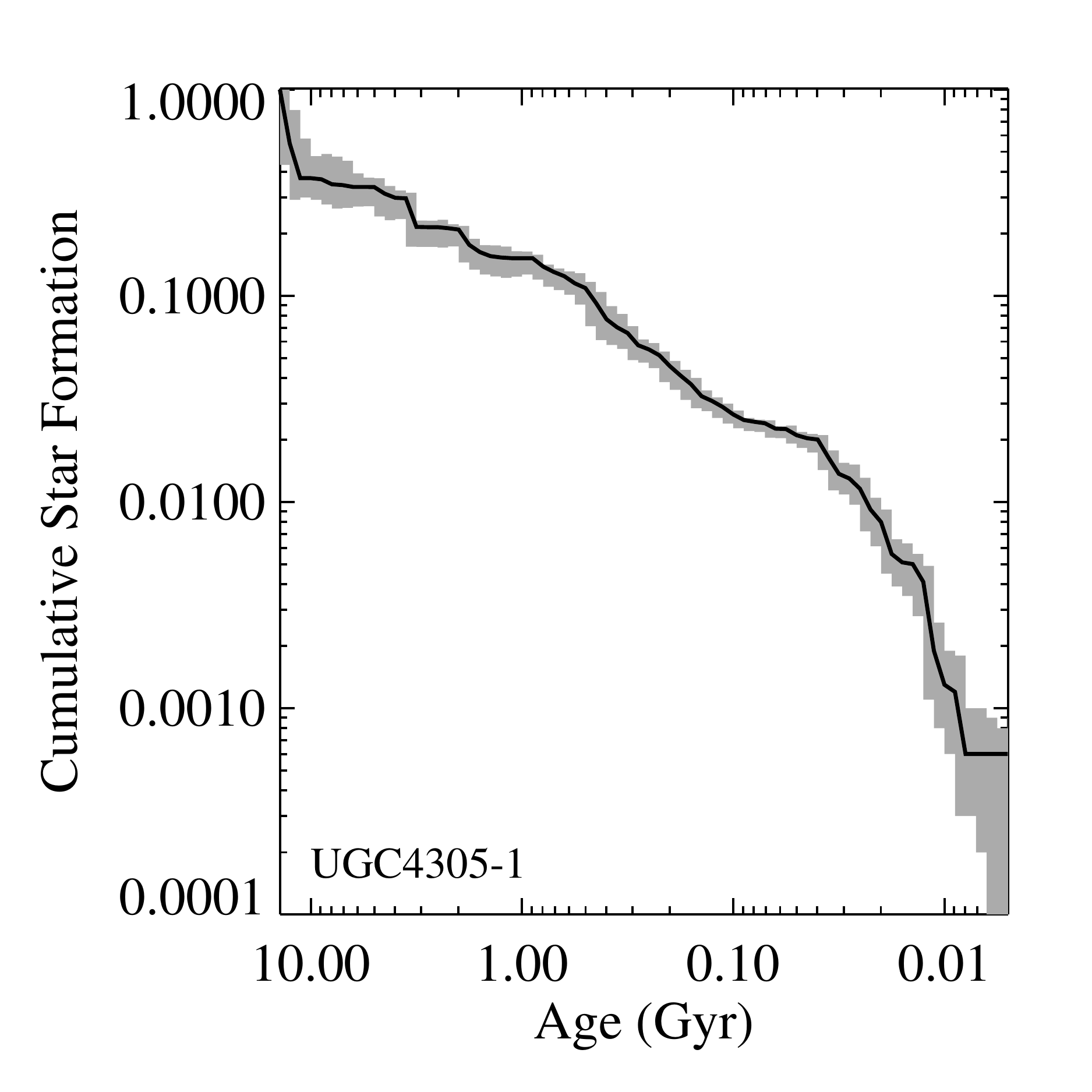}
}
\caption{ Color magnitude diagrams of the WFC3/IR (upper left) and
  optical (upper right) for the target UGC4305-1 within galaxy HoII.
  Lower panels show the star formation history derived from the
  optical data, for both the differential (left, with horizontal
  dotted line indicating the past average SFR) and cumulative (right)
  star formation histories.  The cumulative star formation history is
  calculated from the present back to 14\,Gyrs.  Uncertainties in the
  lower two panels are the 68\% confidence intervals, calculated from
  Monte Carlo tests including random and systematic uncertainties.
  Optical CMDs are restricted to the area covered by the WFC3 FOV. }
\end{figure}
\vfill
\clearpage
 
\begin{figure}
\figurenum{\ref{cmdfig} continued}
\centerline{
\includegraphics[width=3.25in]{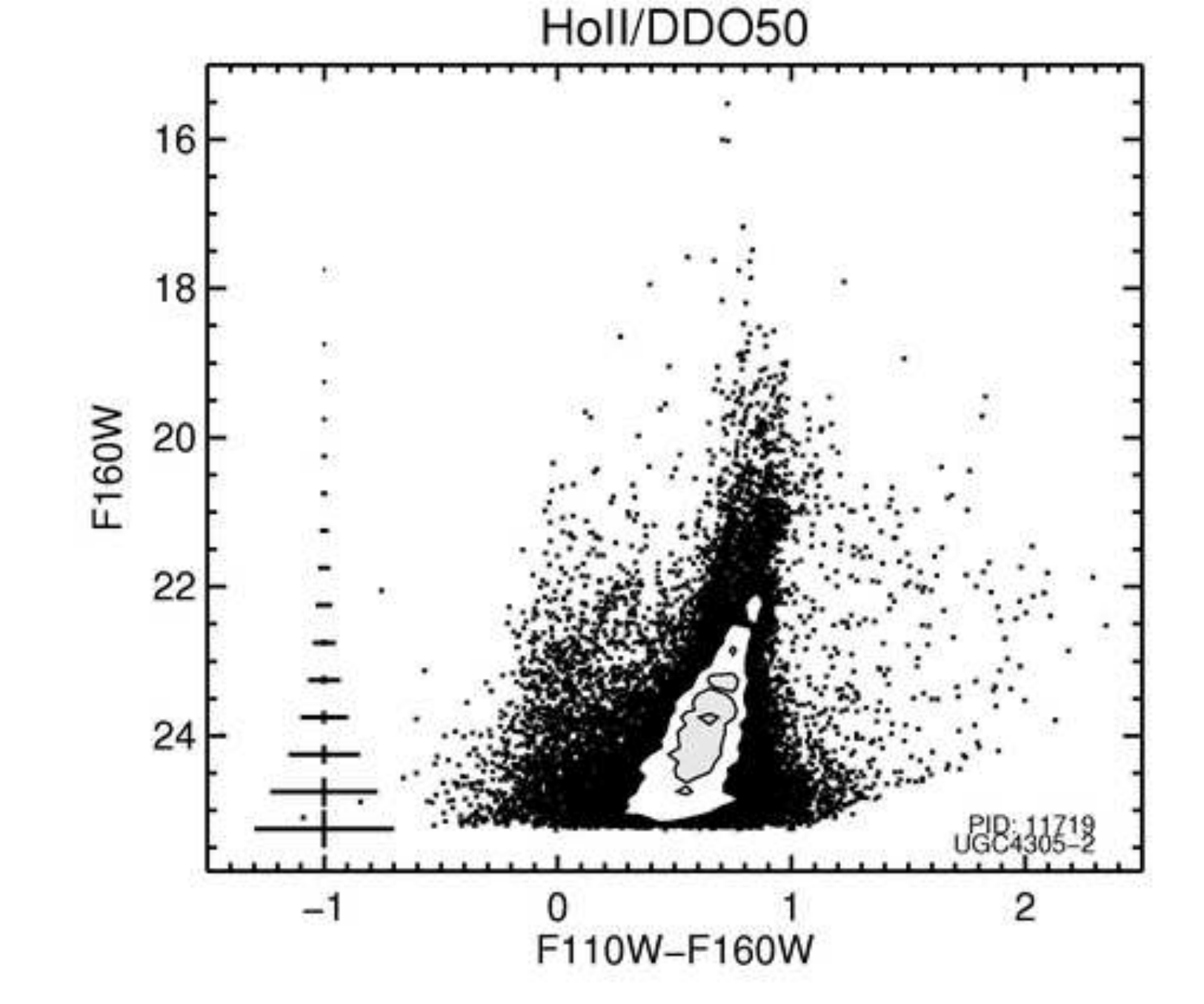}
\includegraphics[width=3.25in]{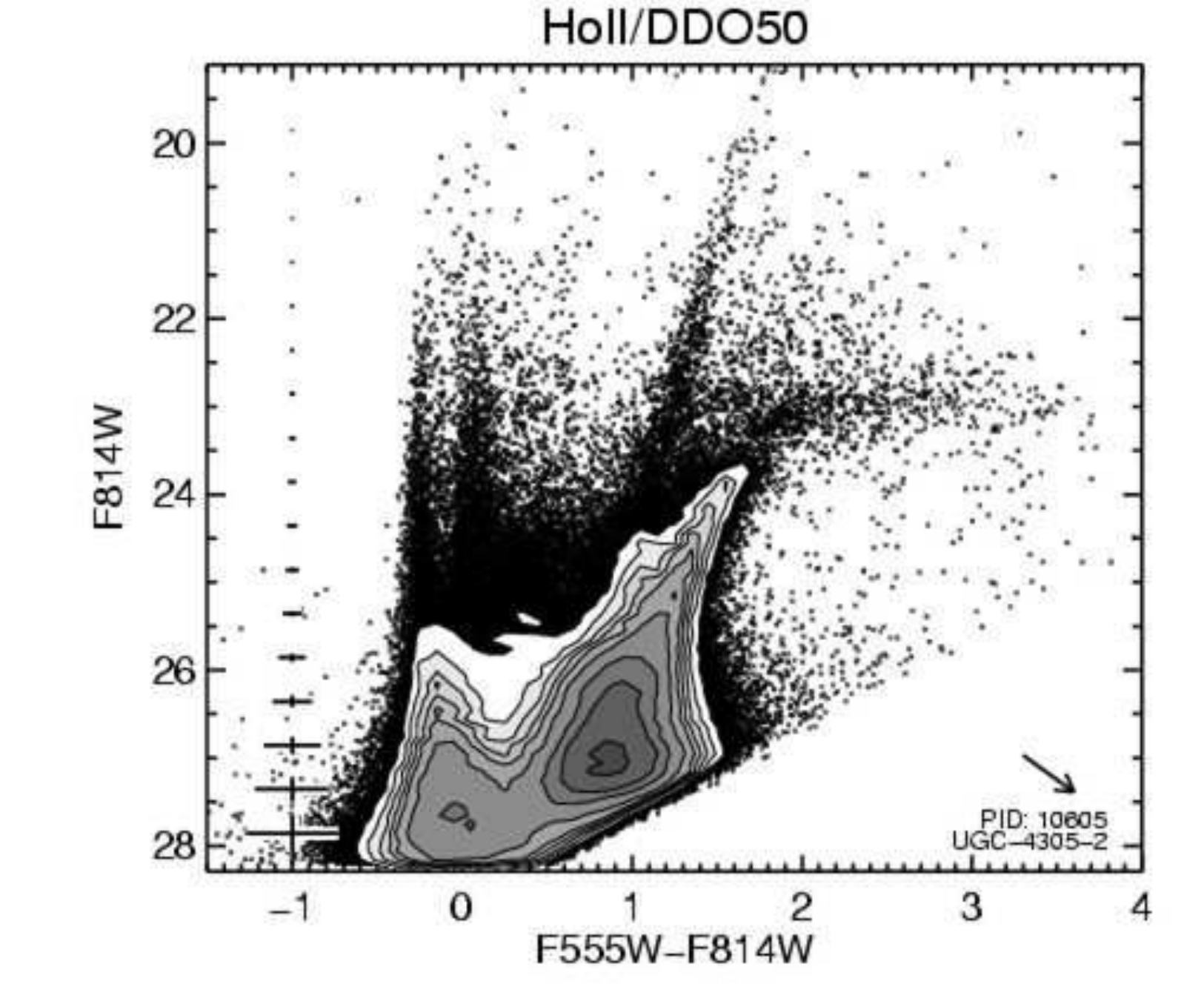}
}
\centerline{
\includegraphics[width=3.25in]{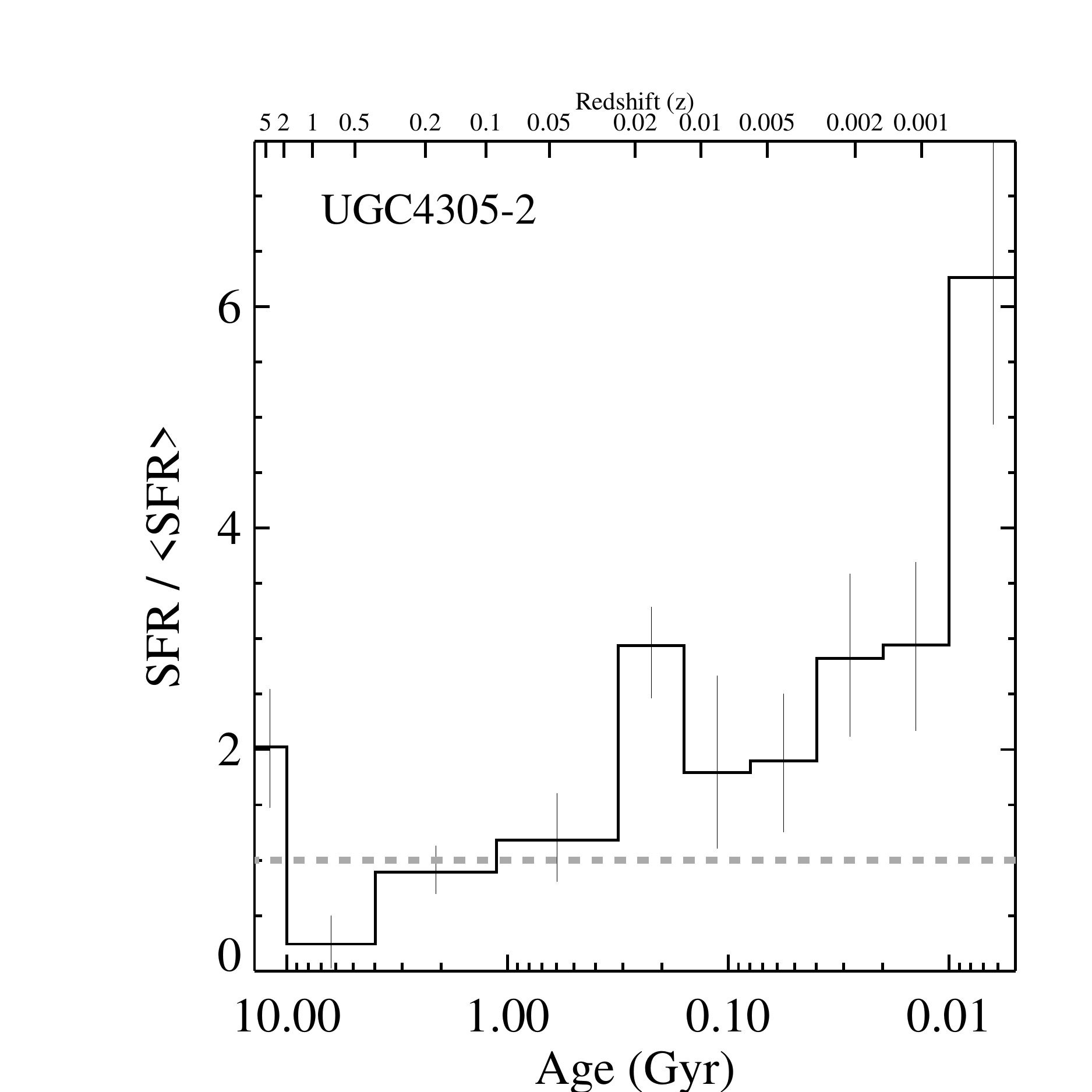}
\includegraphics[width=3.25in]{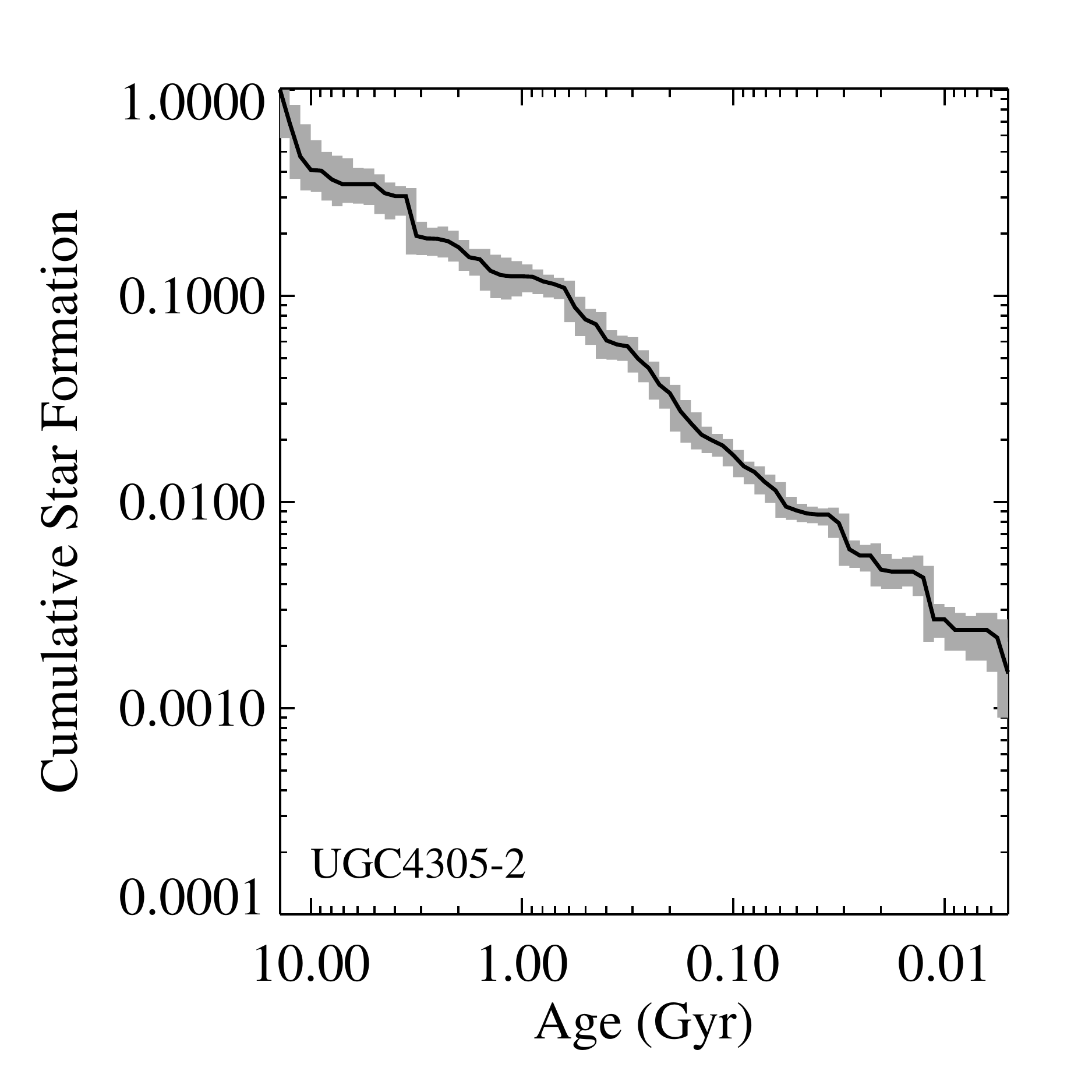}
}
\caption{ Color magnitude diagrams of the WFC3/IR (upper left) and
  optical (upper right) for the target UGC4305-2 within galaxy HoII.
  Lower panels show the star formation history derived from the
  optical data, for both the differential (left, with horizontal
  dotted line indicating the past average SFR) and cumulative (right)
  star formation histories.  The cumulative star formation history is
  calculated from the present back to 14\,Gyrs.  Uncertainties in the
  lower two panels are the 68\% confidence intervals, calculated from
  Monte Carlo tests including random and systematic uncertainties.
  Optical CMDs are restricted to the area covered by the WFC3 FOV. }
\end{figure}
\vfill
\clearpage
 
\begin{figure}
\figurenum{\ref{cmdfig} continued}
\centerline{
\includegraphics[width=3.25in]{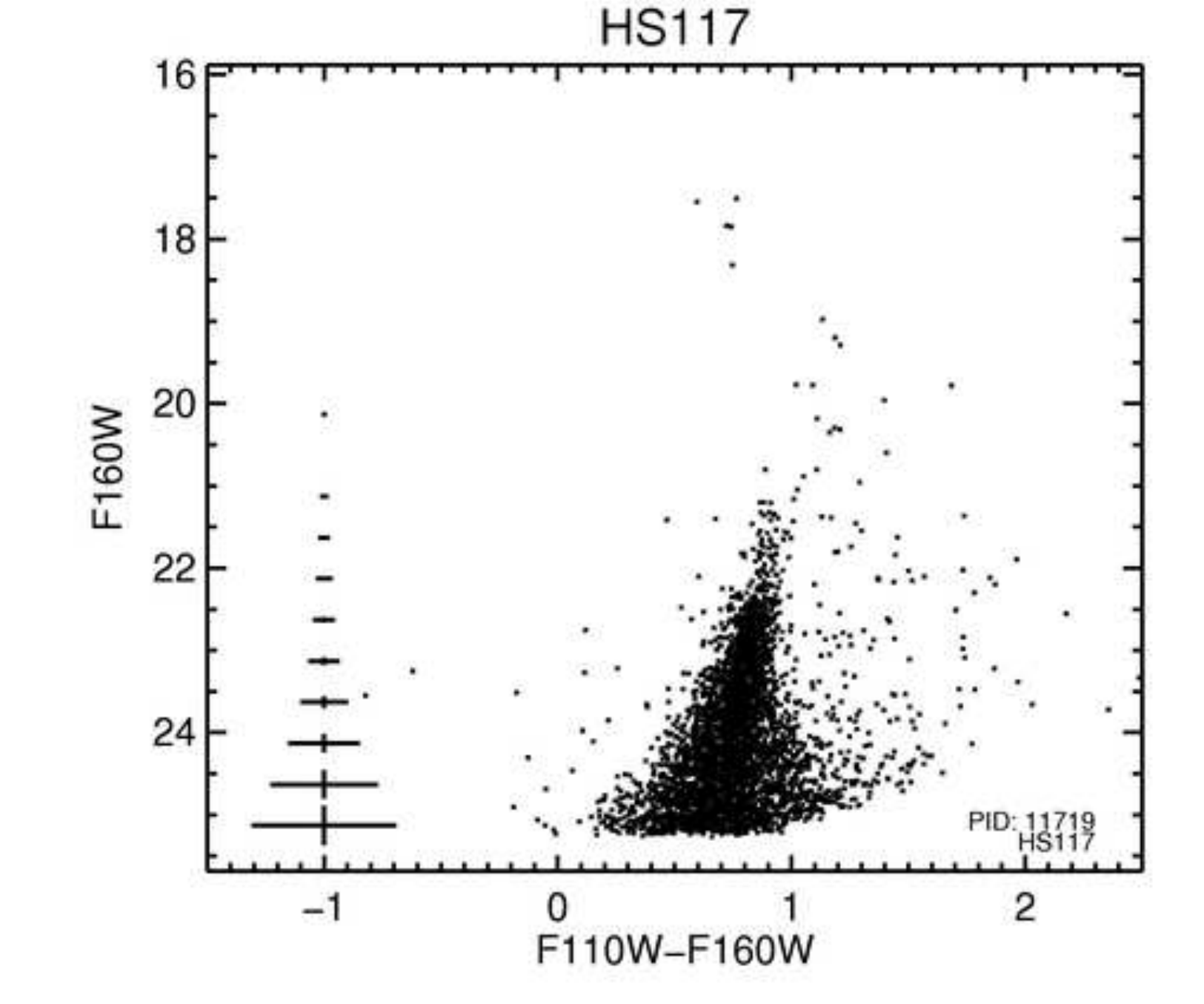}
\includegraphics[width=3.25in]{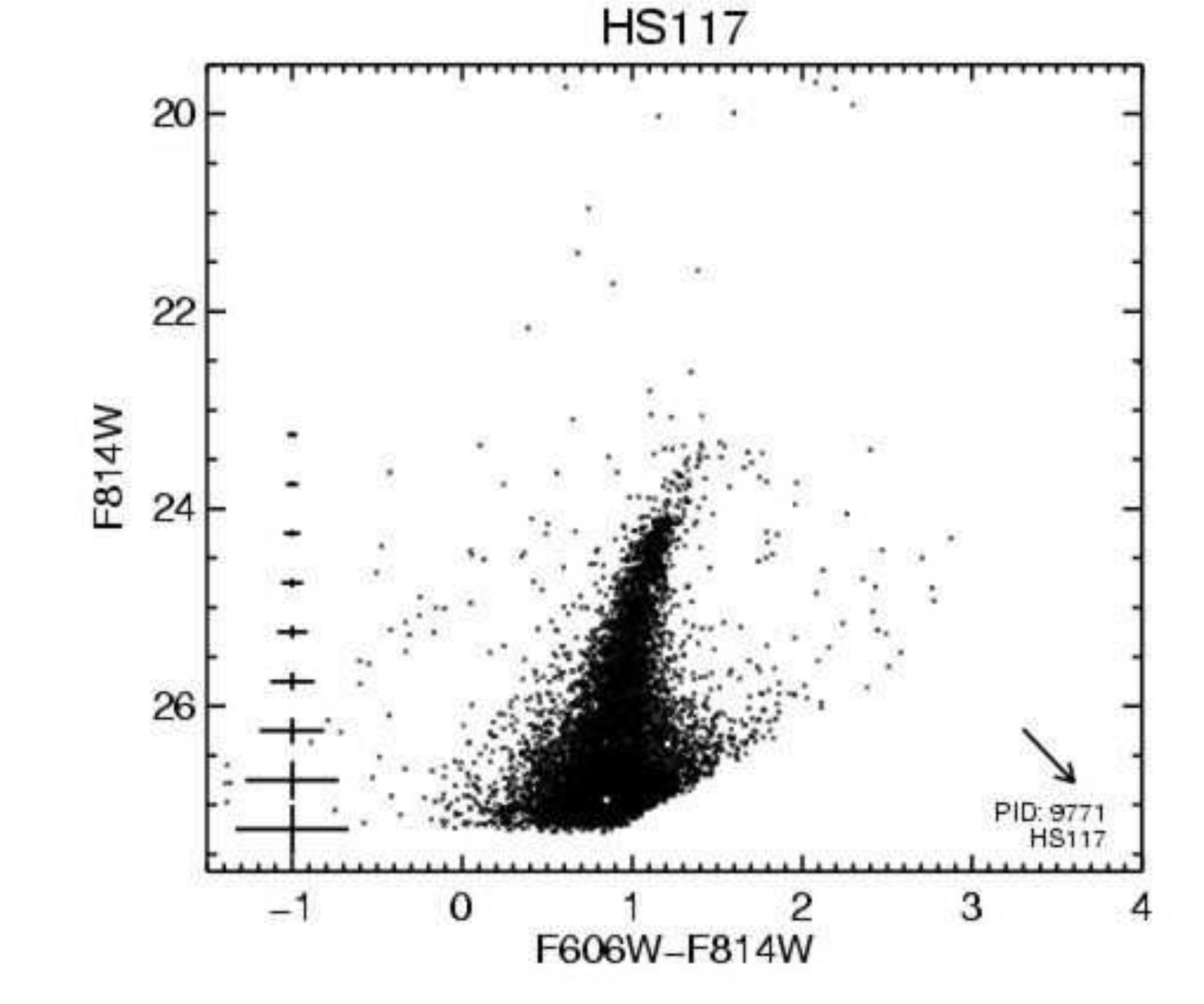}
}
\centerline{
\includegraphics[width=3.25in]{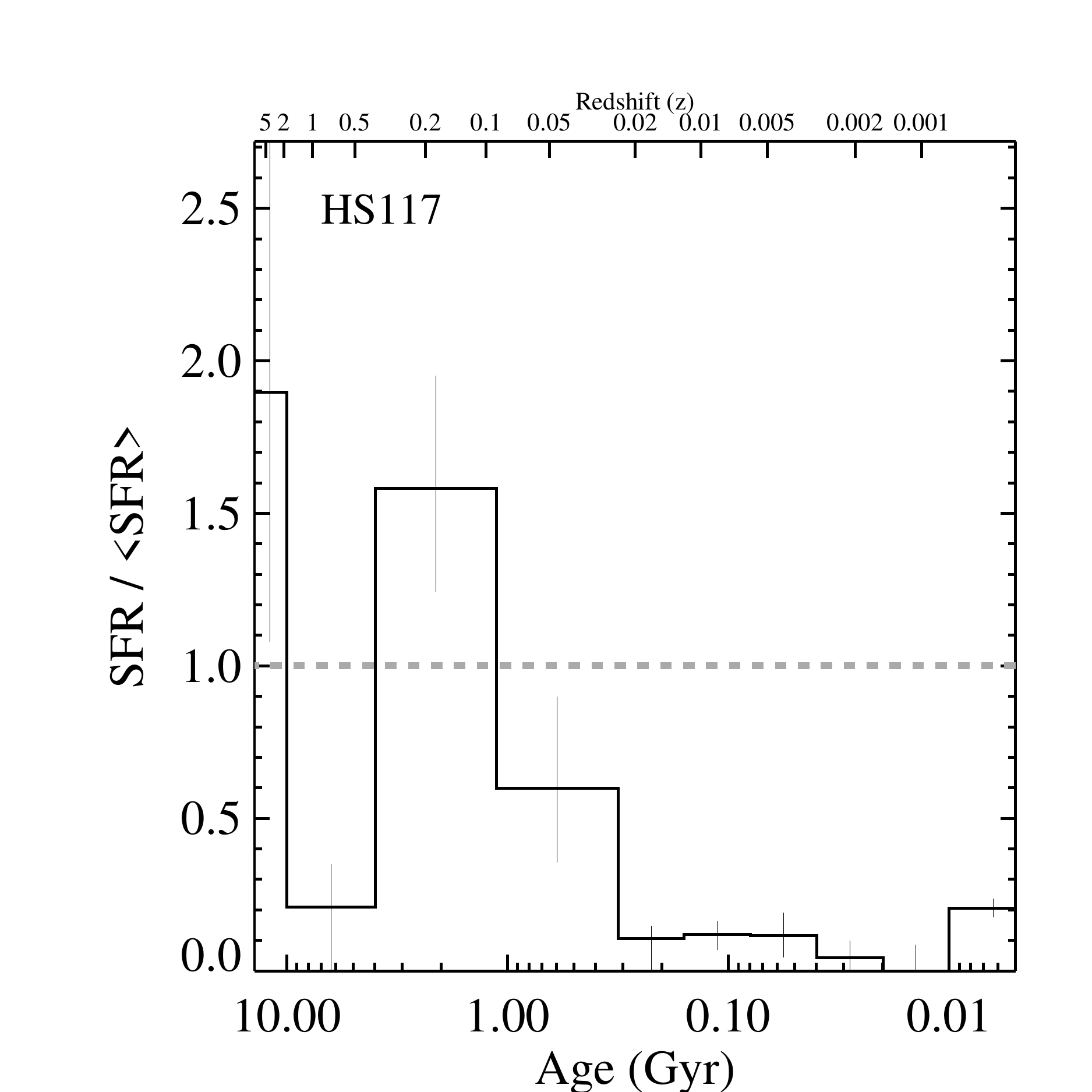}
\includegraphics[width=3.25in]{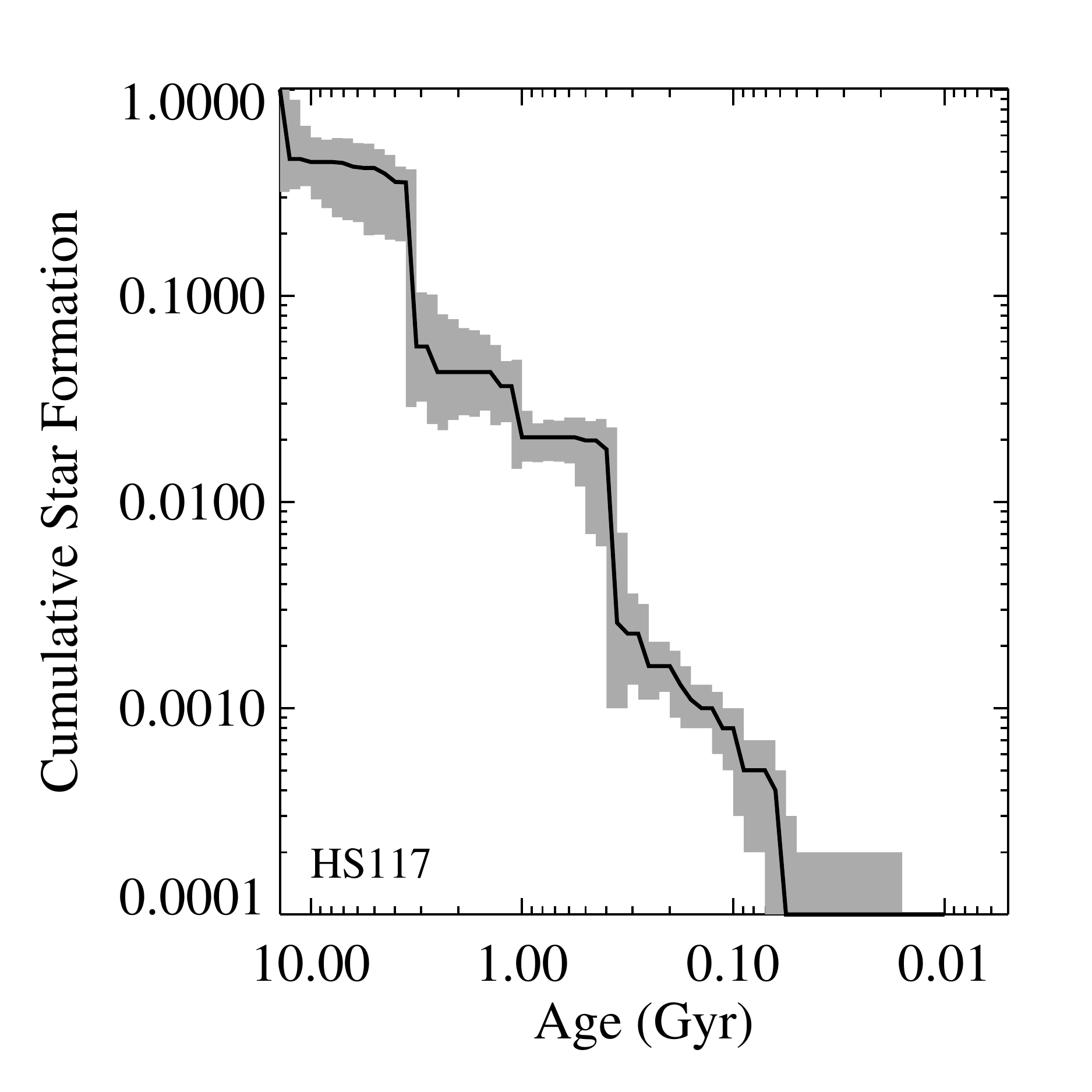}
}
\caption{ Color magnitude diagrams of the WFC3/IR (upper left) and
  optical (upper right) for HS117.  Lower panels show the star
  formation history derived from the optical data, for both the
  differential (left, with horizontal dotted line indicating the past
  average SFR) and cumulative (right) star formation histories.  The
  cumulative star formation history is calculated from the present
  back to 14\,Gyrs.  Uncertainties in the lower two panels are the
  68\% confidence intervals, calculated from Monte Carlo tests
  including random and systematic uncertainties.  Optical CMDs are
  restricted to the area covered by the WFC3 FOV. }
\end{figure}
\vfill
\clearpage
 
\begin{figure}
\figurenum{\ref{cmdfig} continued}
\centerline{
\includegraphics[width=3.25in]{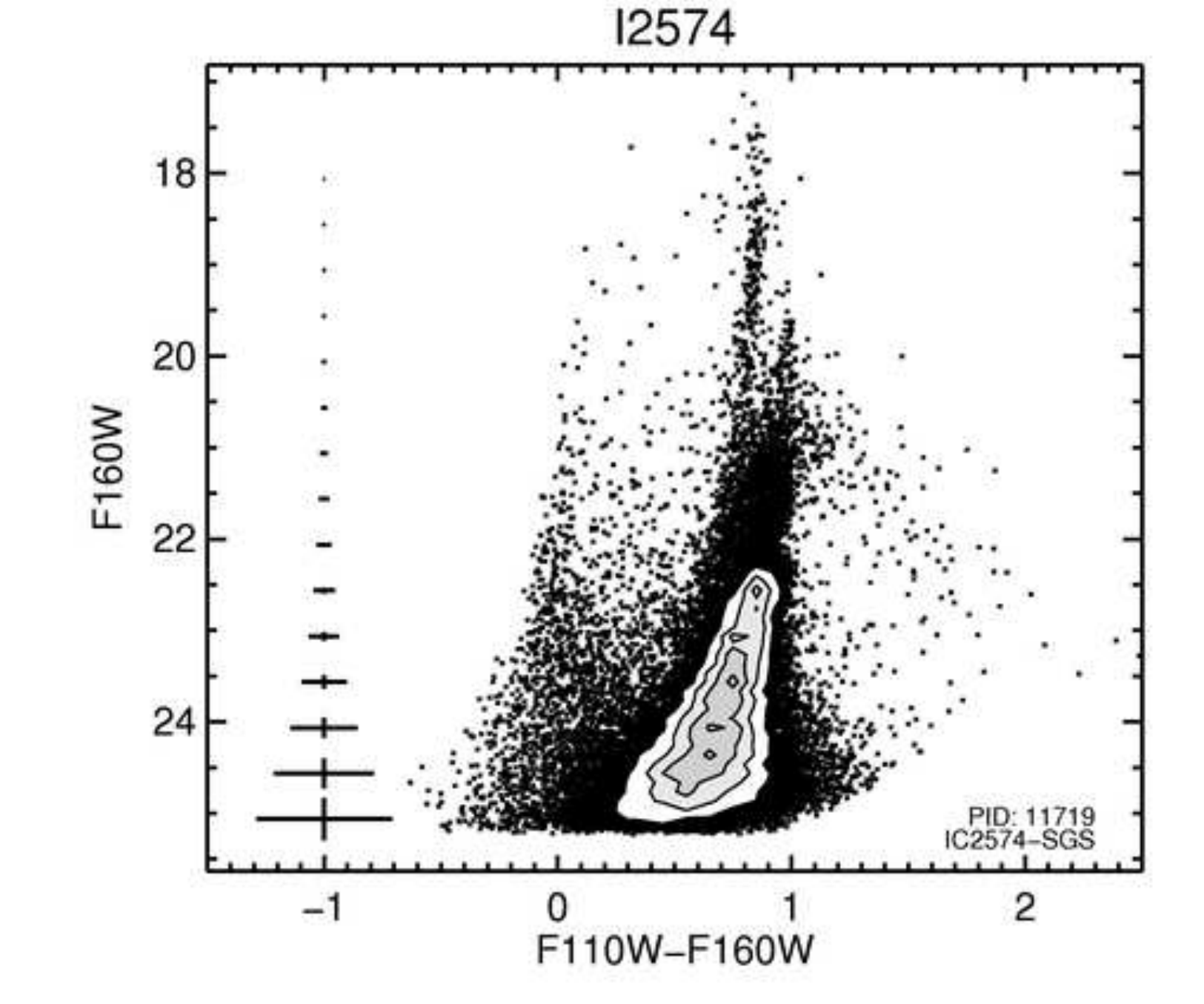}
\includegraphics[width=3.25in]{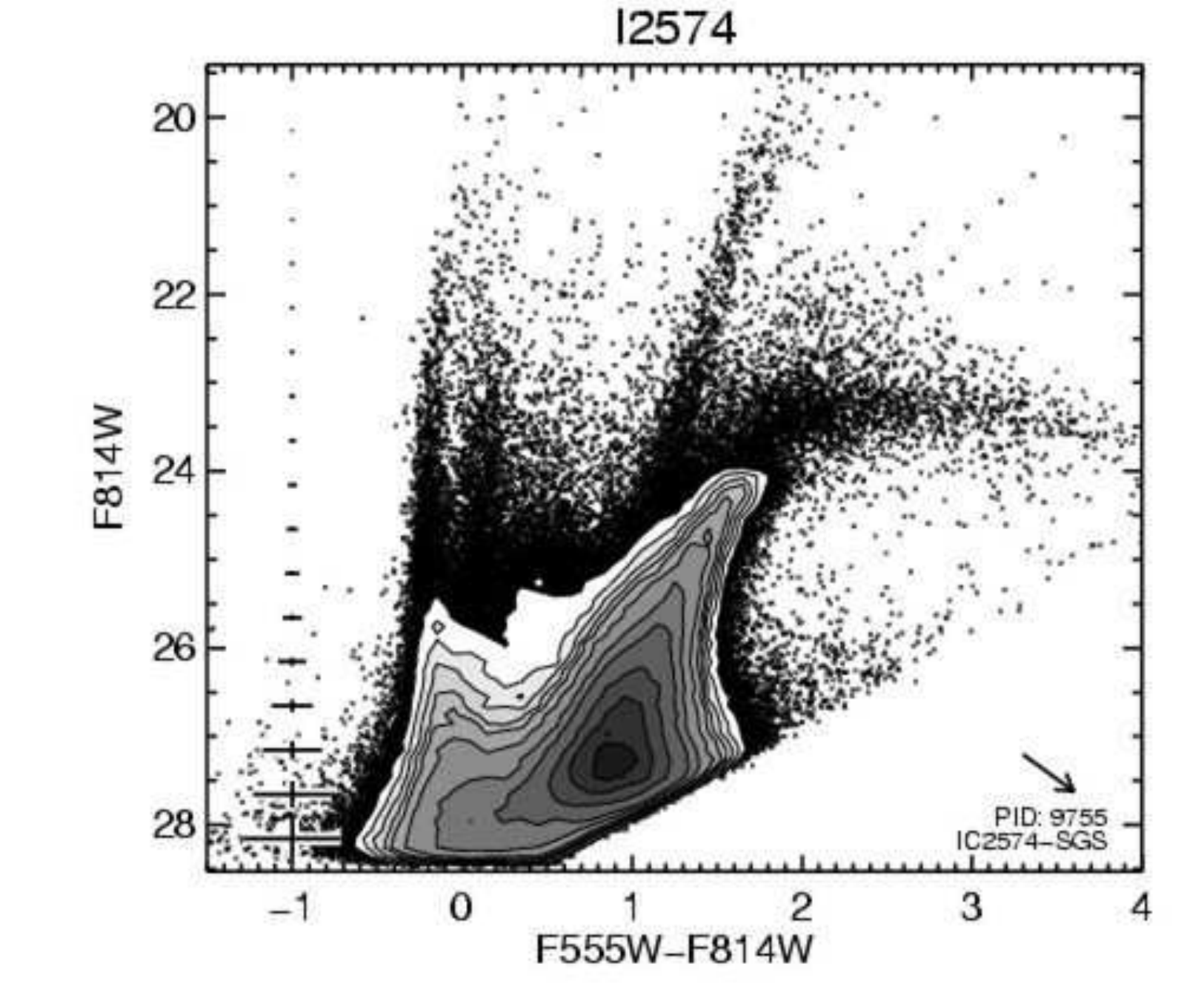}
}
\centerline{
\includegraphics[width=3.25in]{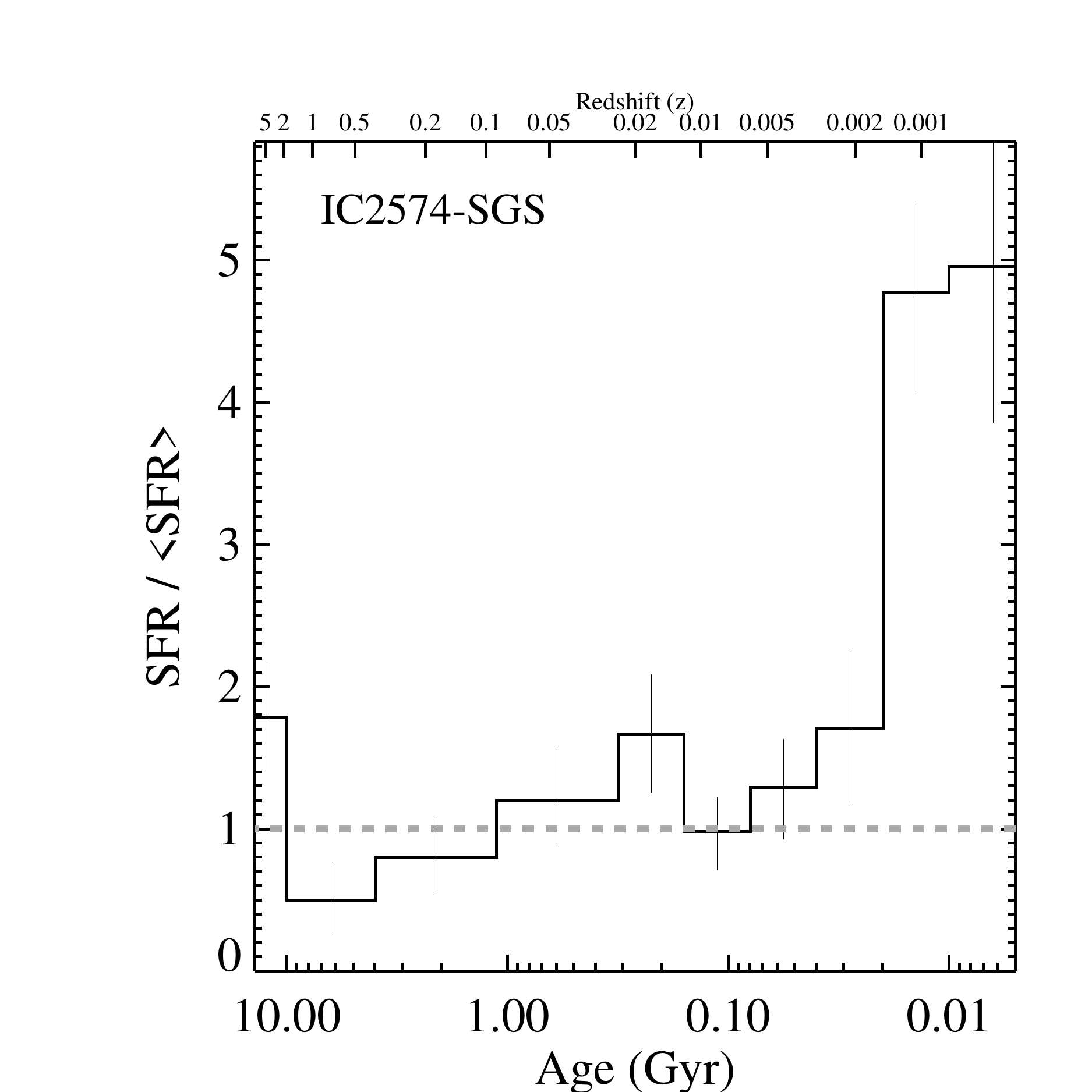}
\includegraphics[width=3.25in]{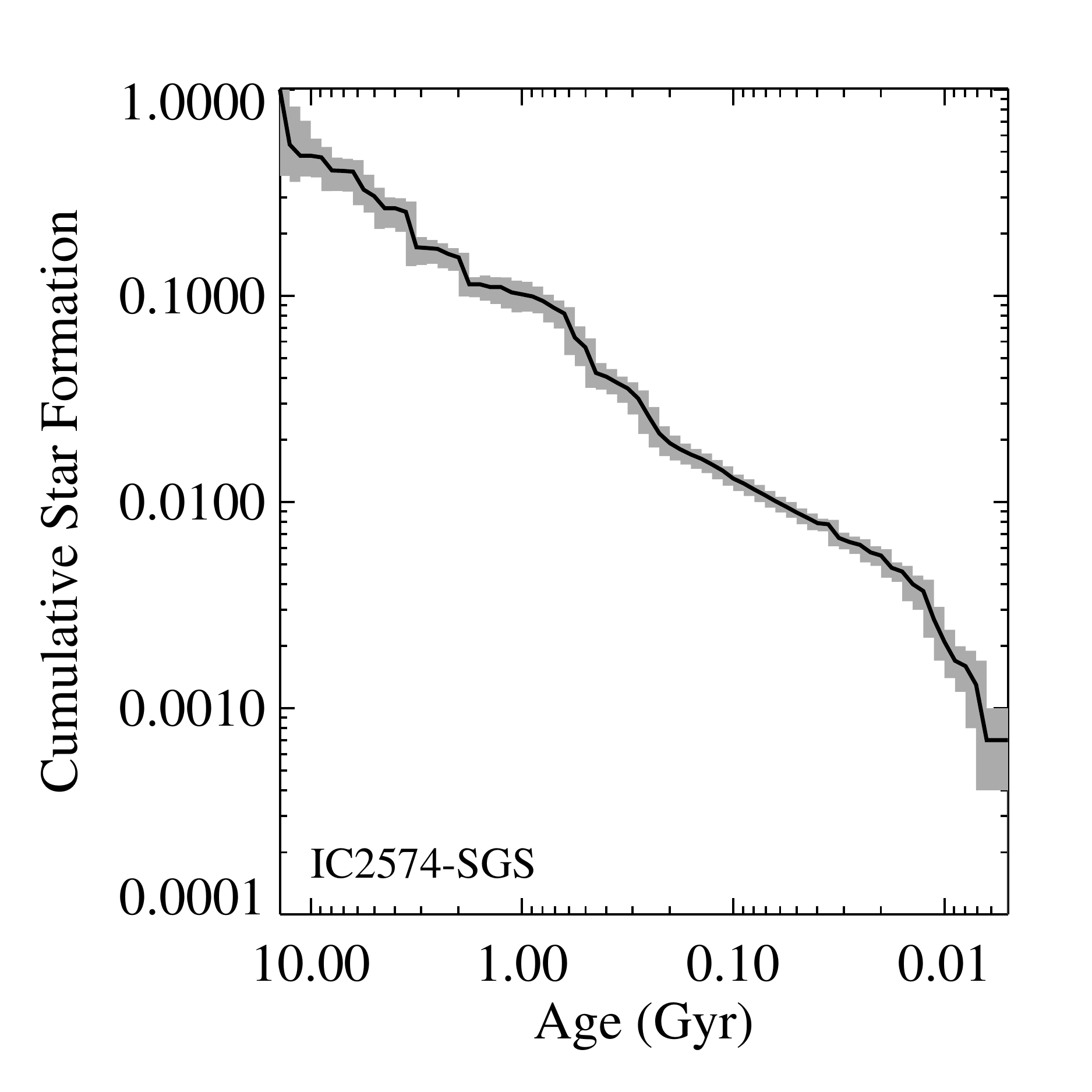}
}
\caption{ Color magnitude diagrams of the WFC3/IR (upper left) and
  optical (upper right) for the target IC2574-SGS within galaxy I2574.
  Lower panels show the star formation history derived from the
  optical data, for both the differential (left, with horizontal
  dotted line indicating the past average SFR) and cumulative (right)
  star formation histories.  The cumulative star formation history is
  calculated from the present back to 14\,Gyrs.  Uncertainties in the
  lower two panels are the 68\% confidence intervals, calculated from
  Monte Carlo tests including random and systematic uncertainties.
  Optical CMDs are restricted to the area covered by the WFC3 FOV. }
\end{figure}
\vfill
\clearpage
 
\begin{figure}
\figurenum{\ref{cmdfig} continued}
\centerline{
\includegraphics[width=3.25in]{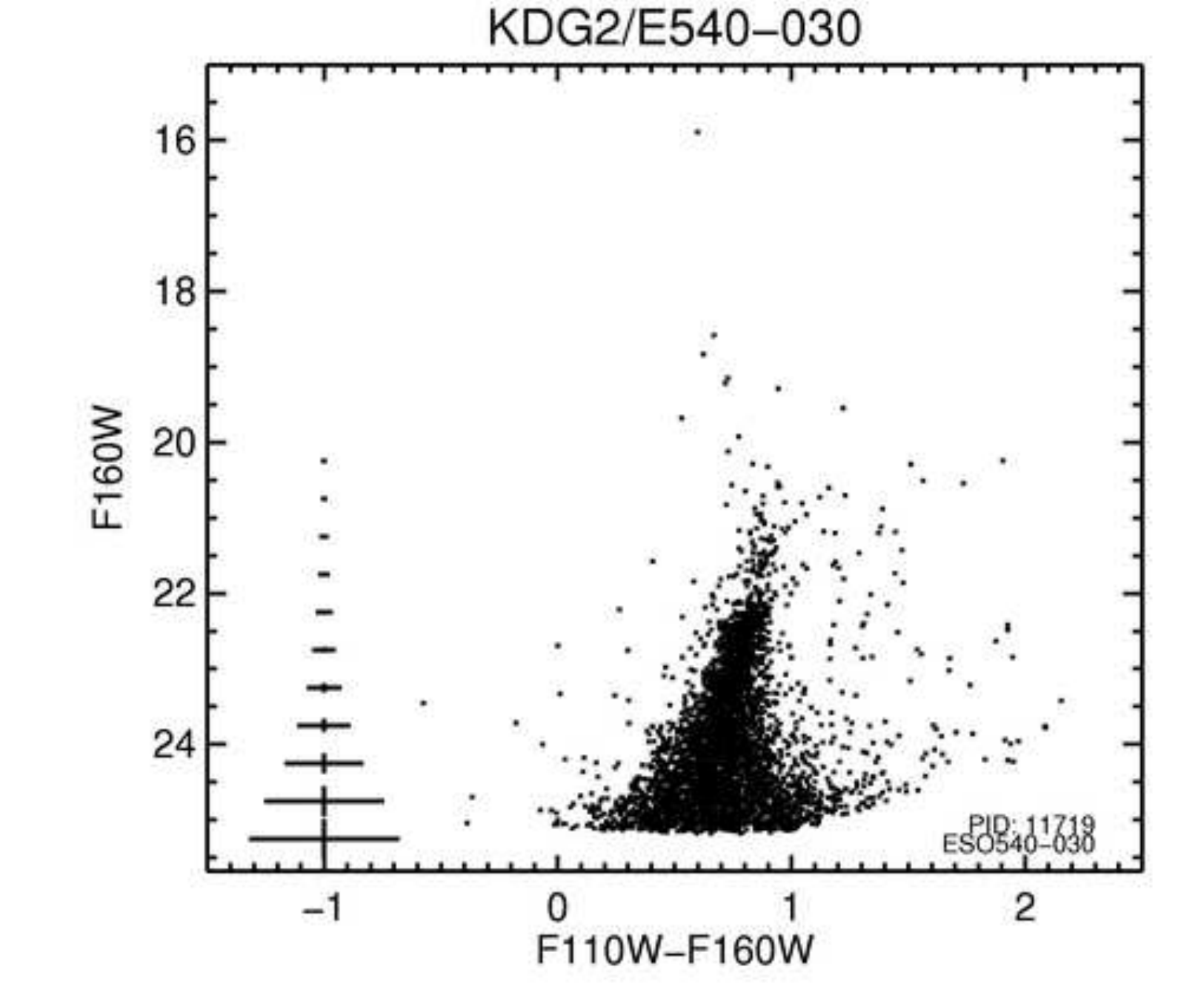}
\includegraphics[width=3.25in]{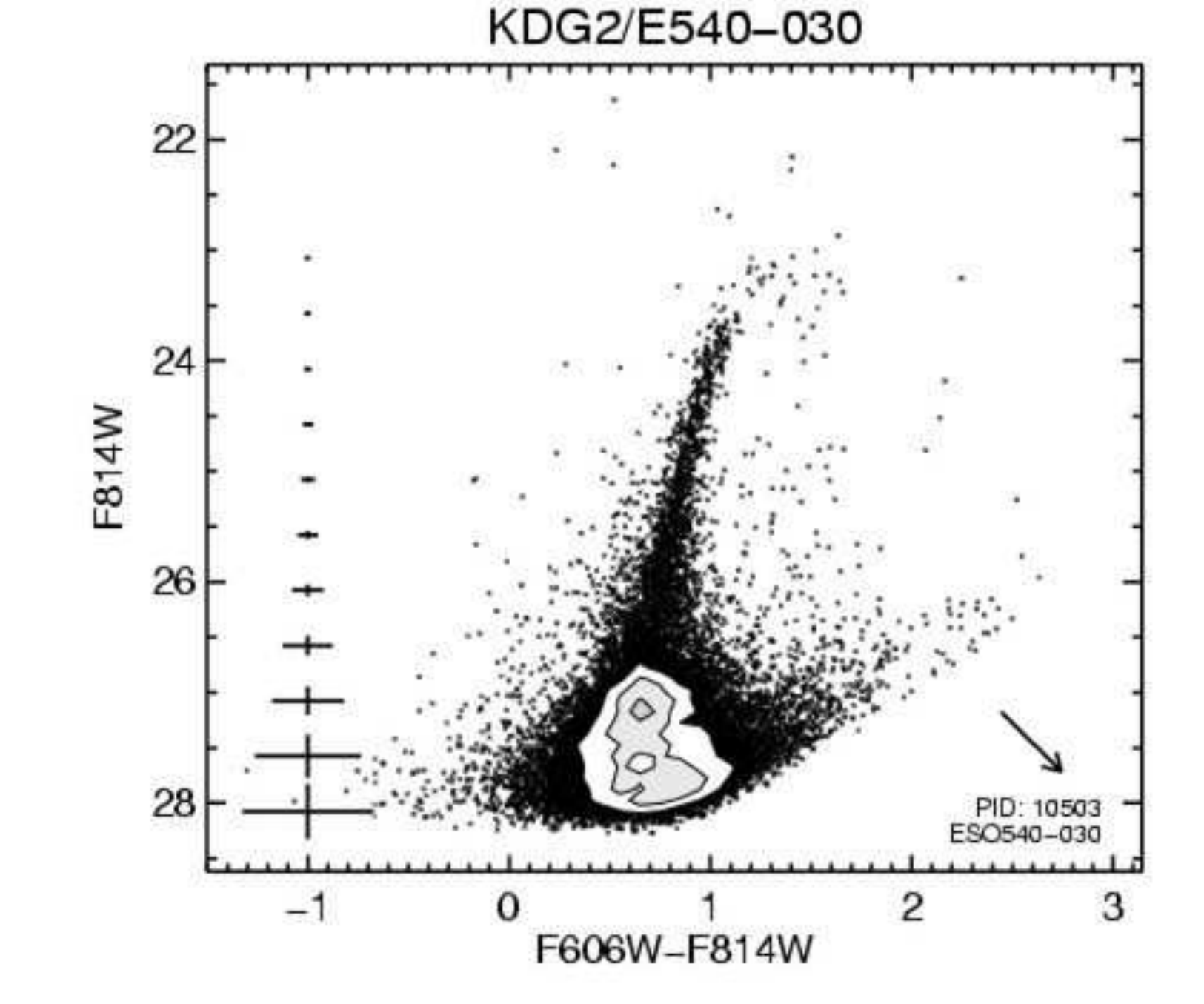}
}
\centerline{
\includegraphics[width=3.25in]{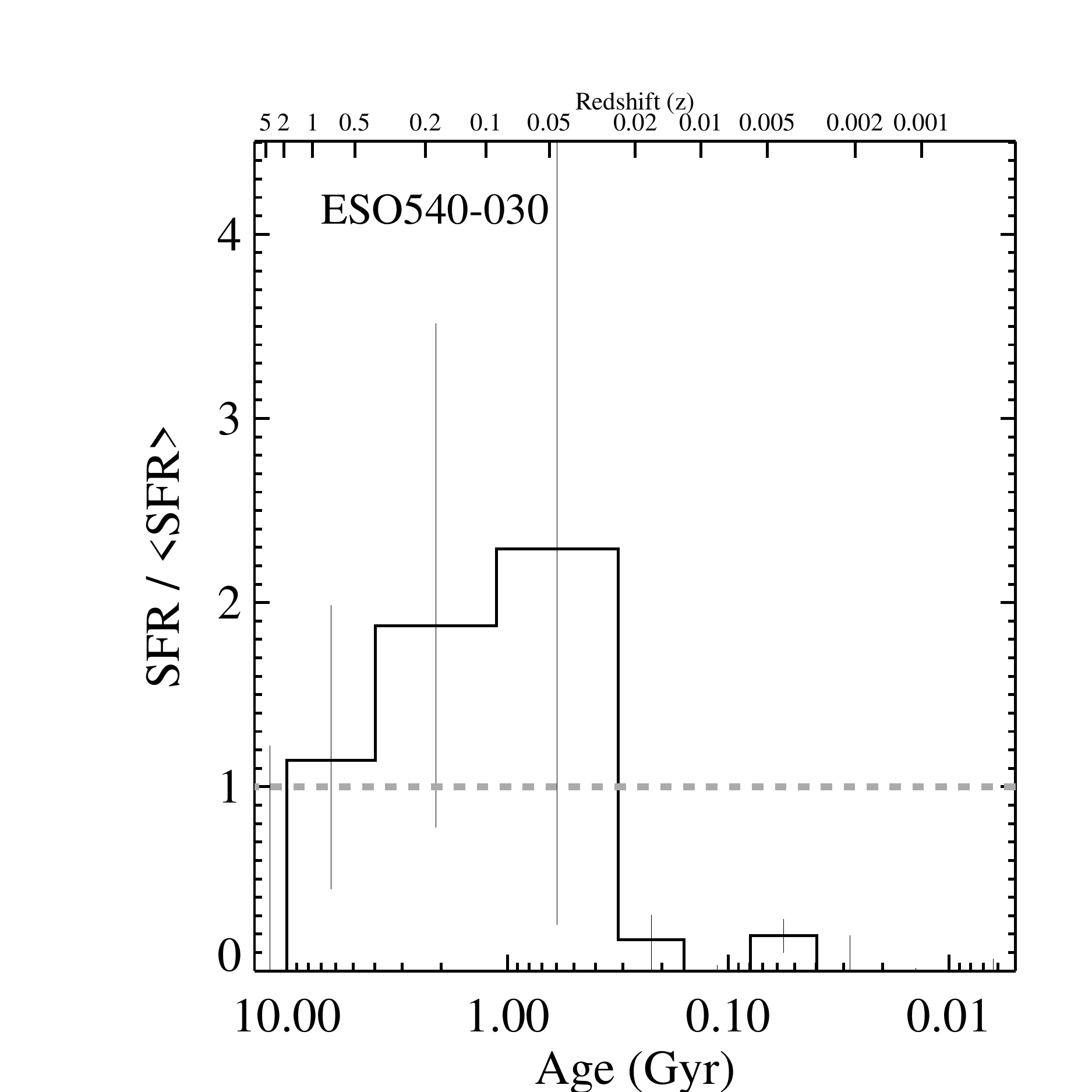}
\includegraphics[width=3.25in]{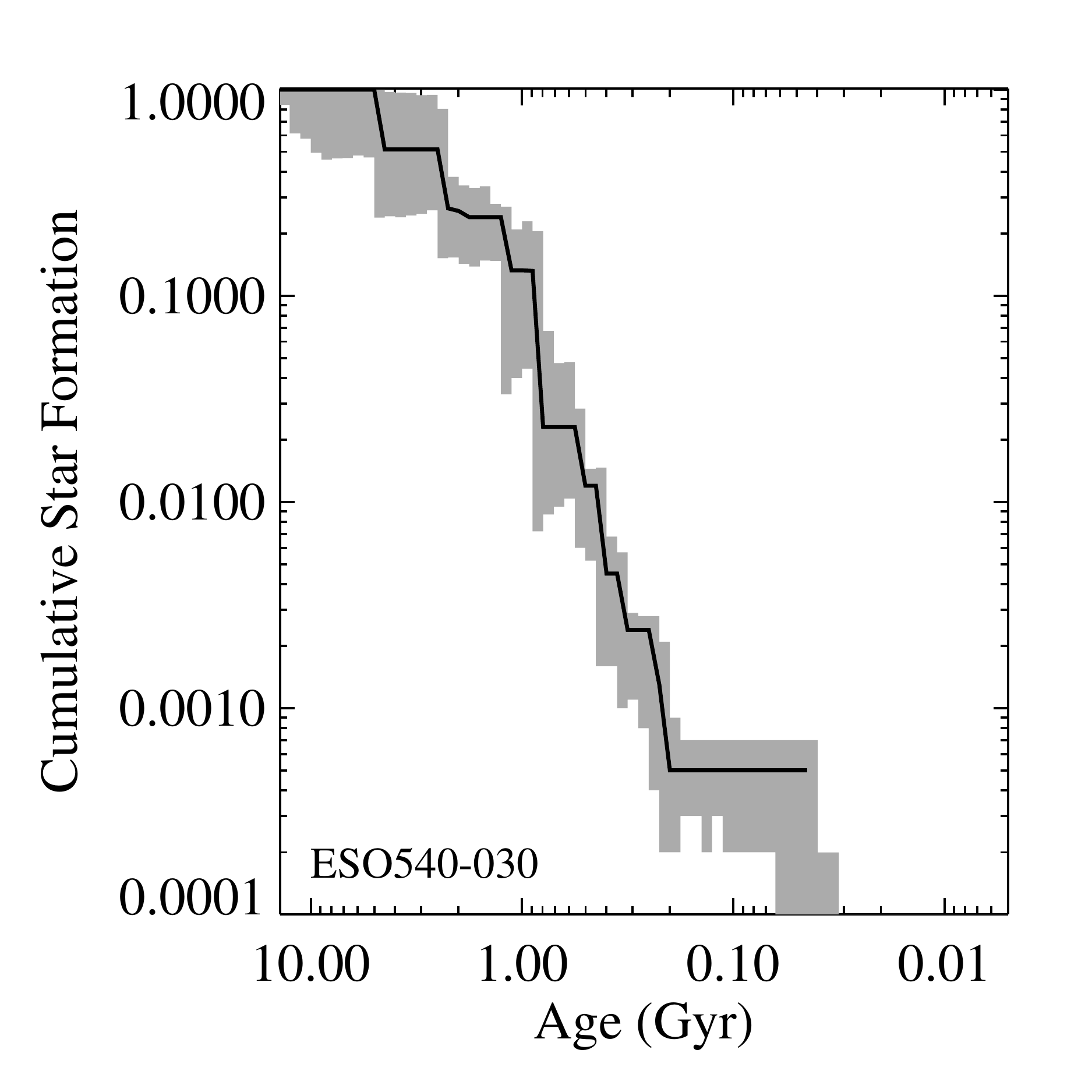}
}
\caption{ Color magnitude diagrams of the WFC3/IR (upper left) and
  optical (upper right) for the target ESO540-030 within galaxy KDG2.
  Lower panels show the star formation history derived from the
  optical data, for both the differential (left, with horizontal
  dotted line indicating the past average SFR) and cumulative (right)
  star formation histories.  The cumulative star formation history is
  calculated from the present back to 14\,Gyrs.  Uncertainties in the
  lower two panels are the 68\% confidence intervals, calculated from
  Monte Carlo tests including random and systematic uncertainties.
  Optical CMDs are restricted to the area covered by the WFC3 FOV. }
\end{figure}
\vfill
\clearpage
 
\begin{figure}
\figurenum{\ref{cmdfig} continued}
\centerline{
\includegraphics[width=3.25in]{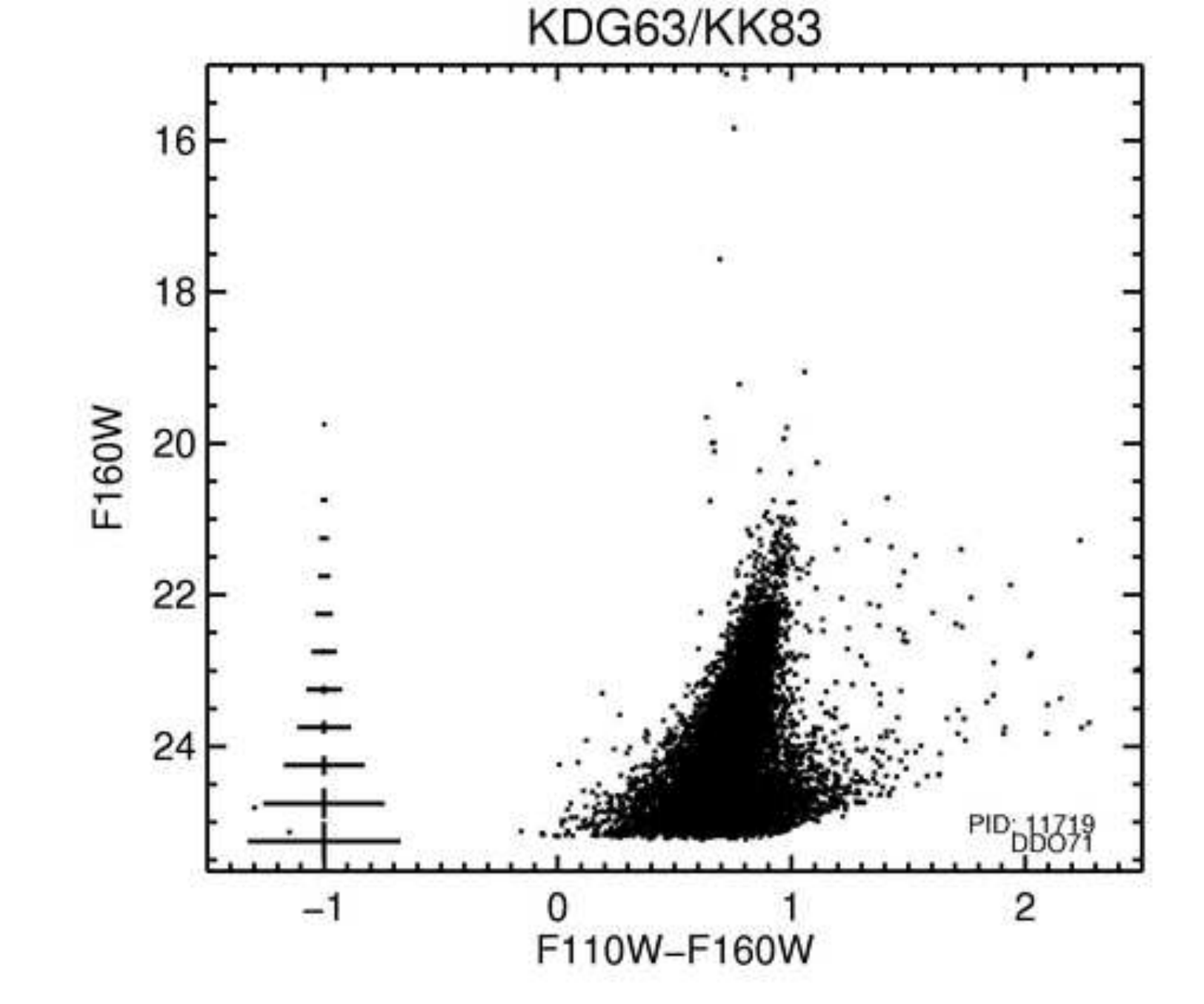}
\includegraphics[width=3.25in]{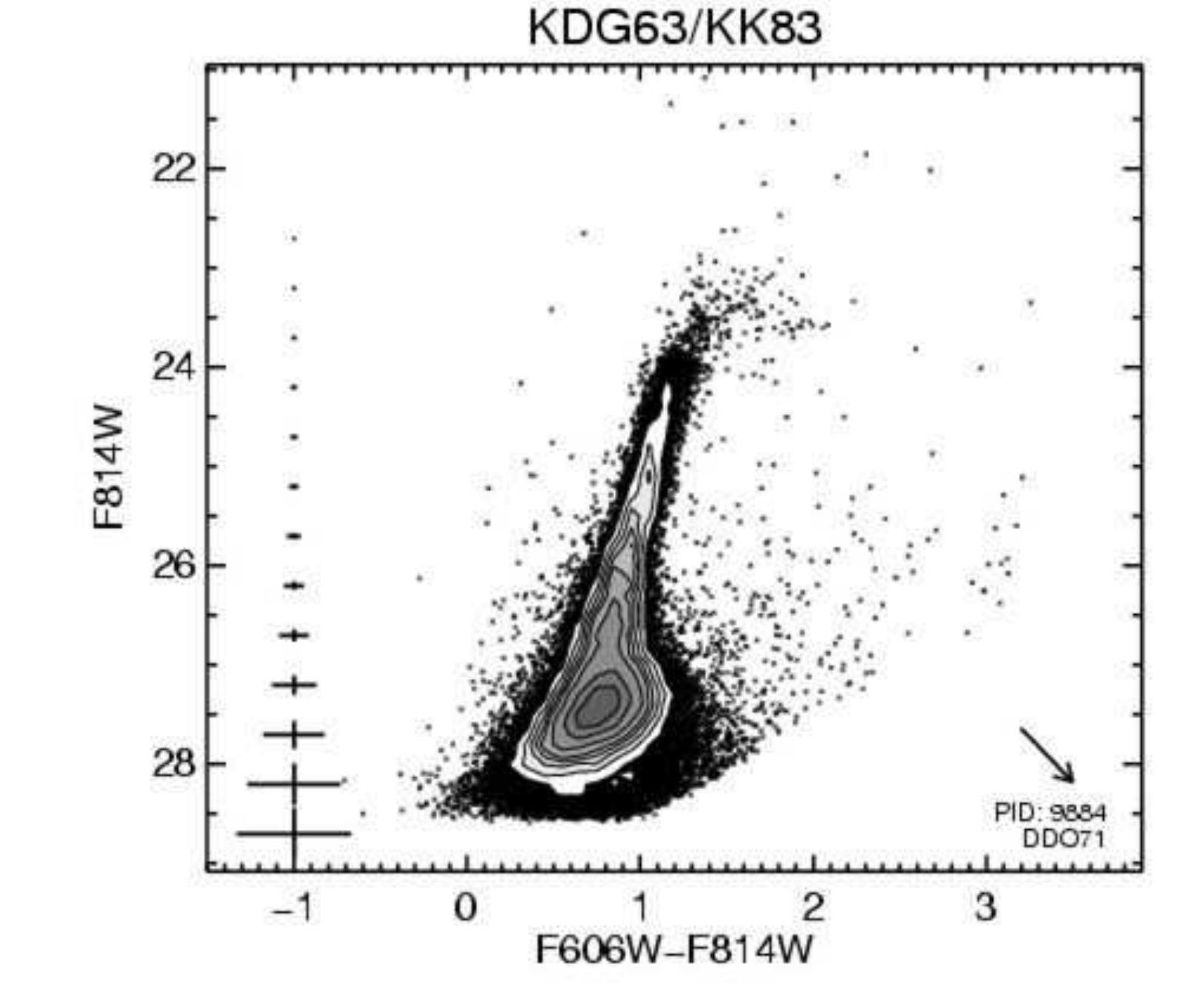}
}
\centerline{
\includegraphics[width=3.25in]{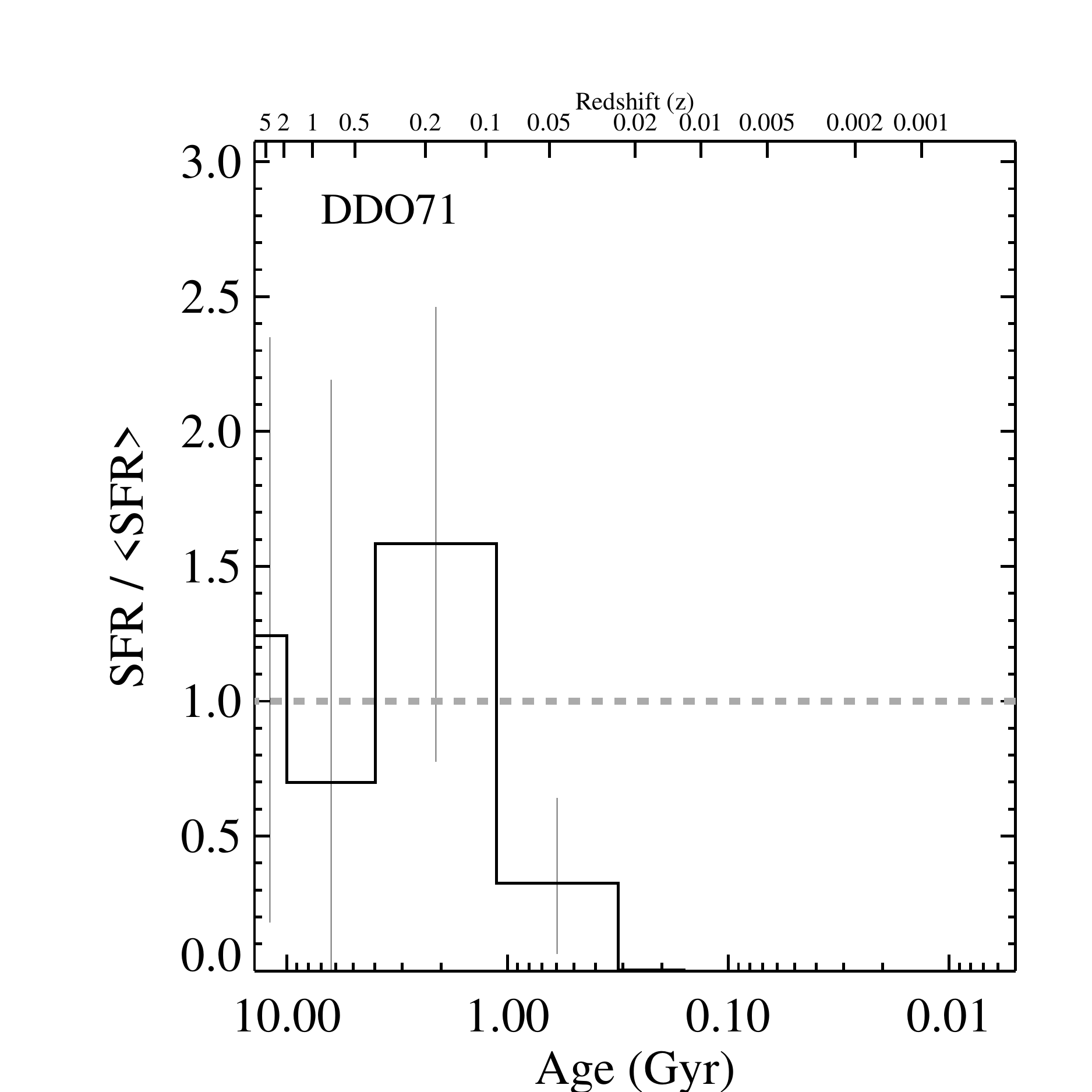}
\includegraphics[width=3.25in]{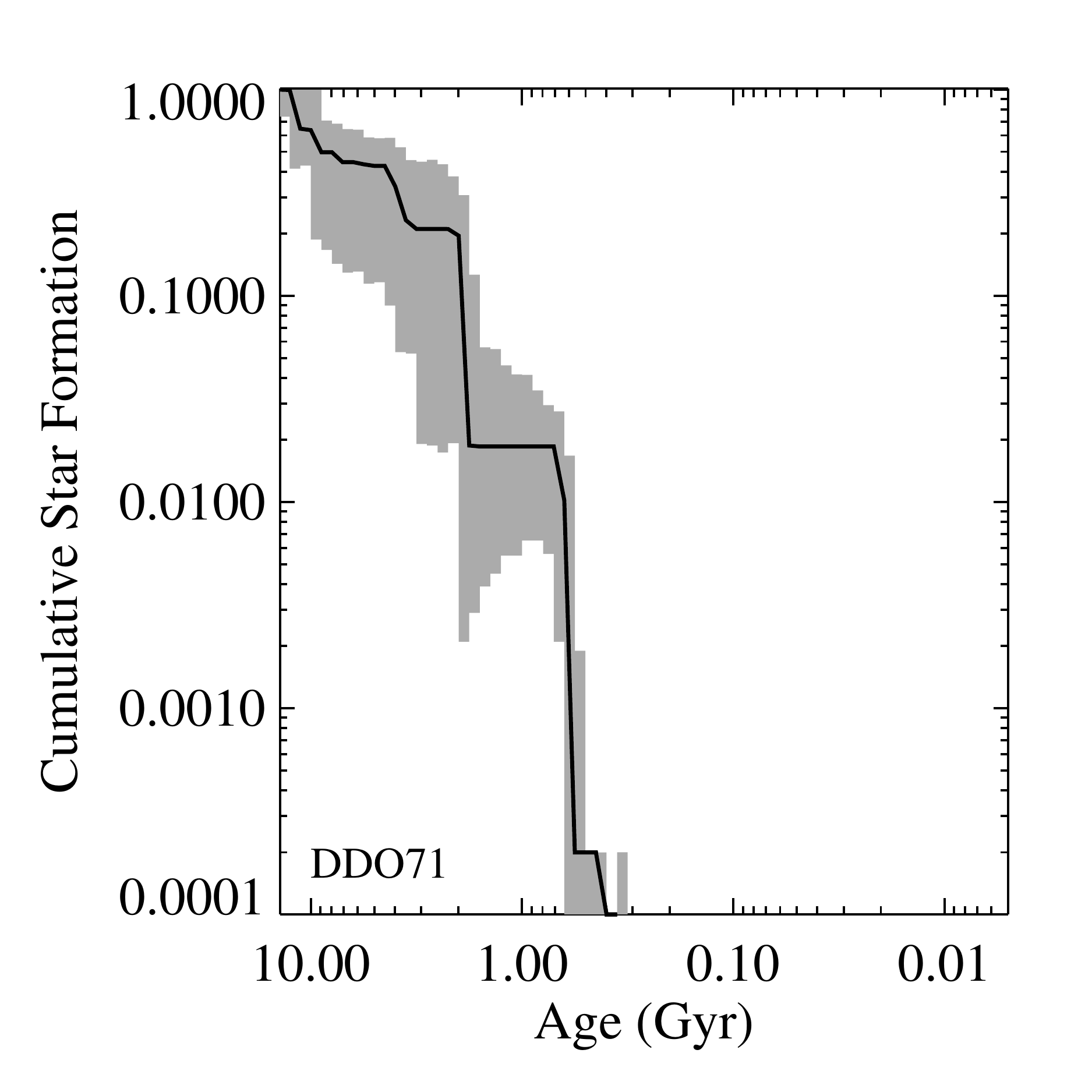}
}
\caption{ Color magnitude diagrams of the WFC3/IR (upper left) and
  optical (upper right) for the target DDO71 within galaxy KDG63.
  Lower panels show the star formation history derived from the
  optical data, for both the differential (left, with horizontal
  dotted line indicating the past average SFR) and cumulative (right)
  star formation histories.  The cumulative star formation history is
  calculated from the present back to 14\,Gyrs.  Uncertainties in the
  lower two panels are the 68\% confidence intervals, calculated from
  Monte Carlo tests including random and systematic uncertainties.
  Optical CMDs are restricted to the area covered by the WFC3 FOV. }
\end{figure}
\vfill
\clearpage
 
\begin{figure}
\figurenum{\ref{cmdfig} continued}
\centerline{
\includegraphics[width=3.25in]{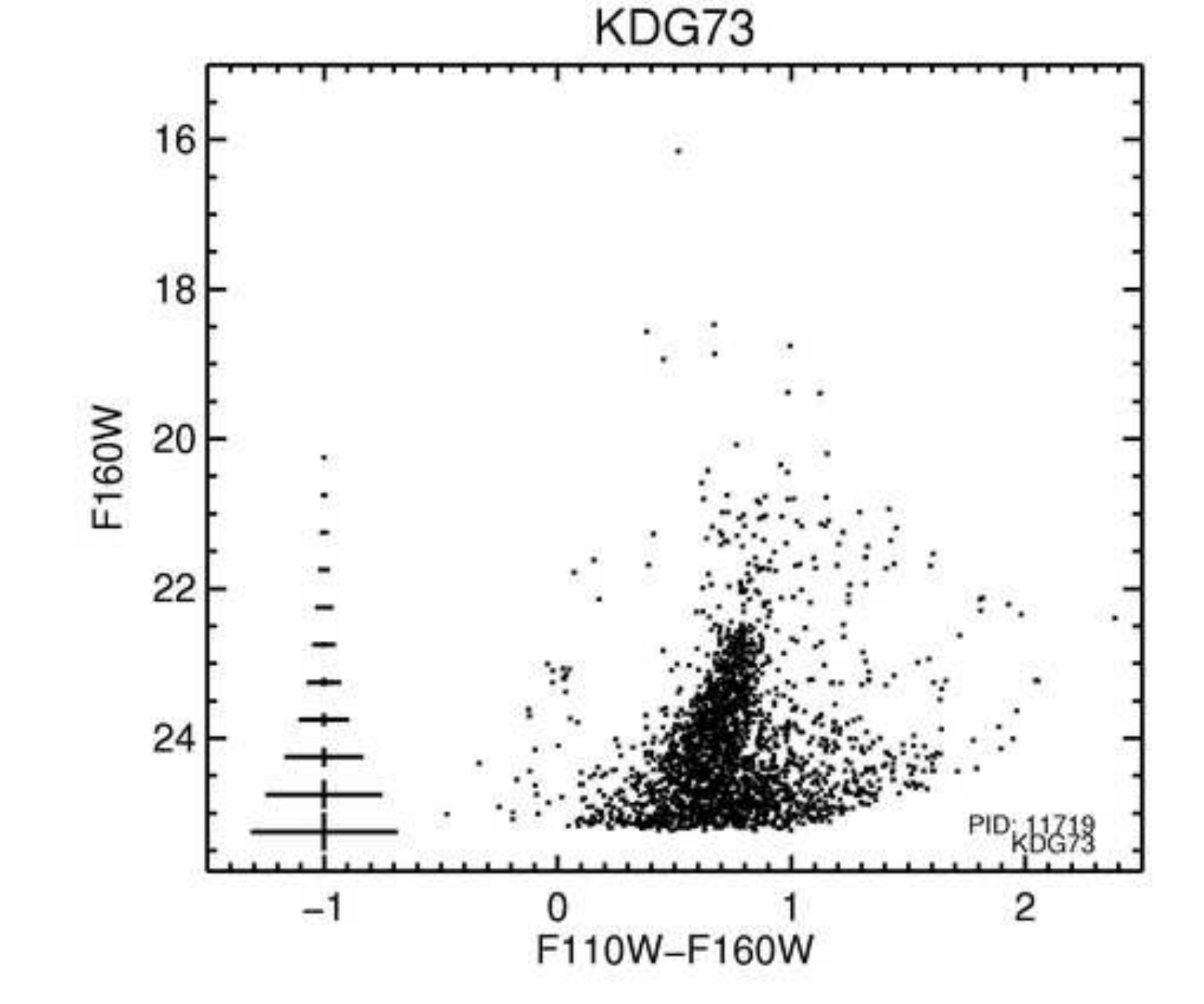}
\includegraphics[width=3.25in]{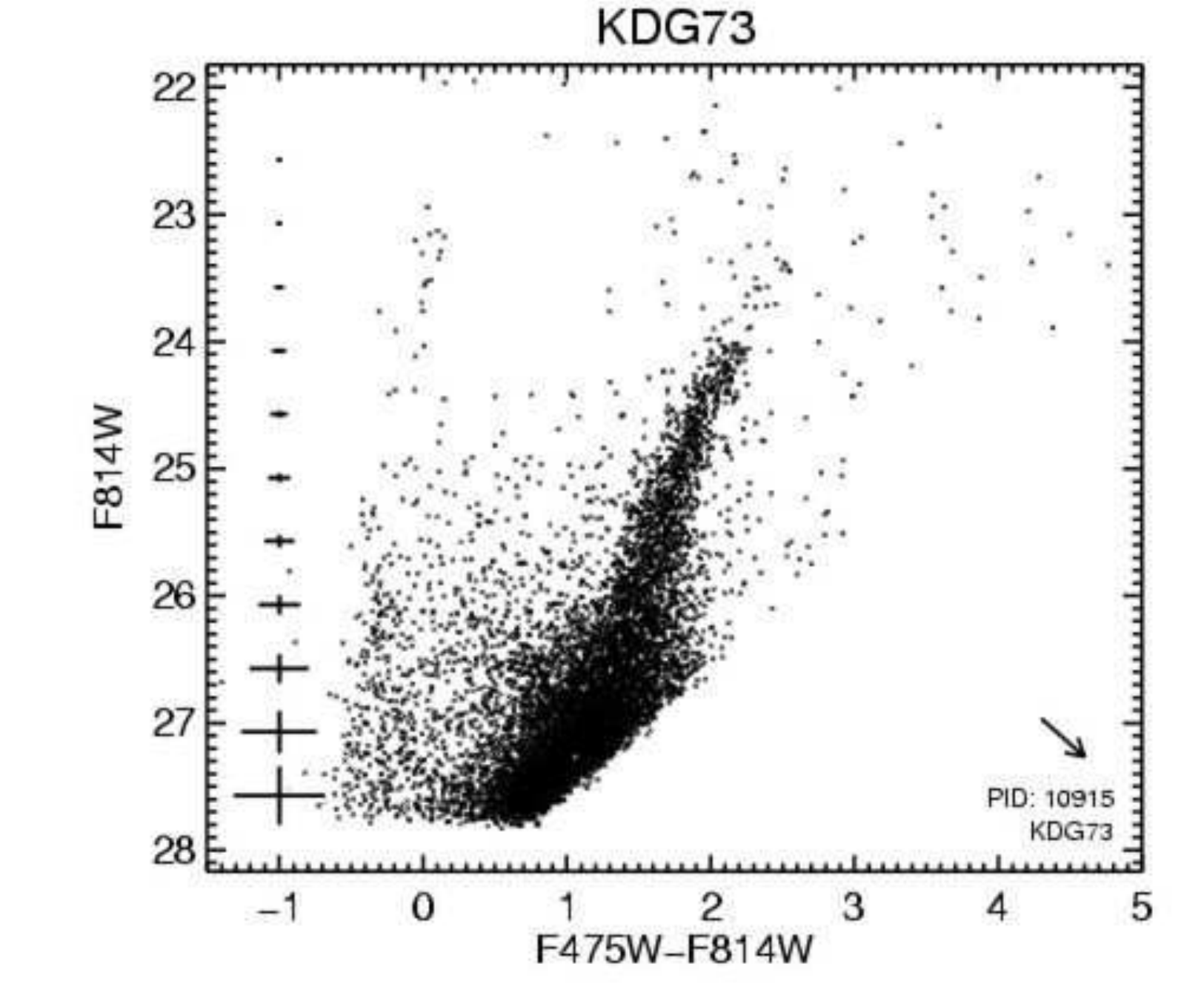}
}
\centerline{
\includegraphics[width=3.25in]{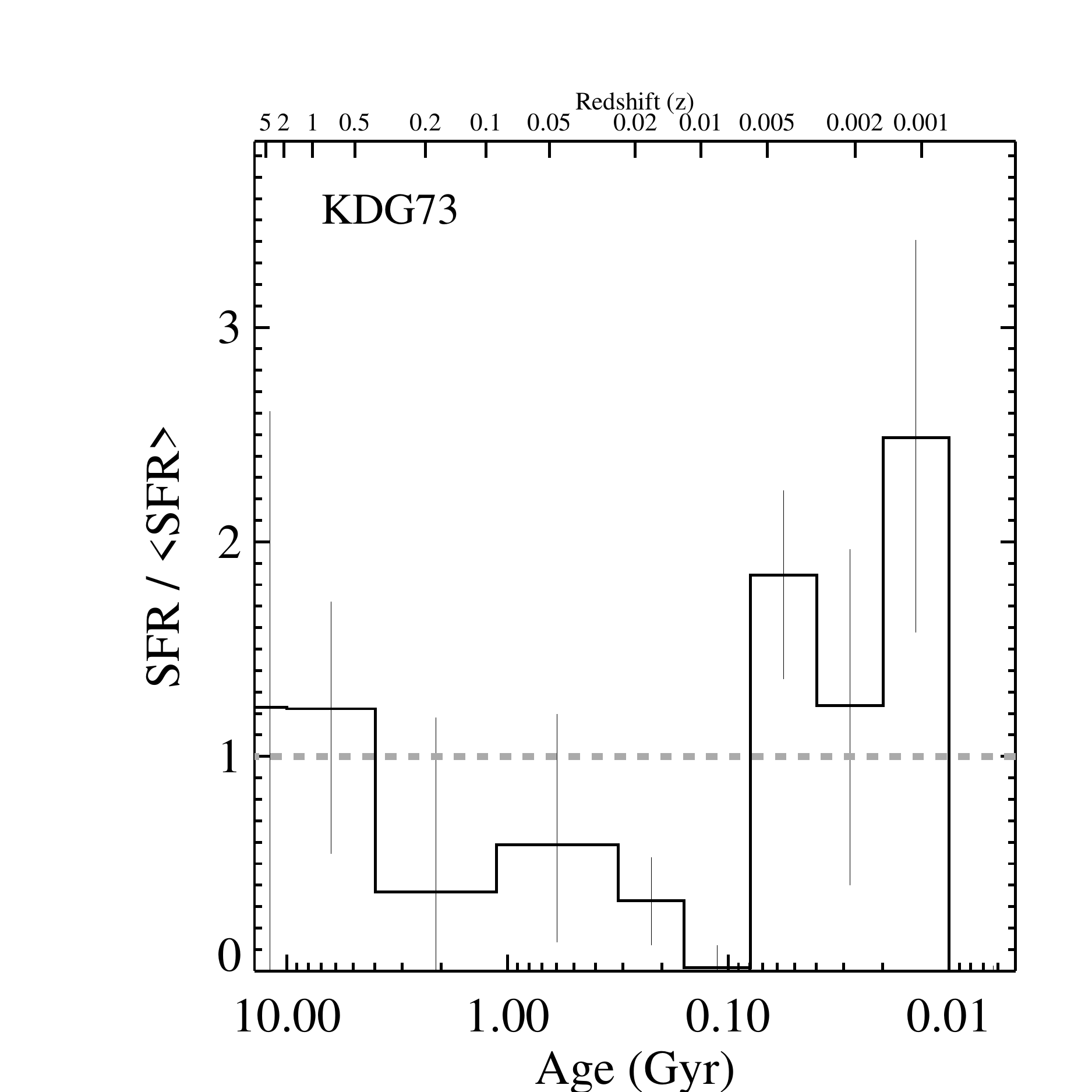}
\includegraphics[width=3.25in]{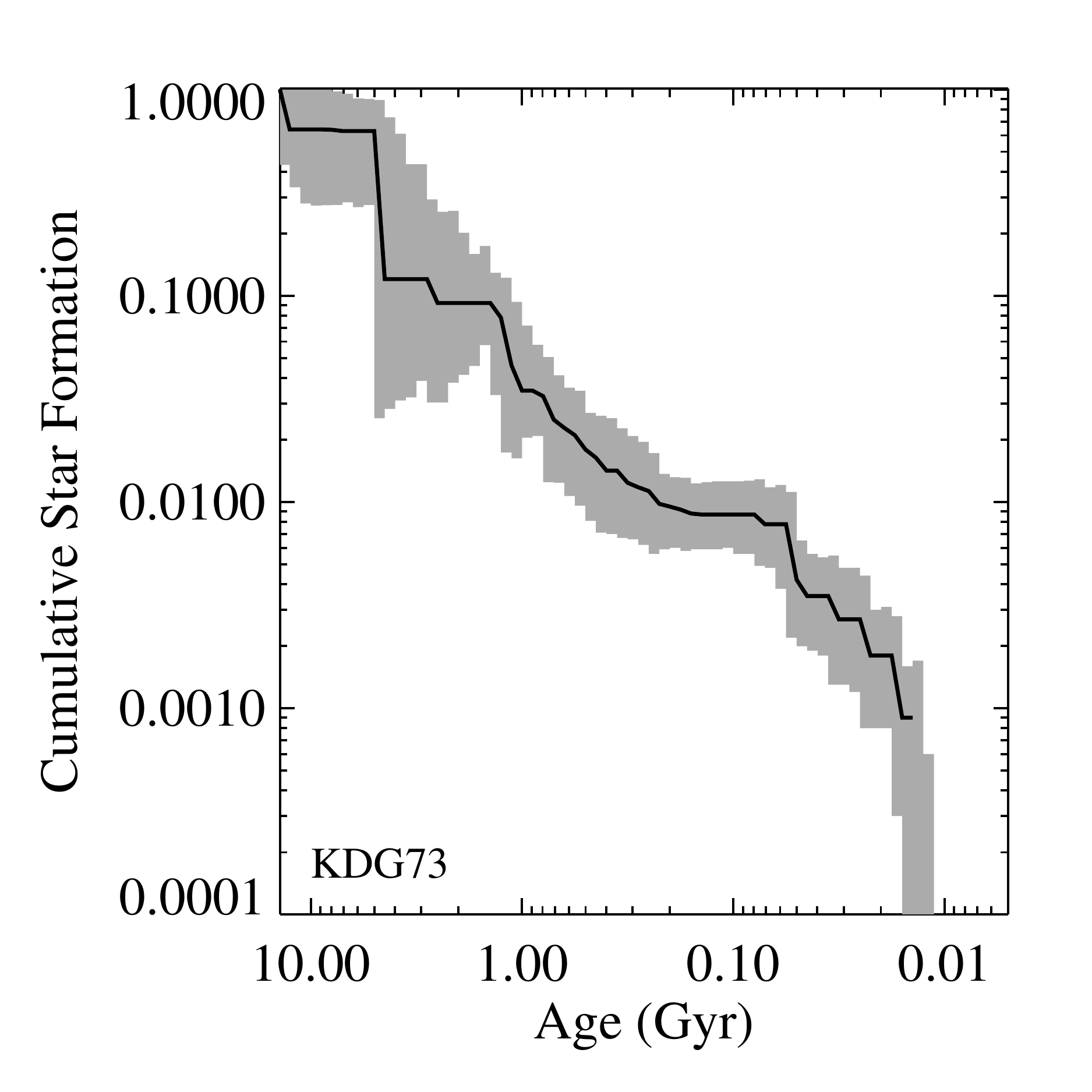}
}
\caption{ Color magnitude diagrams of the WFC3/IR (upper left) and
  optical (upper right) for KDG73.  Lower panels show the star
  formation history derived from the optical data, for both the
  differential (left, with horizontal dotted line indicating the past
  average SFR) and cumulative (right) star formation histories.  The
  cumulative star formation history is calculated from the present
  back to 14\,Gyrs.  Uncertainties in the lower two panels are the
  68\% confidence intervals, calculated from Monte Carlo tests
  including random and systematic uncertainties.  Optical CMDs are
  restricted to the area covered by the WFC3 FOV. }
\end{figure}
\vfill
\clearpage
 
\begin{figure}
\figurenum{\ref{cmdfig} continued}
\centerline{
\includegraphics[width=3.25in]{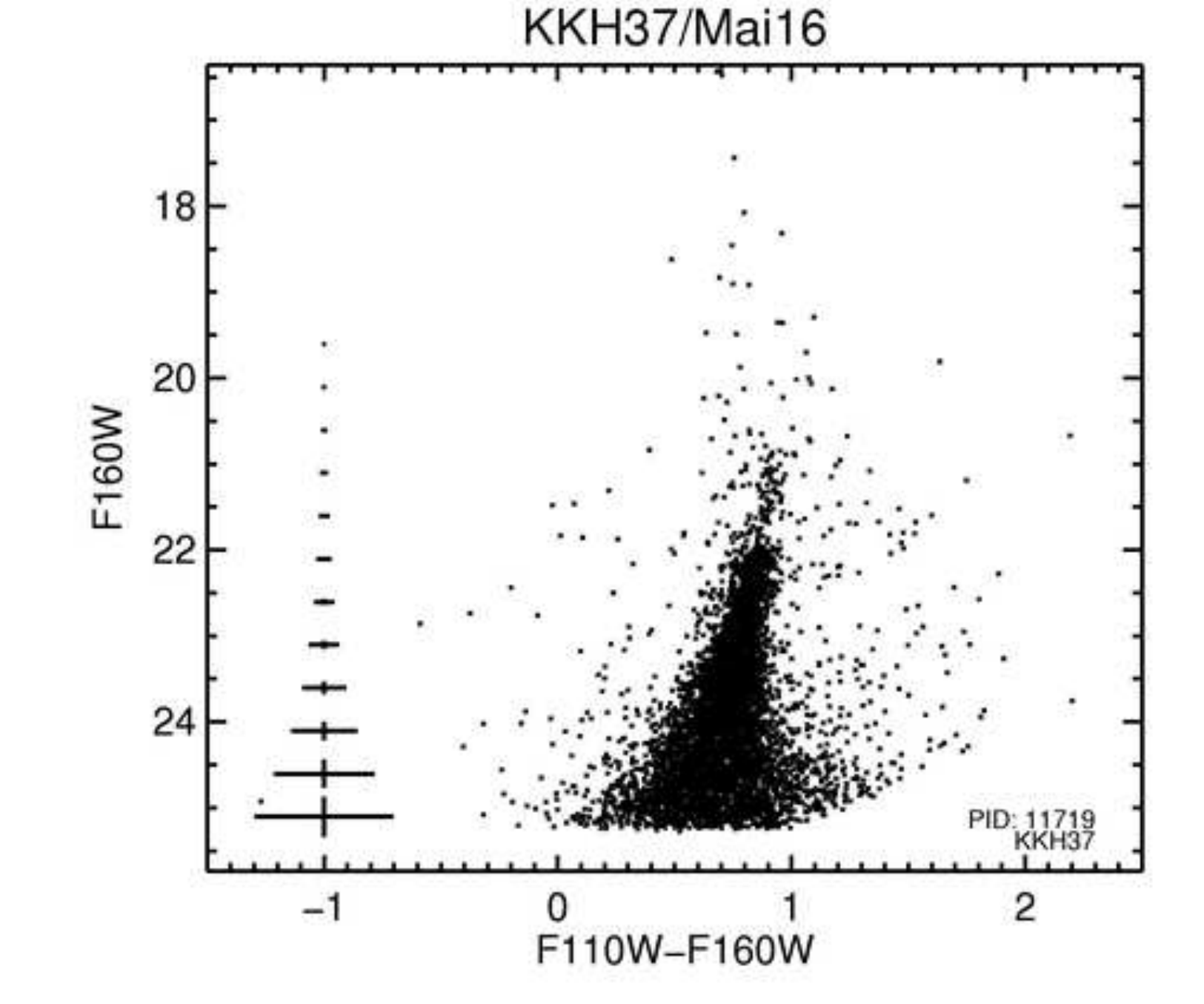}
\includegraphics[width=3.25in]{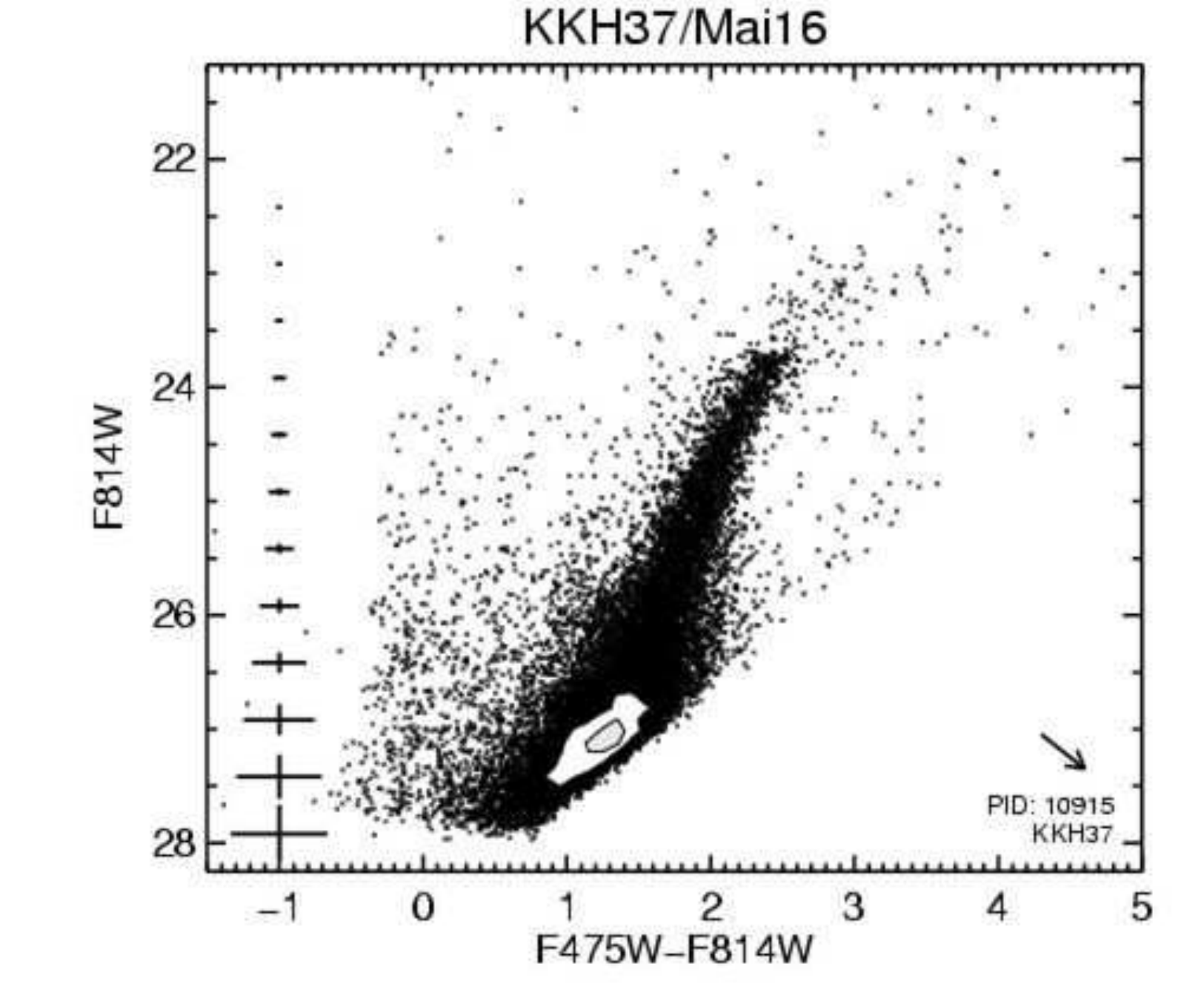}
}
\centerline{
\includegraphics[width=3.25in]{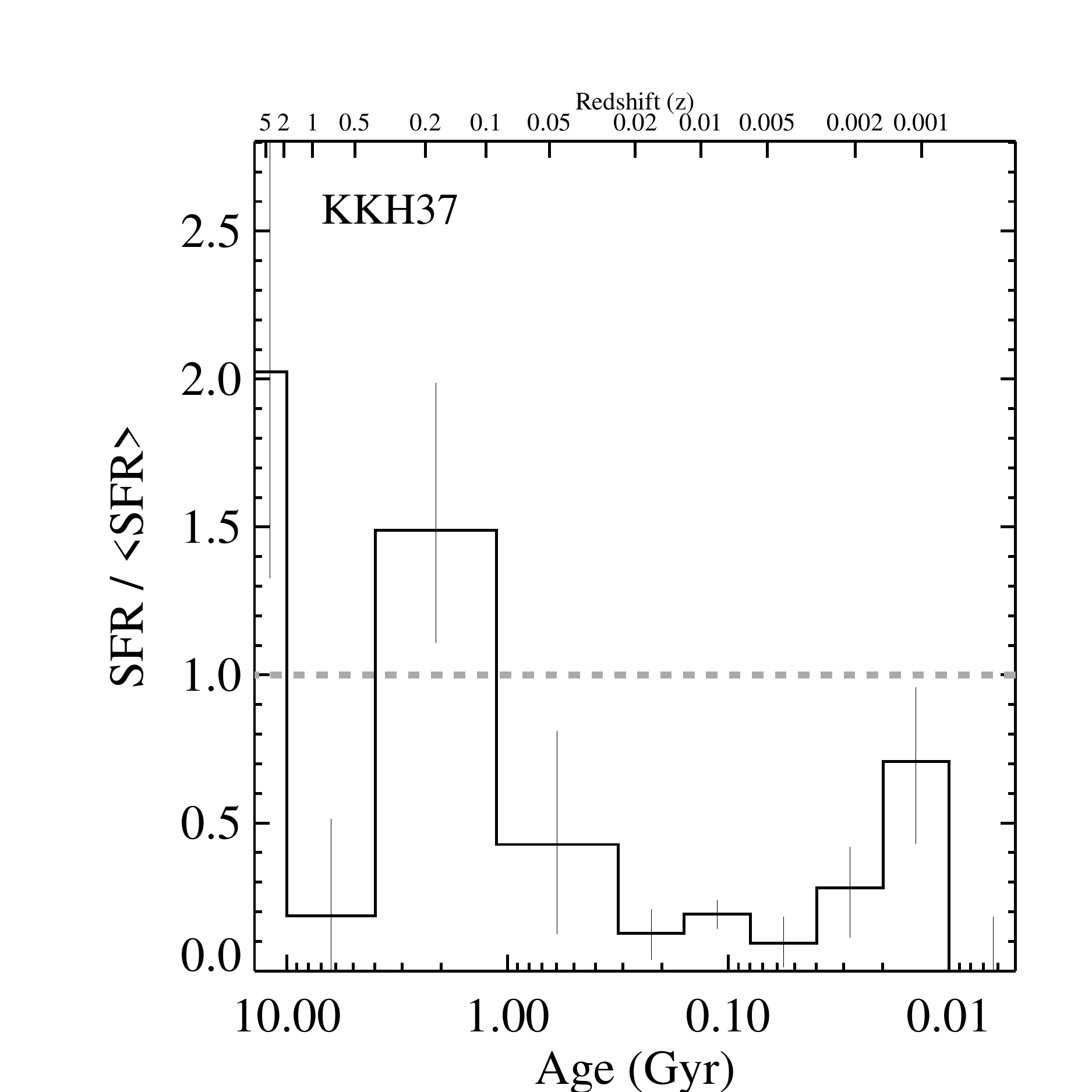}
\includegraphics[width=3.25in]{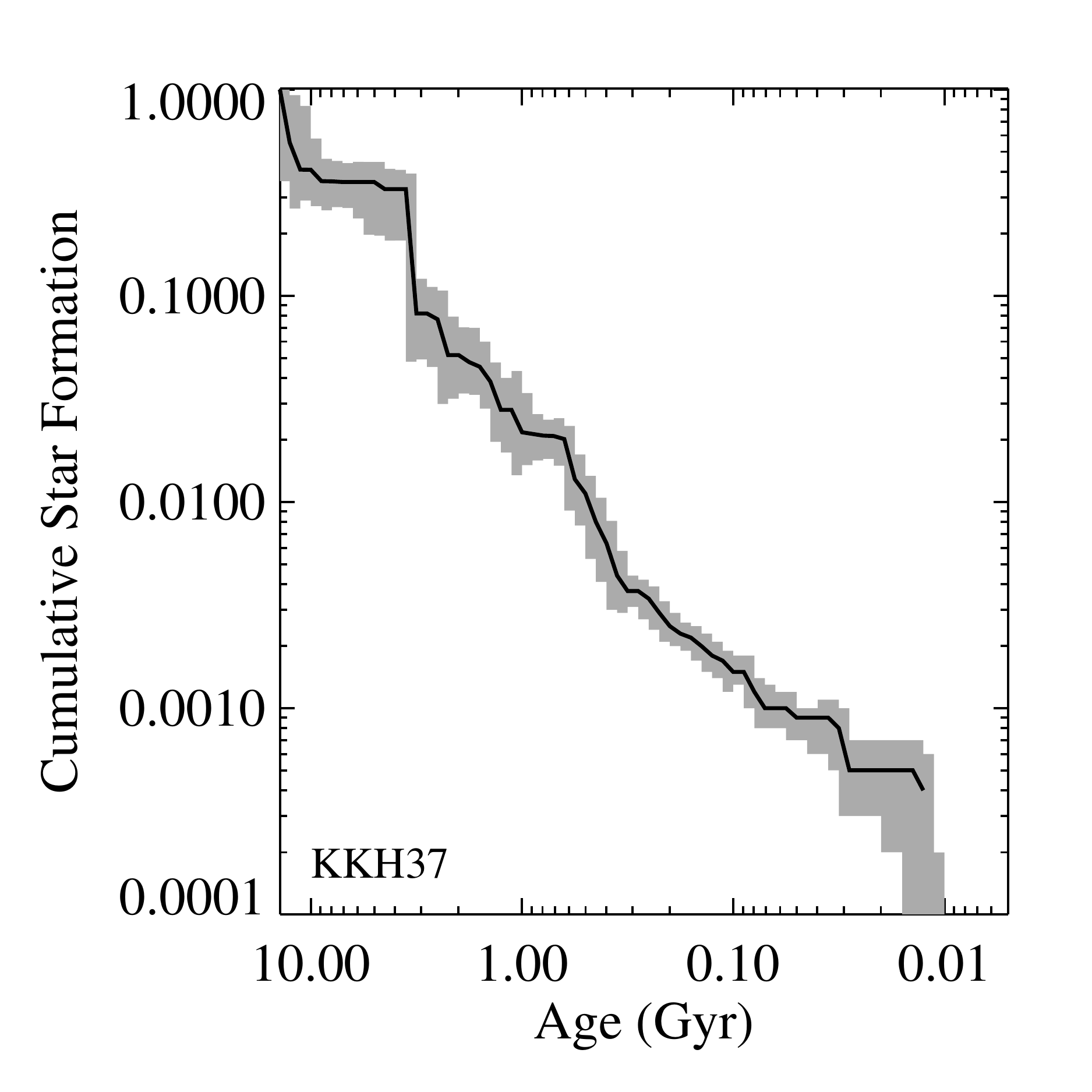}
}
\caption{ Color magnitude diagrams of the WFC3/IR (upper left) and
  optical (upper right) for KKH37.  Lower panels show the star
  formation history derived from the optical data, for both the
  differential (left, with horizontal dotted line indicating the past
  average SFR) and cumulative (right) star formation histories.  The
  cumulative star formation history is calculated from the present
  back to 14\,Gyrs.  Uncertainties in the lower two panels are the
  68\% confidence intervals, calculated from Monte Carlo tests
  including random and systematic uncertainties.  Optical CMDs are
  restricted to the area covered by the WFC3 FOV. }
\end{figure}
\vfill
\clearpage
 
\begin{figure}
\figurenum{\ref{cmdfig} continued}
\centerline{
\includegraphics[width=3.25in]{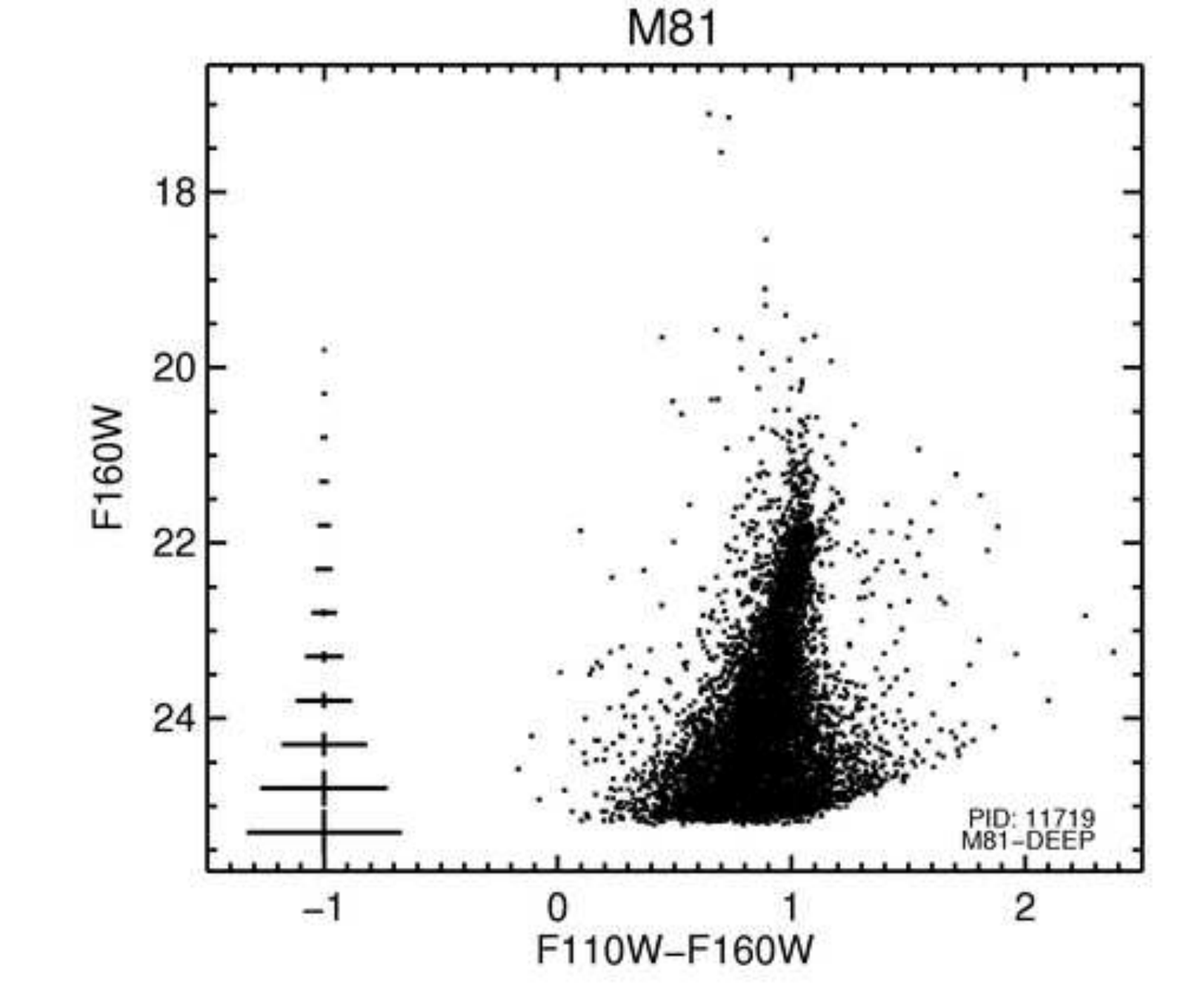}
\includegraphics[width=3.25in]{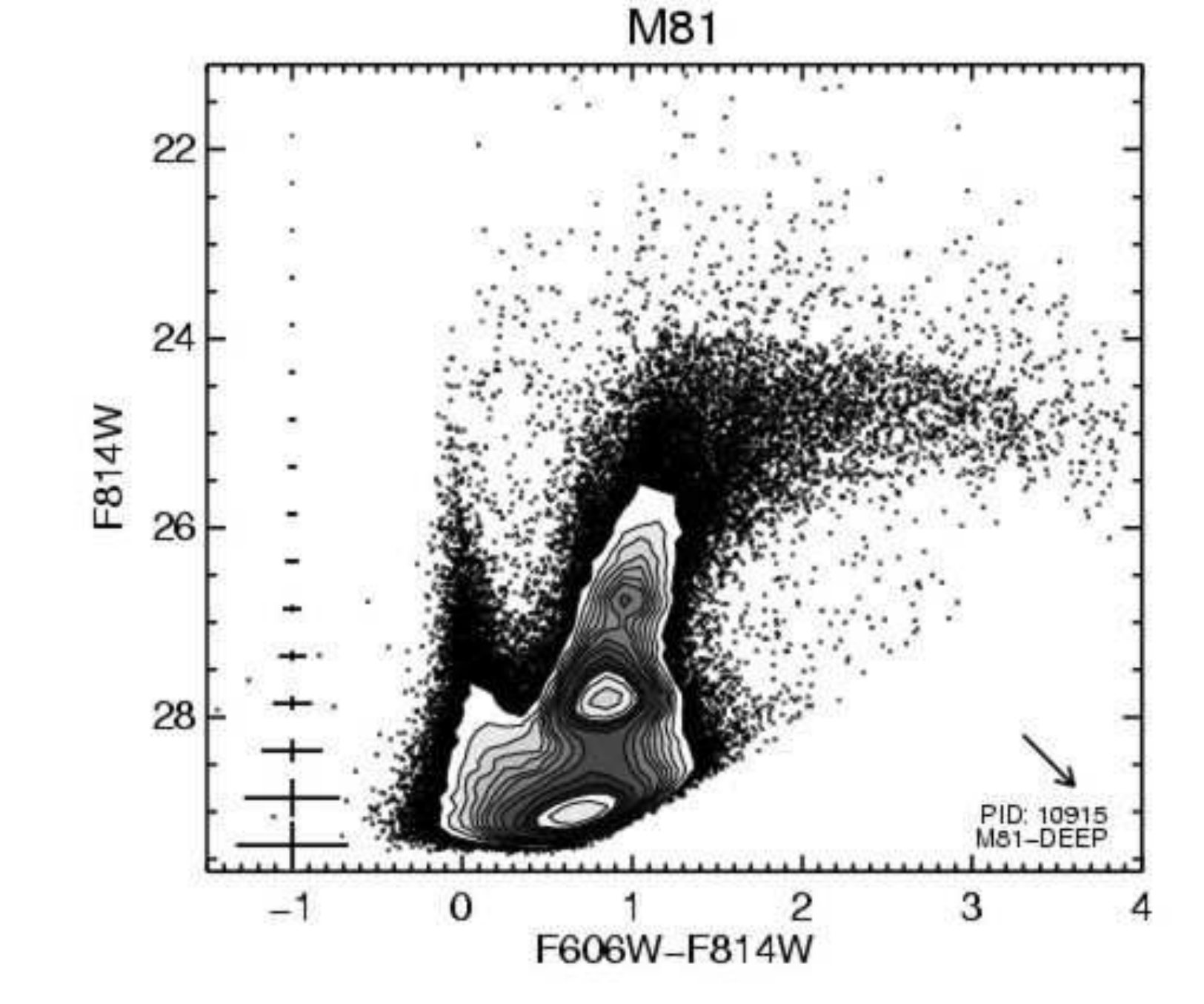}
}
\centerline{
\includegraphics[width=3.25in]{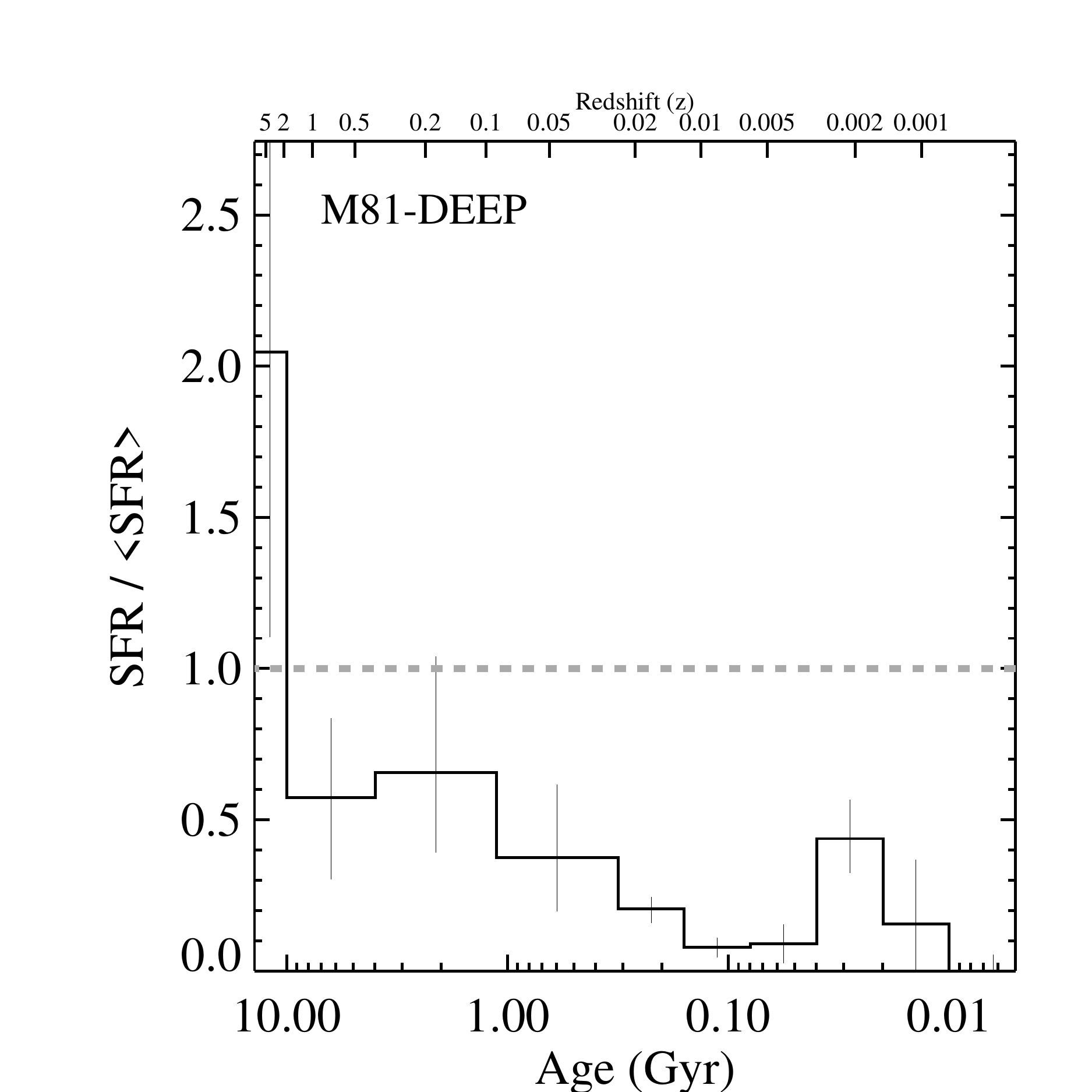}
\includegraphics[width=3.25in]{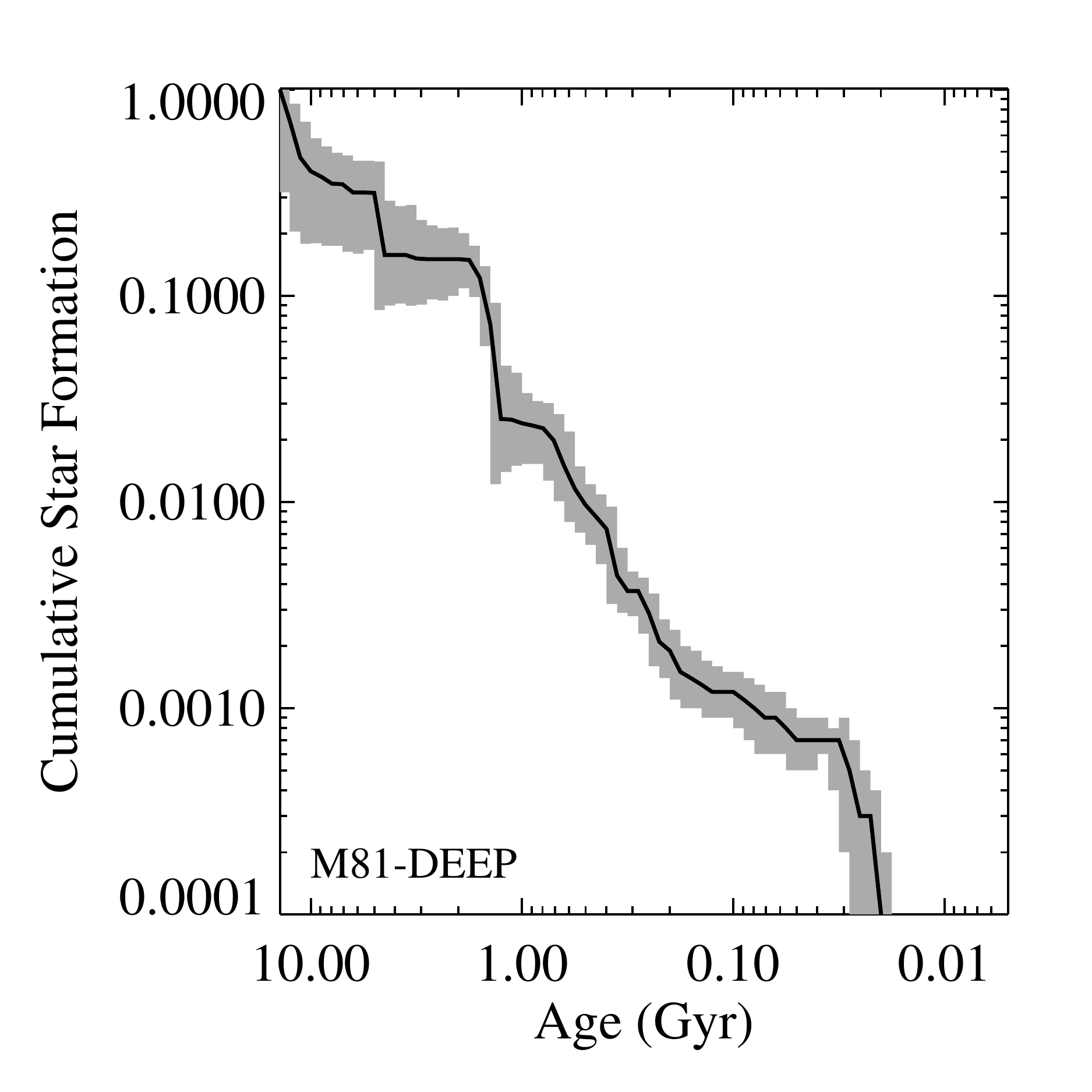}
}
\caption{ Color magnitude diagrams of the WFC3/IR (upper left) and
  optical (upper right) for the target M81-DEEP within galaxy M81.
  Lower panels show the star formation history derived from the
  optical data, for both the differential (left, with horizontal
  dotted line indicating the past average SFR) and cumulative (right)
  star formation histories.  The cumulative star formation history is
  calculated from the present back to 14\,Gyrs.  Uncertainties in the
  lower two panels are the 68\% confidence intervals, calculated from
  Monte Carlo tests including random and systematic uncertainties.
  Optical CMDs are restricted to the area covered by the WFC3 FOV. }
\end{figure}
\vfill
\clearpage
 
\begin{figure}
\figurenum{\ref{cmdfig} continued}
\centerline{
\includegraphics[width=3.25in]{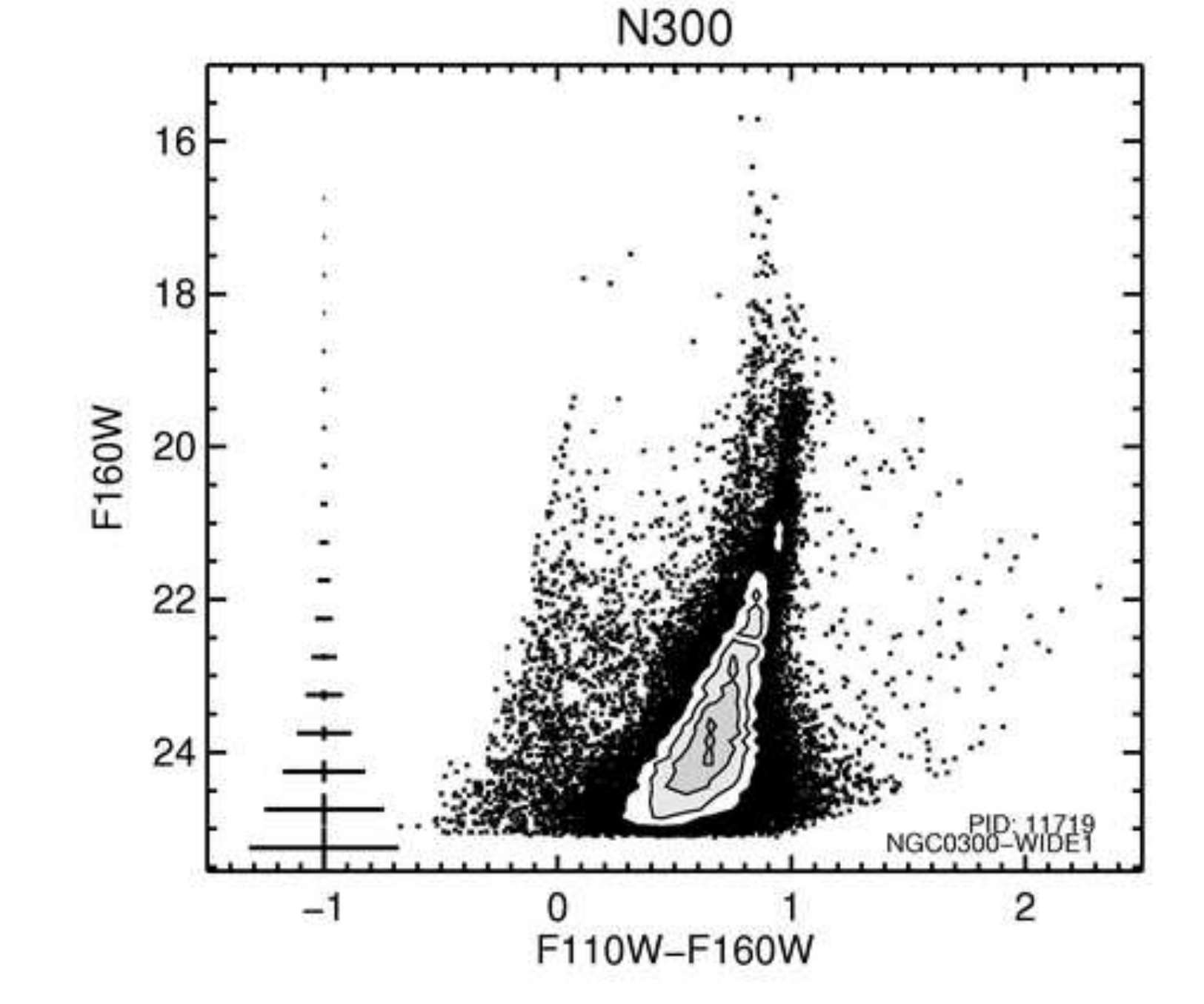}
\includegraphics[width=3.25in]{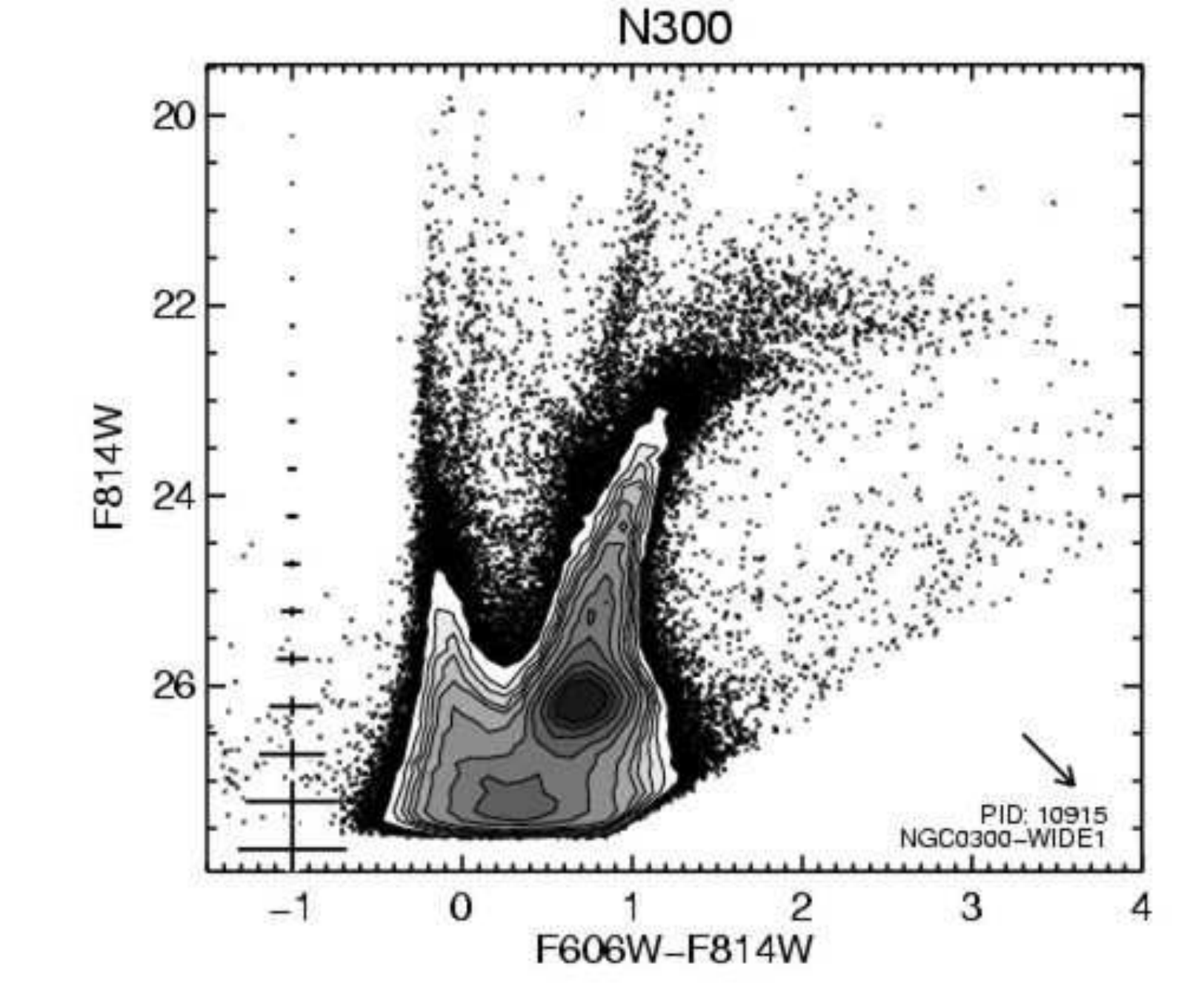}
}
\centerline{
\includegraphics[width=3.25in]{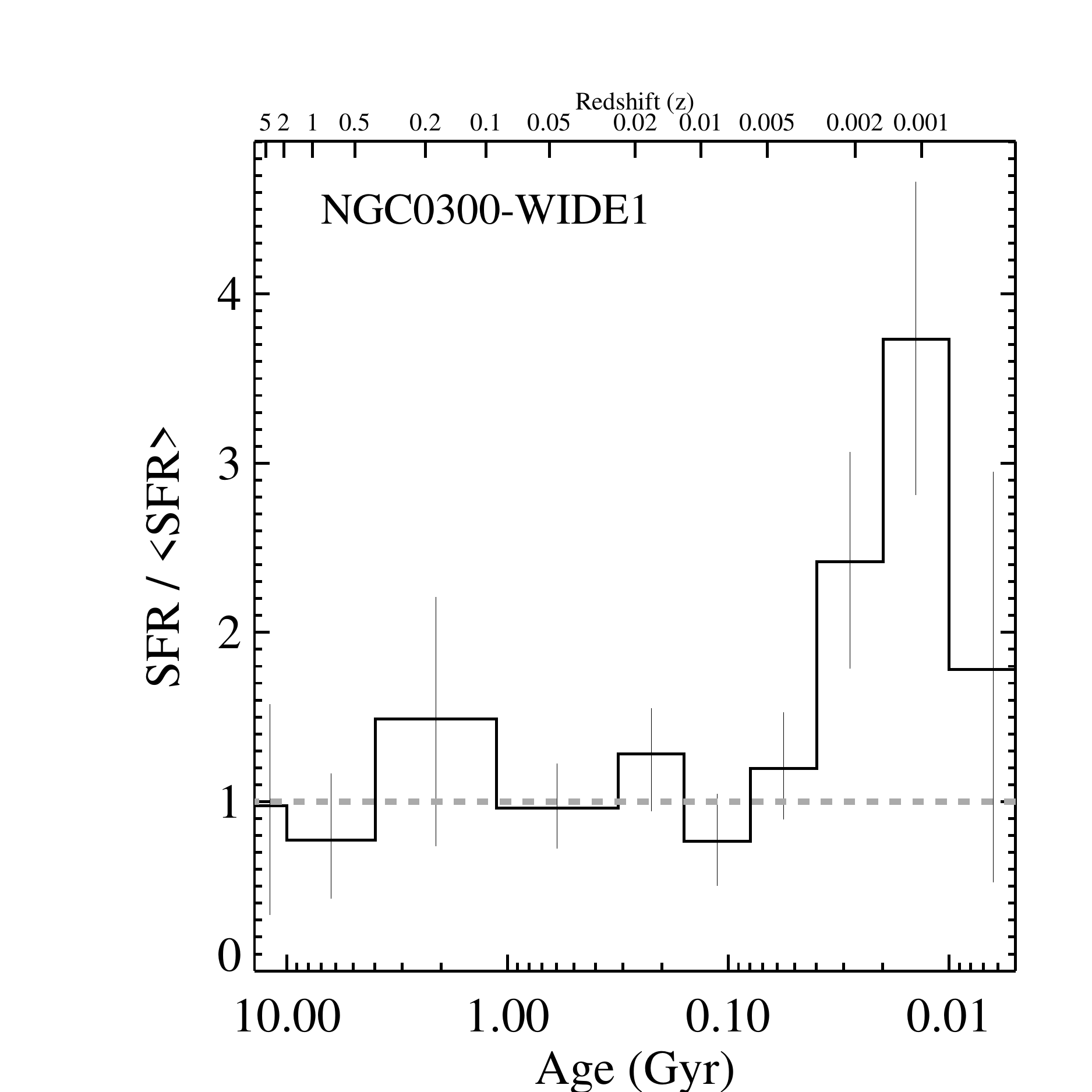}
\includegraphics[width=3.25in]{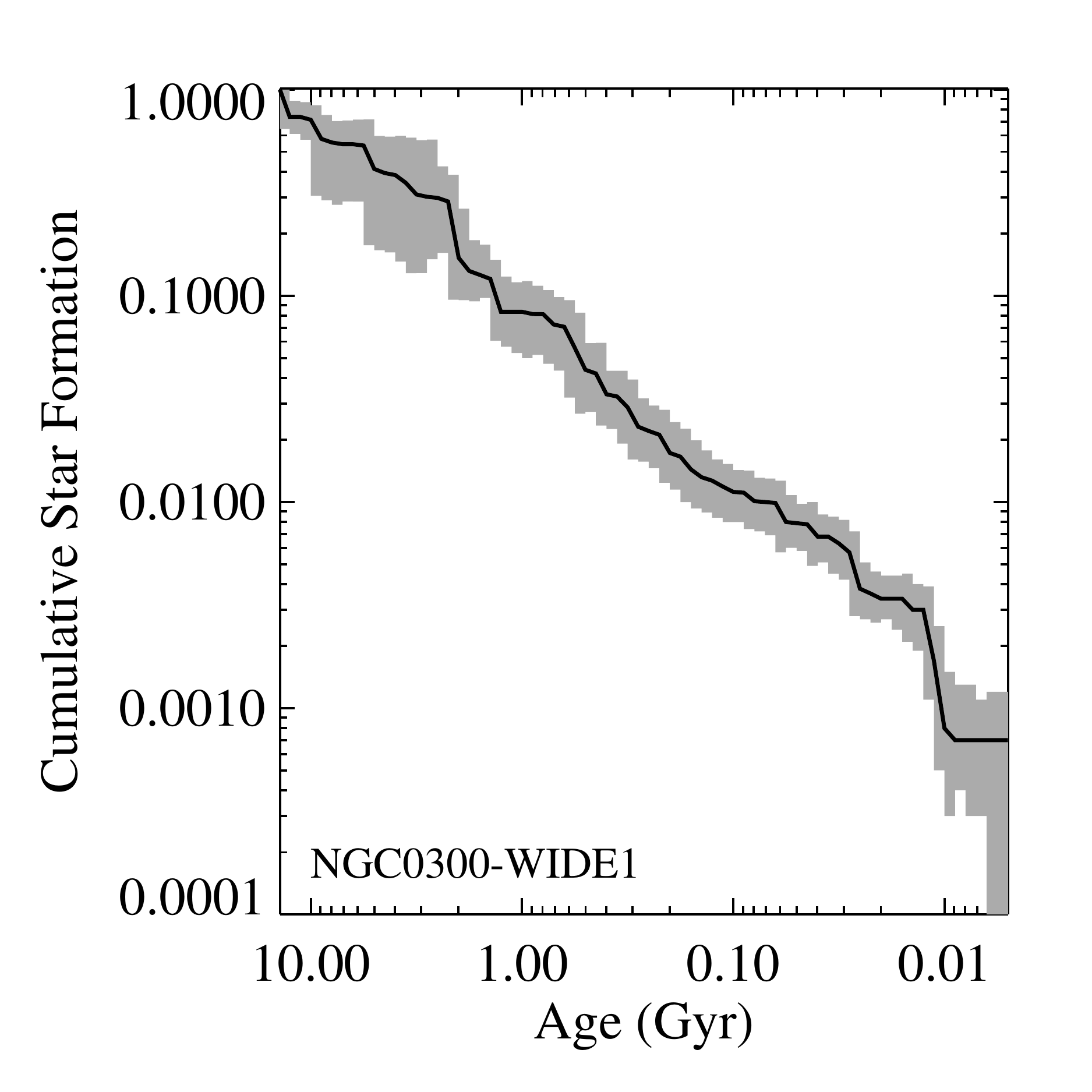}
}
\caption{ Color magnitude diagrams of the WFC3/IR (upper left) and
  optical (upper right) for the target NGC0300-WIDE1 within galaxy
  N300.  Lower panels show the star formation history derived from the
  optical data, for both the differential (left, with horizontal
  dotted line indicating the past average SFR) and cumulative (right)
  star formation histories.  The cumulative star formation history is
  calculated from the present back to 14\,Gyrs.  Uncertainties in the
  lower two panels are the 68\% confidence intervals, calculated from
  Monte Carlo tests including random and systematic uncertainties.
  Optical CMDs are restricted to the area covered by the WFC3 FOV. }
\end{figure}
\vfill
\clearpage
 
\begin{figure}
\figurenum{\ref{cmdfig} continued}
\centerline{
\includegraphics[width=3.25in]{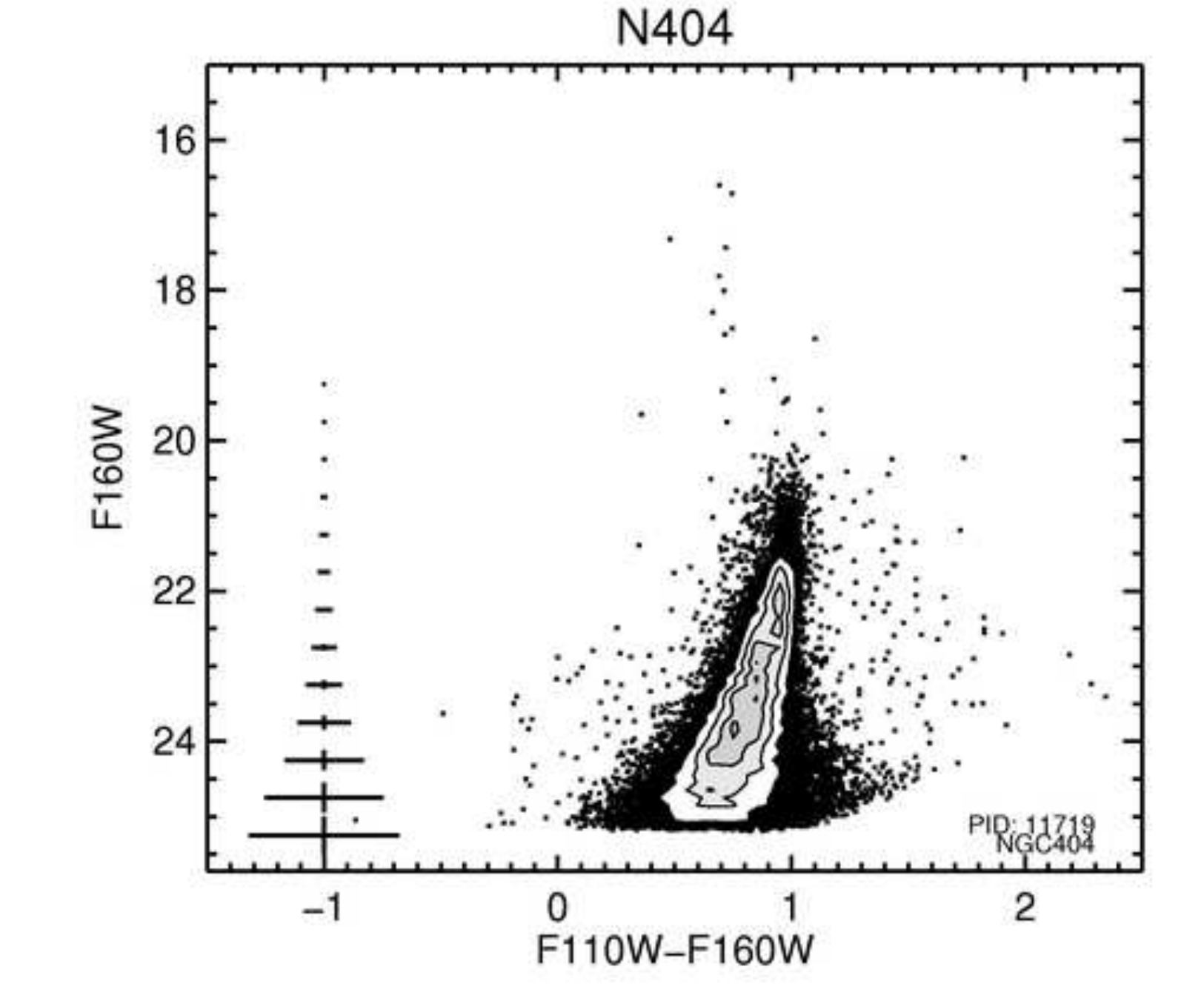}
\includegraphics[width=3.25in]{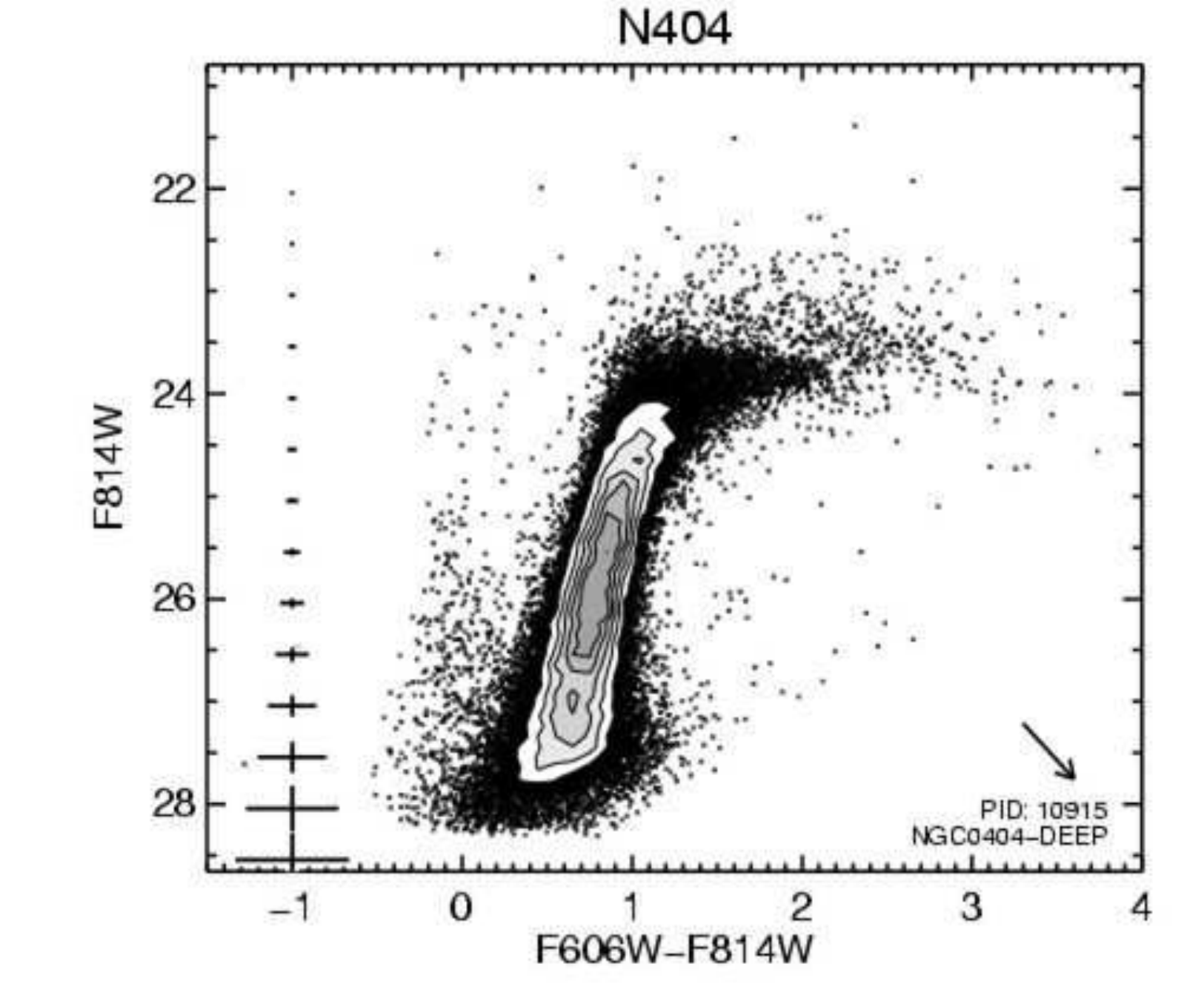}
}
\centerline{
\includegraphics[width=3.25in]{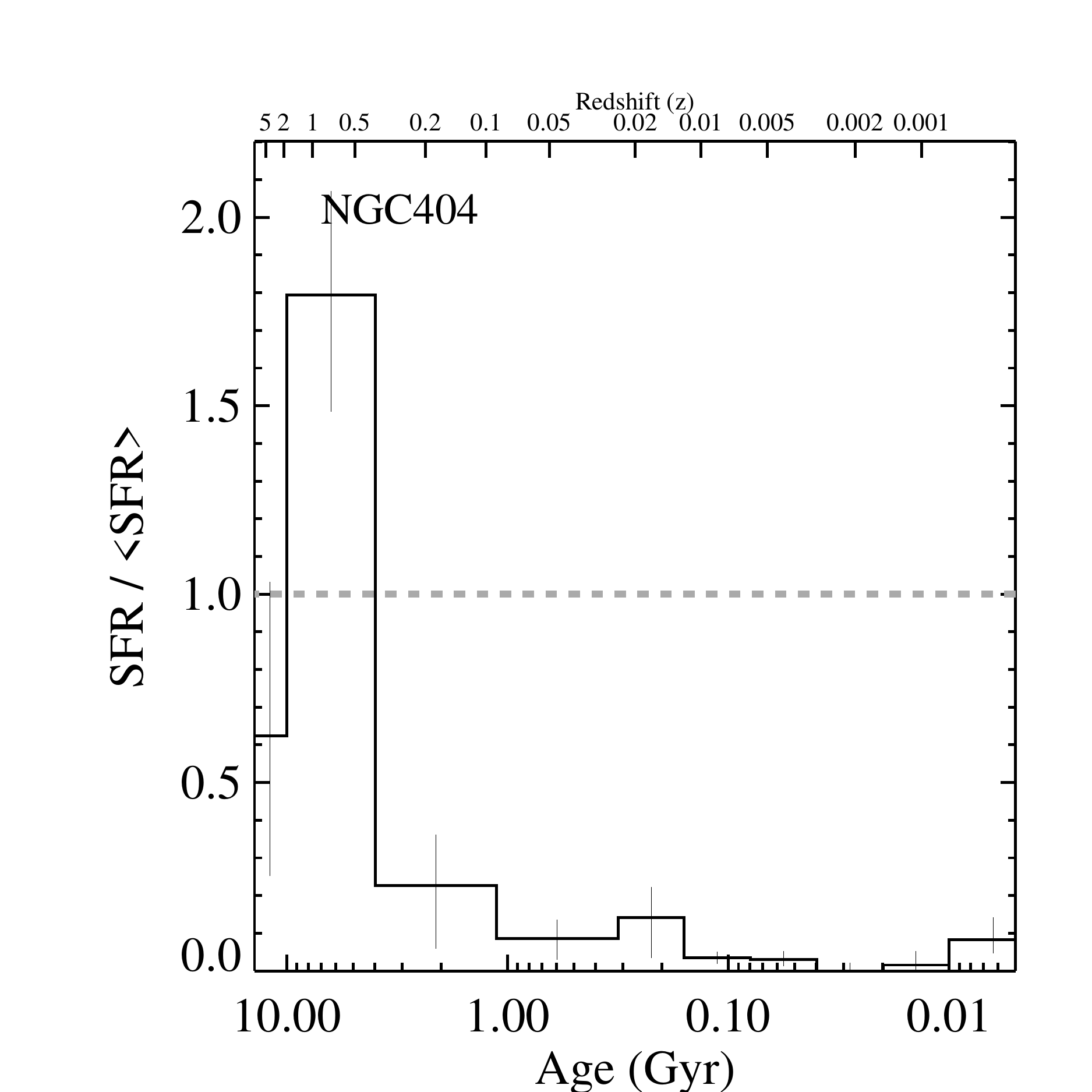}
\includegraphics[width=3.25in]{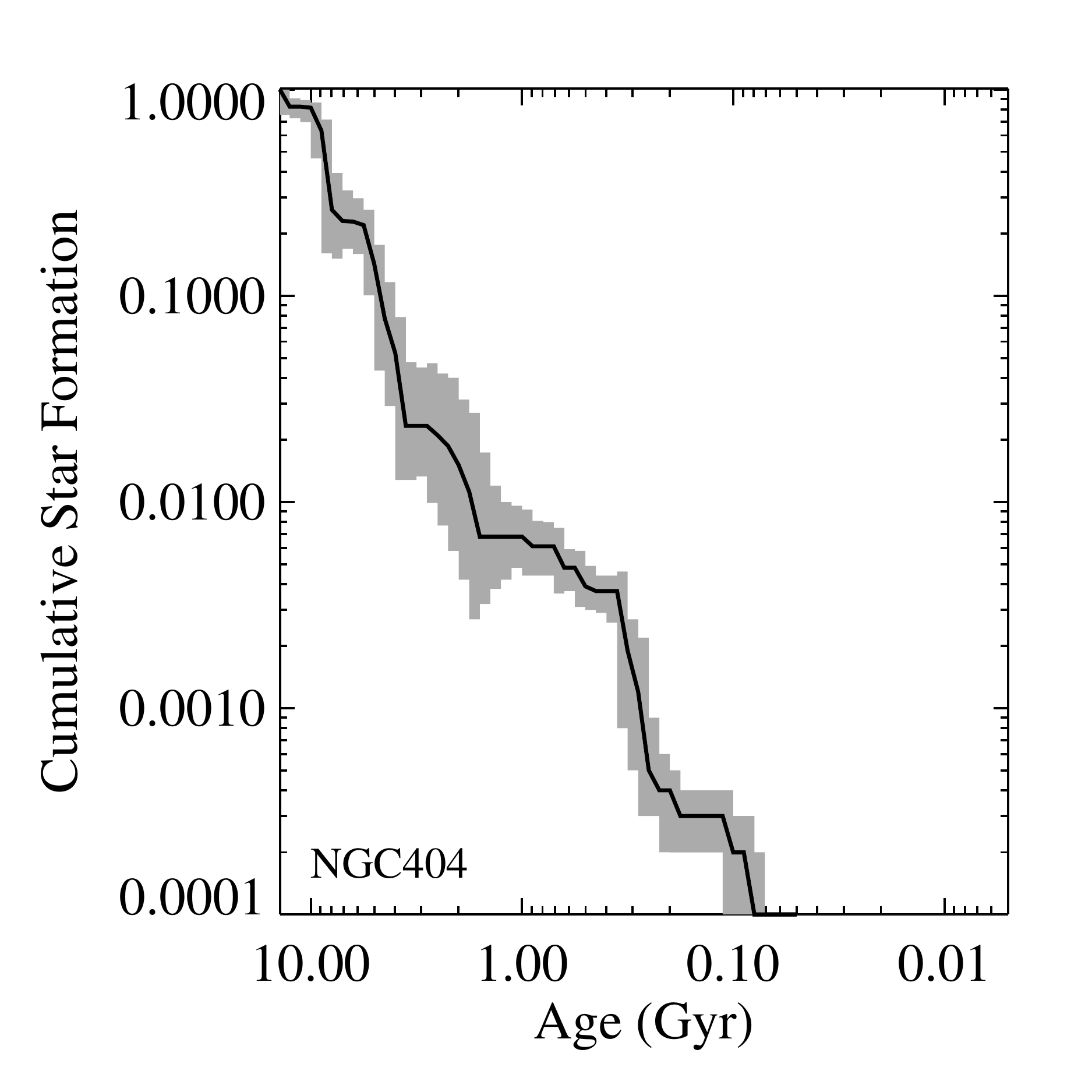}
}
\caption{ Color magnitude diagrams of the WFC3/IR (upper left) and
  optical (upper right) for the target NGC404 within galaxy N404.
  Lower panels show the star formation history derived from the
  optical data, for both the differential (left, with horizontal
  dotted line indicating the past average SFR) and cumulative (right)
  star formation histories.  The cumulative star formation history is
  calculated from the present back to 14\,Gyrs.  Uncertainties in the
  lower two panels are the 68\% confidence intervals, calculated from
  Monte Carlo tests including random and systematic uncertainties.
  Optical CMDs are restricted to the area covered by the WFC3 FOV. }
\end{figure}
\vfill
\clearpage
 
\begin{figure}
\figurenum{\ref{cmdfig} continued}
\centerline{
\includegraphics[width=3.25in]{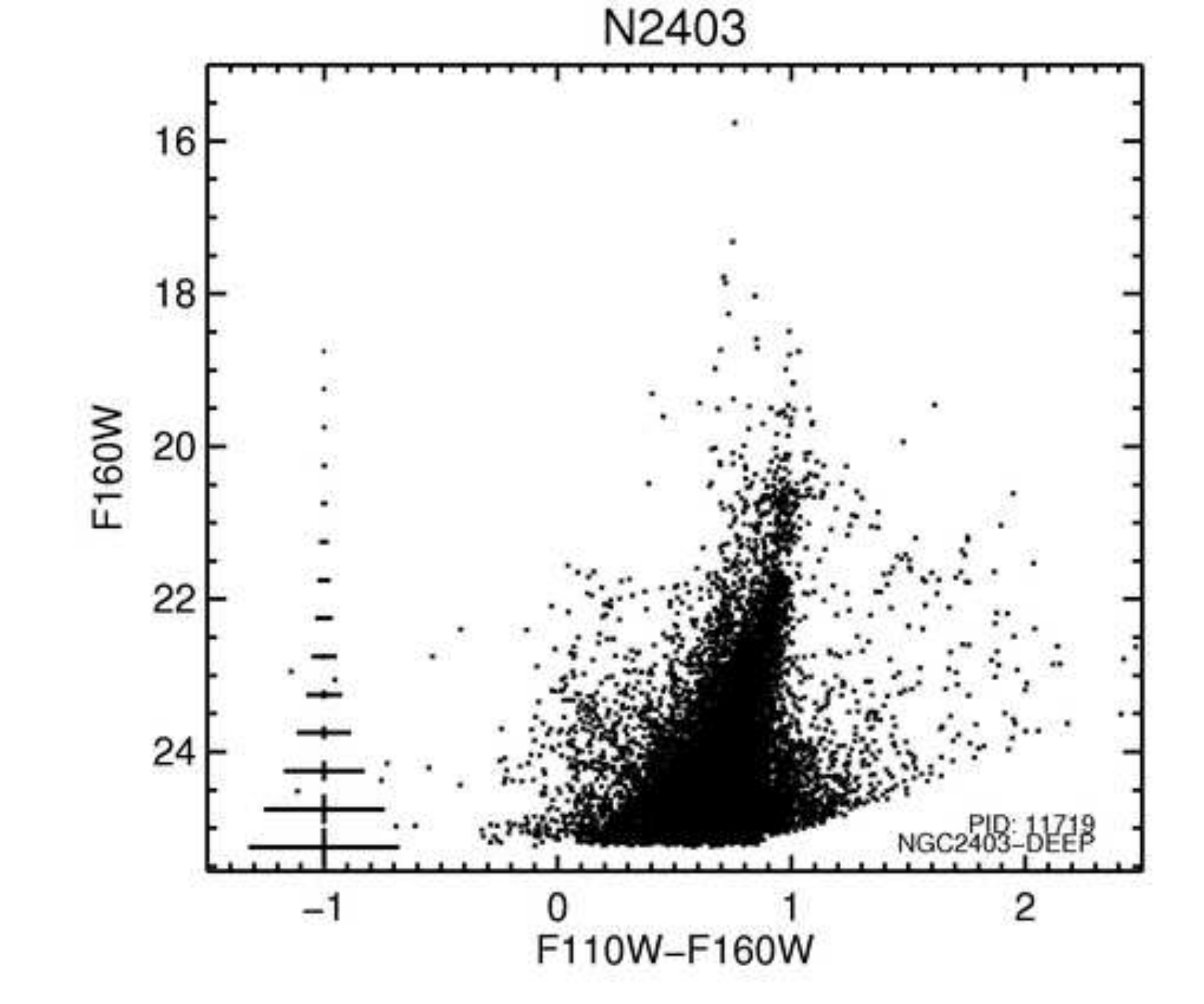}
\includegraphics[width=3.25in]{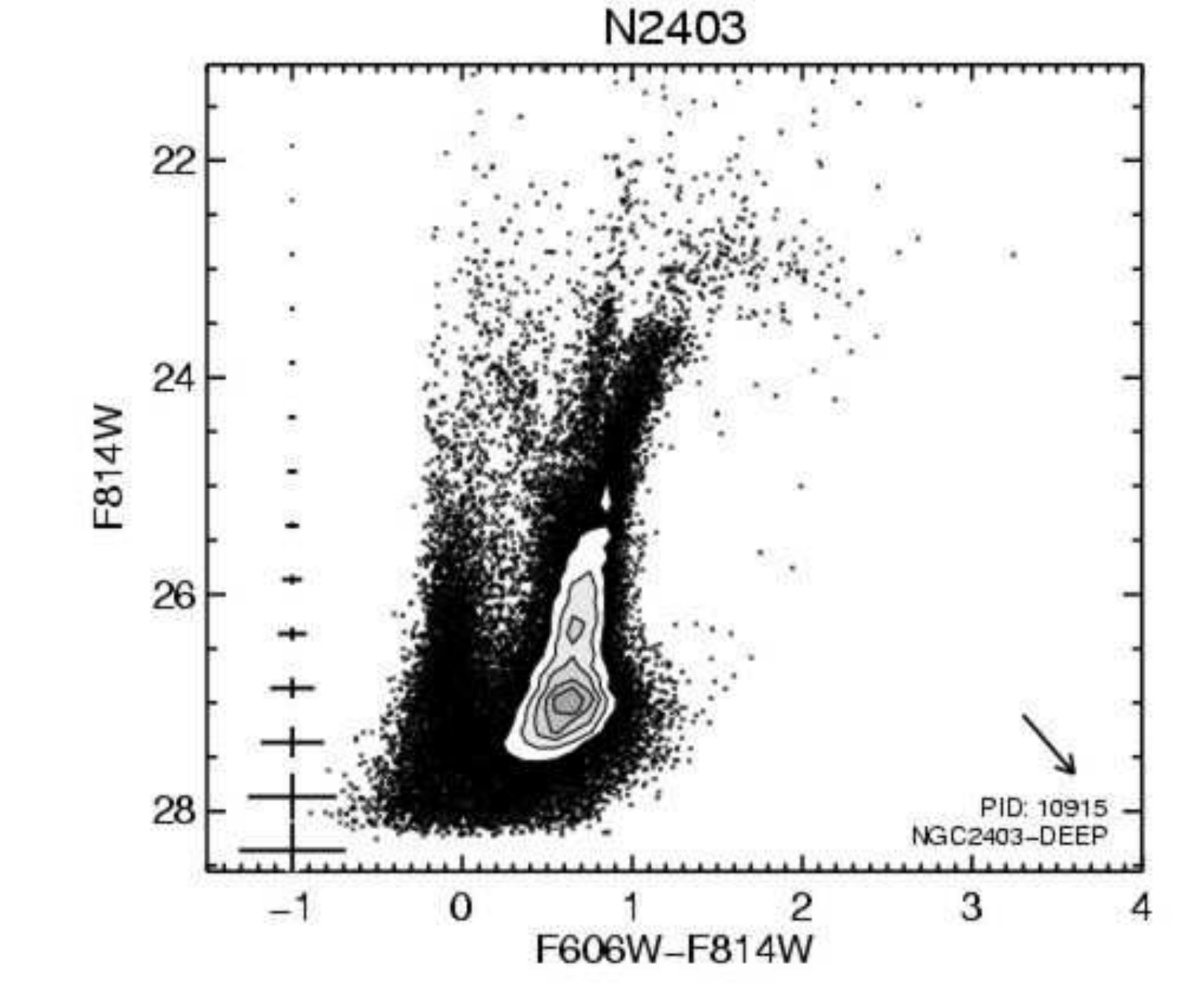}
}
\centerline{
\includegraphics[width=3.25in]{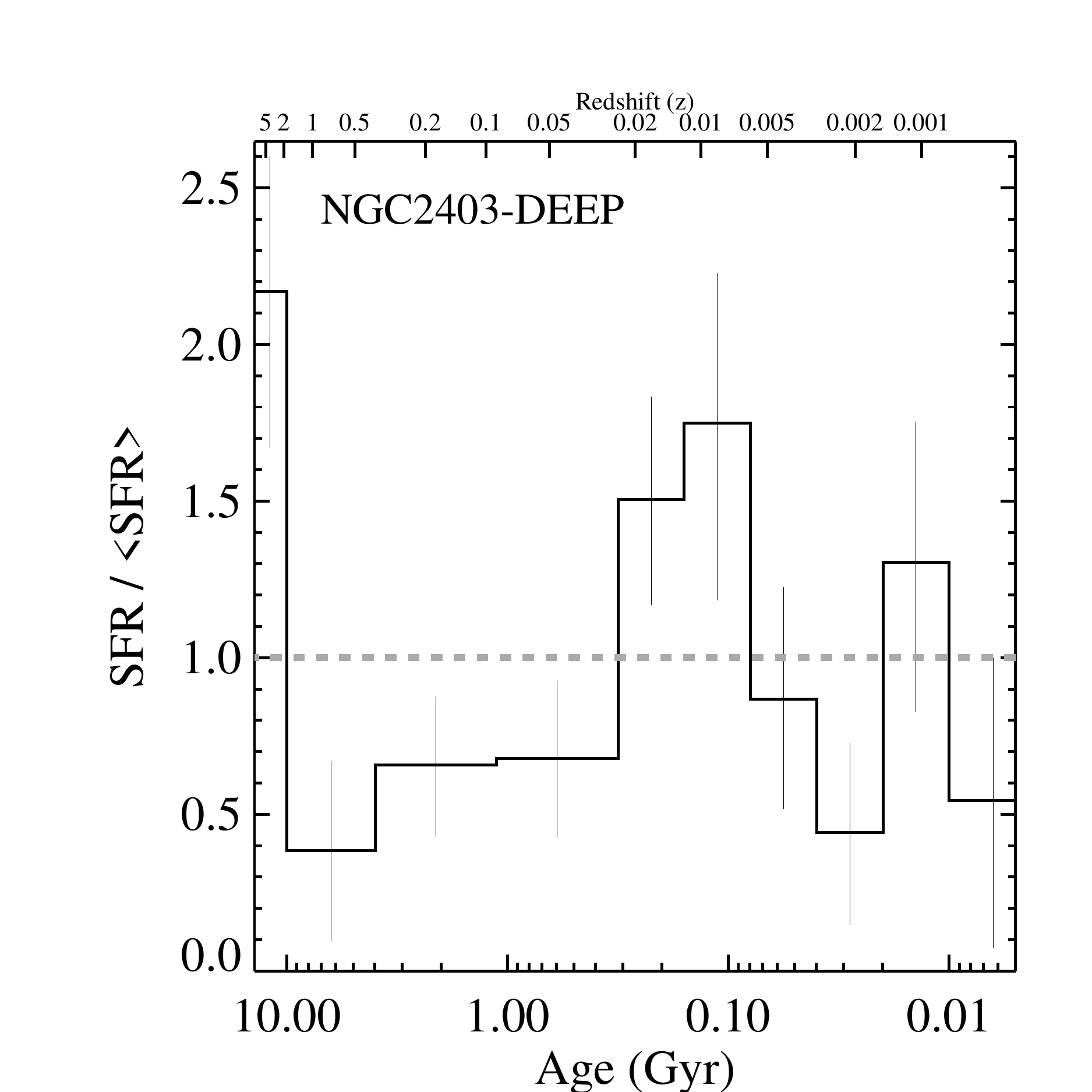}
\includegraphics[width=3.25in]{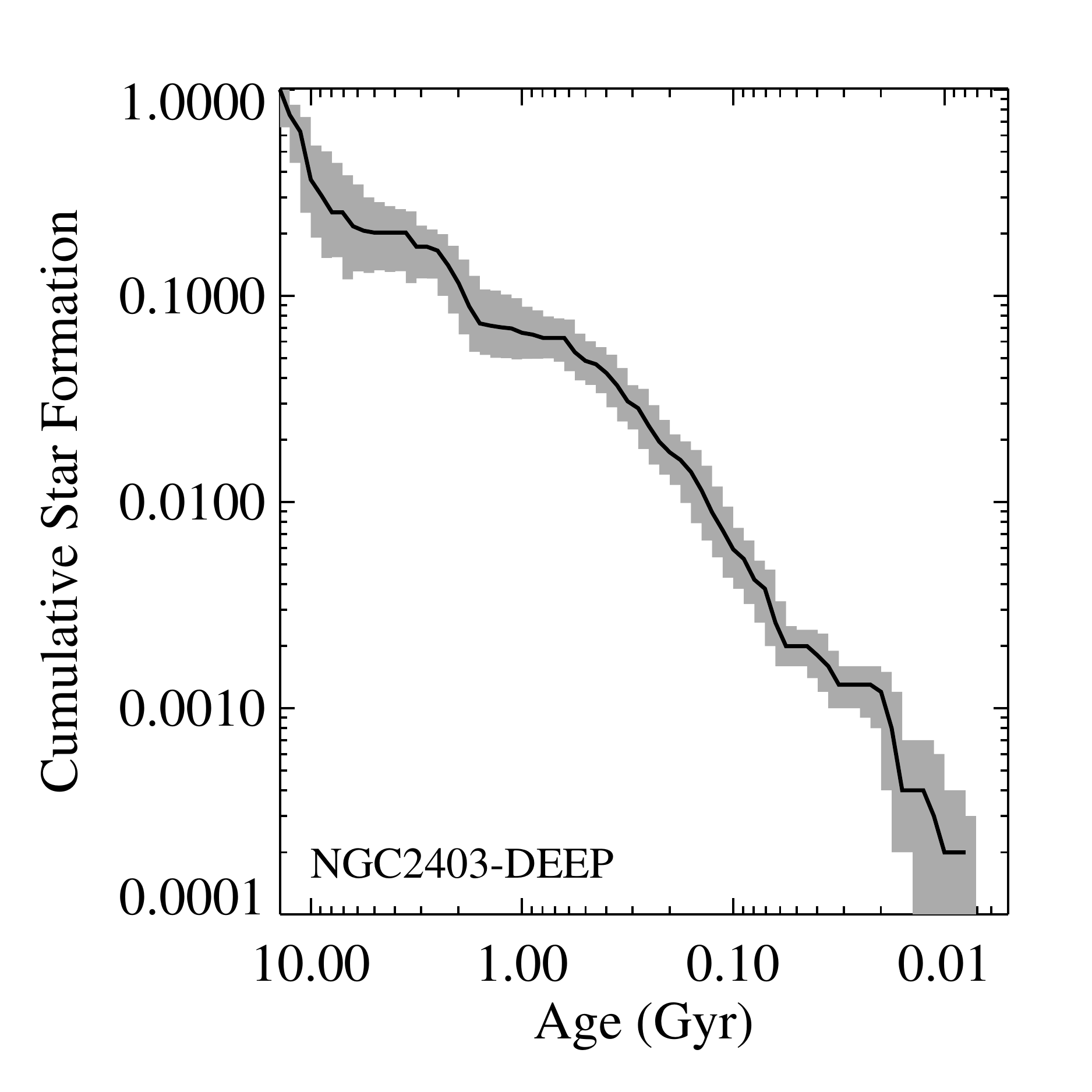}
}
\caption{ Color magnitude diagrams of the WFC3/IR (upper left) and
  optical (upper right) for the target NGC2403-DEEP within galaxy
  N2403.  Lower panels show the star formation history derived from
  the optical data, for both the differential (left, with horizontal
  dotted line indicating the past average SFR) and cumulative (right)
  star formation histories.  The cumulative star formation history is
  calculated from the present back to 14\,Gyrs.  Uncertainties in the
  lower two panels are the 68\% confidence intervals, calculated from
  Monte Carlo tests including random and systematic uncertainties.
  Optical CMDs are restricted to the area covered by the WFC3 FOV. }
\end{figure}
\vfill
\clearpage
 
\begin{figure}
\figurenum{\ref{cmdfig} continued}
\centerline{
\includegraphics[width=3.25in]{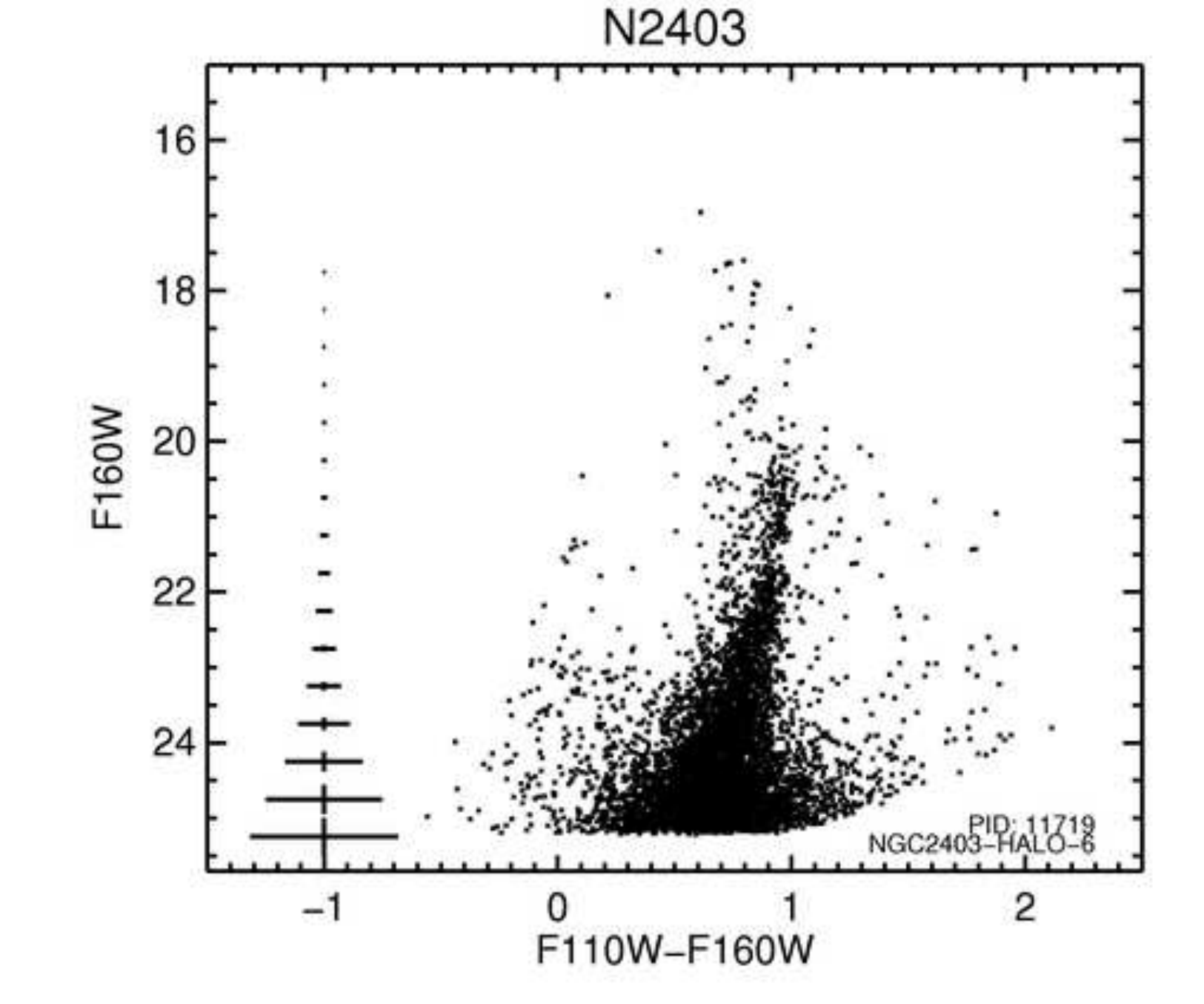}
\includegraphics[width=3.25in]{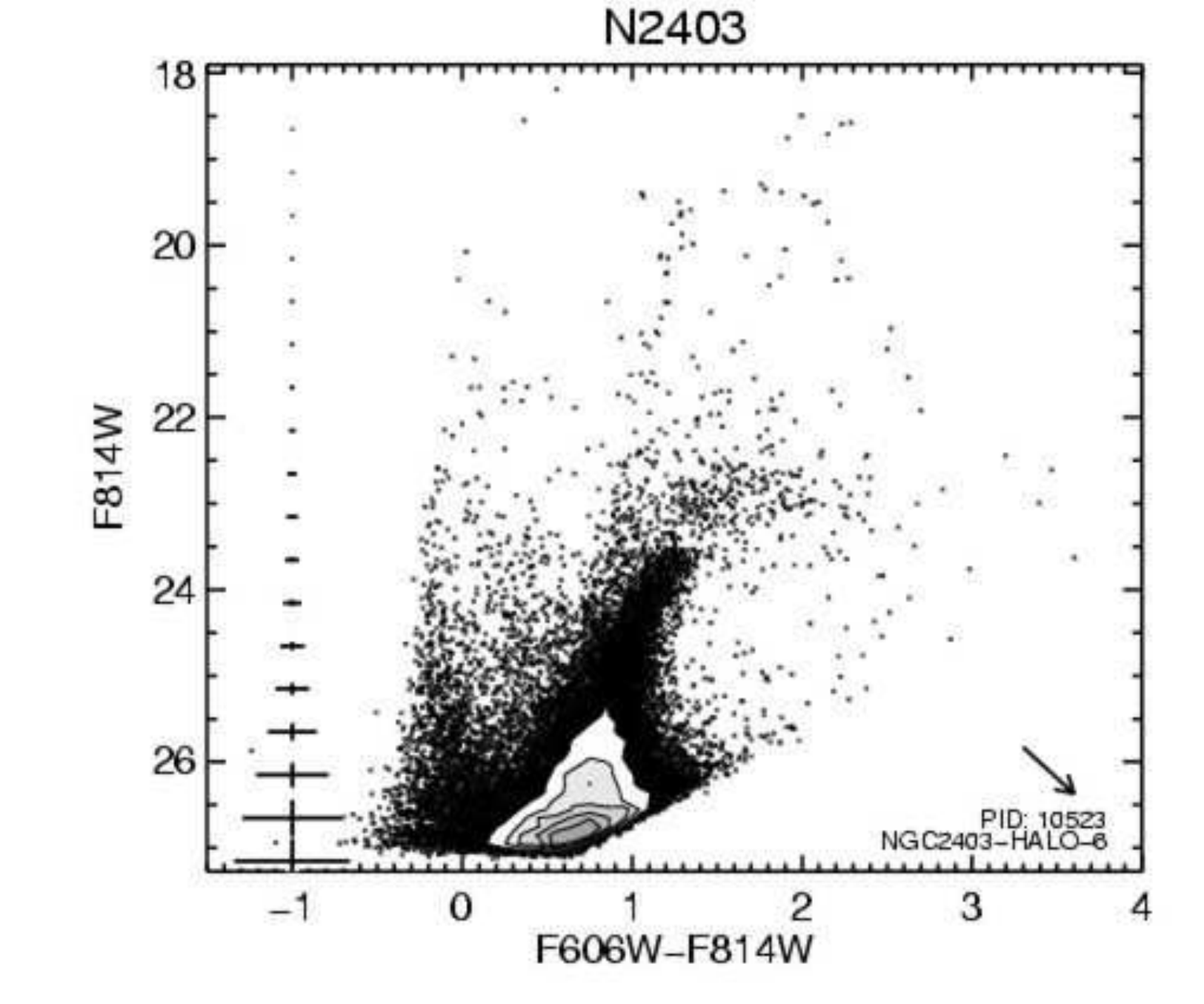}
}
\centerline{
\includegraphics[width=3.25in]{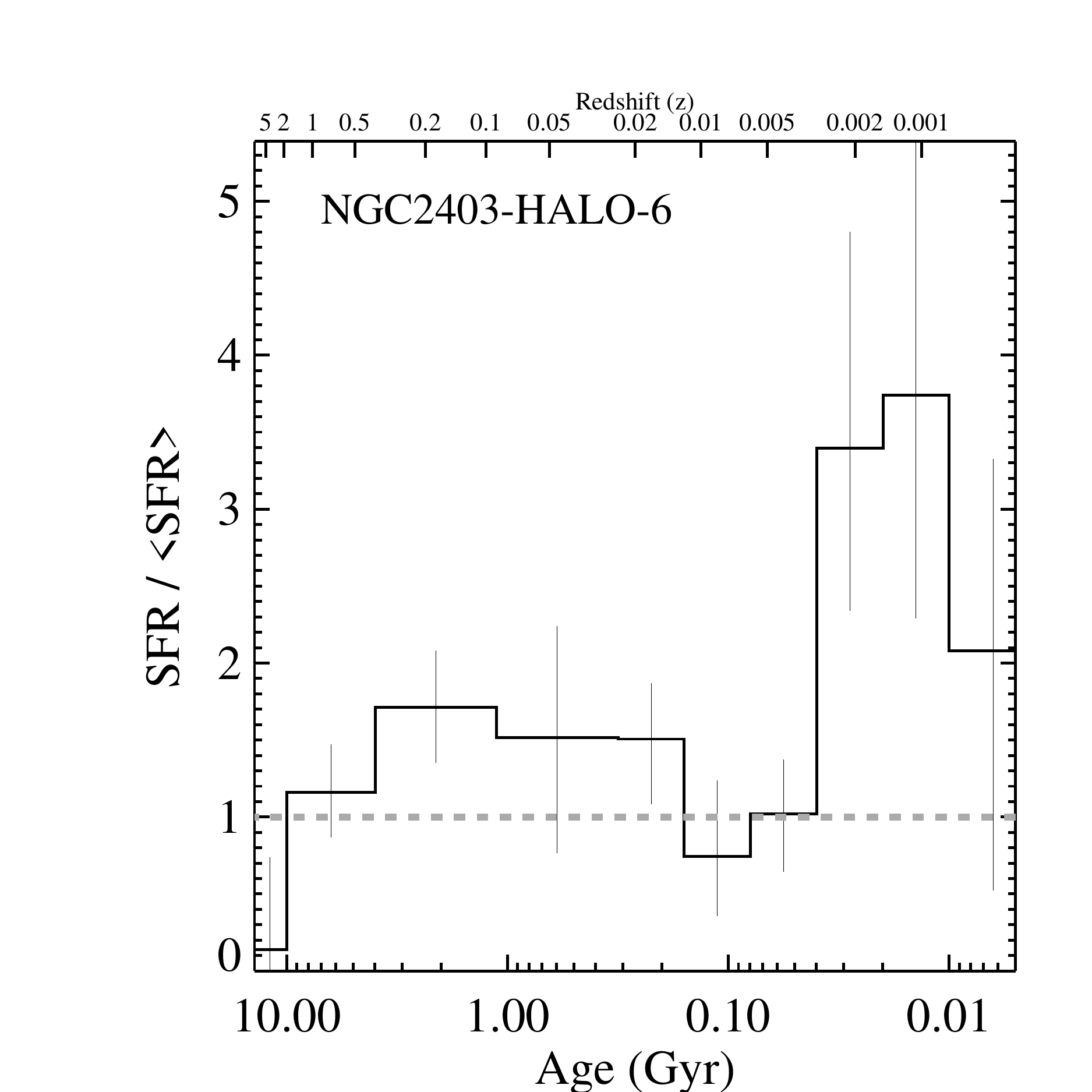}
\includegraphics[width=3.25in]{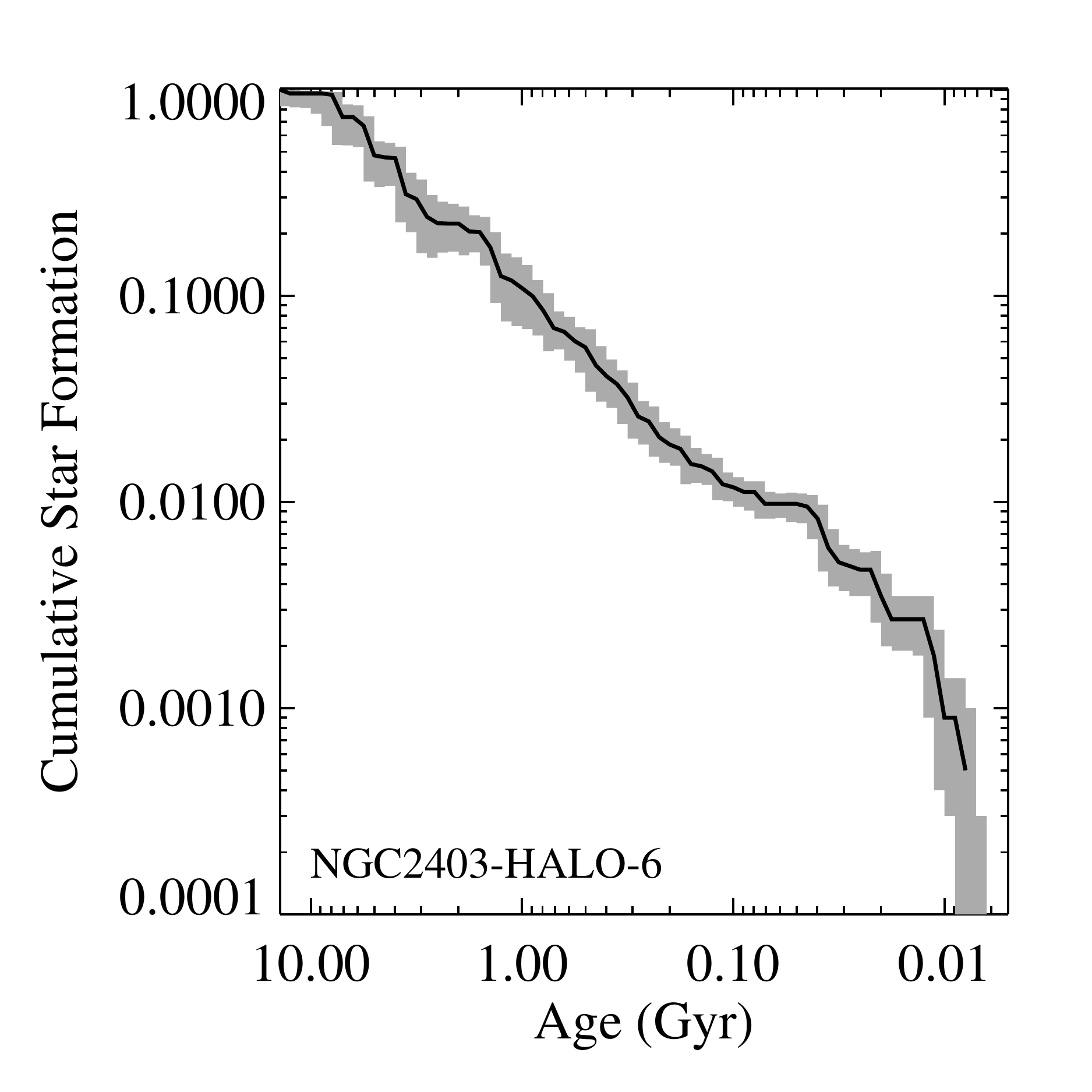}
}
\caption{ Color magnitude diagrams of the WFC3/IR (upper left) and
  optical (upper right) for the target NGC2403-HALO-6 within galaxy
  N2403.  Lower panels show the star formation history derived from
  the optical data, for both the differential (left, with horizontal
  dotted line indicating the past average SFR) and cumulative (right)
  star formation histories.  The cumulative star formation history is
  calculated from the present back to 14\,Gyrs.  Uncertainties in the
  lower two panels are the 68\% confidence intervals, calculated from
  Monte Carlo tests including random and systematic uncertainties.
  Optical CMDs are restricted to the area covered by the WFC3 FOV. }
\end{figure}
\vfill
\clearpage
 
\begin{figure}
\figurenum{\ref{cmdfig} continued}
\centerline{
\includegraphics[width=3.25in]{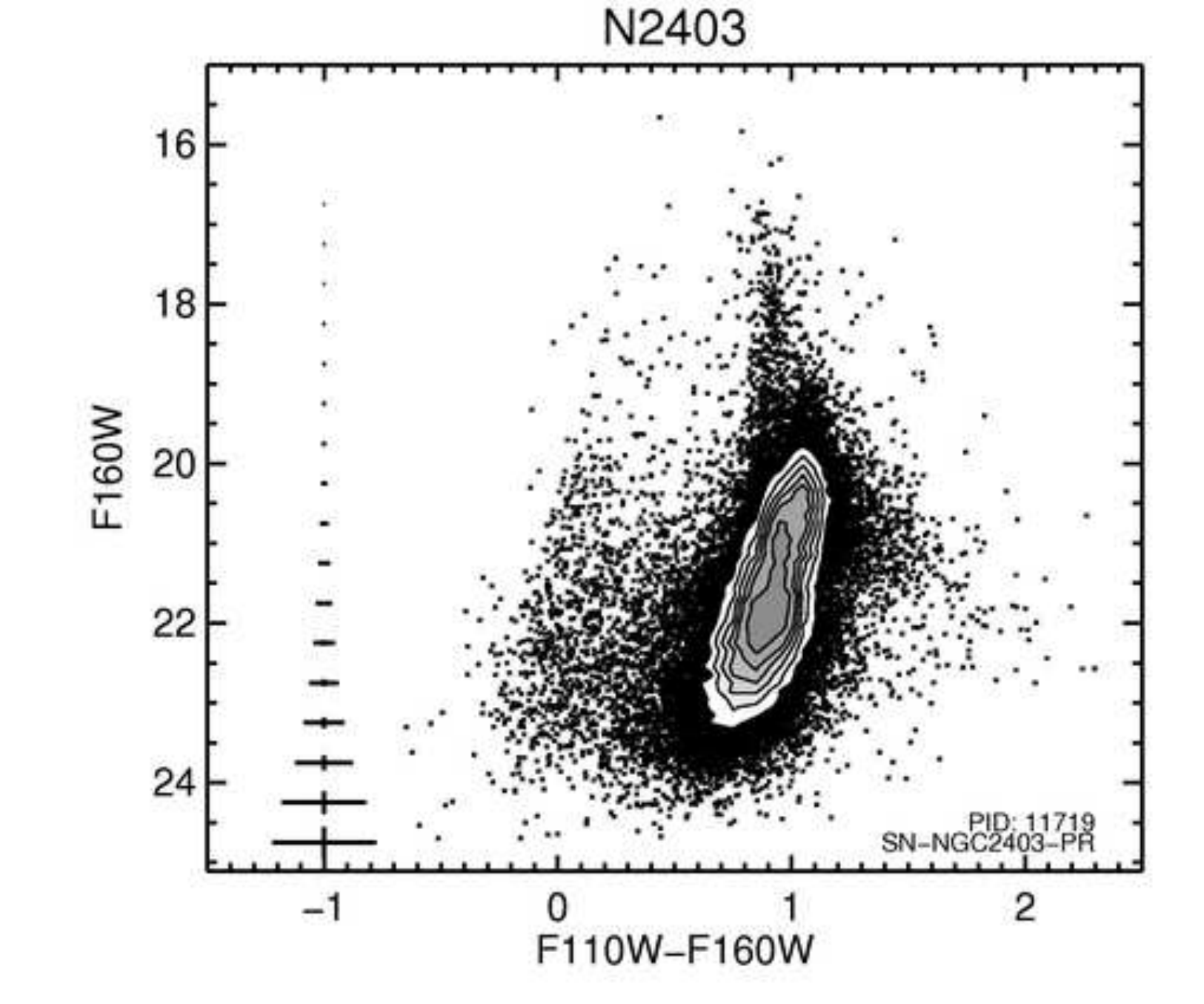}
\includegraphics[width=3.25in]{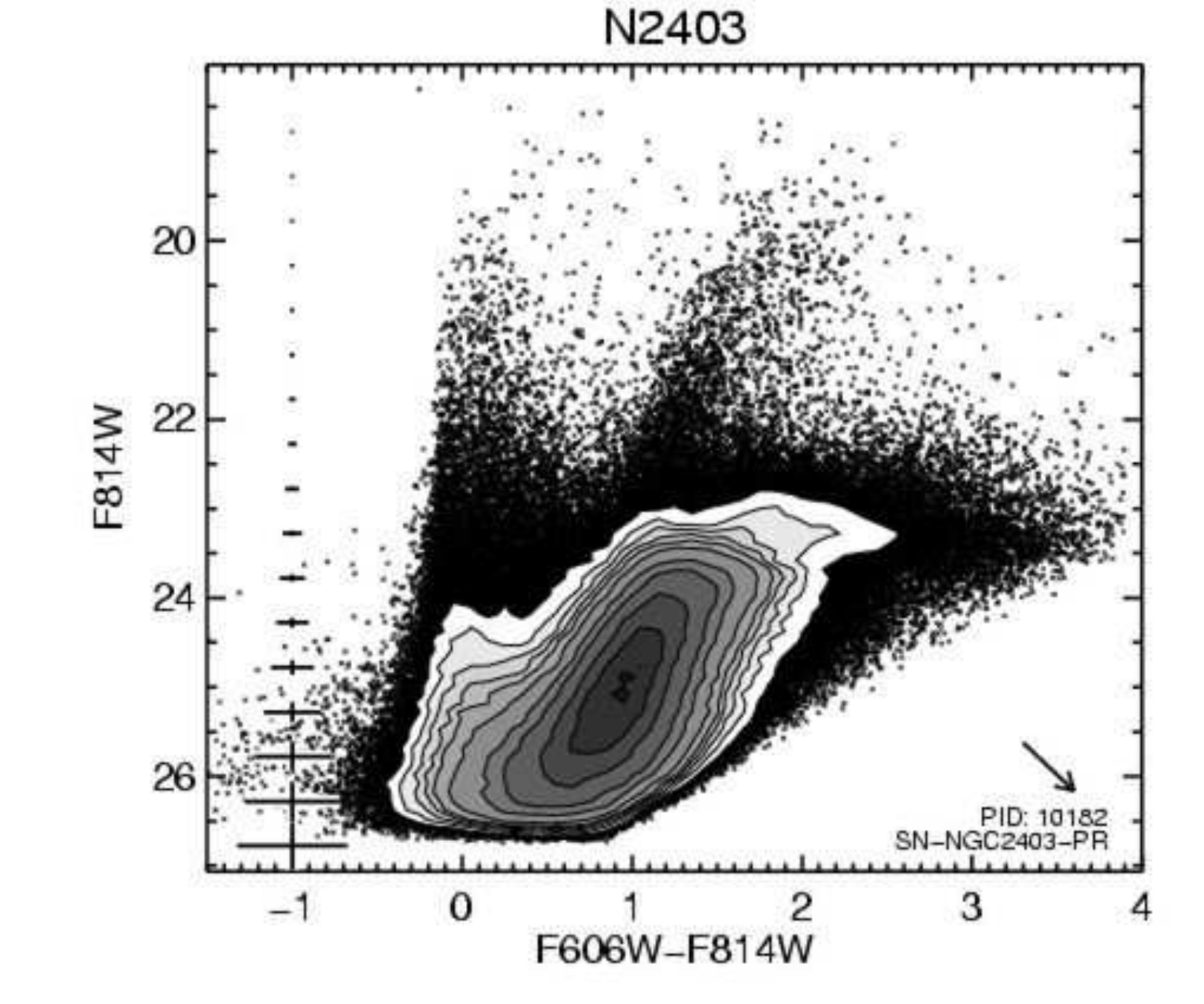}
}
\centerline{
\includegraphics[width=3.25in]{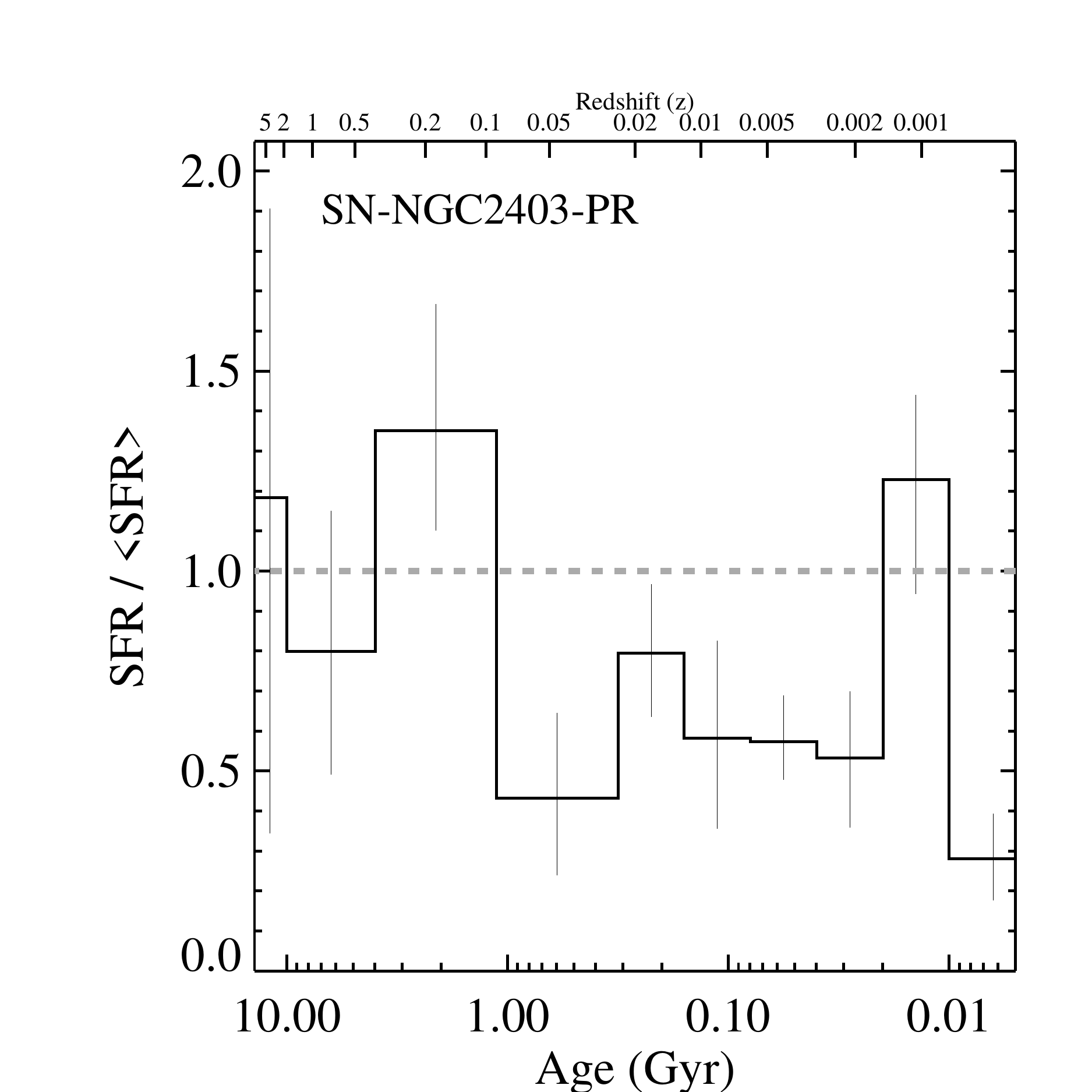}
\includegraphics[width=3.25in]{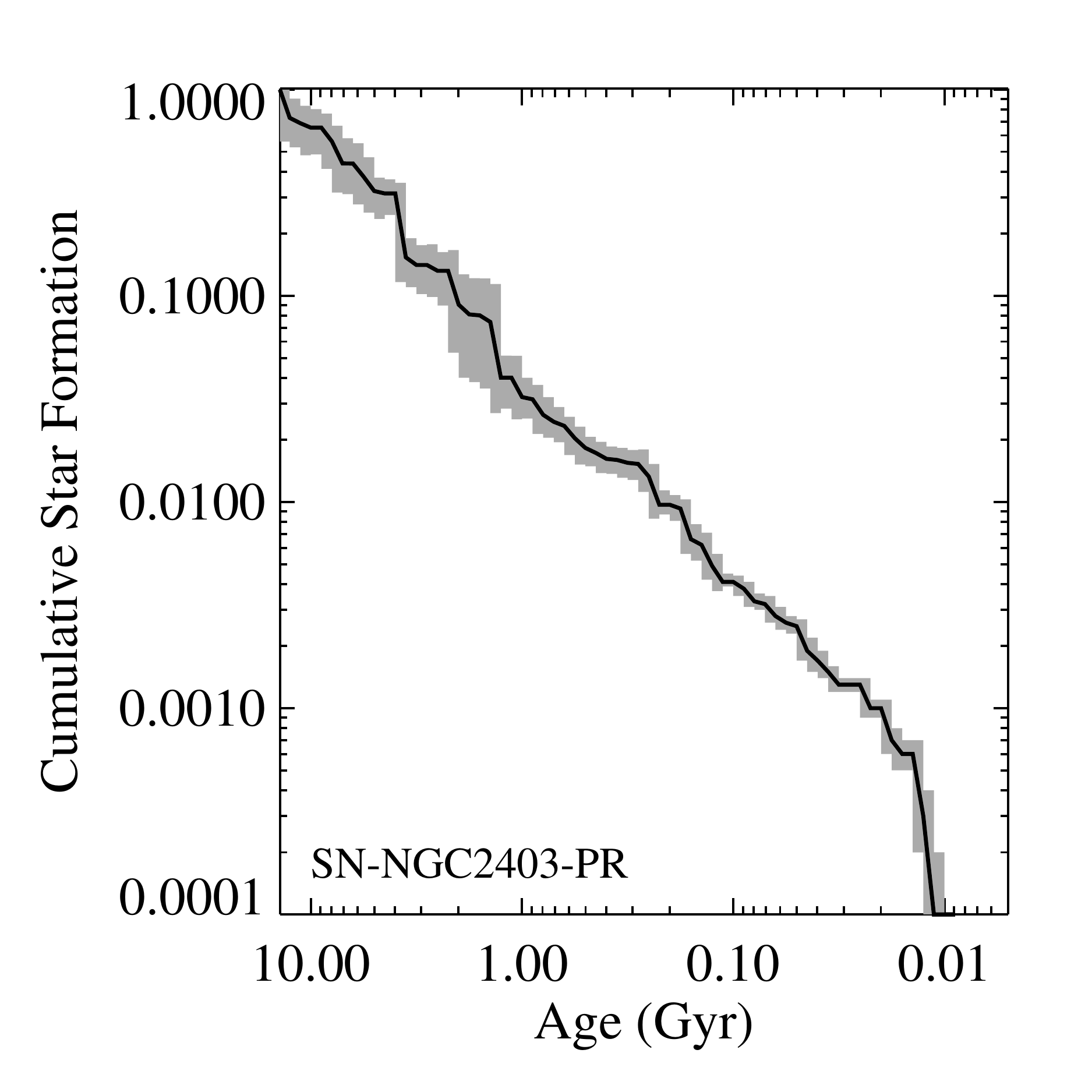}
}
\caption{ Color magnitude diagrams of the WFC3/IR (upper left) and
  optical (upper right) for the target SN-NGC2403-PR within galaxy
  N2403.  Lower panels show the star formation history derived from
  the optical data, for both the differential (left, with horizontal
  dotted line indicating the past average SFR) and cumulative (right)
  star formation histories.  The cumulative star formation history is
  calculated from the present back to 14\,Gyrs.  Uncertainties in the
  lower two panels are the 68\% confidence intervals, calculated from
  Monte Carlo tests including random and systematic uncertainties.
  Optical CMDs are restricted to the area covered by the WFC3 FOV. }
\end{figure}
\vfill
\clearpage
 
\begin{figure}
\figurenum{\ref{cmdfig} continued}
\centerline{
\includegraphics[width=3.25in]{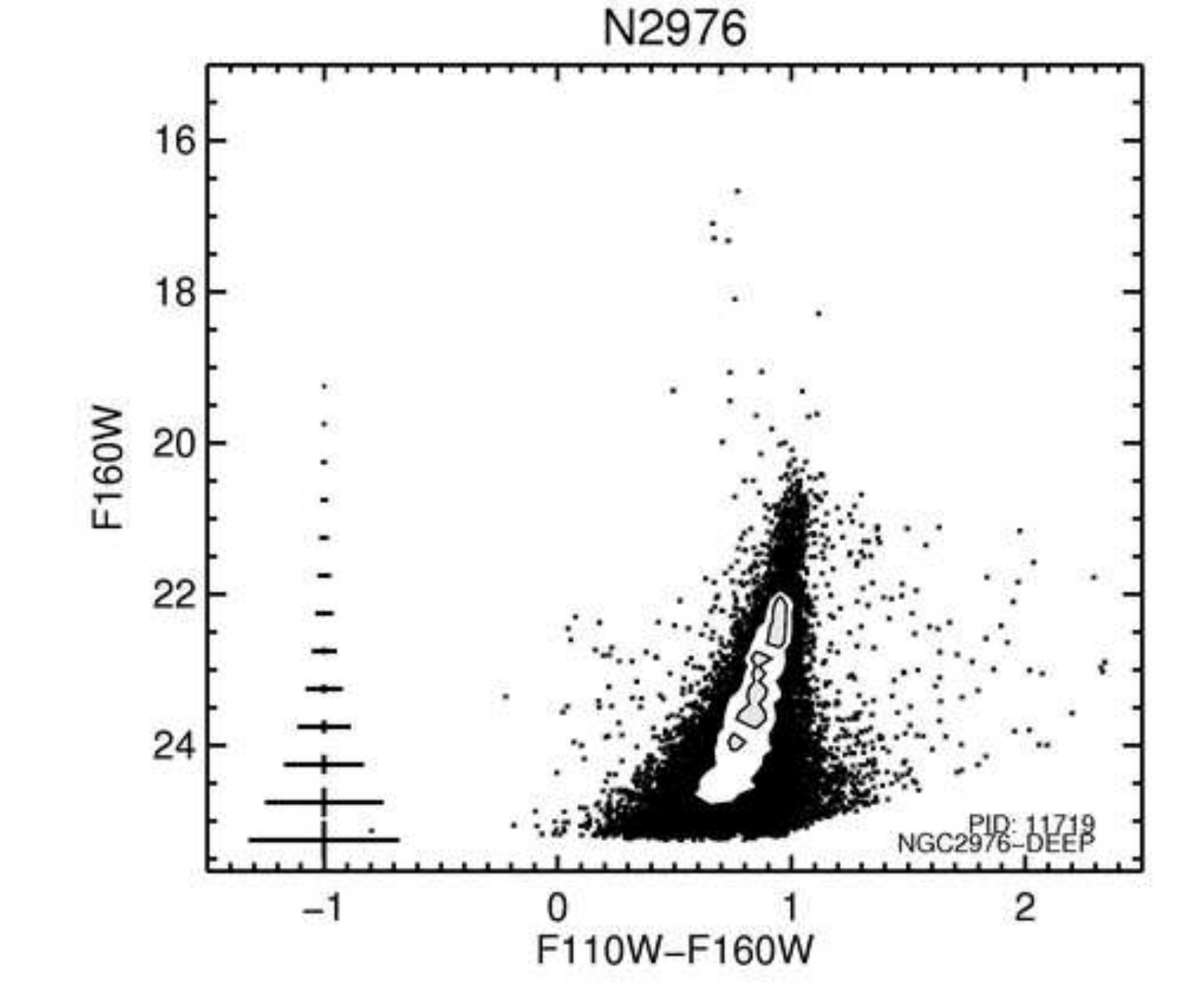}
\includegraphics[width=3.25in]{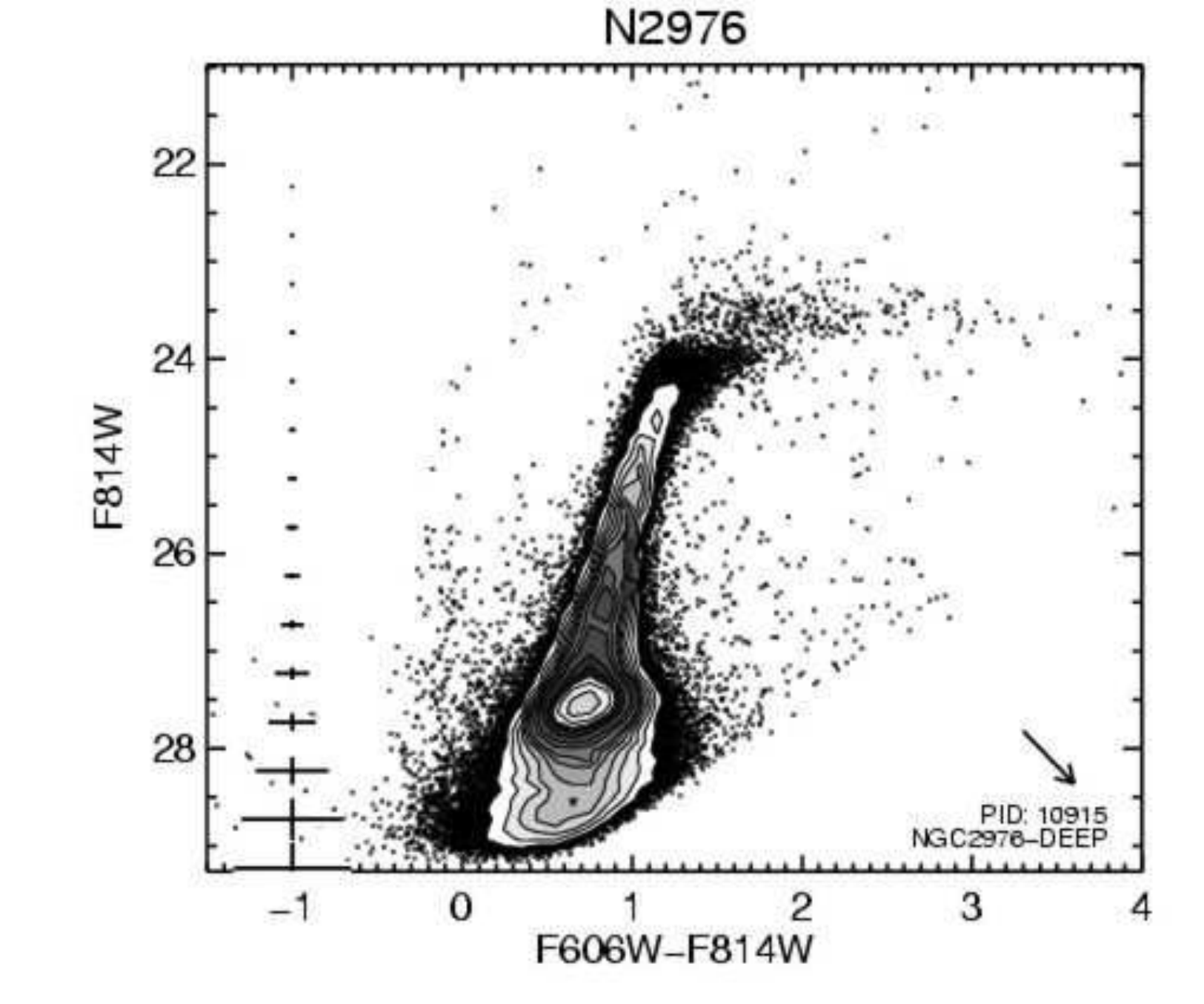}
}
\centerline{
\includegraphics[width=3.25in]{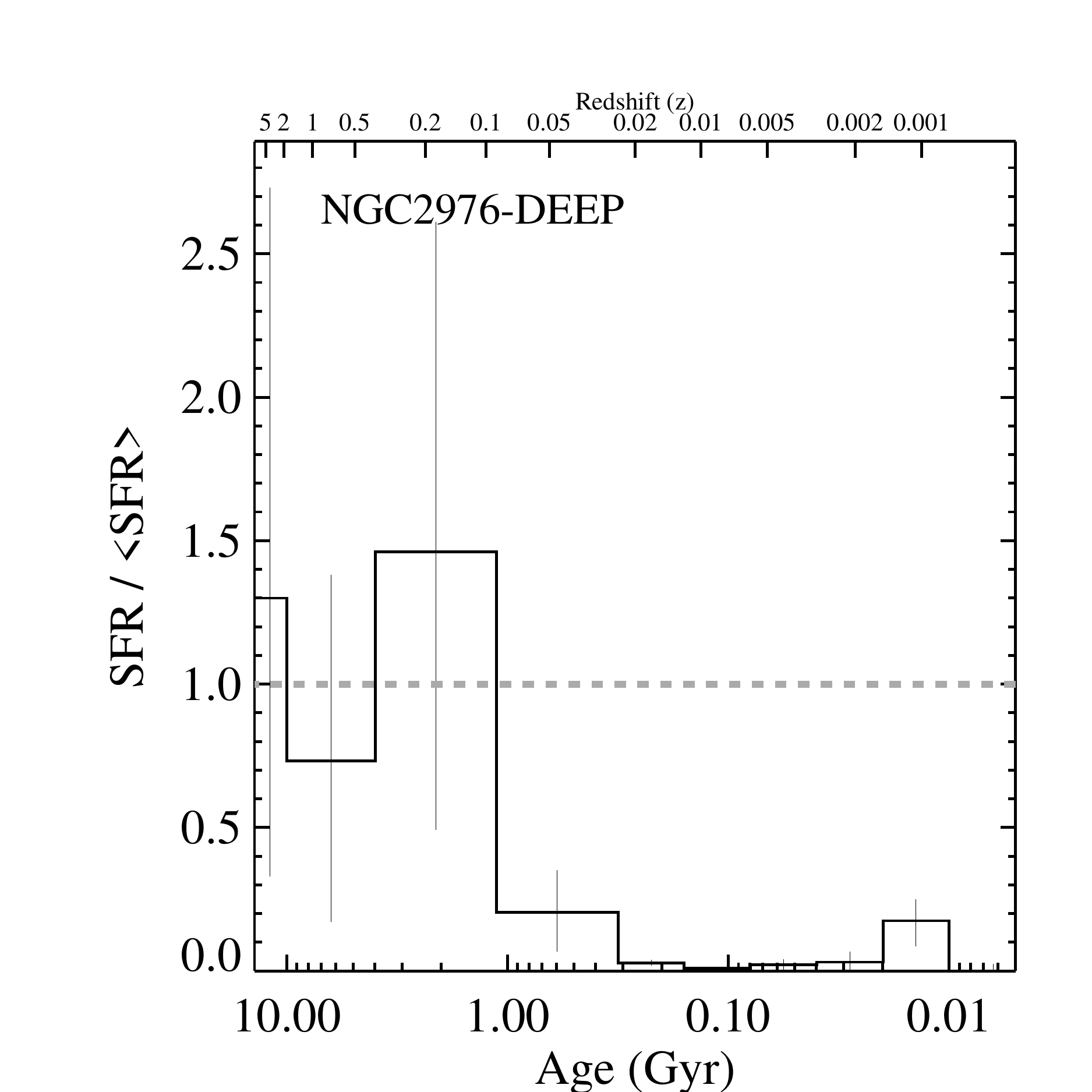}
\includegraphics[width=3.25in]{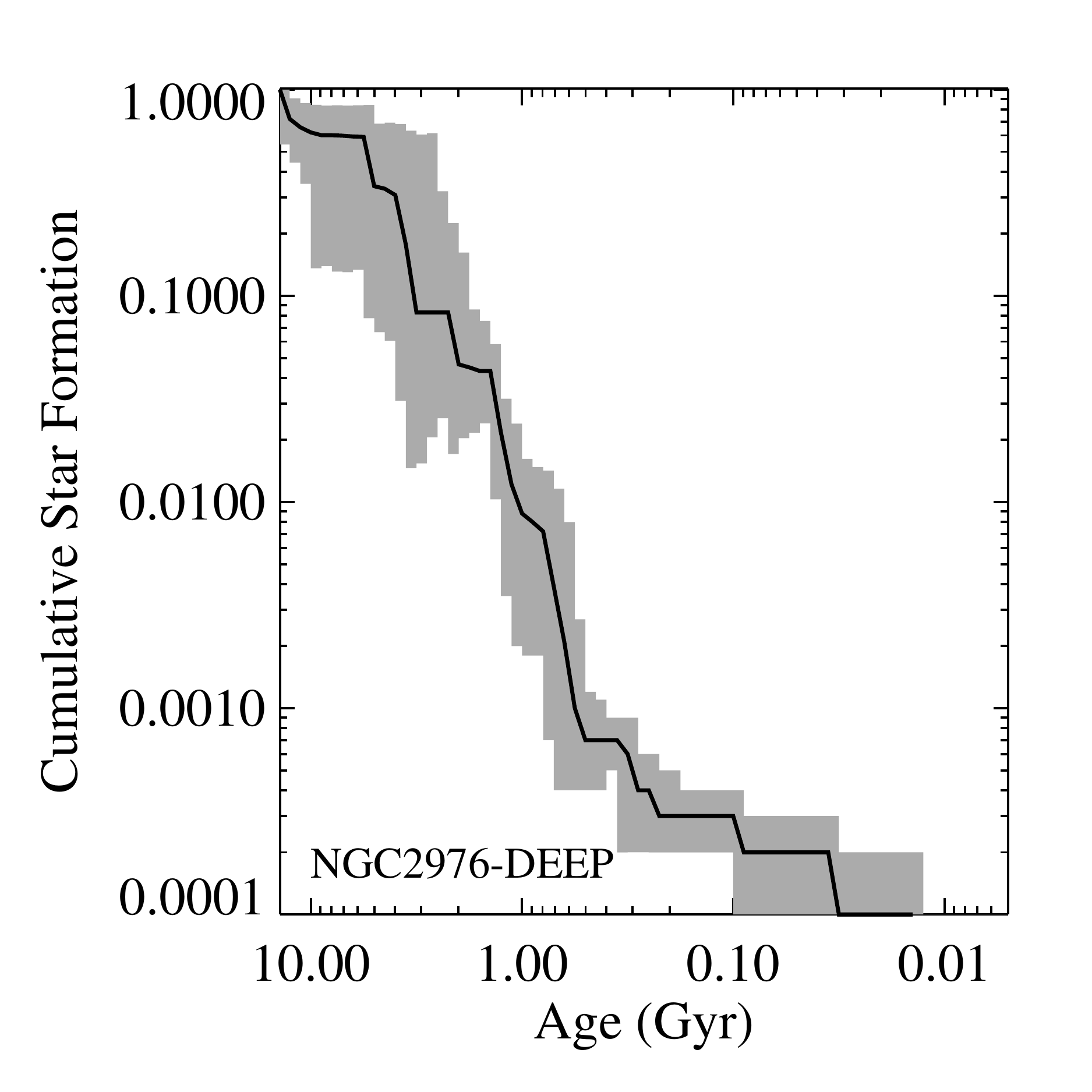}
}
\caption{ Color magnitude diagrams of the WFC3/IR (upper left) and
  optical (upper right) for the target NGC2976-DEEP within galaxy
  N2976.  Lower panels show the star formation history derived from
  the optical data, for both the differential (left, with horizontal
  dotted line indicating the past average SFR) and cumulative (right)
  star formation histories.  The cumulative star formation history is
  calculated from the present back to 14\,Gyrs.  Uncertainties in the
  lower two panels are the 68\% confidence intervals, calculated from
  Monte Carlo tests including random and systematic uncertainties.
  Optical CMDs are restricted to the area covered by the WFC3 FOV. }
\end{figure}
\vfill
\clearpage
 
\begin{figure}
\figurenum{\ref{cmdfig} continued}
\centerline{
\includegraphics[width=3.25in]{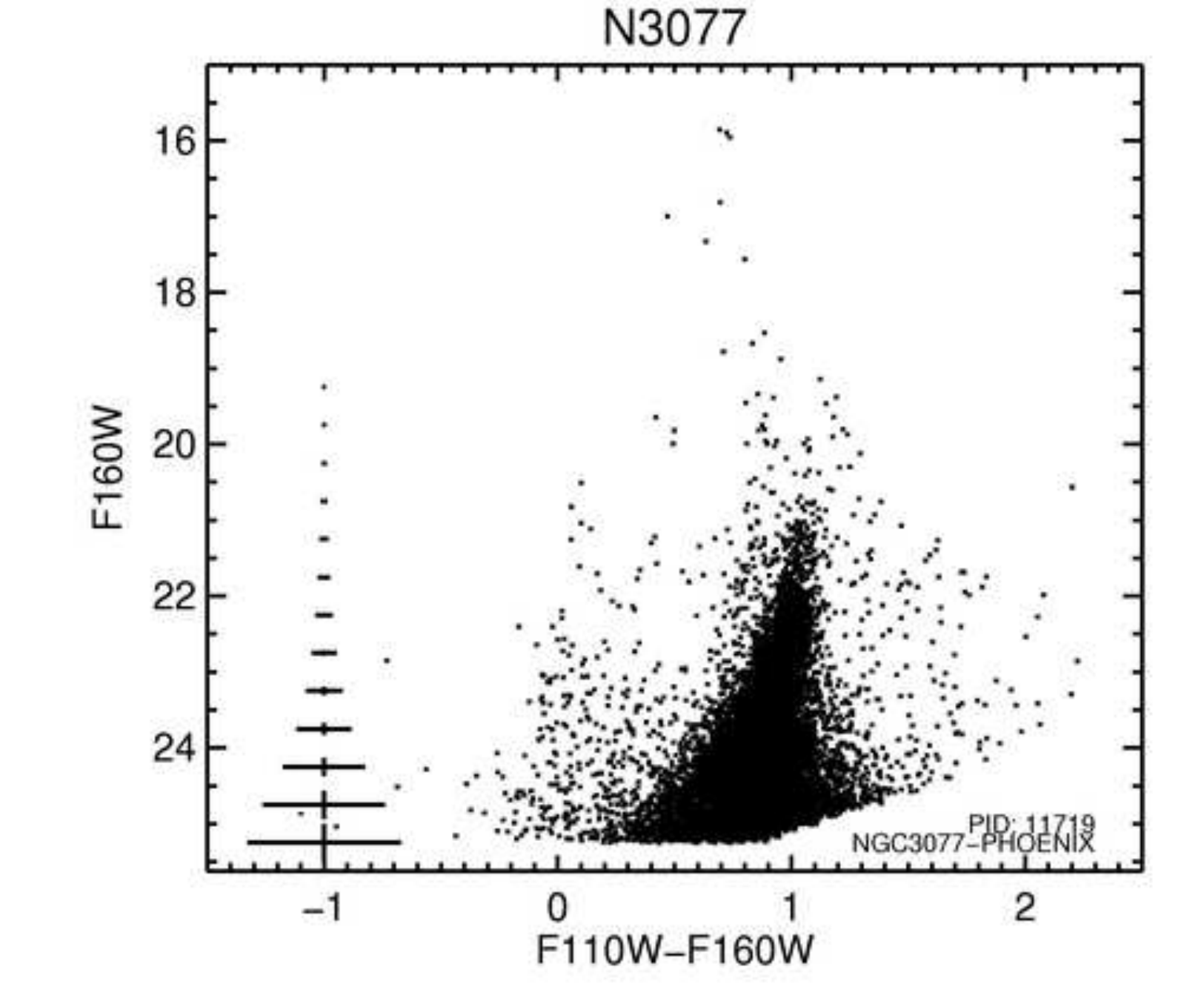}
\includegraphics[width=3.25in]{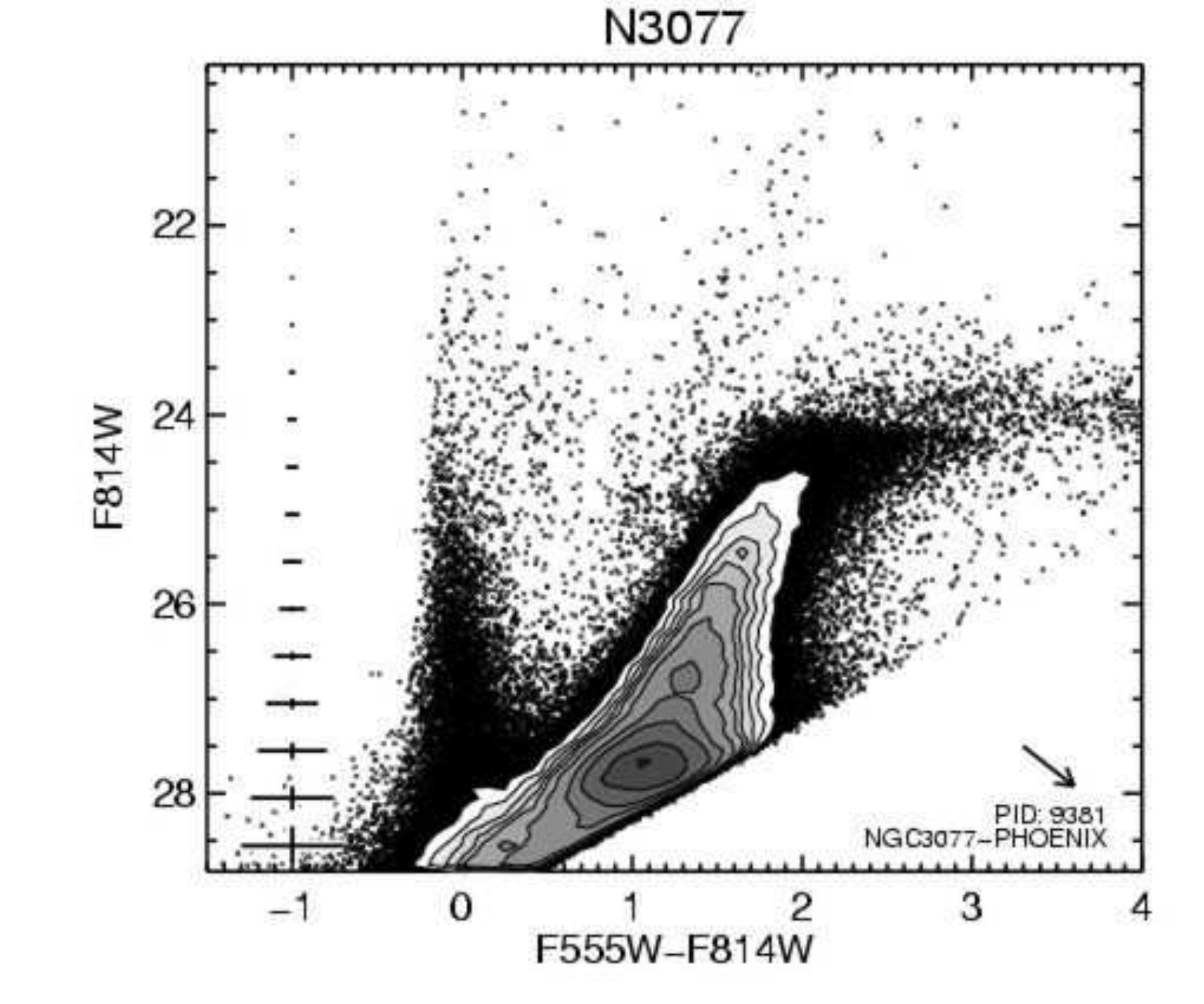}
}
\centerline{
\includegraphics[width=3.25in]{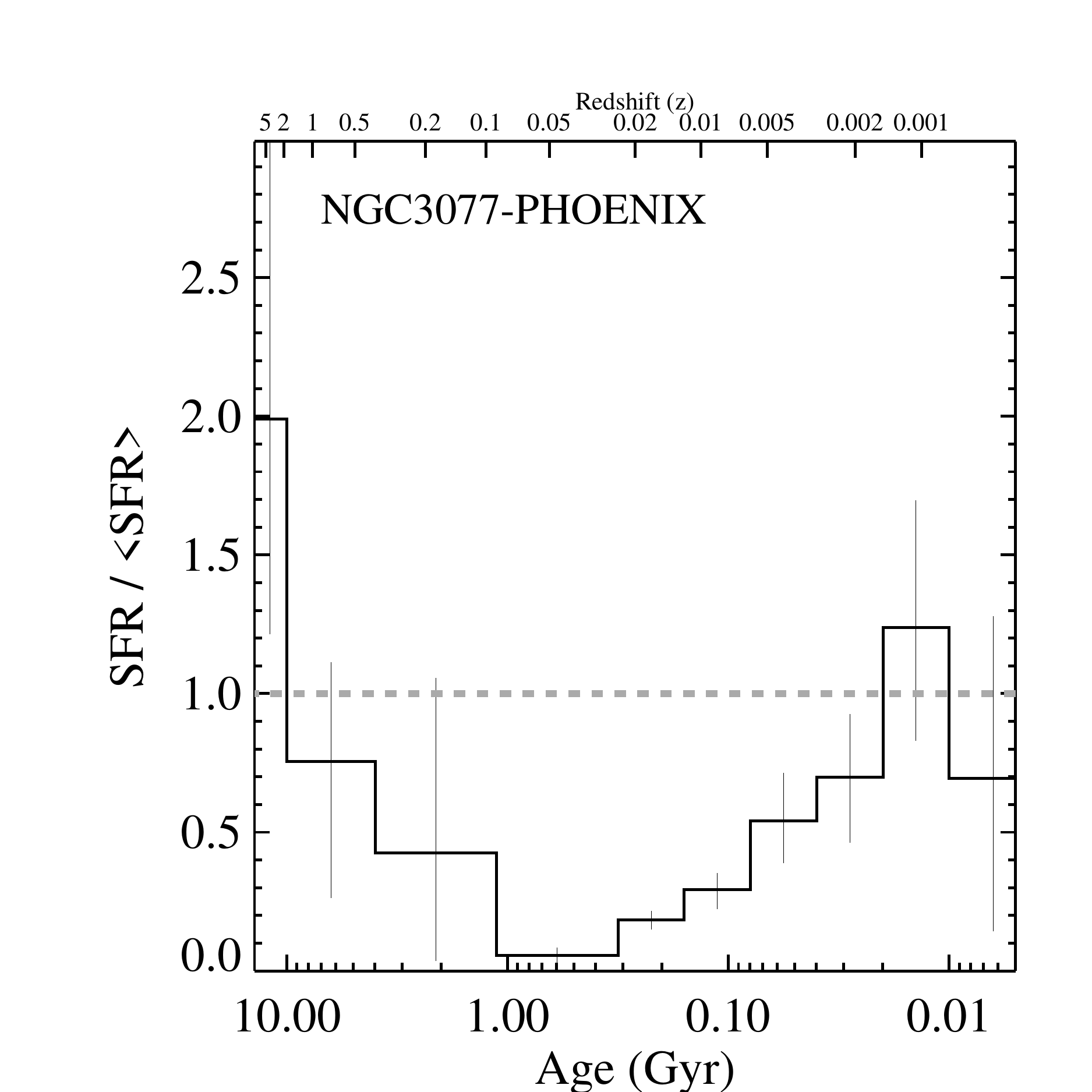}
\includegraphics[width=3.25in]{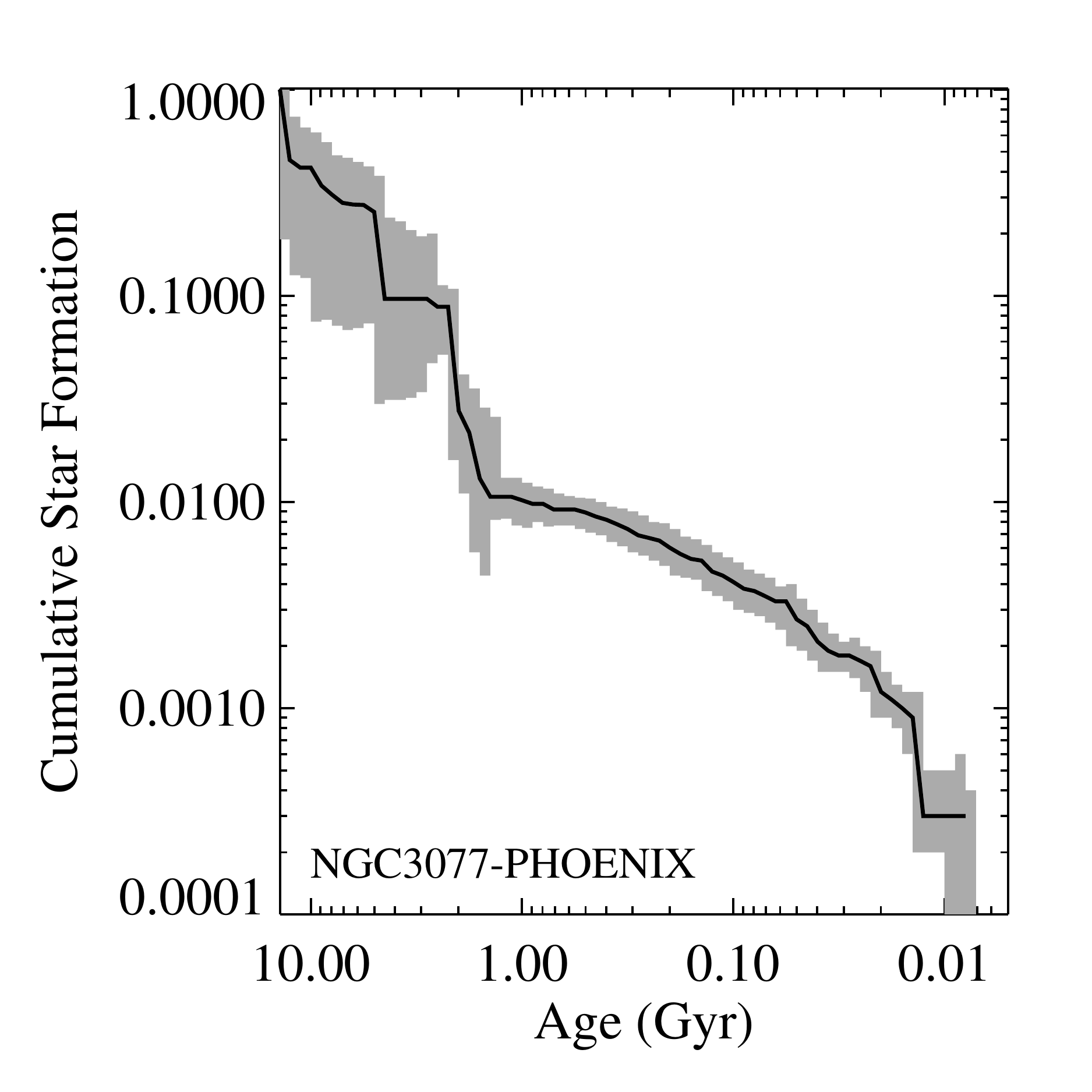}
}
\caption{ Color magnitude diagrams of the WFC3/IR (upper left) and
  optical (upper right) for the target NGC3077-PHOENIX within galaxy
  N3077.  Lower panels show the star formation history derived from
  the optical data, for both the differential (left, with horizontal
  dotted line indicating the past average SFR) and cumulative (right)
  star formation histories.  The cumulative star formation history is
  calculated from the present back to 14\,Gyrs.  Uncertainties in the
  lower two panels are the 68\% confidence intervals, calculated from
  Monte Carlo tests including random and systematic uncertainties.
  Optical CMDs are restricted to the area covered by the WFC3 FOV. }
\end{figure}
\vfill
\clearpage
 
\begin{figure}
\figurenum{\ref{cmdfig} continued}
\centerline{
\includegraphics[width=3.25in]{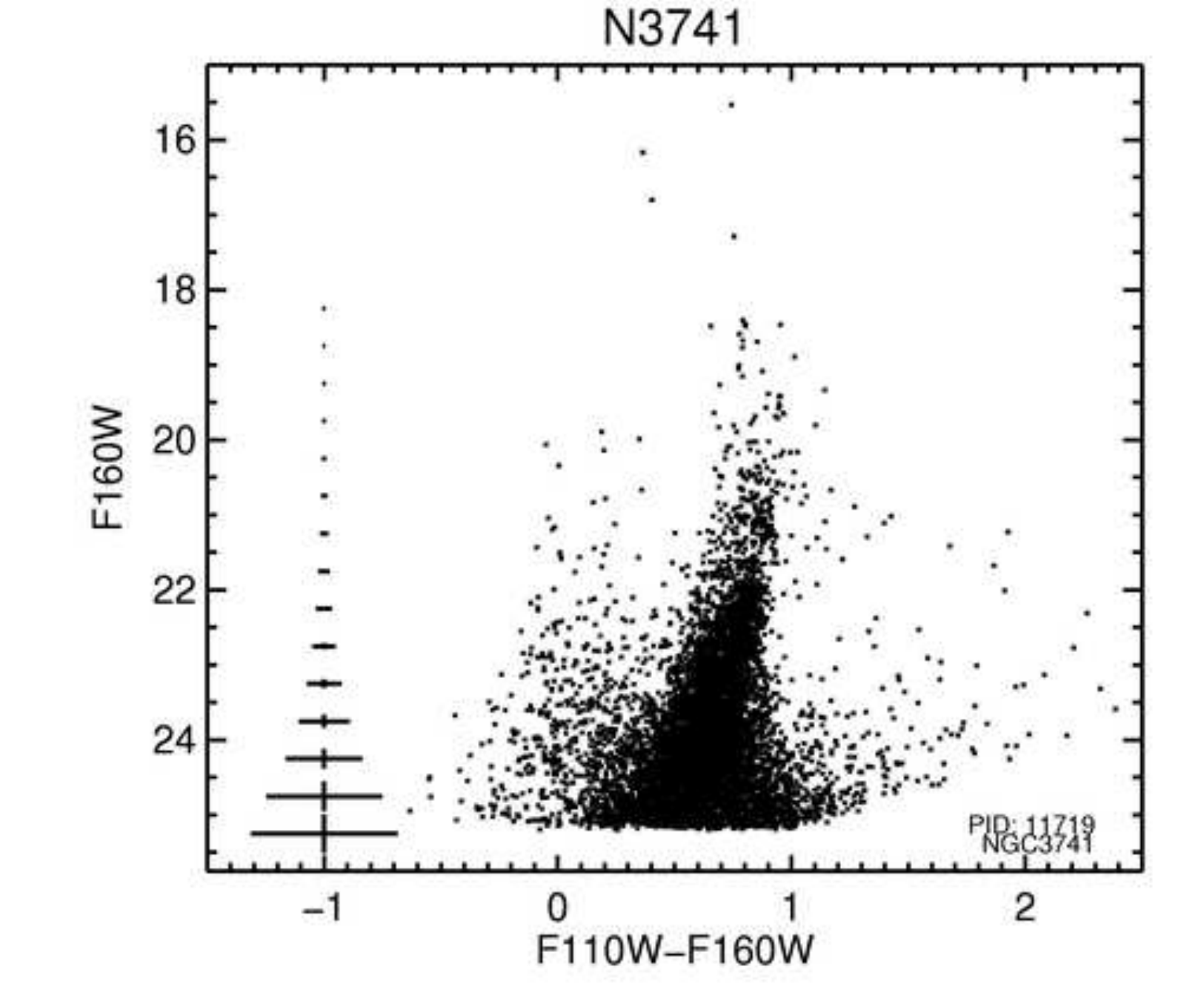}
\includegraphics[width=3.25in]{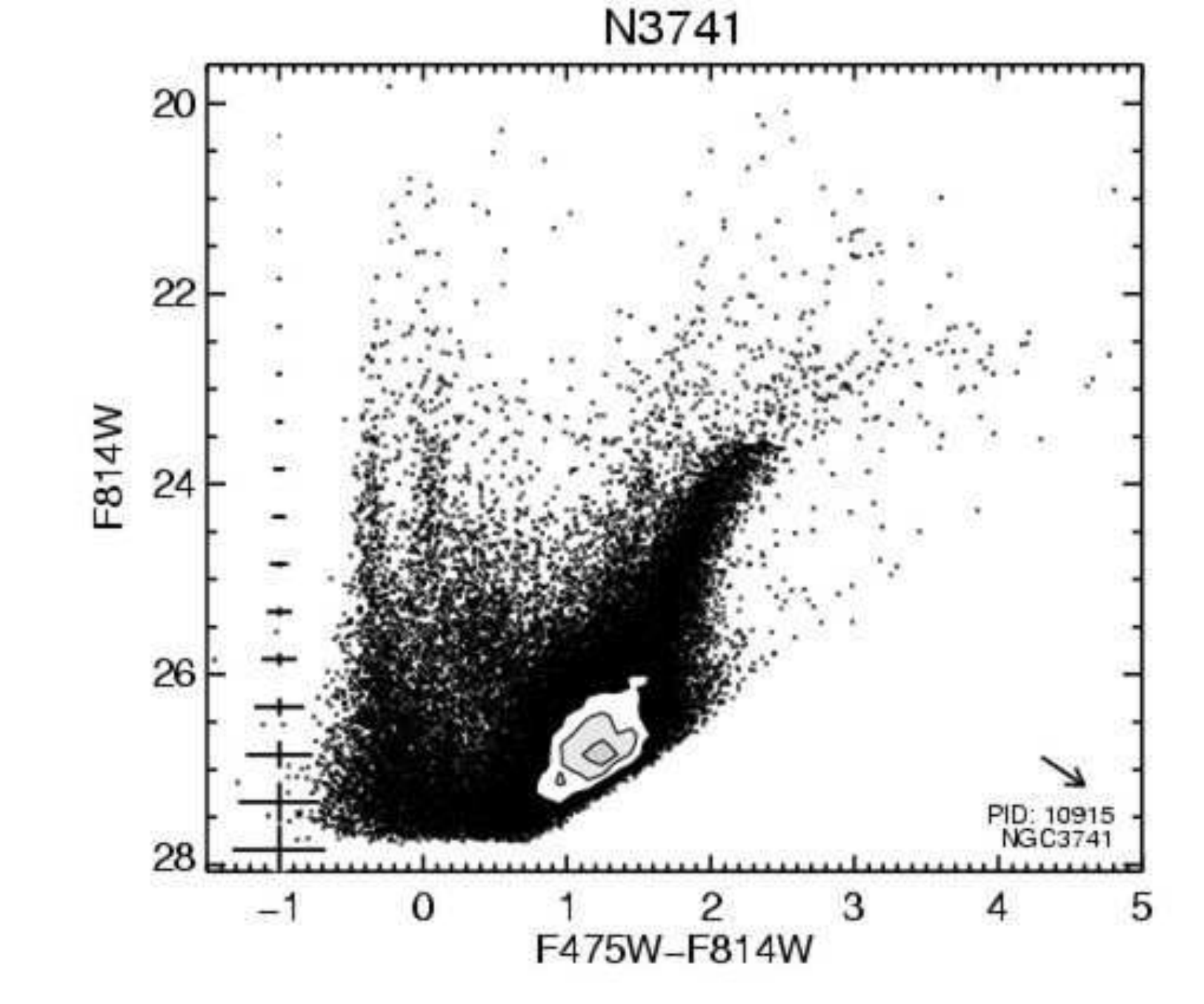}
}
\centerline{
\includegraphics[width=3.25in]{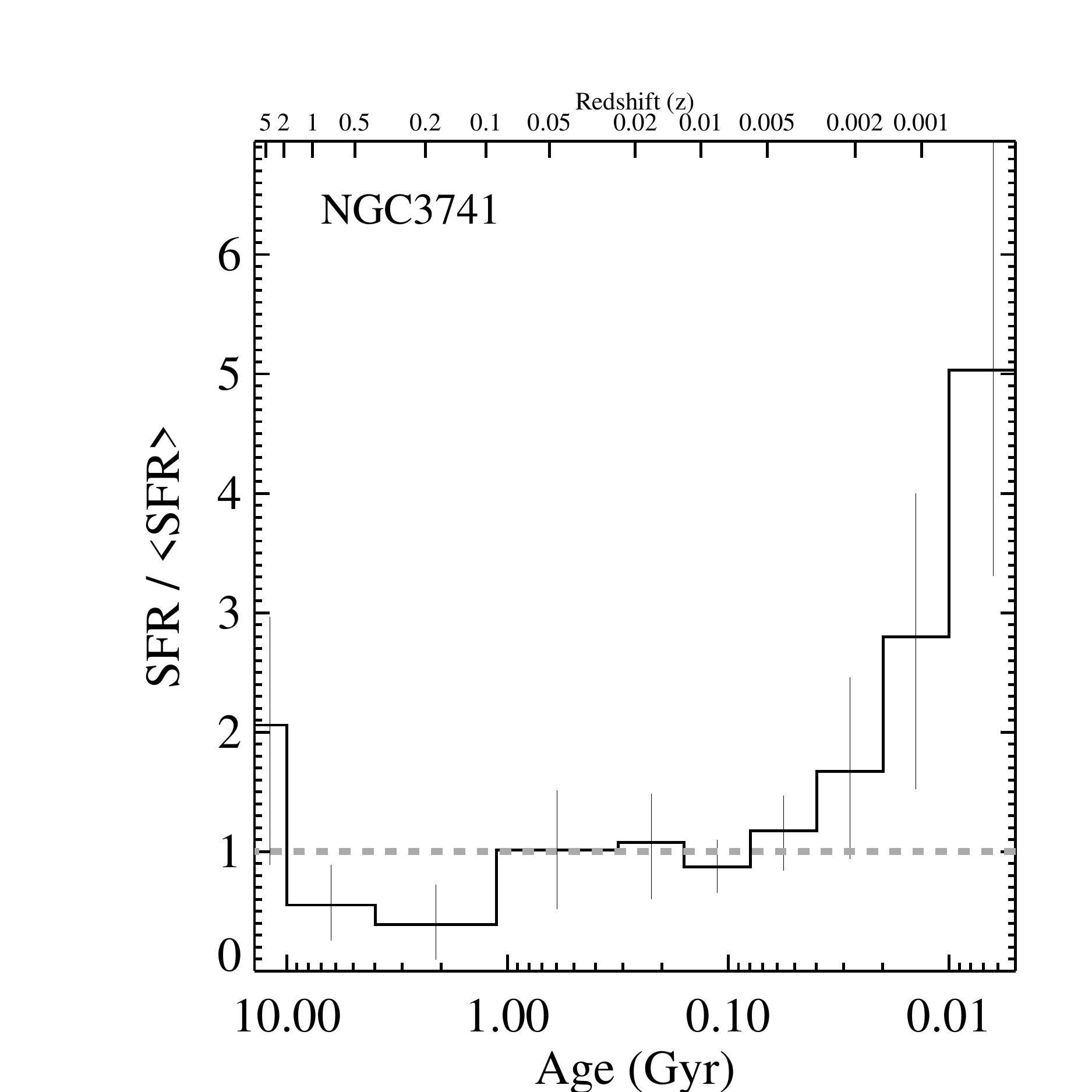}
\includegraphics[width=3.25in]{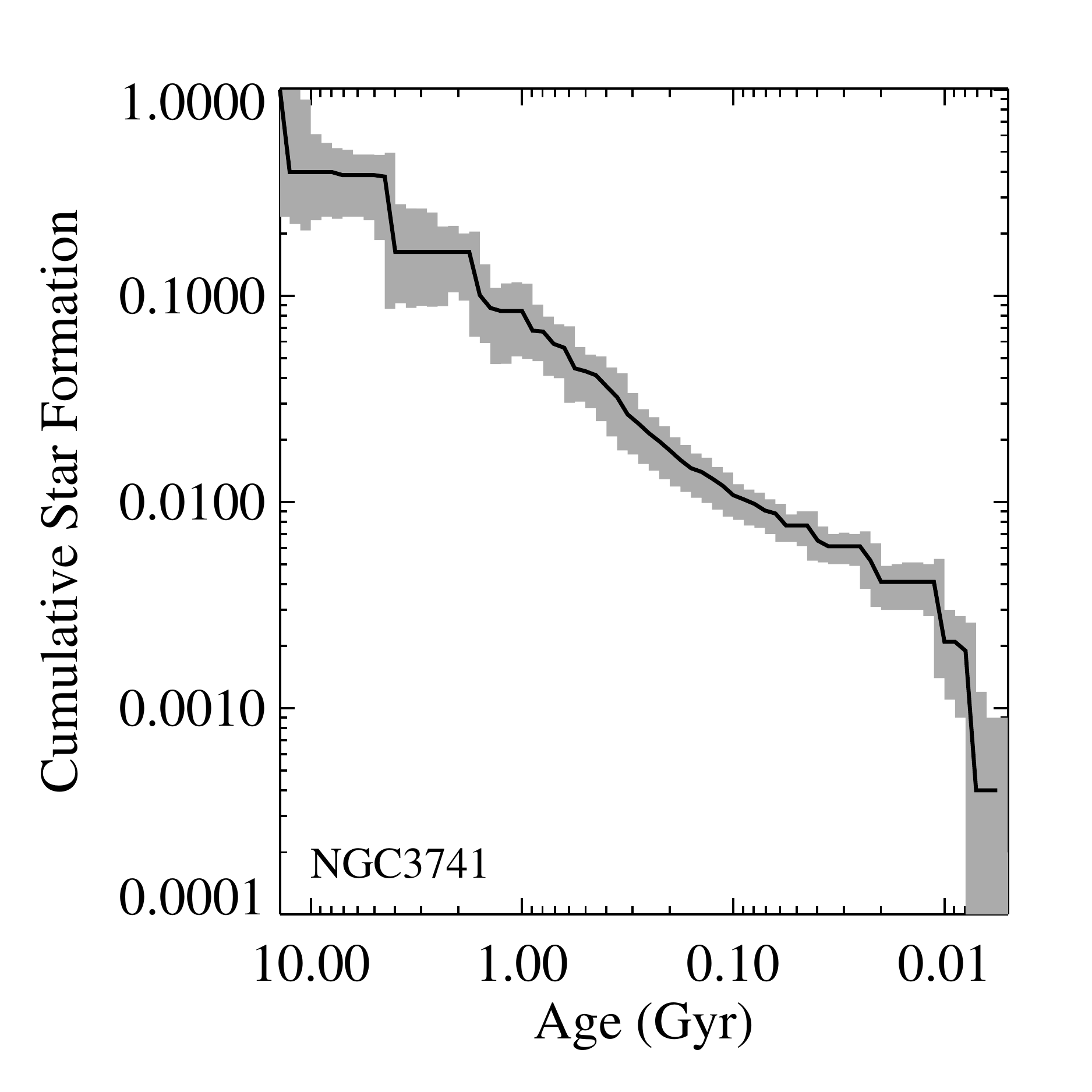}
}
\caption{ Color magnitude diagrams of the WFC3/IR (upper left) and
  optical (upper right) for the target NGC3741 within galaxy N3741.
  Lower panels show the star formation history derived from the
  optical data, for both the differential (left, with horizontal
  dotted line indicating the past average SFR) and cumulative (right)
  star formation histories.  The cumulative star formation history is
  calculated from the present back to 14\,Gyrs.  Uncertainties in the
  lower two panels are the 68\% confidence intervals, calculated from
  Monte Carlo tests including random and systematic uncertainties.
  Optical CMDs are restricted to the area covered by the WFC3 FOV. }
\end{figure}
\vfill
\clearpage
 
\begin{figure}
\figurenum{\ref{cmdfig} continued}
\centerline{
\includegraphics[width=3.25in]{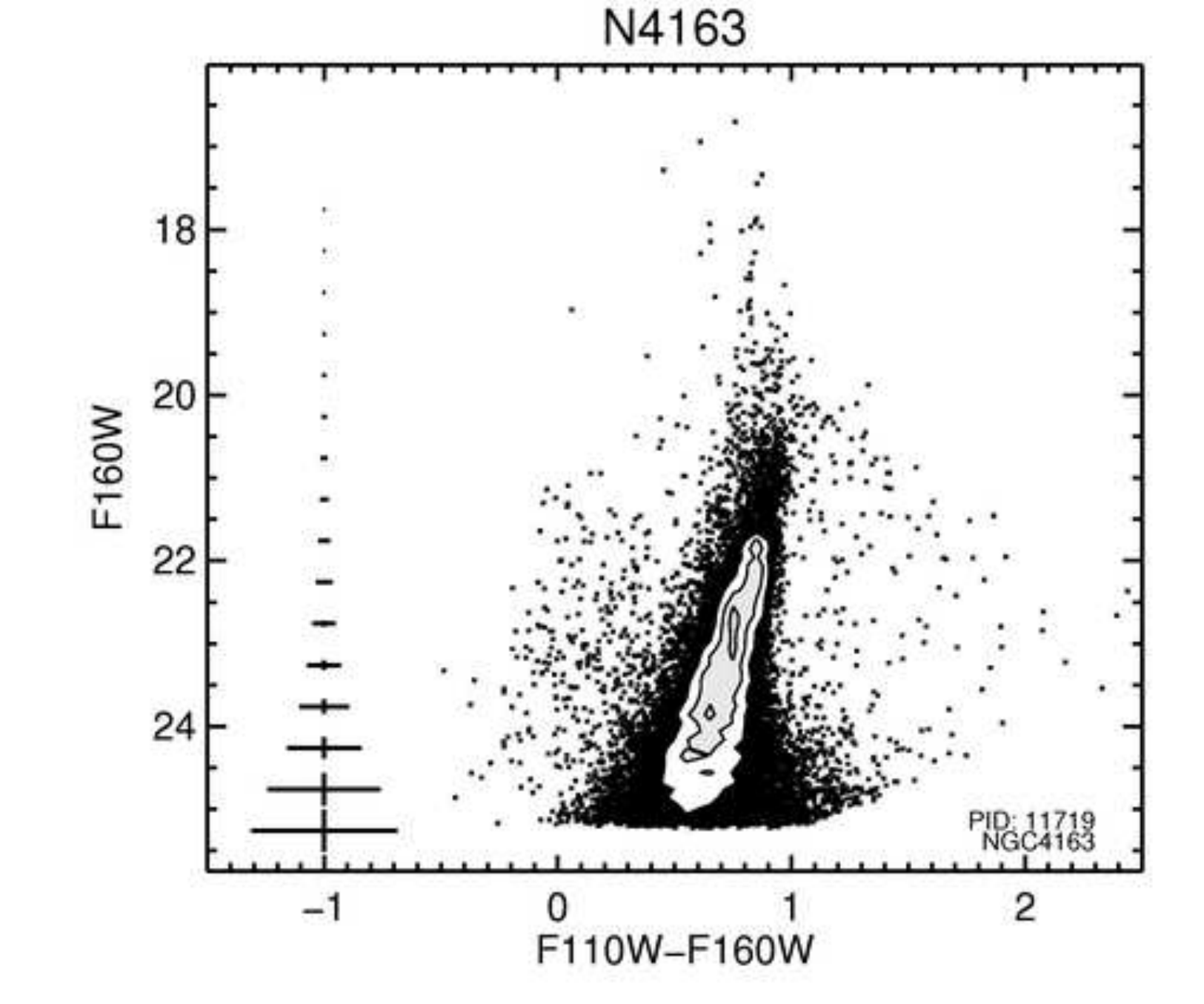}
\includegraphics[width=3.25in]{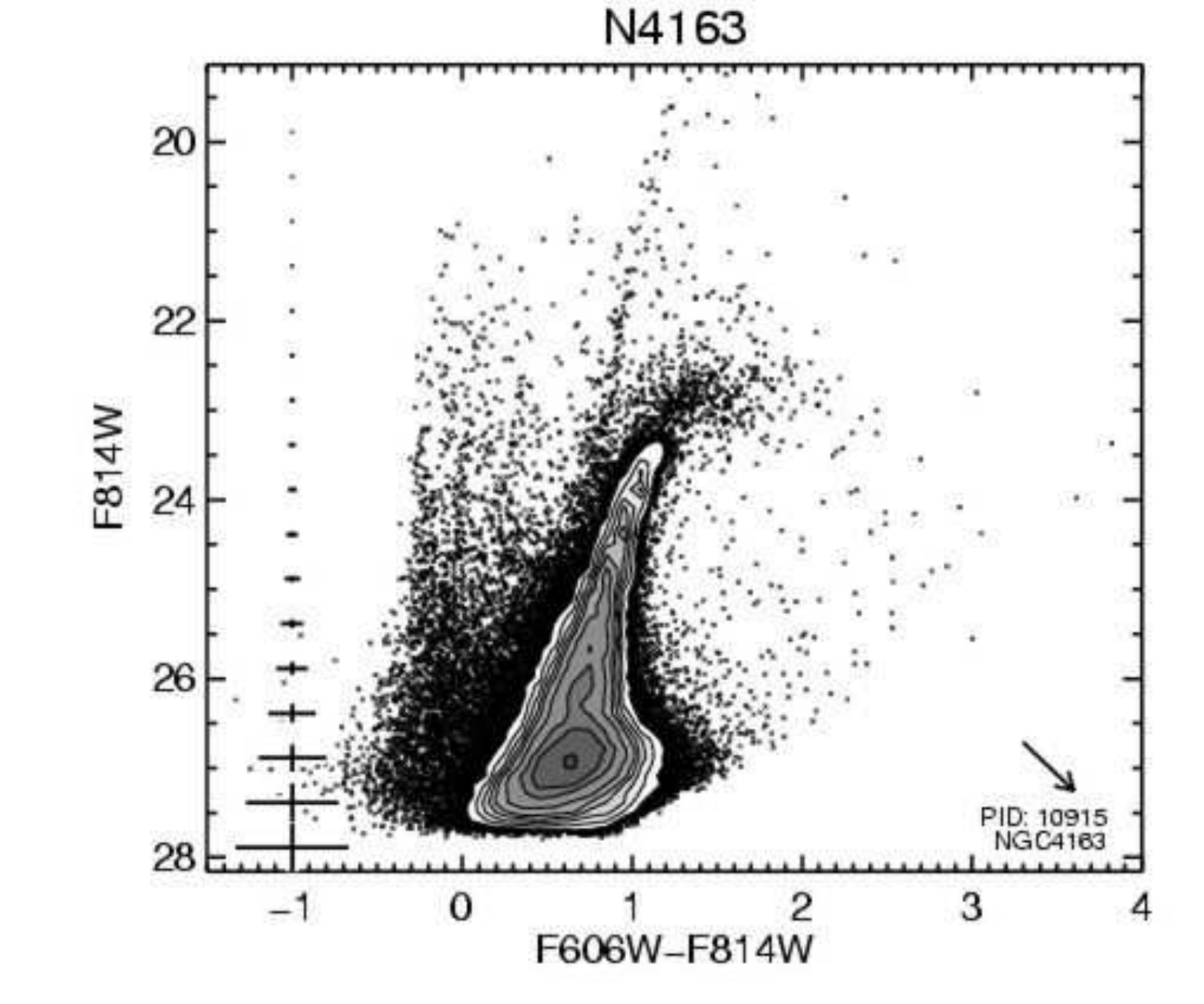}
}
\centerline{
\includegraphics[width=3.25in]{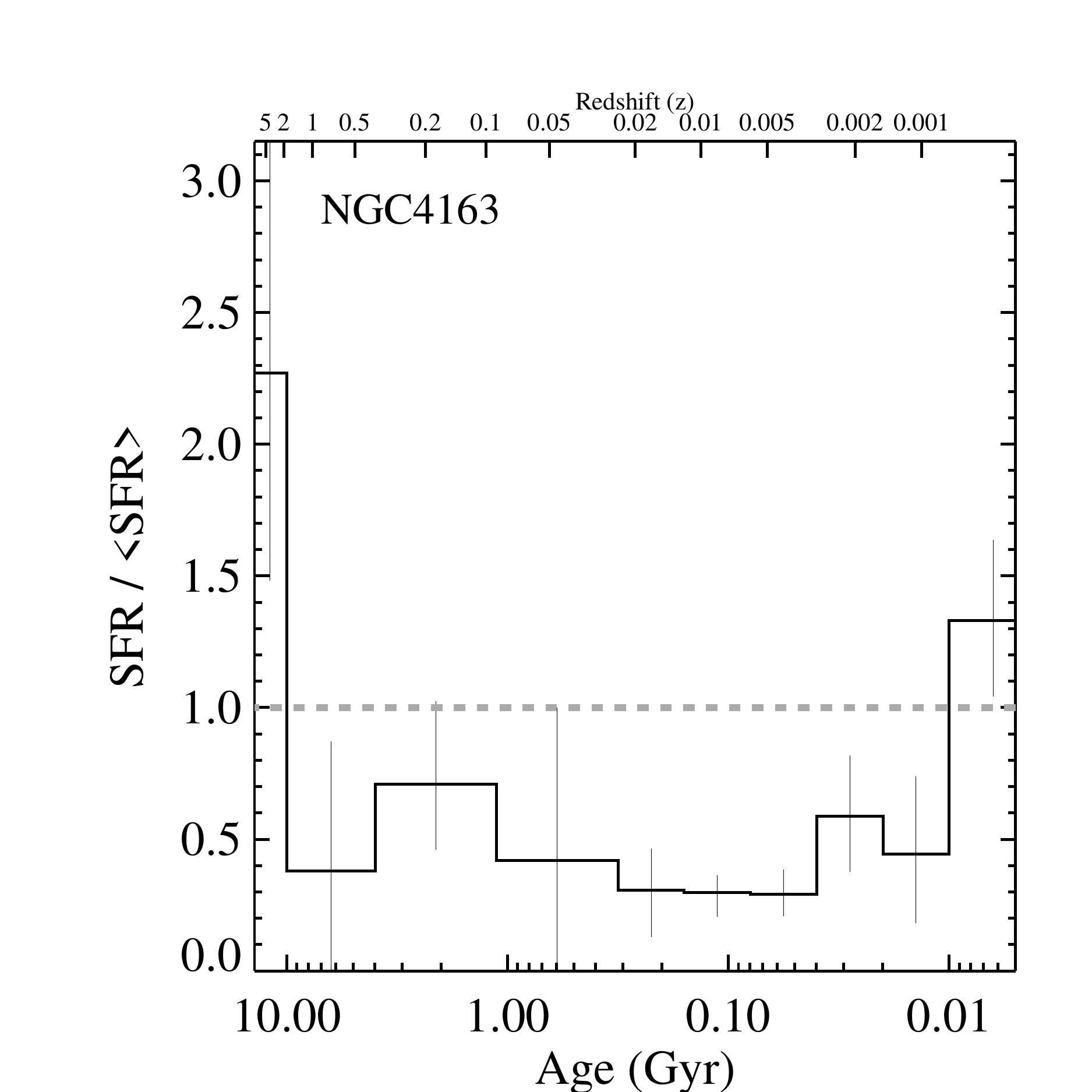}
\includegraphics[width=3.25in]{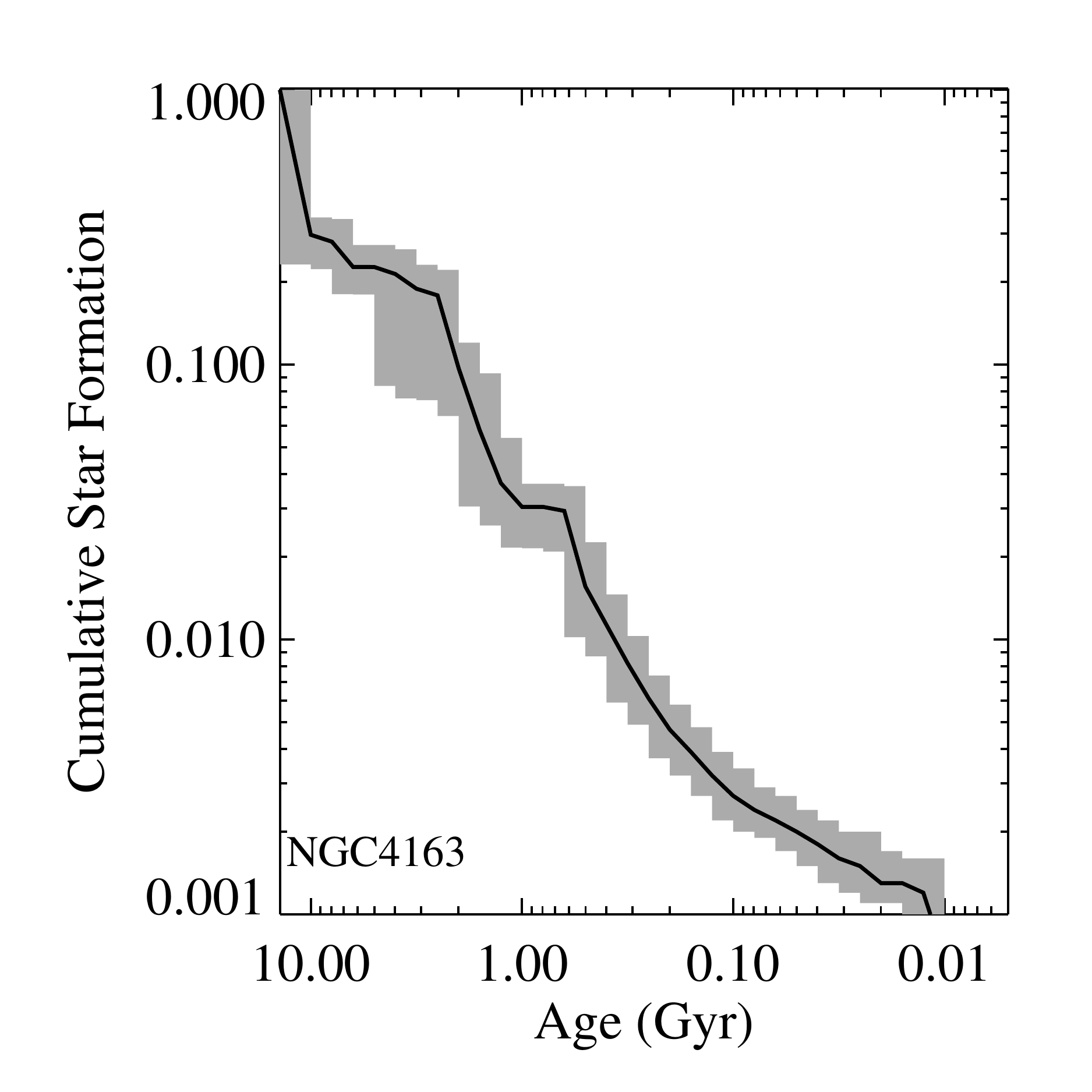}
}
\caption{ Color magnitude diagrams of the WFC3/IR (upper left) and
  optical (upper right) for the target NGC4163 within galaxy N4163.
  Lower panels show the star formation history derived from the
  optical data, for both the differential (left, with horizontal
  dotted line indicating the past average SFR) and cumulative (right)
  star formation histories.  The cumulative star formation history is
  calculated from the present back to 14\,Gyrs.  Uncertainties in the
  lower two panels are the 68\% confidence intervals, calculated from
  Monte Carlo tests including random and systematic uncertainties.
  Optical CMDs are restricted to the area covered by the WFC3 FOV. }
\end{figure}
\vfill
\clearpage
 
\begin{figure}
\figurenum{\ref{cmdfig} continued}
\centerline{
\includegraphics[width=3.25in]{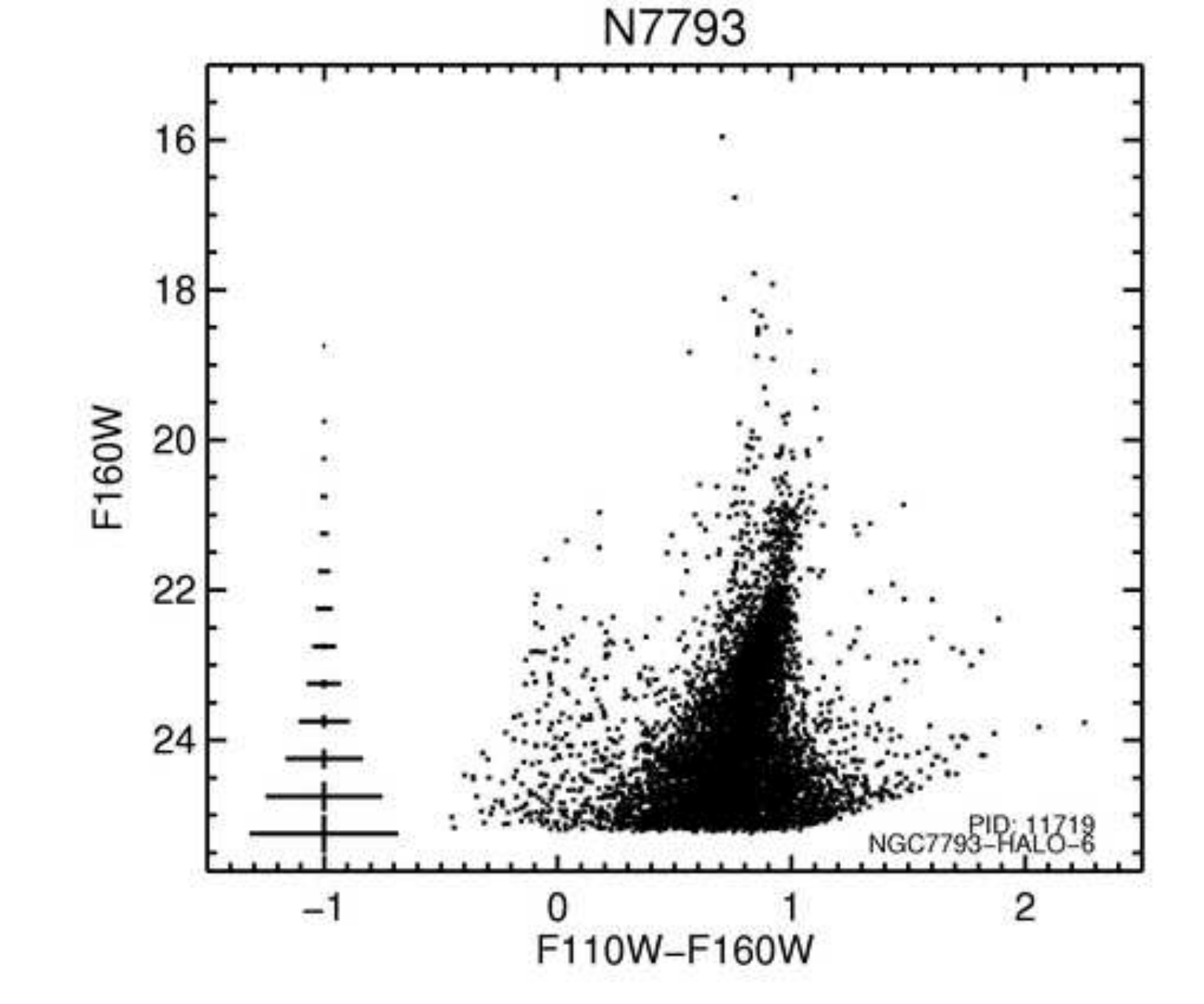}
\includegraphics[width=3.25in]{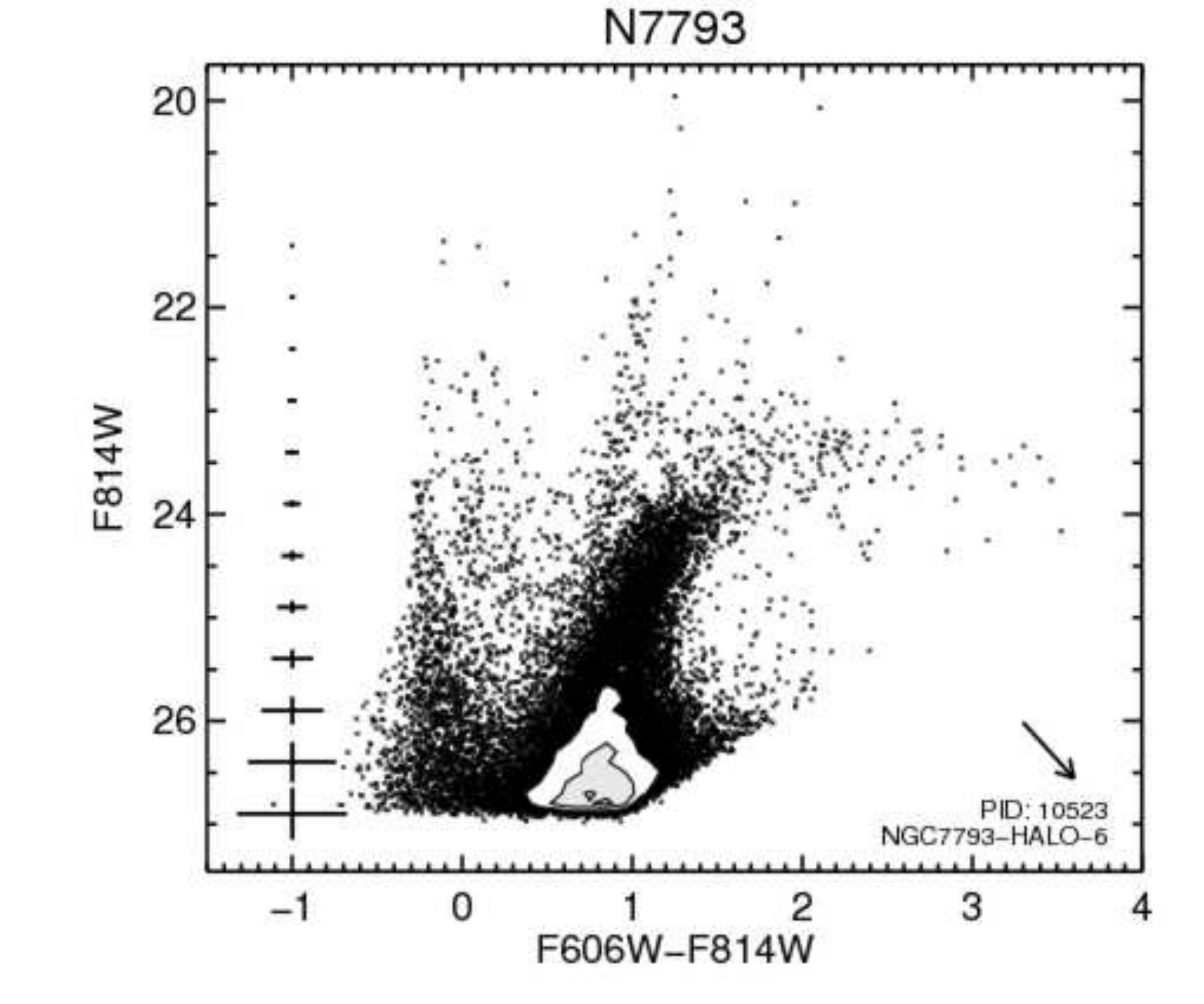}
}
\centerline{
\includegraphics[width=3.25in]{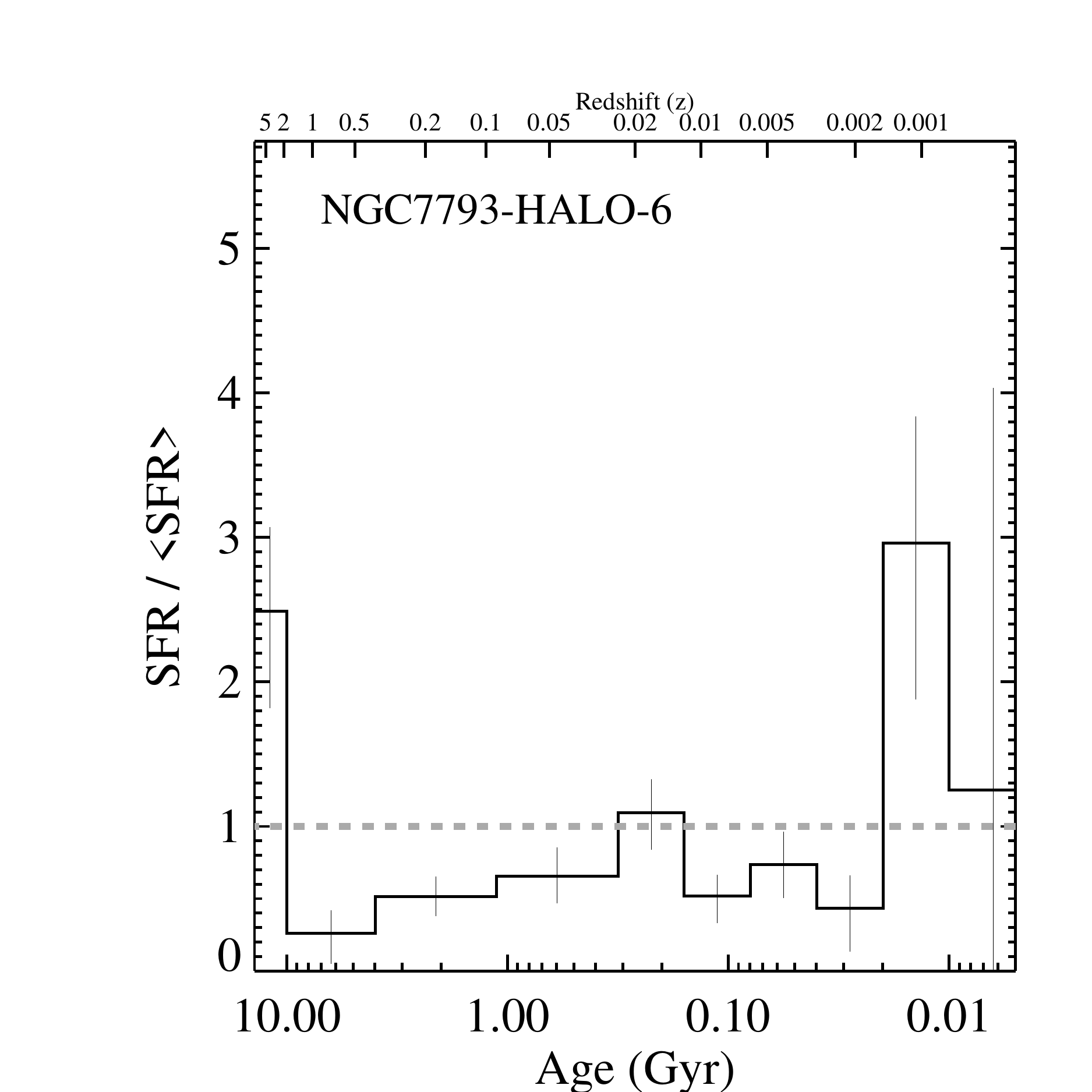}
\includegraphics[width=3.25in]{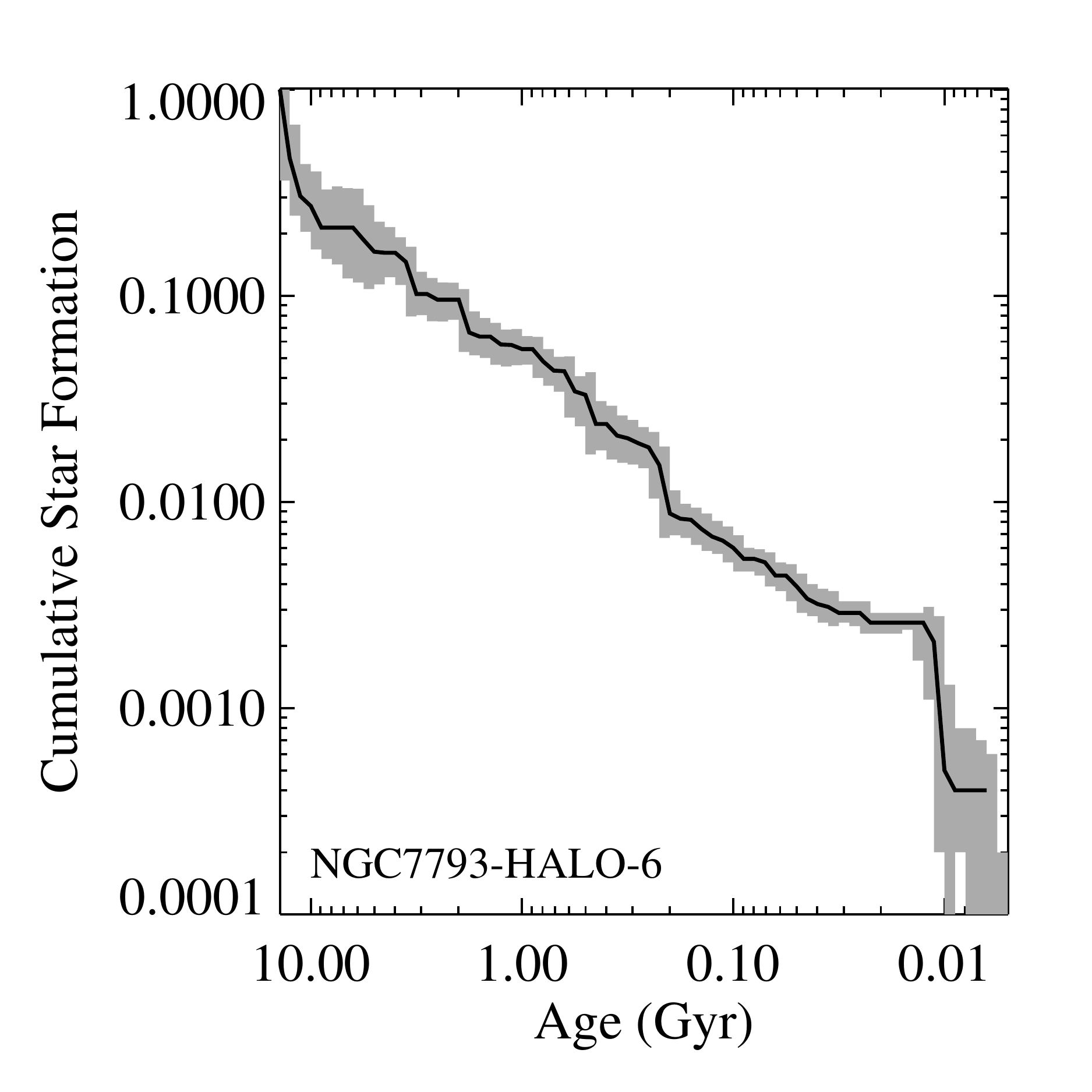}
}
\caption{ Color magnitude diagrams of the WFC3/IR (upper left) and
  optical (upper right) for the target NGC7793-HALO-6 within galaxy
  N7793.  Lower panels show the star formation history derived from
  the optical data, for both the differential (left, with horizontal
  dotted line indicating the past average SFR) and cumulative (right)
  star formation histories.  The cumulative star formation history is
  calculated from the present back to 14\,Gyrs.  Uncertainties in the
  lower two panels are the 68\% confidence intervals, calculated from
  Monte Carlo tests including random and systematic uncertainties.
  Optical CMDs are restricted to the area covered by the WFC3 FOV. }
\end{figure}
\vfill
\clearpage
 
\begin{figure}
\figurenum{\ref{cmdfig} continued}
\centerline{
\includegraphics[width=3.25in]{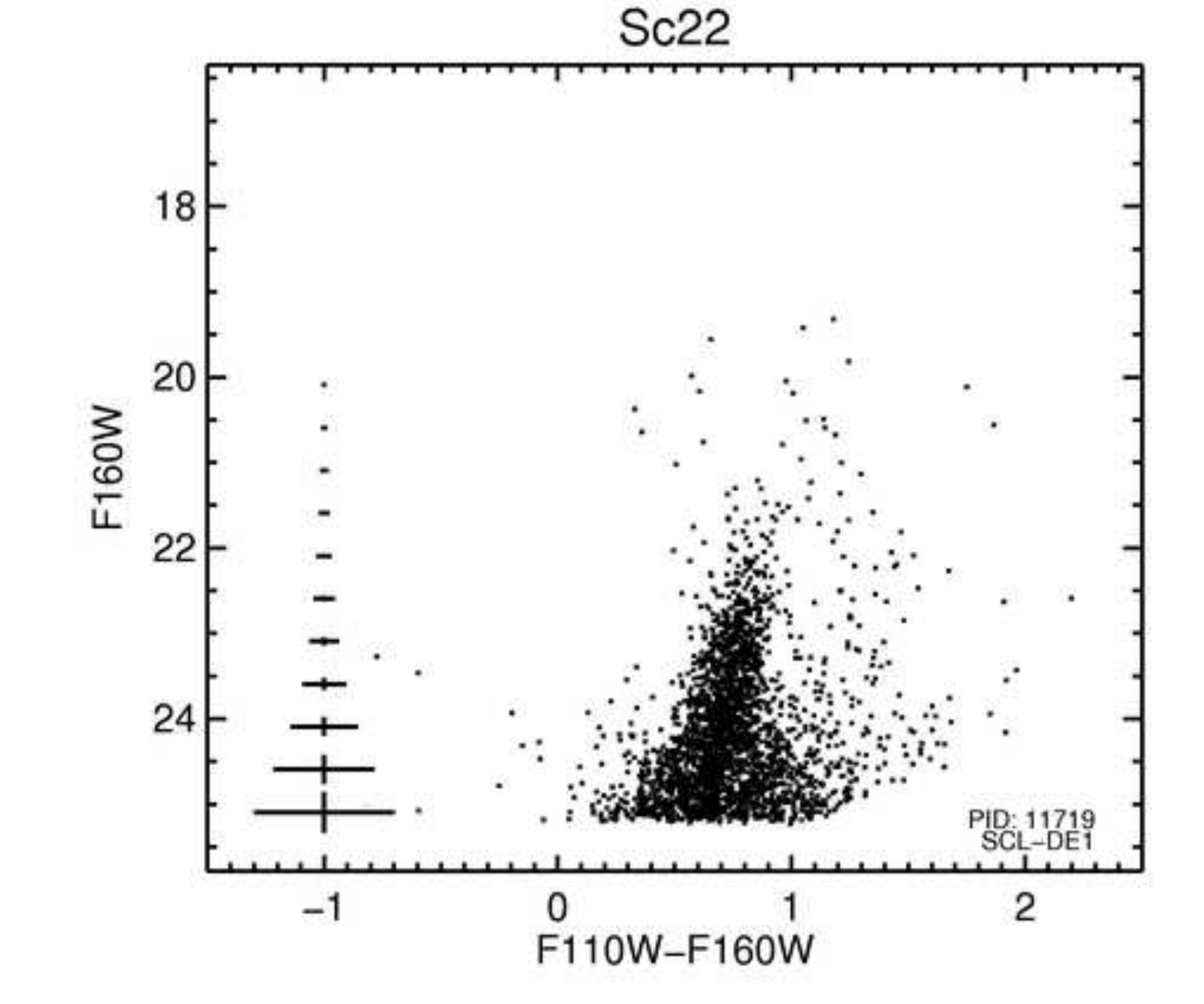}
\includegraphics[width=3.25in]{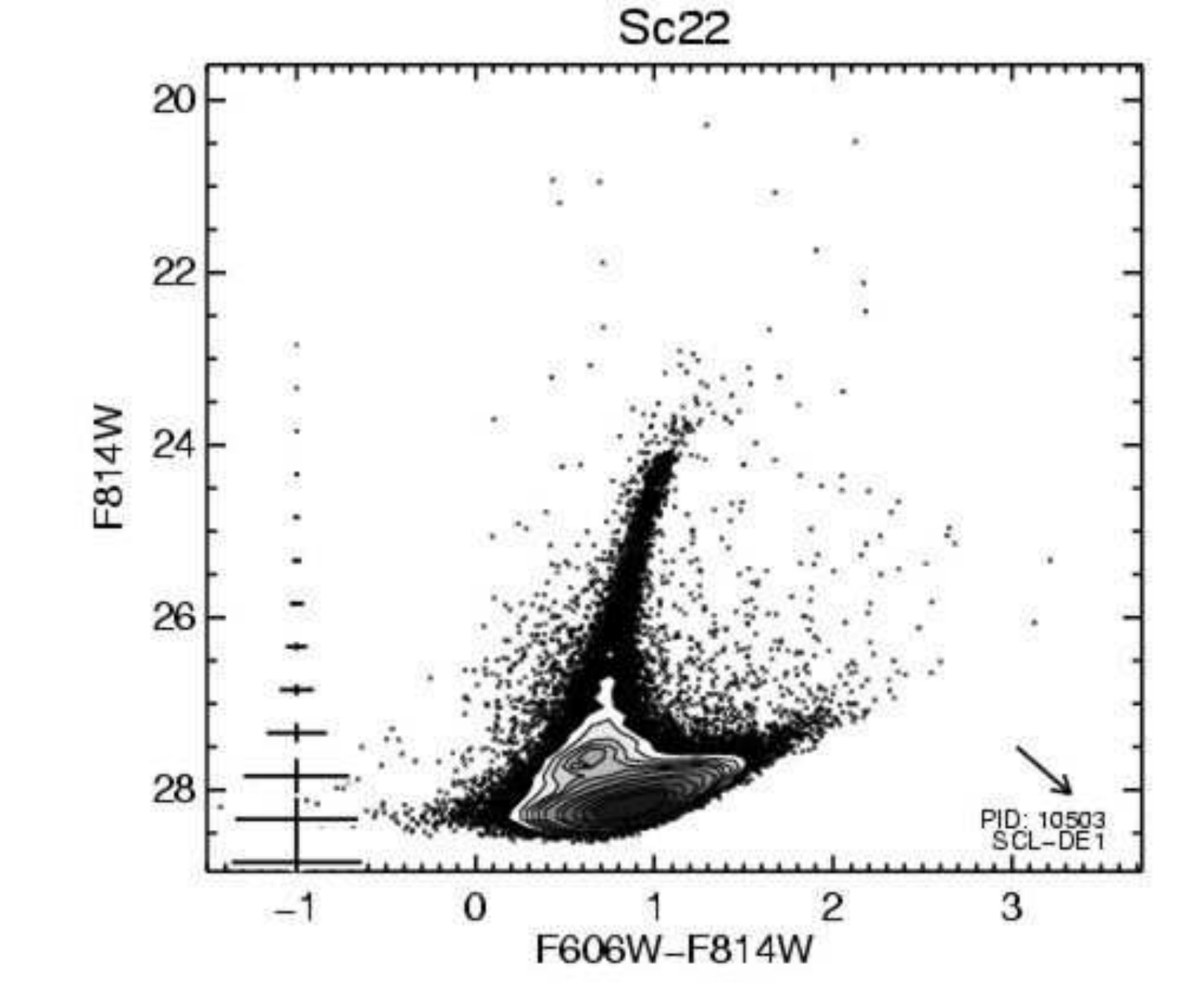}
}
\centerline{
\includegraphics[width=3.25in]{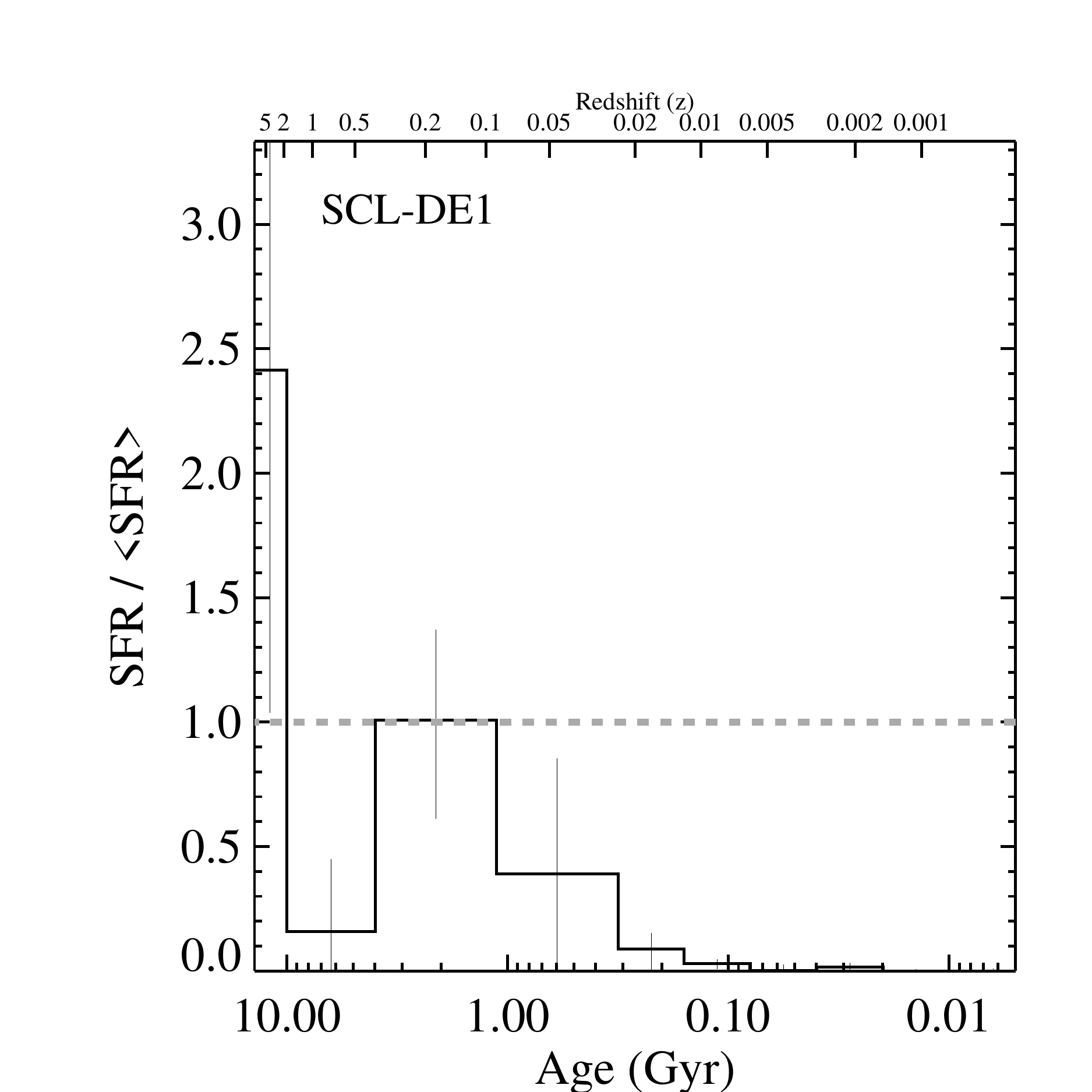}
\includegraphics[width=3.25in]{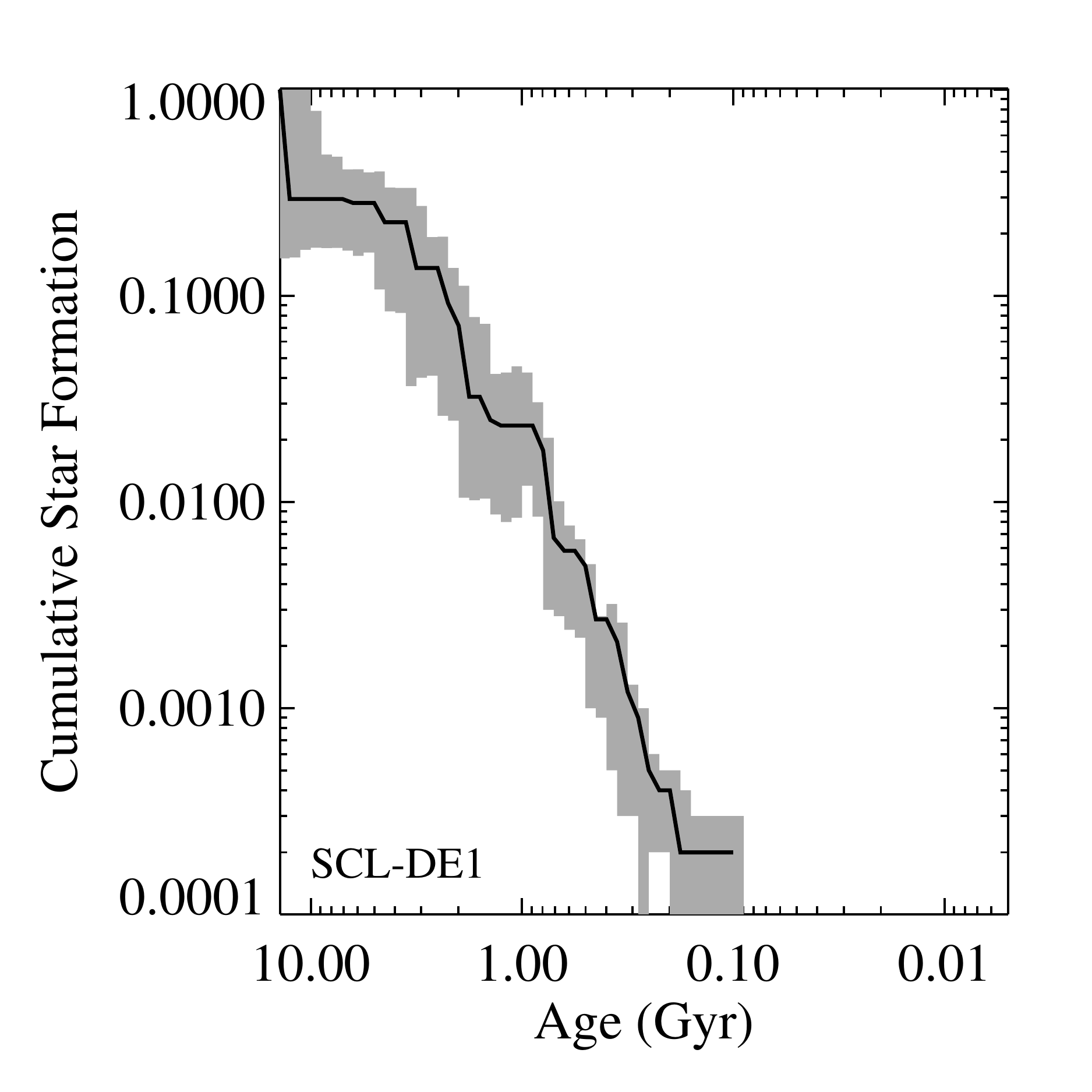}
}
\caption{ Color magnitude diagrams of the WFC3/IR (upper left) and
  optical (upper right) for the target SCL-DE1 within galaxy Sc22.
  Lower panels show the star formation history derived from the
  optical data, for both the differential (left, with horizontal
  dotted line indicating the past average SFR) and cumulative (right)
  star formation histories.  The cumulative star formation history is
  calculated from the present back to 14\,Gyrs.  Uncertainties in the
  lower two panels are the 68\% confidence intervals, calculated from
  Monte Carlo tests including random and systematic uncertainties.
  Optical CMDs are restricted to the area covered by the WFC3 FOV. }
\end{figure}
\vfill
\clearpage
 
\begin{figure}
\figurenum{\ref{cmdfig} continued}
\centerline{
\includegraphics[width=3.25in]{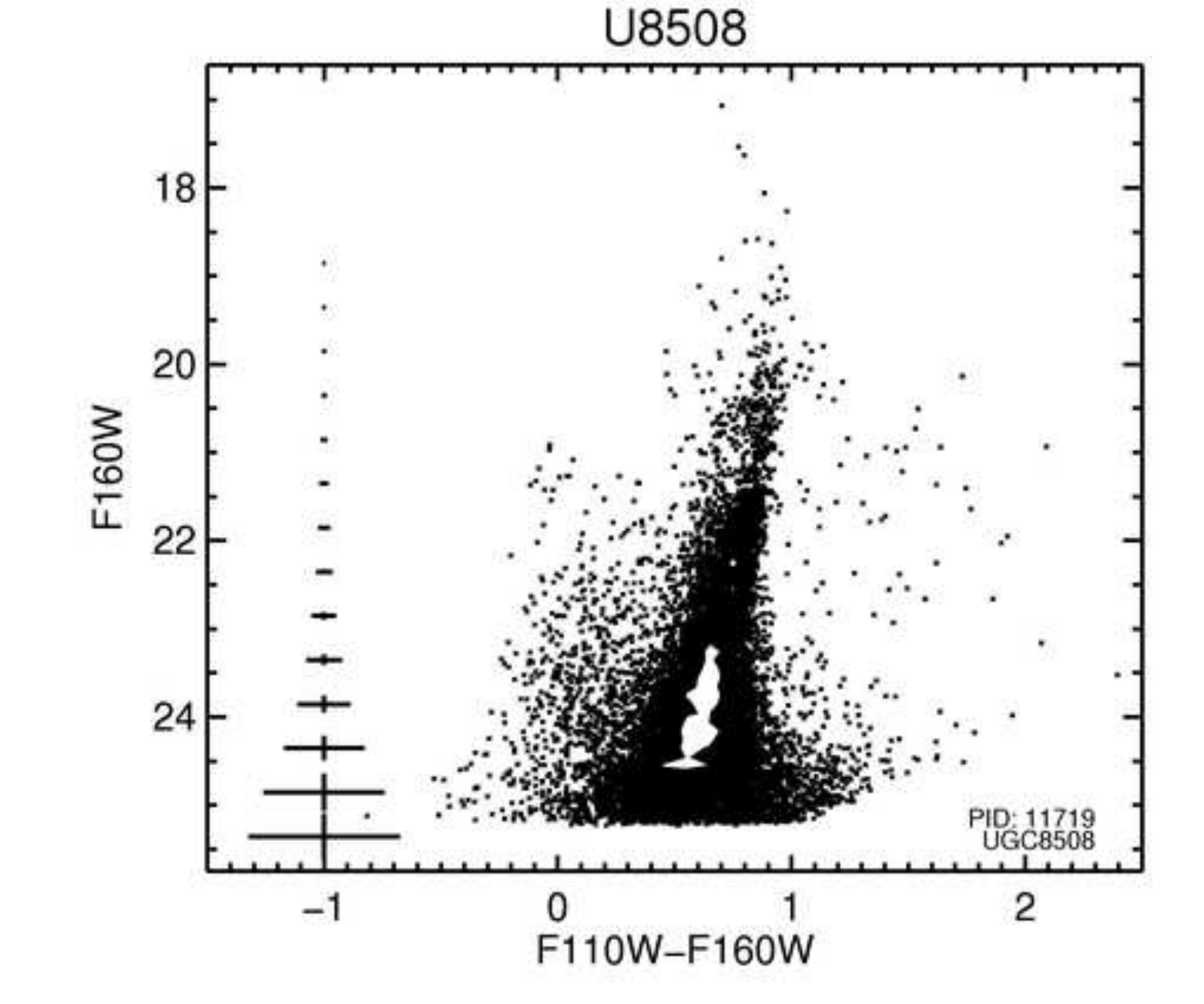}
\includegraphics[width=3.25in]{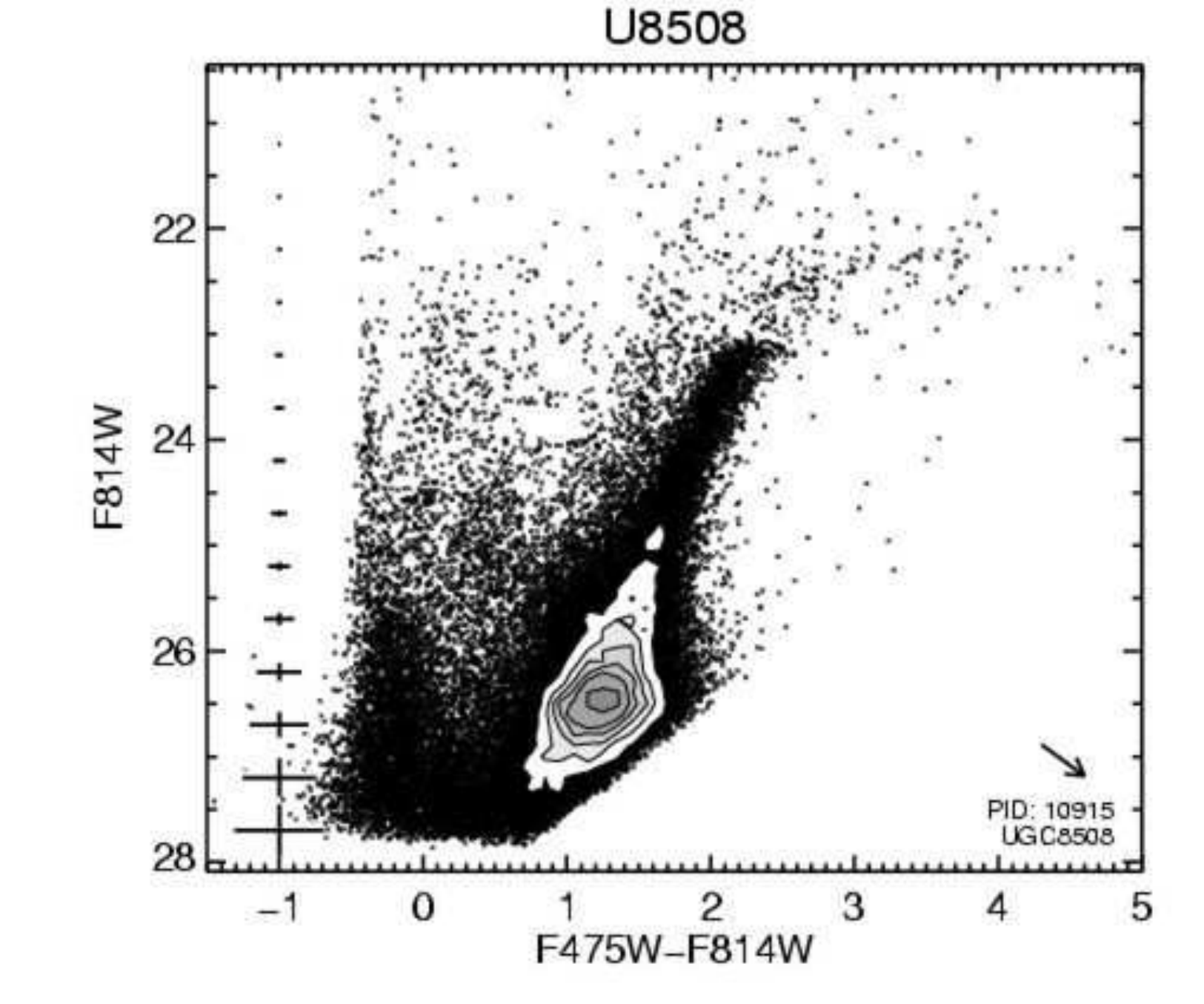}
}
\centerline{
\includegraphics[width=3.25in]{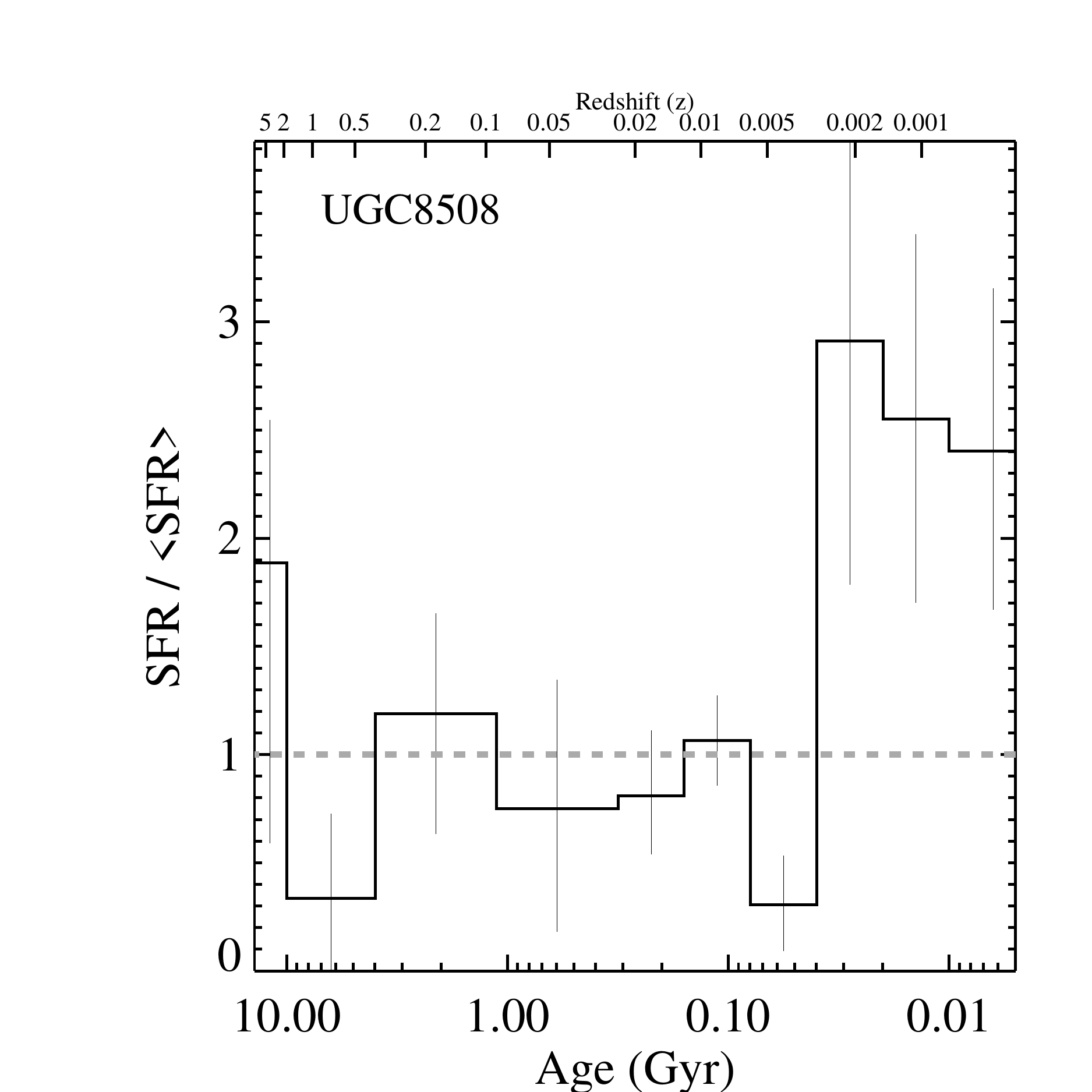}
\includegraphics[width=3.25in]{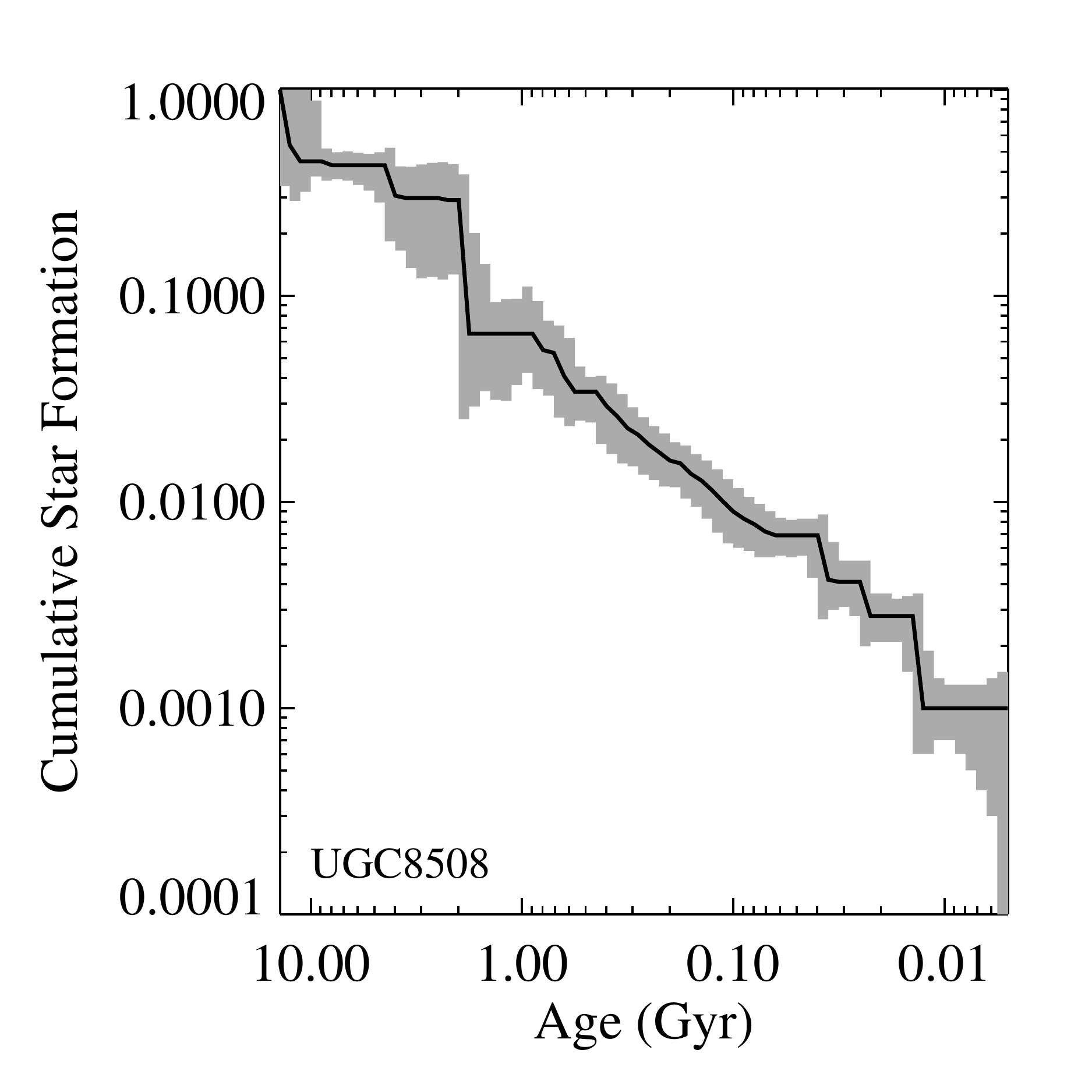}
}
\caption{ Color magnitude diagrams of the WFC3/IR (upper left) and
  optical (upper right) for the target UGC8508 within galaxy U8508.
  Lower panels show the star formation history derived from the
  optical data, for both the differential (left, with horizontal
  dotted line indicating the past average SFR) and cumulative (right)
  star formation histories.  The cumulative star formation history is
  calculated from the present back to 14\,Gyrs.  Uncertainties in the
  lower two panels are the 68\% confidence intervals, calculated from
  Monte Carlo tests including random and systematic uncertainties.
  Optical CMDs are restricted to the area covered by the WFC3 FOV. }
\end{figure}
\vfill
\clearpage
 
\begin{figure}
\figurenum{\ref{cmdfig} continued}
\centerline{
\includegraphics[width=3.25in]{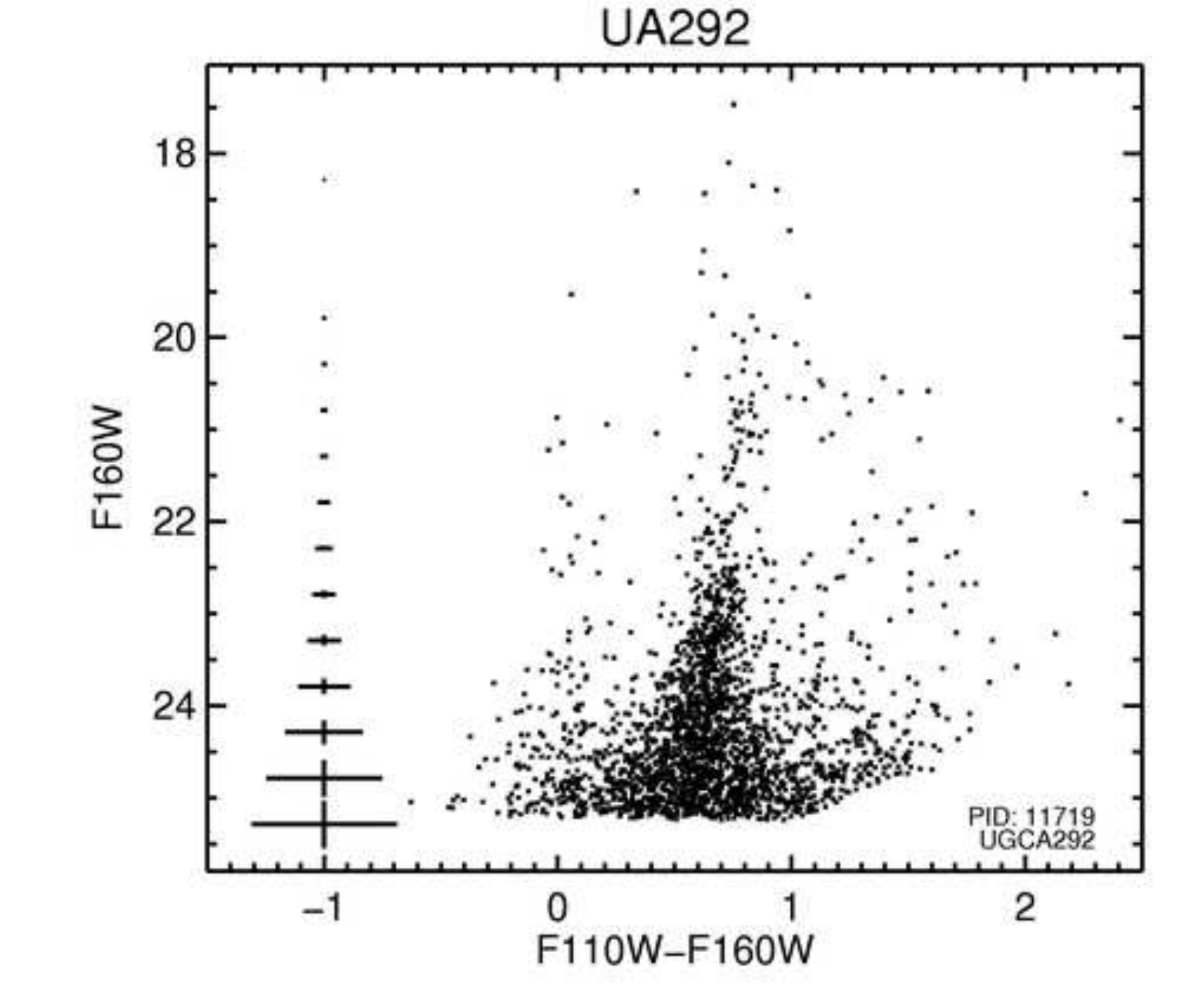}
\includegraphics[width=3.25in]{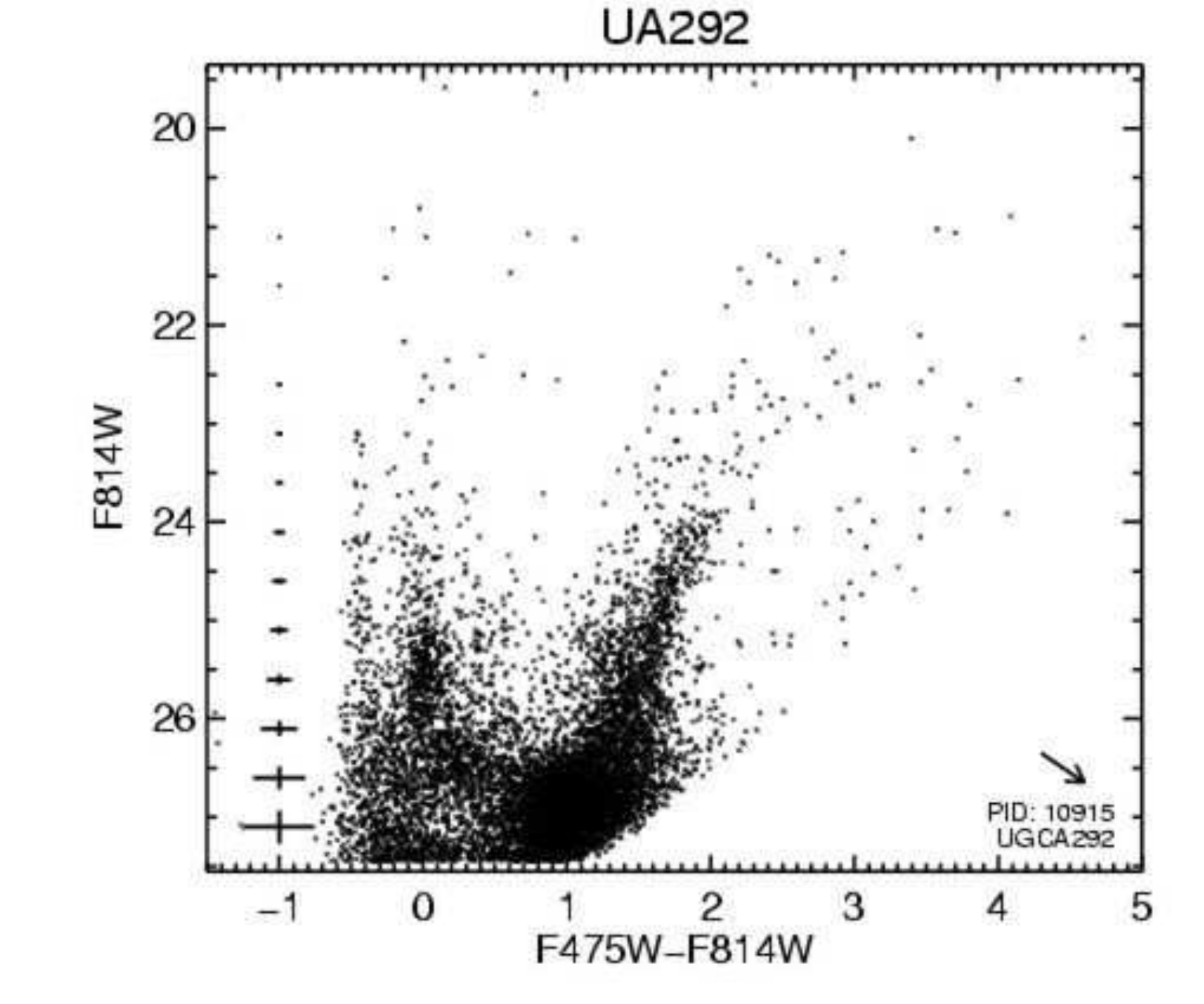}
}
\centerline{
\includegraphics[width=3.25in]{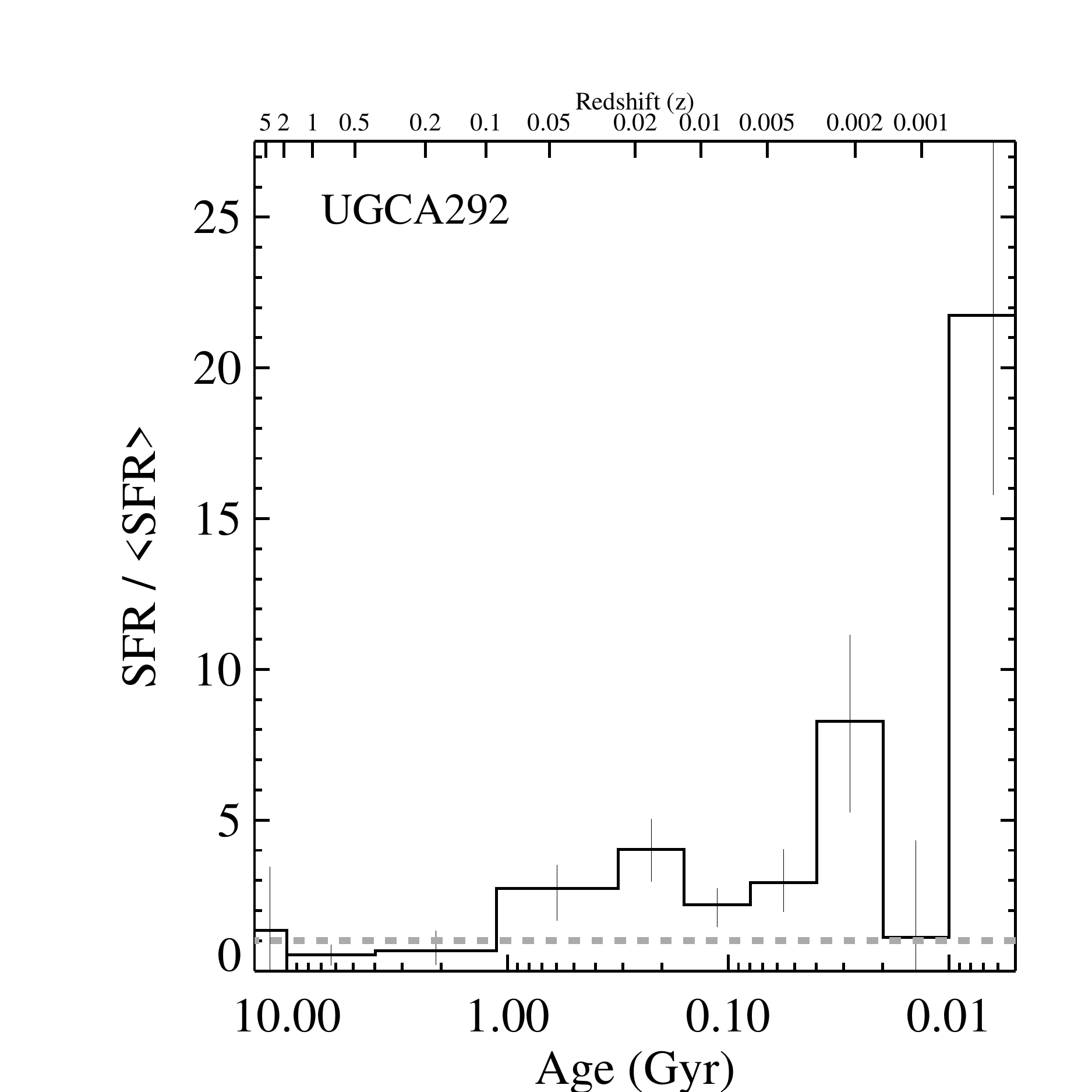}
\includegraphics[width=3.25in]{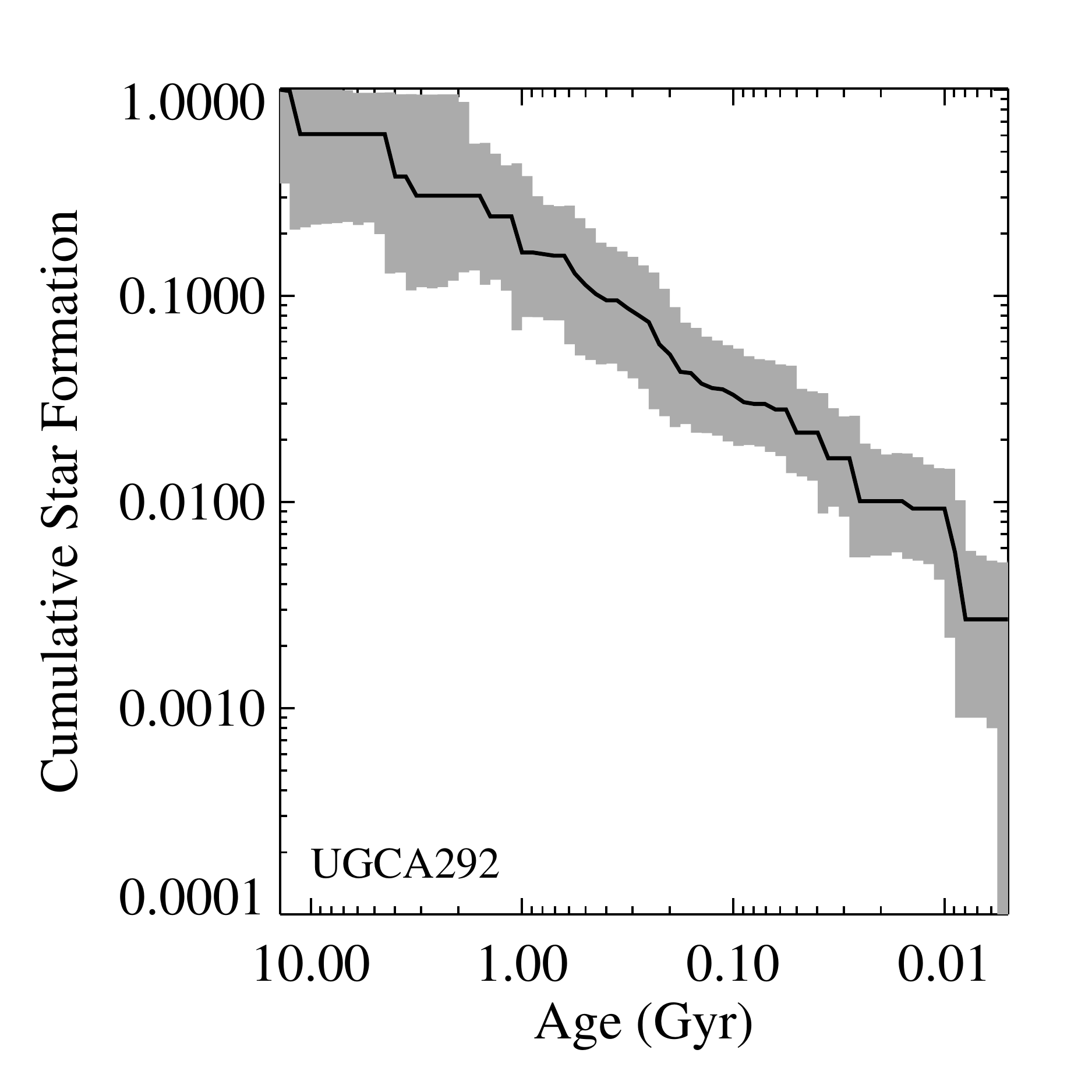}
}
\caption{ Color magnitude diagrams of the WFC3/IR (upper left) and
  optical (upper right) for the target UGCA292 within galaxy UA292.
  Lower panels show the star formation history derived from the
  optical data, for both the differential (left, with horizontal
  dotted line indicating the past average SFR) and cumulative (right)
  star formation histories.  The cumulative star formation history is
  calculated from the present back to 14\,Gyrs.  Uncertainties in the
  lower two panels are the 68\% confidence intervals, calculated from
  Monte Carlo tests including random and systematic uncertainties.
  Optical CMDs are restricted to the area covered by the WFC3 FOV. }
\end{figure}
\vfill
\clearpage

\begin{figure}
\centerline{
\includegraphics[width=3.25in]{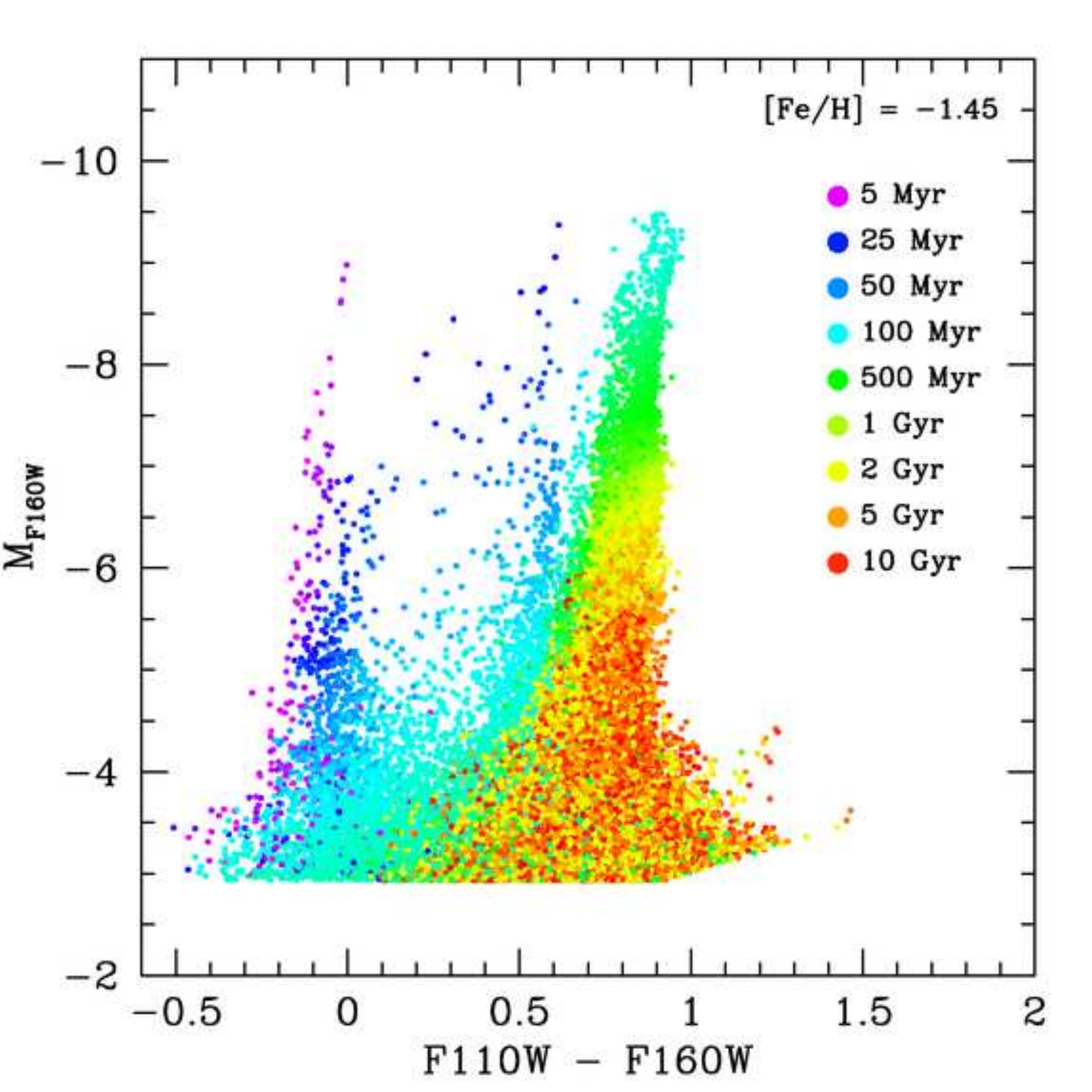}
\includegraphics[width=3.25in]{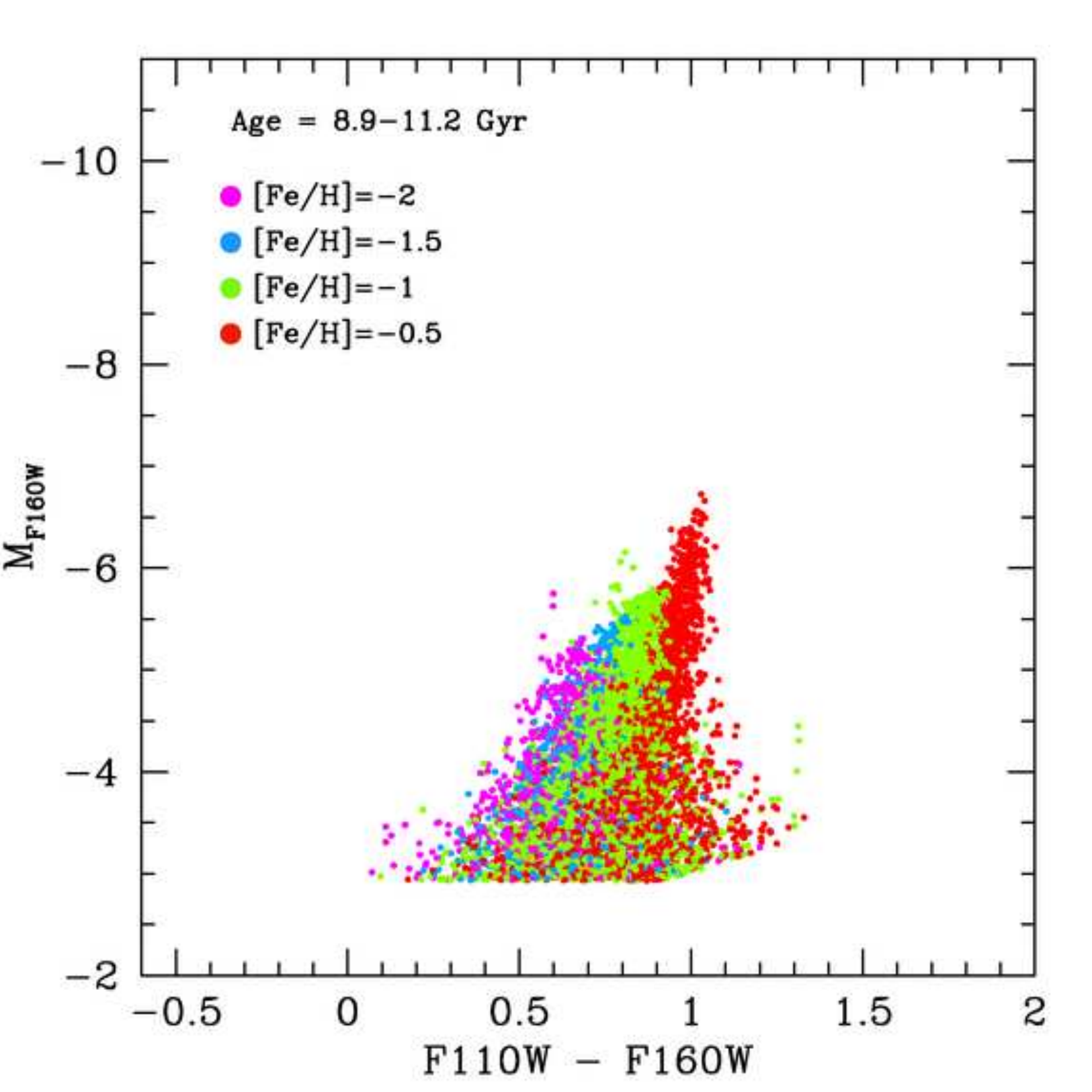}
}
\centerline{
\includegraphics[width=3.25in]{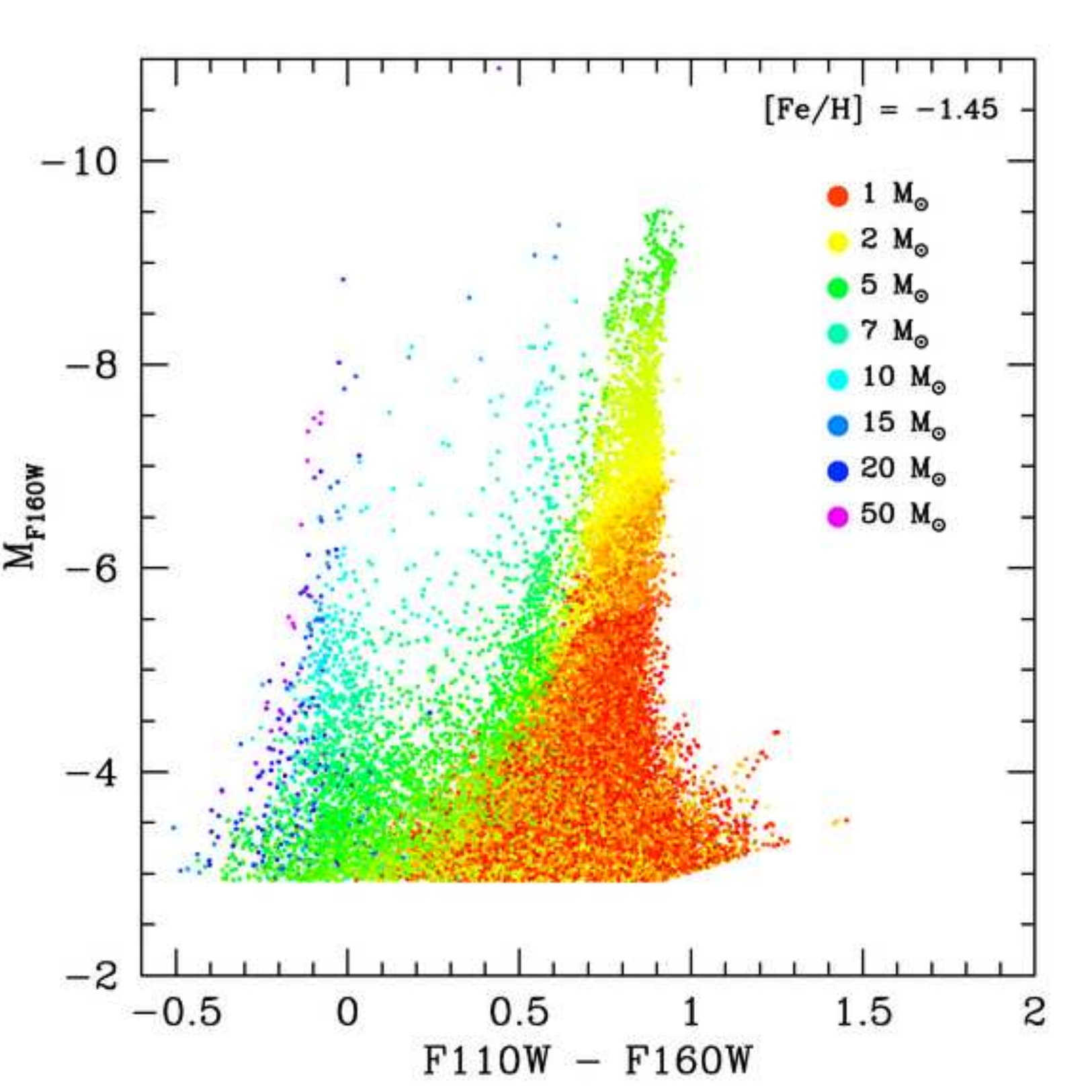}
\includegraphics[width=3.25in]{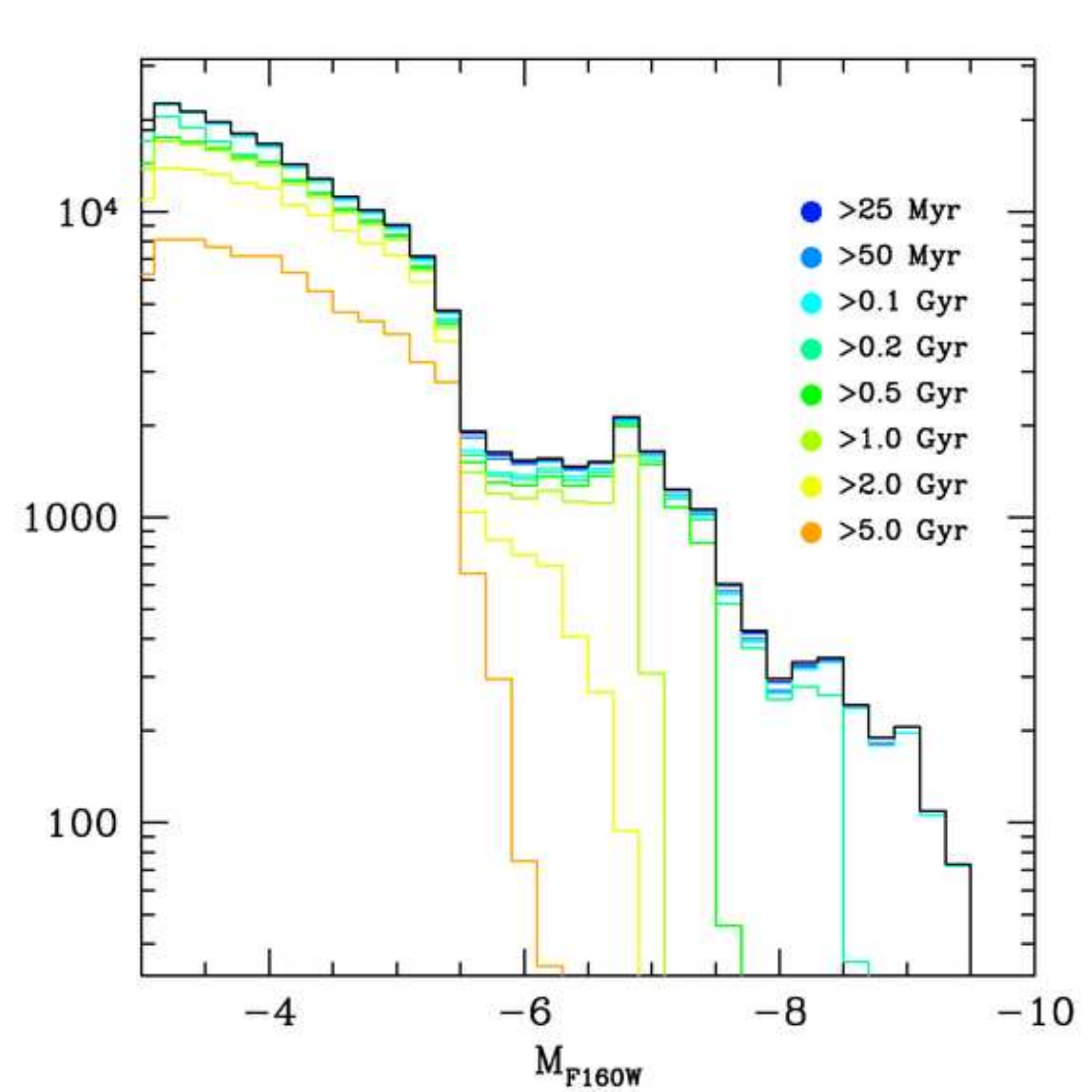}
}
\caption{Simulated CMDs showing the location of populations
  color-coded by age (upper left), initial stellar mass (lower left),
  and metallicity (upper right).  The plots on the left and bottom
  right assume a constant star formation rate and metallicity of
  [Fe/H]=-1.45.  The plot on the upper right assumes a constant star
  formation rate over a narrow age interval (8.9-11.2$\Gyr$), for a
  range of metallicities.  The plot on the lower right shows the total
  luminosity function, along with the contributions of stars older
  than a given stellar age, for stars and ages color coded as in the
  CMD on the upper left.  Stars younger than $2\Gyr$ contribute
  significant numbers of NIR bright stars.  Simulations use Padova
  isochrones with updated AGB models from \citet{girardi2010}, and
  assume photometric errors based on the artificial star tests for the
  M81-DEEP field; as we show in \citet{melbourne2011}, the current
  implementation of these models underestimate the constribution from
  red core Helium burning stars compared to the
  data. \label{fakecmdfig}}
\end{figure}
\vfill
\clearpage

\begin{figure}
\centerline{
\includegraphics[width=6.25in]{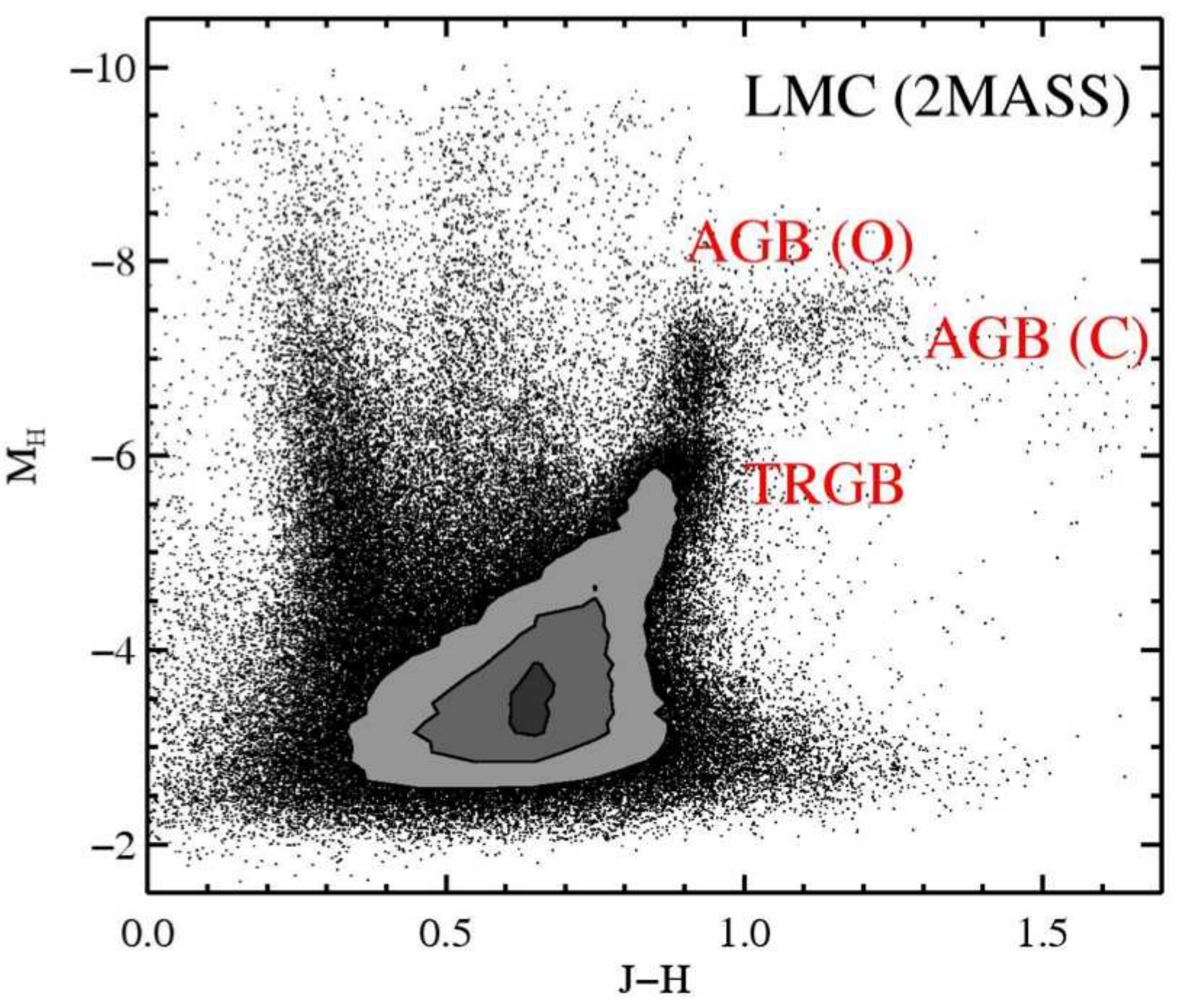}
}
\caption{
$J-H$ vs $M_H$ 2MASS CMD for stars in the LMC.  The sequence extending
up from the TRGB at $M_H=-6$ is dominated by oxygen-rich AGB stars.
The horizontal sequence at $M_H\sim-7.3$ extending redward of
$J-H\sim0.9$ contains mostly carbon-rich AGB stars.  The main sequence
is the bluest vertical sequence at $J-H\sim0.25$.  The vertical sequence
at $J-H\sim0.5$ is dominated by MW foreground stars.
\label{lmcfig}}
\end{figure}
\vfill
\clearpage

\begin{figure}
\centerline{
\includegraphics[width=6.25in]{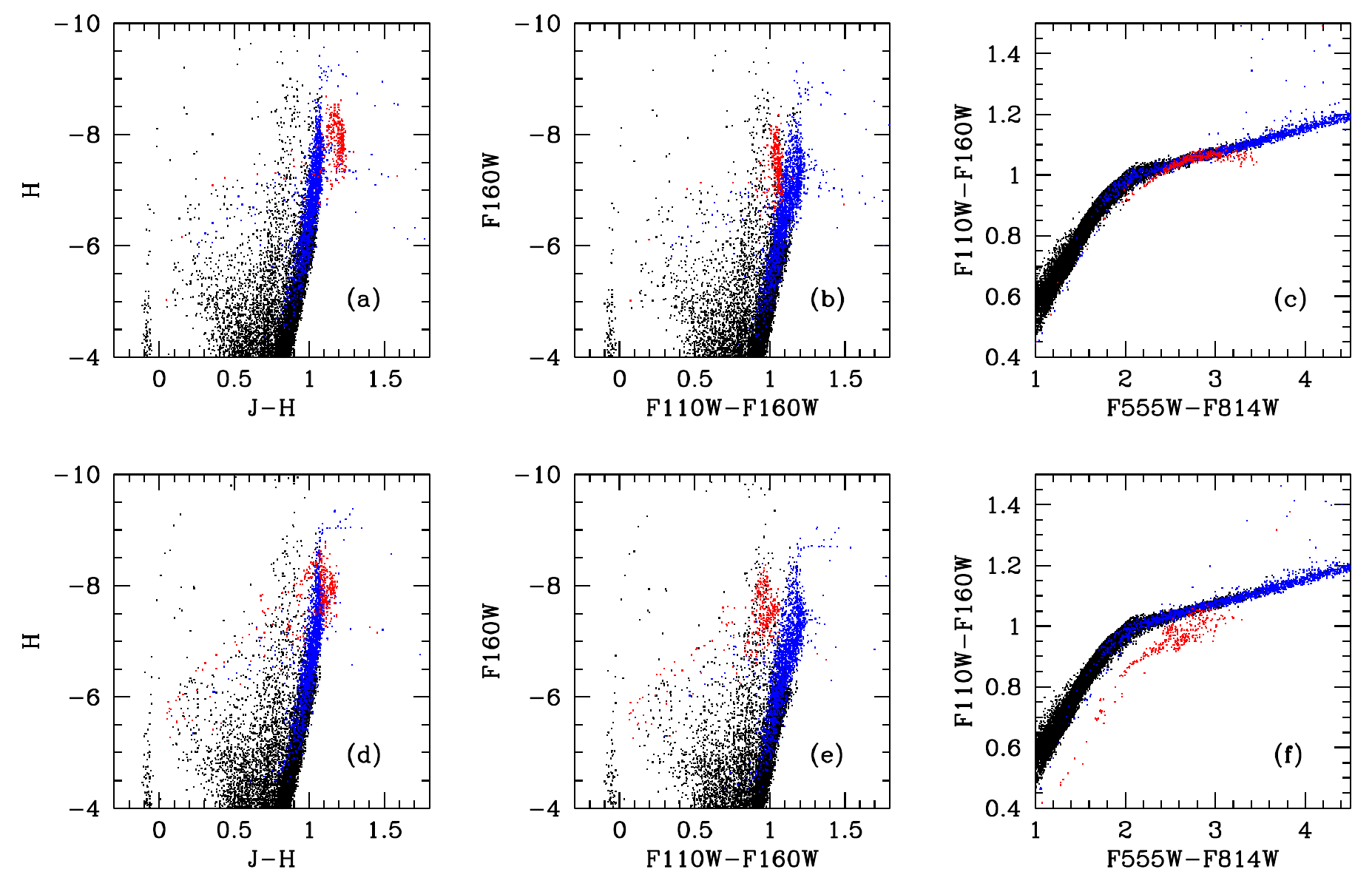}
}
\caption{
Simulated photometry for a model galaxy with a constant SFR and
increasing metallicity (reaching $Z_\odot$ at the present day), made
with the TRILEGAL code \citep{Girardi_etal05, GirardiMarigo07}. The
TP-AGB population is color-coded by blue/red dots for O-rich/C-rich
stars, respectively. The colors and magnitudes of C stars are derived
from the model atmospheres by \citet{loidl2001} (top row, panels a-c)
or by \citet{aringer2009} (bottom row, panels d-f). The colors of cool
O-rich giants come from \citet{fluks} model atmospheres. The left
panels (a and d) show the expected $H$ vs. $J-H$ diagram for 2MASS
filters, where C stars are found (as expected) to be slightly redder
than O-rich stars. In the middle panels (b and e), however, C stars
become bluer than the bulk of O-rich stars for the WFC3/IR $F160W$
vs. $F110W-F160W$ filter set, especially for the \citet{aringer2009}
models. The right panels (c and f) show the optical-IR color-color
plots, comparable to Figure~\ref{colorcolorfig}.  For both models the
the simulations include extended tails of blue and faint TP-AGB stars,
that correspond to the $\sim15$~\% of TP-AGB stars caught in their
low-luminosity dips along thermally-pulsing cycles, as well as the
tails of extremely red stars at the late stages of high mass loss.
\label{carbonfig}}
\end{figure}
\vfill

\begin{figure}
\centerline{
\includegraphics[width=3.25in]{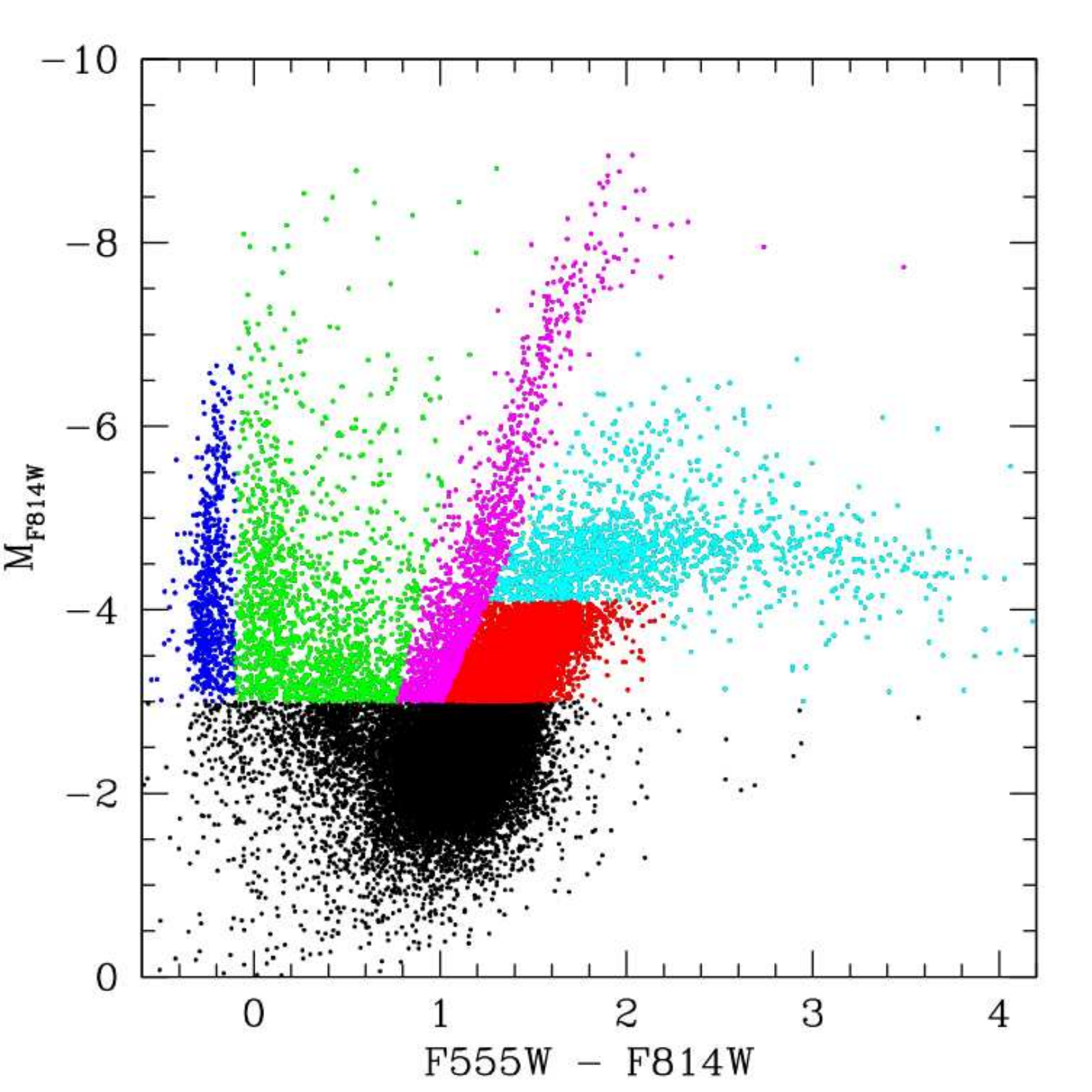}
\includegraphics[width=3.25in]{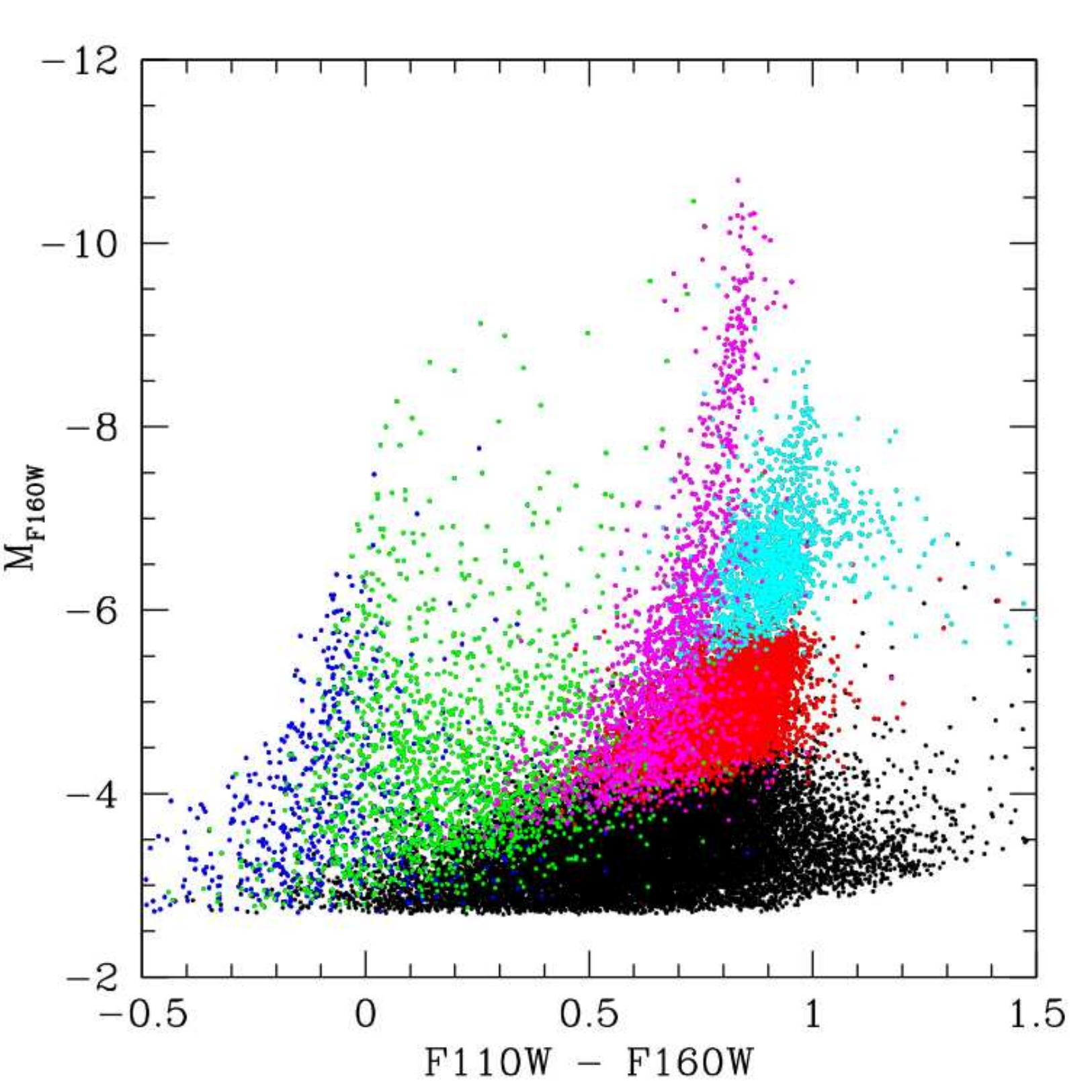}
}
\caption{ Comparison of CMD features in the optical (left) and the NIR
  (right) for IC2574-SGS. Points are color-coded according to their likely
  evolutionary phase (Blue = main sequence; Green = blue core Helium
  burning; Magenta = red core Helium burning; Red = red giant branch;
  Cyan = asymptotic giant branch), as deduced from the optical CMD,
  for stars with $M_{F814W}<-2.5$ that are positionally well-matched
  to stars in the NIR; unmatched or fainter stars are plotted in
  black.  
\label{matchedfig}}
\end{figure}
\vfill
\clearpage

\begin{figure}
\centerline{
\includegraphics[width=6.25in]{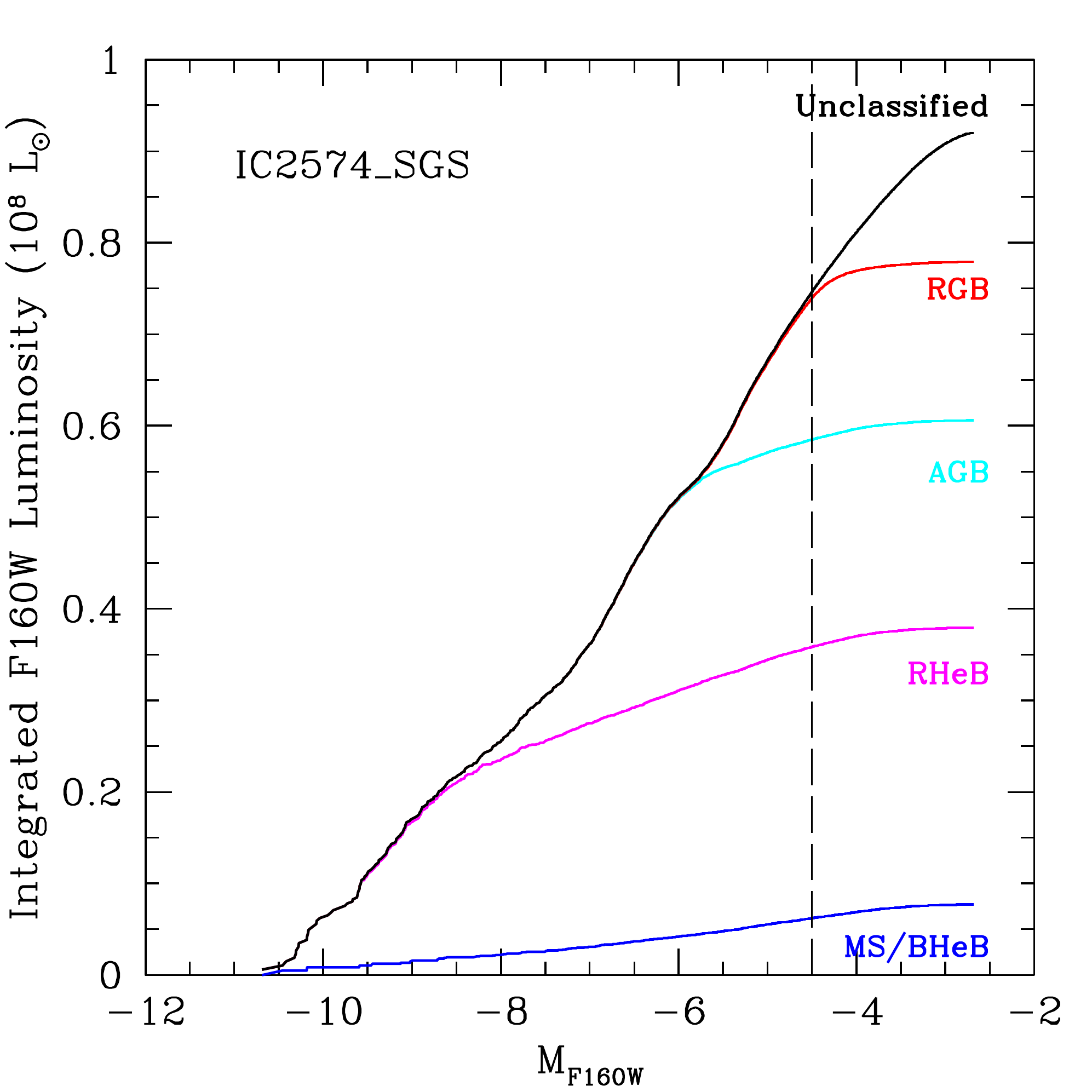}
}
\caption{
Integrated $F160W$ luminosity for IC2574-SGS, integrating from the
bright to the faint end, for stars classified as in
Figure~\ref{matchedfig}.  Each line contains the integrated total
luminosity for stars of a given evolutionary phase or younger, such
that the ``AGB'' line includes the luminosity in RHeB and MS/BHeB
stars as well.  ``Unclassified'' stars are those that were not
reliably matched to stars in the optical catalog.  Note that the
contribution from RHeB stars dominates the NIR luminosity for the
detected stars. The vertical line indicates the approximate
completeness limit of the data; RGB stars are likely to be the
dominant contributor to the luminosity below this limit, which will
reduce the fraction of the luminosity due to RHeB or AGB stars.  Note
that the ``RGB'' classification likely includes some contamination
from AGB and RHeB stars as well, since such stars cannot be reliably
separated within the red giant branch.
  \label{ic2574LFfig}}
\end{figure}
\vfill
\clearpage

\begin{figure}
\centerline{
\includegraphics[width=6.25in]{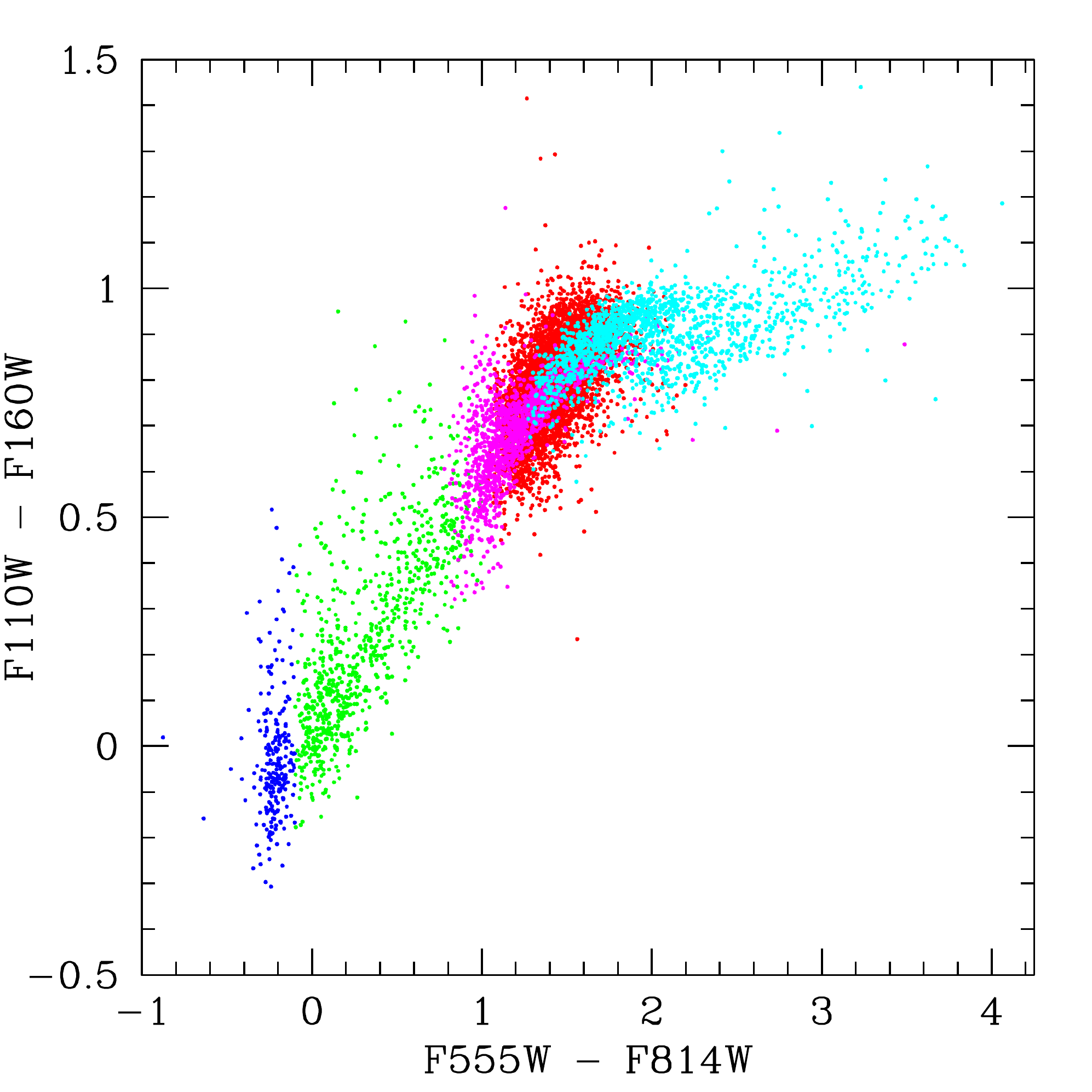}
}
\caption{ Optical vs infrared color for stars within IC2574-SGS,
  color-coded as in Figure~\ref{matchedfig} (MS=dark blue; BHeB=green;
  RHeB=magenta; RGB=red; AGB=cyan).  The AGB stars span a much wider
  range of optical colors, due to increased variability and reddening
  from circumstellar dust.  The NIR colors of the AGB fall in a tight
  sequence for optical colors bluer than $F555W-F814W\!\lesssim\!1.8$,
  but optically redder AGB stars show increased dispersion in their
  NIR colors. Only stars with magnitude uncertainties of less than 0.1
  mag are plotted. \label{colorcolorfig}}
\end{figure}
\vfill
\clearpage

\begin{figure}
\centerline{
\includegraphics[width=3.25in]{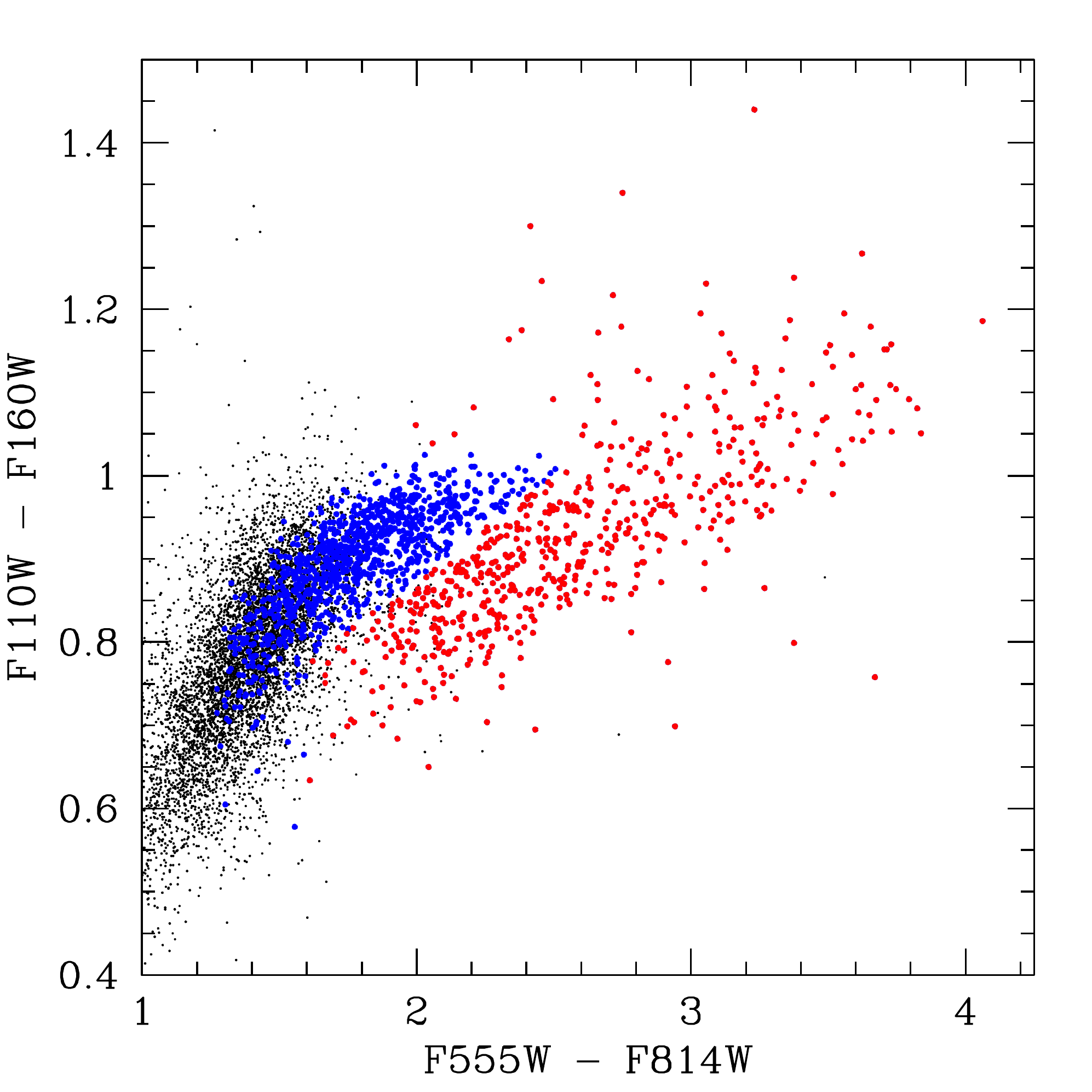}
\includegraphics[width=3.25in]{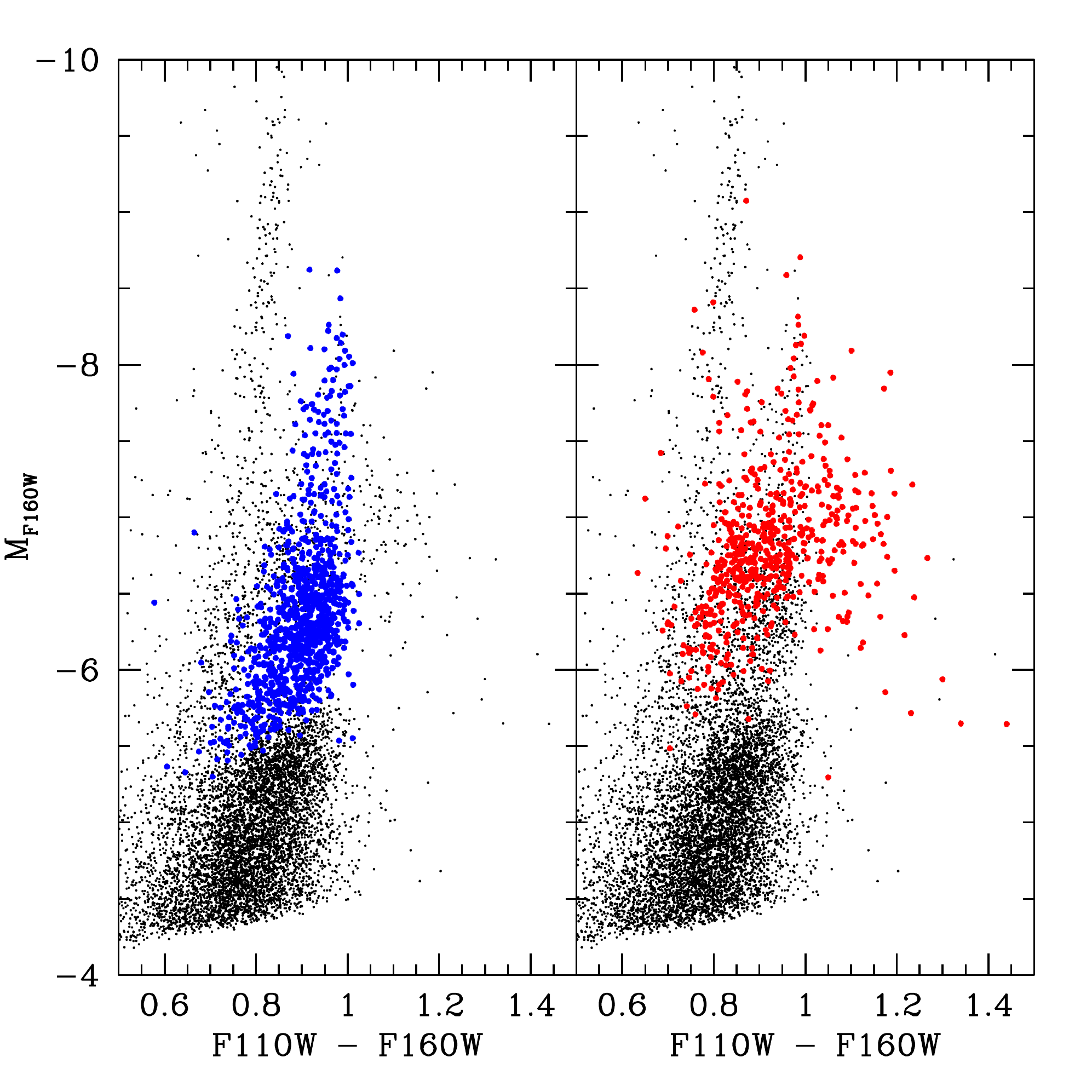}
}
\centerline{
\includegraphics[width=3.25in]{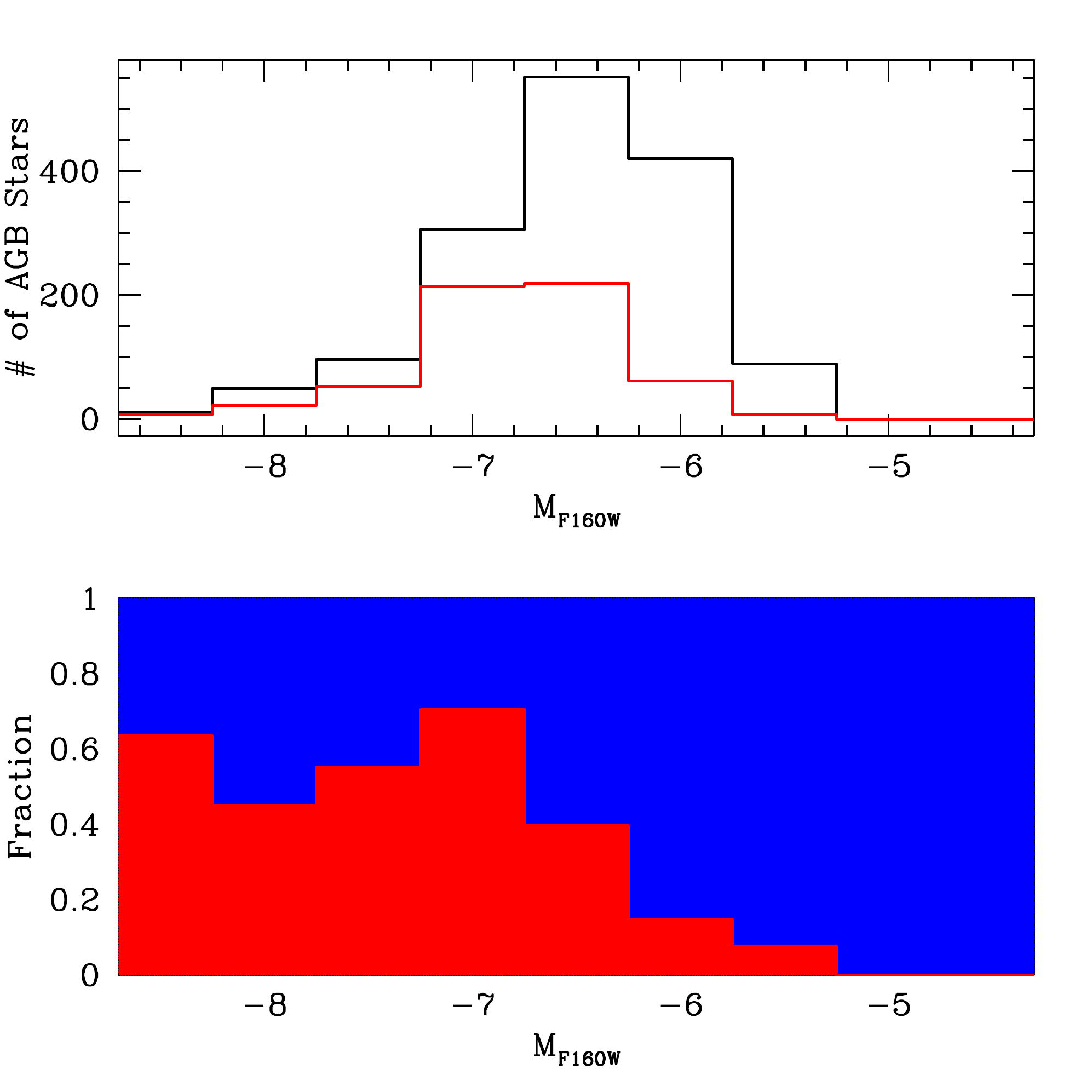}
\includegraphics[width=3.25in]{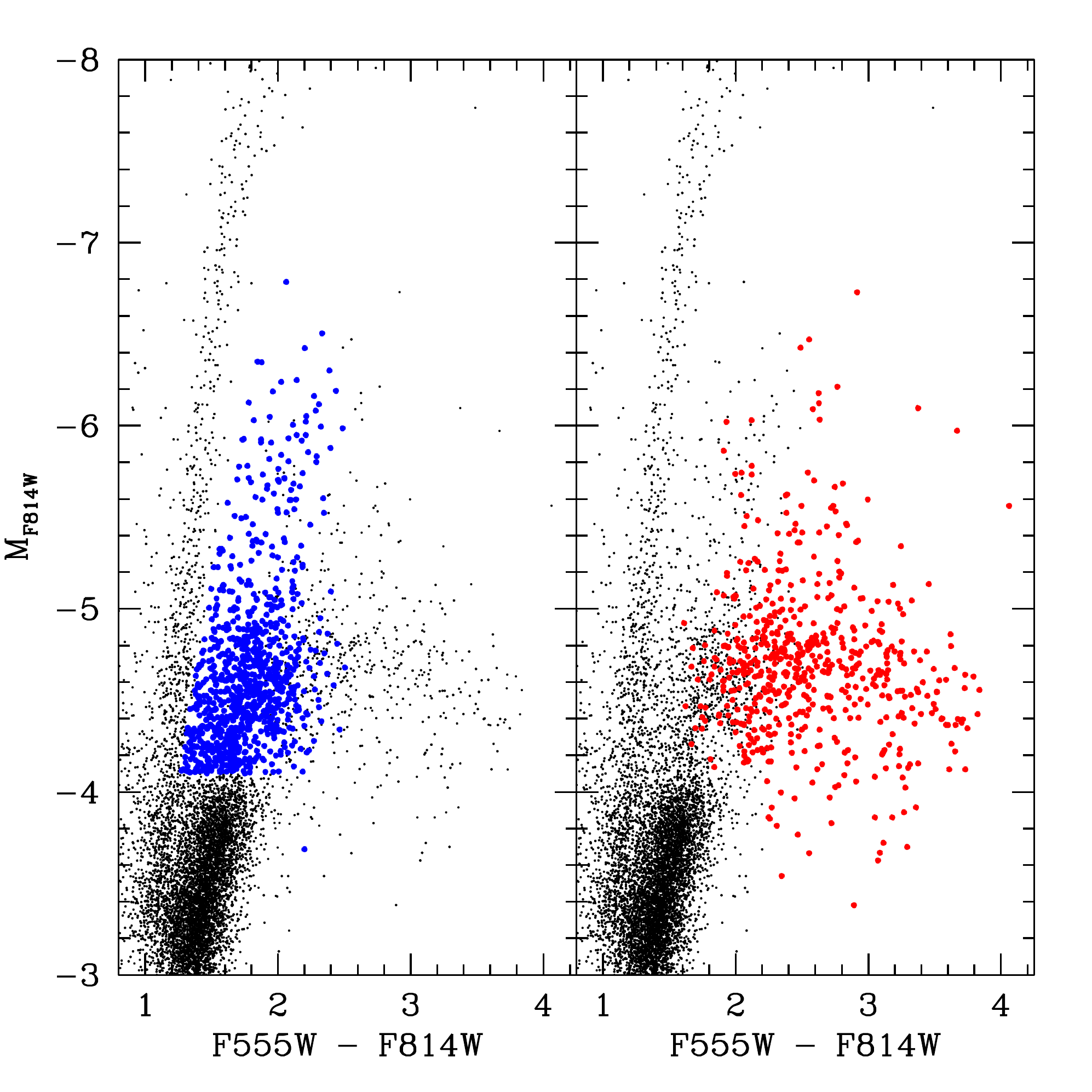}
}

\caption{ 
(Upper Left) Color-color diagram for AGB stars from
Figure~\ref{colorcolorfig}, further split into two groups in an
attempt to isolate the bluer, narrower sequence (blue points) from the
redder population with high dispersion in NIR colors (red points). The
NIR and optical CMDs of these two populations are shown on the upper
right and lower right, respectively.  The high dispersion selection
isolates the reddest carbon stars (by design), and also selects
against low luminosity AGB stars fainter than $M_{\rm{F160W}}=-6.5$
(histograms on lower left). The high dispersion sub-population is also
systematically bluer in $F110W-F160W$ at $F160W$ magnitudes between
-7 and -6.5.
\label{colorcoloragbfig}}
\end{figure}
\vfill
\clearpage

\begin{figure}
\centerline{
\includegraphics[width=2.25in]{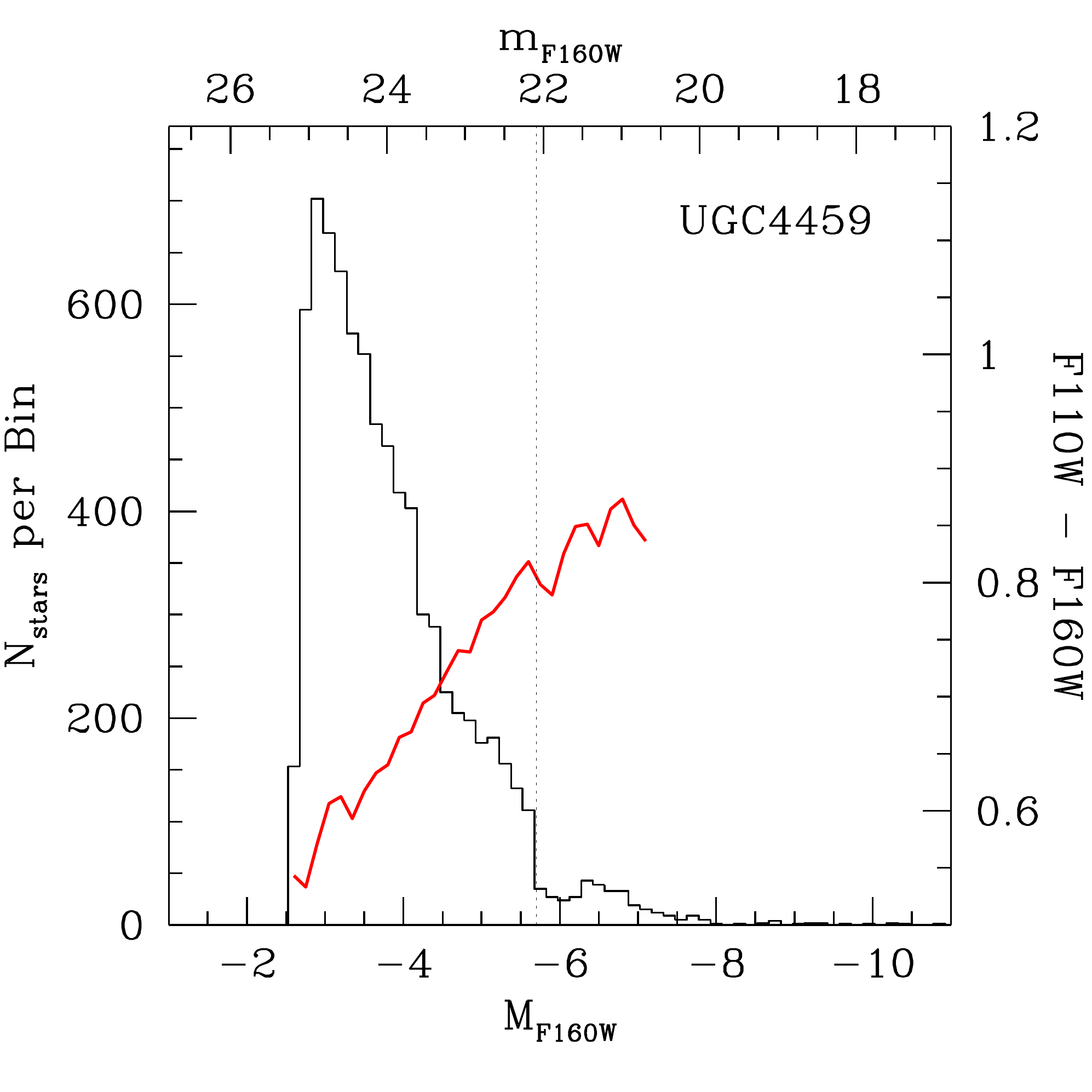}  
\includegraphics[width=2.25in]{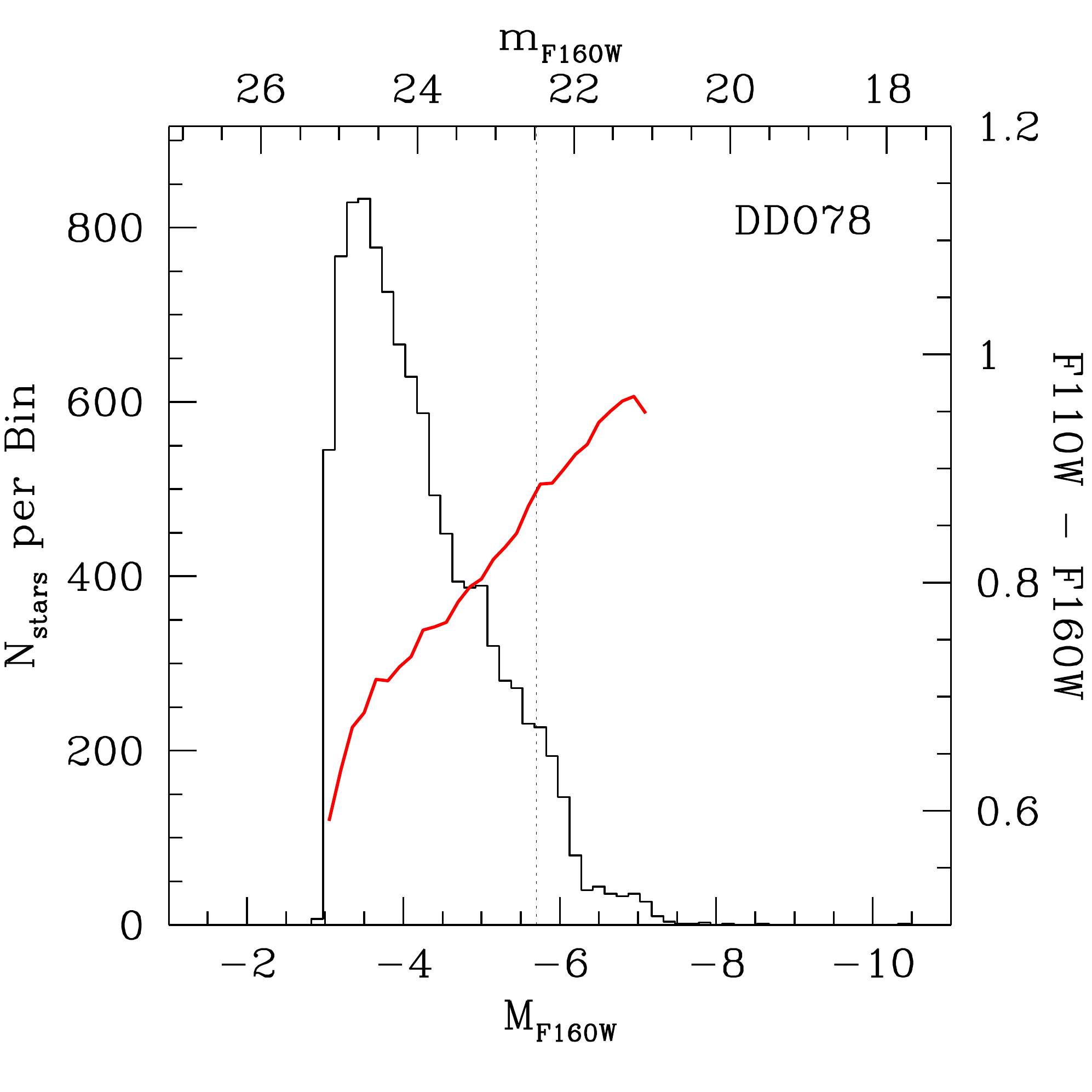}  
\includegraphics[width=2.25in]{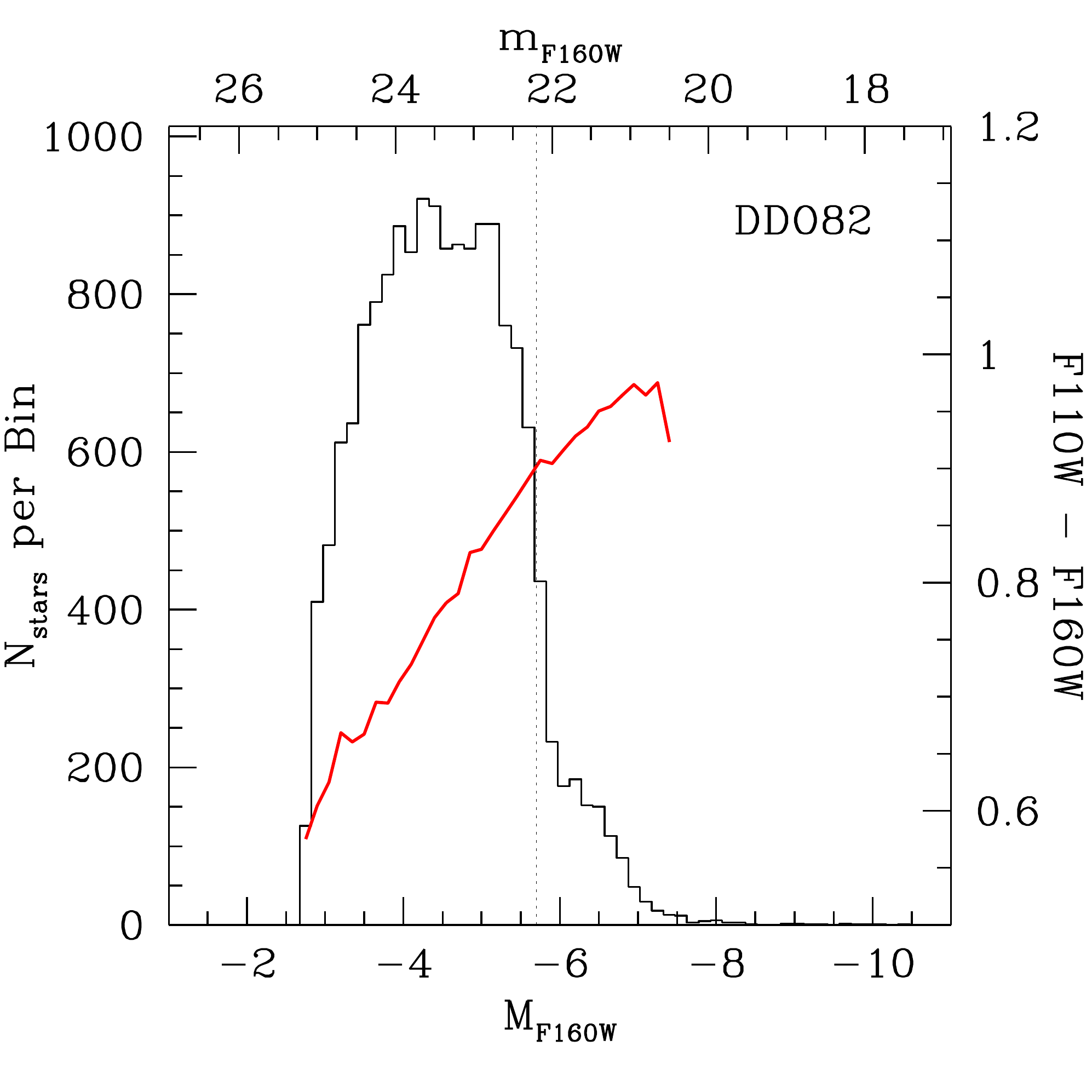}  
}
\centerline{
\includegraphics[width=2.25in]{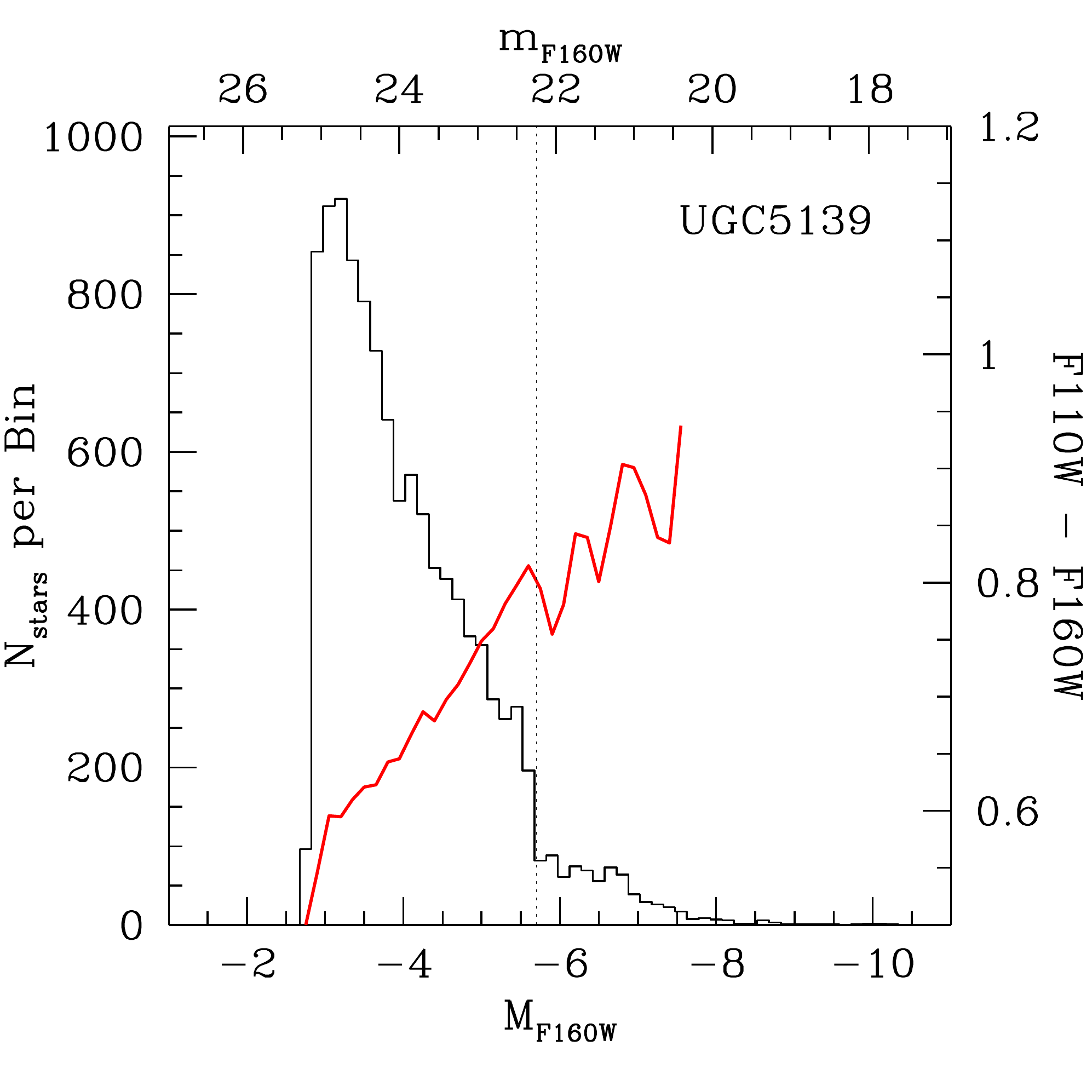}  
\includegraphics[width=2.25in]{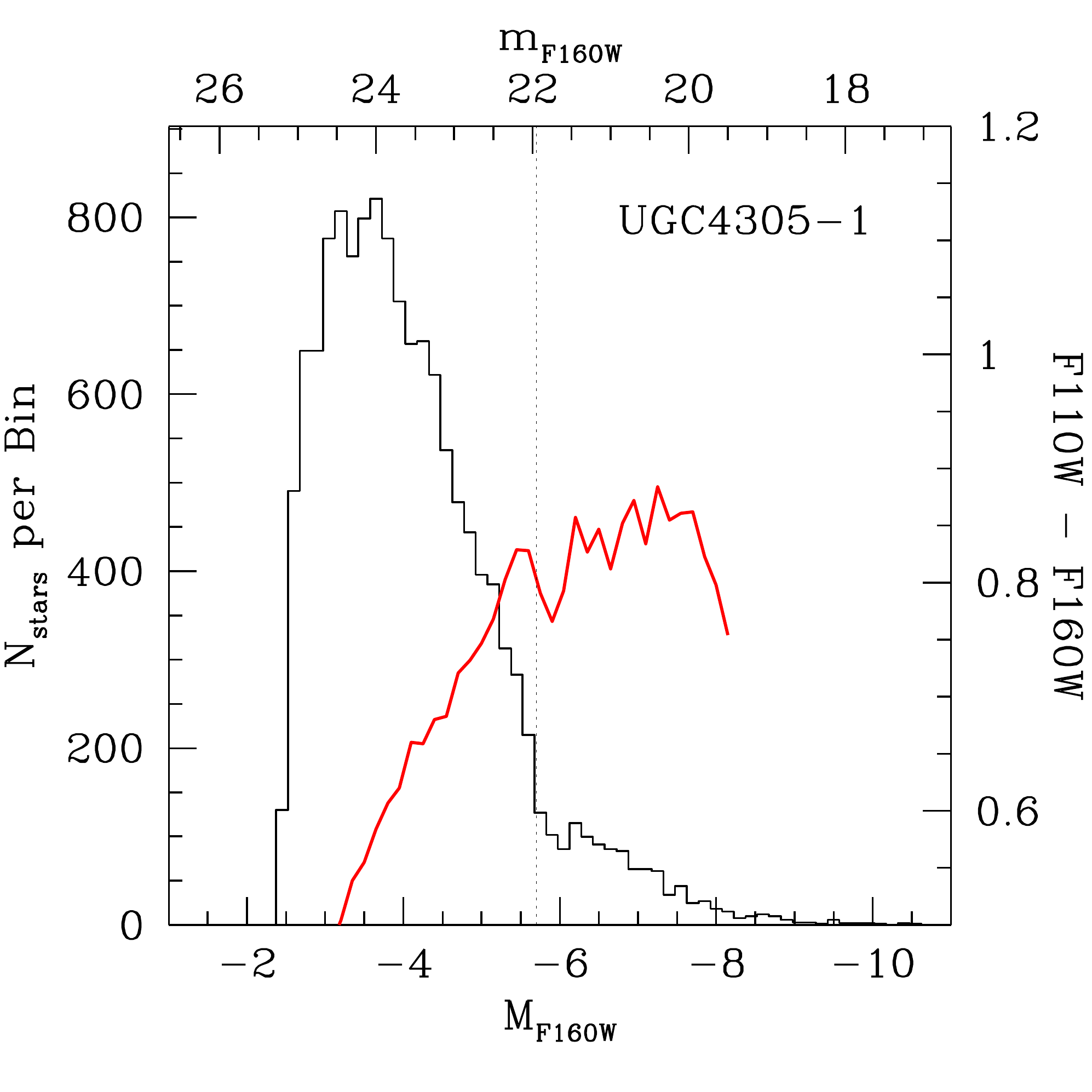}  
\includegraphics[width=2.25in]{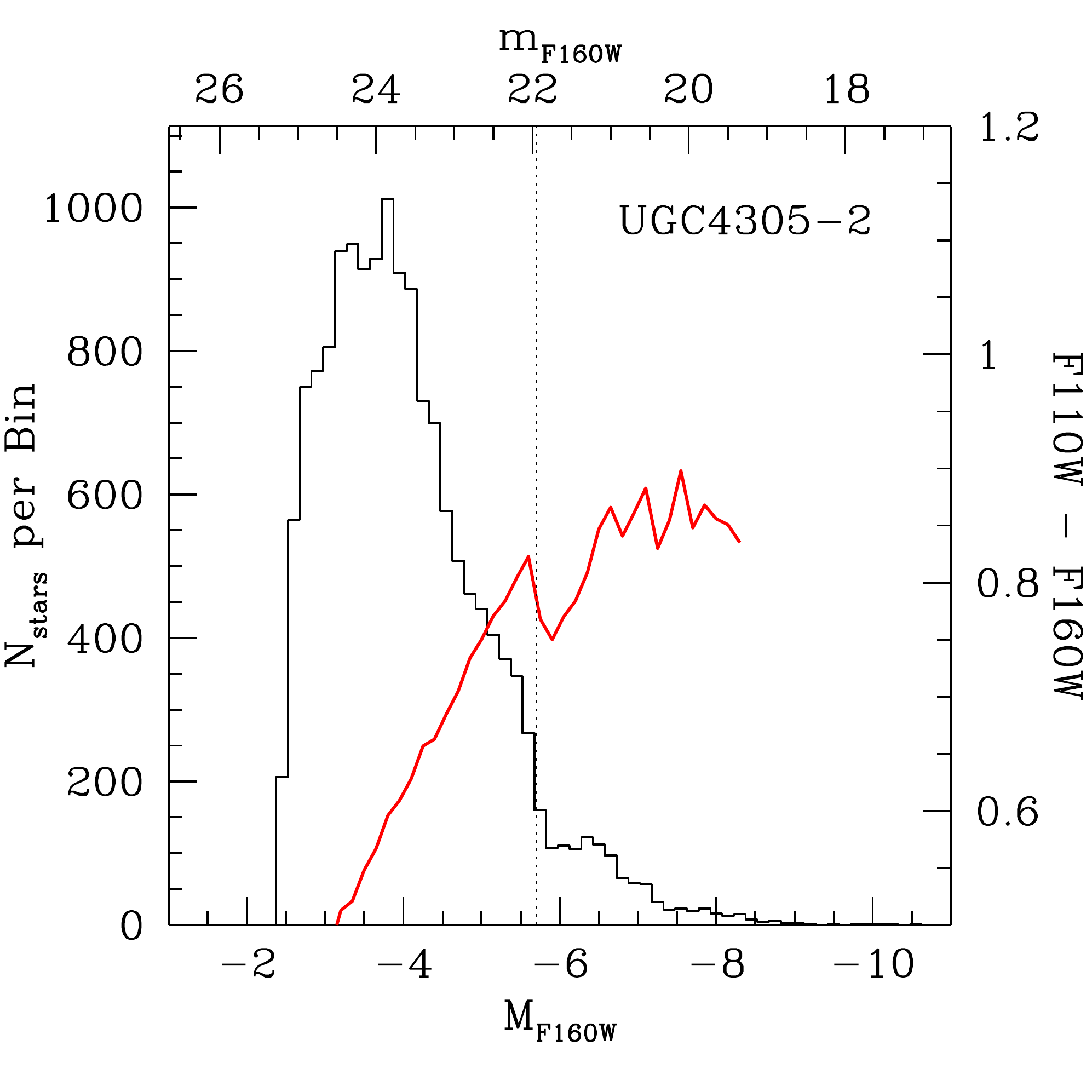}  
}
\centerline{
\includegraphics[width=2.25in]{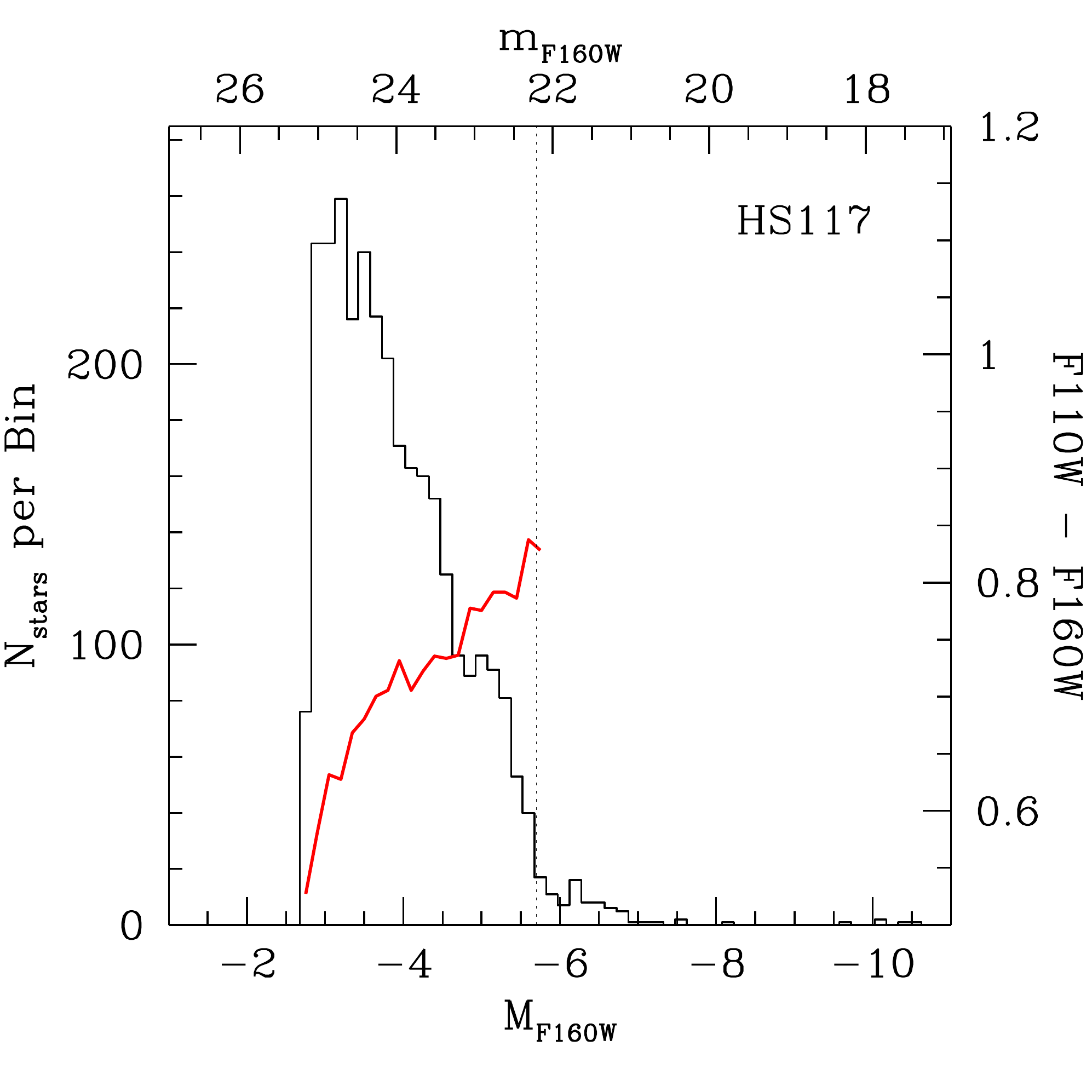}  
\includegraphics[width=2.25in]{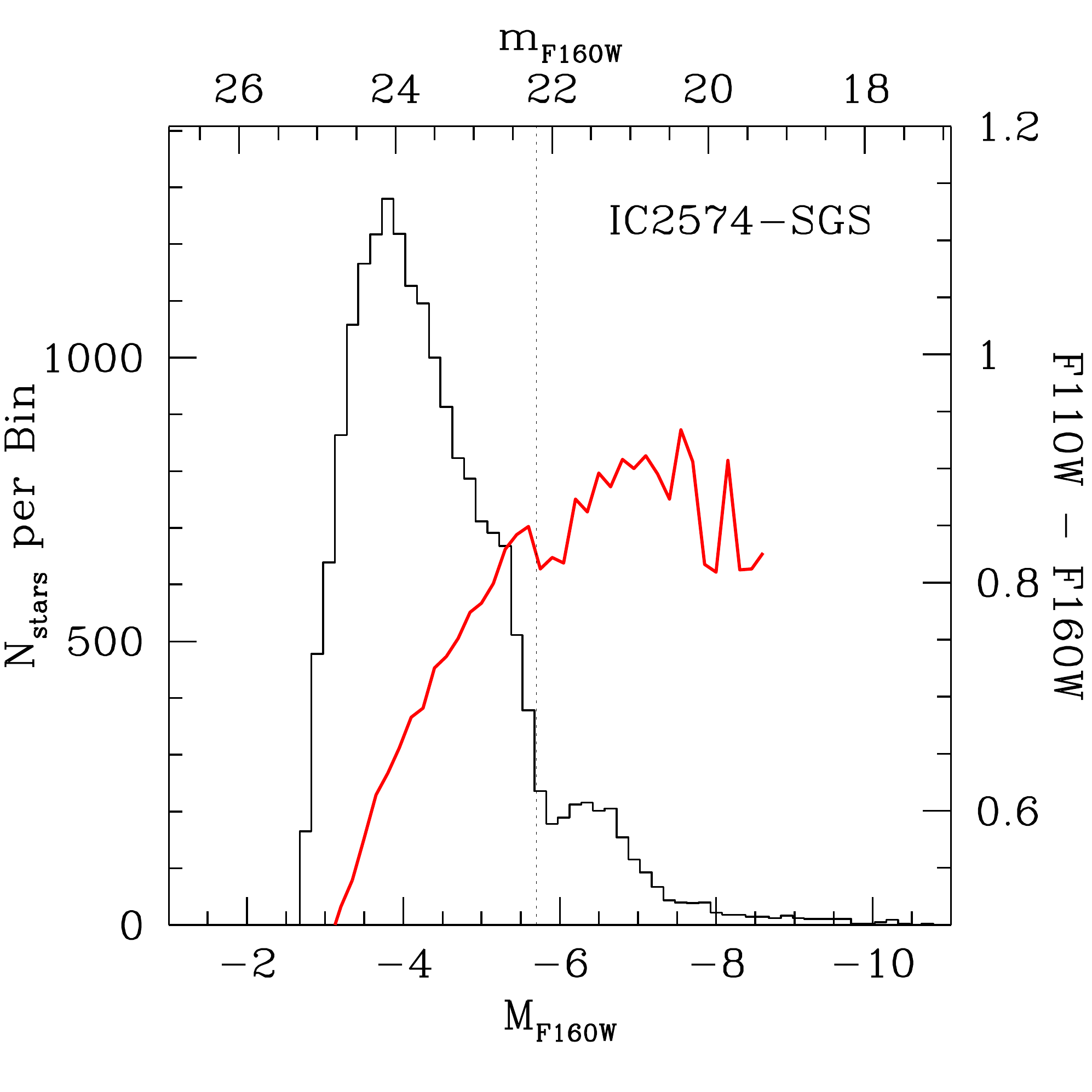}  
\includegraphics[width=2.25in]{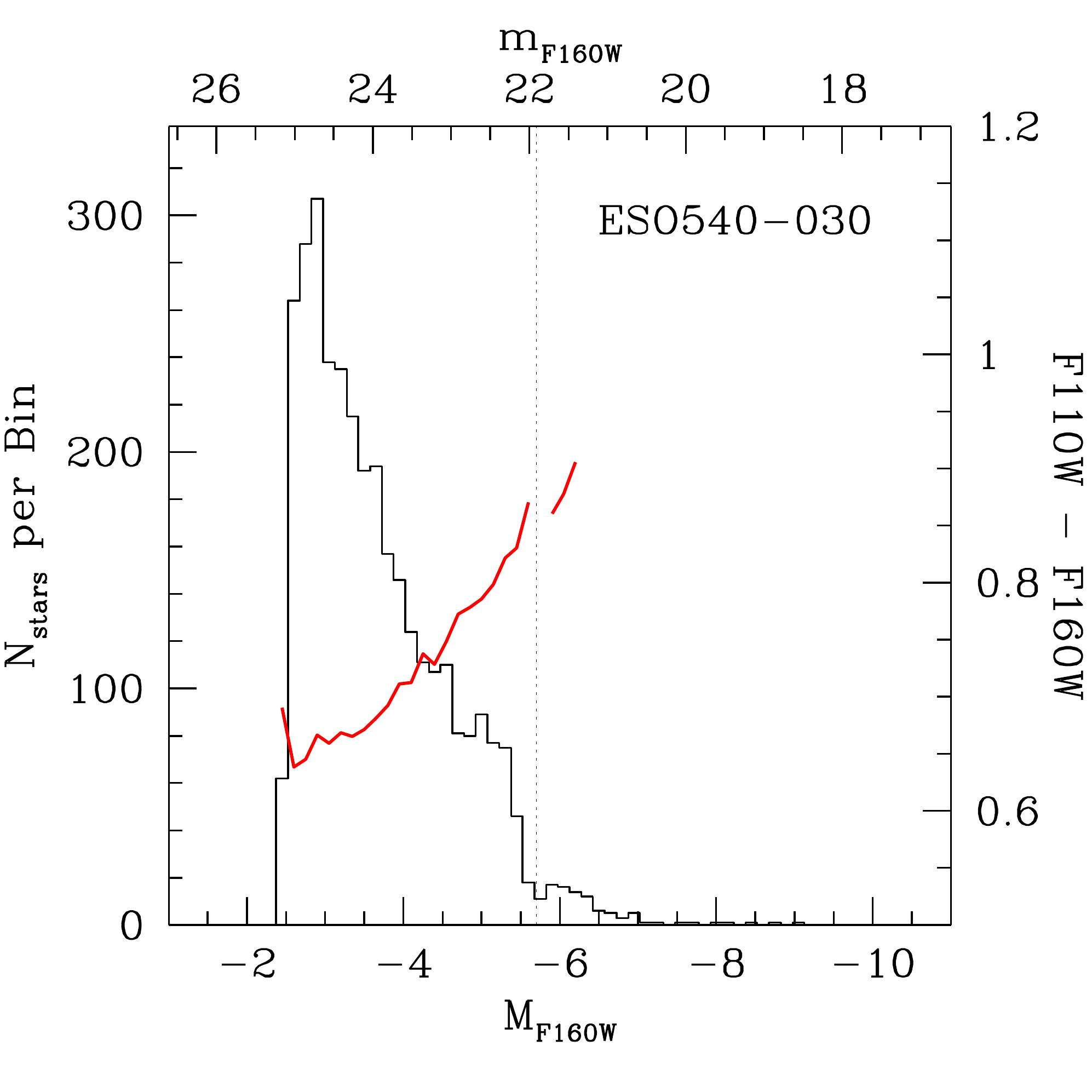}  
}
\caption{ Luminosity functions in $F160W$ (black histogram, left axis)
  and median $F110W-F160W$ color (thick red line, right axis, for bins
  with $>$12 stars) for red stars ( [a] DDO53; [b] DDO78; [c] DDO82;
  [d] HoI; [e] HoII; [f] HoII; [g] HS117; [h] I2574; [i] KDG2; ).  All
  magnitudes have been extinction corrected, and distance moduli are
  as assumed in Table~\ref{sampletable}.  The vertical dotted line
  indicates a fiducial TRGB magnitude of $F160W=-5.7$.
\label{LFfig}}
\end{figure}
\vfill
\clearpage
 
\begin{figure}
\figurenum{\ref{LFfig} continued}
\centerline{
\includegraphics[width=2.25in]{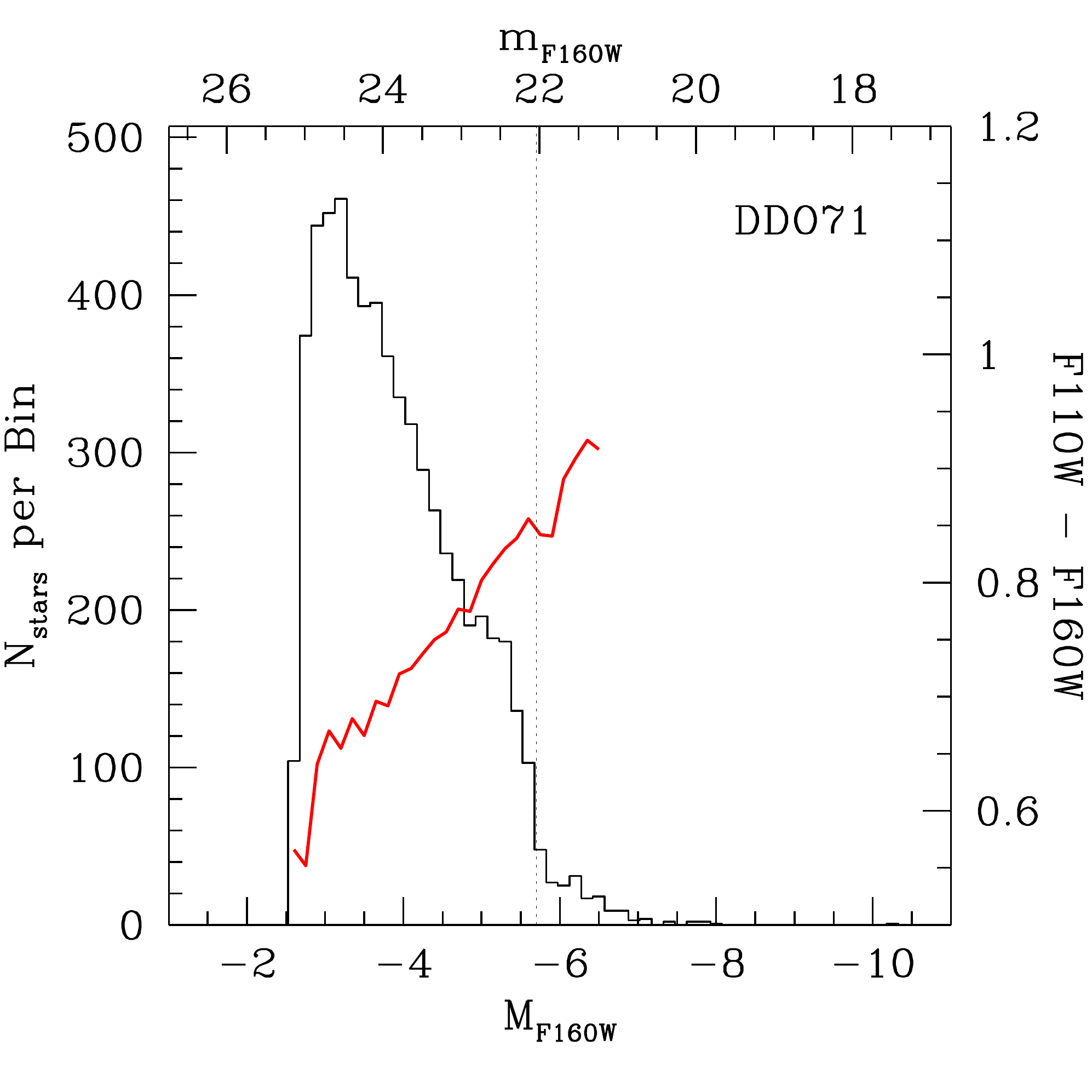}  
\includegraphics[width=2.25in]{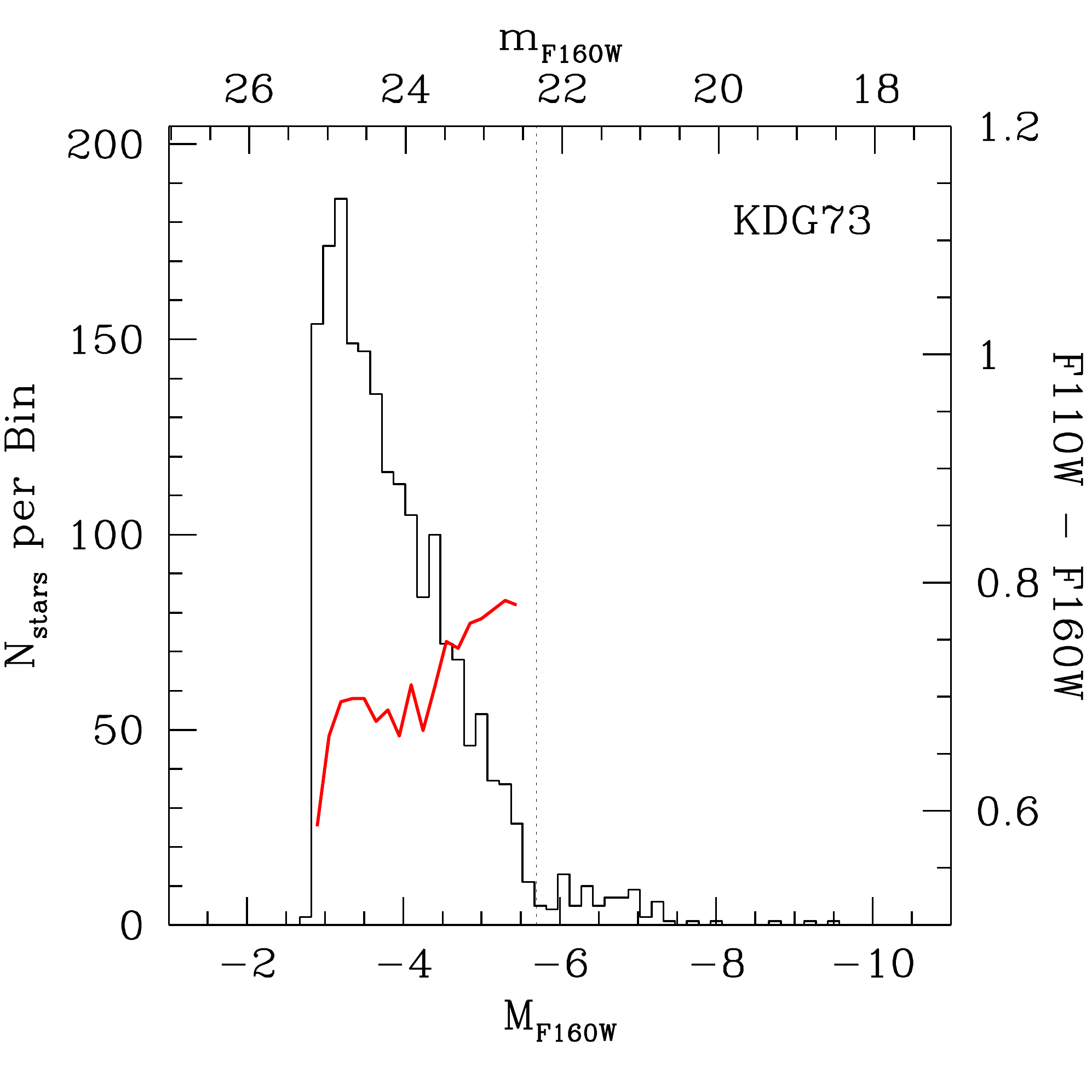}  
\includegraphics[width=2.25in]{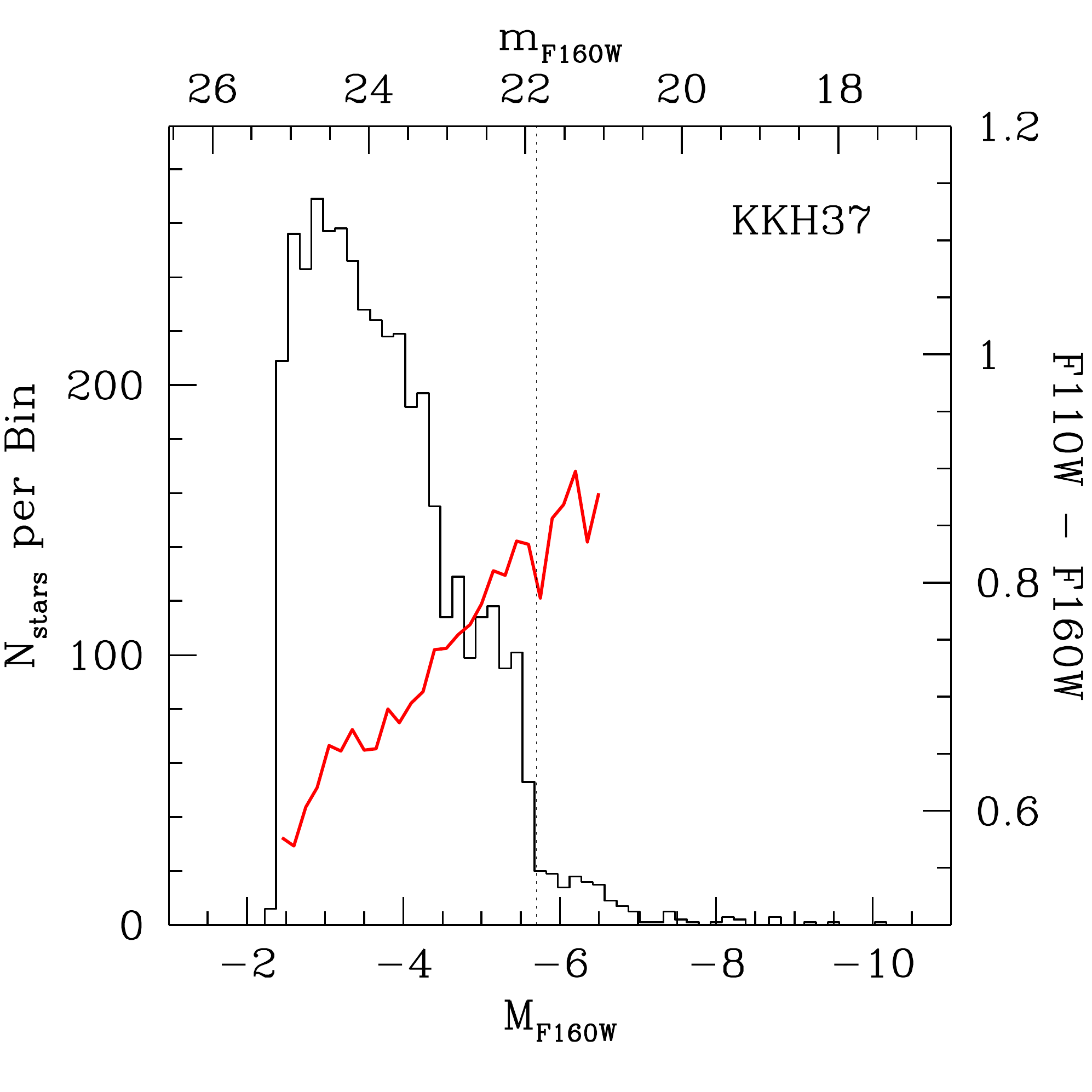}  
}
\centerline{
\includegraphics[width=2.25in]{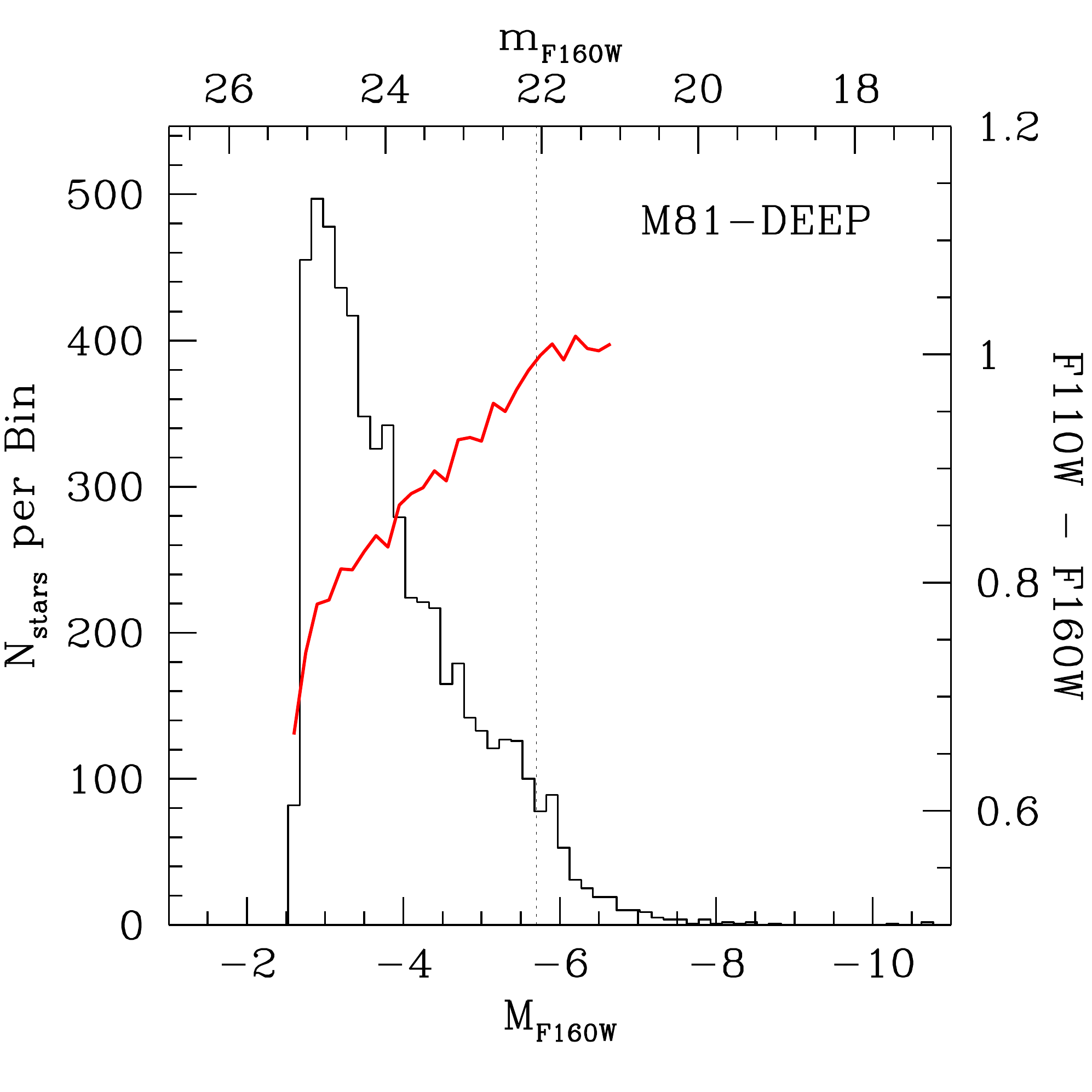}  
\includegraphics[width=2.25in]{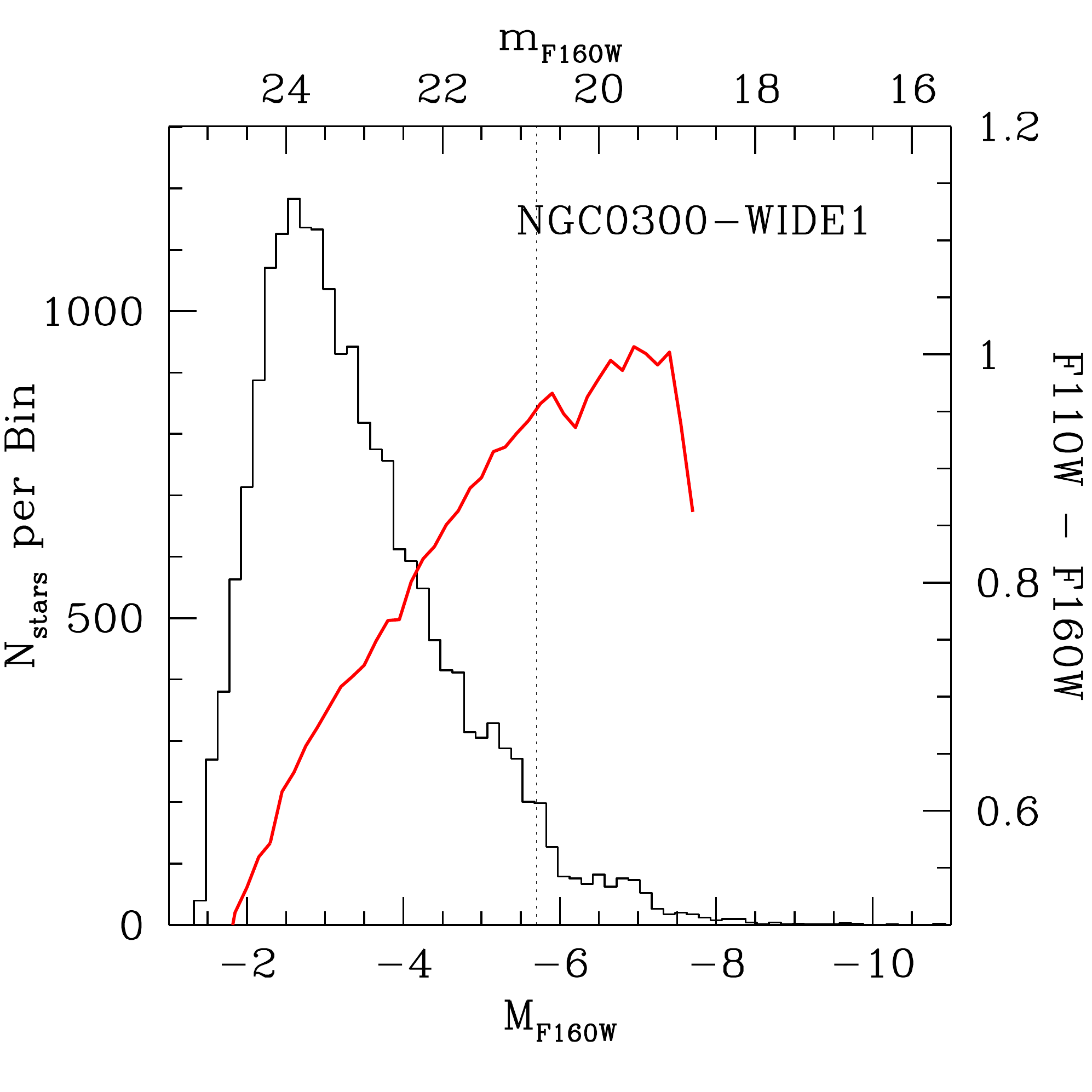}  
\includegraphics[width=2.25in]{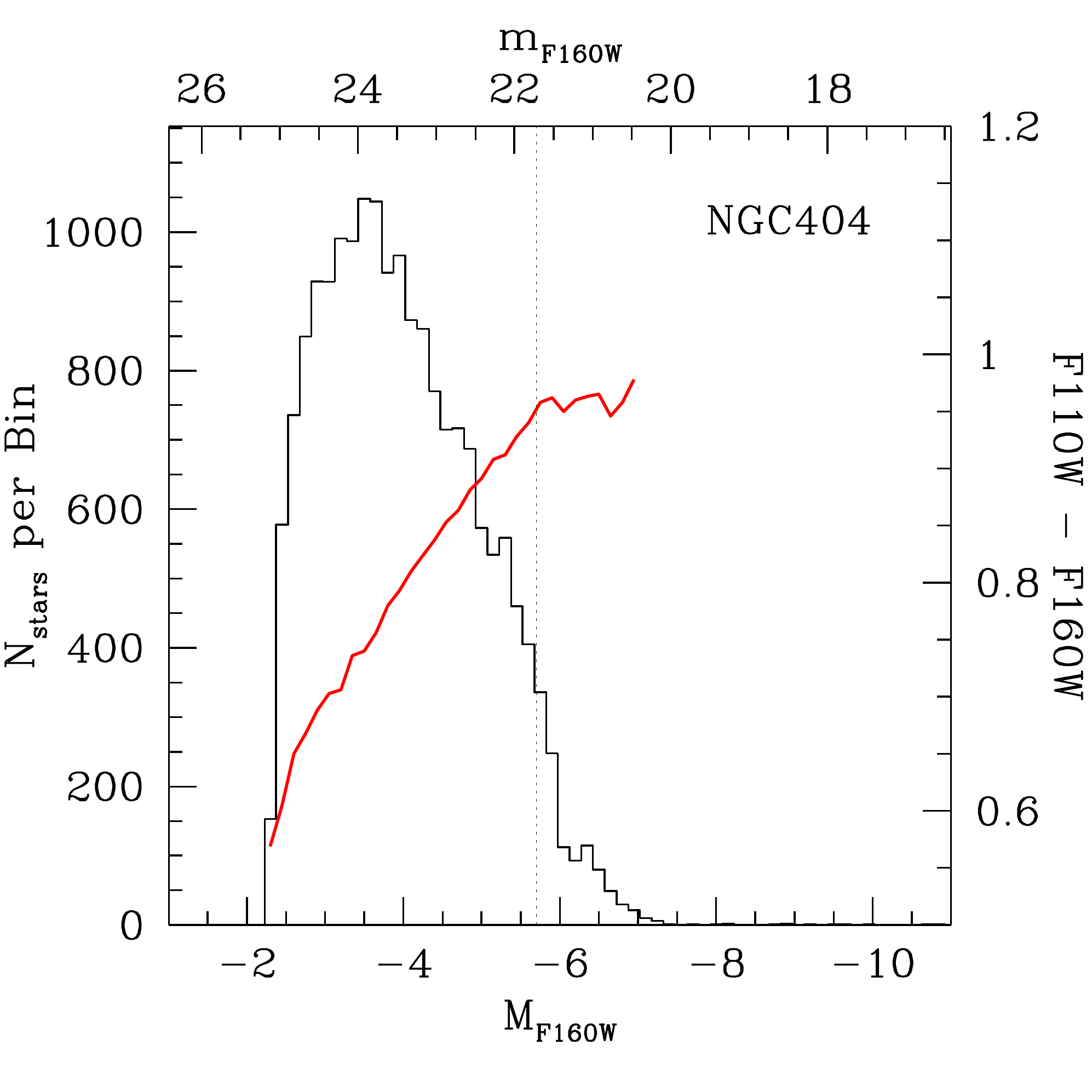}  
}
\centerline{
\includegraphics[width=2.25in]{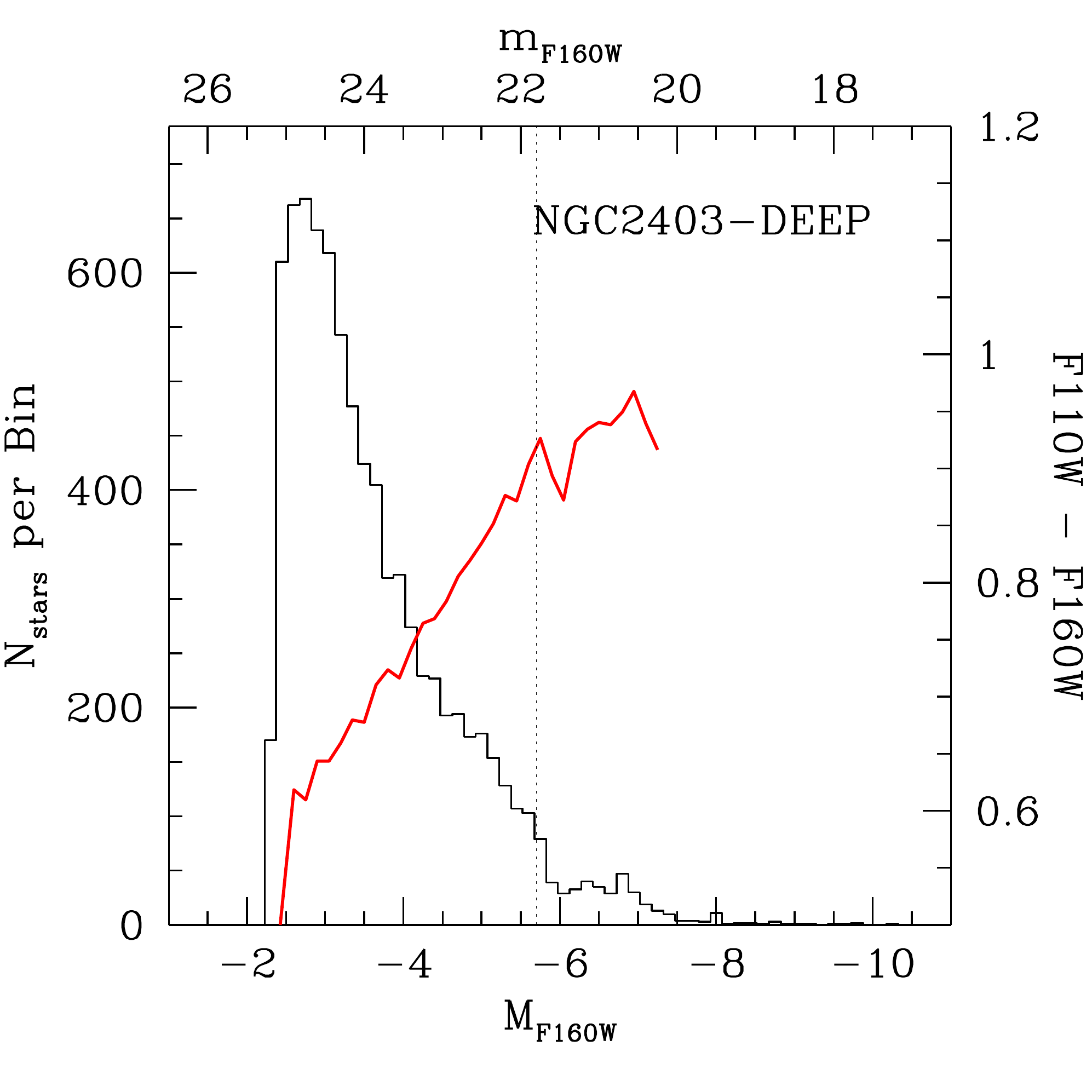}  
\includegraphics[width=2.25in]{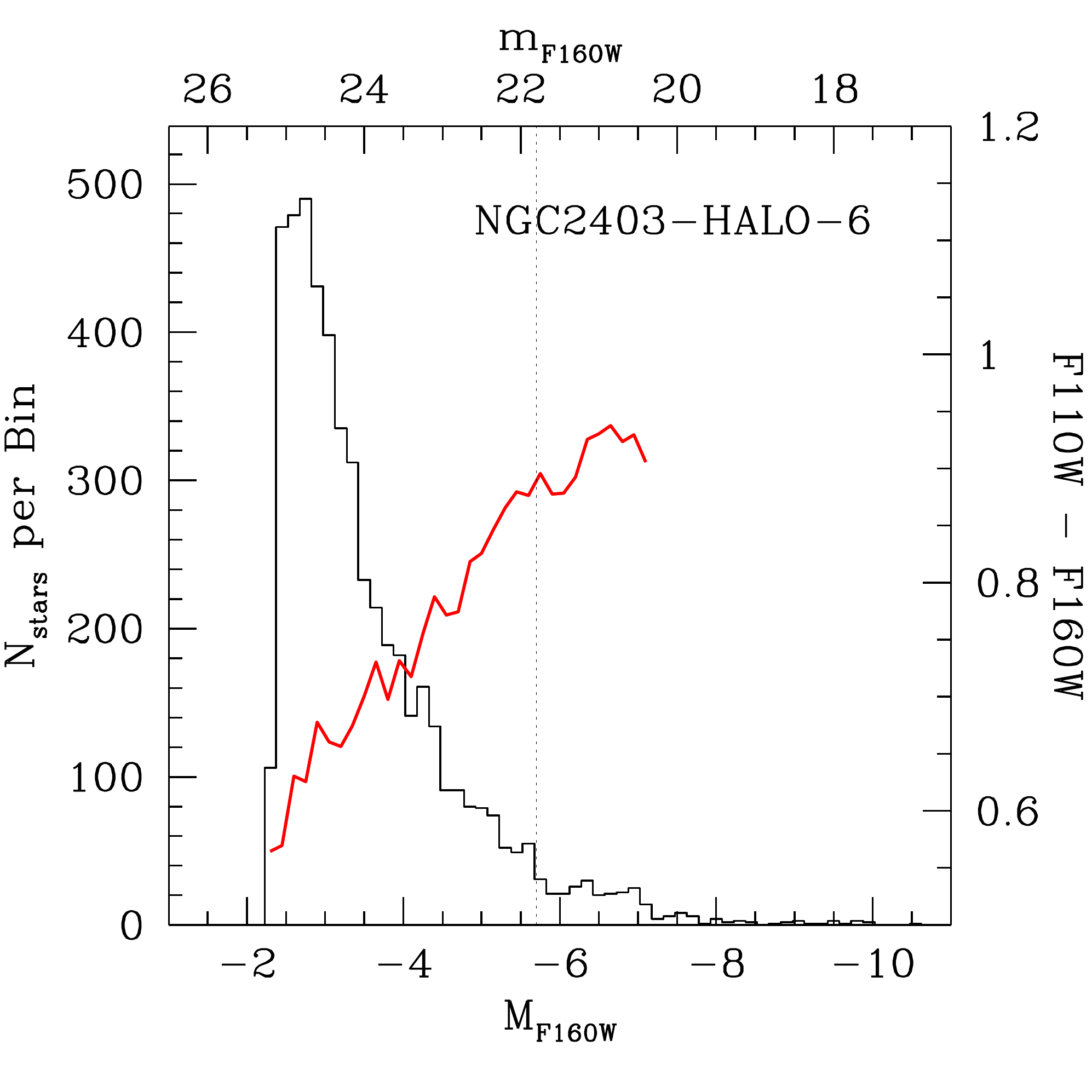}  
\includegraphics[width=2.25in]{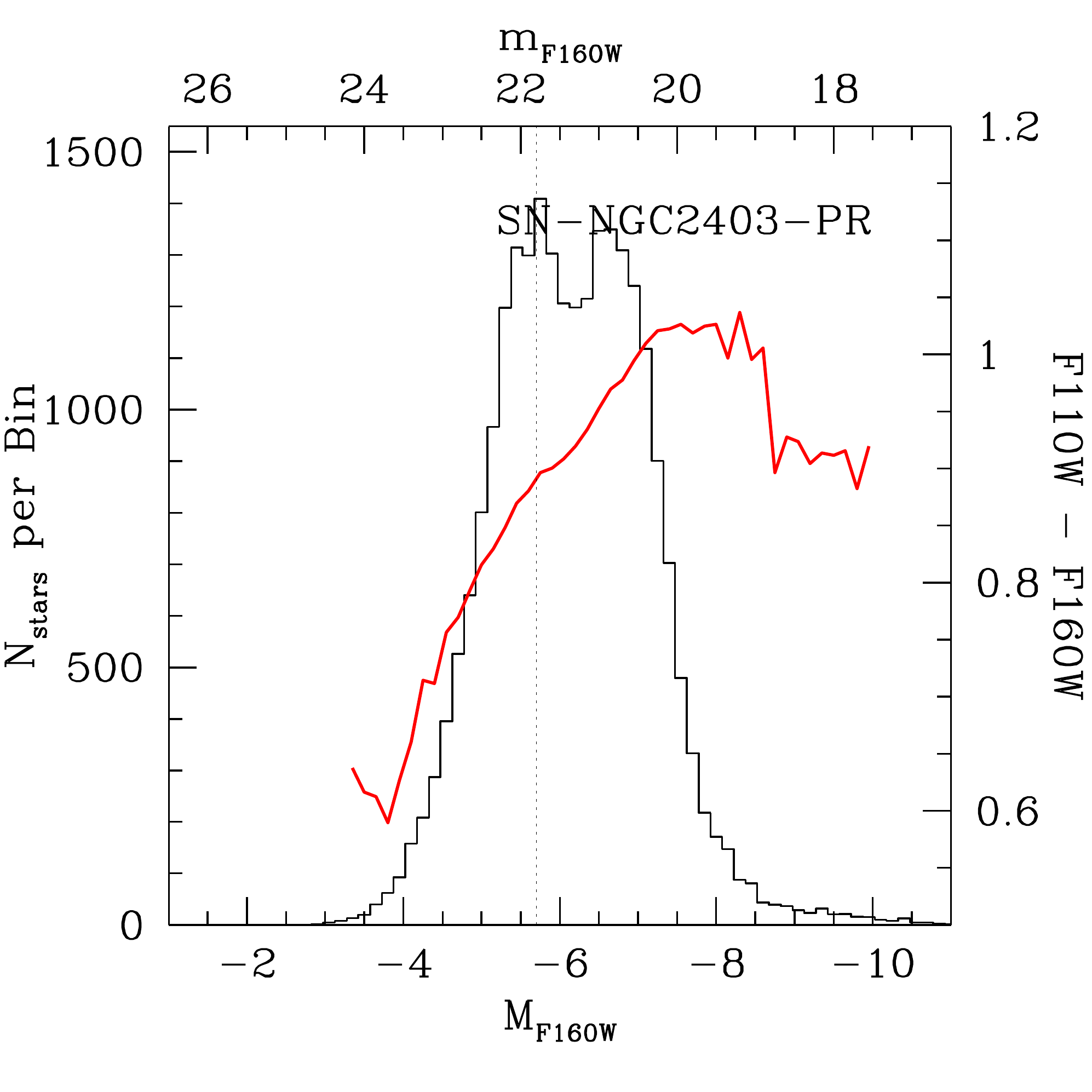}  
}
\caption{ Luminosity functions in $F160W$ (black histogram, left axis)
  and median $F110W-F160W$ color (thick red line, right axis, for bins
  with $>$12 stars) for red stars ( [j] KDG63; [k] KDG73; [l] KKH37;
  [m] M81; [n] N300; [o] N404; [p] N2403; [q] N2403; [r] N2403; ).
  All magnitudes have been extinction corrected, and distance moduli
  are as assumed in Table~\ref{sampletable}.  The vertical dotted line
  indicates a fiducial TRGB magnitude of $F160W=-5.7$.}
\end{figure}
\vfill
\clearpage
 
\begin{figure}
\figurenum{\ref{LFfig} continued}
\centerline{
\includegraphics[width=2.25in]{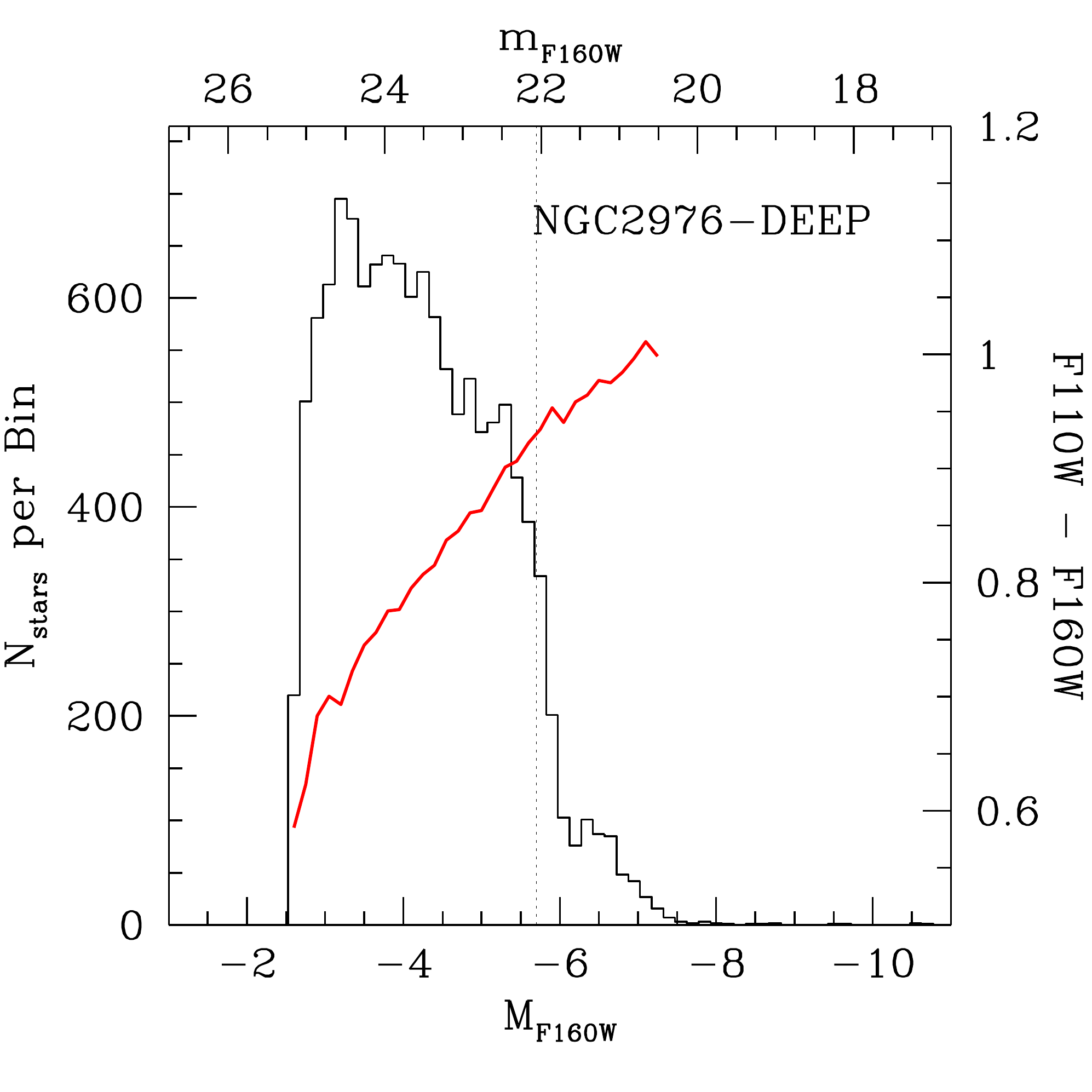}  
\includegraphics[width=2.25in]{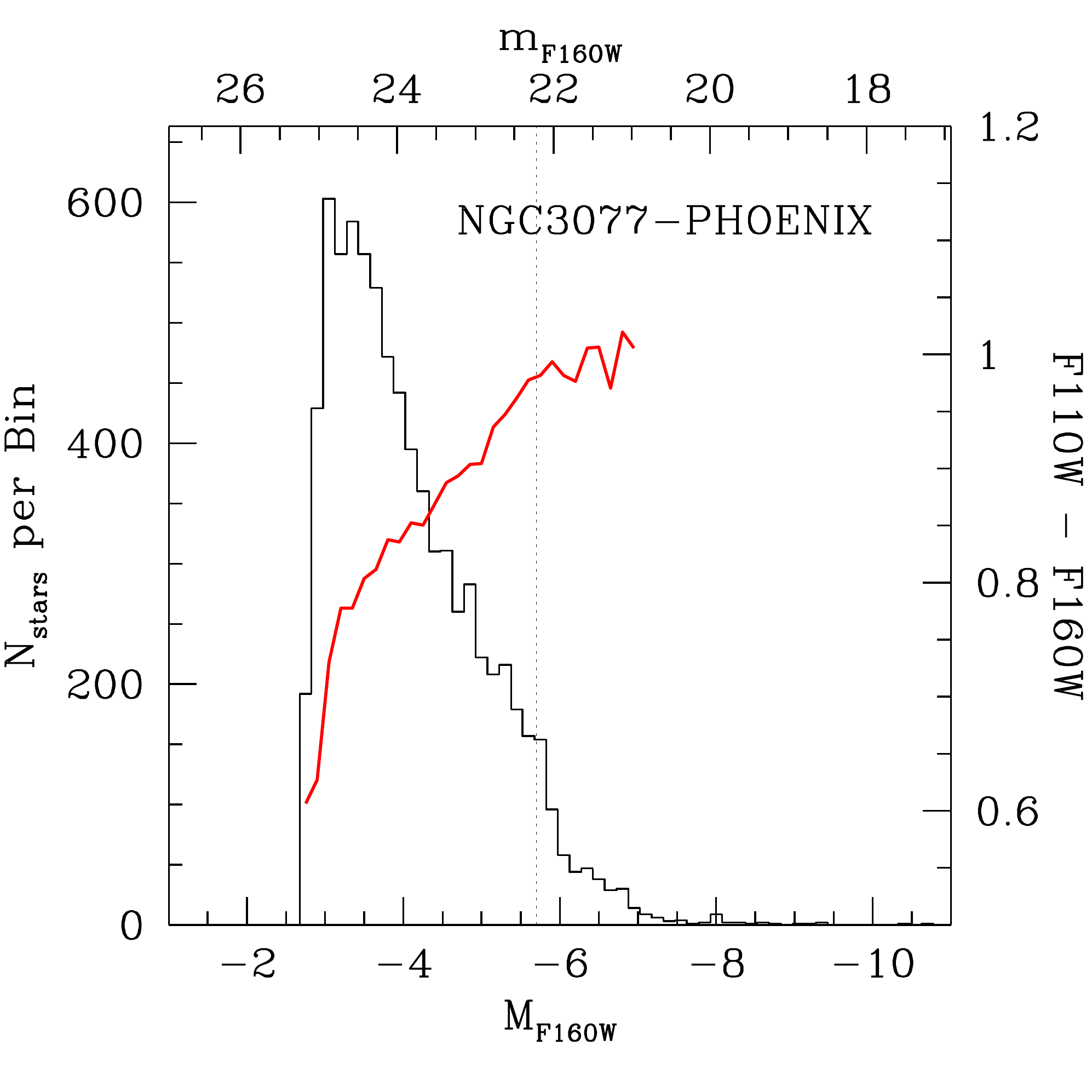}  
\includegraphics[width=2.25in]{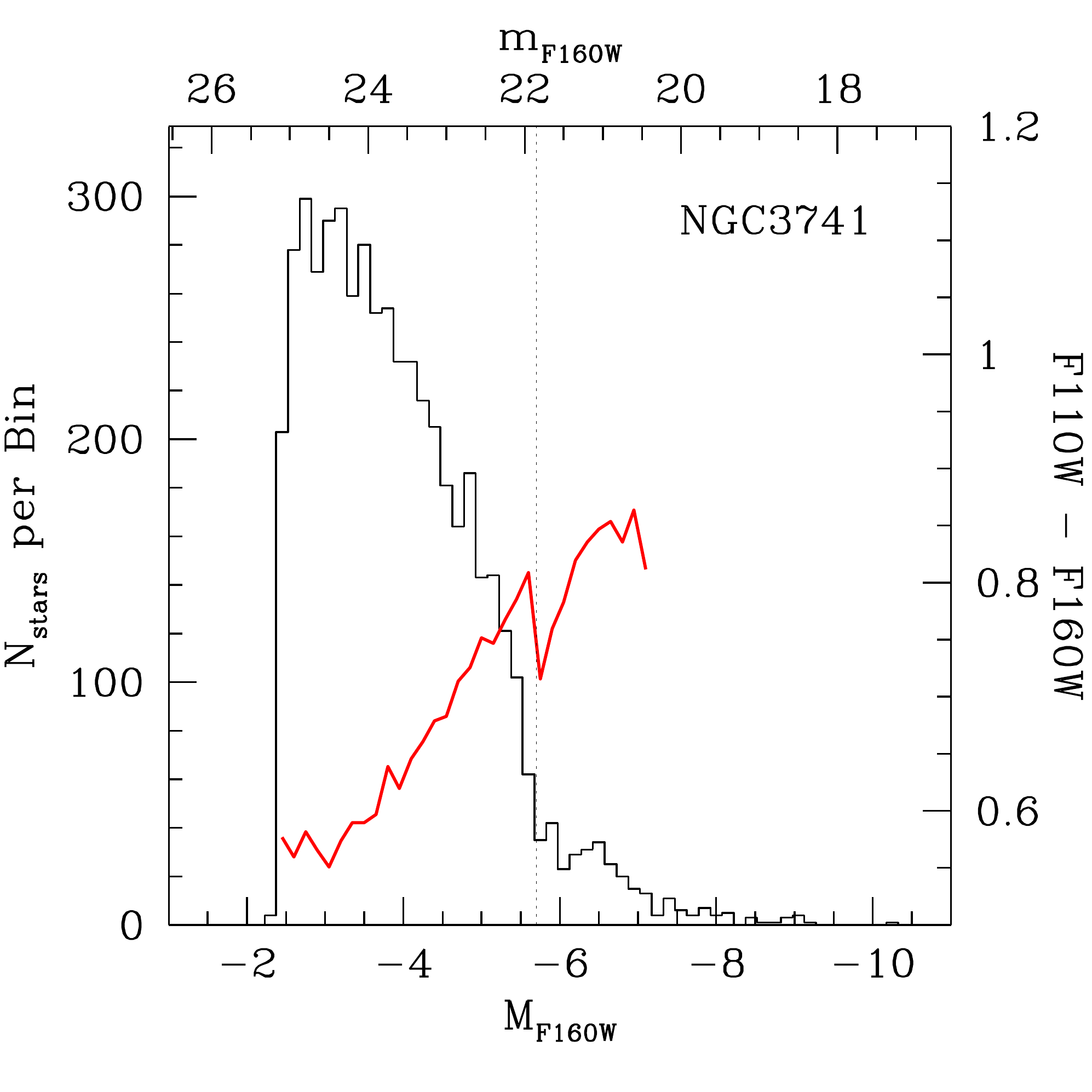}  
}
\centerline{
\includegraphics[width=2.25in]{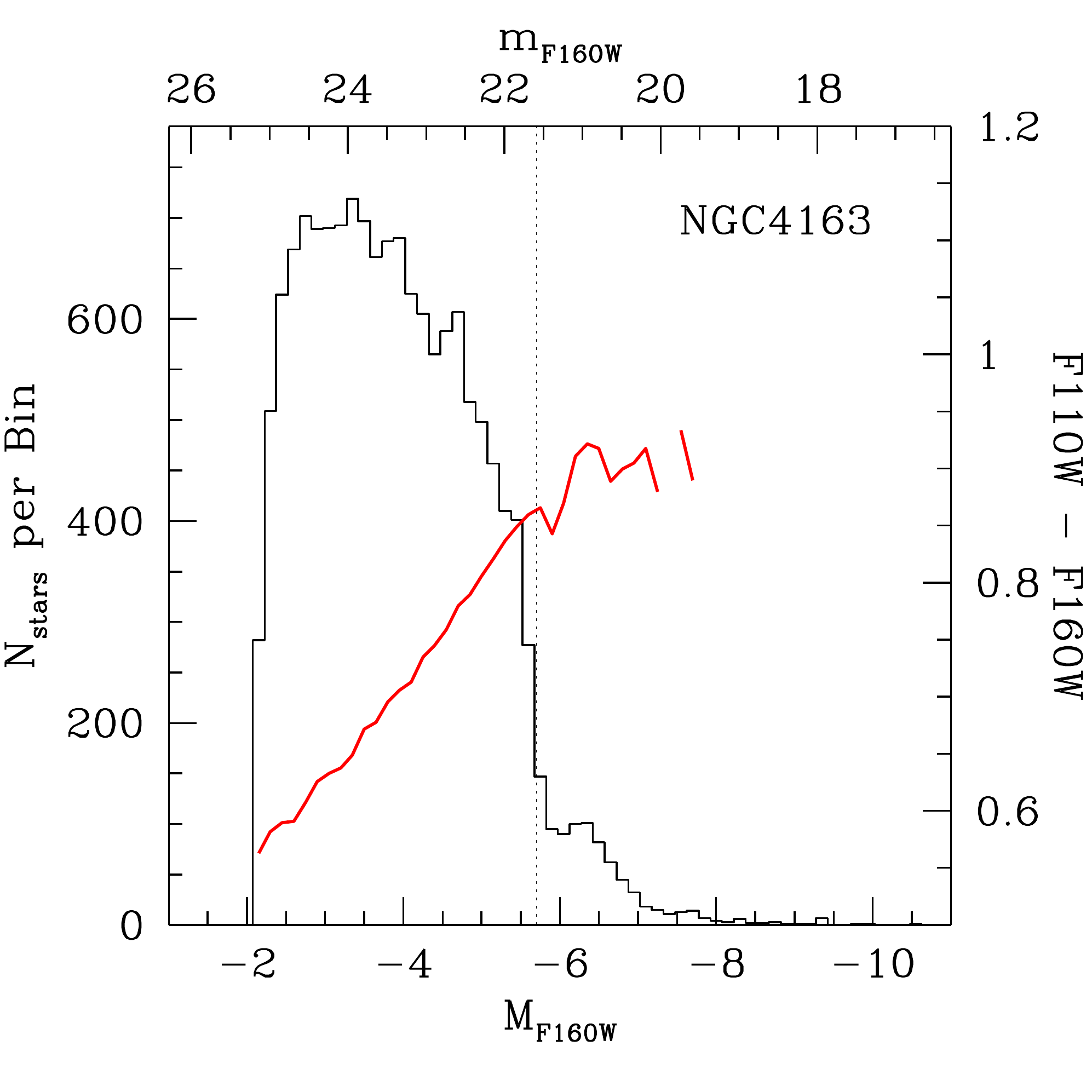}  
\includegraphics[width=2.25in]{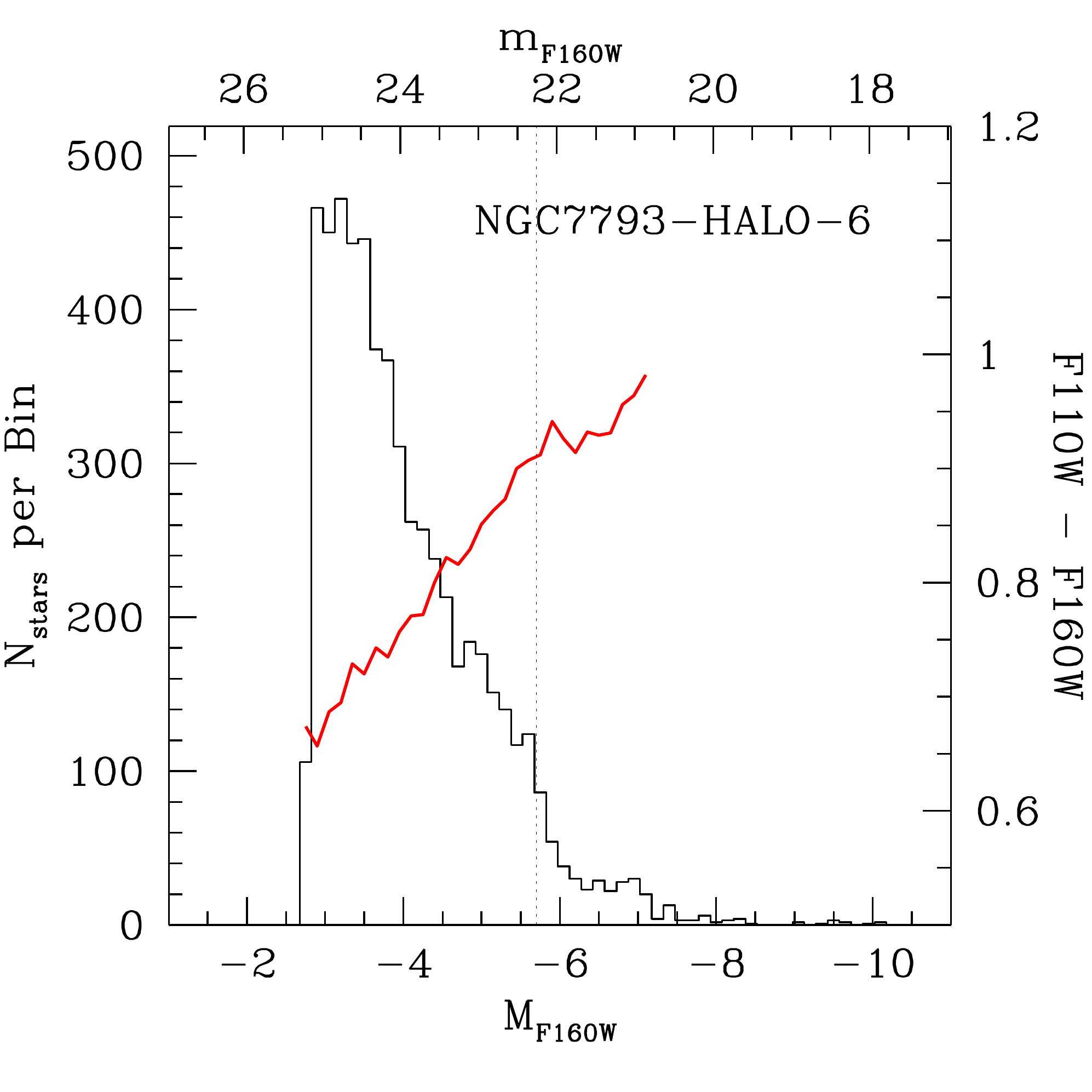}  
\includegraphics[width=2.25in]{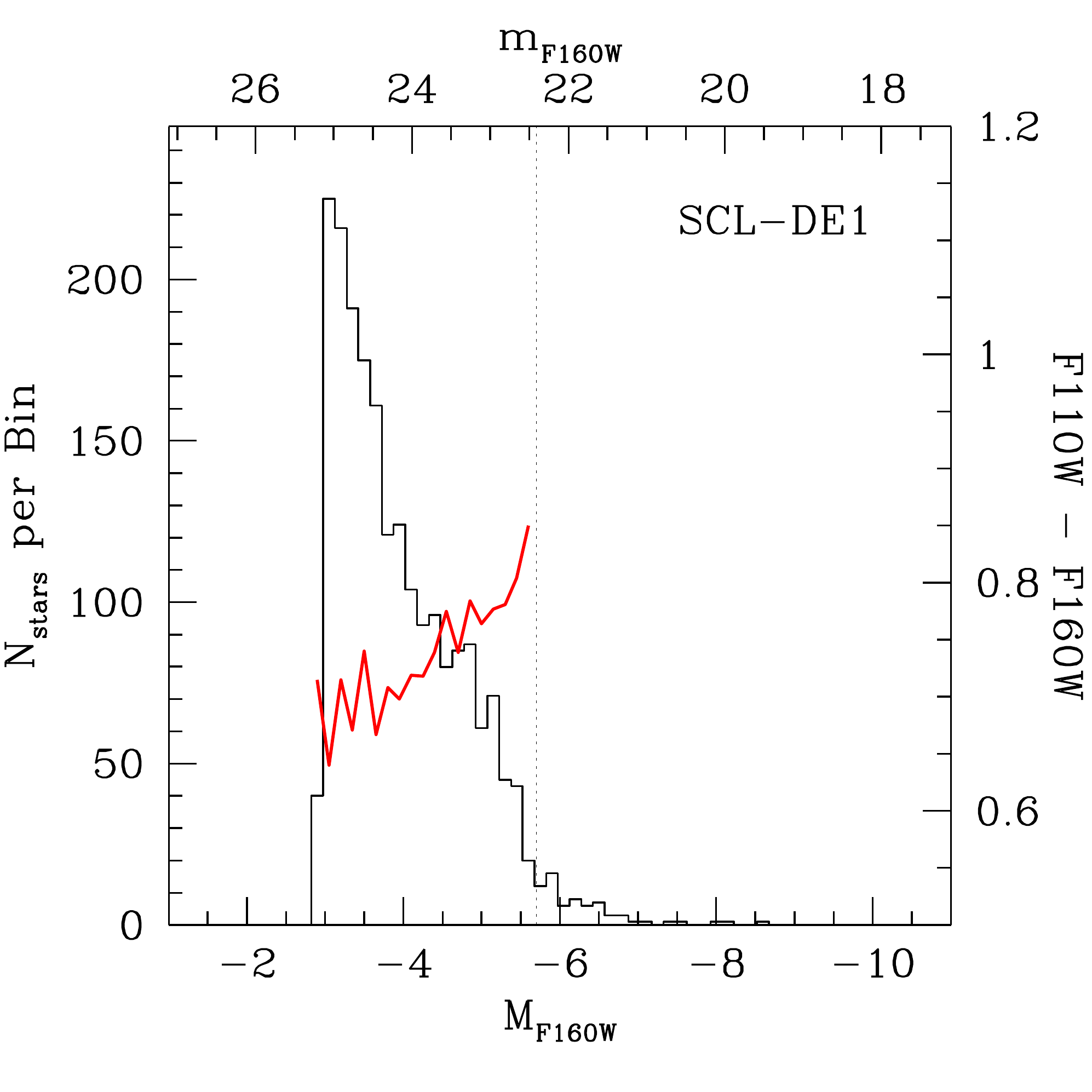}  
}
\centerline{
\includegraphics[width=2.25in]{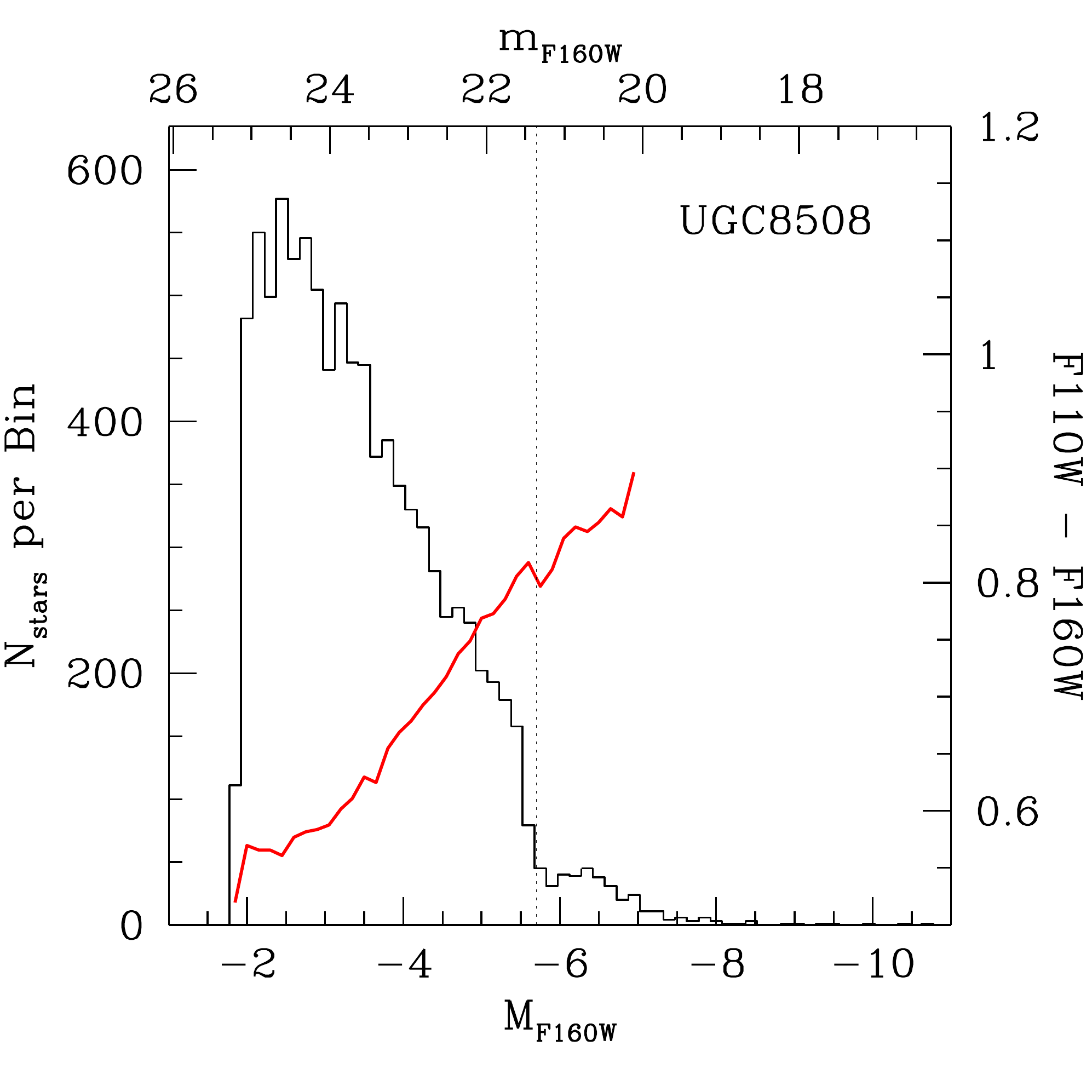}  
\includegraphics[width=2.25in]{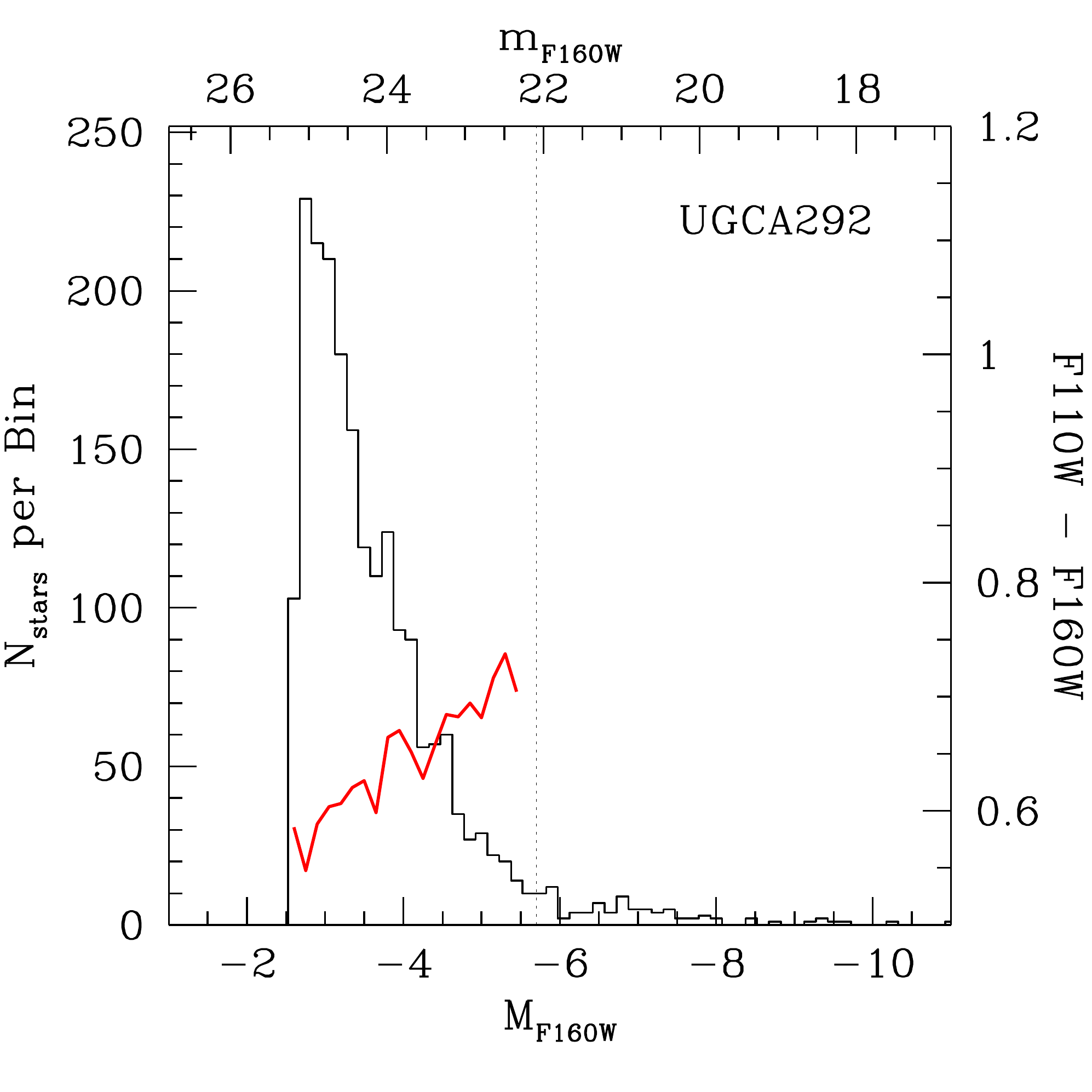}  
}
\caption{
  Luminosity functions in $F160W$ (black histogram, left axis) and
  median $F110W-F160W$ color (thick red line, right axis, for bins
  with $>$12 stars) for red stars ( [s] N2976; [t] N3077; [u] N3741;
  [v] N4163; [w] N7793; [x] Sc22; [y] U8508; [z] UA292; ).  All
  magnitudes have been extinction corrected, and distance moduli are
  as assumed in Table~\ref{sampletable}.  The vertical dotted line
  indicates a fiducial TRGB magnitude of $F160W=-5.7$.}
\end{figure}
\vfill
\clearpage

\begin{figure}
\centerline{
\includegraphics[width=3.25in]{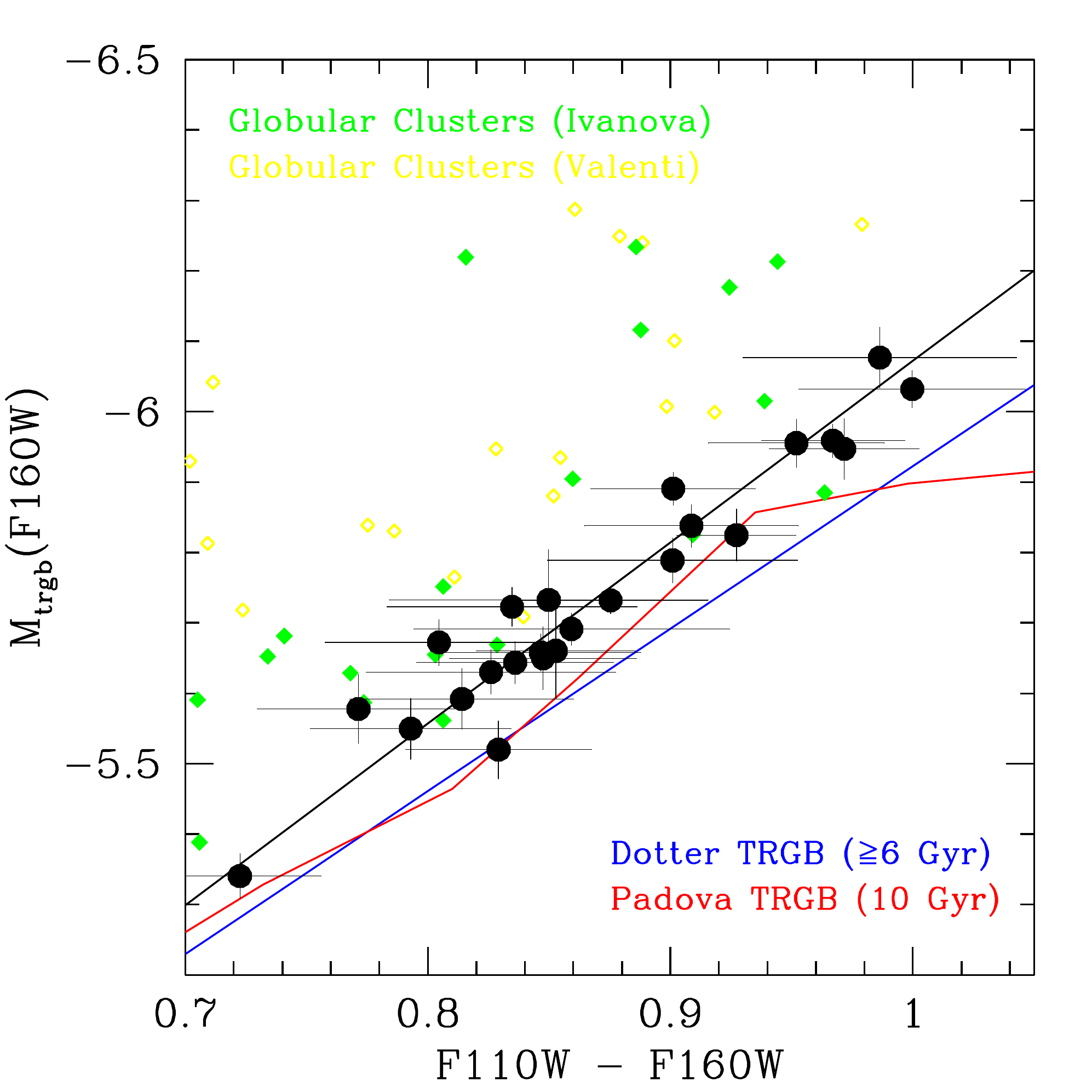}
\includegraphics[width=3.25in]{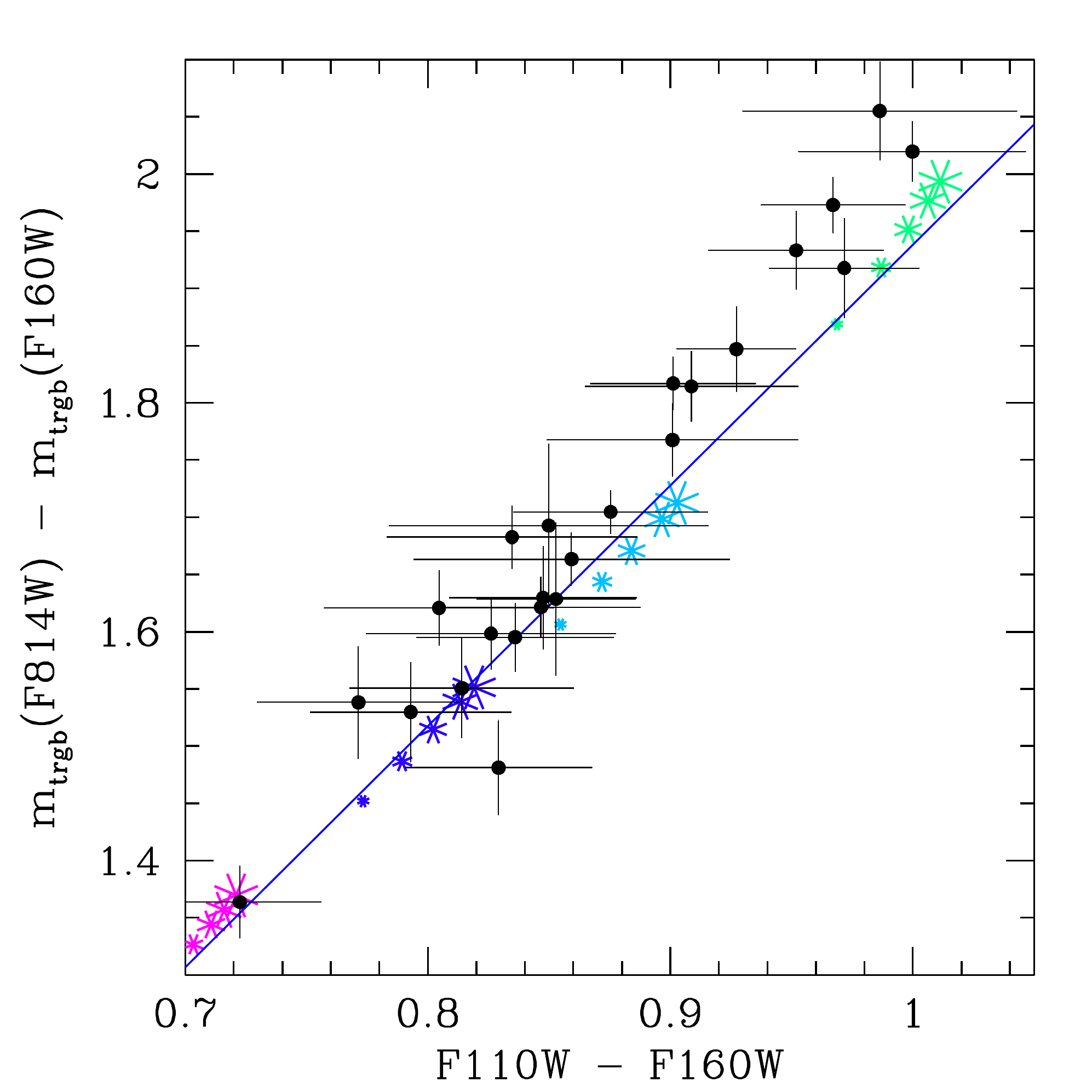}
}

\caption{
(Left) Absolute $F160W$ magnitude of the TRGB as a function of
$F110W-F160W$ color, for data in this sample (solid black circles).
Also shown are loci from the isochrone models of \citet{girardi2008}
for a $10\Gyr$ RGB with a range of metallicities (solid red line;
updated to the WFC3/IR filter set), and a fit to isochrone models
provided by A.\ Dotter (blue solid line; private communication).
Diamonds are data from globular clusters (solid green points are from
Ivanov \& Borissova 2002; open yellow points are from
Valenti et al 2004,2007).  Our measurements are slightly offset
from the models by roughly 0.15 magnitudes brighter in $F160W$, or 0.1
magnitudes bluer in color.  (Right)
Magnitude difference between the optical $F814W$ (from Dalcanton et al
2009) and NIR $F160W$ TRGB, as a function of the measured
$F110W-F160W$ color of the TRGB.  Lines and symbols are the same as in
the left hand plot.  Individual points for the Dotter isochrone TRGB
values are plotted for ages of 6, 8, 10, 12, \& $14\Gyr$ (increasing
asterix sizes) and metallicities of [Fe/H]=-2.49, [$\alpha$/Fe]=0.40,
$Y$=0.245 (magenta), [Fe/H]=-1.99, [$\alpha$/Fe]=0.40, $Y$=0.246
(blue), [Fe/H]=-1.49, [$\alpha$/Fe]=0.20, $Y$=0.246 (cyan), \&
[Fe/H]=-0.99, [$\alpha$/Fe]=0.20, $Y$=0.249 (green).
  \label{trgbfig}}
\end{figure}
\vfill

\begin{figure}
\centerline{
\includegraphics[width=3.5in]{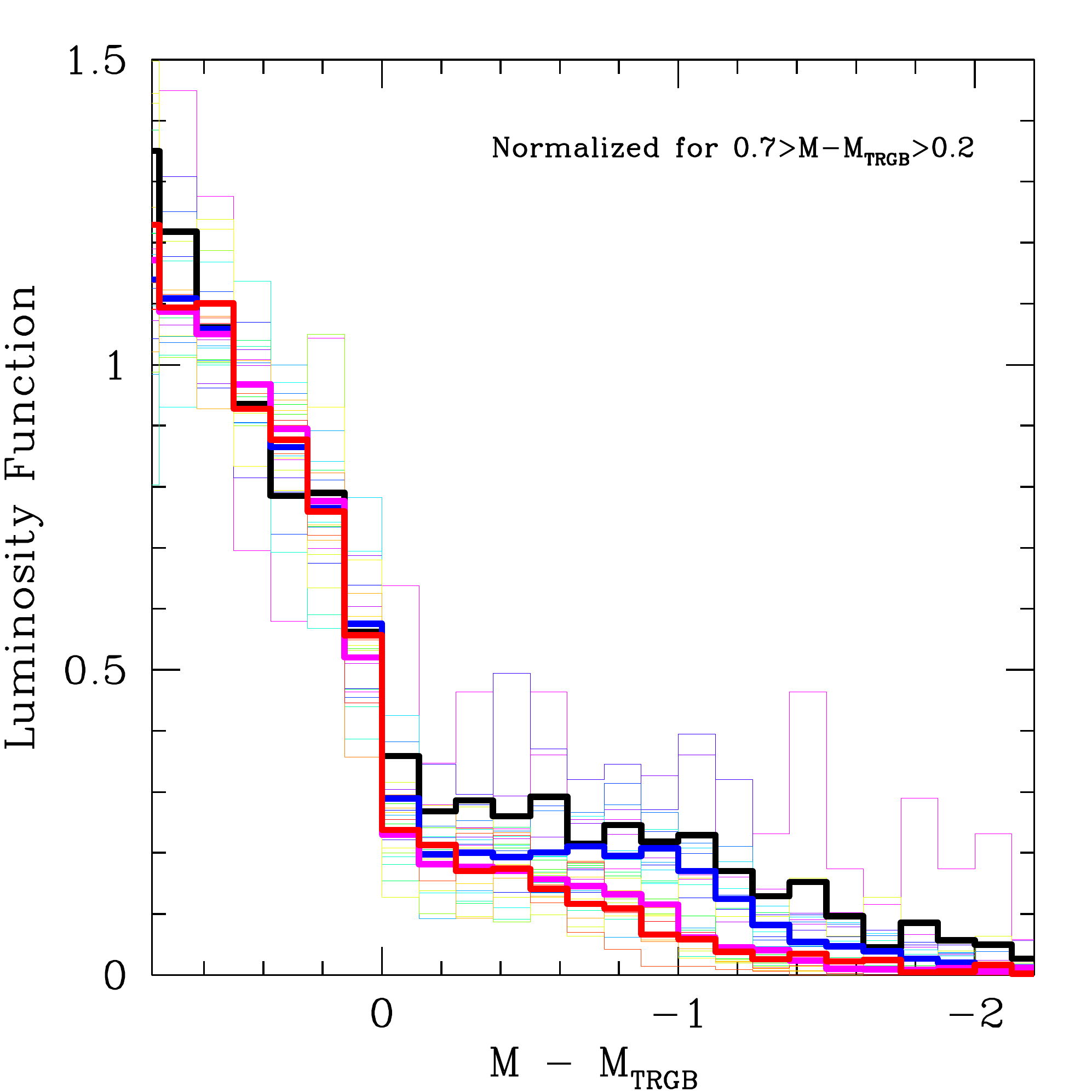}
}
\caption{
RGB luminosity functions for all galaxies, color-coded by their rank
when sorted by the fraction of star formation in the most recent
$1\Gyr$ (light lines, with redder colors indicating older mean stellar
ages). Dark lines show the average luminosity function of galaxies,
sorted by rank into 4 bins of recent star formation, with equal
numbers of galaxies per bin; red, magenta, blue, and black lines go
from lowest to highest fraction of recent star formation ($\langle
f_{0-1\Gyr} \rangle = 0.012, 0.038, 0.071, 0.089$, respectively).  All
luminosity functions are for stars found within 1.5$\sigma$ of the
line fit to the upper 1 magnitude of the RGB, and are normalized to
have the same number of stars within the bins between 0.25 and 0.75
magnitudes fainter than the TRGB.  All magnitudes are relative
to the TRGB magnitude given in Table~\ref{trgbtable}.
\label{allLFfig}}
\end{figure}
\vfill
\clearpage

\begin{figure}
\centerline{
\includegraphics[width=3in]{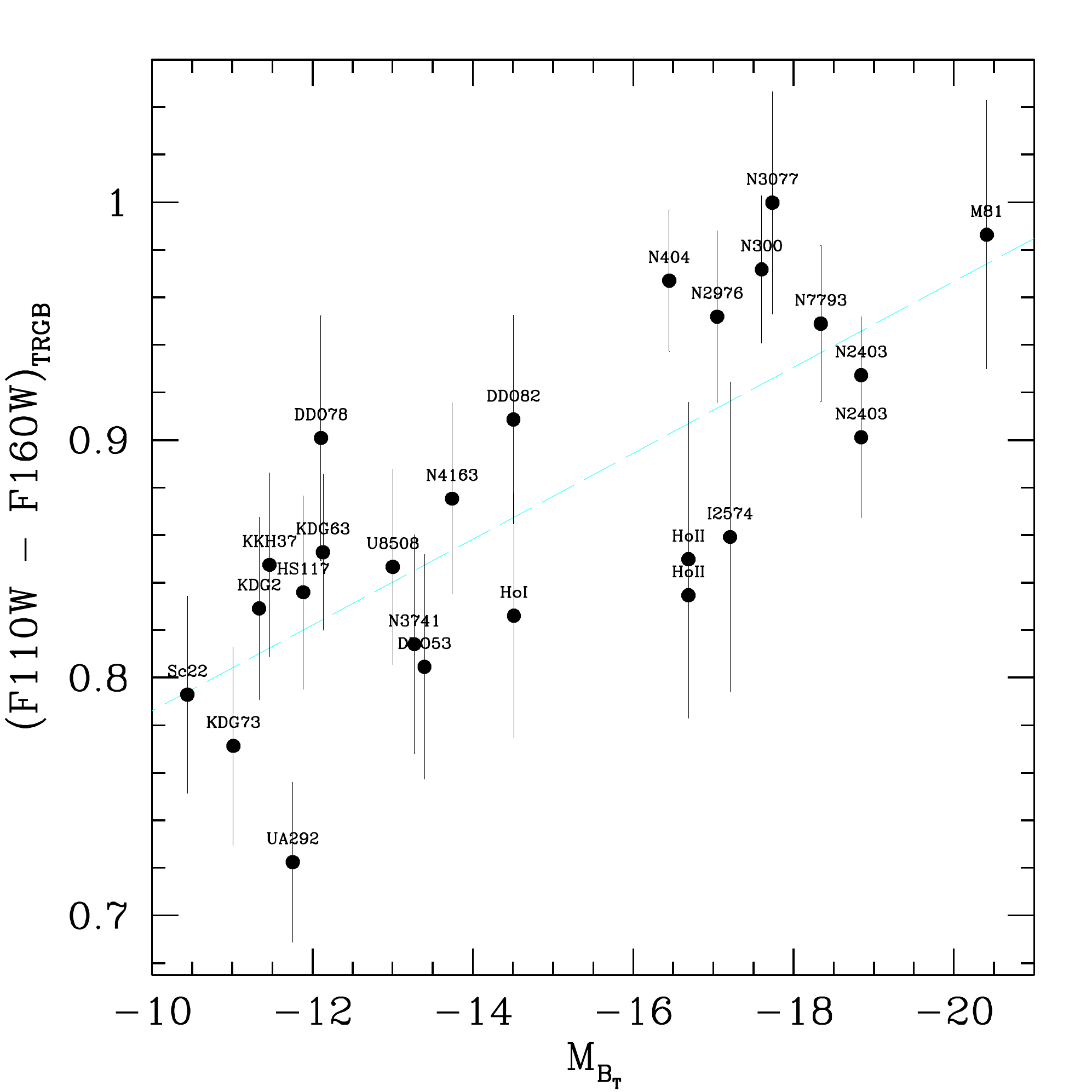}
\includegraphics[width=3in]{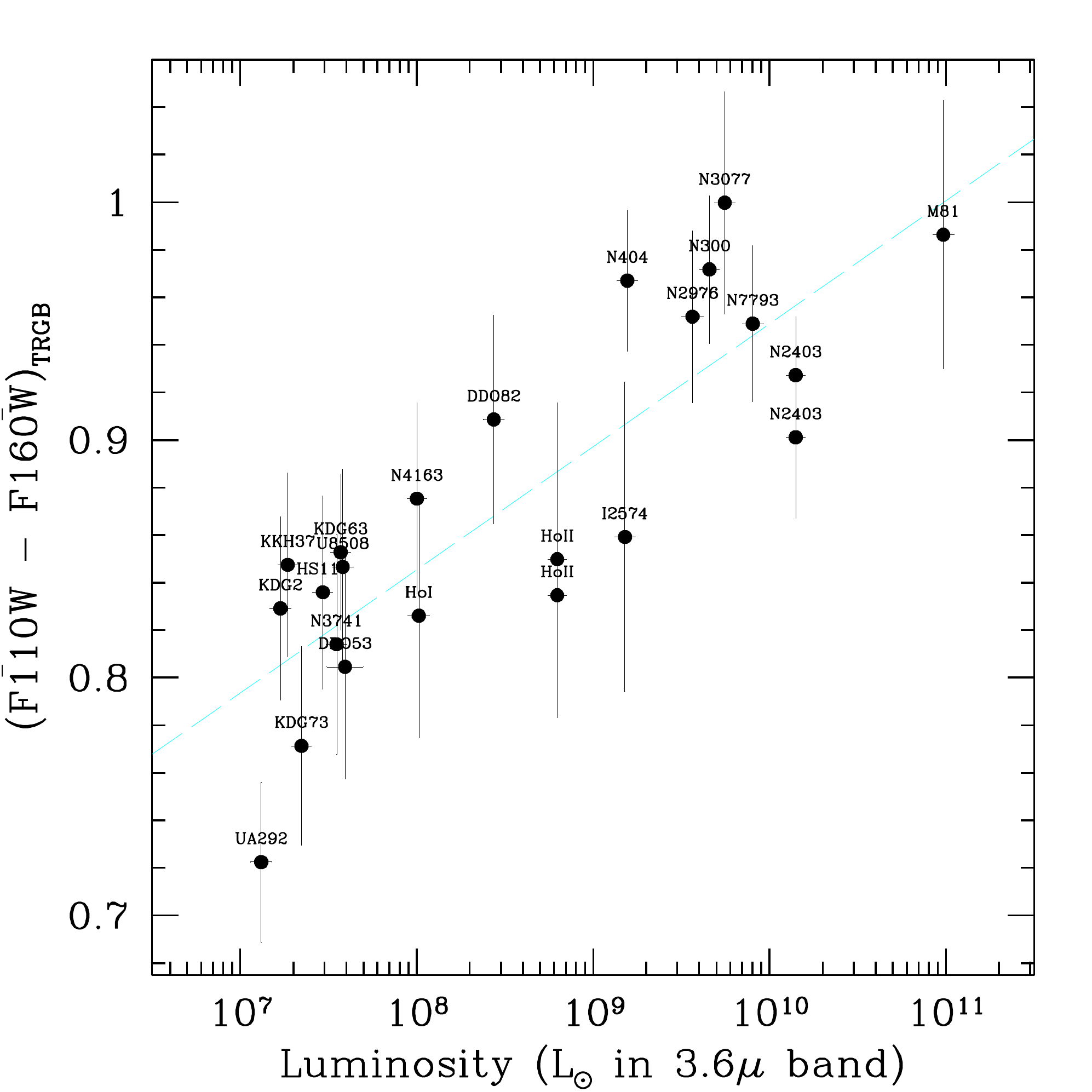}
}
\centerline{
\includegraphics[width=3in]{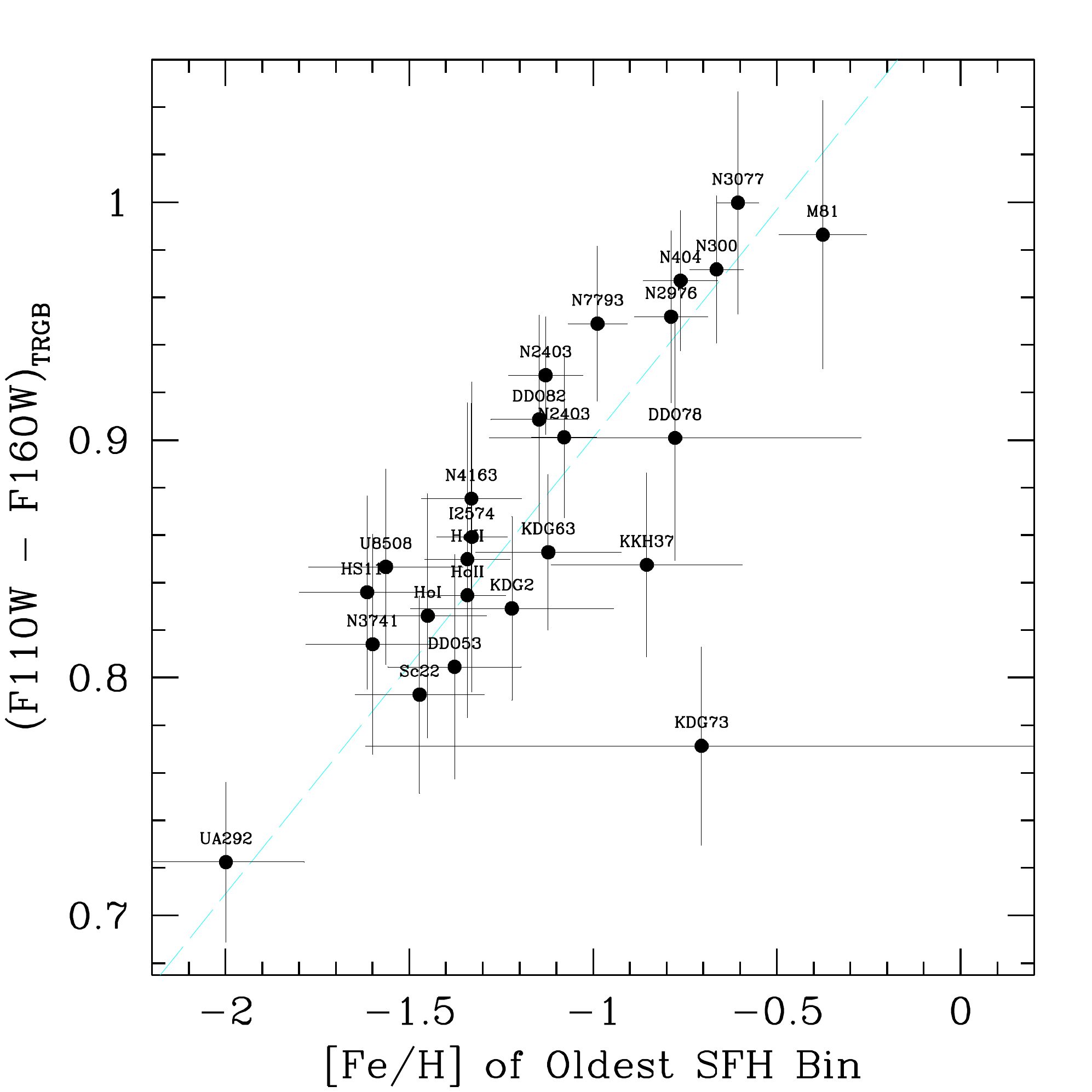}
}
\caption{\footnotesize (Upper Left) $F110W-F160W$ color of the TRGB, plotted
  against the absolute magnitude of the host galaxy in the $B_T$
  filter.  Error bars indicate the characteristic width of the RGB
  near the tip (Table~\ref{trgbtable}).  There is a statistically
  significant correlation between color and galaxy luminosity
  (Spearman rank correlation coefficient of -0.75, giving a less than
  0.001\% chance of occuring by chance), which is likely to be driven
  by the underlying mass-metallicity relationship for galaxies.  Many
  of the galaxies that fall below the mean relationship have high
  recent star formation rates, which may bias the luminosity high (due
  to low mass-to-light ratios) or the RGB color blueward (due to
  contamination from RHeB and AGB stars, or younger mean RGB ages);
  (Upper Right) $F110W-F160W$ color of the TRGB, plotted against the
  solar luminosity of the host galaxy in the Spitzer 3.6$\mu$ bandpass
  \citep{dale2009}.  Again, a strong color-luminosity trend is
  apparent (Spearman rank correlation coefficient of 0.82).  The plot
  excludes DDO78 and Sc22, which are undetected in the Spitzer
  3.6$\mu$ bandpass; (Lower Left) $F110W-F160W$ color of the TRGB as a
  function the [Fe/H] metallicity inferred from the oldest bin of the
  star formation history adopted by \citet{melbourne2011}.  There is a
  strong trend (Spearman rank correlation coefficient of 0.54, giving
  a less than 0.3\% chance of occuring by chance, decreasing to 0.03\%
  if KDG73 is excluded) driven by the fact that the color of the RGB
  is one of the strongest discriminant of metallicities in the CMD;
  Dashed lines indicate the fitting equations given in the text.
\label{magcolorfig}}
\end{figure}
\vfill
\clearpage

\begin{figure}
\centerline{
\includegraphics[width=3.25in]{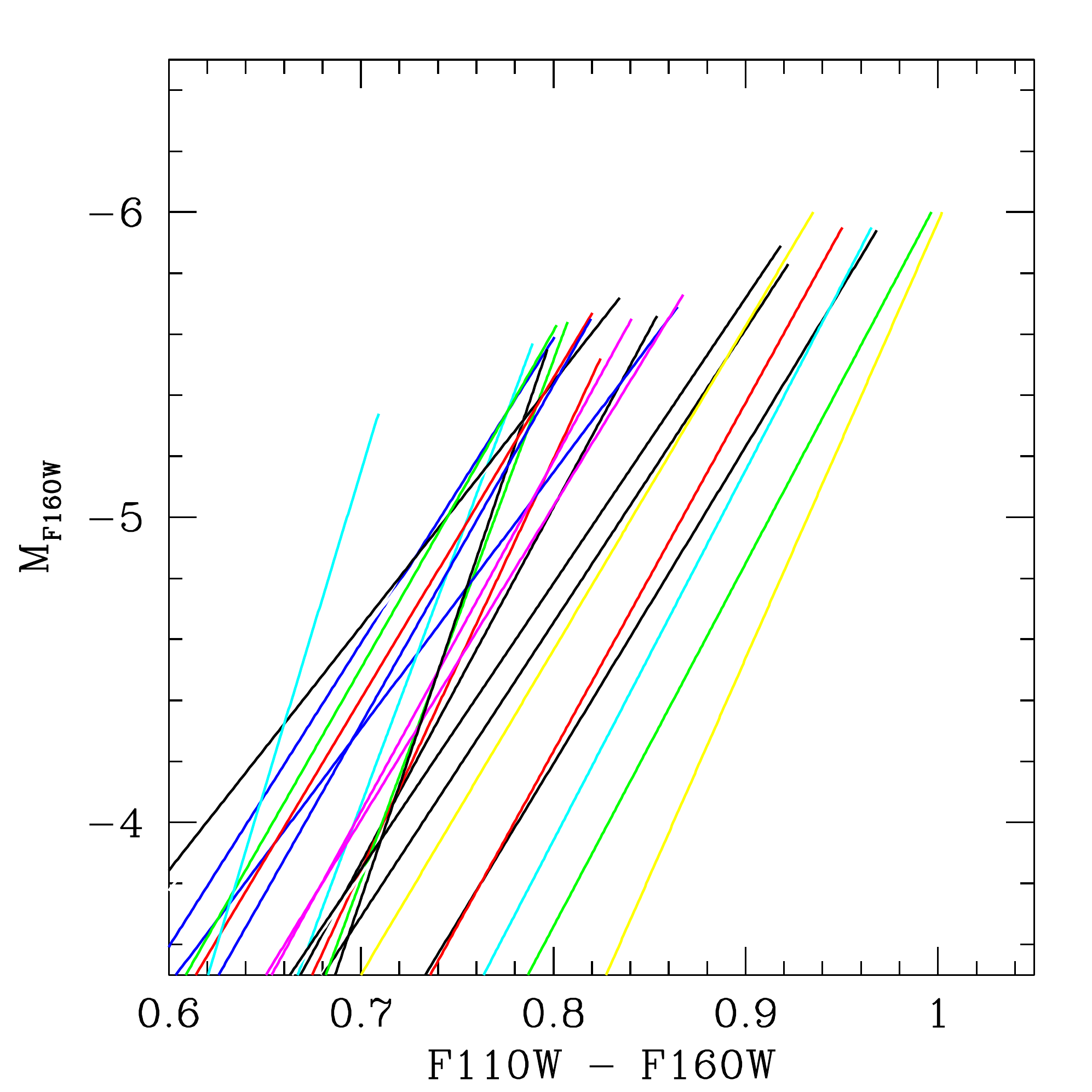}
\includegraphics[width=3.25in]{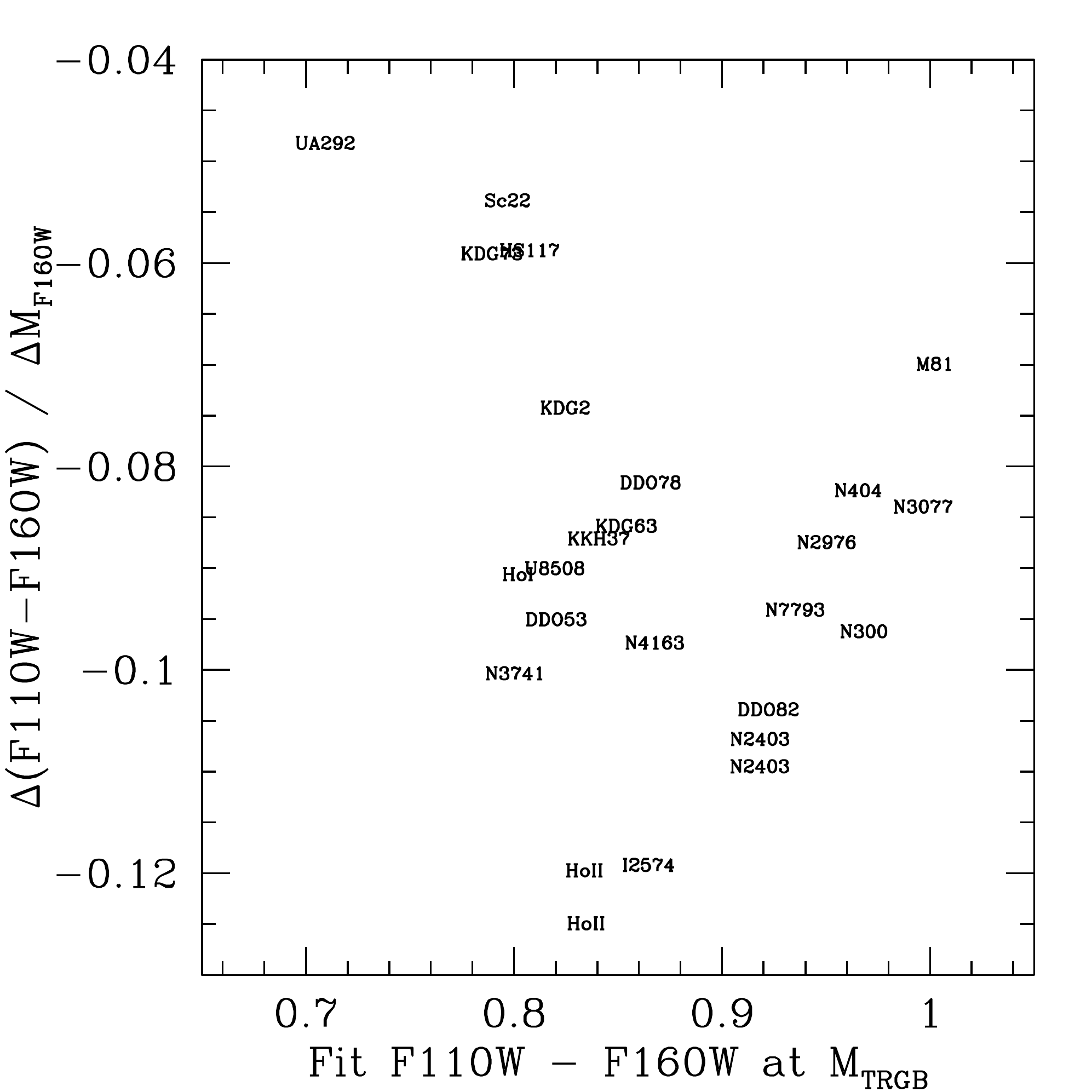}
}

\caption{
(Left) Fits to the median color of the top 2 magnitudes of the RGB,
using data from Figure~\ref{LFfig}.  Colors have been randomly
assigned to individual galaxies.  All quantities have been corrected
for extinction.  (Right) Slopes of the fits shown in the left hand
panel, as a function of the color of the fits at the TRGB.  There are
no obvious systematic trends across the whole sample, suggesting that
the presence complex stellar populations can mask any intrinsic
variation in the shape of the RGB with metallicity.  There are,
however, suggestive trends at the extremes.  First, the three bluest
RGBs make up three of the four steepest RGBs, suggesting a possible
detection of variations in RGB shape at very low metallicity.  Second,
the three fields with extremely prominent RHeB sequences (IC2574 and
two fields in HoII) have the three shallowest RGBs, suggesting that
the presence of RHeB stars are biasing the slope of the RGB by pulling the
base to bluer colors.  These same galaxies are also outliers in the luminosity-metallicity
plots in Figure~\ref{magcolorfig}.
\label{rgbfig}}
\end{figure}
\vfill
\clearpage

\end{document}